\documentclass[a4paper,11pt]{report}
\usepackage[utf8]{inputenc}
\pdfoutput=1

\usepackage{mathrsfs}
\usepackage{tikz}
\usepackage[T1]{fontenc} 
\usepackage[english]{babel}
\usepackage{geometry,amsmath,amsfonts}
\usepackage{slashed}
\usepackage{epsfig}
\usepackage{latexsym}
\usepackage{graphicx}
\usepackage{amssymb}
\usepackage{subfig}
\usepackage{color}
\usepackage{multirow}
\usepackage{bbding}
\usepackage{rotating}
\usepackage{ifthen}
\usepackage{epsfig}
\usepackage{cite}
\usepackage{collref}
\usepackage{hyperref}
\ifx\pdfoutput\undefined
\usepackage[dvips,bookmarks=false]{hyperref}
	
\else
\usepackage{hyperref}	
\fi
\usepackage{hyperref}
\hypersetup{
    colorlinks,
    citecolor=black,
    filecolor=black,
    linkcolor=black,
    urlcolor=black
}

\newcommand{\nn}{\nonumber}
\newcommand{\non}[0]{\nonumber \\}
\newcommand{\bee}[0]{\begin{eqnarray}}
\newcommand{\eee}[0]{\end{eqnarray}}
\newcommand{\comm}[2]{\left[ #1 ,#2 \right]}

\newcommand{\be}[0]{\begin{equation}}
\newcommand{\ee}[0]{\end{equation}}
\newcommand{\brackets}[3]{\left< #1 \left| #2 \right | #3 \right>}
\newcommand{\bra}[1]{\left< #1  \right|}
\newcommand{\ket}[1]{\left| #1 \right>}
\renewcommand{\Re}{\mathfrak{Re}}
\renewcommand{\vec}{\overrightarrow}
\renewcommand{\tilde}{\widetilde}
\newcommand{\de}[0]{\text{d}}
\def\bea{\begin{eqnarray}}   \def\eea{\end{eqnarray}}
\newcommand{\cval}[1]{}
\newcommand{\cmic}[1]{}
\newcommand{\asm}[1]{}
\newcommand{\val}[1]{}
\definecolor{mahtblue}{RGB}{100,100,250}
\newcommand{\Tr}[1]{\mbox{Tr}\left[ #1 \right]}
\newcommand{\hc}{\mathrm{H.c.}}
\newcommand{\eq}{Eq.}
\newcommand{\Ref}{Ref.}

\begin{document}

\thispagestyle{empty}

\enlargethispage{10cm}

\vspace*{-1cm}

\begin{figure}[h]
\vspace{-1cm}
\begin{minipage}[l]{0.24\textwidth}
\includegraphics[height=3cm]{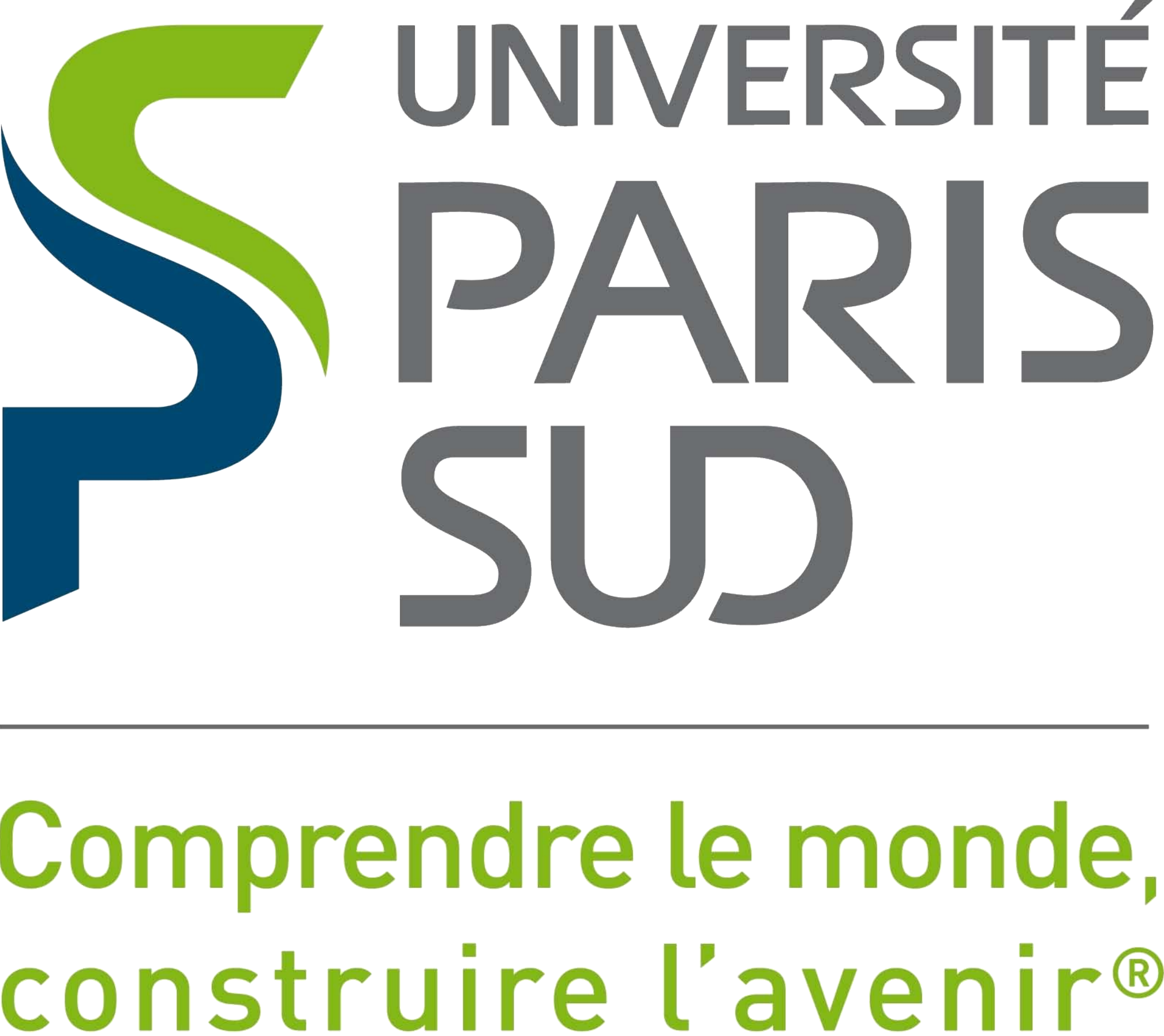}
\end{minipage}
\begin{minipage}[c]{0.24\textwidth}
\hspace{0.06\textwidth}\includegraphics[height=3cm]{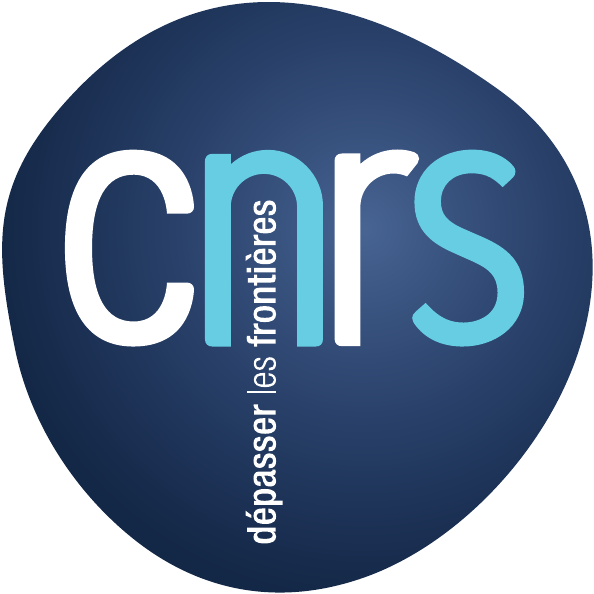}
\end{minipage}
\begin{minipage}[c]{0.24\textwidth}
\includegraphics[height=3cm]{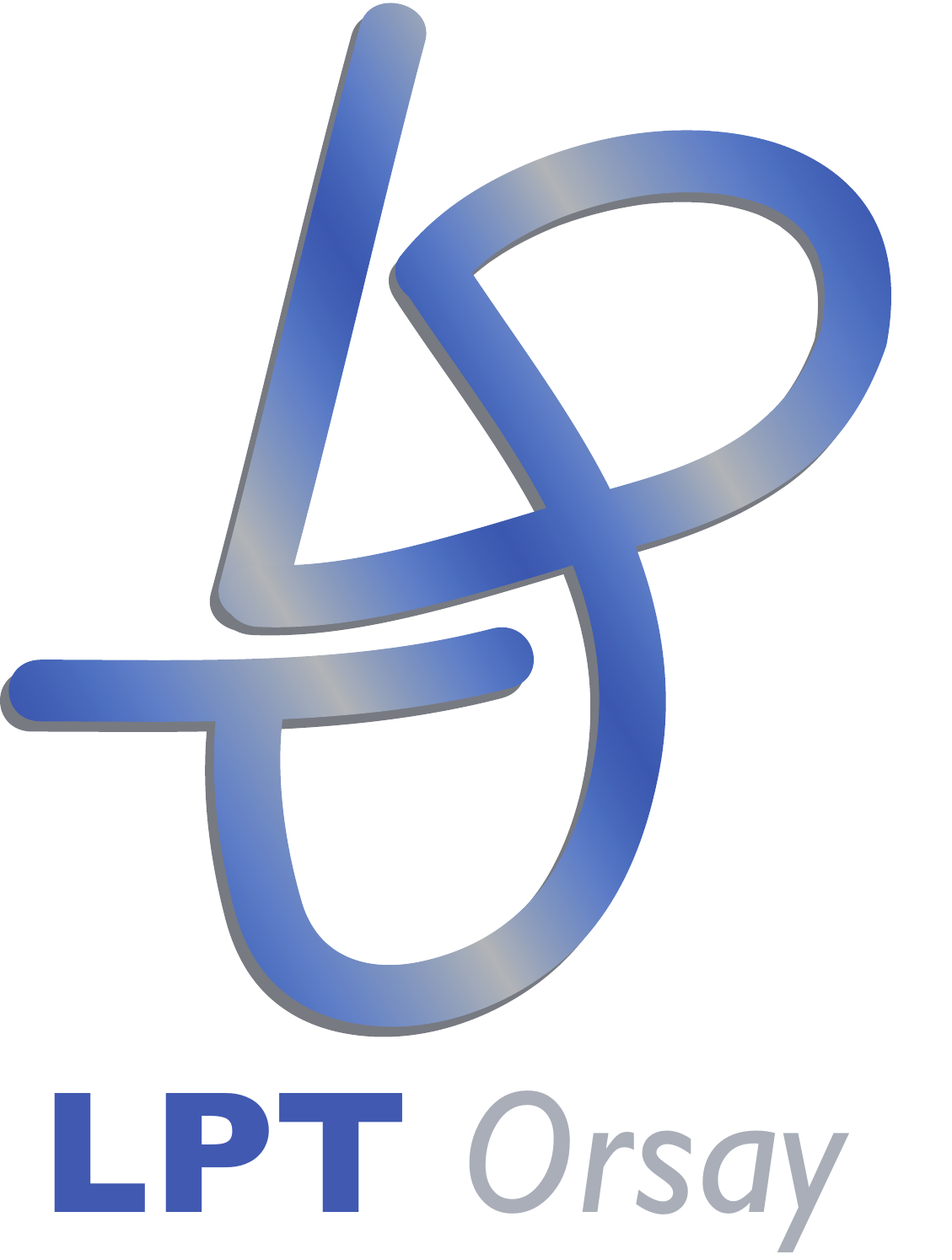}
\end{minipage}
\begin{minipage}[r]{0.24\textwidth}
\includegraphics[height=3cm]{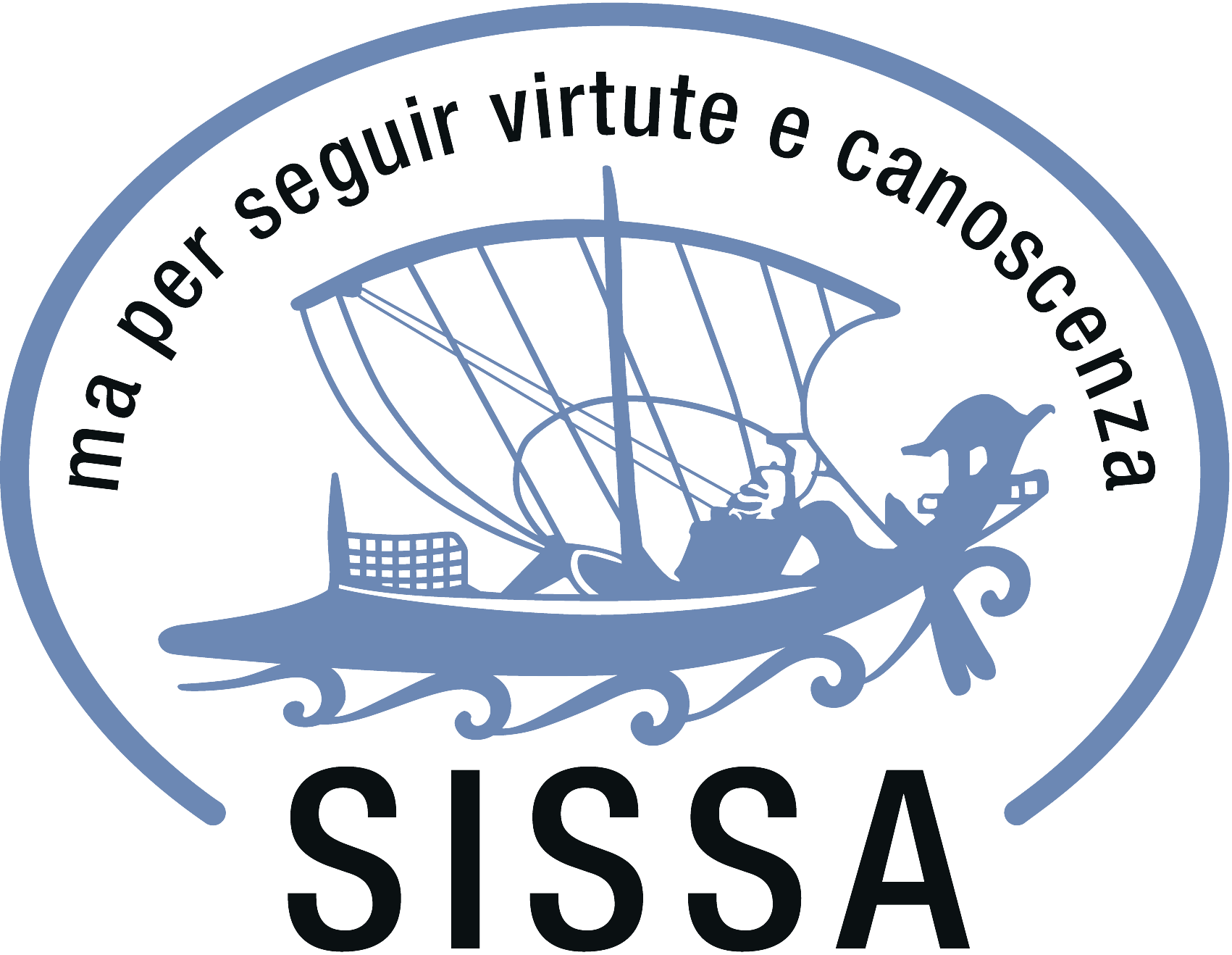}
\end{minipage}
\end{figure}

\begin{center} 
\begin{tabular}{c}
		\\
    {\Large \textsc{Universit\'e Paris-Sud}} \\
    {\large \textsc{\'Ecole Doctorale 564 ``Physique en \^Ile-de-France''}} \\   
		{\normalsize \textsc{Laboratoire Physique Th\'eorique d’Orsay (UMR 8627)}} \\
		\\
    \\
  {\Large \textsc{Scuola Internazionale Superiore di Studi Avanzati}} \\
    {\large \textsc{Area of Physics}} \\   
		{\normalsize \textsc{Astroparticle Physics sector}} \\
		\\
		\\
    \huge \textsc{Ph.D. thesis}\\
    \\
    \large{Defended on September 25th, 2015 by} \\
    \\
    \\
    \Huge{\textbf{{Michele Lucente}}}\\
    \\
		\\
		\\
    \huge\bf{Implication of Sterile Fermions in }\\
		\huge\bf{Particle Physics and Cosmology}\\
    \\
    \\
		\\
		\\
\end{tabular}

\begin{tabular}{llll}
	& \footnotesize\bf{Supervisor}: &  Asm\^aa Abada & \footnotesize{Professor (LPT)} \\
	& \footnotesize\bf{Supervisor}: &  Guido Martinelli & \footnotesize{Professor (SISSA)} \\
	& & &\\
	& \footnotesize\bf\underline{Composition of the jury:}& &\\
	& \footnotesize{President}: & Marie-H\'el\`ene Schune & \footnotesize{Directrice de recherche (LAL)} \\
	& \footnotesize{Referees}: &  Silvia Pascoli & \footnotesize{Professor (IPPP)} \\
  &	&  Thomas Schwetz-Mangold & \footnotesize{Professor (KIT)} \\
  & \footnotesize{Examiners}: &  Marco Cirelli & \footnotesize{Researcher (CNRS)} \\
  &	&  Serguey Petcov & \footnotesize{Professor (SISSA)} \\
\end{tabular}

\vspace{0.8cm}

\begin{figure}[h]
\begin{minipage}[l]{0.5\textwidth}
\includegraphics[height=2.5cm]{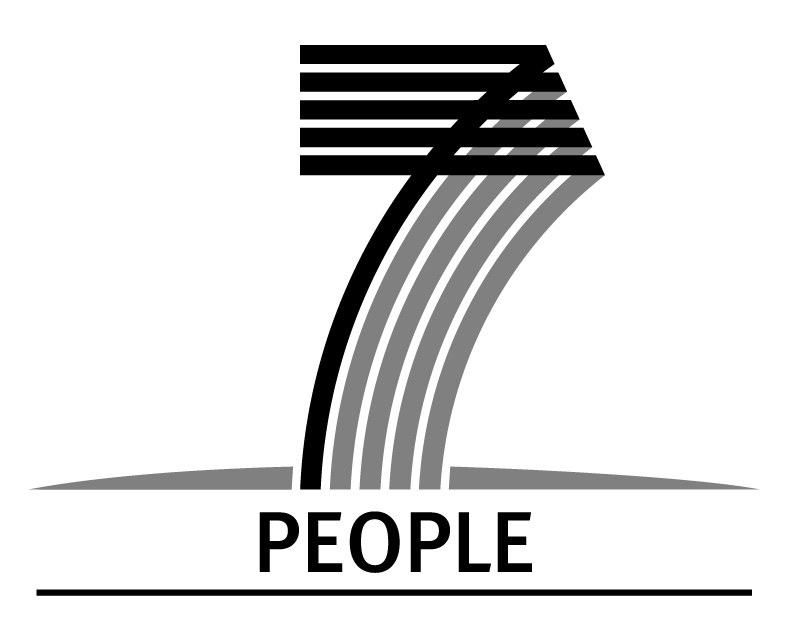}
\end{minipage}
\begin{minipage}[r]{0.5\textwidth}
\includegraphics[height=1.8cm]{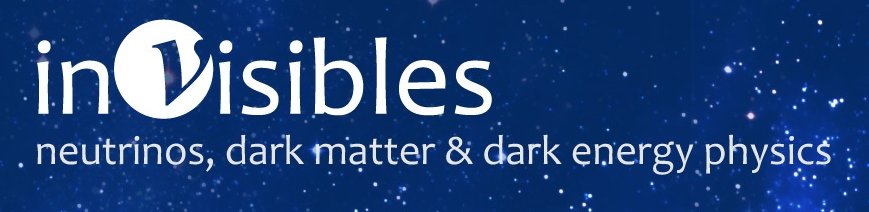}
\end{minipage}
\end{figure}

\end{center}

\thispagestyle{empty}
\phantom{dada}
\newpage

\pagenumbering{roman}

\chapter*{Foreword}

The Ph.D. thesis work summarised in this manuscript was dedicated to studying several aspects of the phenomenology of Standard Model  (SM) extensions by sterile fermions, in particular their impact for particle and astro-particle physics.
An important part of the work is dedicated to a class of SM extensions which allow to explain the smallness of the observed neutrino masses (as well as their mixings) by linking them to the breaking of total lepton number; in the framework of the so-called Inverse seesaw mechanism (ISS), the scale of New Physics can be quite low, and this opens  the door to a rich phenomenology, with an impact on numerous observables, which can be studied in low-energy/high-intensity facilities, colliders and astro-particle experiments. 
The work described in the thesis addresses the r\^ole of these sterile states in providing a satisfactory explanation to three open observational problems of the SM: the generation of neutrino masses and mixings, a viable dark matter candidate, and the dynamical generation of the baryon asymmetry of the Universe.

Motivated by the rich phenomenology of this class of SM extensions, we identified in  Nucl.\ Phys.\ B {\bf 885} (2014) 651 the minimal ISS realisation accounting for the observed neutrino data while at the same time complying with all available experimental and observational constraints.
This study was based on a perturbative approach to the diagonalization of the neutrino mass matrix, which allowed to identify the number of states associated with the different mass scales. A further numerical exploration of the parameter space led to the phenomenological study of the two most minimal realisations. Our study revealed that, depending on the number of additional sterile fermion fields, the ISS can accommodate both a 3-flavour mixing scheme and a 3+more mixing scheme. Interestingly, in the latter scheme, the (light) sterile states can either provide a solution to the neutrino oscillation anomalies or be viable dark matter candidates.

The potential r\^ole of these sterile states as dark matter (DM) candidates led us to carry a dedicated study of the viability of the sterile fermion dark matter hypothesis in a minimal ISS realisation (in which the SM is extended by two right-handed neutrinos and three additional sterile fermion fields), 
JCAP {\bf 1410} (2014) 001.  
From the ISS parameter space complying with all available observational constraints we derived the maximal value of the DM abundance produced via active-sterile neutrino oscillations ($\sim 43\%$ of the observed relic density). Taking into account the effects of entropy injection from the decay of heavier 
pseudo-Dirac pairs, which are present in the spectrum of these minimal ISS realisations, allowed to marginally increase the contribution to the DM abundance; the correct relic abundance can nonetheless be obtained via freeze-in decay processes of the heavy pseudo-Dirac pairs (although this production mechanism is only effective in a limited mass range).

The degeneracy in the sterile neutrino mass spectrum - which is characteristic of low-scale seesaw models with approximate lepton number conservation 
 - can play a relevant r\^ole in cosmology, since it allows to explain the observed baryon asymmetry of the Universe via leptogenesis. In particular, in JCAP {\bf 1511} (2015) no.11,  041 
we focused on the connection between lepton number as an approximate symmetry and low-scale (around the GeV) leptogenesis scenarios. We identified different lepton number violating patterns and their effect on leptogenesis, having also succeeded in isolating the most minimal viable model, which was analytically and numerically studied. 

Laboratory experiments allow to further characterise the sterile states, either by constraining their contributions to a number of SM observables, or by looking for new processes beyond the SM. There are already several experiments actively searching for these states, and several future facilities include searches for sterile fermions in their physics programme. 
In this perspective we performed in JHEP {\bf 1510} (2015) 130 a detailed study of the importance of loop corrections when deriving bounds on active-sterile neutrino mixing from global fits on electroweak precision data, in the context of general Seesaw mechanisms with extra heavy right-handed neutrinos. 

Finally we considered in Phys.\ Rev.\ D {\bf 91} (2015) 11,  113013 new processes (absent in the SM) which can be mediated by sterile states, focusing on rare lepton flavour violating decays of vector bosons (including quarkonia and the $Z$ gauge boson). We computed the relevant Wilson coefficients, and explored the parameter space of a minimal realisation of the ISS, thus determining the maximal allowed branching fractions of the different decay channels.

\tableofcontents

\newpage 

\pagenumbering{arabic}
\chapter{Introduction}
The origin of neutrino masses and the nature of dark matter are two of the most pressing open questions of particle and astroparticle physics. Sterile fermions are an intriguing and popular solution to both these issues.

Sterile fermions generically denote gauge singlet fermionic fields, only capable of interacting with gauge bosons via mixing terms. They  are absent in the Standard Model of particle physics. The general definition of sterile fermions encompasses more particular expressions, such as right-handed or sterile-neutrinos. The term \emph{sterile fermion} will be used in this thesis in its more general meaning reported above; \emph{right-handed neutrino} will be used to refer to a field analogous to the standard left-handed active neutrinos, but with opposite chirality, resulting in a singlet under the standard model gauge group. Finally \emph{sterile} or \emph{heavy neutrino} will be used to refer to a fermionic mass eigenstate, resulting from the diagonalization of a mass matrix that contains the active neutrino mass matrix as a sub-block.

Despite of being gauge singlets, the simple assumption of the existence of right-handed neutrinos -and, more generally, of sterile fermions- can provide a minimal and elegant solution to three observational problems of the SM, namely the origin of neutrino masses and mixing, the nature of dark matter and the origin of the baryon asymmetry of the Universe. 

Neutrino oscillation experiments have established a clear  evidence for two oscillation frequencies ($\Delta m_{ij}^2$) - implying that at least two neutrino states are massive - as well as the basic structure of a 3-flavour leptonic mixing matrix. In contrast with the huge experimental achievements in determining neutrino oscillation parameters, many questions remain to be answered concerning neutrino properties, as  for instance  the neutrino nature (Majorana or Dirac),  the  absolute neutrino mass scale and  the hierarchy of the neutrino mass spectrum, which are not yet determined. Finally, and most importantly, the neutrino mass generation mechanism at work remains to be unveiled as well as the new physics scales that it calls upon.
In order to account for neutrino masses and mixings, many extensions
of the Standard Model (SM) call upon the introduction of sterile fermions. Being gauge singlets, these particles can be stable on cosmological timescales if their mixing with active neutrinos is sufficiently small, and if they are massive they can contribute to the dark matter component of the Universe. They can moreover be coupled to the Standard Model fields via Yukawa terms and can play an important r\^ole in the early Universe, notably in the baryogenesis via leptogenesis mechanism. 

An important feature of sterile fermions is the fact that, being gauge singlets, they can have a Majorana mass term, which is absent in the Standard Model Lagrangian. A Majorana mass term violates all the internal charges of a fermion by two units, and is thus related to fields that are intrinsically neutral, or to fields that are charged under an (unknown) gauge group, broken by an (unknown) Higgs sector.

Sterile fermions are actively searched for in laboratory experiments, but until now only upper bounds on the active-sterile mixing have been established. In particular, their mass scale is unbound from below. It can range from some eV up to the Planck scale. 
For instance, in the simplest implementation of the type-I Seesaw mechanism, in order to account for massive neutrinos with natural neutrino Yukawa couplings, the typical scale of the extra particles is in general very high, 
potentially close to the gauge coupling unification (GUT) scale, thus implying that direct experimental tests of
the Seesaw hypothesis might be  impossible.  
In contrast, low-scale Seesaw mechanisms in which sterile fermions are added to the SM particle content with masses
around the TeV scale or even lower, are very attractive from a
phenomenological point of view since the new states can be
produced in collider and/or low-energy experiments, and their
contribution to physical processes can be sizeable.

In this work we study the implications for the existence of sterile fermions in particle physics and cosmology. We focus on low-scale new physics mechanisms, that can be tested in current and future experiments, and show how the addition of sterile fermions can provide a solution for each of the observational problems of the Standard Model (origin of neutrino masses, dark matter and baryon asymmetry of the Universe). We also address the impact of the new states in laboratory observables, such as lepton flavour violating decays of vector bosons, and their impact on global fits of electroweak precision data.

\chapter{Neutrinos in the Standard Model}
It has been over a century since A.H. Becquerel accidentally discovered radioactivity during a cloudy Parisian day~\cite{pais1986inward}. Since then a huge progress has been made in the understanding of the subnuclear particles and their interactions, a knowledge which is currently incorporated in a theoretical formulation that is the Standard Model (SM) of particle physics~\cite{Glashow:1961tr,Weinberg:1967tq,Salam:1968rm}. Despite being one of the most accurate theories conceived so far,\footnote{Precise measurements for the inverse fine-structure constant $\alpha^{-1}$ inferred by different experiments currently agree within one part in $10^{10}$ in the SM framework~\cite{Mohr:2012tt}.} the SM appears far from being complete. There are fundamental theoretical caveats in the SM, like the flavour puzzle, the hierarchy problem, the strong-CP problem, the gauge coupling unification and the number of families. Furthermore it does not account for gravity, notwithstanding of several arguments suggesting that in a coherent complete theory all fundamental interactions should be quantised~\cite{DeWitt:1962bu, eppley1977necessity,Page:1981aj}. From a phenomenological point of view it does not provide a viable candidate for the Dark Matter (DM) component of the Universe~\cite{Planck:2015xua}, neither a viable mechanism to explain the matter-antimatter asymmetry in the observed Universe~\cite{Farrar:1993hn,Gavela:1993ts}.

In addition to the aforementioned arguments there is an  observation that cannot be accommodated within the SM: the fact that neutrinos are massive and mix. In the first part of this chapter we review how neutrinos are described in the SM and why they are  massless in such a minimal framework. We later discuss the phenomenological consequences of massive neutrinos and compare the massless and massive hypothesis with experimental results, motivating the need to explore extensions of the SM.

\section{The Standard Model and its constraints}\label{SM_constraints}

The Standard Model of particle physics is a relativistic quantum field theory based on a local gauge invariance principle. It is a minimal model, meaning that the matter field content and the gauge symmetry group were postulated in the minimal pattern to agree with observation.

The SM Lagrangian is invariant under Lorentz and gauge transformations, and complies with the renormalizability requirement. In the following sections we review these constraints, pointing out their relation with the (lack of) neutrino masses in the SM.

\subsection{Symmetries}

Given a physical system, a symmetry is defined as the property of  that system of being invariant under some class of transformation acting on its degrees of freedom. The SM has two classes of symmetry: invariance under a global redefinition of the reference frame and the invariance under a local redefinition of the fields based on a precise group of transformations.

\subsubsection{Relativistic invariance}\label{rel:inv}

According to the principle of special relativity, all the laws of physics must retain the same form regardless of the particular inertial reference frame chosen to describe them~\cite{Einstein:1905ve}. This implies that the Lagrangian of a system must be invariant under the transformations generated by the Lorentz group $O(1,3)$, that is the subgroup of the real $4\times 4$ matrices $GL(4,R)$ which  preserve distances in the Minkowski metric
\bee
\Lambda \in O(1,3) \Rightarrow \Lambda^T \eta\ \Lambda = \eta,
\eee
where $\eta$ is the flat metric tensor $\eta=\textrm{diag}(+,-,-,-)$. The transformations of reference frame realisable in Nature are actually the ones that can be deformed continuously into the identity of the Lorentz group, that is the subgroup $SO(1,3)^\uparrow$ of matrices with $\det =1$  and which do not invert the temporal ordering for causally connected events. The Lorentz group must be extended to include also translations in the space-time coordinates, giving rise to the Poincar\'e group.

The Lie algebra of the Lorentz group is defined by:
\begin{eqnarray}\label{commboostrot}
 \comm{J_i}{J_j}&=&i \epsilon_{ijk} J^k,\non 
 \comm{J_i}{K_j} &=& i \epsilon_{ijk} K^k,\non 
\comm{K_i}{K_j}&=&- i \epsilon_{ijk} J^k,
\end{eqnarray} 
where $J_i$ and $K_i$ are the generators of infinitesimal spatial rotations and boosts, respectively.

By defining the linear combinations
\bee
J_i^\pm = \frac{1}{2}\left(J_i\pm i K_i\right),
\eee
which obey the algebra
\bee\label{reduced-lorentz}
\comm{J^\pm_i}{J^\pm_j} = i \epsilon_{ijk} J_k^\pm, \hspace{1cm} \comm{J^\pm_i}{J^\mp_j} =0,
\eee
it is possible to show that the Lorentz group is isomorphic to the direct product of two $SU(2)$ representations
\bee\label{Lorentz:red}
SO(1,3) \cong SU(2)\times SU(2)^*.
\eee
It is thus possible to use a doubled version of the familiar $SU(2)$ labelling in order to classify the irreducible representations of the Lorentz group,
\bee
\left(n,m\right)\hspace{2cm}\text{with } n,m=0,\frac{1}{2},1,\frac{3}{2},2,\dots 
\eee
This means that, apart from the trivial scalar representation $(0,0)$, there exist two distinct fundamental and irreducible representations of the Lorentz group, namely $(1/2,0)$ and $(0,1/2)$, conventionally referred as left and right.

This statement can be made less abstract by choosing an explicit realisation of the algebra~(\ref{commboostrot}) acting on bidimensional spinors,
\begin{eqnarray}\label{Lorentz:fundamental}
 J^i_{R,L} = \frac{\sigma^i}{2}, & \; & K^i_{R,L}=\pm\ i\ \frac{\sigma^i}{2},
\end{eqnarray}
where $\sigma^i$ are the usual Pauli matrices.
These generators define the Lorentz transformations:
\begin{eqnarray}\label{leftrighttransf}
 \Lambda_R= e^{\frac{i}{2} \sigma^j (\omega^j +i \eta^j)}, &\; & \Lambda_L=e^{\frac{i}{2} \sigma^j (\omega^j -i \eta^j)},
\end{eqnarray}
where $\omega^j$ and $\eta^j$ are the real parameters defining rotations and boosts, respectively.

The matrices $\Lambda_{R,L}$ satisfy the identities:
\begin{eqnarray}\label{siglambdasig}
\Lambda_{L,R}^{-1} &=& \Lambda_{R,L}^\dagger,\label{siglambdasig:I}\\
\sigma^2 \Lambda_{L,R} \sigma^2 &=& {\Lambda^{^*}_{R,L}},\label{siglambdasig:II}
\end{eqnarray}
the second one following from $\sigma^2 \sigma^i \sigma^2 = -{\sigma^i}^*$; from the hermitian conjugate of~(\ref{siglambdasig:I}, \ref{siglambdasig:II}) it follows that
\be\label{lambdasigmalambda}
\Lambda^T_{L,R}\ \sigma^2\ \Lambda_{L,R} = \sigma^2.
\ee
These relations are useful to identify the possible Lorentz invariant forms involving bidimensional spinor fields: given a left-handed spinor $\psi_L$ and a right-handed one $\chi_R$, eq.~(\ref{siglambdasig:I}) implies that the combinations
\bee\label{Dirac:inv}
\psi_L^\dagger \chi_R, \hspace{1cm} \chi_R^\dagger \psi_L,
\eee
are Lorentz scalars. In addition, eq.~(\ref{lambdasigmalambda}) implies that the bilinears
\bee\label{Maj:inv}
\psi_L^T \sigma^2 \psi_L, \hspace{1cm} \chi_R^T \sigma^2 \chi_R,
\eee
are Lorentz invariant. 

Equation~(\ref{siglambdasig:II}) implies that the left (right) representation is equivalent to the complex conjugate of the right (left) one, $\Lambda_{R,L}\cong \Lambda_{L,R}^*$. That is, given a bidimensional spinor $\psi_{L,R}$, it is possible to construct another spinor $\sigma^2 \psi_{L,R}^*$ which transforms in the opposite representation:
\be\label{left-right:eq}
\left(\sigma^2 {\psi^*_{L,R}}\right)' =\sigma^2 \Lambda_{L,R}^{^*} \psi_{L,R}^* =\sigma^2 \Lambda_{L,R}^{^*} \sigma^2 \sigma^2 \psi_{L,R}^*= \Lambda_{R,L}\left( \sigma^2 \psi_{L,R}^*\right).
\ee
The expressions in eq.~(\ref{Maj:inv}) can be interpreted as a particular realisation of the ones in eq.~(\ref{Dirac:inv}) in the case in which the left and right-handed spinors are not independent degrees of freedom, i.e. $\psi_L=\sigma^2 \chi_R^*$ or $\chi_R=\sigma^2 \psi_L^*$. 

How are these bidimensional spinors related to the familiar four-dimensional ones, i.e. to the general solutions of the Dirac equation? The reducibility of the Lorentz group, eq.~(\ref{Lorentz:red}), follows from generic group theory arguments, eqs.~(\ref{commboostrot}-\ref{reduced-lorentz}). The physical content of eq.~(\ref{Lorentz:red}) may be obtained by choosing a useful representation of the Dirac algebra, the so called chiral (or Weyl) representation. Defining
\bee
\sigma^\mu=\left(\mathbf{1},\sigma^i\right),\hspace{1cm} \bar{\sigma}^\mu=\left(\mathbf{1},-\sigma^i\right),
\eee
the Dirac matrices in this representation are given by
\bee
\gamma^\mu= \left(\begin{array}{cc}
0&\sigma^\mu\\
\bar{\sigma}^\mu & 0
\end{array}\right).
\eee
Recalling the four-vector notation for the Lie algebra of the Poincar\'e group~\cite{Peskin:1995ev}:
\begin{eqnarray}\label{poincarealgebra}
\comm{P_\mu}{P_\nu}&=&0,\nonumber \\
\comm{M_{\mu \nu}}{P_\sigma}&=&i\eta_{\nu \sigma} P_\mu -i\eta_{\mu \sigma} P_\nu,\nonumber \\
\comm{M_{\mu \nu}}{M_{\rho \sigma}}&=&i\eta_{\nu \rho}M_{\mu \sigma} +i\eta_{\mu \sigma}M_{\nu \rho}-i\eta_{\mu \rho}M_{\nu \sigma}-i\eta_{\nu \sigma}M_{\mu \rho},
\end{eqnarray}
which is connected to the algebra~(\ref{commboostrot}) by:
\begin{eqnarray}
 J^i&=&\frac{1}{2} \epsilon^{i j k} M^{j k},\nonumber\\
K^i&=&M^{0i} \label{boostrotgen},
\end{eqnarray}
and the representation of the tensors $M_{\mu\nu}$ in terms of Dirac matrices acting on four-dimensional spinors
\be
S^{\mu \nu} =  \frac{i}{4} \comm{\gamma^\mu}{\gamma^\nu},
\ee
the generators for infinitesimal rotations and boosts acting on four-dimensional spinors are given by
\bee\label{leftright}
j^i =&\frac{1}{2} \epsilon^{ijk} S^{jk}&=\frac{1}{2} \left( \begin{array}{cc} \sigma^i & 0\\ 0& \sigma^i\end{array} \right), \\
k^i=&S^{0i}&= \frac{i}{2} \left( \begin{array}{cc} -\sigma^i & 0 \\ 0 &  \sigma^i \end{array} \right).
\eee
These relations show explicitly that the upper and lower half of the four-dimensional Dirac spinors transform as invariant subspaces under the Poincar\'e group, with the generators of the respective Lie algebras given in the representation~(\ref{Lorentz:fundamental}). Moreover in this basis the matrix $\gamma^5$ is diagonal
\be
\gamma^5 =i\gamma^0 \gamma^1 \gamma^2 \gamma^3 = \left( \begin{array}{cc} -\mathbf{1} & 0\\ 0  & \mathbf{1} \end{array} \right),
\ee
and thus the projectors $P_R = \left( \frac{\mathbf{1} + \gamma^5}{2} \right)$ and $P_L = \left( \frac{\mathbf{1} - \gamma^5}{2} \right)$ allow to decompose a generic four-component spinor in the explicit form:
\be\label{weyldecomposition}
\psi(x) = \left( \begin{array}{c} \psi_L \\ \psi_R  \end{array} \right) =\underbrace{\left( \begin{array}{c} \psi_L \\ 0 \end{array} \right)}_{ P_L \psi} +\underbrace{\left( \begin{array}{c} 0 \\ \psi_R  \end{array} \right)}_{ P_R  \psi}.
\ee
This shows explicitly that the familiar four-component spinors belong to the representation obtained as direct sum of the two fundamental ones
\bee
\left(\frac{1}{2},0\right) \oplus \left(0,\frac{1}{2}\right).
\eee
The physics is  independent from the chosen representation of the Dirac algebra and the above conclusions are valid without loss of generality, although in a representation different from the Weyl one the reducibility would not be manifest.

In the chiral representation it is also explicit how the invariance of bilinear forms of Dirac spinors is guaranteed in terms of left- and right-handed components. For example the familiar form
\be\label{Dirac-mass}
\bar{\chi} \psi = \left(\begin{array}{cc} \chi^\dagger_L & \chi^\dagger_R  \end{array} \right) 
\left( \begin{array}{cc} 0& \mathbf{1} \\  \mathbf{1} & 0 \end{array}\right) \left( \begin{array}{c} \psi_L \\ \psi_R \end{array} \right) = \chi_R^\dagger \psi_L + \chi_L^\dagger \psi_R,
\ee
only involves the invariants~(\ref{Dirac:inv}). It is natural to ask if a bilinear invariant form containing the invariants~(\ref{Maj:inv}) can be expressed in terms of four-dimensional spinors. The answer is simple and involves eq.~(\ref{left-right:eq}): let us define
\bee\label{2dim-spin}
\psi_{l}=\left( \begin{array}{c} \psi_{L}\\ \sigma^2 \psi_{L}^*  \end{array} \right), \hspace{1cm} \psi_{r}=\left( \begin{array}{c} \sigma^2 \psi_{R}^*\\ \psi_{R}  \end{array} \right),
\eee
which are spinors possessing the correct transformation rules under the Poincar\'e group, but whose left- and right-handed components are not independent degrees of freedom. Taking the analogous of eq.~(\ref{Dirac-mass})  we obtain
\bee\label{Maj-mass}
\bar{\chi}_{l,r} \psi_{l,r} = \left(\begin{array}{cc} \chi^\dagger_{L,R} & \chi^T_{L,R} \sigma^2  \end{array} \right) 
\left( \begin{array}{cc} 0& \mathbf{1} \\  \mathbf{1} & 0 \end{array}\right) \left( \begin{array}{c} \psi_{L,R} \\ \sigma^2\psi_{L,R}^* \end{array} \right) = \chi_{L,R}^\dagger \sigma^2 \psi_{L,R}^* + \chi_{L,R}^T \sigma^2 \psi_{L,R}.\non 
\eee

Although both~(\ref{Dirac-mass}) and~(\ref{Maj-mass}) are Lorentz invariant forms, there is an important difference between them, relevant in the special case $\psi=\chi$. Before addressing it let us introduce, for the sake of clarity and synthesis, the particle-antiparticle conjugation matrix $C$,  i.e. the matrix that gives the correct spinor $\psi^c$ when the r\^oles of particles and antiparticles are interchanged~\cite{Akhmedov:1999uz}:
\begin{eqnarray}\label{maj_relations}
 \psi^c = C {\overline{\psi}}^T,\hspace{1cm}  C= i \gamma^2 \gamma^0,
\end{eqnarray}
with the matrix $C$ satisfying
\begin{eqnarray}\label{cprop}
 C^\dagger =C^T = C^{-1} = -C,&\; & C \gamma_\mu C^{-1} = - \gamma_\mu^T.
\end{eqnarray}
From these relations it is possible to derive the following properties
\begin{eqnarray}\label{cspinors}
\left(\psi^c\right)^c &=&\psi,\non 
\overline{\psi^c} &=& \psi^T C,\nonumber \\
\overline{\psi_1} \psi_2^c &=& \overline{\psi_2^c} \psi_1,\nonumber \\
 \overline{\psi_1} A \psi_2 &=& \overline{\psi_2^c} \left(C A^T C^{-1} \right) \psi_1^c,
\end{eqnarray}
where $\psi,\psi_1,\psi_2$ are four-component spinors and $A$ is a generic $4\times4$ matrix. In the chiral basis the charge-conjugated of a spinor $\psi$ has the explicit form
\bee
\psi^c = C \left( \begin{array}{c} \psi_L \\ \psi_R  \end{array} \right) = \left( \begin{array}{c} i \sigma^2\psi_R^* \\ -i \sigma^2 \psi_L^*  \end{array} \right),
\eee
which indeed possesses the correct transformation properties under the Lorentz group. 

We can now rearrange, with the help of the matrix $C$, the previous information in a more compact form and set up the nomenclature for later use. A four-dimensional spinor of the form~(\ref{weyldecomposition}) with 4 independent degrees of freedom is called a \emph{Dirac spinor}. The bi-dimensional spinors there contained, $\psi_{L,R}$, are called \emph{Weyl spinors}; they can be thought as the fundamental building blocks with which a fermionic theory is composed. A four dimensional spinor with only two independent components, such as the ones in eq.~(\ref{2dim-spin}), is called \emph{Majorana spinor}. It can be also defined as a spinor that respects the condition
\bee\label{phase_mass}
\psi^c = \eta\ \psi,
\eee
where $\eta$ is a global phase factor.

The bilinear~(\ref{Dirac-mass}) can appear in the Lagrangian  associated with a dimensionful constant, playing the r\^ole of a mass term. In particular in the case $\chi=\psi$ the mass term
\bee\label{mDirac}
m\ \bar{\psi} \psi
\eee
is invariant under the redefinition
\bee\label{phases-red}
\psi \rightarrow e^{i\alpha} \psi.
\eee
This implies that if~(\ref{phases-red}) is a symmetry of the massless Lagrangian the addition of~(\ref{mDirac}) does not modify this property, and the global charges associated with this symmetry are conserved. A mass term of the form~(\ref{mDirac}) is called a \emph{Dirac mass term}. On the other hand it is possible to write a mass term that calls upon the structure~(\ref{Maj-mass}), which with the help of the second of the eqs.~(\ref{cspinors}) can be written as
\bee\label{mMaj}
M\ \bar{\psi^c}\ \psi = M\ \psi^T C\ \psi.
\eee  
Contrary to~(\ref{mDirac}) this combination is not invariant under the redefinition~(\ref{phases-red}); in other words whatever is the structure of the rest of the Lagrangian, the term~(\ref{mMaj}) violates the conservation of any global charge associated with $\psi$. A mass term of the form~(\ref{mMaj}) is called a \emph{Majorana mass term}~\cite{Majorana:1937vz}.

\subsubsection{Gauge invariance}

The SM is a gauge theory, meaning that the interactions among fields are not a primary assumption, but they are a natural consequence of certain symmetry requirements. A gauge transformation is a redefinition of the fields of the theory that depends on the space-time coordinates of the field itself; for instance the operation~(\ref{phases-red}) can be seen as a special limit of the transformation
\bee
\psi(x) \rightarrow e^{i\alpha(x)} \psi(x),
\eee
in the case $\partial \alpha/\partial x = 0$. A gauge transformation that does not depend on space-time coordinates is called \emph{global} transformation, in contrast to the \emph{local} character of the generic case.

Having a Lagrangian being invariant under a global transformation of its fields does not in general imply that the same Lagrangian will be invariant once the symmetry is promoted to a local one; this is because under a local gauge transformation, the derivatives of fields do not transform in the same way as the fields, hence, the kinetic term will not be invariant. In order to ensure invariance under a local gauge transformation, it is necessary to enlarge the field content of the theory, adding the so called \emph{gauge fields} which are responsible for the interactions among the original fields~\cite{Yang:1954ek}.

Consider a field theory which is  invariant under the global symmetry:
\be
\psi'(x)=e^{i \alpha_i \Gamma^i}\psi(x)=U\psi(x),
\ee
where $\psi(x)$ represents a generic field, $\psi'(x)$ is the transformed field, $\Gamma^i$ are the Lie generators of the symmetry group $G$ and $\alpha_i $ are the parameters that define the transformation.  The same theory will no longer be invariant if the parameters $\alpha_i$  depend on the space-time coordinate
\be\label{gauge_transf}
\psi'(x)=e^{i \alpha_i(x) \Gamma^i}\psi(x)=U(x)\psi(x),
\ee
since in this case the field derivatives transform differently from the fields themselves:
\be
\frac{\partial \psi'(x)}{\partial x}=\frac{\partial U(x)}{\partial x}\psi(x) + U(x)\frac{\partial \psi(x)}{\partial x} \neq U(x)\frac{\partial \psi(x) }{\partial x}.
\ee
It is possible to recover invariance also under~(\ref{gauge_transf}) by defining a new set of fields $A^i$, one for each generator of the group $G$, and replacing the ordinary derivatives by their covariant version $D_\mu$
\be\label{cov_der}
\partial_\mu \rightarrow D_\mu(x) \equiv \partial_\mu - i g A_\mu^i(x) \Gamma^i.
\ee
The parameter $g$ is the adimensional coupling constant of the gauge group $G$: it is a single free parameter characterising the strength of the interactions mediated by the fields $A^i$ and is to be fixed by comparison with experiments.
In order to recover local invariance it is sufficient to require that under an infinitesimal gauge transformation the fields $A^i$ transform accordingly to
\be\label{gauge_field_transf}
{A_\mu^i(x)}'=A_\mu^i(x) + C^i_{jk} \alpha_j(x) A_\mu^k(x) +\frac{1}{g} \partial_\mu \alpha_i(x),
\ee
where $C^i_{jk}$ are the structure constants of the group $G$. The equations~(\ref{cov_der}) and~(\ref{gauge_field_transf}) imply that under~(\ref{gauge_transf})
\bee
D_\mu (x) \rightarrow U(x) D_\mu(x).
\eee
Thus any Lagrangian that is invariant under the global transformations generated by some generic Lie group $G$ can be made invariant under local transformations of the same group simply by adding a set of new vector fields transforming as in~(\ref{gauge_field_transf}) and replacing the ordinary derivatives by their covariant version~(\ref{cov_der}).

A side-product of this procedure is the introduction of interactions among fields; consider for instance the kinetic term for a fermion field, under the replacement~(\ref{cov_der}) we have
\be
\mathcal{L}_{kin} = i \bar{\psi} \slashed{\partial} \psi \longrightarrow i \bar{\psi}\left( \slashed{\partial} - i g \slashed{A}^i \Gamma^i \right) \psi= \mathcal{L}_{kin} + g \bar{\psi} \slashed{A}^i \Gamma^i \psi,
\ee
where we used the Feynman notation: $\slashed{a}\equiv a_\mu \gamma^\mu$. The interactions among standard model particles  are the product of a gauge invariance. The gauge structure of the SM interactions greatly simplifies the structure of the theory, leaving the nature of the gauge group $G$ and the values of the coupling constant $g$ as the only unknown to be determined.  

The gauge group of the SM, which generates the correct interactions, is the direct product $SU(3)_C \otimes SU(2)_L \otimes U(1)_Y $:
\begin{itemize}
 \item $SU(3)_C$ is the group of the $3\times 3$ unitary matrices with determinant equal to $1$. It has $8$ generators and acts on fields possessing \emph{color} charge, i.e. quark fields. It describes the strong interactions;
\item $SU(2)_L$  is the group of the $2\times2$ unitary matrices with determinant equal to $1$. It has $3$ generators (the Pauli matrices) and it acts on doublets containing left-handed fields;
\item $U(1)_Y$ is the group of phase transformations. Its generator is the hypercharge $Y$. Together with $SU(2)_L$ it describes the electroweak interactions.
\end{itemize}

To complete the construction of the model it is necessary to specify the fields associated with the fermionic particles and their transformation properties under the gauge group. These fields are dubbed \emph{matter fields} (to discriminate them from the aforementioned \emph{gauge fields} necessary to guarantee the gauge invariance of the theory) and are collected in Table~(\ref{matterfields}). We use the following convention: the fields that belong to the fundamental representation of $SU(2)_L$ are collected into doublets, while those that are $SU(2)_L$ singlets are represented alone. The subscripts $(c,y)$ indicate how the fields transform under $SU(3)_C\otimes U(1)_Y$: $c=3$ means that the field is in the fundamental representation of $SU(3)_C$, while $c=1$ means that it is an $SU(3)_C$ singlet; $Y$ is the hypercharge related to the $U(1)_Y$ transformations of the field, i.e. $\psi(x)\rightarrow \psi'(x) = e^{i y \alpha(x) } \psi(x)$. Finally the $L$ and $R$ subscripts denote the left-handed and right-handed chirality of the fields, respectively.

\begin{table}[htb]
\begin{center}
\renewcommand\arraystretch{1.5}
\begin{tabular}{|lcr|}
\hline
$ q_L^{a,i} =\left( \begin{array}{c} u^{a,i}_L  \\ d^{a,i}_L \end{array} \right)_{\left(3,\frac{1}{6}\right)} $ \hspace{2cm} & $ \begin{array}{c} {u^{a,i}_R}_{\left(3,\frac{2}{3}\right)} \\ {d^{a,i}_R}_{\left(3,-\frac{1}{3}\right)} \end{array} $&\hspace{1cm} $\begin{array}{c}
a=1,2,3\\
i=1,2,3
\end{array}$\\
&&\\
$ l_L^\alpha =\left( \begin{array}{c} {\nu}^\alpha_L \\ {e}^\alpha_L \end{array} \right)_{\left(1,-\frac{1}{2}\right)} $ & ${e_R^\alpha}_{\left(1,-1\right)}$ & $\alpha=1,2,3$\\[4ex]
\hline
\end{tabular}
\end{center}
\caption{Fermionic field content of the SM.}
\label{matterfields}
\end{table}

Notice that there exist three ``copies'' of the SM matter fields: the structure depicted in Table~\ref{matterfields} is repeated 3 times as the indices $a$ and $\alpha$ vary. We say that the SM fields belong to three different \emph{generations}; there are no differences among the generations apart for the masses of the particles they contain. The other index $i$ associated to quark fields in Table~\ref{matterfields} is not related to generations but to the gauge group $SU(3)_C$: it simply reminds that the quark fields are charged under this group and belong to an $SU(3)_C$ triplet.

It is possible to use the terminology of the  $SU(2)$ representation theory to label the fields: we can associate to each field the value $I$ so that $I(I+1)$ is the eigenvalue of the $SU(2)$ Casimir  operator over its multiplet; similarly $I_3$ is the eigenvalue of the matrix $\frac{\sigma^3}{2}$. If $Q$ is the electric charge of the field the following relation holds
\be\label{e_charge}
Q=I_3 + Y.
\ee

Notice that the Table~\ref{matterfields} does not contain a right-handed partner for the neutrino fields, since when the SM was constructed there was no evidence suggesting this particle to be required. Further notice that eq.~(\ref{e_charge}) implies that $Y=0$ for a field having {$I_3=Q=0$}, as is the case of a right-handed neutrino, which thus results to be a singlet, neutral under the SM gauge group. Such kind of particle, if it exists, would be coupled to gauge bosons only indirectly via a possible mixing with the left-handed neutrinos.

The construction so far described only allows the description of massless particles. Let us consider the general Lagrangian for a spinor field, from which the Dirac equation is derived,
\be
\mathcal{L}_D=\bar{\psi} (i \slashed{D} - m)\psi.
\ee
By decomposing the spinor in its left- and right-handed part as in~(\ref{weyldecomposition}), and using the orthogonality of the projectors, $P_L P_R =P_R P_L = 0$, it is possible to decompose the mass term as
\be
m \bar{\psi} \psi=m(\overline{\psi_L} \psi_R + \overline{\psi_R}  \psi_L).
\ee
This form is not gauge invariant, because of the different transformation properties of the left- and right-handed SM fields under the $SU(2)_L \otimes U(1)_Y$ group. Moreover a mass term for the gauge bosons
\bee\label{g_boson_mass}
 \frac{1}{2}M_{ij} A^i_\mu A^{j,\mu}
\eee
is not invariant under~(\ref{gauge_field_transf}) as well, suggesting that gauge theories may only accommodate massless gauge bosons. But among the known subatomic interactions only the electromagnetic one manifest a long-range behaviour, while strong and weak interactions are effective at subatomic distances.

The solution to these apparent contradictions is the well known Brout-Englert-Higgs mechanism.

\subsubsection{The Brout-Englert-Higgs mechanism (and the Weinberg-Salam model)}

The Brout-Englert-Higgs (BEH) mechanism~\cite{Higgs:1966ev,Higgs:1964pj,Englert:1964et} shows that the spontaneous symmetry breaking (SSB) of a gauge symmetry implies a nonzero mass term for the gauge bosons associated with the generators of the broken subspace. It can be embedded in the electroweak sector of the SM~\cite{Weinberg:1967tq,Salam:1964ry} in which case it is also responsible for the generation of nonzero fermionic masses. It is important to emphasize that the Lagrangian of the resulting theory is manifestly invariant under the original gauge group, but as a result of the SSB the vacuum state is not.\footnote{The subject of SSB in gauge theories and its historical development is reviewed in~\cite{Bernstein:1974rd}.}

Consider a gauge theory containing a set of Lorentz scalar fields $\phi^i$ charged under a gauge group $G$; we can assume without loss of generality that all the $\phi^i$ are real. Their kinetic term  in the Lagrangian is
\bee\label{kin_higgs}
\frac{1}{2}(D_\mu \phi^i)(D^\mu \phi^i)&=&\frac{1}{2}\left(\partial_\mu \phi^i - i g A_\mu^j\Gamma^j_{ik} \phi^k\right)\left( \partial^\mu \phi^i - i g A^{\mu,s} \Gamma^s_{i r}\phi^r \right)\non 
&=&\frac{1}{2} \left(\partial_\mu \phi^i\right) \left(\partial^\mu \phi^i\right)-\frac{1}{2} g^2 \Gamma^j_{ik}\Gamma^s_{ir} A_\mu^j A^{\mu,s} \phi^k\phi^r+\dots
\eee

Because of the Lorentz scalar nature of the fields it is in general possible that a combination of them acquire a non-zero expectation value on the vacuum state of the theory $\ket{\Omega}$,
\bee
\bra{\Omega} \phi^i (x) \ket{\Omega} = \bra{\Omega} \phi^i (0) \ket{\Omega} = v^i,
\eee
since $\phi^i(x) = e^{-i P^\mu x_\mu} \phi^i (0) e^{i P^\mu x_\mu}$ ($P^\mu$ being the generator of the translations along the $\mu$ direction)  and we can always choose $P^\mu \ket{\Omega} = 0$. In that case it is convenient to separate the vacuum and the dynamical components of the fields
\bee\label{higgs_VEV_red}
\phi^i (x) \equiv v^i + \sigma^i (x), \hspace{1cm} \bra{\Omega} \sigma^i (x) \ket{\Omega} =0.
\eee 
After the redefinition~(\ref{higgs_VEV_red}), the last term in~(\ref{kin_higgs}) generates a mass term for the gauge bosons in the form~(\ref{g_boson_mass}), with
\bee
M_{ij} = g^2 \left(\Gamma^i_{tk} v^k\right)\left(\Gamma^j_{tr} v^r\right).
\eee
Notice that the number of gauge bosons that acquire a non-zero mass depends on the way the scalar fields acquire a vacuum expectation value (VEV): if the orientation of the VEV vector is such that $\Gamma^i_{tk} v^k =0$ for some generator $\Gamma^i$, then the corresponding gauge boson remains massless.

Let us now see how the Higgs mechanism can be embedded in the SM, allowing for a gauge formulation of the theory that includes massive fermions and massive gauge bosons. We enlarge the SM field content by adding  a complex scalar field $\Phi$ (the Higgs field),
\be
{\Phi} = \left( \begin{array}{c} \phi^+ \\ \phi^0 \end{array} \right)_{\left(1,\frac{1}{2}\right)},
\ee
which is an $SU(2)_L$ doublet with hypercharge $Y=1/2$. The SM Lagrangian is modified by the addition of the term
\be
\mathcal{L}_{\Phi} = (D_\mu \Phi)^\dagger (D^\mu \Phi) - V(\Phi^\dagger \Phi),
\ee
where 
\begin{eqnarray}\label{higgspotential}
V(\Phi^\dagger \Phi)& =&-\mu^2 (\Phi^\dagger \Phi)+ \lambda (\Phi^\dagger \Phi)^2 \nonumber \\
&= &- \mu^2 \left(|\phi^+|^2+|\phi^0|^2\right)+\lambda \left(|\phi^+|^2+|\phi^0|^2\right)^2,
\end{eqnarray}
whit $\mu^2,\lambda >0$.

\begin{figure}[htb]
 \begin{center}
\includegraphics[width=0.6\textwidth]{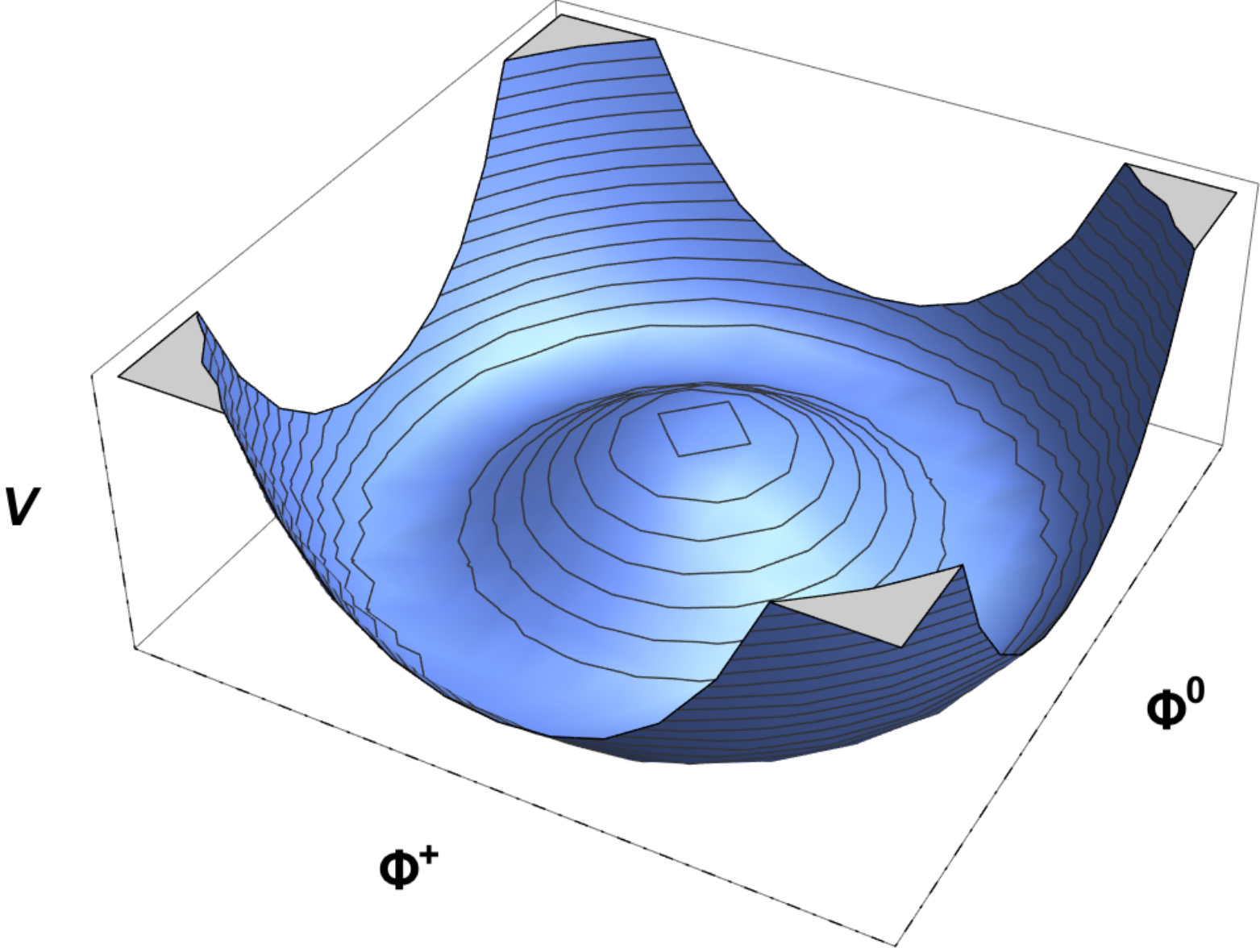}
\caption{Shape of the Higgs field potential.}
\label{higgspotentialgraph}
 \end{center}
\end{figure}

The shape of the potential $V(\Phi^\dagger \Phi)$ as a function of the components of $\Phi$ is shown in Fig.~(\ref{higgspotentialgraph}).

By looking at the stationary points and at the second derivatives of the potential $V$, one can verify that~(\ref{higgspotential}) has a  degenerate absolute minimum determined by the condition
\be\label{solb}
 |\phi^+|^2 +|\phi^0|^2 = \frac{\mu^2}{2\lambda}.
\ee
In terms of energy all the ground configurations described by~(\ref{solb}) are equivalent, but the field $\Phi$ will eventually choose a definite direction in the degenerate minima. That is what spontaneously breaks the gauge symmetry: the Lagrangian is still invariant under $SU(2)_L\otimes U(1)_Y$ local transformations, but the vacuum state is not.

Since the field $\Phi$ is an operator, the condition~(\ref{solb}) is actually referred to the VEV of the field:
\be
\left|\brackets{0}{\phi^+}{0}\right|^2 +\left| \brackets{0}{\phi^0}{0}\right|^2=\frac{\mu}{2\lambda}\equiv \frac{v}{\sqrt{2}}.
\ee
The gauge invariance gives us the freedom to perform an $SU(2)_L\otimes U(1)_Y$ local transformation in order to choose a convenient direction for the Higgs VEV:
\begin{eqnarray}
\Phi(x) &=& \left( \begin{array}{c}  \phi^+(x) \\ \phi^0(x) \end{array} \right) =  \left( \begin{array}{c} \frac{\phi^+_1(x) + i \phi^+_2 (x)}{\sqrt{2}} \\ \frac{\phi^0_1(x) +i \phi^0_2(x)}{\sqrt{2}} \end{array} \right)\nonumber \\
& \stackrel{SU(2)_L}{\longrightarrow}& \left( \begin{array}{c} 0 \\ \frac{\rho(x)+ i \tau(x)}{\sqrt{2}} \end{array} \right)
\stackrel{U(1)_Y}{\longrightarrow} \left( \begin{array}{c} 0 \\ \frac{\sigma'(x)}{\sqrt{2}} \end{array} \right),
\end{eqnarray}
where $\phi_{(1,2)}^{(+,0)},\rho,\tau,\sigma'$ are real fields. In this basis one has
\be
\brackets{0}{\sigma'(x)}{0} = v,
\ee
and one can expand the Higgs field around its minimum
\be
\Phi(x) = \left( \begin{array}{c} 0 \\ \frac{v}{\sqrt{2}} + \sigma(x) \end{array} \right).
\ee
The covariant derivative acting on the Higgs field is
$$
D_\mu \Phi(x) = \left( \partial_\mu - i g W_\mu^i \frac{\sigma^i}{2} - i \frac{g'}{2} B_\mu \right)  \left( \begin{array}{c} 0 \\ v + \frac{\sigma(x)}{\sqrt{2}} \end{array} \right)
$$
\be
=-\frac{i}{2} \left( \begin{array}{c} g v (W^1_\mu -i W^2_\mu) + \frac{g \sigma(x)}{\sqrt{2}} (W^1_\mu - i W^2_\mu ) \\
i \sqrt{2} \partial_\mu \sigma(x) + v(-g W^3_\mu + g' B_\mu) + \frac{\sigma(x)}{\sqrt{2}} (-g W^3_\mu + g' B_\mu ) \end{array} \right),
\ee
and the kinetic term in the Higgs Lagrangian reads
\begin{eqnarray}
(D_\mu \Phi)^\dagger \left(D^\mu \Phi\right) &=& \frac{1}{2} (\partial_\mu \sigma)(\partial^\mu \sigma) + \frac{g^2 v^2 }{4} ( W_\mu^1 {W^1}^\mu + W_\mu^2 {W^2}^\mu )\nonumber \\
&&+\frac{v^2}{4} (g W^3_\mu - g' B_\mu)(g {W^3}^\mu - g' B^\mu) \nonumber \\
&& + \mbox{ cubic } + \mbox{ quartic  terms }\label{higgscovariant},
\end{eqnarray}
which contains the mass terms for the gauge bosons. It is possible to diagonalise the mass terms involving $W^3_\mu$ and $B_\mu$ by defining
\begin{eqnarray}\label{za}
Z_\mu &=& \frac{g W^3_\mu - g' B_\mu}{\sqrt{g^2 +{g'}^2}} \equiv \cos \theta_W W_\mu^3 - \sin \theta_W B_\mu,\\
A_\mu &=& \frac{g' W_\mu^3 + g B_\mu}{\sqrt{g^2 + {g'}^2}} \equiv \sin \theta_W W^3_\mu + \cos\theta_W B_\mu.
\end{eqnarray}
The \emph{Weinberg} or \emph{weak} mixing angle $\theta_W$ is related to the gauge couplings as
\bee
\cos \theta_W = \frac{g}{\sqrt{g^2+{g'}^2}}, \hspace{1cm} \sin \theta_W = \frac{g'}{\sqrt{g^2+{g'}^2}}.
\eee
It is also convenient to define states of definite electric charge as
\be\label{wplusminus}
W^\pm_\mu=\frac{W^1_\mu \mp i W^2_\mu}{\sqrt{2}}.
\ee
With these redefinitions the kinetic term~(\ref{higgscovariant}) reads
\begin{eqnarray}
 (D_\mu \Phi)^\dagger \left(D^\mu \Phi\right) &=& \frac{1}{2} (\partial_\mu \sigma)(\partial^\mu \sigma) + \frac{g^2 v^2 }{4} \left( W_\mu^+ {W^-}^\mu + W_\mu^- {W^+}^\mu \right)\nonumber \\
&&\frac{g^2 v^2}{4 \cos^2 \theta_W} Z_\mu Z^\mu  + \mbox{ cubic } + \mbox{ quartic  terms }.
\end{eqnarray}
Notice that the field $A_\mu$ remains massless, as it is associated to the unbroken $U(1)_{em}$ electromagnetic gauge group with coupling constant
\bee
e=\frac{g g'}{\sqrt{g^2+{g'}^2}}.
\eee
Also the $SU(3)_C$ group remains unbroken by the Higgs mechanism and its gauge bosons are massless. In that case the absence of observation of long range interactions is due to a different mechanism related to strong interactions, the confinement of colour charges~\cite{Wilson:1974sk}.

The nonzero VEV of the Higgs field also accounts for the origin of the masses of the fermionic SM fields, through a gauge and Lorentz invariant Yukawa-like interaction that for the leptonic fields reads as 
\be\label{higgsmassdown}
\overline{l_L} \Phi e_R + h.c.,
\ee
When the Higgs field acquires a VEV, the interaction~(\ref{higgsmassdown}) generates the mass terms for the particles with $I_3=- 1/2$:
\be\label{down_masses}
\left( \begin{array}{cc} \overline{{\nu_e}_L} & \overline{e_L} \end{array} \right) \left( \begin{array}{c} 0 \\ v \end{array} \right) e_R + h.c. = v(\overline{e_L} e_R + \overline{e_R} e_L).
\ee
Defining
\be
\tilde{\Phi} = i \sigma^2 \left( \Phi \right)^*,
\ee
which has $Y=-1/2$ and $I=1/2$, it is possible to construct the gauge and Lorentz invariant term
\be\label{quarkmass}
\overline{q_L} \tilde{\Phi} u_R+h.c.,
\ee
which analogously to~(\ref{down_masses}) generates the mass terms for the fields with $I_3=1/2$.

A direct verification of the BEH mechanism was finally achieved when, on the July 4th 2012, the ATLAS and CMS collaborations announced the independent identification of a new scalar particle with mass around 125 GeV and compatible with a SM-like Higgs boson~\cite{Aad:2012tfa,Chatrchyan:2012ufa}.

\subsection{Renormalizability}

While dealing with the computation of correlation functions in interacting quantum field theories (QFT) it is often impossible to obtain exact results. In these cases the complications of the calculation are overcome by means of numerical simulations if the theory is strongly coupled, or by a perturbative expansion of the results if the coupling constants are sufficiently small. In the latter case divergent quantities commonly appear if the perturbation is taken besides the first non-trivial order (tree level), considering also Feynman diagrams including closed lines (loops). The renormalisation of a theory is the procedure with which these divergent quantities are dealt, in such a way that the final results are finite. The fact that a quantum field theory needs or not to be renormalised is not a priori a criterion for the theory itself, since the necessity of a renormalisation procedure is mostly related to the perturbative approach adapted to perform the computations: in defining a zeroth order theory we are forced to include a set of \emph{bare} parameters, which describe the theory in the absence of interactions. But the bare parameters are unobservable and thus unphysical, since the experiments can only probe the complete theory. The renormalisation procedure consists in expressing all the observables in terms of physical meaningful quantities; in a renormalised theory the aforementioned divergencies only reappear when one tries to establish a connection between physical and bare parameters. 
If it is possible to perform the renormalisation process at all orders in the perturbative expansion, a theory is said to be \emph{renormalizable}.

Consider a general theory in $d$ dimensions, whose Lagrangian includes the interaction term
\be\label{int_ren}
- g f(\phi^r \psi^s),
\ee
where $g$ is a coupling constant and $f(\phi^r \psi^s)$ denotes a combination of $r$ bosonic and $s$ fermionic fields. In any diagram describing the interaction~(\ref{int_ren}) each vertex has $r+s$ legs, but the different momenta are in general not independent. Each diagram is associated with a number of independent integrations over internal momenta, the number equating its number of loops, and each one contributing with $d$ powers of momenta, indicated generically with $q$ in the following. On the other hand each internal line contributes with a propagator of dimension $[q^{-2}]$ in the bosonic case, and $[q^{-1}]$ in the fermionic one. Denoting by $L$ the total number of loops and by $I_{B,F}$ the number of bosonic and fermionic propagators, the \emph{superficial degree of divergence} D of a diagram is
\be\label{degreediv}
D=d L - (2I_B + I_F).
\ee
Let $E_{B,F}$  be the number of external bosonic and fermionic lines, and $n$ the number of vertices: each vertex is associated with a momentum conservation condition, but their combination is constrained by the overall momentum conservation, thus the number of constrains on momenta is $n-1$ and the number of independent momenta is
\be\label{eq:ren2}
L=I_B+I_F - (n-1).
\ee
Each vertex has $r$ bosonic and $s$ fermionic legs, with the internal ones counting twice because they are connected with two vertices. One then has
\begin{eqnarray}
rn &=& E_B+2I_B,\nonumber\\
sn&=& E_F+2I_F,
\end{eqnarray}
from which
\begin{eqnarray}\label{eq:ren3}
2I_B+I_F&=& rn+\frac{sn}{2} - E_B -\frac{E_F}{2}, \nonumber\\
I_B+I_F&=&\frac{rn+sn-E_B-E_F}{2}.
\end{eqnarray}
Replacing~(\ref{eq:ren2}, \ref{eq:ren3}) in~(\ref{degreediv}) we obtain
\be\label{divsup}
D=E_B+\frac{E_F}{2}-d\left( \frac{E_B+E_F}{2} -1\right) + n\left[d\left(\frac{r+s}{2} -1\right) - r -\frac{s}{2} \right].
\ee
Here and in the following it is useful to work in natural units, defined as
\begin{eqnarray}
  c  = 1 &\Rightarrow& \left[ x \right] = \left[ t \right],  \nonumber \\
\hbar  = 1 &\Rightarrow & \left[ E \right] = \left[ t \right]^{-1},
\end{eqnarray}
where $c$ is the speed of light in the vacuum, $\hbar$ is the Planck constant and $x,t,E$ represent  a length, a time and an energy, respectively. In these units the action $S$
\be
S=\int d^d x\ \mathcal{L},
\ee
 is dimensionless, so that
\be
\left[ \mathcal{L} \right] = \left[ E \right]^d,
\ee
or $\left[ \mathcal{L} \right] =d$ if we simply indicate by convention the dimensions as powers in units of energy. Because the kinetic term is $\partial_\mu \phi \partial^\mu \phi$ for a bosonic field, and $i \bar{\psi} \slashed{\partial} \psi$ for a fermionic one, it follows
\begin{eqnarray}
 \left[ \phi \right] = \frac{d-2}{2}, && \left[ \psi \right] = \frac{d-1}{2},
\end{eqnarray}
from which
\be
\left[ g \right] =d \left(1-\frac{r+s}{2} \right) +r+\frac{s}{2},
\ee
and ~(\ref{divsup}) becomes
\be\label{div_naive}
D=E_B+\frac{E_F}{2}-d\left( \frac{E_B+E_F}{2} -1\right) - n\left[g\right].
\ee
The last term of this expression describes how the higher order corrections to a given amplitude diverge as a function of the number of vertices in the diagrams: if $[g]<0$ the $n$th order in the perturbative expansion is more divergent than the $(n-1)$th and the theory is \emph{non-renormalizable}; it does not mean that the theory is not predictive, but the renormalisation must be carried out at each perturbative order, requiring an infinite number of renormalisation conditions to obtain convergent results at all orders. This is usually the case of effective field theories, which inherit their non-renormalizability from the lack of an ultra-violet (UV) completion. Conversely, if only a finite number of renormalisation conditions is necessary to obtain convergent results at all orders in the perturbative expansion, a theory is said \emph{renormalizable}; a renormalizable theory is in principle self-contained and does not formally require any UV completion. A necessary condition for renormalizability is that $[g]\geq 0$, or equivalently
\be\label{ren_cond}
\left[ f(\phi^r \psi^s) \right] \leq d.
\ee
This is however not a sufficient condition, since the counting of the superficial degrees of divergence does not take into account the possibility of nested divergencies, when a divergent non-trivial subgraph makes the whole diagram to diverge more than the na\"ive counting~(\ref{div_naive}).

It has nonetheless been demonstrated that if~(\ref{ren_cond}) holds, then gauge theories are renormalizable provided certain additional conditions (such as the absence of gauge anomalies) are satisfied~\cite{'tHooft:1971fh,'tHooft:1972fi}, and that also spontaneously broken gauge theories are renormalizable~\cite{'tHooft:1971rn}. The SM in $4$ dimensions ($d=4$) is renormalizable, provided that the coupling constants are dimensionless or have positive dimensions in energy.

\section{Neutrino masses in the Standard Model}\label{nu_mass_sm}
We can now address the point of neutrino masses in the SM: we will show why, as a consequence of the SM constraints, neutrinos are  massless and why any signal for nonzero neutrino masses points toward the existence of physics beyond the SM (BSM).

First of all in the SM framework neutrino masses cannot arise from the same mass generation mechanism common to the other elementary fermions, via a Yukawa interaction as in eq.~(\ref{quarkmass}), simply because of the lack of a right-handed neutrino field $\nu_R$, which would have the correct quantum numbers for the purpose. A Dirac mass term as in eq.~(\ref{mDirac}) is thus forbidden.
Nonetheless, we know from Section~(\ref{rel:inv}) that with a single chiral field $\nu_L$ it is possible to construct a Lorentz invariant Majorana mass term as in eq.~(\ref{mMaj}). But a term of this kind suffers from the lack of invariance under the $SU(2)_L\times U(1)_Y$ SM gauge subgroup, and is thus also forbidden. A gauge invariant Majorana mass term for left-handed neutrinos may be generated as a consequence of a SSB mechanism, in a similar way as Dirac mass terms are generated in the SM. However such mechanism would require a Higgs-like scalar field with isospin $I=1$, in order to construct a gauge invariant Yukawa interaction containing the $I=1$ term $\overline{\nu_L^c} \nu_L$. Such a field (a Higgs triplet) is not present in the SM, and so this possibility is also excluded.

To summarise, because of the gauge symmetries and the field content of the theory, and allowing only renormalizable couplings, neutrinos are massless in the SM.

If one relaxes the renormalizability condition and considers the SM as an effective theory valid up to some energy scale, and parametrises the effects of the unknown UV completion as a tower of effective non-renormalizable operators, the first new physics effects are encoded in the collection of allowed dimension 5 operators. Remarkably, there exists a unique 5-dimensional Lorentz and gauge-invariant operator that is possible to construct with the SM fields, the so called Weinberg operator~\cite{Weinberg:1979sa}
\be\label{weinbergmass}
\frac{1}{2} \frac{c_{\alpha \beta}}{\Lambda} \left( \overline{ l_L ^c}_\alpha \tilde{\Phi}^* \right) \left( \tilde{\Phi}^\dagger l_L^\beta \right) + h.c.,
\ee
where $\alpha,\beta=e,\mu,\tau$, $c_{\alpha \beta}$ is a complex symmetric matrix and $\Lambda$ is a constant with the dimensions of energy that is related to the new physics scale. When the Higgs field acquires a nonzero VEV, the operator~(\ref{weinbergmass}) contributes as
\be\label{d5_mass}
\frac{v^2}{2} \frac{c_{\alpha \beta}}{\Lambda} \overline{\nu_L^c}_\alpha \nu_{L \beta} + h.c.,
\ee
that is a Majorana mass term for left-handed neutrinos. It is notable that the first expected effect of physics BSM is just the appearance of non-zero Majorana neutrino masses; in this sense neutrinos are  truly a window to BSM physics.

\section{Leptonic Lagrangian in the Standard Model}\label{nu_lag_sm}
Given the SM field content, the SM Lagrangian is the most general renormalizable Lagrangian which is invariant under the local gauge group and the global Lorentz transformations. Choosing a basis in which the kinetic terms are diagonal, the leptonic part is given by
\begin{eqnarray}\label{lep_lag_sm}
\mathcal{L}_{leptons} &=& \overline{l_L^\alpha} \left(i \slashed{\partial} + \frac{g}{2} \slashed{W}^i \sigma^i - \frac{g'}{2} \slashed{B} \right) l_L^\alpha + \overline{e_R^\alpha} \left(i\slashed{\partial} - g' \slashed{B} \right) e_R^\alpha  \nonumber \\
&& - Y_{\alpha \beta}  \overline{l_L^\alpha} \Phi e_R^\beta - {Y_{\alpha \beta}^\dagger} \overline{e_R^\alpha} \Phi^\dagger  l_L^\beta.
\end{eqnarray}
$Y_{\alpha \beta}$ is the matrix of the Yukawa interactions, which expresses the strength of the couplings between the leptons and the Higgs field. It is a $3 \times 3 $ matrix with entries complex in general, which can be diagonalised through the bi-unitary transformation~\cite{Bilenky:1987ty}
\be\label{biunit_diag}
U^\dagger Y V = \mbox{diag}\left[y_1,y_2,y_3\right],
\ee
where $y_{1,2,3}$ are positive numbers and $U,V$ are unitary matrices. Redefining the lepton fields as
\begin{eqnarray}\label{weak_rot_sm}
l^\alpha_L& =& U_{\alpha \beta} \tilde{l}_L^\beta,\\
e_R^\alpha &=& V_{\alpha \beta} \tilde{e}_R^\beta,
\end{eqnarray}
the leptonic part of the SM Lagrangian is rewritten as
\begin{eqnarray} \label{weakLagrangian}
 \mathcal{L}_{leptons} &=& \overline{\tilde{l}_L^\alpha} \left(i \slashed{\partial} + \frac{g}{2} \slashed{W}^i \sigma^i -\frac{g'}{2} \slashed{B} \right) \tilde{l}_L^\alpha+ \overline{\tilde{e}_R^\alpha} \left(i\slashed{\partial} - g' \slashed{B} \right) \tilde{e}_R^\alpha \nonumber \\
&& - \sum_{\alpha=1}^3 y_\alpha \overline{\tilde{l}_L^\alpha} \Phi \tilde{e}_R^\alpha - \sum_{\alpha=1}^3 y_\alpha \overline{\tilde{e}_R^\alpha} \Phi^\dagger  \tilde{l}_L^\alpha.
\end{eqnarray}
In other words, it is possible to find a simultaneous basis for mass and interaction eigenstates, that we simply indicate with $l_L,e_R$ in the following, while the generation indices can be unambiguously associated with known flavours, $\alpha,\beta=e,\mu,\tau$. The Lagrangian is invariant under the continuous transformations (in the following expressions repeated indices are not meant to be summed):
\be\label{flavour_symm_sm}
\left\{ \begin{array}{rcl}
l_L^\alpha &=& e^{i \theta_\alpha} l_L^\alpha,  \\
e_R^\alpha &=& e^{i \theta_\alpha} e_R^\alpha,\end{array} \right.
\ee
with the $\theta_{e,\mu,\tau}$ parameters not necessarily equal. The Noether current associated to each one of these transformations is
\be
J^\mu_\alpha = - \overline{e}^\alpha \gamma^\mu e^\alpha - \overline{\nu_L}^\alpha \gamma^\mu \nu_L^\alpha.
\ee
The associated conserved charge is a well defined observable; expressing it in the form of a normal product in the operators it contains~\cite{Mandl:1985bg} we obtain:
\begin{eqnarray}\label{flavour_charges_sm}
Q^\alpha &=&- :\int d^3x\ J^0_\alpha(x) :\ =-  \sum_{p,i} \left( a^\dagger_{p,i} a_{p,i} - b^\dagger_{p,i} b_{p,i}+c^\dagger_{p,i} c_{p,i}-d^\dagger_{p,i} d_{p,i}\right)_\alpha \nonumber \\
&=&-\ \left( n_\alpha - \overline{n}_\alpha\right),
\end{eqnarray}
where $a$ and $b$ are the annihilation operators associated with the charged particle and antiparticle of generation $\alpha$, respectively; $c$ and $d$ are the ones associated with the neutrino and the summation is taken over all possible values of momenta $p$ and polarisations $i$. $n$ and $\overline{n}$ are the number operators that count how many particles and antiparticles are respectively present. Thus in the SM the differences
\be
L_\alpha = n_\alpha - \overline{n}_\alpha, \hspace{2cm} \alpha=e,\mu,\tau,
\ee 
between the number of particles and antiparticles in each flavour are constant, and the interactions preserve these quantities.

Expanding the Lagrangian~(\ref{weakLagrangian}) and redefining the gauge fields as in eqs.~(\ref{za} - \ref{wplusminus})  we obtain the following form of the SM Lagrangian involving neutrino fields (repeated indices are summed, hereafter):
\begin{eqnarray}\label{eq:weak_SM}
\mathcal{L}_\nu^{\text{\tiny{SM}}} &=&i \overline{\nu_L^\alpha} \slashed{\partial} \nu_L^\alpha  \nonumber \\
&&- \frac{g}{\sqrt{2}} \left( \overline{\nu_L^\alpha} \slashed{W}^+ e_L^\alpha + \overline{e_L^\alpha} \slashed{W}^- \nu_L^\alpha \right) \nonumber \\
&& - \frac{g}{2\cos{\theta_W}} \overline{\nu_L^\alpha} \slashed{Z} \nu_L^\alpha.
\end{eqnarray} 
The first row is the kinetic term, the second one encodes the charged interactions mediated by the $W^{\pm}$ bosons and the third one accounts for the neutral interactions mediated by the $Z$ boson.

\section{Hypothesis of massive neutrinos and consequences}

We have seen in Section~\ref{nu_mass_sm} that massive neutrinos call for the existence of new physics beyond the SM (BSM). It is thus natural to study the phenomenological consequences of BSM realisations, especially in the light of the recent experimental results that will be reviewed in the next Chapter. We will start by analysing the most direct consequences of the dimension 5 operator~(\ref{d5_mass}), studying how the discussion of Section~\ref{nu_lag_sm} is modified by its presence; we later present a more general situation under the assumption that gauge singlet fermions (e.g. right-handed neutrinos) are added to the SM field content.

\subsection{EFT approach}\label{EFT_approach}
Since the fermionic fields are grassmanian variables and  the charge conjugation matrix $C$ is antisymmetric, the operator
\bee
{\nu_{L}^\alpha}^T\ C\ \nu_{L}^\beta
\eee
is completely symmetric under the exchange of the flavour indices $\alpha,\beta$. Thus the coefficients $c_{\alpha \beta}$ in~(\ref{d5_mass}) are completely symmetric too.

The addition of the operator~(\ref{weinbergmass}) to the Lagrangian~(\ref{lep_lag_sm}) significantly modifies the discussion of Section~\ref{nu_lag_sm}: the transformation~(\ref{biunit_diag}) is still viable, but due to the presence of the neutrino mass matrix
\bee\label{eft_mnu}
m^\nu_{\alpha \beta} =- \frac{v^2}{2} \frac{c_{\alpha \beta}}{\Lambda},
\eee
the same transformation does not in general lead to a diagonal basis for massive neutrinos. Thus, unless the BSM physics is characterised by some unknown symmetry implying that the matrix $c_{\alpha \beta}$ is automatically diagonalised by the transformation~(\ref{weak_rot_sm}), the addition of the effective operator~(\ref{weinbergmass}) to the SM Lagrangian makes it impossible to find a simultaneous basis for the mass and the interaction eigenstates. Moreover the transformation~(\ref{flavour_symm_sm}) is no longer a symmetry of the Lagrangian, and the charges~(\ref{flavour_charges_sm}) are no longer conserved, neither individually nor summed over different flavours. This is due to the Majorana character of the mass term~(\ref{d5_mass}) or, equivalently, to the fact that in~(\ref{weinbergmass}) the operator $l_l$ appears in the combination $~l_l^\alpha l_l^\beta$ (and not as $~\overline{l_l^\alpha} l_l^\beta$, for instance), implying that~(\ref{weinbergmass}) violates the total lepton number by two units. That is not the more general configuration since, as we will see in the next section, massive neutrinos can either conserve or violate the total lepton number, depending on their Dirac or Majorana nature. Notice however that, in order to characterise a massive Dirac neutrino, a right-handed component is required; the right-handed component modifies the above discussion, allowing for the generation of neutrino masses already at the renormalizable level.

\subsection{Impact of sterile fermions on neutrino masses: Majorana, Dirac and pseudo-Dirac states}\label{sec:RH_masses}

Let us consider the general case in which the SM field content is enlarged by the addition of $n$ Weyl fermions $N_i$, singlet under the SM gauge group. This hypothesis includes but is not limited to right-handed neutrinos, since the new sterile fermions can differ among themselves by additional quantum numbers, as for instance a global lepton number. Without loss of generality we can nonetheless assume that the fields $N$ have right-handed chirality, since any fundamental left-handed field can be expressed in term of a right-handed one by means of a charge-conjugation operation, cf.~(\ref{left-right:eq}).

The sterile fermions have the correct quantum numbers to couple to active leptons through a Yukawa term,
\be
Y_{\alpha i} \overline{l_L^\alpha} \tilde{\Phi} N_i+h.c.,
\ee
generating, after the electroweak symmetry breaking (EWSB), a Dirac neutrino mass term. Since they are gauge singlets they can moreover couple via a Majorana mass term. 
The most general mass term, invariant under Lorentz and gauge symmetries, is thus
\be\label{nu_mass_steriles}
-\mathcal{L}_m = \overline{\nu_L^\alpha} {m_{\alpha i}^*} N_i + \frac{1}{2} N_i^T C M^*_{ij} N_j + h.c.
\ee
where $M$ is a complex symmetric mass matrix and $m$ is related to the Higgs VEV and to the Yukawa couplings by
\bee
m_{\alpha i} = \frac{v}{\sqrt{2}}Y_{\alpha i}^*.
\eee
Introducing the basis $n_L^T=(\nu_L^e,\nu_L^\mu,\nu_L^\tau,N_1^c,\dots,N_n^c)$ and with the help of (cf.~(\ref{maj_relations}-\ref{cspinors}))
\begin{eqnarray}
\left(N^T_i C M^*_{ij} N_j \right)^\dagger &=& \left(N^c_j\right)^T C M_{ji} N^c_i, \nonumber \\
\left(\overline{\nu_L^\alpha} {m_{\alpha i}^*} N_i\right)^\dagger &=& (N^c_i)^T C m_{\alpha i} \nu_L^\alpha = \left(\nu_L^\alpha\right)^T C m_{\alpha i} N_i^c,
\end{eqnarray}
the mass term~(\ref{nu_mass_steriles}) can be recast in the more compact form
\bee
-\mathcal{L}_m =\frac{1}{2} n_L^T C \mathcal{M} n_L+h.c.,
\eee 
with the mass matrix $\mathcal{M}$ given by\footnote{The following discussion applies as well if the matrix $\mathcal{M}$ possesses a non-zero block in the (1,1) entry, i.e. when a Majorana mass term for left-handed neutrinos is allowed by the presence of an Higgs isospin triplet, see for instance~\cite{Barbieri:1979ag,Marshak:1980yc,Cheng:1980qt,Magg:1980ut,Lazarides:1980nt,Schechter:1980gr,Mohapatra:1980yp}. We do not consider this possibility in the present discussion.}
\bee\label{general_Seesaw_matrix}
\mathcal{M} = \left(
\begin{array}{cc}
\mathbf{0} & m \\
m^T & M
\end{array}
\right).
\eee
This complex symmetric matrix can be diagonalised with the help of the transformation\cite{Schechter:1980gr}
\bee\label{eq:symm_diagonal}
\mathcal{U}^T \mathcal{M} \mathcal{U}=\hat{\mathcal{M}}=\mbox{diag}\left[\mu_1,\dots,\mu_{3+n}\right],
\eee
where $\mathcal{U}$ is a $(3+n)\times (3+n)$ unitary matrix and $\hat{\mathcal{M}}$ is a diagonal matrix. We can define the basis
\bee\label{mass_basis_rot}
\chi_L = \mathcal{U}^\dagger n_L,
\eee
in terms of which the Lagrangian~(\ref{nu_mass_steriles}) takes the form
\bee\label{diagonal_majorana}
-\mathcal{L}_m =\frac{1}{2} \sum_{k=1}^{3+n}\mu_k \overline{\left(\chi_L^k\right)^c} \chi_L^k + \frac{1}{2} \sum_{k=1}^{3+n}\mu_k^*\ \overline{\chi_L^k} \left(\chi_L^k\right)^c.
\eee
If $\varphi_k$ are the complex arguments of the diagonal elements in $\hat{\mathcal{M}}$, 
\bee\label{complex_masses}
\mu_k = \rho_k e^{i\varphi_k}, \hspace{1cm} \rho_k\ge 0,
\eee 
we can define the mass eigenstates of the system as~\cite{Akhmedov:1999uz}
\bee\label{mass_eigenstates}
\chi_k = \chi_L^k + e^{- i \varphi_k} \left(\chi_L^k\right)^c,
\eee
since in this basis the Lagrangian~(\ref{nu_mass_steriles}) reads
\bee\label{diagonalised_Lagrangian}
-\mathcal{L}_m = \frac{1}{2} \sum_{k=1}^{3+n} \rho_k \overline{\chi_k} \chi_k.
\eee
The masses $\rho_k$ are non-negative quantities, and the fields $\chi_k$ are Majorana states, as it immediately follows from~(\ref{mass_eigenstates})
\bee
\chi_k^c = e^{i\varphi_k} \chi_k.
\eee
This is in agreement with the counting of the degrees of freedom of the system: each Weyl spinor possesses two degrees of freedom, and after diagonalization there is a mass eigenstate for each Weyl spinor, each one should have in turn two degrees of freedom, corresponding to the two possible helicities of a Majorana particle.

There is however a different possibility: suppose that for some symmetry reason the terms $M_{ij}$ in~(\ref{nu_mass_steriles}) are absent. Although there is no limitation on the number $n$ of sterile fields that can be present, it is evident that in this case the rank of the matrix $\mathcal{M}$ is at most\footnote{The $(3\times n)$ matrix $m$ has rank 3 at most. Putting both $m$ and $m^T$ in an echelon form the above statement is straightforward.} 6 and thus it has at most 6 non-zero eigenvalues. We can thus restrict for simplicity to the case $n=3$, keeping in mind that the general case is characterised by the addition of $n-3$ massless states. If no Majorana mass terms are present it is more convenient to diagonalise the mass matrix~(\ref{general_Seesaw_matrix}) in two steps, by rotating first the sub-blocks $m, m^T$ via the matrix
\bee
\mathcal{U'} = \left(\begin{array}{cc}
0 & \mathcal{W}^*\\
\mathcal{V} & 0
\end{array}\right),
\eee
where the unitary matrices $\mathcal{W,V}$ diagonalise $m$ via the biunitary transformation
\bee
\mathcal{W}^\dagger m \mathcal{V} = \hat{m} = \mbox{diag}\left[\zeta_1,\zeta_2,\zeta_3\right].
\eee
The resulting matrix can then be put in a diagonal form by a combination of rotations $\mathcal{R}$ with angle $\pi/4$ in the planes $(i,3+i),\ i=1,2,3$. Notice that the final spectrum is characterised by the eigenvalues
\bee
\mathcal{R}^T\mathcal{U'}^T \mathcal{M'} \mathcal{U'}\mathcal{R}=\hat{\mathcal{M'}}=\mbox{diag}\left[-\zeta_1,\zeta_1,-\zeta_2,\zeta_2,-\zeta_3,\zeta_3\right].
\eee
We can now rotate the states in the mass basis defined in~(\ref{mass_basis_rot}), this time using the matrix $(\mathcal{U'}\mathcal{R})^\dagger$. Since the matrices $\mathcal{W}$ and $\mathcal{V}$ only act on the $\nu_L$ and $N$ fields, respectively, we can define for convenience a new basis for them
\bee
\left\{ \begin{array}{rcl}
{N'}^c &=& \mathcal{V}^\dagger N^c,\\
\nu_L' &=& \mathcal{W}^T \nu_L,
 \end{array}\right.
\eee
such that the mass basis resulting from the previous diagonalization procedure can be expressed as
\bee
\left\{ \begin{array}{lcr}
\chi_L^{k,-} &=& \frac{{N'}_k^c-\nu_{L,k}'}{\sqrt{2}},\\
\chi_L^{k,+} &=& \frac{{N'}_k^c+\nu_{L,k}'}{\sqrt{2}},
\end{array}\right.
\eee
with $k=1,2,3$ running over the different mass eigenvalues.
We can now repeat the construction~(\ref{complex_masses}, \ref{mass_eigenstates}) to define the mass eigenstates, noticing however that in this case the masses $\rho_k$ appear in pairs with degenerate moduli and phase factors that are opposite in sign, $e^{i \varphi_k^+}=-e^{i \varphi_k^-}$. The mass eigenstates associated with the degenerate mass $\rho_k$ result, from the previous diagonalization procedure,
\bee
\left\{ \begin{array}{lcr}
\chi_k^+ &=& \chi_L^{k,+} + e^{-i \varphi_k} \left(\chi_L^{k,+}\right)^c,\\
\chi_k^- &=& \chi_L^{k,-} - e^{-i \varphi_k}\left(\chi_L^{k,-}\right)^c,
\end{array}\right.
\eee
and the Lagrangian~(\ref{diagonalised_Lagrangian}) is characterised by the degenerate mass terms
\bee
-\mathcal{L}_m &\ni &\frac{1}{2} \rho_k \left(\overline{\chi_k^+} \chi_k^+ + \overline{\chi_k^-} \chi_k^-\right) \non
&=& \frac{1}{2}\rho_k\left(e^{-i \varphi_k} \overline{{N'}^c_k} {\nu'}_{L,k}^c+e^{-i \varphi_k} \overline{{\nu'}_{L,k}} {N'}_k +e^{i \varphi_k} \overline{{N'}_k} {\nu'}_{L,k}+e^{i \varphi_k}\overline{{\nu'}_{L,k}^c} {N'}^c_k \right)\non
&=& \rho_k \overline{\nu_D^k}\nu_D^k,
\eee
where
\bee
\nu_D^k = \nu^k_L + e^{-i \varphi_k} N^k,
\eee
is a Dirac state possessing a left- and a right-handed component, which are independent. The degrees of freedom of the system are now arranged in a different fashion: the 6 initial Weyl spinors, each one possessing two degrees of freedom, are arranged to form 3 Dirac spinors, each one characterised by 4 degrees of freedom (left- and right-handed helicities for particle and antiparticles states). 

An important point of the above discussion is that we have shown that a Dirac state can be seen as the combination of two Majorana states, that are degenerate in mass and with opposite eigenvalues under the particle-antiparticle conjugation operation. Also notice that the Majorana or Dirac character of the mass eigenstates is related to the symmetries of the Lagrangian: if there exist a nontrivial assignment of lepton charges to the fields in $n_L$ such that the terms in~(\ref{nu_mass_steriles}) preserve this lepton number, then the symmetry must be preserved regardless of the chosen basis. In the previous example, where the $M_{ij}$ terms are absent, we can assign $L=1$ to the fields $\nu_L, N$ and easily verify that the mass matrix $m$ preserves this number. After the diagonalization the mass terms are rearranged in the form~(\ref{diagonal_majorana}), where each single term clearly violates all the internal numbers of the fields $\chi_L$ by two units. However, since the conservation of $L$ does not depend on the chosen basis, the Lagrangian~(\ref{diagonal_majorana}) must preserve this number, although the symmetry is no longer manifest. The mechanism through which the symmetry is implemented is precisely the one described above: the individual states can violate the number $L$, but their combined effect preserve it. An interesting situation is realised when an assignment of lepton numbers preserving $L$ is not possible, but the mass matrix~(\ref{general_Seesaw_matrix}) is characterised by a certain hierarchy, with the terms that violate the total lepton number suppressed with respect to the conserving ones. In this case we indeed expect a violation of the total lepton number and thus Majorana mass eigenstates; however this violation must be a weak effect with respect to the lepton conserving interactions and must vanish when the lepton violating terms are set to zero. The would-be Dirac states in the limiting $L$-conserving scenario are composed by 2 degenerate Majorana states, each one violating $L$ in  exactly the opposite way with respect to its companion; when small $L$-violating terms are considered, the degeneracy in the pairs is lifted and the states become two truly Majorana particles. However their masses are still almost degenerate and their lepton number violating (LNV) interactions still compensate among them, apart for a small resulting amount proportional to the size of the $L$-violating mass terms in~(\ref{nu_mass_steriles}). Such kind of particles are thus named pseudo-Dirac states, meaning that they are Majorana particles whose combined effect resembles a Dirac state.

\subsection{Neutrino oscillations}

As stated in the previous sections any signal for non-zero neutrino masses calls for the existence of physics BSM; it is thus essential to look for experimental evidences related to this hypothesis. As we will show in Section~\ref{sec:nu_mass_exp}, neutrinos are extremely light particles, and any direct measurement of their masses based on kinematic observations is compatible with massless particles in the limit of experimental uncertainties. It is thus necessary to look for different signatures of massive neutrinos, and a powerful guideline is given by the observation made in Section~\ref{EFT_approach}, that the appearance of a neutrino mass matrix~(\ref{eft_mnu}) makes it in general impossible to find a simultaneous diagonal basis for neutrino mass and flavour eigenstates, implying that leptonic flavours are not preserved in the neutrino propagation~\cite{Pontecorvo:1967fh}.

Let us adopt an effective approach to the problem, assuming that a neutrino mass matrix~(\ref{eft_mnu}) is present in the low-energy Lagrangian, disregarding for the moment any possible underlying neutrino mass generation mechanism. Adding the effective operator~(\ref{weinbergmass}) to the Lagrangian~(\ref{lep_lag_sm}) we obtain, after electroweak symmetry breaking, the following leptonic mass terms
\be
-\mathcal{L}_m =  \overline{e'_R}_\alpha M_l^{\alpha \beta} {e'_L}_\beta +\frac{1}{2}{\nu'_L}^T_\alpha C m_\nu^{\alpha \beta} {\nu'_L}_\beta  + h.c. ,
\ee
with $\alpha=e,\mu,\tau$ and
\bee
M_l^{\alpha \beta} = \frac{v}{\sqrt{2}} Y^\dagger_{\alpha \beta}, \hspace{1cm} m_\nu^{\alpha \beta}= -\frac{v^2}{2} \frac{c_{\alpha \beta}}{\Lambda}.
\eee
In the basis defined in~(\ref{lep_lag_sm}) the charged current interactions are diagonal
\be\label{eq:charged_int}
\mathcal{L}_W = -\frac{g}{\sqrt{2}}  \overline{e'_\alpha}\  \slashed{W}^- {\nu'_L}_\alpha + h.c.
\ee
and we have seen in Section~\ref{nu_lag_sm} that, if $m_\nu=0$, the same is true in the mass basis too. This conclusion is not valid anymore as long as $m_\nu\neq 0$, since the independent mass matrices need to be diagonalised by the transformations
\begin{eqnarray}
M_l \rightarrow U_R M_l U_L^\dagger = \hat{M_l}, \nonumber \\
m_\nu \rightarrow U_\nu^T m_\nu  U_\nu = \hat{m_\nu},
\end{eqnarray}
\begin{eqnarray}\label{redefinitions}
{e_{L}}_\alpha &=&  {U_{L}}_{\alpha \beta} {e'_L}_\beta,\nonumber \\
{e_{R}}_\alpha &=&  {U_{R}}_{\alpha \beta} {e'_R}_\beta,\nonumber \\
{\nu_{L}}_i &=&  {U_\nu^\dagger}_{i \alpha} {\nu'_L}_\beta.
\end{eqnarray}
The leptonic kinetic terms and the neutral weak interaction current are invariant under the transformation\footnote{This is no longer true if the matrix $U_\nu$ is not unitary.}~(\ref{redefinitions}), while the charged weak interaction current becomes 
\be\label{mixed_charged_curr}
\mathcal{L}_W = -\frac{g}{\sqrt{2}}  \overline{e_\alpha}\  \slashed{W}^- \left(U_L^\dagger U_\nu\right)_{\alpha i} {\nu_L}_i + h.c.
\ee
The charged current is characterised by the presence of a leptonic mixing matrix $U=U_L^\dagger U_\nu$ and the interactions are not diagonal in the mass basis~(\ref{redefinitions}). Notice that it is in principle possible to choose a basis in which either charged leptons or neutrinos are simultaneous mass and interaction eigenstates. However, because of the tiny neutrino mass values, it is practically impossible to experimentally distinguish among the different neutrino mass eigenstates, while the same operation is very simple for charged leptons. For such a reason it is customary to work in a basis where charged leptons are diagonal in the mass and interactions basis, while neutrinos are not. We recall this convention by appending to the charged states a unique flavour index generically represented by a greek letter, while we distinguish among neutrino interaction eigenstates $\nu_\alpha,\ \alpha = e,\mu,\tau$ and neutrino mass eigenstates $\nu_i,\ i=1,2,3$. By choosing  such a basis, the flavour of a neutrino is determined by the superposition of states that couples with a charged lepton of a given flavour in the charged current vertex,
\be\label{interactionbymass}
\nu_\alpha =  U_{\alpha j} \nu_i.
\ee
Following this definition we can infer the flavour of a neutrino by looking at the charged lepton entering the common interaction vertex. A neutrino produced at a certain space-time point (identified for convenience with the origin of the reference frame) by the interaction~(\ref{mixed_charged_curr}) corresponds to a linear superposition of the (kinematically accessible) mass eigenstates, each one possessing a proper four-momentum defined by the kinematic of the process. Its propagation is described by the equation
\be
\ket{\nu_\alpha (x)} =  U_{\alpha i} e^{- i p_i x} \ket{\nu_i},
\ee
where $x = \left(t,\textbf{x}\right)$ is a space-time point and $p_i = \left(E_i, \textbf{p}_i\right) = \left(\sqrt{m_i^2 + |\textbf{p}_i|^2 }, \textbf{p}_i \right)$ is the momentum of the mass eigenstate $\ket{\nu_i}$.
Consequently the probability amplitude of observing a neutrino with flavour $\ket{\nu_\beta}$  at the point $x$  is
\begin{eqnarray}\label{oscillation_amp}
\mathcal{A}_{\nu_\alpha \rightarrow \nu_\beta} (x) &=&\left< \nu_\beta | \nu_\alpha(x) \right> = U_{\beta j}^*  U_{\alpha i} e^{- i p_i x} \left<\nu_j| \nu_i\right> \nonumber \\
&=&  U_{\beta i}^* U_{\alpha i} e^{-i p_i x},
\end{eqnarray}
where we used $\left< \nu_i|\nu_j \right> = \delta_{ij}$. Thus the probability is given by
\begin{eqnarray}\label{oscillationssimpl}
P_{\nu_\alpha \rightarrow \nu_\beta} (x) &=& \left|\mathcal{A}_{\nu_\alpha \rightarrow \nu_\beta} (x)\right|^2  = \left|  U_{\beta i}^* U_{\alpha i} e^{-i p_i x} \right|^2=  |U_{\beta i }|^2 |U_{\alpha i}|^2\nonumber \\
&+& \sum_{i<j} 2\ \Re \left[ U_{\beta j} U_{\alpha j}^* U_{\beta i}^* U_{\alpha i} e^{i (p_j - p_i) x} \right],
\end{eqnarray}
that is, a constant term plus a periodic function of the space-time point, and is not vanishing also for $\beta \neq \alpha$. The previous formula can be simplified considering that neutrinos are very light and highly relativistic particles, for which is $\ell \simeq t$, where $\ell$ is the distance from the production point in natural units and $t$ is the time passed from the production. One thus has 
\begin{eqnarray}\label{osc_per}
\left(p_j - p_i\right) x &=& \left(E_j - E_i\right) t - \left(\left|\textbf{p}_j\right| - \left|\textbf{p}_i\right|\right) \ell \simeq \left[  \left(E_j - E_i\right) -  \left(\left|\textbf{p}_j\right| - \left|\textbf{p}_i\right|\right) \right] \ell\nonumber \\
& \simeq & \frac{(m_j^2 - m_i^2) \ell}{2E},
\end{eqnarray}
where we have used the approximations $\left|\textbf{p}_i\right| \simeq E_i -\frac{m_i^2}{2E_i}$ and  $m_i/2E_i \simeq m_i/2E$, where $E$ is the neutrino energy in the assumption of vanishing masses.

Thus if neutrinos are massive particles and the leptonic mixing matrix is non-trivial, the individual lepton flavours are not conserved (in the neutral sector), and the probability of observing a neutrino of a given flavour is a periodic function of the distance $\ell$ between the production and the detection points, i.e. the neutrino flavour oscillates. This is a relevant example of how the operator~(\ref{weinbergmass}) violates the conservation of the individual lepton flavour numbers. We will analyse further examples of lepton flavour violating (LFV) processes in the following sections; neutrino oscillations stand out among  the other LFV processes since the amount of flavour violation is amplified by the neutrino propagation, making the effect easily experimentally detectable.

\subsubsection{Quantum field theory treatment}

The above derived formula for the probability of the transition $\nu_\alpha \rightarrow \nu_\beta$ is suitable to describe a vast majority of the possible experimental configurations. It is however an approximate expression derived assuming several simplifications~\cite{Lipari:2001is}. First of all neutrinos are produced and detected in localised space-time regions: this implies that they cannot be described by plane waves, but a wave packet approach must be adopted. Moreover, because of their different masses, the neutrino mass eigenstates have different velocities,
\be\label{velocities}
v_i = \frac{p_i}{E_i} \simeq 1 - \frac{m_i^2}{2 E^2}.
\ee
A neutrino emitted in a certain process can be described by a wave packet with size $\delta_x$, determined by the resolution within which the neutrino production point is known. The wave packet can be approximated with a gaussian distribution with width $\delta_x$, and consequently the neutrino momentum distribution is a gaussian with width $\sigma_p = 1/\sigma_x$ and centred at $p_i$. If a neutrino flavour eigenstate produced at a certain point is the superposition of three different mass eigenstates, the corresponding wave packets move with different group velocities~(\ref{velocities}) and tend to separate over long distances. The superposition between the states $\nu_i$ and $\nu_k$ remains significant only if the separation between the centres of their wave packets is smaller than $\Delta x \sim 2 \sigma_x$, i.e. for values of $\ell$ smaller than
\be
\ell \lesssim \ell_{coh}^{ik} = \sqrt{2} \frac{2 \sigma_x}{|v_j - v_k|} \simeq \frac{4 \sqrt{2} E^2 \sigma_x}{|m_i^2 - m_k^2|}.
\ee

If a rigorous analysis is performed considering the localisation of the production and detection points, the decorrelation between different wave packets and adopting a fully field theoretical approach, a more general formula can be obtained~\cite{Kayser:1981ye,Zralek:1998rp}
\begin{eqnarray}
P_{\nu_\alpha \rightarrow \nu_\beta} (\ell) &=& |U_{\beta j}|^2 |U_{\alpha j}|^2  \\
&& + \sum_{i<j} 2\Re \left[  U_{\beta j} U_{\alpha j}^* U_{\beta i}^* U_{\alpha i} e^{i 2\pi \frac{\ell}{\ell_{osc}^{jk}}} \right] e^{- \left( \frac{\ell}{\ell_{coh}^{jk}} \right)^2} e^{- 2 \pi^2 \xi^2 \left(\frac{\sigma_x}{\ell_{osc}^{jk}}\right)^2},\nonumber 
\end{eqnarray}
where  $\ell_{osc}^{jk} = 4\pi E/|m_j^2 - m_k^2|$, $\sigma_x$ is the combination of both the spatial uncertainties relative to the production and the detection points and $\xi$ is a dimensionless parameter of  order unity depending from the production process. In all the experimental configurations characterised by $\ell \ll \ell_{coh}$ and $\sigma_x \ll \ell_{osc}$ the formula reduces to the simpler form of eq.~(\ref{oscillationssimpl}).

\subsection{Matter effects on neutrino oscillations}

The above treatment describes the oscillation of neutrinos in vacuum. In most of the interesting configurations, neutrinos actually propagates in a matter environment (Earth mantle, solar and supernovae interior, for instance). The surrounding matter affects the propagation of neutrinos via processes of forward coherent scattering that can strongly modify the transition probabilities with respect to the vacuum case. These effects can be described using an effective potential that is added to the neutrino vacuum Hamiltonian~\cite{Wolfenstein:1977ue,Barger:1980tf,Mikheev:1986gs,Mikheev:1986wj,Langacker:1986jv,Bilenky:1987ty,Krastev:1988yu,Kuo:1989qe}. The effective potential for the scattering mediated by a $Z^0$ boson reads~\cite{Lipari:2001is}
\begin{eqnarray}
V_{\nu_\mu e} &= V_{\nu_\tau e} = V_{\nu_e e}^{Z^0} =& - \frac{\sqrt{2}}{2} G_F N_e,\nonumber \\
V_{\nu_\mu p} &= V_{\nu_\tau p} = V_{\nu_e p} =&+  \frac{\sqrt{2}}{2} G_F N_p,\nonumber \\
V_{\nu_\mu n} &= V_{\nu_\tau n} = V_{\nu_e n} =&  - \frac{\sqrt{2}}{2} G_F N_n,
\end{eqnarray}
where $G_F$ is the Fermi electroweak constant and $N_{e,p,n}$ are the densities of electrons, protons and neutrons in the medium, respectively. Differently from the other flavours, the electron neutrinos receive a contribution from the forward coherent scattering mediated by the charged bosons $W^\pm$, resulting in the effective potential
\be
V_{\nu_e e} = V_{\nu_e e}^{Z^0} + V_{\nu_e e}^W = -\frac{\sqrt{2}}{2} G_F N_e + \sqrt{2} G_F N_e.
\ee
The relevant quantity driving the oscillations of neutrinos in matter is the difference of potential among different flavours:
\be\label{eq:vnue}
V\equiv V_{\nu_e} - V_{\nu_{\mu}} =V_{\nu_e} - V_{\nu_{\tau}} = + \sqrt{2} G_F N_e,
\ee
that is proportional to the electron density in the medium $N_e$.
Antineutrinos are subject to a similar treatment but their potentials are opposed in sign,
\be
V_{\overline{\nu}_\alpha} = - V_{\nu_\alpha}.
\ee
Given that in all the relevant configurations the medium is not charge-symmetric (it contains electrons but not antielectrons) the difference between the potentials for neutrinos and antineutrinos can enhance possible CP-violating effects in neutrino oscillations.

The quantum mechanical dynamics at work for neutrino oscillations in matter is the same as the one for the vacuum case, the difference relies on the fact that the effective potential modifies the neutrino mass eigenstates and eigenvectors, affecting the flavour evolution on neutrino propagation. The effective neutrino Hamiltonian in the flavour basis reads
\be\label{matterhamiltonian}
\mathcal{H}_\nu = U \left( \begin{array}{ccc} E_1 &0 &0 \\ 0&E_2 &0 \\ 0&0&E_3 \end{array} \right) U^\dagger +\left( \begin{array}{ccc} V_{\nu_e}&0&0 \\ 0&V_{\nu_\mu}&0\\0&0&V_{\nu_\tau} \end{array} \right),
\ee
where $U$ is the unitary matrix connecting the states in the flavour and the mass basis. It formally describes the mass to interaction basis change. Because of the tiny neutrino masses it is possible to approximate $t\simeq \ell$ ($\ell$ being the propagation distance) in the neutrino evolution, and  the Hamiltonian~(\ref{matterhamiltonian}) governs the Schr\"oedinger equation
\be
i \frac{d}{d\ell} \nu_\alpha = \mathcal{H}_\nu \nu_\alpha.
\ee
Notice that it is always possible to subtract a constant term from a Hamiltonian, and the matter effects can be described without loss of generality by the matrix $\mathcal{H}_{\nu}^{m} = \left( V,0,0 \right)$,  where $V$ is given in eq.~(\ref{eq:vnue}).

In most cases neutrinos propagate in non homogeneous media, resulting in the evolution equation
\begin{eqnarray}
i \frac{d}{d\ell} \left( \begin{array}{c} \nu_e \\ \nu_\mu \\ \nu_\tau \end{array} \right) &=& \left[ \mathcal{H}_0 + \mathcal{H}^m (\ell) \right]  \left( \begin{array}{c} \nu_e \\ \nu_\mu \\ \nu_\tau \end{array} \right) \\
& =& \left\{ \frac{1}{2E} U \left[\begin{array}{ccc} m_1^2 &0&0 \\ 0&m_2^2 &0 \\ 0&0&m_3^2 \end{array} \right] U^\dagger + \left[ \begin{array}{ccc} V(\ell) &0&0 \\ 0&0&0 \\0&0&0 \end{array} \right] \right\}  \left( \begin{array}{c} \nu_e \\ \nu_\mu \\ \nu_\tau \end{array} \right), \nonumber
\end{eqnarray}
where $V(\ell)= \sqrt{2} G_F N_e(\ell)$ and where the approximation $E_i\simeq E+m_i^2/2E$ has been used. This equation can be analytically or numerically solved once the matter density distribution is known, taking as initial condition a pure flavour eigenstate\cite{Bouchez:1986kb,Petcov:1987zj,Petcov:1988wv,Krastev:1988ci,Petcov:1989du,Krastev:1991hc,Abada:1992hw}.

\chapter{Signals from the BSM realm: neutrino masses, dark matter and baryon asymmetry of the Universe} 

Beside some theoretical caveats, the SM described in Section~\ref{SM_constraints} cannot account for at least three observational problems, that are: neutrino masses and mixing, the lack of a dark matter (DM) candidate and the observed baryon asymmetry of the Universe (BAU). These observations lead to call for extensions of the SM and to consider BSM realisations accounting for the aforementioned problems.

It is remarkable that each one of these observational results (and often more than one at the same time) can be successfully addressed by one of the simplest extensions of the SM: the addition of fermionic gauge singlets to its field content.
The simple motivation for such a minimal BSM framework is the fact that neutrinos are massive like the other elementary particles of the SM, which all have a right-handed component, so in order to have a Dirac mass term for neutrinos one has to consider the possibility of having right-handed neutrinos (i.e. sterile fermions) as well. If present, their phenomenological effect has to be probed using different experimental strategies.

The present chapter summarises the observational evidences for non-zero neutrino masses, for the existence of a non-baryonic matter component in the Universe and the determinations of the observed matter-antimatter asymmetry, together with the present experimental and cosmological status.

\section{Evidence of nonzero neutrino masses from oscillation experiments}\label{sec:nu_mass_exp}

Neutrinos are weakly interacting particles with very tiny masses, so light that it is not possible, in the limits of experimental uncertainties, to disentangle among the massive and massless hypothesis using direct kinematical methods. On the other hand the mass differences among different neutrino eigenstates are small enough such that the coherence between them is preserved in the propagation over ordinary lengthscales, making experimentally feasible to observe the phenomenon of flavour oscillations from known neutrino sources.

All the oscillation experiments follow the same guideline: they use a detector to measure the ratios of the different flavours composing a neutrino flux after it has propagated some distance, and compare the results with the flavour composition expected at the origin. If the flavour compositions differ then flavour numbers are not preserved in Nature. By varying the energy of the neutrino flux and the distance between the source and the detection points, it is possible to compare the predictions obtained using the transition probability~(\ref{oscillationssimpl}) with observations, determining if the violation of flavour numbers is due to neutrino oscillations or if different hypothesis need to be taken into account.

In this context the leptonic mixing matrix is assumed to be unitary and is usually parametrised as~\cite{Agashe:2014kda} 
\begin{eqnarray}\label{upmns}
U_\text{PMNS} &=& \left( \begin{array}{ccc} 1&0&0 \\ 0 & c_{23} & s_{23} \\ 0&- s_{23}& c_{23} \end{array} \right) \left( \begin{array}{ccc}c_{13} & 0&s_{13} e^{-i\delta} \\ 0&1&0 \\ - s_{13} e^{i\delta} &0&c_{13} \end{array} \right) \left( \begin{array}{ccc} c_{12} & s_{12} & 0 \\ - s_{12} & c_{12} & 0 \\ 0&0&1 \end{array} \right) \nonumber \\
&& \times \mbox{diag}(1,\  e^{i \frac{\alpha_{21}}{2}},\  e^{i \frac{\alpha_{31}}{2}} ) \nonumber \\
&=& \left( \begin{array}{ccc} c_{12} c_{13} & s_{12} c_{13} & s_{13} e^{-i \delta} \\ -s_{12}c_{23} - c_{12}s_{13}s_{23}e^{i\delta} & c_{12} c_{23}-s_{12}s_{13}s_{23}e^{i \delta} & c_{13} s_{23} \\ s_{12}s_{23}-c_{12}s_{13}c_{23}e^{i\delta} & -c_{12}s_{23}-s_{12}s_{13}c_{23}e^{i\delta} & c_{13}c_{23} \end{array} \right)\nonumber \\
&& \times \mbox{diag}(1,\  e^{i \frac{\alpha_{21}}{2}},\  e^{i \frac{\alpha_{31}}{2}} ),
\end{eqnarray}
where $c_{ij}\equiv \cos \theta_{ij}$ and $s_{ij}\equiv \sin \theta_{ij}$. The CP-violating imaginary part of the mixing matrix is parametrised by the $\delta,\alpha_{ij}$ phases. The phases $\alpha_{ij}$ are related to the complex phases of the mass matrix eigenvalues, cf.~(\ref{complex_masses}), and are physical degrees only if neutrinos are Majorana particles, while they can be rotated away in the Dirac case. Notice however that they cancel out in the oscillation formula~(\ref{oscillation_amp}) and cannot thus be probed in neutrino oscillation experiments, which do not distinguish between the Dirac or Majorana hypothesis for neutrinos. The label PMNS stands for Pontecorvo-Maki-Nakagawa-Sakata after~\cite{Pontecorvo:1967fh,Maki:1962mu}.

Recall that oscillation physics only depends on the neutrino mass squared differences and not on the absolute neutrino mass scale, and the neutrino oscillation experiments are sensitive to the parameters $\Delta m_{ij}^2=m_i^2-m_j^2$  where $m_i,\ i=1,2,3$ are the three mass eigenvalues (see eqs.~(\ref{oscillationssimpl}, \ref{osc_per})). Given that with 3 different masses there are 2 independent mass squared differences, $\Delta m_{12}^2$ and $\Delta m_{13}^2$, in the absence of information on the absolute mass scale the oscillation data can be equivalently explained by two sets of solutions, characterised by the 2 possible orderings of the known mass differences. For the sake of definiteness, the following convention is usually adopted to label the neutrino mass eigenstates
\begin{itemize}
 \item $\nu_3$ is the mass eigenstate with the largest mass difference in modulus, \\$\left|\Delta m_{3 i}^2\right|>\left|\Delta m_{21}^2\right|$ for $i=1,2$;
\item $\nu_{1,2}$ are the other two mass eigenstates, $\nu_1$ being the lightest, $\Delta m_{21}^2$>0.
\end{itemize}
Accordingly, the free parameters to be determined from oscillation data are the values $\Delta m_{21}^2$, $\left|\Delta m_{3 1}^2\right|$ and the sign of $\Delta m_{3 1}^2$. The solution with $\Delta m_{3 1}^2>0$ is labeled ``Normal Hierarchy'' (NH), the one with $\Delta m_{3 1}^2<0$ ``Inverted Hierarchy'' (IH).

Notice that under $CP$ conjugation the leptonic mixing matrix is replaced by its complex conjugate, $U\rightarrow U^*$, and the amount of $CP$-violation in the leptonic sector is determined by the imaginary part of $U$. From the oscillation probability, eq.~(\ref{oscillationssimpl}), it follows that the relevant $CP$-violating quantities are the coefficients
\bee
J_{ij}^{\alpha \beta} = - \text{Im} \left[U_{\alpha i}U_{\alpha j}^* U_{\beta i}^* U_{\beta j}\right],
\eee
which are zero for $\alpha=\beta$ or $i=j$. It can be shown that, for a $3\times 3$ unitary mixing matrix, the values of $J_{ij}^{\alpha \beta}$ for the 9 non-zero combinations of indices are equal, and the amount of $CP$ violation is determined by a single parameter, the Jarlskog invariant~\cite{Jarlskog:1985ht}. It can be parametrised, using~(\ref{upmns}), as
\bee
J=J_{12}^{e\mu} = c_{13}^2 s_{13}s_{12}c_{12}s_{23}c_{23}\sin \delta.
\eee
Thus $CP$-violating interactions are only possible if all the three mixing angles and the phase $\delta$ are different form zero.

We recall in the following the main investigated neutrino sources and the corresponding experiments~\cite{Lipari:2001is}, and summarise the results obtained from global fits to neutrino data.

\subsection{Atmospheric neutrinos}

Cosmic rays~\cite{Blasi:2013rva} are charged particles and nuclei of extraterrestrial origin, reaching the Earth with a rate of approximatively 1 particle/(cm$^2$ sec sr). They are mainly composed by protons and charged nuclei, and produce a shower of particles when they interact with the upper layers of the atmosphere
\be
p + A_{air} \rightarrow p, n, \pi^\pm, \pi^0, K^{\pm}, \dots.
\ee
The atmospheric neutrino flux is mainly originated by the chain of pion decays~\cite{Kajita:2000mr}
\be\label{pion_decay_chain}
\begin{array}{rcl}
\pi^+ &\rightarrow & \mu^+ +  \nu_\mu \\
&& \downarrow\\
&& e^+ +\nu_e + \overline{\nu}_\mu,
\end{array}
\ee
and charge conjugate channels

The overall cosmic radiation originates from different astrophysical sources responsible for the primary cosmic rays flux, plus secondary particles produced form the interactions of the primary flux with the interstellar gas, and its precise determination is not an easy task. Nonetheless it is possible to derive robust predictions relative to the neutrino atmospheric flux. Firstly, since cosmic rays are trapped in the galactic magnetic field for millions of years, erasing any spatial dependence from their sources, the resulting flux is isotropic and uniform in time. The produced neutrino flux is expected to be uniform in time and up-down symmetric with respect to the Earth surface, due to the Earth's sphericity. Moreover, as a consequence of the production chain~(\ref{pion_decay_chain}), the muon neutrino flux, $\Phi(\nu_\mu)$, is expected to be as twice as the electron neutrino one, $\Phi(\nu_e)$, in the absence of neutrino oscillations. Notice however that at higher energies relativistic effects make the muons travelling longer in the atmosphere before decaying, and thus the ratio $\Phi(\nu_\mu)/\Phi(\nu_e)$ increases with neutrino energies.

The properties of the atmospheric neutrino flux offer an ideal ground to test the neutrino oscillation hypothesis because atmospheric neutrinos travel for very different distances from the production point to the detector depending on their incoming angle, ranging from tens of kilometres for the ones produced at the zenith until ~6000 km for the ones passing through the Earth.

Notable results in the study of atmospheric neutrino oscillations came from Frejus~\cite{Daum:1994bf}, Kamiokande and Super-Kamiokande (SK)~\cite{Kajita:1998bw,Wendell:2010md}, Nusex~\cite{Aglietta:1988be}, Macro~\cite{Ambrosio:2004ig} and Soudan-2~\cite{Kafka:2006qy}. The SK collaboration in particular firstly  claimed evidence for atmospheric neutrino oscillations in 1998. A running research program is currently carried out by SK~\cite{Wendell:2014dka}, Hyper-Kamiokande (HK)~\cite{Kearns:2013lea}, MINOS~\cite{Adamson:2014vgd}, ICARUS~\cite{Arneodo:2001tx}, ANTARES~\cite{AdrianMartinez:2012ph}, IceCube~\cite{Aartsen:2013jza} and Baikal-GVD~\cite{Avrorin:2013sla}.

The atmospheric neutrino experiments exhibit a dominant dependence on the parameters $\theta_{23}$ and $\Delta m_{31}^2$, and a subdominant dependence on $\theta_{13}$ and $\delta$.

\subsection{Solar neutrinos}

\begin{figure}[htb]
\begin{center}
 \includegraphics[width=0.7\textwidth]{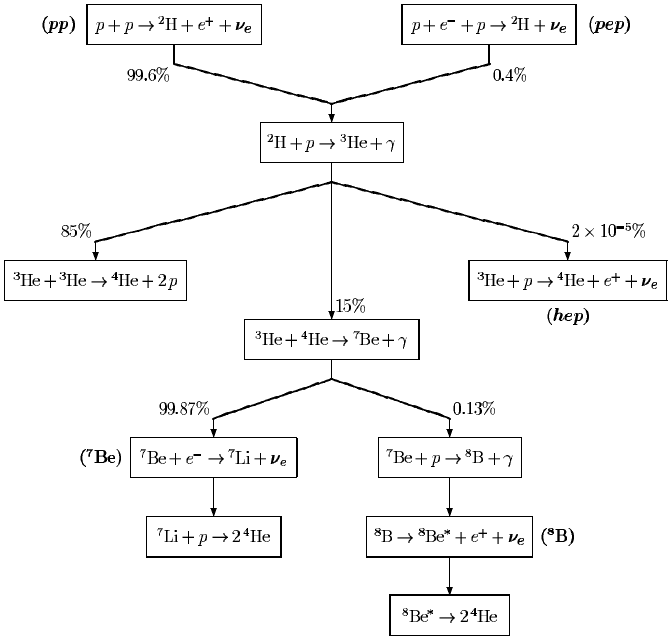}
\end{center}
\caption{Intermediate channels for the solar nuclear fusion reaction (\ref{eq:nu_solar}). The probability for each branch calculated in the standard solar model, as well as the different neutrino sources (in parenthesis) are reported. Figure taken from~\cite{Lipari:2001is}.}
\label{fig:sol_chann}
\end{figure}

Solar neutrinos are generated in the nuclear processes that release the thermic energy in the Sun, which can be summarised in the reaction
\be\label{eq:nu_solar}
4 p + 2 e^- \rightarrow {}^4\text{He} + 2 \nu_e.
\ee
The above nuclear fusion process can actually occur in different intermediate channels, reported in Fig.~\ref{fig:sol_chann}, and the solar neutrino spectrum results from the superposition of the different spectra~\cite{clayton1983principles}.
The $Q$ value of the reaction, defined as the difference between the  initial and the final state masses, is (neglecting the neutrino masses)
\be 
Q=4 m_p + 2 m_e - m_{He} = 26.73 \mbox{MeV},
\ee
and is carried away by the final state particles in the form of kinetic energy. While the Helium nuclei contribute to the thermal energy of the Sun, the neutrinos easily escape generating a flux which can be estimated on the Earth to be~\cite{bahcall1989neutrino}
\be
\Phi_\nu \simeq \frac{1}{4\pi d_\odot^2} \frac{2 L_\odot}{(Q-\left< E_\nu \right>)} \sim 6 \times 10^{10} \mbox{ cm}^{-2} s^{-1},
\ee
where $L_\odot = 3.842\times 10^{33}$erg is the solar luminosity, $\left< E_\nu \right> \simeq 0.3$ MeV  is the average neutrino energy at the end of a cycle (\ref{eq:nu_solar}) and $d_\odot \simeq 1.495 \times 10^{13}$ cm is the Sun-Earth distance. 

\begin{figure} [htb]
\begin{center}
\includegraphics[angle=270,width=0.7\textwidth]{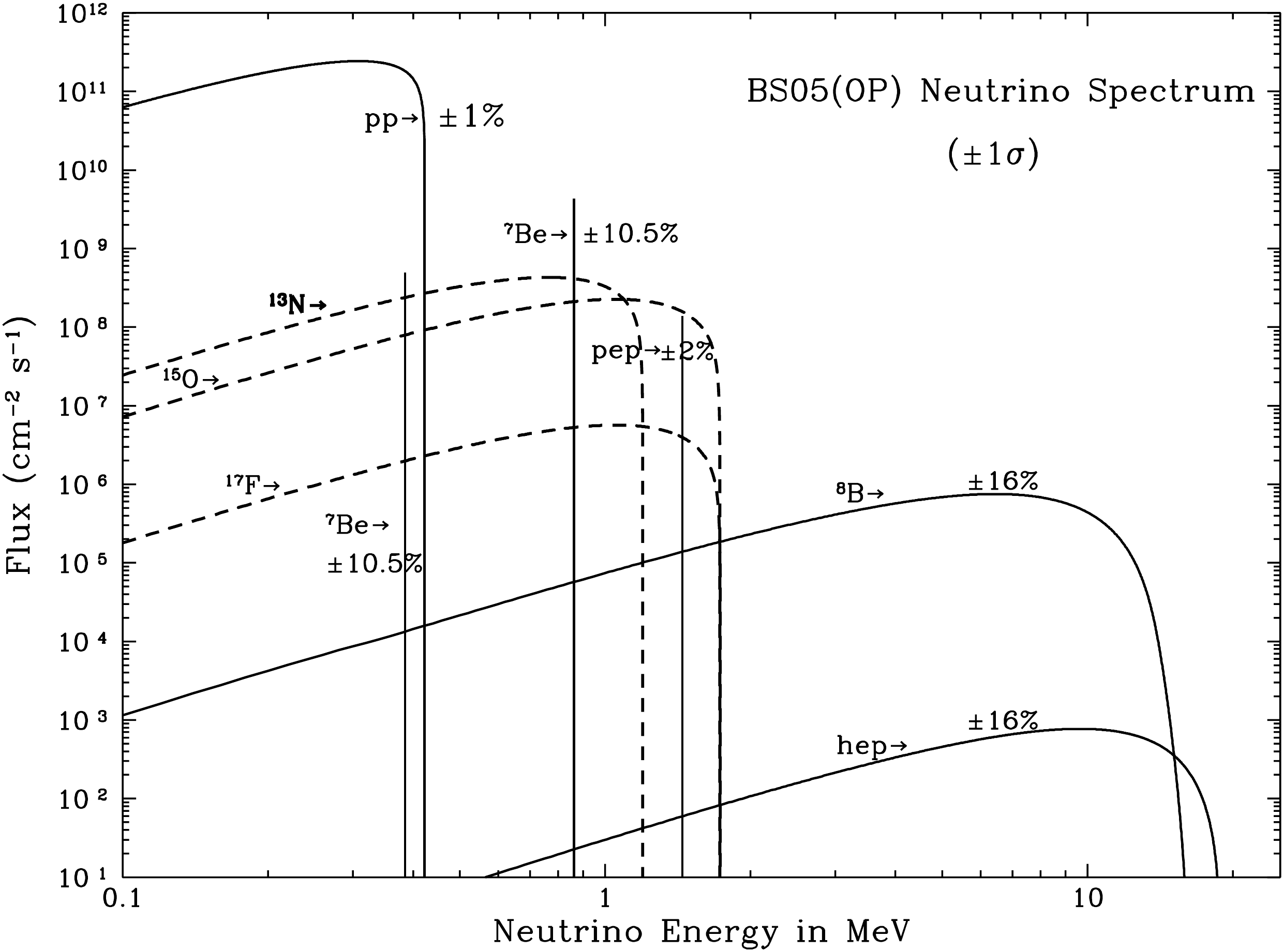}
\end{center}
\caption{Energy spectrum of the solar neutrinos according to the solar model BS05. Figure taken from~\cite{Bahcall:2004pz}.\label{fig:energyspectrum}}
\end{figure}

In order to test the neutrino oscillation hypothesis it is important to know the energy spectrum of the neutrino flux reaching the Earth expected in the absence of oscillations. That requires a precise knowledge of the internal structure of the sun and of the underlying nuclear reactions, provided by the \emph{Standard Solar Model} (SSM)~\cite{Bahcall:2000nu}. In this model the Sun is approximated by a sphere in hydrodynamical equilibrium, with the gravitational attraction balanced by the thermal pressure generated by nuclear reactions. The internal parameters of the Sun such as the density, the composition, the temperature profiles and the rate of nuclear reactions are computed starting from an initial configuration, with parameters determined by the solar mass and by an initial composition taken to match the observed abundances of elements on the Sun's surface, but with the helium fraction as a free parameter. The system is then evolved and the current neutrino flux is computed. The solar neutrino energy spectrum predicted in the BS05 solar model is reported in Fig.~\ref{fig:energyspectrum}. The most important information on solar neutrino oscillations comes from the most energetic ${}^8$B and \emph{hep} channels, while only few experiments have a sufficiently low threshold to be sensitive to the other sources.

Notable results in the measurement of the intensity and flavour composition of the solar neutrino flux were obtained by Chlorine~\cite{Cleveland:1998nv}, Gallex/GNO~\cite{Hampel:1998xg,Kaether:2010ag}, SAGE~\cite{Abdurashitov:1999zd,Abdurashitov:2009tn}, Kamiokande~\cite{Hirata:1991ub}, Super-Kamiokande~\cite{Fukuda:2002pe,Abe:2010hy}, SNO~\cite{Ahmad:2001an,Aharmim:2011vm} and Borexino~\cite{Bellini:2008mr,Bellini:2011rx}. The Chlorine experiment at the Homestake Gold Mine was at the origin of the so called solar neutrino problem, detecting approximately one third of the solar neutrino flux expected from the SSM. The problem was successfully solved in the context of neutrino oscillations, taking into account the important matter effects due the propagation of neutrinos in the solar environment~\cite{Mikheev:1986gs,Mikheev:1986wj}. 

An ongoing research program on solar neutrinos is currently pursued by SK~\cite{Renshaw:2014awa}, HK, Borexino~\cite{Smirnov:2014hbh}, ICARUS and KamLAND~\cite{Gando:2014wjd}.

Solar neutrino experiments are mainly sensitive to the parameters $\theta_{12}$ and $\Delta m_{21}^2$, with a subdominant dependence on $\theta_{13}$.

\subsection{Reactor neutrinos}

Commercial nuclear plants are conceived to convert the energy released in nuclear fission reactions into electric energy. Their fuel is usually constituted by ${}^{238}$U enriched in ${}^{235}$U, which is a fissile isotope. The fission reaction can be summarised as
\be\label{fission}
n + {}^{235}\text{U} \rightarrow X_1 + X_2 + 2n,
\ee
where $X_{1,2}$ are the fragments of the ${}^{235}$U, which are unstable because of their excess in neutrons and which reach stability by a succession of beta decays, with an average of 6 per reaction, emitting one electron anti-neutrino in each of them. The neutrons emitted in the nuclear fission can be captured by other ${}^{235}$U nuclei originating an analogous fission process and sustaining a chain reaction. The overall neutrino flux can be estimated knowing that each fission reaction~(\ref{fission}) releases 204 MeV,
\bee
\Phi(\overline{\nu_e}) \simeq 6 \left(\frac{\text{Plant power}}{204 \text{ MeV}}\right).
\eee
Once its energy distribution is known, the intense flux of reactor antineutrinos can be used to test the oscillation hypothesis~\cite{Bemporad:2001qy}. This determination requires a precise knowledge of the statistical distribution of beta decays following the reaction~(\ref{fission}) which depends on the reactor structure and fuel composition, with the last parameter slowly evolving in time. Notice for instance that only a fraction of $\sim 25\%$ of the total flux has an energy above the threshold for detection via inverse neutron $\beta$ decay ($\sim 1.8$ MeV).

Important results in the study of reactor antineutrino fluxes have been obtained by Chooz~\cite{Apollonio:1999ae}, Palo Verde~\cite{Boehm:2001ik}, KamLAND~\cite{Araki:2004mb,Gando:2010aa}, Double Chooz~\cite{Abe:2012tg}, Daya Bay~\cite{An:2012eh} and RENO~\cite{Ahn:2012nd}. Notably, the Daya Bay collaboration determined in 2012 a non-zero value for the neutrino mixing angle $\theta_{13}$ (the so-called Chooz-angle)  at a 5.2 $\sigma$ level; the realised non-zero value of all the 3 angles in the 3 neutrino mixing paradigm is a necessary condition for CP-violating effects in neutrino oscillations to take place. 

The results from the above listed collaborations agree well among themselves within the 3-flavour oscillation paradigm for baselines longer than $\sim 100$ m. Agreement also held between shorter baselines, but in 2011 a reevaluation of the neutrino fluxes expected from nuclear reactors led to an expected flux larger of about 3\% than the previously quoted results~\cite{Mueller:2011nm,Huber:2011wv}. When reanalysed in terms of this new predictions, the observed reactor antineutrino fluxes all exhibit a short-baseline deficit with respect to the expected values. If interpreted as an effect of neutrino oscillations, this deficit requires a further mass splitting larger than the solar and atmospheric ones~\cite{Mention:2011rk}, requiring the existence of a fourth (light) neutrino state.

An ongoing research program focused on reactor neutrinos is currently pursued by KamLAND~\cite{Gando:2013nba}, Double Chooz~\cite{Abe:2014bwa}, Daya Bay~\cite{Zhan:2015aha}, RENO~\cite{Seon-HeeSeofortheRENO:2014jza} and JUNO~\cite{An:2015jdp}.

Medium baseline reactor experiments are sensitive to $\Delta m_{31}^2$ and $\theta_{13}$, while long baseline experiments can probe the ``solar'' parameters, $\theta_{12}$ and $\Delta m_{21}^2$, with a subdominant dependence on $\theta_{13}$.

\subsection{Accelerator neutrinos}

All the above described neutrino sources are ``just-there'' sources, meaning that they produce a neutrino flux that is not expressly designed for experimental searches. On the contrary, accelerator neutrino fluxes are designed for investigation purposes. This means that neutrinos from accelerator come in intense, focused and highly energetic beams whose energy and flavour composition can be modulated in the limits of the experimental technical constraints.

Neutrino beams~\cite{Kopp:2006ky} are produced in the decays of charged mesons obtained by hitting a nuclear target with an highly energetic proton beam. The most commonly used are muon neutrino beams obtained from the decays
\begin{eqnarray}
\pi^\pm & \rightarrow & \mu \ \nu\ \phantom{\nu} \mbox{ (B.R. } \sim 100  \%),\nonumber \\
K^\pm & \rightarrow & \mu \ \nu\ \phantom{\nu} \mbox{ (B.R. } = 63.4  \%),\nonumber \\
K_L^\pm & \rightarrow &\pi\ \mu \ \nu \mbox{ (B.R. } = 27.2  \%).
\end{eqnarray}
The charged decay products are stopped by some shield, while neutrinos easily propagate towards the detector. The composition of the flux (neutrinos or antineutrinos) can be selected by focusing the mesons of desired charge through magnetic horns. To obtain different neutrino flavours other techniques are used, notably beam dumps. The detectors  can be located at distances ranging form some tens of meters from the neutrino source (short baseline experiments) up to thousands of kilometres (long baseline experiments).

Given that neutrino beams are usually conceived to observe the disappearance of muon neutrinos and the appearance of electron/tau neutrinos due to oscillations, the most  important challenge in such experiments is to monitor the beam contamination from electron neutrinos generated in the on-flight decays of muons.\footnote{There have been proposals to produce enhanced electron neutrino beams from the channel $K_L \rightarrow \pi e \nu_e\ (\text{B.R.}\sim 38.8\%)$~\cite{Mori:1978nr,Camerini:1980zf,Camerini:1980vd,Ong:1985xv} but they have not been realised.}

Remarkable results from the study of accelerator neutrinos came from NOMAD~\cite{Astier:2001yj}, K2K~\cite{Ahn:2002up,Ahn:2006zza} , CHORUS~\cite{Eskut:1997ar,Ludovici:2001tn}, LSND~\cite{Athanassopoulos:1997pv,Aguilar:2001ty},  MiniBooNE~\cite{AguilarArevalo:2008qa}, MINOS~\cite{Michael:2006rx,Adamson:2013whj,Adamson:2013ue}, T2K~\cite{Abe:2013hdq,Abe:2014ugx} and OPERA~\cite{Agafonova:2015jxn}, the last collaboration having successfully observed the oscillation $\nu_\mu \rightarrow \nu_\tau$. 

Most of the above results fit the 3-flavour neutrino oscillation paradigm, except for the ones of LSND and MiniBooNe. The LSND collaboration (characterised by a baseline of 30 m)  reported an excess of electron neutrino events incompatible with the known mass differences obtained from the analysis of solar and atmospheric neutrinos.  MiniBooNe was designed to probe the same mass splitting range via neutrino oscillations; the collaboration did not observe any excess in a first run~\cite{AguilarArevalo:2007it}, while subsequent results reported an anomaly compatible with the LSND results~\cite{AguilarArevalo:2008rc,AguilarArevalo:2010wv}.

An ongoing long baseline research program on accelerator neutrinos is currently carried out by ICARUS, OPERA, MINOS, NO$\nu$A~\cite{Jediny:2014lda}, T2K and LBNF/DUNE~\cite{Adams:2013qkq}. On the short-baseline side there are MINER$\nu$A~\cite{Aliaga:2013uqz} and MicroBooNE~\cite{Soderberg:2009rz}. The last one represents the evolution of the LSND and MiniBooNe experiments, and is expected to probe the related neutrino anomaly.

Long-baseline accelerator experiments looking for $\nu_\mu$ disappearance can probe the ``atmospheric'' parameters, $\theta_{23}$ and $\Delta m_{31}^2$, while LBL $\nu_e$ appearance experiments are sensitive to $\theta_{13}$,  $\theta_{23}$ and $\delta$.

\subsection{Global results}\label{sec:global_fit}

Apart from accelerator (LSND and MiniBoone) and reactor short-baseline anomalies, the combination of the results obtained on solar, atmospheric, reactor and accelerator neutrinos is very well accommodated by the oscillation hypothesis in a 3-flavour framework~\cite{Capozzi:2013csa,Forero:2014bxa,Gonzalez-Garcia:2014bfa} and with 3 different neutrino masses.

\begin{table}\centering
  \begin{footnotesize}
    \begin{tabular}{|l|cc|cc|}
      \hline
      & \multicolumn{2}{c|}{Normal Ordering}
      & \multicolumn{2}{c|}{Inverted Ordering}
      \\
      \hline
      & bfp $\pm 1\sigma$ & $3\sigma$ range
      & bfp $\pm 1\sigma$ & $3\sigma$ range
      \\
      \hline
      \rule{0pt}{4mm}\ignorespaces
      $\sin^2\theta_{12}$
      & $0.304_{-0.012}^{+0.013}$ & $\left[0.270 , 0.344\right]$
      & $0.304_{-0.012}^{+0.013}$ & $\left[0.270 , 0.344\right]$
      \\[1mm]
      $\theta_{12}/^\circ$
      & $33.48_{-0.75}^{+0.78}$ & $\left[31.29 , 35.91\right]$
      & $33.48_{-0.75}^{+0.78}$ & $\left[31.29 , 35.91\right]$
      \\[3mm]
      $\sin^2\theta_{23}$
      & $0.452_{-0.028}^{+0.052}$ & $\left[0.382 , 0.643\right]$
      & $0.579_{-0.037}^{+0.025}$ & $\left[0.389 , 0.644\right]$
      \\[1mm]
      $\theta_{23}/^\circ$
      & $42.3_{-1.6}^{+3.0}$ & $\left[38.2 , 53.3\right]$
      & $49.5_{-2.2}^{+1.5}$ & $\left[38.6 , 53.3\right]$
      \\[3mm]
      $\sin^2\theta_{13}$
      & $0.0218_{-0.0010}^{+0.0010}$ & $\left[0.0186 , 0.0250\right]$
      & $0.0219_{-0.0010}^{+0.0011}$ & $\left[0.0188 , 0.0251\right]$
      \\[1mm]
      $\theta_{13}/^\circ$
      & $8.50_{-0.21}^{+0.20}$ & $\left[7.85 , 9.10\right]$
      & $8.51_{-0.21}^{+0.20}$ & $\left[7.87 , 9.11\right]$
      \\[3mm]
      $\delta_\text{CP}/^\circ$
      & $306_{-70}^{+39}$ & $\left[0 , 360\right]$
      & $254_{-62}^{+63}$ & $\left[0 , 360\right]$
      \\[3mm]
      $\frac{{\Delta m^2}_{21}}{10^{-5}~{\text{eV}^2}}$
      & $7.50_{-0.17}^{+0.19}$ & $\left[7.02 , 8.09\right]$
      & $7.50_{-0.17}^{+0.19}$ & $\left[7.02 , 8.09\right]$
      \\[3mm]
      $\frac{{\Delta m^2}_{3\ell}}{10^{-3}~{\text{eV}^2}}$
      & $+2.457_{-0.047}^{+0.047}$ & $\left[+2.317 , +2.607\right]$
      & $-2.449_{-0.047}^{+0.048}$ & $\left[-2.590 , -2.307\right]$
      \\[3mm]
      \hline
    \end{tabular}
  \end{footnotesize}
  \caption{Results from a global fit to the three-flavour oscillation parameters from the NuFIT collaboration~\cite{Gonzalez-Garcia:2014bfa}. The label for the atmospheric mass difference reads ${\Delta m}_{3\ell} \equiv
    {\Delta m}_{31}$ for NH and ${\Delta m}_{3\ell} \equiv {\Delta m}_{32}$ for IH. The fit is performed by leaving the reactor fluxes as free parameters.}
  \label{tab:nufit}
\end{table}

The current experimental parameters determined by the NuFIT collaboration~\cite{Gonzalez-Garcia:2014bfa} are summarised in Table~\ref{tab:nufit}, where the best fit points together with the 1$\sigma$ and 3$\sigma$ ranges are reported for the normal and inverted hierarchies. The only parameter completely undetermined at the $3\sigma$ level is the CP-violating phase $\delta_\text{CP}$; notice however that indications towards the value $\delta_\text{CP}\sim  \frac{3}{2} \pi$ appear from the global fit, although with a low statistical significance at current time.

\section{Limits on neutrino masses}

The oscillation experiments have successfully established the massive character of neutrinos, but they are only sensitive to the neutrino mass squared differences and cannot provide information on the absolute mass scale. Regarding this quantity only upper bounds exist, which are summarised in the present section.

\subsection{End-point searches}\label{sec:end_point}
Since the kinematics of a three-body decay depends on the masses of the involved particles, any $\beta$ decay process can be used in principle to infer the value of the absolute neutrino mass scale. Given the decay $N_i \rightarrow N_f+ e+ \nu_e$, where $N_{i,f}$ are the initial and final state nuclei, the  maximum energy carried away by the electron is determined by the $Q$ value of the reaction $Q_\beta=m_i - m_f - m_e - m_\nu$ for the case in which the (massive) neutrino is produced at rest. Neglecting the recoil of the nucleus we have
\bee
E_e^{\text{max}} = Q_\beta - m_{\nu_e}.
\eee
Considering that the electron neutrino $\nu_e$ is a linear superposition of neutrino mass eigenstates, the differential distribution of electron energies as a function of neutrino masses is given by~\cite{Osipowicz:2001sq}
\bee\label{beta_distr}
\frac{\de N}{\de E_e} &=& C \times F(Z,E_e)\ p\ E_e (Q_\beta -  E_e)\non 
&&\sum_{i} \left|U_{ei}\right|^2 \sqrt{(Q_\beta-E_e)^2-m_i^2}\Theta\left[Q_\beta-E_e-m_i \right], 
\eee
where $C$ is a constant related to the weak current structure and to the nuclear matrix element, $F(Z,E)$ is the Fermi function which accounts for the Coulomb interaction between the electron and nucleus and $p$ is the electron momentum. Notice that both $C$ and $F$ are independent from the neutrino mass. The step function $\Theta$ guarantees that only kinetically accessible final states account for the spectrum. Equation~(\ref{beta_distr}) has two important phenomenological consequences. The first one is that the endpoint of the reaction is shifted towards the value $Q_\beta - m_{\nu_l}$ with respect to the massless case, where $\nu_l$ is the lightest neutrino state ($\nu_1$ for the normal hierarchy case and $\nu_3$ for the inverted one). The second consequence is the appearance of kinks in the electron spectrum at energies $E_e^i \sim Q_\beta - m_i$ with size related to the mixing element $|U_{ei}|^2$. If the experimental energy resolution is sensitive to the size of the kinks, the experiment can determine individual neutrino masses; in the opposite case the incertitude in the determination of $E_e$ justifies the expansion of the distribution in eq.~(\ref{beta_distr}) in the small parameters $m_i/(Q_\beta-E_e)$. In this regime, and under the assumption of a unitary leptonic mixing matrix, the neutrino mass dependence in eq.~(\ref{beta_distr}) can be approximated  
\bee
\sum_{i} \left|U_{ei}\right|^2  \sqrt{(Q_\beta-E_e)^2-m_i^2}\simeq   \sqrt{(Q_\beta-E_e)^2-\sum_{i} \left|U_{ei}\right|^2 m_i^2}, 
\eee
and the experiment is sensitive to the effective electron neutrino mass
\bee\label{eq:mnueff}
m_{\nu_e}^{eff} = \sqrt{\sum_{i} \left|U_{ei}\right|^2 m_i^2}.
\eee
From eq.~(\ref{beta_distr}) the number of $\beta$-decays near the endpoint of the electron spectrum is proportional to $Q_\beta^{-3}$, thus the most interesting reactions are characterised by a low $Q_\beta$ value. For this reason tritium (${}^3$H) is a suitable isotope for the study of the $\beta$-decay endpoint via the reaction
\be
{}^3\text{H} \rightarrow {}^3\text{He}^+ +e^- + \overline{\nu_e},
\ee
characterised by a $Q$ value of 18.6 keV. Moreover ${}^3$H has a simple single-electron configuration that makes it simpler the computation of the Fermi function. The current upper bound on the effective electron neutrino mass has been established by the Mainz and Troitsk experiments~\cite{Kraus:2004zw,Aseev:2011dq}
\bee\label{nueff_bound}
m_{\nu_e}^{eff} < 2.05 \text{ eV at 95\% C.L.} 
\eee

These results are expected to be improved by the next generation of experiments: KATRIN~\cite{Osipowicz:2001sq} and MARE~\cite{Nucciotti:2010tx} with a planned sensitivity of 0.35 eV (3-years running) and 0.2 eV, respectively.

\subsection{$m_{\nu_\mu}^{eff}$ and $m_{\nu_\tau}^{eff}$ mass limits}
Analogously to the definition (\ref{eq:mnueff}), it is possible to define an effective mass for the other neutrino flavours, $m_{\nu_\mu}^{eff}$ and $m_{\nu_\tau}^{eff}$, by the replacement $U_{ei} \rightarrow U_{\mu i}$ and $U_{\tau i}$, respectively~\cite{Agashe:2014kda}.

A limit on the effective mass $m_{\nu_\mu}^{eff}$ can be extracted from the pion decay
\bee
\pi^+ \rightarrow \mu^+ + \nu_\mu,
\eee
by measuring the muon energy, since the kinematics of the process gives
\bee
m_{\nu_{\mu}}^2 = m_\pi^2 + m_\mu^2 - 2 m_\pi E_\mu.
\eee
The current bound on the effective muon neutrino mass is~\cite{Assamagan:1995wb}
\be\label{eq:mmunueff}
m_{\nu_\mu}^{eff} < 170 \mbox{ keV at 90 \% C.L.}
\ee
A limit on $m_{\nu_\tau}^{eff}$ can be obtained by measuring the missing energy in the decays
\bee
\tau^- &\rightarrow& 2 \pi^- + \pi^+ + \nu_\tau,\non 
 \tau^- &\rightarrow& 3 \pi^- + 2\pi^+ + \nu_\tau,
\eee
resulting in the upper bound~\cite{Barate:1997zg}
\be\label{eq:mtaueff}
m_{\nu_\tau}^{eff} < 18.2 \mbox{ MeV at 95 \% C.L.}
\ee

Notice that, in the three-flavour paradigm, the above referred quantities are constrained to be orders of magnitude smaller than the bounds (\ref{eq:mmunueff}, \ref{eq:mtaueff}), due to the combination of the values of the mixing matrix elements, Table~\ref{tab:nufit}, and the upper bound on the neutrino mass scale (\ref{nueff_bound}).

\subsection{Neutrinoless double beta decay}\label{sec:0betanunu}
\begin{figure}[htb]
\begin{center}
 \includegraphics[width=0.6\textwidth]{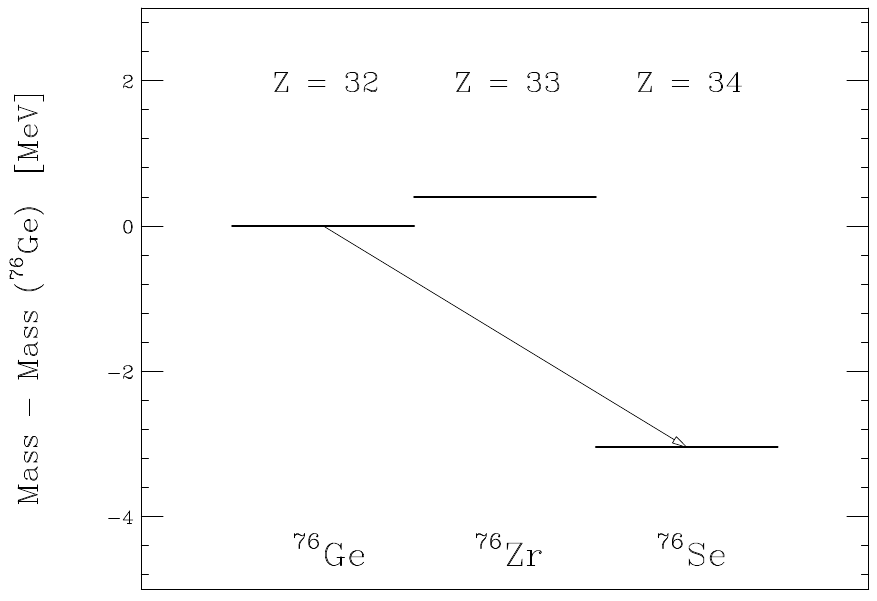}
\end{center}
\caption{Energy levels for the $A=76$ nuclei. Figure taken from~\cite{Lipari:2001is}.}
\label{dblevels}
\end{figure}

The double beta ($2\beta$) decay is a second order weak process characterised by the transition
\be\label{dbd}
\mathcal{N}(A,Z) \rightarrow \mathcal{N}(A,Z+2) + 2 e^- + 2 \overline{\nu_e}.
\ee
Being a second order process in the weak coupling, this process is relevant when the single beta decay is kinematically forbidden, as is the case for instance of the nuclei ${}^{48}$Ca, ${}^{76}$Ge, ${}^{82}$Se, ${}^{96}$Zr, ${}^{100}$Mo, ${}^{116}$Cd, ${}^{130}$Te, ${}^{136}$Xe, ${}^{150}$Nd~\cite{Agashe:2014kda}, see Fig.~\ref{dblevels} for the $A=76$ case.

If neutrinos are Majorana particles they can mediate a variation of the $2\beta$-decay process, the neutrinoless double beta ($0\nu2\beta$) decay process~\cite{Furry:1939qr}
\be\label{0nudbd}
\mathcal{N}(A,Z) \rightarrow \mathcal{N}(A,Z+2) + 2 e^-.
\ee
This process violates the conservation of the total lepton number by two units and is characterised by a clear experimental signature, since the two final electrons carry away the total Q value of the reaction, resulting in a peak over the continuous $2\beta$-decay background.

The $0\nu 2\beta$ decay process requires a chirality flip and a particle-antiparticle identification, thus its amplitude is proportional to the Majorana neutrino masses.  Other mediators than Majorana neutrinos can in principle contribute to the amplitude, however it has been demonstrated that if $0\nu 2\beta$-decay is possible then the same underlying physics generates a Majorana mass term for neutrinos~\cite{Schechter:1981bd,Takasugi:1984xr}. Thus $0\nu 2\beta$ experiments are a powerful tool to probe the Majorana hypothesis for massive neutrinos.

The contribution of a single Majorana neutrino to the $0\nu 2 \beta$-decay amplitude is proportional to the combination~\cite{Blennow:2010th} 
\be 
A_i \propto m_i U_{ei}^2 M^{0\nu 2 \beta}(m_i)\,,
\ee
where $M^{0\nu 2 \beta}(m_i)$ is the nuclear matrix element that characterises the process. The latter is a function of the neutrino mass $m_i$ and depends on the nucleus that undergoes the $0\nu 2 \beta$ transition. It can be satisfactorily approximated by the analytic expression
\be 
M^{0\nu 2 \beta}(m_i) \simeq M^{0\nu 2 \beta}(0) \frac{p^2}{p^2-m_i^2},
\ee
where  $p^2 \approx - (125 \mbox{ MeV})^2$ is the virtual momentum of the neutrino, whose exact value depends on the nucleus. From the experimental results described in Section~\ref{sec:end_point} we know that $m_i \ll |p|$, the contribution of active neutrinos to the $0\nu2\beta$ amplitude is proportional to the combination
\bee
m_{2\beta} = \left|\sum_i U_{ei}^2 m_i\right|.
\eee
Since the above expression depends on the square of the complex mixing matrix elements, a cancellation between the contributions of different mass eigenstates is in general possible.

\begin{figure}[htb]
\begin{center}
 \includegraphics[width=0.70\textwidth]{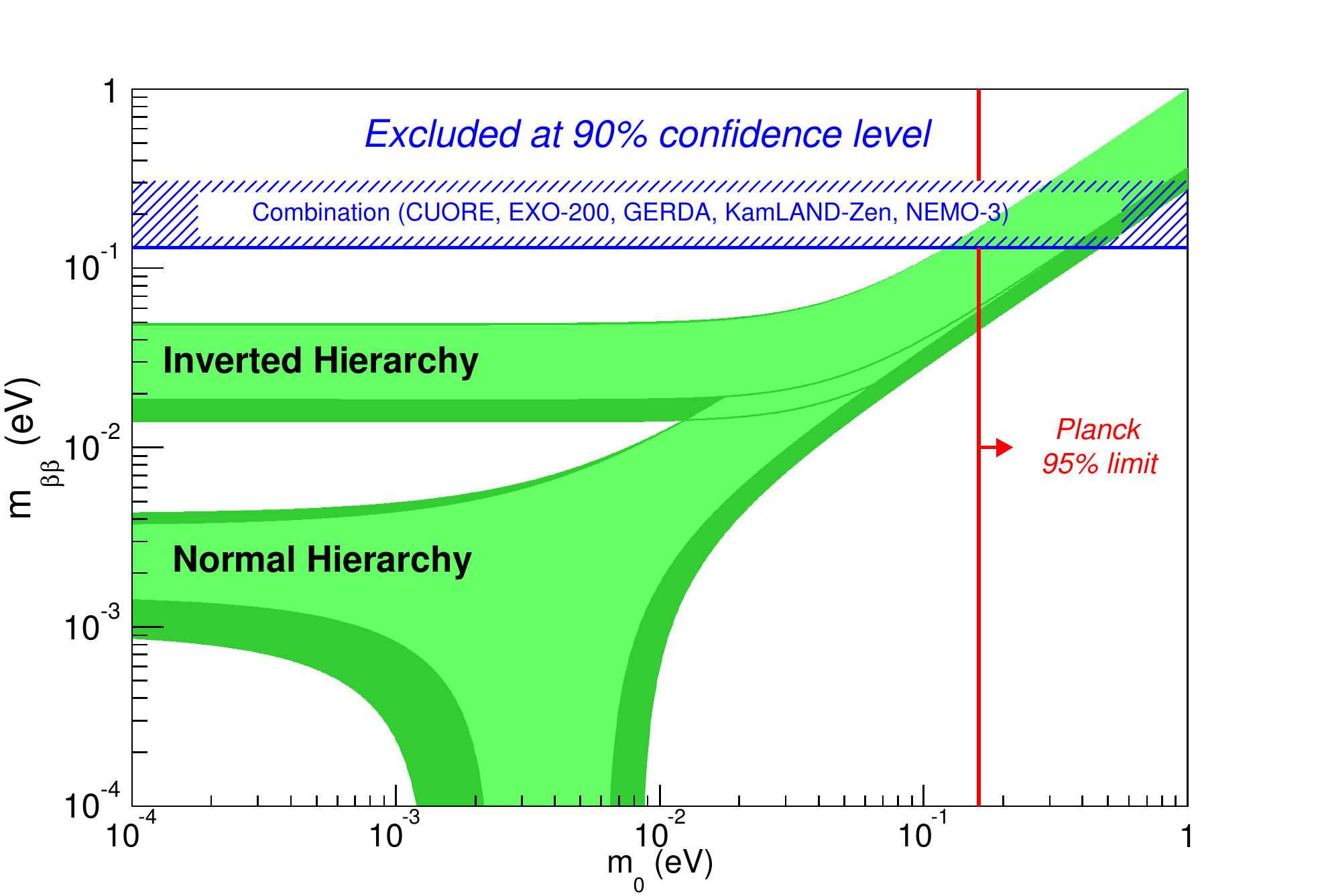}
\end{center}
\caption{Current bounds on the $0\nu2\beta$ mass $m_{2\beta}$ from different experiments together with the allowed regions determined from the neutrino oscillation parameters in the 3-flavour mixing scheme. Figure taken from~\cite{Guzowski:2015saa}.}
\label{0nubblimit}
\end{figure}

Numerous experiments worldwide are looking for $0\nu2\beta$-decay events using different experimental techniques and nuclei. Currently only upper bounds on the parameter $m_{2\beta}$ are available, while a controversial claim made by some member of the Heidelberg-Moscow collaboration~\cite{KlapdorKleingrothaus:2004wj} was not confirmed by subsequent measurements. The most stringent bounds on $0\nu2\beta$ are published by CUORICINO~\cite{Andreotti:2010vj}, KamLAND-Zen~\cite{Gando:2012zm}, GERDA~\cite{Agostini:2013mzu}, NEMO-3~\cite{Arnold:2013dha}, EXO-200\cite{Albert:2014awa} and CUORE~\cite{Alfonso:2015wka}. A combined analysis gives the limit~\cite{Guzowski:2015saa}
\bee
m_{2\beta} < [0.130, 0.310] \text{ eV},
\eee
where the spread is due to the incertitude on the calculation of the different nuclear matrix elements. The current limit on $m_{2\beta}$ together with the region allowed by the known $U_\text{PMNS}$ and mass parameters is reported in Fig.~\ref{0nubblimit}.

In addition to the above referred, an experimental program on $0\nu 2\beta$-decay is carried on by MAJORANA~\cite{Efremenko:2015kya}, LUCIFER~\cite{Cardani:2014vga}, AMoRE~\cite{Bhang:2012gn}, MOON~\cite{Ejiri:2010zz}, SNO+~\cite{Grullon:2015ica}, XMASS~\cite{Hiraide:2015cba}, DAMA-LXe~\cite{Bernabei:2006rf} and SuperNEMO~\cite{Bongrand:2015yfa}. 

\subsection{Supernova bounds}

\subsubsection{Supernova theory}

Supernovae are classified among two main classes: Type Ia supernova explosions take place when a white dwarf star accretes mass from a companion star, while Type II supernovae characterise the end life of stars with masses in the range~\cite{Raffelt:2002tu}
\be
8\ M_\odot \lesssim M \lesssim 60\ M_\odot,
\ee
where $M_\odot$ is the solar mass. Such stars have a sufficient mass to produce heavy elements with increasing atomic number, with the series that ends with ${}^{56}$Fe, which is the most bounded nucleus in Nature. When no nuclear fusion processes are energetically accessible, the hydrodynamical equilibrium between the thermal pressure and gravity is lost, and the star starts to collapse.  At the beginning of the collapsing phase the star is characterised by an onion-like structure, with an iron core surrounded by several shells composed by elements of decreasing atomic number. The gravitational collapse contracts the core which heats, causing iron photodissociation via~\cite{Giunti:2007ry}
\be \label{photofe}
\gamma+{}^{56}\mbox{Fe}\rightarrow 13\ \alpha + 4\ n.
\ee
Each one of these reactions absorbs $124$ MeV and, together with the electron capture by nuclei and protons
\be \label{ecapture}
e^- + \mathcal{N}(A,Z)\rightarrow \nu_e +\mathcal{N}(A,Z-1),
\ee   
they reduce the total kinetic energy and the electron degeneracy pressure, while the gravitational pressure increases because of the in-falling matter. The processes in eq.~(\ref{ecapture}) generate a first burst of neutrinos, the neutronization burst. When the mass of the core reaches the Chandrasekhar limit ($\approx 1.46\ M_\odot$) the pressure of degenerate electrons can no longer balance the gravitational force, and the core starts to collapse.
As more matter from the outer layers falls into the core, the  reactions~(\ref{photofe}) and~(\ref{ecapture}) proceed faster, resulting in the formation of a proto-neutron star in the core of the collapsing star.
When the density of the inner core reaches the nuclear density (about $10^{14}$ g cm${}^{-3}$), the pressure from  degenerate nucleons abruptly stops  the collapse, preventing new matter to fall into the  proto-neutron star. As a consequence, the free falling matter is ``bounced'', generating a shock wave that propagates through the outer iron core towards the surface of the collapsing star. At this stage the proto-neutron star has an un-shocked core with a radius of the order of 10 km, and a shocked mantle.

Neutrinos are thought to play a central r\^ole in the supernova explosions since they can revitalise the outgoing shock wave. During its propagation the shock wave  dissipates energy by the photodissociation of the free falling matter, that is abruptly stopped. Numerical simulations show that the energy loss would cause a stall of the shock at about 100-300 km from the core~\cite{Colgate:1966ax}, while the outer material will continue to fall through the shock, increasing the core mass. If nothing else happens, this sequence of events does not lead to a supernova explosion, because when the mass of the core exceeds the neutron star stability limit ($\approx 2 M_\odot$) a black hole forms.

If the shock wave stalls, a supernova explosion can be achieved only if something revitalises it. One of the mechanisms that is thought to renew the energy of the shock is the energy deposition from the huge flux of neutrinos that are thermally produced in the proto-neutron star~\cite{Bethe:1984ux} (other sources of energy are concurrently possible, for example convection behind the shock or acoustic power generated by oscillations of the accreting proto-neutron star). If the shock is successfully revived, the result is a delayed supernova explosion, about 0.5 s after the bounce. 

The core of the proto-neutron star has a temperature of about 40  MeV; inside it thermal neutrinos of all flavours are produced via ``flavour blind'' processes~\cite{Giunti:2007ry}, i.e. electron-positron pair annihilations
\be 
e^-+e^+ \rightarrow \nu + \overline{\nu},
\ee  
electron-nucleon bremsstrahlung
\be 
e^\pm + N \rightarrow e^\pm + N + \nu + \overline{\nu},
\ee 
nucleon-nucleon bremsstrahlung
\be 
N+N \rightarrow N+N+\nu+\overline{\nu},
\ee 
plasmon decay
\be  
\tilde{\gamma} \rightarrow \nu + \overline{\nu},
\ee 
and photoannihilation in matter
\be 
\gamma + e^{\pm} \rightarrow e^\pm + \nu + \overline{\nu}.
\ee 
There are also processes sensitive to the flavour, which are electron capture
\be 
e^- + p \rightarrow n + \nu_e,
\ee  
and positron capture
\be 
e^+ + n \rightarrow p + \overline{\nu}_e.
\ee
From numerical simulations~\cite{Bethe:1984ux} the electron neutrino and antineutrino fluxes are estimated to be of the order
\be\label{eq:supernova_flux} 
L_{\nu_e} \approx L_{\overline{\nu}_e} \approx   10^{52} \mbox{ erg s}^{-1},
\ee 
while the fluxes of $\nu_\mu,\overline{\nu}_\mu, \nu_\tau,\overline{\nu}_\tau$ are slightly smaller, since they are produced in less processes.

Due to the high density of the core, neutrinos cannot free-stream. A neutrinosphere is defined as the sphere whose radius coincides with the collision mean free path of a neutrino. The neutrinosphere radii $R_\nu$ are estimated to be between 50 and 100 km, in a region of the proto-neutron star mantle  with density of about $10^{11}$ g cm${}^{-3}$, from where neutrinos can escape by free-streaming. Thus, thermal neutrinos can be considered as emitted from a neutrinosphere within the mantle. Since the medium in which neutrinos move is mainly composed by protons, neutrons and electrons, the neutrino scattering cross section has different values depending on the (anti)-neutrino flavour, so  that there are three energy-dependent neutrinospheres, one for $\nu_e$, one for $\overline{\nu}_e$ and one for $\nu_\mu,\overline{\nu}_\mu,\nu_\tau,\overline{\nu}_\tau$, the last ones collectively denoted as $\nu_x$ in the following. Estimates of the time-averaged thermal neutrino energies give $\left<E_{\nu_e}\right>\approx$ 10 MeV, $\left<E_{\overline{\nu}_e}\right>\approx$ 15 MeV, $\left<E_{\nu_x}\right>\approx$ 20 MeV, with a total number of neutrinos of the order of $10^{58}$.

A simple parametrisation of the rate at which the energy from thermal neutrinos is absorbed by a gram of matter at a large distance $R_m$ from the center is~\cite{Bethe:1984ux}
\be\label{netrate}
\dot{E}=k(T_\nu)\left[\frac{L_\nu}{4\pi R_m^2}-\left(\frac{T_m}{T_\nu}\right)^2 a\ c\ T^4_m\right] \mbox{ erg }\mbox{ g}^{-1} \mbox{ s}^{-1},
\ee
where $T_{m}$ is the matter temperature in MeV, $T_\nu$ the neutrinosphere temperature, $k(T_\nu)$ the neutrino absorption coefficient in cm${}^{-2}\mbox{g}^{-1}$ (which depends on the neutrino temperature)  and $a\ T_m^4$ is the energy for unit volume of a ``blackbody'' neutrino gas
\be
a=\frac{7}{16}\cdot 1.37\cdot 10^{26} \mbox{ ergs }\mbox{cm}^{-3}\mbox{ MeV}^{-4}. 
\ee
In~(\ref{netrate}) the first term is the energy gain of matter due to neutrino interactions, while the second one is the energy loss due to electron and positron capture at temperature $T_m$, which are processes that release neutrinos.

The most important absorption processes are the charged weak current interactions:
\bee \label{energytrasfer}
\nu_e+n\rightarrow p+e^-,\non
\overline{\nu}_e+p\rightarrow n+e^+.
\eee  
By writing $L_\nu$ (see eq. (\ref{eq:supernova_flux})) in terms of the neutrinosphere temperature
\be 
L_\nu = \pi R_\nu^2 a c T_\nu^4,
\ee
the relation~(\ref{netrate}) can be rewritten in the useful way
\be
\dot{E}=k(T_\nu) a c T_\nu^4\left[\left(\frac{R_\nu}{2R_m}\right)^2-\left(\frac{T_m}{T_\nu}\right)^6\right],
\ee
which tell us that there is an energy gain if
\be
\frac{T_m}{T_\nu}<\left(\frac{R_\nu}{R_m}\right)^\frac{1}{3}.
\ee 

The capture of about $5-10 \% $ of the neutrino thermal flux could revive the shock wave~\cite{Burrows:2002bz}, so that the supernova explosion takes place around $0.5$ s after the bounce. Notice however that most of the one-dimensional (i.e. spherically symmetric) numerical simulations fail in reproducing the explosion~\cite{Giunti:2007ry}, because they do not account for convection, rotation effects and asymmetries that can enhance the energy deposition behind the shock, and a truly multidimensional treatment of the problem has to be taken into account~\cite{Janka:2012wk}. 

\subsubsection{SN1987A}

On the 23rd February 1987 the radiations from a relatively close-by supernova, labeled SN1987A, reached the Earth. SN1987A was located at 168,000 light-years from the Earth in the Large Magellanic Cloud (LMC), a satellite galaxy of the Milky Way. It is the closest observed supernova since 1604 (SN1604). 

The event offered a unique opportunity to test the theory of neutrino emission in supernova explosions. Two independent detectors observed a neutrino burst in coincidence with the SN1987A explosion: Kamiokande~\cite{Hirata:1987hu} reported 11 events in the energy range [7.5, 36] MeV on a time interval of 13 seconds, while IMB~\cite{Bionta:1987qt} reported 8 events in the energy range [20, 40] MeV over a time interval of 6 seconds. The inferred flux and energy distribution of the neutrino burst were found to be in agreement with the theoretical expectations.

From the time at which one detects a neutrino burst originating at a distance $D$ from the Earth it is possible to infer neutrino masses, since the neutrino velocity depends on the neutrino mass
\be 
v_{\nu} = \frac{p_{\nu}}{E_\nu} \simeq 1 - \frac{m_\nu^2}{2E_\nu^2},
\ee
and neutrinos of different energies will reach the Earth with a relative time delay
\be 
\Delta t = t_2 - t_1 = \frac{D}{v_2} - \frac{D}{v_1} \simeq D m_\nu^2 \left( \frac{1}{2E_2^2} - \frac{1}{2E_1^2} \right).
\ee
No such a delay was observed in the SN1987A events~\cite{Bilenky:2002aw}, resulting in a model independent upper bound of~\cite{Schramm:1987ra}
\bee
m_\nu < 30 \text{ eV at 95\% C.L.}
\eee
A more recent analysis that takes into account the subsequent developments in the understanding of supernova explosions gives the upper bound~\cite{Loredo:2001rx}
\bee
m_\nu < 5.7 \text{ eV at 95\% C.L.}
\eee 

\subsection{Cosmological bounds}

\subsubsection{Relic neutrino abundance}

Neutrinos in the early Universe interacted with the thermal bath via the weak interaction reactions~\cite{Giunti:2007ry}
\bee
\nu+\overline{\nu} &\leftrightarrow& e^+ + e^-,\non 
\stackrel{\left(-\right)}{\nu} +\ e^\pm &\leftrightarrow & \stackrel{\left(-\right)}{\nu} +\ e^\pm, 
\eee
with an interaction rate of the order
\bee
\Gamma \simeq G_F^2 T^5,
\eee
where $T$ is the temperature. The rate thus drops rapidly with the temperature, and considering the expansion rate of the Universe in the radiation dominated epoch,
\bee
H \approx \frac{T^2}{M_P},
\eee
where $M_P$ is the Planck mass, it is possible to estimate the neutrino decoupling temperature from the condition $\Gamma \simeq H$, which gives $T_\nu^{\text{dec}}\simeq 2$ MeV. A more refined analysis taking into account the flavour effects gives the decoupling temperatures\cite{Dolgov:2002wy}
\bee
T_{\nu_e}^{\text{dec}} = 1.34 \text{ MeV}, \hspace{1cm} T_{\nu_{\mu,\tau}}^{\text{dec}} = 1.5 \text{ MeV}.
\eee
Below these temperatures neutrinos freeze-out with a relativistic distribution and their number density remains constant in a comoving volume. It is possible to infer the current relic neutrino temperature from the cosmic microwave background (CMB) photons, with which neutrinos were in thermal equilibrium before decoupling; CMB  has been extensively studied by the COBE~\cite{Smoot:1992td}, WMAP~\cite{Spergel:2003cb} and Planck~\cite{Ade:2015xua} collaborations. However, the two temperatures are currently different, since the CMB photons received an entropy injection when the temperature decreased below the electron mass, making the process
\be 
\gamma \gamma \leftrightarrow e^+ e^-
\ee
to proceed only from the right to the left hand side. After the thermal decoupling, the entropy of positrons and electrons was transferred to photons, which remain the only relativistic particle in thermal equilibrium. The entropy density is given by the sum over the relativistic degrees of freedom in thermal equilibrium,
\bee
s=\frac{2 \pi^2}{45} g_s T^3 \left( \sum_{\stackrel{\text{bosonic}}{\text{d.o.f.}}} + \frac{7}{8} \sum_{\stackrel{\text{fermionic}}{\text{d.o.f}}}\right)
\eee
and is conserved during the annihilation process; thus the ratio between the photon temperatures before and after the $e^+ e^-$ annihilation is given by
\bee
\frac{T_\gamma^\text{after}}{T_\gamma^\text{before}}  =\left(\frac{11}{4}\right)^\frac{1}{3} = \frac{T_\gamma^\text{now}}{T_\nu^\text{now}},
\eee
where the last equality comes from the fact that neutrino and photon temperatures were the same before leptons annihilation. From the knowledge of the current CMB temperature~\cite{Fixsen:2009ug}, $T_\gamma^\text{now} = 2.725$ K, the temperature of the cosmic neutrino background (C$\nu$B) is
\bee
T_\nu^\text{now} = 1.945 \text{ K} = 1.68\times 10^{-4} \text{ eV},
\eee
which corresponds to an average density for each degree of freedom of
\be 
n_\nu^{d.o.f} = \frac{1}{(2\pi \hbar)^3} \int d^3 p \frac{1}{e^{|p|/T_\nu} +1} \simeq  56 \text{ cm}^{-3}.
\ee
The total density of neutrinos is determined by the number of degrees of freedom (d.o.f.) that were in thermal equilibrium at the decoupling epoch: if neutrinos are Majorana particles, the number of d.o.f. is 2 for each flavour, corresponding to the two possible helicity states. If neutrinos are Dirac particles there are in principle 4 degrees of freedom for each flavour, however only the left-handed neutrinos and right-handed antineutrinos were in thermal equilibrium at decoupling, resulting again in a number of 2 degrees for each flavour. Thus the density of relic neutrinos does not depend on the nature (Dirac or Majorana) of massive neutrinos. 

The neutrino contribution to the energy density of the Universe strongly depends on the neutrino masses. Notice that since $\sqrt{\Delta m_{12}^2} > T_\nu^\text{now}$ at least two mass eigenstates are non relativistic today. The contribution of a relativistic neutrino degree of freedom is
\be 
\rho_\nu^{\text{d.o.f, rel}} = \frac{1}{(2\pi \hbar)^3} \int d^3 p \frac{|p|}{e^{|p|/T_\nu} +1} \simeq  2.97 \times 10^{-2}\text{ eV cm}^{-3},
\ee
which is a small correction with respect to the contribution of a non relativistic degree of freedom
\be
\rho_\nu^{\text{d.o.f, non-rel}} = n_\nu m_\nu \geq n_\nu \sqrt{\Delta m_{12}^2} \simeq 4.87 \times 10^{-1} \text{ eV cm}^{-3}.
\ee
Neutrinos thus contribute to the matter component of the Universe, while a negligible radiation component is present only if the value of the lighter mass eigenstate does not exceed the temperature $T_\nu^\text{now}$. 

The matter component of the Universe has been accurately determined by the Planck collaboration~\cite{Ade:2015xua}
\bee
\Omega_m =\frac{\rho_m}{\rho_c} = 0.308 \pm 0.012,
\eee
where $\rho_c=\frac{3}{8 \pi G} H^2 = 10.54\ h^2 \text{ keV/cm}^3$ is the critical density of the Universe. Taking into account the current value of the Hubble parameter, $H_0 = h_0\ 100\ \text{km\ (sec Mpc)}^{-1} = (67.8 \pm 0.9) \text{km\ (sec Mpc)}^{-1}$~\cite{Ade:2015xua}, it is possible to extract an upper bound on the sum of the neutrino masses by imposing that the neutrino component does not exceeds $\Omega_m$
\bee\label{eq:nu_abundance}
\Omega_\nu = \frac{\sum_i m_i}{h^2\ 94.1 \text{ eV}} < \Omega_m,
\eee
leading to the constrain
\bee
\sum_i m_i < 13.32 \text{ eV}.
\eee    

\subsubsection{CMB limits}\label{sec:CMB_nu}
The cosmic microwave background (CMB) is composed by the photons that decoupled from the primordial plasma at the time of recombination, when at the temperature $T\approx 0.25$ eV the electrons and protons formed bound states, making the Universe transparent to the electromagnetic radiation\cite{Kolb:1990vq}. From the study of the CMB radiation it is possible to extract useful information regarding the thermal fluctuations in the primordial Universe, which can in turn constrain the parameters of the standard cosmological model, the $\Lambda$CDM model~\cite{Bahcall:1999xn}.

The CMB radiation has been studied by different collaborations~\cite{Smoot:1992td,Spergel:2003cb,Ade:2015xua} and is characterised by an extreme level of isotropy, with a blackbody temperature of {$\overline{T} = (2.72548 \pm 0.00057) $ K~\cite{Fixsen:2009ug}} and fluctuations of the order $\delta T/T \approx 10^{-5}$, which are correlated with the asymmetries in the matter distribution at the time of recombination. Given that the temperature of recombination is of the same order of the upper bound on neutrino masses from $\beta$-decay searches, eq.~(\ref{nueff_bound}), massive neutrinos can potentially affect the related CMB observables.

The CMB anisotropies can be parametrised by the observable
\bee
\frac{\delta T(\hat{n})}{\overline{T}},
\eee
where $\hat{n}$ is a direction in the sky. The relevant observables are the two-point correlations in the CMB map,
\bee\label{eq:CMB}
\left< \frac{\delta T(\hat{n}')}{\overline{T}} \frac{\delta T(\hat{n})}{\overline{T}}\right> = \sum_{l=0}^\infty \frac{2l+1}{4\pi} C_l P_l\left(\hat{n}' \cdot \hat{n}\right),
\eee
where the correlations are conveniently expanded as a series in the Legendre polynomials $P_l$,  the nature of the anisotropies being encoded in the coefficients $C_l$, which probe the angular scales $\theta=\pi/l$. The absolute neutrino mass scale affects the CMB anisotropies in both a direct and an indirect way~\cite{Lesgourgues:2012uu}. Indirectly, heavier neutrinos impact the background evolution by delaying the onset of the matter-radiation equality epoch, and by increasing the non relativistic matter density at later times; directly, massive neutrinos affect secondary anisotropies by enhancing the gravitational potential wells when they become non-relativistic. The CMB data alone are not very sensitive to neutrino masses because their effect can be cancelled by varying other cosmological parameters, such as for instance the Hubble expansion rate $H$. CMB measurements can however give strong constraints when   combined with independent observations that fix one or several observables in order to break the degeneracy in the parameter space, for instance by a direct measurement of $H$. In this way,
the upper bound on the sum of neutrino masses from the Planck collaboration is~\cite{Ade:2015xua}
\bee
\sum_i m_i < 0.23 \text{ eV at 95\% C.L.}
\eee

\subsubsection{Structure formation bounds}\label{sec:LSS}
The strongest bounds on neutrino masses are currently extracted from the observation of the large scale structure (LSS) of the Universe. Notice however that the obtained results are model dependent and rely on complex hydrodynamical simulations accounting for the non-linear evolution of the matter component of the Universe. Beside the effects on the CMB anisotropies described in the previous subsection, neutrinos affect the structure formation because they decouple with a relativistic momentum distribution (and thus with a large velocity dispersion) and they tend to erase the matter density inhomogeneities by escaping from the related gravitational potential wells. The quantitative importance of this effect depends on the absolute value of the neutrino masses, since heavier neutrinos have a smaller free-streaming length. The net effect is twofold: in the linear regime massive neutrinos induce a step-like suppression of the matter power spectrum, while in the non-linear regime they delay the collapse time~\cite{Bond:1980ha}.

A powerful tool to trace the matter distribution at cosmological scales of the order $(1 - 80) h^{-1}$ Mp is the Lyman-$\alpha$ (Ly-$\alpha$) forest~\cite{Narayanan:2000tp}, which is a series of absorption lines in the spectra of distant quasi stellar objects (QSO). The absorption is caused by the hydrogen present in the intergalactic medium (IGM) on the line of sight between the QSO and the Earth, in the Ly-$\alpha$ transition between the hydrogen atomic levels $1s \rightarrow 2p$. The ``forest'' structure is due to redshift, since the wavelength at which the Ly-$\alpha$ transition takes place depends on the distance of the IGM from the Earth; by measuring the intensities of absorption in the Ly-$\alpha$ forest it is possible to trace the IGM distribution on large scales and infer from this the underlying matter distribution.

From the combination of CMB, Ly-$\alpha$ and baryon acoustic oscillations (BAO)~\cite{Dawson:2012va} data, the following bound on the sum of neutrino masses has been obtained~\cite{Palanque-Delabrouille:2014jca}
\bee
\sum_i m_i < 0.14 \text{ eV at 95\% C.L.}
\eee
Notice that the analysis~\cite{Palanque-Delabrouille:2014jca} tends to favour the NH ordering against the IH one.

\section{The Dark Matter component of the Universe}
There exist several independent cosmological observations which are difficult (if not impossible) to reconcile with the hypothesis that the matter component of the Universe is completely constituted by the known luminous matter. These observations are all consistently reconciled with the theoretical expectations   under the assumption that, in addition to the standard baryonic matter, the matter component of the Universe encompasses a dark component, which does not interact electromagnetically.\footnote{An alternative way to account for the aforementioned observations is to modify the gravitational dynamic on galactic scales. This option is known as Modified Newtonian Dynamics (MoND)~\cite{Milgrom:1983ca,Sanders:2002pf}.} This component is called dark matter (DM) and is five times more abundant than the ordinary baryonic matter. In the present section we review the observations that call for the presence of DM and the phenomenological constraints that exclude any SM particle as a viable DM candidate.

\subsection{Galaxy cluster velocity dispersion}
The first indications for the existence of a dark matter component arose in 1933 from the measurement of the velocity dispersion of 800 galaxies in the Coma cluster~\cite{Zwicky:1933gu}. By assuming that the gravitational potential is completely determined by the observed luminous matter, the virial theorem gave an expected velocity dispersion of 80 km/s, while the measured value resulted in the interval 1500 to 2000 km/s. The discrepancy can be reconciled under the assumption that the average matter density in the cluster is at least 400 times larger than the value inferred from the observation of luminous matter.
These results have been confirmed by modern observations on galaxy clusters~\cite{Allen:2011zs}.

\subsection{Galaxy rotation curves}
On galactic scales a strong evidence for the existence of DM comes from the observation of the galaxy rotation curves, namely the velocity rotation of luminous matter (stars and gas) as a function of the distance from the galactic center. From Newtonian dynamics, the velocity rotation of a body subject to gravitational attraction at a distance $R$ from the galaxy center is given by the relation
\bee
v(R) = \sqrt{\frac{G_N M(R)}{R}},
\eee
where $M(R)$ is the mass contained inside a sphere of radius $R$ and $G_N$ is the Newton constant. From the above relation, the rotation velocity should drop as $v \propto 1/R$ when the radius $R$ exceeds the radius where the luminous matter is contained. However a pioneering study of the Andromeda galaxy showed in 1970 that the rotation velocity in the outer-luminosity region is actually constant around the value $v \approx $ 200 km/s~\cite{Rubin:1970zza}. This rotation curve can be explained by assuming that the function $M(R)$ grows linearly despite the fact that the contribution of luminous matter is negligible in that region, $M(R)\propto R$, implying the presence of a DM halo that encompasses the visible galaxy. 
Subsequent studies on the rotation curves of galaxies confirmed the above referred results, leading to the conclusion that the luminous matter accounts at most for the 10\% of the total matter in the halo~\cite{Rubin:1980zd,Battaner:2000ef}.

\subsection{Gravitational lensing and the Bullet Cluster}
According to the theory of general relativity~\cite{Einstein:1916vd}, the propagation of photons follows the space-time curvature due to the presence of energy; the bending of light due to the gravitational field of the Sun was firstly observed during the solar eclipse of 1919, with a result that was in agreement with the predictions of general relativity~\cite{Dyson1920.0009}. While the Sun causes only a mild deflection of the light reaching the Earth, more massive and distant objects can bend it so strongly that  light from a luminous source lying behind them reaches the Earth via multiple paths; this is the case, for instance, of black holes, galaxies and clusters of galaxies. They effectively act as lenses, bending and focusing the light during its propagation~\cite{Bartelmann:1999yn}.

Gravitational lensing is a powerful and robust tool, since it only relies on the presence of matter in some spacial region, without requiring any assumption on the dynamics of the system. It is usually classified in weak and strong lensing, although a generally applicable definition of the two regimes is not possible. Strong lensing refers to the deflection generated by a well defined gravitational source, that results in the multiple images of a single luminous body and which provides information on the mass distribution of the deflector. On the contrary, weak lensing is related to a study of statistical nature,  in which the analysis of multiple luminous sources provides information on the distribution of matter between the sources and the observer.

\begin{figure}[htb]
\begin{center}
\subfloat{\includegraphics[width=0.45\textwidth]{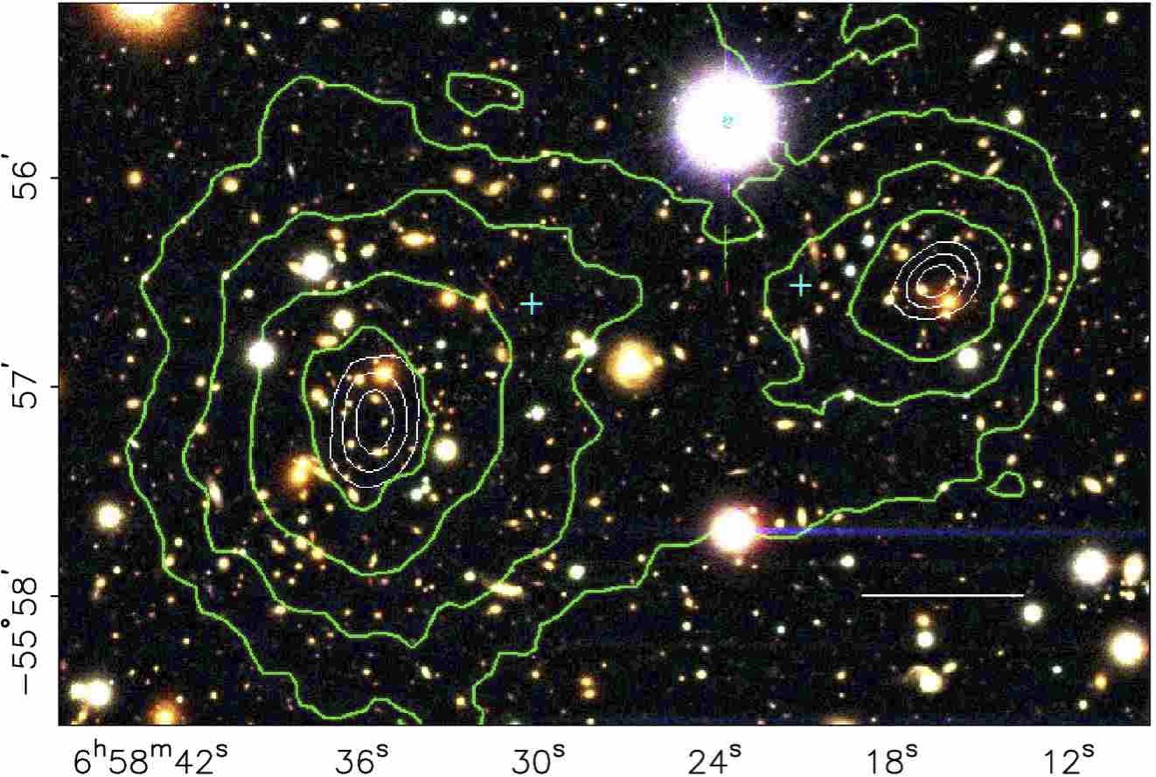}}\hspace*{0.5cm}
\subfloat{\includegraphics[width=0.45\textwidth]{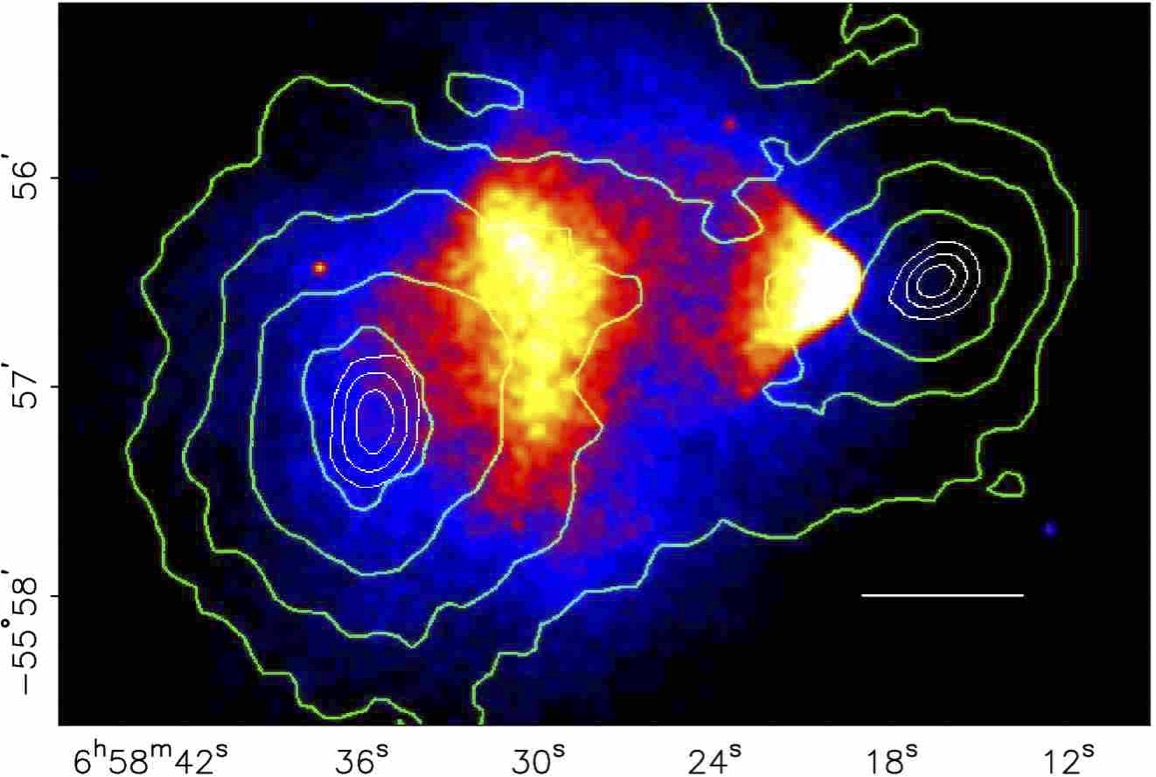}}
\end{center}
\caption{
On the left panel, a colour image of the Bullet cluster from the Magellan telescopes. The white bar indicates 200 kpc at the distance of the cluster. The green contours are the mass profiles reconstructed by weak lensing. On the right panel, an X-ray Chandra image of the same cluster. Figures taken from~\cite{Clowe:2006eq}.}
\label{fig:bullet}
\end{figure}

The presence of DM in an amount compatible with kinematic estimates has been confirmed by means of gravitational lensing observations in galaxies~\cite{Maoz:1993ix,Griffiths:1996pu}, clusters of galaxies~\cite{Tyson:1990yt,Mellier:1998pk} and on larger scales~\cite{Wittman:2000tc}. One of the most compelling observations that favour the particle nature of DM comes from the weak lensing measure of the matter distribution in the Bullet Cluster~\cite{Clowe:2003tk,Clowe:2006eq}. It is a system composed by two merging galaxy sub-clusters, whose luminous matter is dominated by baryons observable in the X-ray spectrum; the associated X-ray image shows the presence of bow shock in the emitting plasma  of the smaller sub-cluster, meaning that the latter is currently moving away from the more massive one. From the analysis of the relative velocities it is possible to infer that the two sub-clusters passed through each other around $10^{8}$ yr ago. The specificity of the Bullet Cluster relies on the fact that the relative motion of the sub-clusters lies in the sky plane, and the clean observation of the shock bow allows to determine the relative velocity and geometry of the merging plasma. The cross-section of the galaxies during the merging is negligible and the two populations remain collisionless in the process. This offers an unique opportunity to test the DM particle hypothesis by means of weak lensing: if the DM is composed by collisionless particles, the gravitational potential will trace the distribution of the collisionless galaxies, while in the absence of DM the dominant matter component coincides with the X-ray emitting plasma and the gravitational potential will trace its emission. The mass profiles reconstructed from weak lensing are shown in Fig.~\ref{fig:bullet} and are in agreement with the collisionless particle hypothesis, giving a robust model-independent evidence for the existence of DM.

\subsection{Cosmic microwave background}\label{sec:CMB_DM}
Although the above described observations are strong evidences for the existence of DM, they probe the matter density in the Universe only locally, in a limited spatial region. A further observable that provides evidence for the existence of a DM component in the Universe is the CMB which, contrary to the other probes, is sensitive to the global DM matter density, since at the decoupling epoch the Universe was characterised by an extreme degree of homogeneity.

The two-point correlation function of the CMB, eq. (\ref{eq:CMB}), is sensitive to both the total amount of matter in the Universe and to the ratio between baryonic and dark matter~\cite{Samtleben:2007zz}. The total matter density $\Omega_m$ determines the matter-radiation equality epoch: a higher matter density pushes this epoch to earlier times, allowing more time for the gravitational enhancement of primordial inhomogeneities before the CMB decoupling. On the other hand, the relative abundance of baryonic matter density  $\Omega_b$ with respect to the dark matter one,  $\Omega_c$, determines the relative change in the gravitational potential wells due to baryon accretion: if the baryonic fraction increases the baryons that fall in a potential well  modify its potential in a stronger way, and the outgoing photons will be more redshifted. The Planck collaboration has determined with high precision these parameters in the $\Lambda$CDM cosmological model, resulting in the values\cite{Ade:2015xua}:
\bee
\Omega_m h^2 &=& 0.1426 \pm 0.0020,\\ 
\Omega_b h^2 &=& 0.02226 \pm 0.00023,\\
\Omega_c h^2 &=&  0.1186 \pm 0.0020.\label{eq:planckDM}
\eee
Other than being a further evidence for the existence of a DM component in the Universe, the result (\ref{eq:planckDM}) is an important constraint on extensions of the SM of particle physics aiming at providing a particle candidate for the DM component.

\section{Dark matter hypothesis and the SM}
The discovery of a DM component of the Universe opens the question of its nature. The observations discussed in the previous section are all in agreement with the hypothesis of having the DM composed by a population of massive particles that do not interact electromagnetically; the weak lensing observations favour in particular the particle nature of DM over other alternative hypothesis. Moreover, the results from numerical simulations of the evolution of large scale structures of the Universe, performed under the assumption of a particle nature for DM, are all in good agreement with the observed structure~\cite{Springel:2005nw,Springel:2008cc}. These results motivate the study of the compatibility of possible particle candidates with the DM hypothesis.

In order to be a viable DM candidate, a particle must have negligible electromagnetic interactions, be massive and stable on cosmological timescales. In the SM, there is only one particle  with all these requirements,  the active neutrino. Since the first evidences for nonzero neutrino masses, the hypothesis of neutrinos as the DM constituent received a strong attention~\cite{Primack:2001ib}. There are however at least two reasons that exclude SM neutrinos as the dominant DM component. The first one comes from the cosmological abundance of massive neutrinos, eq.~(\ref{eq:nu_abundance}): in order to account for the total DM abundance, eq.~(\ref{eq:planckDM}), the sum of neutrino masses would have to be
\bee
\sum_i m_i \simeq 11.16 \text{ eV}.
\eee
This value strongly exceeds the upper bound on the absolute neutrino mass scale from laboratory experiments, eq.~(\ref{nueff_bound}), implying that neutrinos can only be a sub-dominant DM component. In addition to that, neutrinos decouple relativistically with a large velocity dispersion, and represent a Hot Dark Matter (HDM) component, meaning that their free streaming length $\lambda$ at decoupling was of the order of the cosmic horizon. The DM free streaming length strongly affects the subsequent structure formation, since the inhomogeneities in the matter density are erased at scales smaller than $\lambda$. This implies that in HDM cosmologies large structures, such as superclusters of galaxies, form first and later fragment into smaller structures. HDM cosmologies are at odds with observation, and the hypothesis of having the SM neutrinos as the dominant DM component is ruled out~\cite{Bond:1980ha,Bond:1983hb,White:1984yj}. In fact, LSS observations can be used to put an upper bound on the contribution of the SM neutrinos to DM, as already discussed in Section~\ref{sec:LSS}.

A possibility that would not require any physics beyond the SM is the hypothesis that DM is composed by standard astrophysical objects characterised by a very low luminosity, such as black holes, neutron stars, brown dwarfs or nomad planets, that are collectively denoted as Massive Compact Halo Objects (MACHOs). If MACHOs  compose the Milky Way halo, they can be detected by means of the gravitational microlensing they would cause on the stars in nearby galaxies~\cite{Paczynski:1985jf,Griest:1990vu}. An analysis of 5.7 years on 11.9 million stars in the LMC has excluded the hypothesis that MACHOs compose the totality of the Milky Way halo at the 95 \% C.L.~\cite{Alcock:2000ph}, while a subsequent analysis on the same galaxy has put an upper bound of 8 \% on the maximum halo fraction that can be composed by MACHOs~\cite{Tisserand:2006zx}.

We finally mention the hypothesis that DM may be composed by SM particles, forming stable exotic bound states of quarks (quark nuggets) that may arise in the early Universe if the QCD phase transition was of first order~\cite{Witten:1984rs}. This is an intriguing possibility, since it does not require any new physics; however recent lattice computations suggest that the QCD phase transition is a continuous cross-over~\cite{Bhattacharya:2014ara}.

In conclusion, the existence of DM and its properties call for the existence of BSM physics.

\section{The baryon asymmetry of the Universe}
The fact that the observed Universe is matter-antimatter asymmetric is a robust observation~\cite{Steigman:1976ev,Cohen:1997ac}: all astrophysical bodies appear to be composed by matter, while antimatter is observed in a limited amount. One could assume that the Universe is globally matter-antimatter symmetric, but with spatially separated regions dominated by one of the two kinds. However this kind of cosmologies are in contradiction with observation~\cite{Steigman:1976ev}. In the standard Big-Bang picture the current Universe  derives from an extremely hot initial state where all the SM interactions were in thermal equilibrium. In this phase of its evolution, the Universe is expected to contain an equal amount of particles and antiparticles. We thus infer that at some intermediate phase in the evolution of the Universe a net asymmetry between baryons and antibaryons was created, and that this asymmetry has survived until present times. This process is commonly called baryogenesis.

It has been shown that the baryogenesis process requires three necessary conditions~\cite{Sakharov:1967dj}:
\begin{itemize}
\item the baryon number $B$ is not conserved;
\item the $C$ and $CP$ symmetries must be broken;
\item the Universe must deviate from the thermal equilibrium condition at some epoch.
\end{itemize}
These are known as Sakharov's conditions~\cite{Sakharov:1967dj}. The first one is clear: if the baryon number is an integral of the system no baryon asymmetry can be generated starting from $B=0$. The second condition contains two requests. Firstly the interactions must be different for particles and antiparticles. Suppose in fact that, following the first condition, there exist some process $\ket{i} \rightarrow \ket{f}$  that violates the baryon number by an amount $\delta B$, where $\ket{i} (\ket{f})$ represents the initial (final) state in the scattering. Then if $C$ is conserved the following condition for the amplitude of the process $\mathcal{M}_{i\rightarrow f}$ holds
\bee\label{eq:Cinv}
\mathcal{M}_{i\rightarrow f} = \mathcal{M}_{\overline{i} \rightarrow \overline{f}},
\eee
where the bars stand for the charge conjugated states $\ket{\overline{i}},\ket{\overline{f}}$. The process $\mathcal{M}_{\overline{i} \rightarrow \overline{f}}$ violates the baryon number by an amount $-\delta B$, and eq.~(\ref{eq:Cinv}) implies that it  is as probable as the charge conjugated one, thus no net baryon asymmetry can be globally generated. The violation of the $CP$ conjugation symmetry is  related, from the $CPT$ conservation theorem~\cite{Lueders:1992dq}, to the violation of the $T$ conjugation symmetry. If that is not the case, then
\bee\label{eq:Tinv}
\mathcal{M}_{i\rightarrow f} = \mathcal{M}_{f\rightarrow i}.
\eee
Again, since the process $\ket{f} \rightarrow \ket{i}$ violates the baryon number by an amount $-\delta B$, eq.~(\ref{eq:Tinv}) implies that no baryon asymmetry can be globally generated.
The $CPT$ theorem further implies that particles and antiparticles have the same mass. Thus in a thermal equilibrium condition, the phase space distribution for baryons and antibaryons is the same
\bee
f(\textbf{p},m_B) =  \frac{1}{1+e^{\frac{|\textbf{p}|^2 + m_B^2}{T}}}  = f(\textbf{p},m_{\bar{B}}),
\eee
and no asymmetry in the occupation numbers of baryons and antibaryons can be created. This is the reason why a mechanism for baryogenesis is required even in the hypothesis that the Big Bang initial conditions provided the Universe with a net baryon asymmetry: since the Universe has been in thermal equilibrium for most of its evolution, that initial asymmetry would have been washed out by now.

The above considerations are all qualitative; in order to test any baryogenesis mechanism a quantitative definition of the BAU is necessary. It is convenient to define it as
\bee
Y_{\Delta B} = \frac{n_b-n_{\bar{b}}}{s},
\eee
where $n_b,n_{\bar{b}}$ are the number densities of baryons and antibaryons, respectively, and $s$ is the entropy density. This definition is useful because, in the absence of baryon number violation, the quantity $Y_{\Delta B}$ remains constant during the expansion of the Universe. An alternative definition normalised to the photon density is
\bee
\eta_{\Delta B} = \frac{n_b-n_{\bar{b}}}{n_\gamma}.
\eee
There are two independent observations that probe the BAU at different epochs, Big Bang Nucleosynthesis (BBN) and CMB, and remarkably they both give compatible results. We review them in the following sections.

\subsection{Big Bang Nucleosynthesis}
\begin{figure}[htb]
\begin{center}
\includegraphics[width=0.55\textwidth]{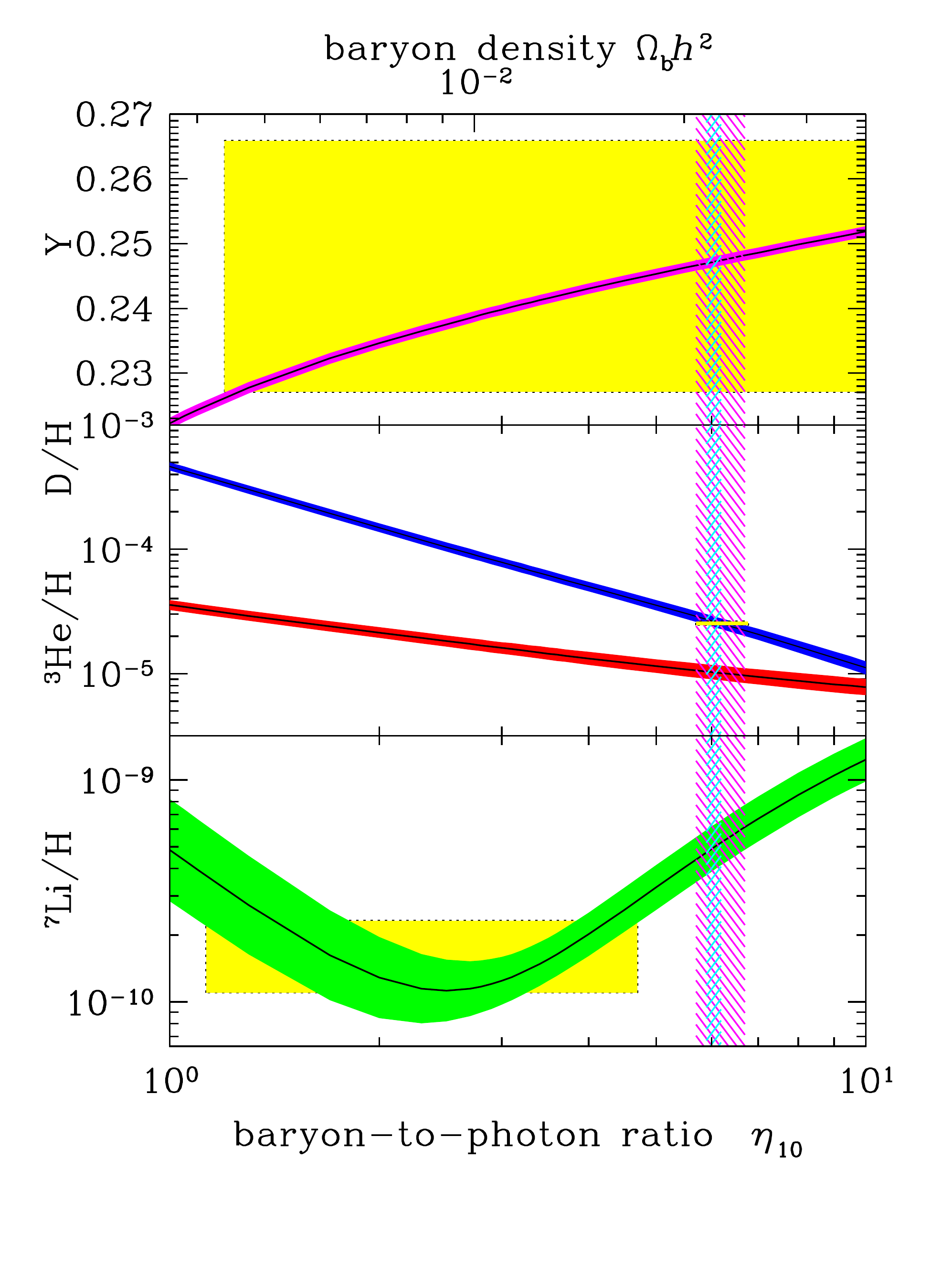}
\end{center}
\caption{The abundances of light elements ${}^2$H, ${}^3$H, ${}^3$He, ${}^4$He and ${}^7$Li as a function of the BAU parameter $\eta_{10} \equiv 10^{10} \eta_{\Delta B}$ at 95\% C.L.. The yellow boxes indicate the observed abundances, the wider vertical band is the BBN concordance region at 95 \% C.L., the narrow vertical band is the value inferred from CMB. Figure taken from~\cite{Fields:2014uja}.}
  \label{fig:BAU}
\end{figure}

Big Bang Nucleosynthesis~\cite{Gamow:1946eb,Alpher:1948ve,Peebles:1966zz,Wagoner:1966pv,Wagoner:1972jh,Steigman:2007xt,Iocco:2008va,Kolb:1990vq} is the process that led to the formation of the lightest chemical elements in the primordial plasma. After the QCD phase transition (around a temperature of 150 MeV) it became energetically favourable for the free quarks to be bounded in baryons. The relative abundances of elements were set by nuclear statistical equilibrium, which implies a population dominated by (anti)protons and (anti)neutrons with a completely negligible fraction of heavier elements. Notice that weak interactions were keeping baryons in thermal equilibrium at that epoch but, since the temperature was already smaller than the proton and neutron masses, baryons and antibaryons annihilated among them.   If the Universe was characterised by a baryon asymmetry $Y_{\Delta B}>0$ $(Y_{\Delta B}<0)$ all the antiparticles (particles) would eventually disappear. The value of $Y_{\Delta B}$ is an input parameter in BBN and, since  it affects the relative abundances of light elements, it can be constrained by observation. The baryons left in the plasma were kept in thermal equilibrium by the weak interaction reactions
\bee\label{eq:BBNeq}
n + \nu_e  &\leftrightarrow & p + e^-,\non 
p + \overline{\nu_e} &\leftrightarrow & n + e^+,\non 
n & \leftrightarrow & p + e^- + \overline{\nu_e},
\eee
which set the relative abundances of protons and neutrons to the equilibrium value
\bee
\frac{n_n^{eq}}{n_p^{eq}} = e^{-\frac{\Delta m}{T}},
\eee
where $\Delta m \equiv m_n- m_p = 1.29$ MeV is the mass difference between the neutron and the proton. At the temperature $T_D \sim 0.7$ MeV the reactions (\ref{eq:BBNeq}) departed from thermal equilibrium and the neutron to proton abundance froze-out at the value $e^{-\frac{\Delta m}{T_D}}\simeq 0.16$. From that moment two processes started to reduce the abundance of free neutrons: spontaneous neutron decay
\bee
n & \rightarrow & p + e^- + \overline{\nu_e},
\eee 
and deuterium production
\bee\label{eq:deuterium}
n + p \rightarrow {}^2 \text{H} + \gamma.
\eee
The amount of produced deuterium depends on the value of $Y_{\Delta B}$: although the temperature is already below the deuterium binding energy $B_D \simeq 2.2$ MeV, the tail of the photon phase space distribution may contain sufficiently energetic photons to make the reaction (\ref{eq:deuterium}) to proceed in the inverse direction (photodissociation). The smaller is $Y_{\Delta B}$ the higher is the number of photons per baryon, and the higher is the photodissociation rate at fixed temperature. The photodissociation process becomes ineffective at a temperature $T_N$ such that
\bee\label{eq:TN}
e^{-\frac{{B_D}}{T_N}} \sim \eta_{\Delta B},
\eee
that is when the deuterium and baryon abundances become comparable. When the ${}^2$H formation becomes effective the production of the intermediate states ${}^3$H and ${}^3$He starts and, to an excellent approximation, all the neutrons that have not yet decayed are eventually bounded in  ${}^4$He nuclei, which is the most bounded element in the series (the production of ${}^3$H, ${}^3$He and ${}^4$He is not effective in the absence of deuterium since it would require  three body reactions that are statistically disfavoured). The helium mass fraction can thus be estimated as
\bee
Y_p \simeq \frac{4 n_{{}^4\text{He}}}{n_B} = \frac{4 n_n^N/2}{n_n^N+n_p^N} = \frac{2 \left(n_n^N/n_p^N\right)}{1+\left(n_n^N/n_p^N\right)},
\eee
where
\bee
\frac{n_n^N}{n_p^N} = e^{-\frac{\Delta m}{T_D}} e^{-{\frac{t\left(\tiny{T_N}\right)}{\tau_n}}},
\eee
is the ratio between neutron and proton abundances at temperature $T_N$, $t(T_N)$ being the time elapsed at temperature $T_N$ and $\tau_n$ denotes the neutron lifetime. Notice however from eq.~(\ref{eq:TN}) that the value of $T_N$ depends only logarithmically on the parameter $\eta_{\Delta B}$, thus the measurement of $Y_p$ weakly constrains the BAU, while it strongly depends on $T_D$ and thus on the expansion rate of the Universe. For this reason the quantity $Y_p$ is usually considered a chronometer of the BBN process. The observables that strongly depend on $Y_{\Delta B}$ are the residual abundances of light nuclei that are not converted into ${}^4$He at the end of the process, that is the abundances of ${}^2$H, ${}^3$H and ${}^3$He. Indeed, they are the result of the competition between the two body processes for production and destruction, whose rates strongly depend on the baryon abundance. These observables are thus considered baryometers of the BBN process. The prediction for the abundances of light elements as a function of the BAU, computed and reported in the PDG~\cite{Agashe:2014kda}, is shown in Fig.~\ref{fig:BAU}. 

The analysis gives the concordant range for the BAU:
\bee\label{eq:BAU_BBN}
5.7 \times 10^{-10} \leq \eta_{\Delta B} \leq 6.7 \times 10^{-10}\ \left(\text{at 95 \% C.L.}\right).
\eee
Notice from Fig.~\ref{fig:BAU} that the observed ${}^7$Li abundance does not agree with the BBN prediction. The discrepancy is at the $[2.3 , 5.3]\ \sigma$ level, depending on the adopted analysis~\cite{Cyburt:2008kw}, and constitutes the so-called \emph{Lithium problem}. It is not clear if the discrepancy is due to systematic errors in the observation, to uncertainties in the nuclear inputs, to underestimated processes that may reduce the Lithium abundance during the stellar evolution or to new physics at work.

\subsection{Cosmic Microwave Background}
The study of the CMB anisotropies provides a precise measurement of the baryon to photon ratio at the decoupling epoch via the determination of the parameter $\Omega_b$, which is related to the BAU via the relation~\cite{Cyburt:2015mya}
\bee
\eta_{\Delta B} = \frac{\rho_c}{\left<m\right> n_\gamma^0}\Omega_b,
\eee
where $n_\gamma^0$ is the present photon number density and $\left<m\right>$ is the mean mass per baryon, which is slightly lower than the proton one due to the Helium binding energy. We already discussed the effects of $\Omega_b$ on the CMB angular power spectrum in Section~\ref{sec:CMB_DM}. The analysis performed by the Planck collaboration gives the value~\cite{Ade:2015xua}
\bee
\eta_{\Delta B} = \left(6.10 \pm 0.04\right) \times 10^{-10},
\eee
which is in remarkable agreement with the value derived from BBN, eq.~(\ref{eq:BAU_BBN}). Notice that this observation is a probe of the BAU at an epoch characterised by a temperature {$T\sim$ eV}, and is thus complementary to the BBN one that probes the BAU at  $T\sim$ MeV.

We notice here that the CMB angular power spectrum also depends on the parameter $Y_p$, that sets the number of free electrons between helium and hydrogen recombination, that in turns determines the mean free path of photons due to Thomson scattering. Thus the parameters $(\Omega_b,H,Y_p)$ are directly probed by the CMB observation: since they are correlated in BBN, it is possible to test the BBN scenario from CMB~\cite{Cyburt:2015mya}.

\section{BAU and the Standard Model}
Having established the presence of a small but finite BAU at the BBN epoch makes it necessary to determine the mechanism at its origin. The first question to answer is whether the SM can account for this asymmetry. Qualitatively, it complies with Sakharov's conditions. The $C$ and $CP$ symmetries and the baryon number are violated by weak interactions. The violation of the $C$ symmetry relies on the chiral structure of the $SU(2)_L$ gauge group, with the weak current given by the sum of a vector component (odd under $C$) and an axial one (even under $C$)~\cite{Peskin:1995ev}. The $CP$ violation is  related to the presence of a physical phase in the Lagrangian, the $\delta_{CKM}$ phase in the Cabibbo-Kobayashi-Maskawa quark mixing matrix~\cite{Agashe:2014kda}.
Baryon number is conserved in the SM at the perturbative level, however non-perturbative  effects violate the sum of the baryon and lepton numbers $B+L$, while preserving their difference $B-L$~\cite{Belavin:1975fg,'tHooft:1976up,'tHooft:1976fv}. Indeed, the ground state of a $SU(N)$ gauge theory is not unique, but is composed by an infinite series of vacua, degenerate in energy and which cannot be deformed into each other by a gauge rotation. In the SM a transition between two different $SU(2)$ vacua induces a violation of the $B+L$ charge. However, at zero temperature, the transition can only happen via tunnelling and is characterised by the amplitude
\bee
\Gamma \simeq e^{-\frac{16 \pi^2}{g^2}},
\eee
which is completely negligible given the value of the electroweak gauge coupling $g$. Tunnelling is however not the only way to connect inequivalent vacua: it has been shown that there exists a configuration of the Higgs and of the electroweak gauge boson fields that is associated to a saddle point in the energy separating two inequivalent vacua~\cite{Klinkhamer:1984di}. These solutions, called sphalerons\footnote{From the classical Greek $\sigma \phi \alpha \lambda \epsilon \rho o \varsigma$, ``ready to fall''.}, are unstable. The sphaleron energy is of the order 
\bee
E_{sph}(T) \simeq \frac{8 \pi v(T)}{g},
\eee
where $v(T)$ is the Higgs VEV at temperature $T$ (and thus $E_{sph}(0)\sim$ 10 TeV); if the energy is high enough, the system can move from a vacuum to an inequivalent one by passing through a sphaleron configuration.  It is unclear if a coherent sphaleron configuration of the fields can be generated in high energy collisions, while it is known that thermal effects can dramatically enhance the sphaleron transition amplitude~\cite{Kuzmin:1985mm}, because thermal fluctuations populate high energy configurations and because the Higgs VEV $v(T)$ decreases with increasing temperature. Sphaleron transitions are in thermal equilibrium in the early Universe for temperatures $T$ such that~\cite{Kuzmin:1985mm,D'Onofrio:2014kta}
\bee\label{eq:sphal_eq}
130 \text{ GeV} \lesssim T \lesssim 10^{12} \text{ GeV}.
\eee
Thus, in the above defined range of temperatures, any $B+L$ asymmetry is effectively erased by sphalerons, while they do not affect the $B-L$ charge.

Regarding the third Sakharov's condition, the standard cosmological model provides several departures from thermal equilibrium~\cite{Kolb:1990vq}, as for instance the BBN process, the CMB decoupling or the kinematical decoupling when the temperature drops below the mass of a given particle. Among these there is one departure that can generate a net baryon asymmetry, which is the electroweak phase transition if it is of first order. In this case there exists a temperature $T_c$ at which the Higgs potential has two degenerate minima: one at $v\neq 0$ that will evolve into the zero temperature (``true'') vacuum state, and one at $v=0$ that will evolve into a local maximum. Around the temperature $T_c$ the Higgs field can tunnel between the two energetically equivalent minima, however when the temperature decreases the true minimum becomes energetically more favourable. The symmetry breaking takes place via the formation of bubbles in the primordial plasma where  $v\neq 0$ (bubble nucleation), that are surrounded by an unbroken phase where $v=0$. The bubbles expand, eventually filling the whole Universe; as a bubble expands the field configuration on the bubble wall rapidly changes, resulting in a departure from thermal equilibrium~\cite{Trodden:1998ym}.

In the SM framework a net baryon asymmetry would be indeed generated if the EWPT was of first order. The mechanism is the following~\cite{Farrar:1993sp,Farrar:1993hn}: in the unbroken phase the sphaleron transitions are effective and set the number of quarks and antiquarks to be equal. As a result of $CP$ violation, however, they possess different transmission and reflection rates among the bubble wall, leading to an excess of quarks transmitted inside the bubble, where the sphaleron rate is strongly suppressed by the large value of the Higgs VEV. The excess is thus preserved, while the antiparticle excess outside the bubble is rapidly erased by sphaleron transitions. The final baryon asymmetry is determined by two parameters: the order of the phase transition and the amount of $CP$ violation quantified by the Jarlskog invariant (which depends on the value of $\delta_{CKM}$). Also assuming that the first criterion was met, it has been shown that the amount of $CP$ violation in the SM is not enough to account for the observed value of $\eta_{\Delta B}$~\cite{Gavela:1993ts,Huet:1994jb,Gavela:1994ds,Gavela:1994dt}. Moreover the order of the phase transition depends on the Higgs boson mass $m_h$, and a strong first order is expected only for $m_h \lesssim 70$ GeV~\cite{Bochkarev:1987wf,Kajantie:1995kf}, a value much smaller than the observed one, $m_h \simeq 125$ GeV. Indeed lattice simulations performed after the determination of the Higgs boson mass confirm that the EWPT is a smooth crossover~\cite{D'Onofrio:2014kta}.

Thus the value of the BAU is a third, firm observation that calls for BSM physics.

\chapter{Neutrino mass generation mechanisms and phenomenology}\label{sec:nu_mass_gen}
As discussed in the previous chapters the observation that neutrinos are massive and mix requires new physics beyond the SM. The determination of the mechanism responsible for these effects is one of the main open questions in modern particle physics, and several possible solutions have been proposed. In the first part of the present chapter we will review the main classes of neutrino mass and mixing generation mechanisms. We will then focus on mechanisms based on the existence new fermions that are singlet under the SM gauge group (sterile fermions), on their phenomenological consequences and constraints. The second part of this chapter is devoted to the detailed analysis of an interesting mechanism, the Inverse Seesaw.

\section{Quarks and leptons: similarities and differences}
The notable distinction among quarks and leptons is that the former are subject to strong interactions, while the latter are not. On the other hand  the $SU(3)_C$ gauge group is not broken and, as far as we know, its dynamics is not related to the mass generation and mixing of the elementary particles (this is not true for baryons and mesons, that are QCD bound states).

Both quarks and leptons are structured in three generations, meaning that their field content is repeated three times, with the only difference among the particles of different generations given by a different Yukawa coupling (flavour structure), that results in a difference in the corresponding masses and mixing properties. Looking at the Table~\ref{matterfields}, the most striking difference in the flavour structure of quarks and leptons in the SM (in its original formulation) is given by the absence of right-handed neutrino fields, which implies that while quarks are massive and mix, neutrinos are massless and leptons do not mix. Given that neutrino oscillation experiments clearly established that neutrinos are massive and mix, a first clue on the neutrino mass generation mechanism can be obtained by comparing the flavour phenomenology in both quark and lepton sectors.

\begin{figure}[htb]
\begin{center}
\includegraphics[width=0.8\textwidth]{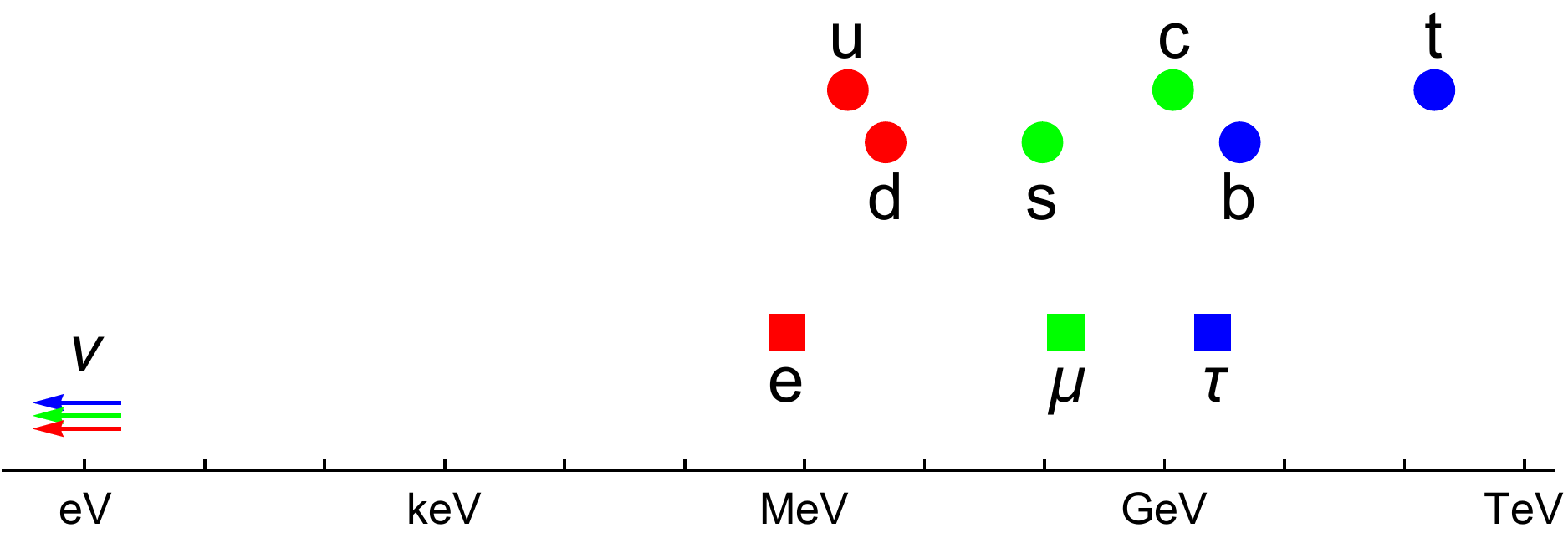}
\end{center}
\caption{Logarithmic distribution of the mass values of the SM fermions, for the first (red), second (green) and third (blue) generations. Quark masses are denoted by circles, while charged leptons by squares. For neutrino masses only the upper bound value (\ref{nueff_bound}) is available. Numerical data from~\cite{Agashe:2014kda}.}
  \label{fig:SMmasses}
\end{figure}

We begin by comparing the masses in the two sectors, reporting the mass values of the SM fermions on a logarithmic scale in Fig.~\ref{fig:SMmasses}, from which the difference between the quark and the lepton sectors is evident.
While the \emph{up} and \emph{down} quarks in the same generation have comparable masses, 
neutrino masses seem to be unrelated with the values of their $SU(2)_L$ partners, the charged leptons. The difference among the heavier neutrino and the electron is at least of 6 orders of magnitude, which grow to 9 when the $\tau$ lepton is considered.

\begin{figure}[htb]
\begin{center}
\includegraphics[width=0.55\textwidth]{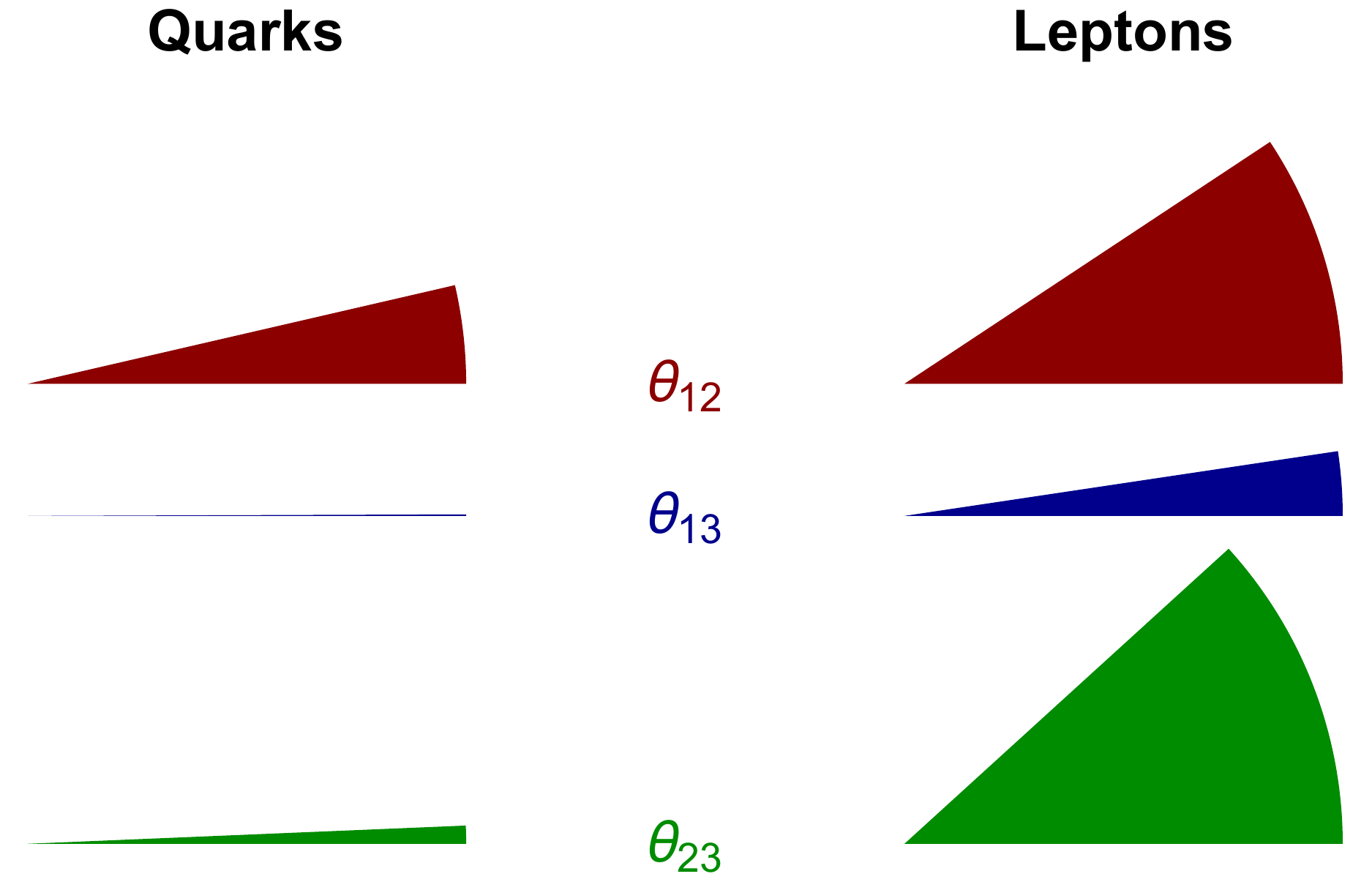}
\end{center}
\caption{Comparison of the mixing angles in the quark and leptonic sectors. Numerical values from~\cite{Agashe:2014kda,Gonzalez-Garcia:2014bfa}.}
  \label{fig:SMmixing}
\end{figure}

The mixing among the generations is intriguing. Comparing the mixing angles (defined by adopting the parametrisation in eq.~(\ref{upmns}) for the mixing matrices)  of the two sectors (see Fig.~\ref{fig:SMmixing}), one can see
 that quarks are weakly mixed in comparison to leptons, having only one sizeable angle $\theta_{12}^q \simeq 13^\circ$, which is of the same order of the smallest leptonic mixing angle, $\theta_{13}^l \simeq 9^\circ$. 
The difference in the mixing structure translates in a potential quantitative difference in the amount of $CP$-violation in the two sectors. The Jarlskog invariant of the quark sector is experimentally determined, since all the parameters of the $CKM$ matrix are (almost) known\cite{Agashe:2014kda}
\bee
J^q = \left(3.06^{+0.21}_{-0.20}\right)\times 10^{-5}.
\eee
The value of the same parameter in the leptonic sector is currently unknown since a determination of the $\delta$ phase in the $PMNS$ matrix~(\ref{upmns}) is still missing. We can nonetheless estimate the allowed range for the Jarlskog invariant from the  values of the known leptonic mixing parameters. Considering for definiteness the best fit values in the NH solution~\cite{Gonzalez-Garcia:2014bfa}, we obtain
\bee
J^l_\text{NH}(\theta_{12}^\text{bf},\theta_{13}^\text{bf},\theta_{23}^\text{bf},\delta) = 3.3 \times 10^{-2}\ \sin \delta.
\eee
Thus, depending on the value of $\delta$, the amount of $CP$ violation in the leptonic sector can be up to 3 orders of magnitude larger than the one in the quark sector.

Any neutrino mass generation mechanism must account for the large hierarchy between the neutrino and the charged lepton masses, and for the  large mixing in the leptonic sector. The simplest mechanism consists in the addition of RH neutrino fields to the SM, generating nonzero neutrino masses via the Higgs mechanism in the same way as the masses of the up-type quarks are generated, thus resulting in massive neutrinos of Dirac nature. This mechanism requires Yukawa couplings smaller than $\mathcal{O}(10^{-11})$ and,  although it is phenomenologically viable, it  does not provide a dynamical explanation for the smallness of neutrino masses. Moreover, the new RH neutrino fields are SM gauge singlets, and thus nothing prevents them to acquire a Majorana mass term: depending on the latter mass scale, the phenomenology can be very rich.

\section{Neutrino mass generation mechanisms}
Several mechanisms have been conceived to address the observation of the tiny neutrino masses~\cite{Mohapatra:1998rq}. A viable minimal option is to extend the field content of the SM, such that the interactions of the new fields can account for massive neutrinos, resulting in the operator~(\ref{d5_mass}) when they are integrated out. The suppression of the neutrino mass scale can be related to a suppression of the $c_{\alpha \beta}$ coefficients, as it is the case for models with an extended Higgs sector if the new VEV is suppressed with respect to the electroweak one~\cite{Cheng:1980qt}. Alternatively, $c_{\alpha \beta}$ can be suppressed if neutrino masses are generated at loop level~\cite{Cheng:1980qt,Zee:1980ai,Babu:1988ki,Ma:2006km}. Another possibility is offered by models with an approximated $B-L$ symmetry, since in the limit in which the symmetry is restored neutrinos are massless. A small violation of the symmetry is then connected to small neutrino masses; this is the case, for instance, of supersymmetric models with $R$-parity violation~\cite{Ellis:1984gi,Ross:1984yg,Romao:1999up,Abada:1999ai,Hirsch:2000ef,Abada:2001zh}. On the other hand, the Weinberg operator~(\ref{d5_mass}) could contain $\mathcal{O}(1)$ coefficients, with the new fields lying at an energy scale $\Lambda \gg v$. In this case the neutrino mass will be suppressed by the small value of the ratio $v/\Lambda$. This is the so called \emph{Seesaw mechanism}, which can be implemented in three basic ways, namely by the addition of fermionic gauge singlet fields (Type-I Seesaw)~\cite{Minkowski:1977sc,GellMann:1980vs,Yanagida:1979as,Glashow:1979nm,Mohapatra:1979ia}, of an $SU(2)_L$ triplet of scalar fields (Type-II Seesaw)~\cite{Barbieri:1979ag,Marshak:1980yc,Magg:1980ut,Lazarides:1980nt,Schechter:1980gr,Cheng:1980qt,Mohapatra:1980yp,Wetterich:1981bx} or of an $SU(2)_L$ triplet of fermionic fields (Type-III Seesaw)~\cite{Foot:1988aq,Ma:1998dn}.

A less minimal option to account for massive neutrinos is to enlarge the SM gauge group, by considering for instance an additional $U(1)_{B-L}$ gauge symmetry~\cite{GellMann:1980vs,Yanagida:1979as,Witten:1979nr,Mohapatra:1979ia,Mohapatra:1980qe,Chikashige:1980ht,Mohapatra:1980yp}, which is an anomaly-free combination of charges in the SM. This group also appears in left-right symmetric models, where it comes together with an additional $SU(2)_R$ gauge symmetry~\cite{Pati:1974yy,Mohapatra:1974hk,Senjanovic:1975rk,Marshak:1979fm,Davidson:1978pm}. Theoretically appealing extensions of the SM gauge group are the grand unified theories (GUT), where the SM gauge structure derives from the breaking of a single gauge group which is unbroken at some high-energy scale~\cite{Mohapatra:1986uf,ross1990grand}. Notable examples of  possible GUT groups are $SU(5)$~\cite{Georgi:1974sy}, $SO(10)$~\cite{Georgi:1974my,Fritzsch:1974nn,Witten:1979nr} and $E_6$~\cite{Gursey:1975ki,Achiman:1978vg,Shafi:1978gg,Gursey:1981kf}, which contains $SU(5)$ and $SO(10)$ as subgroups. The $SO(10)$ group is especially attractive, since it contains both $U(1)_{B-L}$ and $SU(2)_R$ as subgroups, and it requires the existence of right-handed neutrino fields.

It is worth to mention mechanisms providing a suppression of the neutrino mass scale in the context of extra dimensions~\cite{ArkaniHamed:1998vp,Dienes:1998sb,Grossman:1999ra}. Neutrinos can acquire small Dirac masses, with the suppression of the Yukawa couplings related to the localisation of the right-handed neutrino fields in the extra space, or small Majorana masses if the lepton number is broken by a field localised in the extra space. Finally small neutrino masses are also provided in some string theory realisations~\cite{Ibanez:2001nd,Beasley:2008kw}.

\section{Phenomenology of sterile fermions}
As discussed in Section~\ref{sec:RH_masses}, the addition of sterile fermions to the SM field content naturally leads to massive neutrinos. However, the Weinberg operator~(\ref{d5_mass}) is common to any BSM realisation that provides Majorana massive neutrinos, and the experimental data on neutrino masses and mixing only constrain the value of the $c_{\alpha \beta}$ coefficients. Thus the data on active neutrinos can only constrain the parameters of a given model, but they are not sufficient to unveil the underlying neutrino mass generation mechanism. In order to do that, it is necessary to consider other manifestations of the new physics, that are encoded in the higher dimensional operators ($d\geq 6$) of the effective field theory expansion. These new physics effects are connected with the phenomenology of the new states that are present in the model: if they are gauge singlets, they  can couple to the SM gauge bosons only indirectly via mixing effects. Thus the phenomenology of mechanisms based on fermionic gauge singlets is completely encoded in the mixing matrix $\mathcal{U}$ appearing in eq.~(\ref{eq:symm_diagonal}) and in the value of the masses of the heavy sterile fermions. These parameters are determined by the mass matrix $\mathcal{M}$. In order to investigate the phenomenology of the mechanisms, it is convenient to diagonalise the mass matrix in two steps, firstly by putting it in a block diagonal form via the transformation~\cite{Kanaya:1980cw,Schechter:1981cv}
\bee\label{eq:block_diag}
\Xi ^T \mathcal{M} \Xi = \left( \begin{array}{cc} m_l & 0 \\
 0 & M_h
 \end{array}\right),
\eee
where $\Xi$ is a $(3+n)$-dimensional unitary matrix, $n$ being the number of sterile fermions. The dimensions of the blocks $m_l$ and $M_h$ can be arbitrarily arranged, and it is convenient to collect   in $m_l$ the ``light'' mass eigenstates and in $M_h$ the ``heavy'' ones, where light and heavy are to be intended in comparison with the energy of the considered process. In the following discussion we fix for definiteness the dimensions of $m_l$ and $M_h$ to be $(3\times 3)$ and $(n\times n)$, respectively, that is we assume that the only light states are the ones determined by neutrino oscillation data, keeping however in mind that the results can be easily adapted if $k$ further light states are present, replacing 3 with $3+k$. The matrix $\Xi$ can be parametrised as the exponential of an anti-hermitian matrix~\cite{Antusch:2009pm}
\bee\label{eq:xi_par}
\Xi =  \text{exp} \left(\begin{array}{cc} 0 & \Theta \\ -\Theta^\dagger & 0 \end{array}\right) = \left(\begin{array}{cc} 
1- \frac{1}{2} \Theta \Theta^\dagger & \Theta \\ 
- \Theta^\dagger & 1-\frac{1}{2}\Theta^\dagger \Theta \end{array}\right) + \mathcal{O}(\Theta^3),
\eee
where $\Theta$ is a $(3 \times n)$ matrix. The submatrices in the block rotated mass matrix~(\ref{eq:block_diag}) can be diagonalised by two unitary rotations, $\hat{m}_\nu = U^T m_l U, \hat{M}_s = V^T M_h V$, where $U,V$ are unitary matrices and the hat denotes a diagonal matrix. Thus the unitary rotation $\mathcal{U}$ in eq.~(\ref{eq:symm_diagonal}) can be expressed as
\bee\label{eq:non_unit_U}
\mathcal{U} = \Xi \left(\begin{array}{cc} U & 0 \\ 0 & V \end{array} \right) = \left(\begin{array}{cc} 
\left(1- \frac{1}{2} \Theta \Theta^\dagger\right) U & \Theta V\\
-\Theta^\dagger U & \left(1-\frac{1}{2}\Theta^\dagger \Theta\right)V
\end{array}\right) + \mathcal{O}(\Theta^3).
\eee
By defining the neutrino mass basis $\{\chi^i\}$ as in eq.~(\ref{mass_basis_rot}), the weak charged current~(\ref{eq:charged_int}) reads
\be\label{eq:weak_charged_SF}
\mathcal{L}_W = -\frac{g}{\sqrt{2}}  \overline{e_\alpha}\  \slashed{W}^-\mathcal{U}_{\alpha i}\  {\chi^i_L}_\alpha + h.c.,
\ee
while the neutral current between neutrinos~(\ref{eq:weak_SM}) reads
\bee\label{eq:weak_neutral_SF}
\mathcal{L}_Z^{\nu \nu} = - \frac{g}{2\cos{\theta_W}} \overline{\chi_L^i}\mathcal{U}^\dagger_{i \alpha}\ \slashed{Z}\ \mathcal{U}_{\alpha j} \chi_L^j,
\eee
where we used the convention $\slashed{a} = a_\mu \gamma^\mu$.
There are thus two important phenomenological consequences. The first one is that a charged lepton of flavour $\alpha$ is coupled to all the fermions $\chi^i$, with a strength proportional to the mixing element $\mathcal{U}_{\alpha i}$, giving rise to non universal weak interactions. The second one is that two fermions $\chi^i,\chi^j$ are coupled between them with a coupling proportional to the combination
\bee
C_{ij} \equiv \sum_{\alpha = e,\mu,\tau} \mathcal{U}_{\alpha i}^* \  \mathcal{U}_{\alpha j}.
\eee
Notice that since the sum is performed only on the first 3 rows of the mixing matrix, the coefficients $C_{ij}$ are in general different from zero also for $i\neq j$, giving rise to non diagonal interactions.

By denoting with $\chi^i$, $i=1,2,3$, the three mass eigenstates of the active neutrinos, as  defined in Section~\ref{sec:global_fit}, the $(3\times 3)$ upper-left block of $\mathcal{U}$ corresponds to the PMNS matrix
\bee
N_\text{PMNS} = \left(1- \frac{1}{2} \Theta \Theta^\dagger\right) U + \mathcal{O}(\Theta^3).
\eee
We denote the $(3 \times 3)$ leptonic mixing matrix by $N_\text{PMNS}$ to account for the fact that it is in general non-unitary, the deviation from unitarity being parametrised by the matrix $\Theta$. The same matrix determines the strength of the interactions between the charged leptons and the other mass eigenstates,
\bee\label{eq:act_ster_mix}
\mathcal{U}_{\alpha I} = \left(\Theta V\right)_{\alpha I} + \mathcal{O}(\Theta^3), \hspace{1cm} I\geq 4.
\eee
Therefore the manifestations of new physics other than neutrino masses and mixing are determined by the matrix $\Theta$: in the limit $\Theta=0$ the PMNS matrix is unitary, the active leptons are only coupled with the three neutrino eigenstates and the coefficients $C_{ij}$ are diagonal. 

There exist experimental constraints on both the deviation from unitarity of the PMNS matrix and on the $\mathcal{U}_{\alpha I}$ elements, that will be reviewed in the following sections.

\subsection{Unitarity of the leptonic mixing matrix}
Any deviation of the PMNS matrix from unitarity can have observables effects, notably in a violation of the flavour universality in weak interaction processes~\cite{Antusch:2006vwa,FernandezMartinez:2007ms,Antusch:2008tz}. In the presence of mixing, the charged current~(\ref{eq:weak_charged_SF}) implies that the amplitude for any tree-level decay process involving a neutrino and a charged lepton in the final state depends on the coupling
\bee
\mathcal{A} \left(X \rightarrow \ell_\alpha \nu_i\right) \propto -\frac{g}{\sqrt{2}}\mathcal{U}_{\alpha i}.
\eee
If the neutrino mass is not resolved, the experiment can only measure the total width for the decay in a given flavour $\alpha$, given by the incoherent sum over all the possible final states, that is all fermions that are kinematically accessible ($m_{\nu_i} + m_{\ell_\alpha} < M_X$)
\bee\label{eq:non_univ_lnu}
\Gamma \left(X \rightarrow \ell_\alpha \nu_\alpha\right) \propto \frac{g^2}{2} \sum_i^\text{kin}\left|\mathcal{U}_{\alpha i}\right|^2.
\eee 
If the mass eigenstates of the mass matrix are such that they are all kinematically accessible, the sum in eq.~(\ref{eq:non_univ_lnu}) equals 1 due to the unitarity of the matrix $\mathcal{U}$ and the SM prediction is recovered. In the opposite case, the sum is smaller than one and the flavour universality in the weak interactions is violated in general
\bee
\sum_i^\text{kin}\left|\mathcal{U}_{\alpha i}\right|^2 \neq \sum_i^\text{kin}\left|\mathcal{U}_{\beta i}\right|^2, \hspace{1cm} \text{for } \alpha \neq \beta.
\eee
It is important to notice that the deviation from universality in a given process is a function of the $Q$ value of the reaction, since the upper bound in the sum depends on the quantity $M_X- m_{\ell_\alpha} $. It is in this sense that the block diagonalization in light and heavy states of eq.~(\ref{eq:block_diag}) is phenomenologically meaningful.

Weak decay processes that are sensitive to a deviation from flavour universality are for instance the $W$ boson leptonic decays, $W \rightarrow \ell_\alpha \nu_\alpha$, and the 3-body charged lepton decays, $\ell_\alpha \rightarrow \ell_\beta \nu_\alpha \nu_\beta$. Further interesting processes are the leptonic and semileptonic decays of mesons, $M \rightarrow \ell_\alpha \nu_\alpha$, $M \rightarrow N \ell_\alpha \nu_\alpha$, where $M,N$ are mesons. Since the branching ratios of these processes are affected by large uncertainties related to the hadronic matrix elements, more suitable observables to constrain new physics scenarios are the ratios of the above decays~\cite{Abada:2012mc,Abada:2013aba}
\bee
R^M_{\alpha \beta} &=& \frac{\Gamma \left(M  \rightarrow \ell_\alpha \nu_\alpha\right)}{\Gamma \left(M \rightarrow \ell_\beta \nu_\beta\right)},\non 
R^M_{\alpha \beta}\left(N\right) &=& \frac{\Gamma \left(M  \rightarrow N \ell_\alpha \nu_\alpha\right)}{\Gamma \left(M \rightarrow N \ell_\beta \nu_\beta\right)},
\eee
where the hadronic uncertainties cancel out to a very good approximation. Also notice that  the Fermi constant $G_\mu$, measured in the muon decay process $\mu \rightarrow e \nu_\mu \nu_e$, differs from the SM definition $G_F$ in the presence of the current~(\ref{eq:weak_charged_SF}). The total width of the process is related to the SM one by
\be 
\Gamma_{\mu\rightarrow e \nu \nu} = \Gamma_{\mu\rightarrow e \nu \nu}^{\mbox{\tiny{\text{SM}}}}  \sum_{i,j}^\text{kin} |U_{\mu i}|^2 |U_{e j}|^2,
\ee
implying
\be 
G_\mu^2 = G_F^2 \sum_{i,j}^\text{kin} |U_{\mu i}|^2 |U_{e j}|^2.
\ee

Analogously to the charged case, the modified weak neutral current~(\ref{eq:weak_neutral_SF}) can cause a deviation from the SM predictions if some of the eigenstates of the mass matrix is not kinematically allowed in a given process. For instance the width of the invisible decay of the $Z$ boson is proportional to
\bee
\Gamma_\text{inv} \left(Z\right) \equiv \sum_{i,j} \Gamma \left(Z \rightarrow \nu_i \nu_j\right) \propto \frac{g^2}{4 \cos^2 \theta_W} \sum_{i,j}^\text{kin} \left|C_{ij}\right|^2,
\eee
with the sum that deviates form the SM prediction if some masses are such that {$m_i + m_j > M_Z$}.

Finally a deviation from unitarity of the leptonic mixing matrix can be manifest at loop level in processes that are absent at tree-level, notably in lepton flavour violating decays of charged leptons (cLFV), $\ell_\alpha \rightarrow \ell_\beta \gamma$~\cite{Petcov:1976ff,Bilenky:1977du,Cheng:1980tp,Abada:2008ea}, $\ell_\alpha \rightarrow \ell_\beta \ell_\gamma \ell_\delta$~\cite{Ilakovac:1994kj} and in the $\mu \rightarrow e$ conversion in nuclei, $\mu N \rightarrow e N$~\cite{Deppisch:2005zm,Dinh:2012bp,Alonso:2012ji}. Sterile fermions in any range of mass contribute to these processes by propagating in the loops, and since a GIM mechanism is present the final amplitude is a function of the mass differences between the different states. For instance the branching ratio for the process $\mu \rightarrow e \gamma$ is given by
\be\label{Petcov:1976ff,Bilenky:1977du,Cheng:1980tp}
\text{Br}(\mu \to e \gamma) = \frac{3 \alpha}{32 \pi}\left|\sum_{i} \mathcal{U}_{\mu i}^* \mathcal{U}_{e i} G\left(\frac{m_i^2}{M_W^2}\right)\right|^2 ,
\ee
where $G$ is a loop function 
\bee
G(x) = \frac{10-43x+78x^2-49x^3+4x^4+18x^3\log(x)}{3(x-1)^4}.
\eee
If all the masses $m_i$ were equal the unitarity of the mixing matrix would imply an exact cancellation of the different contributions when summed over $i$. The function $G(x)$ has the limiting values $G(0) = 10/3$ and $\lim_{x\rightarrow \infty} G(x) = 4/3$, thus in a configuration in which $\mathcal{O}(m_l) \ll M_W$ and $\mathcal{O}(M_h) \gg M_W$ the branching ratio can be approximated as
\be \label{Petcov:1976ff,Bilenky:1977du,Cheng:1980tp}
\text{Br}(\mu \to e \gamma) \simeq \frac{3 \alpha}{32 \pi}\left|2 \left(NN^\dagger\right)_{e \mu} \right|^2 ,
\ee
which is proportional to the deviation form unitarity of the PMNS matrix.

The deviation from unitarity of the PMNS matrix is strongly constrained by experimental data. In the assumption that the masses of the heavy degrees of freedom are large enough such that they can be safely integrated out, which is accurate if they lie above the electroweak energy scale, the current bounds at 90\% C.L. are~\cite{Antusch:2014woa}
\be
\left| NN^\dagger \right| = \left( \begin{array}{ccc} 0.9979 - 0.9998 & < 10^{-5} & < 0.0021 \\
< 10^{-5} & 0.9996 - 1.0 & < 0.0008 \\
< 0.0021 & < 0.0008 & 0.9947 - 1.0 \end{array}\right)\,.
\ee
Thus in the parametrisation~(\ref{eq:non_unit_U}) the $\mathcal{O}(\Theta^3)$ terms can be safely neglected.

\subsection{Direct searches of sterile fermions}
The observables discussed in the previous section are indirect effects related to the presence of sterile fermions.  These particles can be looked for also directly in laboratory searches~\cite{Atre:2009rg}. For both the Dirac and Majorana case a sterile fermion can be produced in the leptonic decay of a meson via the current~(\ref{eq:weak_charged_SF}), $M \rightarrow \ell_\alpha \chi_i$~\cite{Shrock:1980vy}. Being a two-body process  the kinematic is completely determined  by the masses of the involved particles, with the energy of the lepton given by
\bee
E_\ell = \frac{m_M^2+m_\ell^2-m_i^2}{2 m_M}.
\eee
Thus a sterile fermion would manifest as a monochromatic line in the energy spectrum of the charged lepton, with the position related to the sterile fermion mass and the intensity proportional to the mixing element $\mathcal{U}_{\alpha i}$. Especially interesting channels are the pion and kaon decays, because of their large leptonic branching ratios. Sterile fermions can also be produced in three-body decays, as for instance in $\beta$-decay processes, in which case they manifest as kinks in the energy spectra of the other decay products, as discussed in Section~\ref{sec:end_point}. While the above referred channels are relevant if the sterile fermion is lighter than the mass of the decaying particle, the region of heavier masses can be studied in accelerator experiments by looking at the products of the sterile fermion decay. If kinematically allowed, sterile fermions can be produced in accelerators via $W$ and $Z$ mediated processes, cf. eqs.~(\ref{eq:weak_charged_SF}, \ref{eq:weak_neutral_SF}), with a branching fraction proportional to the quantities $|\mathcal{U}_{\alpha i}|^2$ and $|C_{ij}|^2$. The sterile fermion subsequently decays in SM particles by means of the same weak currents, with a decay width proportional to the same couplings. From the non-observation of the decay products experimental bounds on the mixing of the sterile fermions can be put, as a function of the sterile fermion mass.

The above discussed processes are especially phenomenologically relevant since they constrain the mixing element of a single sterile fermion. Contrary to that, in processes where the sterile particles appear as virtual states cancellations among the different contributions can arise. The constraints related to the  latter class of processes  apply to the combination of masses and mixing for the whole mass spectrum.

If the sterile fermions are Majorana states they can mediate LNV processes, as for instance the $0\nu\beta \beta$ decay discussed in Section~\ref{sec:0betanunu}. Relevant processes are also the LNV tau decays, $\tau^- \rightarrow \ell_\alpha^+ M_1^- M_2^-$, and the LNV meson decays, $M_1^+ \rightarrow \ell_\alpha^+ \ell_\beta^+ M_2^-$, since their matrix elements are resonantly enhanced if the sterile fermion mass is  $\mathcal{O}(0.1, 1)$ GeV, resulting in very stringent bounds for the mixing elements in this mass range~\cite{Atre:2009rg}.

Finally in the very low mass range, $m_i \lesssim 1$ eV, a sterile fermion can manifest in neutrino oscillation experiments, modifying the results expected from the three-flavour mixing pattern~(\ref{upmns}). The non observation of these effects puts upper bounds on the mixing elements~\cite{Declais:1994su,Agafonova:2013xsk,Antonello:2013gut,An:2014bik,Timmons:2015lga}.

\section{Lepton number violation and the new physics scale}\label{sec:LNV_scale}
The discussion performed so far is valid for any mechanism in which the SM field content is extended by the addition of sterile fermions. Nonetheless, the exact phenomenology of a given model depends on the possible values of the masses and mixing elements for the new states. 

The equation~(\ref{eq:block_diag}) with the matrices $\Xi$ and $\mathcal{M}$ parametrised as in (\ref{eq:xi_par}, \ref{general_Seesaw_matrix}) implies, at the lowest order in $\Theta$, the relations~\cite{Ibarra:2010xw}
\bee\label{eq:Seesaw_pert_theta}
\Theta^* &\simeq & m\ M^{-1},\non 
m_l &\simeq & -m\ \Theta^\dagger - \Theta^* m^T + \Theta^* M \Theta \simeq -\Theta^* M \Theta^\dagger \simeq -m\ M^{-1} m^T, \non 
M_h &\simeq & M +\Theta^T \Theta^* M + M \Theta^\dagger \Theta
.
\eee
These equations imply the following relation among the neutrino mass scale $m_l$ and the couplings $\mathcal{U}_{\alpha I}$ between the charged leptons and the heavy states  $\chi_I$
\bee
\left|\sum_I \mathcal{U}_{\alpha I}\ \mathcal{U}_{\beta I}\ m_I\right| \simeq \left| m_l^{\alpha \beta} \right|.
\eee
An important discriminating factor in the phenomenological discussion is the mechanism responsible for the suppression of the neutrino mass scale: if the smallness of the ratio $m_\nu/v$ is only due to the smallness of the ratio $v/\Lambda$ in~(\ref{d5_mass}), the large value of the new physics scale $\Lambda$ necessarily implies a strong suppression of the higher dimensional operators, $d\ge 6$, in the effective theory expansion 
\be\label{eq:eft_expansion}
\mathcal{L}_{eff} = \mathcal{L}_{SM} + \frac{c_5}{\Lambda}\mathcal{O}^{d=5} + \frac{c_6^i}{\Lambda^2}\mathcal{O}_i^{d=6} + \dots
\ee
Thus in this framework new physics effects other than neutrino masses are difficult to observe. This is what happens for instance in the Type-I Seesaw mechanism, where the relations~(\ref{eq:Seesaw_pert_theta}) imply
\bee
\Theta^* &\simeq & \frac{v}{\sqrt{2}}Y^*\ M^{-1},\\ 
m_l &\simeq &   -\frac{v^2}{2}Y^* M^{-1} Y^\dagger.
\eee
If the submatrices $m$ and $M$ in~(\ref{general_Seesaw_matrix}) do not have any substructure and barring accidental cancellations between the (a priori independent) entries of the matrices $Y$ and $M$,
the smallness of the ratio $\mathcal{O}(m)/\mathcal{O}(M)$ required to accommodate neutrino masses necessarily implies a suppression of the active-sterile mixing $\Theta V$. 

As pointed out in~\cite{Abada:2007ux}, the phenomenology is different if the suppression of the five dimensional operator in~(\ref{eq:eft_expansion}) is not related to a suppression of the higher-dimensional operators. This is notably the case of mechanisms characterised by an approximate lepton number conservation: the five-dimensional operator~(\ref{weinbergmass}) violates lepton number by two units, and its coefficient is necessarily zero if the Lagrangian preserves the total lepton number. On the other hand the $d > 5$ operators in~(\ref{eq:eft_expansion}) can violate or preserve the total lepton number, and they do not necessarily vanish when the symmetry is restored. Hence new physics effects are not necessarily suppressed by the small value of neutrino masses. Examples of mechanisms of this kind are the linear~\cite{Akhmedov:1995ip,Akhmedov:1995vm} and the inverse Seesaw (ISS)~\cite{Mohapatra:1986bd,GonzalezGarcia:1988rw,Deppisch:2004fa}, where  pairs of fermionic singlets $(\nu_R, s)$ with lepton number $L=1$ are added to the SM. In the ISS the submatrices $m$ and $M$ in the mass matrix~(\ref{general_Seesaw_matrix}) read, in the basis $(\nu_L,\nu_R^c,s)$,
\bee
m &=& \left(\begin{array}{cc} d & 0 \end{array}\right),\non 
M &=& \left( \begin{array}{cc} 0 & n \\
 n^T & \mu \end{array}\right),
\eee
where $d,n$ are complex matrices and $\mu$ is a complex symmetric matrix. The matrix $d$ arises from the Yukawa couplings between the left- and right-handed neutrino fields $\nu_L$ and $\nu_R$ after the EWSB, while the matrix $n$ is related to the new physics energy scale $\Lambda$. The matrix $\mu$ is the only entry in the mass matrix that violates the total lepton number and  the hierarchy $\mathcal{O}(\mu) \ll \mathcal{O}(d) < \mathcal{O}(n)$ is assumed. The relations~(\ref{eq:Seesaw_pert_theta}) give in this case
\bee\label{eq:ISS_mass_mix}
\Theta^* &=& \left(\begin{array}{lr} -\frac{v}{\sqrt{2}} Y^* \left(n^T\right)^{-1} \mu\ n^{-1}\hspace{0.5cm} & \frac{v}{\sqrt{2}} Y^* \left(n^T\right)^{-1} \end{array}\right),\non 
m_l &=&  -\frac{v^2}{2} Y^* \left(n^T\right)^{-1} \mu\ n^{-1} Y^\dagger. 
\eee
Thus the suppression of the neutrino mass scale is related to the smallness of the lepton number violating parameter $\mu$. In the limit $\mu\rightarrow 0$ neutrinos are massless, but the matrix $\Theta$ does not vanish: new physics effects are not suppressed by the small value of the neutrino mass scale.

The scale of the LNV terms in the Lagrangian strongly affects the phenomenology of a given neutrino mass generation mechanism, and in the absence of direct experimental inputs the value of this scale is unbound. In the rest of this work we will concentrate on mechanisms characterised by an approximate lepton number conservation, since they allow to accommodate neutrino data for relatively low values of the new physics scale, $\mathcal{O}$(TeV) or lower, making it possible to test this kind of models in current and future laboratory experiments.

\section{Looking for the minimal inverse Seesaw realisation}\label{Sec:intro}

In view of the strong potential of low-scale Seesaw mechanisms, in this section we consider the inverse Seesaw (ISS)~\cite{Mohapatra:1986bd,GonzalezGarcia:1988rw,Deppisch:2004fa} which requires the addition of both a number  $\#\nu_R\ne 0$ of right-handed (RH) neutrino fields
and a number $\# s\ne 0$ of extra sterile fermions to the SM field content\footnote{In the case where $\# s=0$, one recovers the type I  Seesaw realisation which could account for neutrino masses and mixings provided that the number of right-handed neutrinos is at least $\#\nu_R=2$.}. As discussed in Section~(\ref{sec:LNV_scale}),  a distinctive feature of the ISS is that an additional dimensionful  
parameter ($\mu$) allows to accommodate the smallness of the active neutrino
masses $m_\nu$ for a low Seesaw scale, and still with natural Yukawa
couplings ($Y^\nu\sim {\mathcal{O}}(1)$), cf. eq.~(\ref{eq:ISS_mass_mix}).  In turn, this allows for 
sizeable mixings between the active and the additional
sterile states. Such features are in clear contrast with, for instance, the canonical type I Seesaw~\cite{Minkowski:1977sc,GellMann:1980vs,Yanagida:1979as,Glashow:1979nm,Mohapatra:1979ia},
where $\mathcal{O}(1)$ Yukawa couplings require the mass of the right-handed neutrinos to be close to the GUT scale, 
$M_R \sim 10^{15}$~GeV, thus leading to extremely small active-sterile mixings.

Any type I Seesaw realisation requires the introduction of $N$ gauge singlet Weyl fermions $w$ that can thus couple via a Majorana mass term $\sim M_{ij} w^c_i w_j$. Both the number $N$ and the energy scale $M$ are in principle free parameters that can be fixed by neutrino data. It is thus natural to ask what is the minimal number of fermionic singlets  $N$ required to successfully generate neutrino masses and mixings in agreement with observation.  It was shown in~\cite{Donini:2011jh} that the choice $N=1$, although containing in principle enough parameters, fails in fitting the neutrino oscillation data, while the choice $N=2$ is the minimal one that is phenomenologically viable. 
Minimal models, in the framework of low-scale Seesaw mechanisms characterised by an approximate lepton number conservation,  have been addressed in~\cite{Gavela:2009cd}, where $N=2n$ Weyl fermions were added to the SM field content, with a  lepton number assignment allowing them to be cast into two groups of $n$ elements with opposite lepton number charges. It was found that the minimal phenomenologically viable model is the one with $n=1$, which can be the mechanism at work  if all the (gauge invariant) lepton number violating interactions are allowed.  In this situation the tree level neutrino masses derive from the sum of two terms which are differently suppressed by the high-energy scale - and which  depend on the  two sets of Yukawa couplings that are present (lepton number violating and conserving ones) - while the coefficients of the LFV dimension 6 operators only  depend,  to a first approximation, on the lepton number conserving Yukawas. 
The situation is different in the case of the inverse Seesaw scenario, where LNV Yukawas are not allowed and the dimension 5 and 6 effective operators have the same high-energy suppression~\cite{Gavela:2009cd}. In this case the minimal phenomenologically viable model is the one with $n=2$, that is $N=4$. 

Usually, in the  inverse Seesaw scenario, where a LNV parameter $\mu$ is present,  an equal number of singlet Weyl fermions with opposite lepton number is added to the SM field content, i.e. $N=n+n$.  After the diagonalization of the neutral mass matrix, one ends up with three active neutrinos (at least  two massive in order to accommodate neutrino data) and  $n$ pseudo-Dirac pairs with mass differences of the order of the LNV parameter $\mu$. Notice that in this scenario  the scale $\mu$ does not correspond to the mass of any new physical state (after diagonalization). 
In this analysis,  we will consider  the inverse Seesaw scenario in which  we relax  the previous assumption, by adding $N=n+n'$ Weyl fermions with opposite lepton number,  with $n$  not necessarily coinciding with $n'$. 
We will show that when $n\ne n'$, the LNV scale $\mu$ can indeed correspond to the mass of a physical (almost sterile) state, i.e.,  a light sterile neutrino.

Since both RH neutrinos and sterile states are gauge singlet, there is no requirement on their (field) number from anomaly cancellation.
Moreover, in view of the presence of two independent mass scales 
(the mass of the RH neutrinos and the Majorana mass of the sterile states), associated to gauge singlet fermions, it is natural to investigate which is the 
minimal content of the ISS extension of the SM successfully accounting for neutrino data, while at the same time complying with all available experimental and observational constraints. 

We thus embed the inverse Seesaw  mechanism into the SM, considering
models with an arbitrary non-vanishing (and different) number of RH neutrinos and of additional sterile states, in order to establish which class of models provides a  minimal 3-flavour and   3 +~more-mixing schemes. The latter class of realisations (configurations) may 
offer an explanation to the  reactor anomalies or, depending on the mass scales, a solution for the Dark Matter (DM) problem, in the form of a sterile neutrino DM candidate (see Chapter~\ref{sec:DMMISS}). 
In a first stage, we do not impose a particular mass  scale for the (RH) Majorana states nor the hierarchy of the associated light spectrum; 
likewise, we do not specify a mass range for the sterile fields. 

Our study has allowed to identify two
classes of minimal ISS realisations that can successfully account for neutrino data:  
the first leads to a 3-flavour mixing scheme, and requires only two scales (that of light neutrino masses, $m_\nu$, and the mass of the RH neutrinos, $M_R$); 
the second corresponds to a 3~+~1-mixing scheme, and calls for an additional scale 
($\mu$ $\in[m_\nu,M_R]$).  
For each of these minimal classes, we carried  a numerical analysis taking into account all possible bounds associated to the presence of sterile fermions (which 
constrain the mixings between active and sterile neutrinos for different mass regimes). 
We  also provide predictions regarding the hierarchy of the light neutrino spectrum (normal or inverted) and the effective mass in neutrinoless double beta decay, for each of the minimal realisations identified. 

The rest of the chapter is organised as follows:  in Section~\ref{Sec:towards}, we 
briefly review the inverse Seesaw mechanism and define the framework; we also determine the generic class of frameworks leading to 3- and to 3 +~more-mixing schemes as well as their generic features concerning the different mass scales. 
In Section~\ref{Sec:constrains}, we  consider all the different constraints 
from neutrino data, electroweak observables and laboratory measurements 
 applied  in the analysis. Section~\ref{Sec:analysis} 
is devoted
to the phenomenological analysis of the minimal ISS framework leading to the 3-flavour and to the 3~+~1-mixing schemes.   
For completeness, some  technical details concerning the computation
are included in the Appendix~\ref{sec:min_ISS}.

\subsection{Towards the minimal inverse Seesaw realisation}\label{Sec:towards}
In this analysis we consider the inverse Seesaw mechanism~\cite{Mohapatra:1986bd,GonzalezGarcia:1988rw,Deppisch:2004fa} for the generation of  neutrino masses and lepton mixings,  with a minimal field content. 
We work in the framework of the SM extended by one or more generations of right-handed neutrinos $\nu_R$ and additional fermionic singlets~$s$. 

\subsubsection{The one generation case} \label{largeM}
We first consider the illustrative one generation case. 
In the basis $n_L \equiv \left( \nu_L,\nu_R^c,s \right)^T$,  the  neutrino mass term  reads:
\be
-\mathcal{L}_{m_\nu} =\frac{1}{2} n_L^T\ C\ {M}\ n_L + \text{h.c.},
\ee
where $C\equiv i \gamma^2 \gamma^0$ is the charge conjugation matrix and  the matrix ${M}$ is given by
\be\label{isszeromatrix}
{M} = \left( \begin{array}{ccc} 0 & d & 0 \\ d & m & n \\ 0 & n & \mu \end{array} \right).
\ee
We assume that there is no term mixing the left-handed neutrino with  the fermionic singlet $s$ ($ \sim \overline{\nu_L^c} s$). 
In the above, $d$ corresponds to the Dirac mass term. The matrix $ {M}$ also includes  a Majorana mass term  for the RH neutrino,
\be\label{majmassright}
- \frac{m^*}{{2}} \nu_R^T C \nu_R + \text{h.c.}\,.
\ee
The values of $m$ and $\mu$ in Eq.~(\ref{majmassright})  are arbitrary. However, accommodating neutrino masses of $\mathcal{O}(\text{eV})$ implies that both  must be very small in the case of the inverse Seesaw framework.
Assigning  a leptonic charge to both $\nu_R$ and $s$, with lepton number
$L=+1$~\cite{Mohapatra:1986bd,GonzalezGarcia:1988rw,Deppisch:2004fa}
 (such that the Dirac mass term $-d^* \overline{\nu_L} \nu_R + \text{h.c.}$ preserves the leptonic number),  the terms  
$\nu_R^T C \nu_R$ and $s^T C s$ 
violate total leptonic number $L$ by two units. 
Small values of $m$ and 
 $\mu$  are  natural in the sense of 't~Hooft~\cite{'tHooft:1979bh} since in the limit where 
$m, \mu\to 0$, the total lepton number symmetry is restored. 
 In the following, we  assume for simplicity  that $\mu$ and $m$ are of the same order of magnitude.

In order to obtain the {tree-level} neutrino mass spectrum and the leptonic mixing, we diagonalise the matrix ${M}$ as \cite{Schechter:1980gr}
 \be\label{diagonalization}
U^T M U = \mbox{diag}(m_0,m_1,m_2)\,,
\ee
where $U$ is a unitary matrix, and $m_{0,1,2}$  
correspond to the physical neutrino masses. The mixing matrix is obtained from
\be\label{diagmsquare}
\mbox{diag}(m_0^2,m_1^2,m_2^2) = \left( U^T M U \right)^\dagger \left( U^T M U \right)  = U^\dagger M^\dagger M U\, ,
\ee
so that the matrix $U$ diagonalising $M^\dagger M$ is the same as  the one in Eq.~(\ref{diagonalization}).

We determine the neutrino spectrum {\it perturbatively}: 
the perturbations correspond to taking into account the tiny effects of the lepton number violating 
diagonal entries, 
\be\label{DeltaM}
\Delta M = \mbox{diag}(0,m,\mu)\,.
\ee 
The lightest neutrino mass arises from perturbative corrections\footnote{We denote by $(n)$ superscript perturbative corrections of order $n$.} to the zeroth order $m_0=0$ eigenvalue;
the two other states are pseudo-Dirac heavy neutrinos, massive and degenerate.

Concerning $m_0$,
the second order corrections ${m_0^2}^{(2)}$ (the first order one gives vanishing contributions) 
 are given by
\begin{equation}\label{1genneutrinomass(0)}
{m_0^2}^{(2)} \,=\, \frac{|d|^4 |\mu |^2}{\left(|d|^2+|n|^2\right)^2} \, ,
\end{equation}
which reduces to the usual inverse Seesaw expression once one assumes $|d| \ll |n|$. 
The first order corrections to ${m_{1,2}^2}^{(0)}= |d|^2+|n|^2$ lift the degeneracy: 
\be\label{1genneutrinomass(1,2)}
\begin{array}{cc}
 {m_1^2}^{(1)} = -\frac{\left|\mu ^* n^2+m |d|^2+m |n|^2\right|}{\sqrt{|d|^2+|n|^2}}\,, 
 & {m_2^2}^{(1)}= \frac{\left|\mu ^* n^2+m |d|^2+m |n|^2\right|}{\sqrt{|d|^2+|n|^2}}\,.
\end{array}
\ee
The corresponding eigenvectors allowing to build the leptonic mixing matrix can be found in  Appendix~\ref{AppendixA}.
Notice that in this approach,  the only assumption on the magnitude of the physical parameters, i.e.  
\bea\label{condition}
|m|, \,|\mu| \ll |d|,\,|n|\ ,\quad (n\ne 0)
\eea
 is driven (and justified) 
by the naturalness criterium. Notice that when $n\to 0$, one recover the simple realisation of the usual type I Seesaw, which is not the scenario we consider in this study. 

\subsubsection{Minimal Inverse Seesaw realisations}\label{miss}
In this section, we build the minimal ISS framework complying with experimental observations. The latter lead to the following requirements: 
\begin{itemize}
\item there are $3$ generations of neutrino fields with $SU(2)_L \otimes U(1)_Y$ gauge interactions 
($\# \nu_L=3$);
\item  there are at least $3$ non-degenerate light mass eigenstates. 
\end{itemize}

We extend the one generation matrix  of Eq.~(\ref{isszeromatrix})
to the case of several 
generations of $\nu_R$ and $s$ fields,  so that $M$ now  reads
\be\label{generalmatrix}
M= \left( \begin{array}{ccc} 0 & d & 0 \\ d^T & m & n \\ 0 & n^T & \mu \end{array} \right)\,,
\ee
 $d,m,n,\mu$ now being complex matrices. 
 Since $M$ is symmetric (due to the Majorana character of the fields), it follows that $m$ and $\mu$ are also symmetric matrices.
 
 A possible choice in Eq. (\ref{generalmatrix}) is to set the matrix $n=0$, such that the singlets $s$ decouple. In this case, the model reduces in practice to the type I Seesaw model, already compatible with low-energy data. We will conduct our analysis always assuming the (perturbativity) condition  Eq. (\ref{condition}) and thus considering the matrix $n\ne0$ and its entries always such that $|m|, \,|\mu| \ll |d|,\,|n|$.

In the following, we denote by $\# \nu_L, \#\nu_R$  and $\# s$  (with $\# \nu_R\ne 0$ and $\# s\ne 0$) the number of generations of left-handed, right-handed  and sterile fields, respectively. 
The Dirac mass matrix $d$ arises from the  Yukawa couplings to the Higgs boson $(\tilde{\Phi} =i  \sigma^2 \Phi)$,
\be\label{yukawa}
Y_{\alpha \beta} \, \overline{l_L}^\alpha \,\tilde{\Phi }\, \nu_R^\beta + \text{h.c.}\,,
\ee
where $Y$ is a complex matrix, $l_L^\alpha$ denotes the left-handed (LH)  leptonic doublet, 
\be
l_L^\alpha \,=\, \left( \begin{array}{c} \nu_L^\alpha \\ e_L^\alpha \end{array} \right)\,,
\ee
$\alpha$ and $\beta$ being generation indices. After electroweak symmetry breaking (EWSB), 
the matrix $d$ is given by 
\be\label{diracterms}
d_{\alpha \beta} = \frac{v}{\sqrt{2}} \,Y^*_{\alpha \beta}\,, 
\ee
and its dimension  is 
\be
\mbox{ dim } d = \left( \# \nu_L \times \# \nu_R \right).
\ee
The matrix $n$ describes the lepton number conserving interactions involving  
$\nu^c_R$ and $s$ fields, and its dimension is
\be
\mbox{ dim } n = \left( \# \nu_R \times \# s \right).
\ee
Finally, the dimension of the (symmetric) Majorana mass matrices $m$ and $\mu$ are  given by
\be
\mbox{ dim } m = \left( \# \nu_R \times \# \nu_R \right)\,, \quad
\mbox{ dim } \mu = \left( \# s \times \# s \right)\,.
\ee

Being gauge singlets, and since there is no direct evidence for their existence,  the number of additional fermionic 
singlets $\# \nu_R$ and $\# {s}$ is unknown. 
In the following we determine their {\it minimal} values when accommodating either a 3-flavour  or a 3 + 1 (or more) -flavour mixing schemes.
The different possibilities are summarised in Table~\ref{massspectrum}.

\begin{table}[htb]
 \begin{tabular}{|c|c|c||c|c|c||c|c|}
\hline
\hspace*{-3mm}
\begin{tabular}{c}
\#   {\footnotesize new} \\
{\footnotesize fields}
\end{tabular} \hspace*{-3mm}
& $\# \nu_R$ & $\# s$ &
$  \# {m^2_i}^{(0)} =0$
&
$  \# {m^2_i}^{(1,2)} \neq0$
&\begin{tabular}{c}
\# {\footnotesize of} \\ {\footnotesize non-deg. }\\ {\footnotesize light} $m_i$\end{tabular} &  
\hspace*{-3mm} \begin{tabular}{c} {\footnotesize oscillation}  \\ {\footnotesize data:}\\ $\Delta m^2$ \end{tabular} \hspace*{-3mm} 
&  \hspace*{-3mm}
\begin{tabular}{c} {\footnotesize oscillation} \\ 
{\footnotesize data:}\\  $\Delta m^2$ \& $U_\text{\scriptsize PMNS} $\end{tabular} 
\hspace*{-3mm}
\\
\hline
2&1&1&3&1&2&\XSolidBrush &\XSolidBrush  \\
\hline
3&1&2&4&2&3&\Checkmark (s) &\XSolidBrush  \\
\hline
3&2&1&2&1 &2&\XSolidBrush &\XSolidBrush \\
\hline
4&1&3&5&3&4&\Checkmark (a)&\XSolidBrush \\
\hline
4&2&2&3&2&3&\Checkmark (s) &\Checkmark \\
\hline
4&3&1&1&1&1&\XSolidBrush &\XSolidBrush \\
\hline
5&2&3&4&3&4&\Checkmark (a) &\Checkmark \\
\hline
5&3&2&2&2&2&\XSolidBrush &\XSolidBrush \\
\hline
6&3&3&3&3&3&\Checkmark (s) &\Checkmark\\
\hline
 \end{tabular}
\caption{Tree-level neutrino mass spectra for different choices of the number of additional fields, $\nu_R$ and $s$, and different properties of the light neutrino spectrum (see text for details and for description of used symbols). We limit the table to the case where the maximum number of additional singlet fields is six.}
\label{massspectrum}
\end{table}

The first three columns of Table~\ref{massspectrum}   indicate 
the total number of additional  fermionic singlets $\# \nu_R+ \#s$,  $\# \nu_R$ and  $ \#s$, respectively. 
The fourth column contains the number of  massless eigenstates at zeroth order 
(in the absence of accidental cancellations between the a priori independent entries {of the mass matrix}). 
Always in the absence of accidental cancellations,
the fifth column displays how many  massless eigenstates acquire mass once higher order  corrections from perturbations  
are taken into account (see Appendix \ref{AppendixA}): although massive, these states remain light
since the corresponding masses are proportional to  entries of $m$ and $\mu$ 
(this can be inferred from the one generation illustrative case, see Eq.~(\ref{1genneutrinomass(0)})). 
It is important to notice that states which are already massive at zeroth order have masses proportional to the $d$ and $n$ matrix entries. Finally, the sixth column contains information on the number of non-degenerate light mass eigenstates predicted by each of the different ISS configurations considered.

The last two columns provide information on the phenomenological viability of the different ISS realisations. Firstly, 
neutrino oscillation experiments require at least two independent oscillation frequencies ($\Delta m_{ij}^{{2}}$); if 
there are less than 3 different light masses, the model is then excluded by observation, and this is denoted by a  \XSolidBrush\ .
Models with 3 different light masses can generate the correct neutrino mass spectrum and are marked with a \Checkmark~(s) 
in the seventh column of the table. 

Interestingly,  models with 4 different light masses could potentially explain the (anti)neutrino anomalies reported by the short baseline experiments LSND~\cite{Aguilar:2001ty} and MiniBooNE~\cite{AguilarArevalo:2007it,AguilarArevalo:2010wv,Aguilar-Arevalo:2013pmq}, the Gallium anomaly  in radioactive 
source experiments~\cite{Acero:2007su,Giunti:2010zu} and the reactor antineutrino anomalies~\cite{Mueller:2011nm,Huber:2011wv,Mention:2011rk}.
Such configurations, leading to a 3 + 1-mixing scheme (see for example~\cite{Giunti:2012tn}) are 
indicated by a \Checkmark~(a) in the seventh column of Table~\ref{massspectrum}. 

For all cases with a viable mass spectrum - either (s) or (a) - we have then verified if the observed mixing pattern could be successfully reproduced. Should this be the case,  a \Checkmark is present in the eighth column of the table. 

\medskip
As can be seen from the information summarised on Table~\ref{massspectrum}, the  
simplest model\footnote{In our study, the  first scenario (ISS(1,1))  would have corresponded to the $n=1$  scenario  in~\cite{Gavela:2009cd}, provided  the entry $(1,3)$ of Eq. (\ref{isszeromatrix}) was different from zero.} which could accommodate the observed neutrino spectrum is the one with $ \left( \# \nu_R=1, \# s=2 \right) $, which will be here denoted as ISS(1,2). 
It predicts $4$ light eigenstates, two of which are massive; provided that the latter are non-degenerate, one could have two independent mass squared differences (corresponding to the solar and atmospheric mass differences).  Notice however that this model cannot provide the observed leptonic mixing matrix $U_\text{PMNS}$. This is a consequence 
of having one of its light mass eigenstates dominated by sterile components, and as such it cannot be identified with a SM active neutrino. A similar problem is present for the ISS(1,3) configuration, which although in principle accommodating the correct neutrino mass spectrum fails to provide the observed mixings. 

The scenarios $ \left( \# \nu_R=2, \# s=1 \right) $ and $ \left( \# \nu_R=3, \# s=1 \right) $ could in principle accommodate neutrino data (masses and mixing) in the limiting case where sterile fields decouple, i.e. the matrix $n\to  0$ in Eq. (\ref{generalmatrix}). We further emphasise here that we are not in this situation (of a type I Seesaw with 2 or 3 right-handed neutrinos), and these two scenarios do not comply with neutrino data. In the case of  $ \left( \# \nu_R=2, \# s=1 \right) $,  the corresponding mass spectrum contains one massless active neutrino, one light active while the third active one is too heavy to explain solar and atmospheric oscillation frequencies. 
A similar situation occurs for the $ \left( \# \nu_R=3, \# s=1 \right) $ case, where one has only one light active neutrino and two (too) heavy active ones.  

From this simple analysis and in view of  Table~\ref{massspectrum}, the first realisation  of the inverse Seesaw (with $\#s\ne 0$) possibly accommodating neutrino data is $ \left( \# \nu_R=2, \# s=2 \right) $, which we define to be the minimal one under the previous assumption of Eq. (\ref{condition}), 
 hereafter denoted by ISS(2,2) realisation. Notice that this solution  corresponds to the minimal model found in~\cite{Gavela:2009cd} in the case where no lepton number violating Yukawa couplings are allowed. 
This ISS(2,2) scenario  does not provide an explanation for the reactor anomaly;  the next (to minimal) ISS realisation 
which could explain such anomaly is the one with $ \left( \# \nu_R=2, \# s=3 \right) $,
 which we denote  by ISS(2,3) configuration.

Before addressing in detail the phenomenology of each minimal framework above identified, we will briefly comment 
on some aspects intrinsic to all ISS realisations.

\subsubsection{Different neutrino mass scales}\label{generaliss}

As a function of the number of generations for the sterile fields ($\#s\ne 0, \#\nu_R\ne 0$), 
the model always exhibits $\#\nu_L + (\#s-\#\nu_R)$ light mass eigenstates. These states would be massless at zeroth order, and their masses arise from higher order corrections (in perturbation) due to the block-diagonal matrix which now generalises $\Delta M$, see Eq.~(\ref{DeltaM}).
In addition, the full spectrum contains heavy states with masses  $\sim \mathcal{O}(n_{i,j})+\mathcal{O}(d_{i,j})$, which form 
$\#\nu_R$ pseudo-Dirac pairs with mass differences $\sim \mathcal{O}(\mu_{i,j}), \ \mathcal{O}(m_{i,j})$. 
In the limit where lepton number is conserved (i.e. $\Delta M=0$) these states become Dirac particles.

The low-energy phenomenology of these models is determined by two quantities: the scale of the Lepton Number Violating  parameters $\mu$ and  the ratio between the scale of the Dirac mass terms $d$ and that of the $n$ mass matrix, denoted by $k$.
To understand the key r\^ole of these  quantities, let us consider again the illustrative one-generation model 
(i.e. $\# \nu_L = \# \nu_R = \# s =1$) of 
Section~\ref{largeM}. The active neutrino mass of Eq.~(\ref{1genneutrinomass(0)}) can be rewritten as
$m_\nu = |\mu| k^2/(1+k^2)$, with $k=|d|/|n|$. 
In the realistic case of several generations, $d,n,\mu$ are matrices, and  these considerations loosely apply to the order of magnitude of their entries.
The ratio $k$ is directly related to  deviations from unitarity of the leptonic mixing matrix, as shown in Appendix \ref{AppendixA}, Eq.~(\ref{1genneutrinoeigenvec}).
Constraints on the non-unitarity of the PMNS matrix impose that $k$ should not be too large; 
as we will discuss in the section devoted to the numerical analysis,  
solutions in agreement with experimental data can be found if, and only if,
$\mathcal{O}(d)/\mathcal{O}(n) \lesssim 10^{-1}$. 
These features are shared by the different realistic extensions presented in Table~\ref{massspectrum}.
 
 The mass spectrum of the ISS models is thus characterised by either 2 or 3 different mass scales (as illustrated in Fig.~\ref{frize}):

\begin{figure}[htb]
 \begin{center}
  \includegraphics[width=0.8\textwidth]{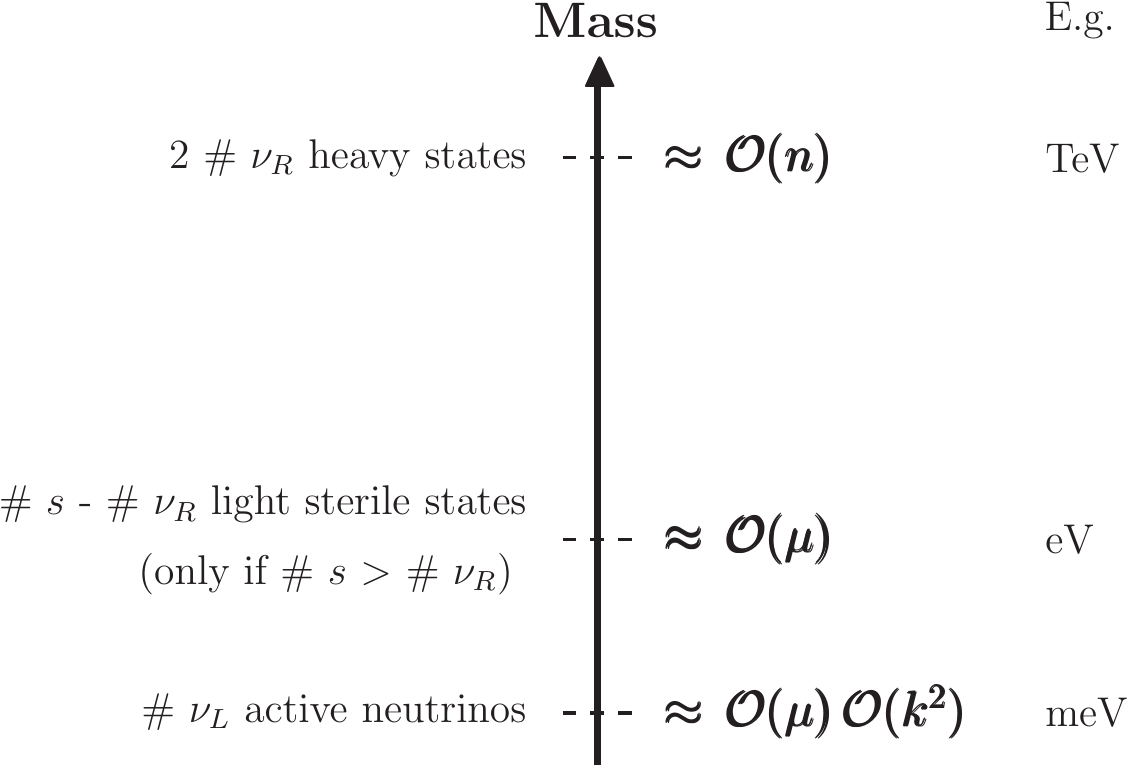}
\caption{Pictorial representation of typical scales for the neutrino mass spectrum in several ISS realisations.}
 \label{frize}
 \end{center}
\end{figure}

\begin{itemize}
\item the one of the light active neutrinos  $\sim \mathcal{O}(\mu) \mathcal{O}(k^2)$;
\item the scale corresponding to the heavy states, roughly $\mathcal{O}(d)+\mathcal{O}(n)\approx \mathcal{O}(n)$;
\item an intermediate scale of order $\mathcal{O}(\mu)$  corresponding to $\#s-\#\nu_R$ sterile light states (only present when $\#s>\#\nu_R$).

\end{itemize}

\subsubsection{Removing unphysical parameters}

The relevant leptonic terms of a general inverse Seesaw  Lagrangian can be written in the following compact  form, 
\be \label{lag22}
\mathcal{L}_\text{leptonic} \,= \,\mathcal{L}_\text{kinetic} + \mathcal{L}_\text{mass} + 
\mathcal{L}_\text{CC}+ \mathcal{L}_\text{NC}+ \mathcal{L}_\text{em}\,,
\ee
where
\begin{eqnarray}
\mathcal{L}_\text{kinetic}&=&  i \,\overline{e_L}^\alpha \,\slashed{\partial} \,
\delta_{\alpha,\beta} \,e_L^\beta + 
i \,\overline{e_R}^\alpha \,\slashed{\partial} \,\delta_{\alpha,\beta} \,e_R^\beta + 
i \,\overline{\nu_L}^\alpha \,\slashed{\partial} \,\delta_{\alpha,\beta} \,\nu_L^\beta+ 
i \,\overline{\nu_R}^i \,\slashed{\partial} \,\delta_{i,j} \,\nu_R^j + 
i \,\overline{s}^a \,\slashed{\partial} \,\delta_{a,b} \,s^b\,, \nonumber \\
\mathcal{L}_\text{mass}&=& 
- \overline{e_R}^\alpha \,{\mathfrak{m}}_{\alpha,\beta} \,e_L^\beta - 
\overline{\nu_R}^i \,d^T_{i,\alpha} \,\nu_L^\alpha -  
\overline{\nu_R}^i \,m_{i,j} \, {\nu_R^c}^j - 
\overline{\nu_R}^i \,n_{i,a} \,s^a - 
\overline{s^c}^a \,\mu_{a, b} \,s^b + \text{h.c.} \,, \nonumber \\
 \mathcal{L}_\text{CC}&=& 
 \frac{g}{\sqrt{2}} \,\overline{e_L}^\alpha \,\slashed{W}^- \,\delta_{\alpha,\beta} \,\nu_L^\beta + \text{h.c.}\,, 
 \nonumber \\
\mathcal{L}_\text{NC} &=& 
\frac{g}{\cos{\theta_W}} \,\left\{ \frac{1}{2} \,\left[ \overline{\nu_L}^\alpha \,\gamma_\mu \,
\delta_{\alpha,\beta}  \,\nu_L^\beta - \overline{e_L}^\alpha \,\gamma_\mu \,
\delta_{\alpha,\beta}\, e_L^\beta \right] - \sin^2{\theta_W} \,J_\mu^\text{em} \right\} 
\,Z^\mu\,, \nonumber \\
\mathcal{L}_\text{em} &=& e \,J_\mu^\text{em} \,A^\mu.
\end{eqnarray}
In the above equation  $\alpha,\beta=1,2,3$, $i,j=1,\dots,\# \nu_R$ and $a,b=1,\dots,\# s$. 
The total number  $n_u$ of physical and non-physical parameters in the mass matrices present in the  
Lagrangian of Eq.~(\ref{lag22}) is equal to 
\be
n_u=18 + 6\ \# \nu_R + \# \nu_R (\# \nu_R + 1) + \# s (\#s + 1) + 2\ \# \nu_R\ \# s\,,
\ee
and detailed in Table~\ref{numorder22}.

\begin{table}[htb]
\begin{center}
\begin{tabular}{|c|c|}
\hline
{\footnotesize Matrix} & {\footnotesize Total number of parameters} \\
\hline
${\mathfrak{m}}$ & $18$ \\
$d$ & $6\times \# \nu_R$\\
$n$ & $2\times \# \nu_R\times\# s$ \\
$m$ & $\# \nu_R\times (\# \nu_R + 1)$ \\
$\mu$ & $\# s \times(\#s + 1)$ \\
\hline
{\footnotesize Total} & $18 + \# \nu_R (7 + \# \nu_R +2\ \# s) + \# s (\# s+1)$\\
\hline
\end{tabular}
\end{center}
\caption{Total number of physical and non-physical parameters in the Lagrangian of Eq.~(\ref{lag22}).}
\label{numorder22}
\end{table}

In order to determine the actual number of  physical parameters, one has to identify all  independent transformations under which the Lagrangian of Eq.~(\ref{lag22}) is invariant. One finds four classes of transformations with the following unitary matrices: 
\begin{enumerate}
\item $U^{L}$ ($3 \times 3$):
\noindent\be\label{sym1.22} \hspace*{-8mm}
e_L^\alpha \rightarrow  U^L_{\alpha,\beta}e_L^\beta\,,
\quad
{\mathfrak{m}}_{\alpha,\beta} \rightarrow {\mathfrak{m}}_{\alpha,\gamma} 
{U^L}^\dagger_{\gamma,\beta}\,,
\quad
\nu_L^\alpha \rightarrow U^L_{\alpha,\beta} \nu_L^\beta\,,
\quad
d^T_{i,\alpha} \rightarrow d^T_{i,\beta} {U^L}^\dagger_{\beta,\alpha}\,;
\ee

\item  $U^{R}$ ($3 \times 3$):
\be\label{sym2.22} \hspace*{-8mm}
e_R^\alpha \rightarrow U_{\alpha,\beta}^R e_R^\beta\,,
\quad
{\mathfrak{m}}_{\alpha,\beta} \rightarrow U^R_{\alpha,\gamma} {\mathfrak{m}}_{\gamma,\beta}\,;
\ee

\item $U^{\nu_R}$ ($\# \nu_R \times \# \nu_R$):
\be\label{sym3.22} \hspace*{-8mm}
{\nu_R^c}^i  \rightarrow  U^{\nu_R}_{i,j} {\nu_R^c}^j\,,
\quad
m_{i,j}  \rightarrow  {U^{\nu_R}}^*_{i, k} m_{k,l} {U^{\nu_R}}^\dagger_{l,j}\,,
\quad
d^T_{i,\alpha}  \rightarrow  {U^{\nu_R}}^*_{i, j} d_{j,\alpha}^T\,,
\quad
n_{i,a}  \rightarrow  {U^{\nu_R}}^*_{i, j} n_{j,a}\,;
\ee

\item $U^s$ ($\# s \times \# s$):
\be\label{sym4.22} \hspace*{-8mm}
s^a  \rightarrow  U^s_{a,b} s^b\,,
\quad
\mu_{a,b}  \rightarrow {U^s}^*_{a, c} \mu_{c,d} {U^s}^\dagger_{d, b}\,,
\quad
n_{i,a} \rightarrow  n_{i, b} {U^s}^\dagger_{b,a}\,.
\ee

\end{enumerate}
The number of parameters defining the transformations of Eqs.~(\ref{sym1.22} - \ref{sym4.22}) 
is $n_t=18 +(\# \nu_R)^2 + (\# s)^2$, as shown in Table~\ref{numparsym22}, so that the 
number of physical parameters $n_p$ thus reduces to 
\be \label{nphys}
n_p\,=\,n_u-n_t\,=\,7\ \# \nu_R +\# s + 2 \# \nu_R \ \# s\,.
\ee

\begin{table}[htb]
\begin{center}
\begin{tabular}{|l|c|}
\hline 
{\footnotesize Matrix} & {\footnotesize Number of free parameters}\\
\hline
$U^L$ & $9$ \\
$U^R$ & $9$ \\
$ U^{\nu_R}$ & $(\# \nu_R)^2$ \\
$U^s$ & $(\# s)^2$\\
\hline 
{\footnotesize Total} & $18 +(\# \nu_R)^2 + (\# s)^2$\\
\hline
\end{tabular}
\end{center}
\caption{Number of parameters defining the transformations of Eqs.~(\ref{sym1.22} - \ref{sym4.22}).}
\label{numparsym22}
\end{table}

Since $\mathcal{L}_\text{kin}$ is invariant under 
 each of the transformations of  Eqs.~(\ref{sym1.22} - \ref{sym4.22}), we can use 
 the latter to redefine the fields and cast the  mass matrices only in terms of physical parameters. 
 For instance,  with the transformations of Eqs.~(\ref{sym1.22}, \ref{sym2.22}), one  can  choose a basis in which the charged leptonic matrix ${\mathfrak{m}}$ is real and diagonal,  and similarly for the symmetric Majorana mass matrices $m$ and $\mu$ (in this case using Eqs.~(\ref{sym3.22}, \ref{sym4.22})). 
Finally,  one can eliminate three phases from the Dirac mass matrix $d$  while  keeping ${\mathfrak{m}}$ real, 
via a combination of transformations of Eq.~(\ref{sym1.22}) and Eq.~(\ref{sym2.22}).
In this simple example, there are exactly $n_p$ free parameters, as summarised in Table~\ref{physpar}.

\begin{table}[htb]
\begin{center}
\begin{tabular}{|c|c|c|c|}
\hline
{\footnotesize Matrix} & {\footnotesize \# of moduli} & {\footnotesize \# of phases} &  
{\footnotesize Total} \\
\hline
{\footnotesize Diagonal and real ${\mathfrak{m}}$ }& $3$ & $0$ & $3$ \\
{\footnotesize $d$ with three real entries} & $3\ \# \nu_R$ & $3\ \# \nu_R-3$ & $6\ \# \nu_R-3$ \\
{\footnotesize Real and diagonal} $m$ & $\# \nu_R$ & $0$ & $\# \nu_R$ \\
 $n$ & $\# \nu_R \ \# s$ & $\# \nu_R \ \# s$ & $2 \ \# \nu_R \ \# s$ \\
{\footnotesize Real and diagonal $\mu$} & $\# s$ & $0$ & $\# s$\\
\hline
{\footnotesize Total} & \multicolumn{3}{ |c| }{$7\ \# \nu_R +\# s + 2 \# \nu_R \ \# s$}\\
\hline 
\end{tabular}
\end{center}
\caption{Example of a basis in which all unphysical degrees of freedom have been rotated away. 
}
\label{physpar}
\end{table}

\section{Effects of fermionic gauge singlets and constraints on the ISS parameters}\label{Sec:constrains}

In addition to succeeding in accommodating neutrino oscillation data, models with sterile fermions 
are severely constrained, since the mixings of the sterile neutrinos with the active
left-handed states might induce contributions to several observables, leading to conflict with experimental 
data. The mixings of the sterile neutrinos with the active
left-handed states  imply a departure from unitarity of the $3\times 3$ 
$U_\text{PMNS}$ matrix, which 
can have an impact on several observables, inducing deviations from the SM predictions.
Bounds on the non-unitarity of the PMNS were derived in~\cite{Antusch:2008tz}, using 
Non-Standard Interactions (NSI). These bounds are especially relevant in our analysis when 
the masses of the sterile states are heavier than the GeV, but some are still lighter than 174 GeV.

If the sterile states are sufficiently light and  have large mixings with the active neutrinos
(as for example in the inverse Seesaw~\cite{Mohapatra:1986bd}
, 
the $\nu$MSM \cite{Asaka:2005an} and the 
low-scale  type I Seesaw~\cite{Gavela:2009cd,Ibarra:2010xw,Ilakovac:2009jf,Alonso:2012ji,Dinh:2012bp}), 
then the deviations from unitarity  
of the PMNS mixing matrix can be sizeable, and lead to 
(tree-level) corrections to the $W \ell \nu$ vertex. This will have a significant impact to 
several observables, such as corrections to the invisible $Z$ decay width~\cite{Akhmedov:2013hec}, 
significant contributions to lepton flavour 
universality (LFU) violation observables~\cite{Shrock:1980vy,Abada:2012mc,Abada:2013aba}, 
and new contributions to numerous low-energy rare decays.

Another important  constraint concerns charged lepton flavour violation (cLFV)  
since the modified $W \ell \nu$ vertex gives rise to cLFV processes, typically at rates higher than the current bounds unless the active-sterile mixings are  small~\cite{Mohapatra:1986bd,GonzalezGarcia:1988rw,Deppisch:2004fa,Ilakovac:1994kj}.  
In the case of $\mu \to e \gamma$ decays, the rate induced by the presence of the sterile states is given by~\cite{Petcov:1976ff,Bilenky:1977du,Cheng:1980tp}:
\be \label{Petcov:1976ff,Bilenky:1977du,Cheng:1980tp}
\text{Br}(\mu \to e \gamma) = \frac{3 \alpha_{\text{em}}}{32 \pi}\left|\sum_{i} U_{\mu i}^* U_{e i} G\left(\frac{m_i^2}{M_W^2}\right)\right|^2 ,
\ee
where the index $i$ runs over all neutrino states,  $U$ is 
the leptonic mixing matrix obtained after diagonalization of the mass matrix and $G$ is the associated loop function.
The current bound on this branching ratio is $\text{Br}(\mu \to e \gamma) < 5.7 \times 10^{-13}$ at 90\% C.L., as reported 
very recently by the MEG experiment~\cite{Adam:2013mnn}. This will prove to be 
the most relevant LFV bound in most of our scenarios with light sterile neutrinos.
\vskip 0.5cm
Constraints arising from  neutrinoless double beta ($0\nu 2 \beta$) decay bounds can be particularly relevant, 
since in the ISS  the heavy sterile states also contribute to the process.
The effective neutrino mass $m^{\nu_e}_{\text{eff}}$, to which  the amplitude 
of the $0\nu 2 \beta$ process is proportional,  can receive further corrections with respect to the standard expression, 
$\sum_{i=1}^3 U_{e,i}^2 m_{\nu_i}$. 
Since the heavy Majorana states mix to form pairs of pseudo-Dirac states, their contribution is proportional to their mass difference weighted by the $\nu_e$-sterile mixing.  Each Majorana state thus contributes 
to the amplitude of a $0\nu 2 \beta$ decay as~\cite{Blennow:2010th} 
\be 
A_i \propto m_i U_{e,i}^2 M^{0\nu 2 \beta}(m_i)\,,
\ee
where $M^{0\nu 2 \beta}(m_i)$ is the nuclear matrix element that characterises 
the process. The latter is a function of the neutrino mass $m_i$ and depends on the nucleus that 
undergoes the $0\nu 2 \beta$ transition. It can be satisfactorily approximated by the analytic expression
\be 
M^{0\nu 2 \beta}(m_i) \simeq M^{0\nu 2 \beta}(0) \frac{p^2}{p^2-m_i^2},
\ee
where $p^2 \simeq - (125 \mbox{ MeV})^2$ is the virtual momentum of the neutrino. 
We will conduct a detailed analysis of the impact of two minimal  ISS  realisations, the ISS(2,2) and ISS(2,3),  
on the effective electron neutrino mass in Sections~\ref{nu0bb} and \ref{nu0bb-bis}.

\medskip
Moreover, if the typical scale of the new states is sufficiently light, they can be produced in collider or low-energy experiments, thus being subject to further constraints~\cite{BhupalDev:2012zg}.
Robust laboratory bounds arise from direct searches for sterile neutrinos, which can be produced in meson decays such as $\pi^\pm \to \mu^\pm \nu$, with rates that depend on their mixing with the active neutrinos. Therefore, negative searches for monochromatic lines in the muon spectrum can be translated into bounds on the active-sterile 
mixing~\cite{Kusenko:2009up,Atre:2009rg}.
\medskip

All the above mentioned bounds will be taken into account in our subsequent numerical analysis of the two minimal ISS realisations.

\section{Phenomenological analysis}\label{Sec:analysis}

Although it is possible to derive analytical expressions for the neutrino mass eigenvalues and leptonic mixing matrix (see Appendices~\ref{AppendixA} and \ref{AppendixB}), 
these expressions are  lengthy and involved, and do not easily convey the general features and behaviour of the ISS configurations investigated. 
We thus conduct a numerical analysis for each of the minimal ISS(2,2) and ISS(2,3) realisations. 
In order to unveil some interesting features, we performed a scan of the parameter space (corresponding 
to all the entries of the mass matrix; in our analysis we will not address the effect of CP violating phases, both Dirac and Majorana). This also allows to numerically compute interesting quantities, as for instance
the effective mass in $0\nu2\beta$ decay amplitude.
All the  constraints listed in Section~\ref{Sec:constrains} were implemented. We proceed to discuss the results
in the following sections.

\subsection{The ISS(2,2) realisation}\label{22realisation}
Some aspects of this model have already been studied, in particular CP violation and Non Standard 
Interactions~\cite{Malinsky:2009df,Malinsky:2009dh}. 
We have determined the neutrino spectrum and the leptonic mixing matrix using  a perturbative approach, whose details are summarised in Appendix~\ref{AppendixB}.
At second order in the perturbative expansion, the light neutrino  spectrum is given by:
\be\label{22issmasses-t}
{m_1^2}^{(2)}\,=\,0,\quad
{m_2^2}^{(2)}\,=\,\frac{b-\sqrt{b^2+4 c}}{2},\quad
{m_3^2}^{(2)}\,=\,\frac{b+\sqrt{b^2+4 c}}{2}\,,
\ee
where the parameters $b$ and $c$ are defined in terms of the entries of the (2,2) mass matrix; these expressions are 
lengthy, as explained in Appendix~\ref{AppendixB}. 
Notice that $b$ and $c$ do not depend on the submatrix $m$ of the mass matrix of Eq.~(\ref{generalmatrix}).

Having one massless eigenstate (to all orders in perturbation theory) is a feature of this minimal 
ISS(2,2) model (see also Table~\ref{massspectrum}). 
The expressions of Eq.~(\ref{22issmasses-t}) 
allow to easily understand why the ISS(2,2) model strongly prefers the normal 
hierarchy scheme. 
In order to accommodate an inverted hierarchy, i.e. $m_2^2 \simeq m_3^2 \simeq 10^{-3}\ \text{eV}^2$ and 
$m_3^2  - m_2^2 \simeq 10^{-5}\ \text{eV}^2$, one would be led  to comply with 
$10^{-6} \text{ eV}^4 +4c \simeq 10^{-10} \text{ eV}^4$. This amounts to an extreme fine-tuning.  
Although some solutions can indeed be found (see the numerical studies of the following section), 
it should be stressed that accommodating a NH spectrum also requires a certain amount of fine-tuning.

Even if useful when addressing  the issue of the hierarchy of the light neutrino spectrum, the analytical expressions we have  
derived for the neutrino masses and leptonic mixings cannot be used to extract general features, nor to 
infer the  magnitude of the fundamental scales of the 
ISS model (i.e. the magnitude of the entries of the matrices $\mu$, $m$, ...).
To do so, we performed numerical scans of the ISS(2,2)  parameter space, the result of which we proceed to report. 

\subsubsection{Mass hierarchy}

As discussed in Section~\ref{Sec:towards} and illustrated in Fig.~\ref{frize},  the low-energy phenomenology of a ISS(2,2) model is 
determined by two scales: 
that of the LNV parameter $\mu$, and the ratio $k$ between the magnitude of the entries of the $d$ and $n$ matrices, 
see Appendix~\ref{AppendixA}.

 In our numerical analysis, we randomly scan over all parameters: the entries of  the $d$ and $n$ submatrices are varied such that the obtained mixing matrix $U_\text{PMNS} $ is in agreement with oscillation data (global fits to both hierarchies, normal and inverted~\cite{GonzalezGarcia:2012sz}) and the interval of variation for the entries of $\mu$ is chosen to ensure that the largest neutrino squared mass value $ \sim 2.4 \times 10^{-3} \text{ eV}^2$. 
 While scanning over the parameter space, we always make sure that Eq. (\ref{condition}) is fulfilled,  assuming $\mu$ and $m$ to be  of the same order of magnitude. Moreover,  we take all parameters to be real (leading to vanishing Dirac and Majorana phases, and hence no contributions to leptonic electric dipole moments). 

\begin{figure}[htb]
 \begin{tabular}{ll}
\includegraphics[width=0.45\textwidth]{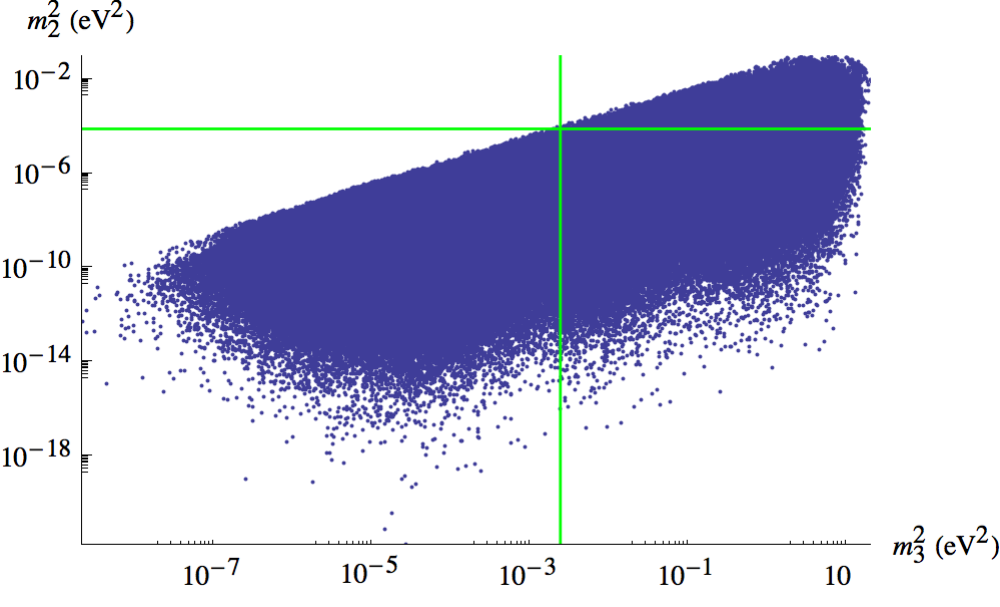}
\hspace*{2mm}&\hspace*{2mm}
\includegraphics[width=0.45\textwidth]{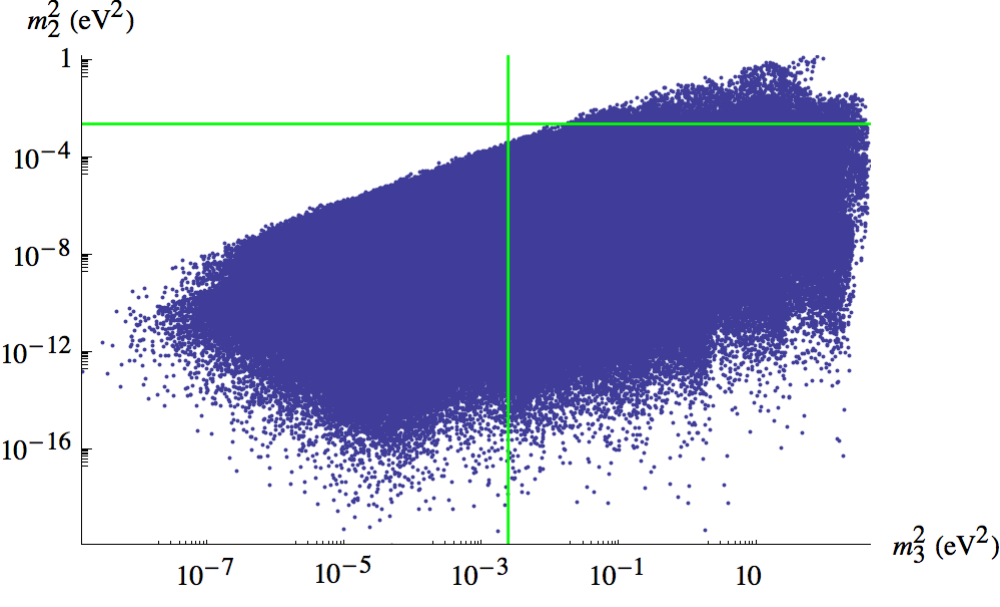}
\end{tabular}
\caption{Squared masses of the active neutrinos for the ISS(2,2) model (the lightest neutrino is massless). All points displayed fulfil the experimental constraints on the PMNS entries for the NH (left) and IH (right) schemes. The green lines denote the experimental best fit values \cite{GonzalezGarcia:2012sz} in the NH or IH schemes.
 The scan details are summarised in the text.}
\label{m1m2sq1.22}
\end{figure}

In Figure~\ref{m1m2sq1.22}, 
we collect the  values of the squared masses  $ m_{i}^2$ imposing that all the obtained mixing angles $\theta_{ij}$  are in agreement with oscillation data (for both cases of hierarchy, NH and IH).  
Leading to this figure, we varied for the left (right) panel the entries of each submatrix (see Eq.~(\ref{22massmatrix})) as $d_{i,j} \in  [10^{6},10^{8}] \text{ eV}$, $n_{i,j} \in  [10^{7},10^{9}] \text{ eV}$ ($n_{i,j} \in  [10^{8},10^{10}] \text{ eV}$), and  $m_{i,j},\mu_{i,j} \in  [10^{-3},10] \text{ eV}$ ($m_{i,j},\mu_{i,j} \in  [10^{-1},10^2] \text{ eV}$).

The best fit values for the mass eigenvalues resulting from the global analysis of the oscillation experiments~\cite{GonzalezGarcia:2012sz} are indicated in Fig.~\ref{m1m2sq1.22} by horizontal and vertical lines. 
This example clearly illustrates the analytical result found in Section \ref{22realisation} (as well as in Appendix \ref{AppendixB}):  
the ISS(2,2) model favours a normal hierarchical scheme - the inverted hierarchy requiring in this case an extreme fine tuning of the parameters, see Eq. (\ref{22issmasses-t}).  This  can be seen on the right panel of Fig.~\ref{m1m2sq1.22},  as no solutions can be encountered for an IH scheme (corresponding to 
$\Delta{m^2_{32}}\sim 10^{-5} \text{eV}^2$ together with $m_2^2\sim m_3^2\sim 10^{-3} \text{eV}^2$). 
Moreover, as can be seen on the left panel of Fig.~\ref{m1m2sq1.22}, for the NH scheme, finding solutions for the light neutrino masses in agreement with data is possible although difficult.

\subsubsection{Constraints from unitarity}\label{NSI}
The non-observation of NSI in the leptonic sector as induced by the deviation from unitarity of the $U_\text{PMNS}$ matrix due to the presence of additional fermions puts stringent constraints~\cite{Antusch:2008tz} on the ISS parameter space.

The non-unitarity effects can be quantified by 
\be\label{nsi}
\epsilon_{\alpha \beta}\equiv \left| \sum_{i=4}^7 U_{\alpha,i} \,U_{i,\beta}^\dagger \right|\,=\,\left|\delta_{\alpha,\beta} - \left(N\,N^\dagger\right)_{\alpha,\beta}\right|\,, 
\ee
where $N$ is the  $3 \times 3$ submatrix encoding the  mixings between the active neutrinos and  the charged leptons, i.e. the PMNS matrix. 
Depending on the mass regime for the sterile fermions (above or below the EW scale)  the constraints on $ \left(NN^\dagger\right)$ are different~\cite{Antusch:2008tz}.
We thus identify the following mass regimes for our sample of ISS(2,2)  mass matrices:
\begin{itemize}
\item no (or only some) sterile states are above 1 GeV - implying that not all the extra states can be indeed integrated out; 
the NSI constraints of~\cite{Antusch:2008tz} do not apply in this case;
\item all sterile states are heavier than 1 GeV, but do not necessarily lie above the EW scale, $\Lambda_\text{EW} \sim$174 GeV; 

\item all sterile states are heavier than $\Lambda_\text{EW}$.

\end{itemize}
When appropriate, we thus compute the amount  of non-unitarity from Eq.~(\ref{nsi}), and apply the corresponding bounds, to further constrain the ISS parameter space.

Notice that in the ISS models the non-unitarity effects are proportional to the ratio $\mathcal{O}(d)/\mathcal{O}(n)$ (see for example the neutrino mass eigenvector expression for the one-generation model 
(Eq.~\ref{1genneutrinoeigenvec})).

\begin{figure}[htb]
 \begin{tabular}{ll}
\includegraphics[width=0.45\textwidth]{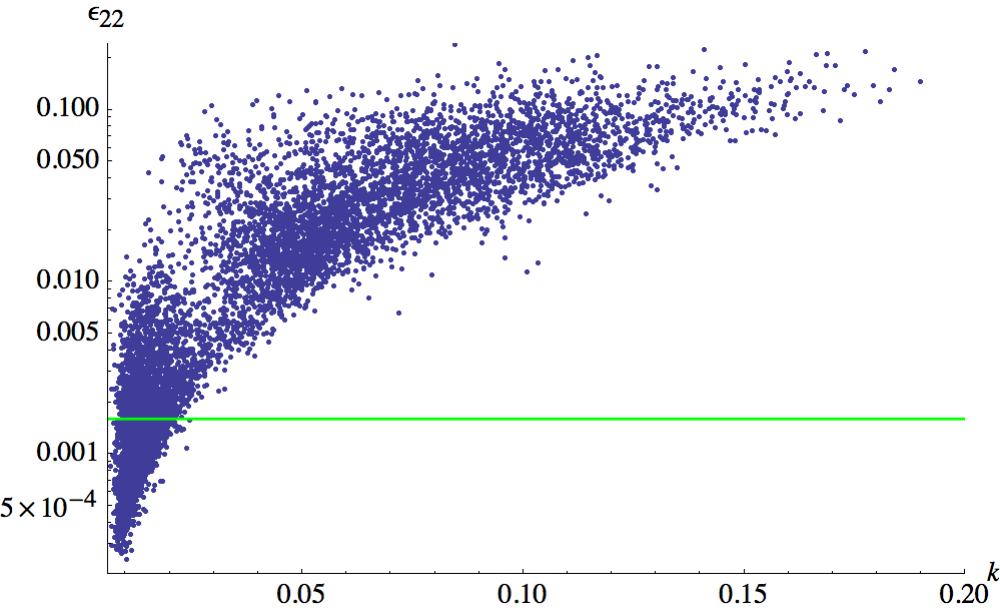}
\hspace*{2mm}&\hspace*{2mm}
\includegraphics[width=0.45\textwidth]{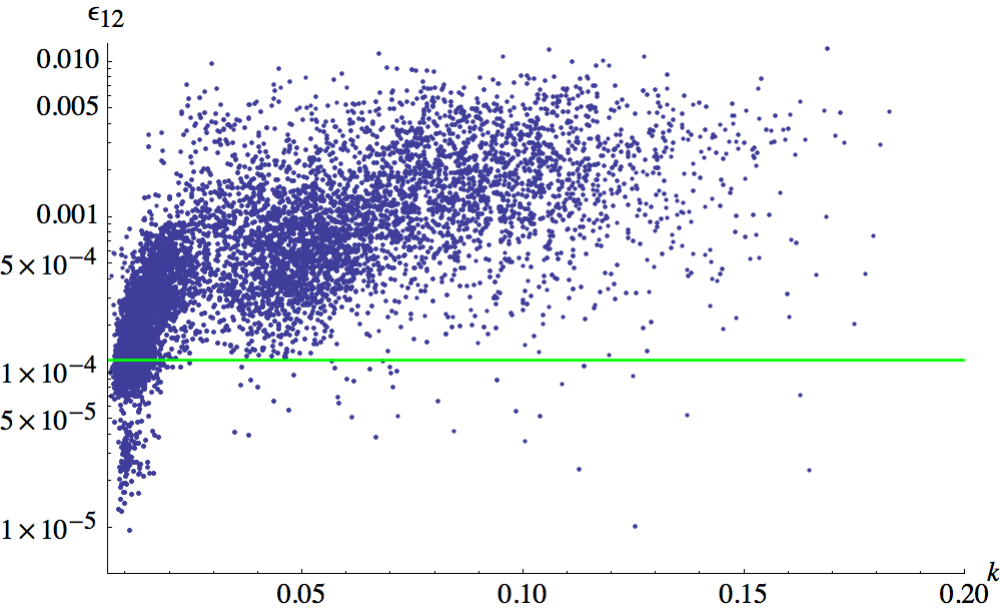}\end{tabular}
\caption{Examples of $  \epsilon =\left|\mathbf{1}- \left(N\,N^\dagger\right)\right|$ entries, as  a function of an effective factor $k$ (see Eq.~(\ref{k})).
On the left, $ \epsilon_{22}$, for a  
mass regime in which the sterile neutrino masses are  between 1 GeV and $\Lambda_\text{EW}$; on the right, $ \epsilon_{12}$, in the regime where all sterile states are heavier than  $\Lambda_\text{EW}$.
The green lines indicate the corresponding upper bounds~\cite{Antusch:2008tz}. All points comply with oscillation data in the NH scheme.
 The scan details are summarised in the text.}
\label{nsi22gev}
\end{figure}

We display on Fig.~\ref{nsi22gev} the most constraining deviations from unitarity parametrised by 
$ \epsilon =| {\mathbf{1}- }\left(NN^\dagger\right)|$, see Eq.~(\ref{nsi}), as a function of  an effective factor  $k$ 
 generalising the one  introduced in Section~\ref{generaliss}, which is defined as (see Eq.~(\ref{22massmatrix}) in Appendix~\ref{AppendixB}):
\be\label{k}
k\,=\,\frac{\left(d_{1,1}+d_{2,1}+d_{3,1}+d_{1,2}+d_{2,2}+d_{3,2}\right)/6}{\left( n_{1,1}+n_{2,2}\right)/2}\,.
\ee

Each point is generated with random values for the entries of the $d,n$ submatrices 
- but allowing the entries of each submatrix to vary at most over two orders of magnitude -,
 and such that the mass matrix would generate a PMNS matrix and a neutrino mass spectrum in agreement with experimental constraints (in the NH scheme). 
Leading to this figure (left  and right panels) , we varied  the entries of each submatrix (see Eq.~(\ref{22massmatrix})) as $d_{i,j} \in  [10^{3}, 1.7 \times 10^{11}] \text{ eV}$, $n_{i,j} \in  [5.5 \times  10^4,1.6 \times 10^{13}] \text{ eV}$ and  $m_{i,j},\mu_{i,j} \in  [5 \times 10^{-6}, 100] \text{ eV}$.

As can be seen from both panels of Fig.~\ref{nsi22gev}, NSI constraints significantly reduce the number of 
otherwise phenomenologically viable solutions for the ISS(2,2) model. 

\subsubsection{LFV constraints: Br(${\mu \to e \gamma}$)}
The presence of sterile fermions  may impact several observables in particular   LFV processes, with rates potentially larger than current bounds. We focus here on the radiative muon decay $\mu
\to e \gamma$, searched for by the MEG experiment~\cite{Adam:2013mnn} and which 
provides the most stringent constraint on the branching ratio of Eq.~(\ref{Petcov:1976ff,Bilenky:1977du,Cheng:1980tp}).

 In Fig.~\ref{Petcov:1976ff,Bilenky:1977du,Cheng:1980tp22}, we display this observable as a function of the mass of the lightest sterile state, $m_4$. The investigated parameter space (the same as the one leading to the previous figures) leads to contributions typically below the future experimental sensitivity. However, for $m_4$  heavier than $\sim 1$ GeV, one might observe a cLFV signal of the ISS(2,2) at MEG.

\begin{figure}[htb]
 \begin{center}
\includegraphics[width=0.65\textwidth]{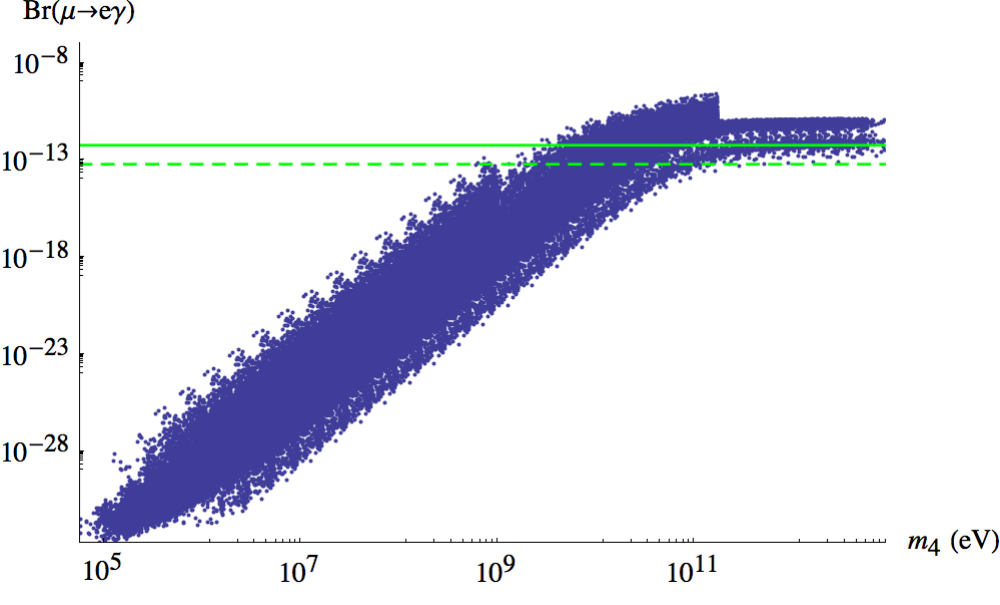}
\caption{Br(${\mu \to e \gamma}$) as a function of the mass of the lightest sterile state, $m_4$. 
The green full (dashed) horizontal lines denote MEG's current upper bound~\cite{Adam:2013mnn} (future sensitivity~\cite{Baldini:2013ke}). All points comply with oscillation data in the NH scheme and unitarity constraints. Scan details as in Fig.~\ref{nsi22gev}.}
\label{Petcov:1976ff,Bilenky:1977du,Cheng:1980tp22}
 \end{center}
\end{figure}

\subsubsection{Lepton number violating parameter space}
From the numerous numerical scans we conducted, certain features of the ISS(2,2) model became apparent: 
\begin{itemize} 
\item Low-energy neutrino data (i.e. masses and mixings)
can be accommodated if the entries in each of the submatrices of Eq.~(\ref{generalmatrix}) are allowed 
a strong  hierarchy  - varying at least over 2 orders of magnitude.
\item The model leads to a strongly hierarchical light neutrino mass spectrum, with   the second lightest neutrino mass being strongly suppressed with respect to the heaviest one (the first state being massless). 
\end{itemize}

The size of the LNV parameters (i.e. the entries of the $\mu$ submatrix - recall that the LNV matrix $m$ does not enter in the expression for the lightest neutrino mass eigenvalues, 
as derived in a perturbative approach - see for instance, Eq.~(\ref{1genneutrinomass(0)})) is bounded from below by PMNS matrix constraints, and from above by the naturalness requirement.
The lower limit is due to the fact that, to a good approximation, the entries of $d$ must be at least one order of magnitude smaller than those of  $n$ (in order to accommodate oscillation data).
In order to fulfil solar and atmospheric mass squared differences, and given that one typically has 
$k<10^{-1}$ (see Eq.~({\ref{1genneutrinomass(0)})), it follows that
\be 
|\mu| \ \gtrsim \ k^{-2}\times  8\times 10^{-3} \text{ eV} \ \gtrsim \ 8\times 10^{-1} \text{ eV}\,.
\ee

We have checked that the latter condition is indeed valid in the ISS(2,2) model;  
the lower values for the $\mu$ submatrix entries, 
in agreement with both $U_\text{PMNS}$ data  and neutrino mass squared differences 
are:
$\min|\mu_{i,i}| \sim 0.13$ eV, $\min|\mu_{i\neq j}| \sim 5 \times 10^{-6}$ eV.
The upper bound on the LNV parameters comes from  't Hooft naturalness criterium, even though
a clear definition regarding the naturalness of a small dimensionful parameter breaking some SM accidental 
symmetries does not exist. 
In this study, we have posited a "naturalness" upper limit of 100 eV on the entries of the submatrix $\mu$. 
This translates into a lower bound on the factor $k$ ({since $m_\nu \approx k^2 \mu$}).
\subsubsection{Neutrinoless double beta decay}\label{nu0bb}
When applied to the ISS(2,2) model, the effective neutrino mass $m^{\nu_e}_\text{eff}\ $ determining the 
amplitude of the neutrinoless double beta decay rate is given by (see Section~\ref{Sec:constrains})~\cite{Blennow:2010th}:
\bee \label{22bbdecay}
m_\text{eff}^{\nu_e}&\simeq&\sum_{i=1}^7 U_{e,i}^2 \,p^2 \frac{m_i}{p^2-m_i^2}\simeq \left(\sum_{i=1}^3 U_{e,i}^2\, m_{\nu_i}\right) \non 
&&+ p^2 \left(- U_{e,4}^2 \,\frac{|m_4|}{p^2-m_4^2}+U_{e,5}^2\, \frac{|m_5|}{p^2-m_5^2}-U_{e,6}^2 \,\frac{|m_6|}{p^2-m_6^2}+U_{e,7}^2 \,\frac{|m_7|}{p^2-m_7^2}\right)\,,
\eee
where $p^2 \simeq - (125 \mbox{ MeV})^2$ is the virtual momentum of the neutrino. 
From the analytical expressions derived in Appendix~\ref{perturbative}, one can see that  
in the limit $\mu_{i,j},\,m_{i,j} \rightarrow 0$,  one has 
$m_5\rightarrow m_4,\, m_7 \rightarrow m_6,\, U_{e,4}^2\rightarrow U_{e,5}^2,\,U_{e,6}^2 \rightarrow 
U_{e,7}^2$,  and thus the extra contribution vanishes.

Our predictions for the effective electron neutrino mass are collected in Fig.~\ref{bbdecay.jpg}, 
and displayed as a function of the mass of the  lightest sterile state, $m_4$. By defining   an "average" effective sterile mass, 
$m_s= \frac{m_4+m_5+m_6+m_7}{4}$,  three distinct mass regimes for $m_s$ can be identified from 
Fig.~\ref{bbdecay.jpg},

\begin{itemize}
\item 
$m_s \ll |p|$: in this regime the effective mass goes to zero, since from Eq.~(\ref{22bbdecay}) one 
approximately has
\be\label{lightsteriles}
m_\text{eff}^{\nu_e}\,=\,p^2 \sum_{i=1}^7 U_{e,i}^2 \, \frac{m_i}{p^2-m_i^2}\,\simeq 
\sum_{i=1}^7 U_{e,i}^2 \,m_{_i}\,,
\ee
and one can write 
\be
\sum_{i=1}^7 U_{\alpha,i}^2 \,m_{_i} \,=\, \sum_{i=1}^7 U_{\alpha,i} \,m_{_i} \,U_{i,\alpha}^T 
\,= \,M_{\alpha, \alpha}\,, 
\ee
where $M$ denotes the full neutrino mass matrix. 
\item $m_s \approx |p|$: the contribution of the pseudo-Dirac states becomes more important, and can induce sizeable effects to $m_\text{eff}^{\nu_e}$.
\item $m_s \gg |p|$: in this regime the heavy states decouple, and the contributions to $m_\text{eff}^{\nu_e}$ only arise from the 3 light neutrino states.
\end{itemize}

Notice that the values of $m_\text{eff}^{\nu_e}$ displayed in Fig.~\ref{bbdecay.jpg} correspond to conservative
(maximal) estimations; since in our scan all parameters are taken to be real, 
no cancellation due to possible (Majorana) phases can take place, and thus reduce the contributions of the ISS(2,2) model.
It is important to stress that all points leading to Fig.~\ref{bbdecay.jpg} comply with all available low-energy constraints discussed in Section~\ref{Sec:constrains}. 
The MEG bound on Br($\mu \to e \gamma$)~\cite{Adam:2013mnn} and the constraints from laboratory experiments~\cite{Atre:2009rg} are particularly important, and the latter are in fact responsible for the exclusion of a significant amount of points found (corresponding to the grey regions) in  Fig.~\ref{bbdecay.jpg}.

\begin{figure}[htb]
 \begin{center}
\includegraphics[width=0.7\textwidth]{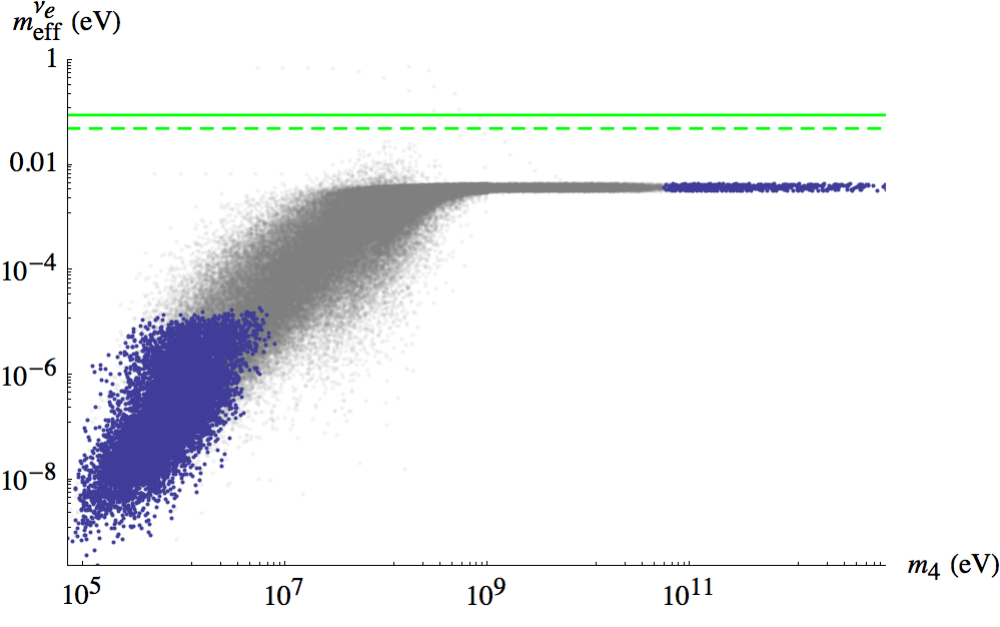}
\caption{Effective electron neutrino mass, $m_\text{eff}^{\nu_e}$, as a function of the lightest sterile mass $m_4$. The green full and dashed horizontal lines denote the current upper bound and the expected future sensitivity \cite{GomezCadenas:2011it}; blue points pass all imposed constraints (oscillation data, NSI, Br($\mu \to e \gamma$) and laboratory direct searches), while grey points are excluded by laboratory bounds. Scan details as in Fig.~\ref{nsi22gev}.
}
\label{bbdecay.jpg}
 \end{center}
\end{figure}

\subsection{The ISS(2,3) realisation}\label{23realisation}
We now address the phenomenology of the next-to-minimal configuration,  the ISS(2,3), where two generations of RH neutrinos and three sterile states are added to the SM content. In view of the degree of complexity of the analytical 
expressions derived for the simpler ISS(2,2), in this case we directly base our analysis  on a numerical approach.

\subsubsection{Allowed mass hierarchies}
Concerning the neutrino spectra, the crucial difference of the ISS(2,2) and the ISS(2,3) configurations is that the latter contains {\it four} light states, one being dominantly sterile-like. Its mass typically lies below the GeV (in the analysis we have explored the interval $[0, 100] \ \text{keV}$ for all the entries of the $\mu$ submatrix); recall that the four remaining states are heavy,  pseudo-Dirac pairs.
As can be seen in Table~\ref{massspectrum}, and similar to what occurred for the ISS(2,2), the lightest neutrino is also 
massless in the ISS(2,3) configurations. Thus, bounds on squared mass differences also translate into bounds for the masses themselves. 

Our study reveals that the ISS(2,3) model is not as  fine-tuned 
as the ISS(2,2) one. Allowing the entries 
of each submatrix of Eq.~(\ref{generalmatrix}) to vary over one order of magnitude leads to abundant solutions in agreement with low-energy neutrino data. Concerning the hierarchy of the light neutrino spectrum, we have verified that both NH and IH spectra are possible in the explored ISS(2,3) parameter space, although IH tends to be only marginally allowed, as is illustrated on 
Fig.~\ref{m1m2sq1.23}. For the left panel (NH), the parameters were varied as
$d_{i,j} \in  [10^{6},10^{7}]$~eV, $n_{i,j} \in  [10^{7},10^{8}]$~eV, $m_{i,j},\,\mu_{i,j} \in  [10^{-1},10]$~eV, while leading to the 
right plot (IH) we considered $d_{i,j} \in  [10^{6},10^{7}]$~eV, $n_{i,j} \in  [10^{8},10^{9}]$~eV, 
$m_{i,j},\,\mu_{i,j} \in  [10,10^3]$~eV.

\begin{figure}[htb]
 \begin{tabular}{cc}
\includegraphics[width=0.45\textwidth]{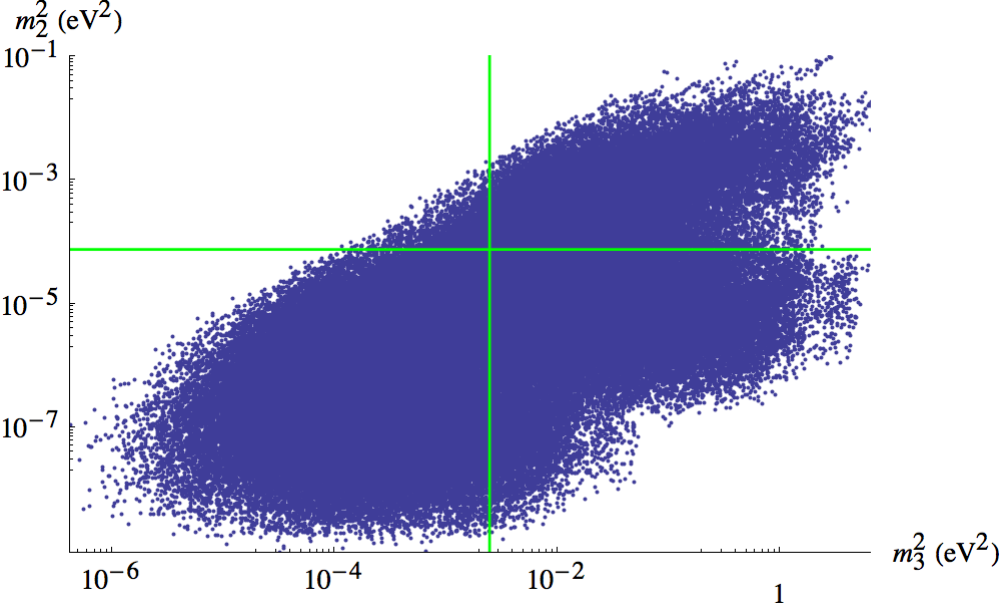}
\hspace*{2mm}&\hspace*{2mm}
\includegraphics[width=0.45\textwidth]{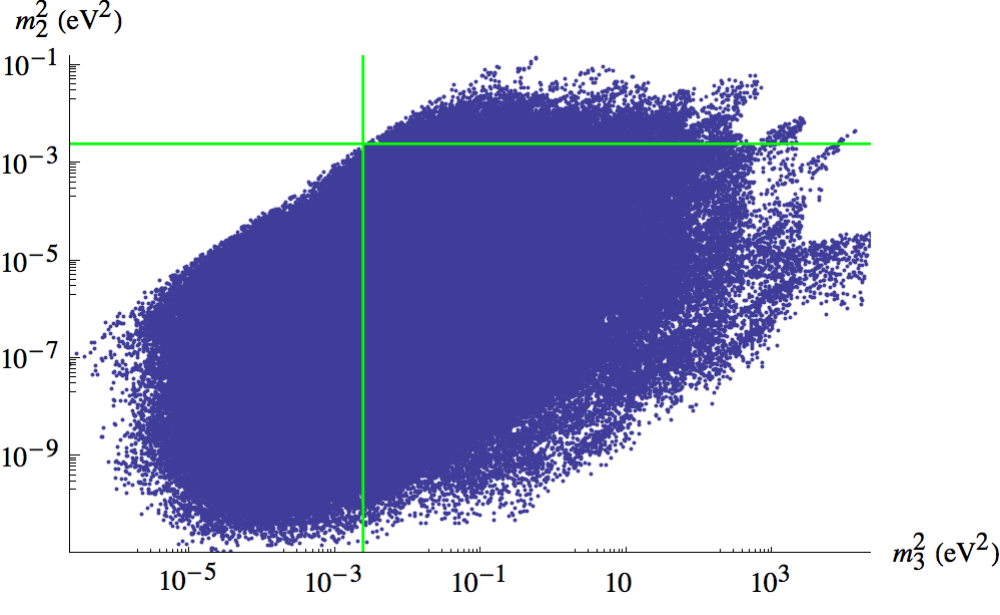}
\end{tabular}
\caption{Squared masses of the active neutrinos for the ISS(2,3) model (the lightest neutrino is massless). All points displayed fulfil the experimental constraints on the PMNS entries in the NH (left) and IH (right) schemes. The green lines denote the experimental best fit values \cite{GonzalezGarcia:2012sz} in the NH or IH schemes.
 The scan details are summarised in the text.}
\label{m1m2sq1.23}
\end{figure}

\subsubsection{Constraints from non-unitarity}
Similar to what was previously discussed for the ISS(2,2) configuration, the 
constraints coming from the non-observation of NSI (see Section~\ref{Sec:constrains}) also apply to 
ISS(2,3) models. We conducted here an analogous study:  the formulae and notations are simple generalisations of those introduced in Section \ref{NSI}, the only difference being that 
in the present case the index $i$ in Eq.~(\ref{nsi}) runs over the states that are integrated out ($\gtrsim 1$ GeV), 
i.e., $i=5,\dots,8$. 
Moreover and since we are interested in a potential "Warm" DM  candidate, we consider realisations of the ISS(2,3) model in which only the lightest sterile state lies below $100$ keV (i.e. $\mu\in[0,100]$ keV). 

In Figure~\ref{nsi23gev} we display two examples of deviations from unitarity as parametrised by 
$\epsilon_{\alpha \beta}\equiv \left| \sum_{i=5}^8 U_{\alpha,i} \,U_{i,\beta}^\dagger \right|$ as a function of  an effective  factor $k$. 
We notice that 
the relative density of points in the figure confirms that the ISS(2,3) allows for both spectra, although with a clear preference for NH. As in the previous ISS(2,2) model, we again verify that 
NSI constraints significantly reduce the number of viable solutions for a ISS(2,3) configuration. 
Leading to this figure, we varied  the entries of each submatrix  as $d_{i,j} \in  [10^{3}, 1.7 \times 10^{11}] \text{ eV}$, $n_{i,j} \in  [4.3 \times 10^4,4.8 \times 10^{14}] \text{ eV}$ and  $m_{i,j},\mu_{i,j} \in  [2\times10^{-2}, 10^5] \text{ eV}$. 

\begin{figure}[htb]
 \begin{tabular}{cc}
\includegraphics[width=0.45\textwidth]{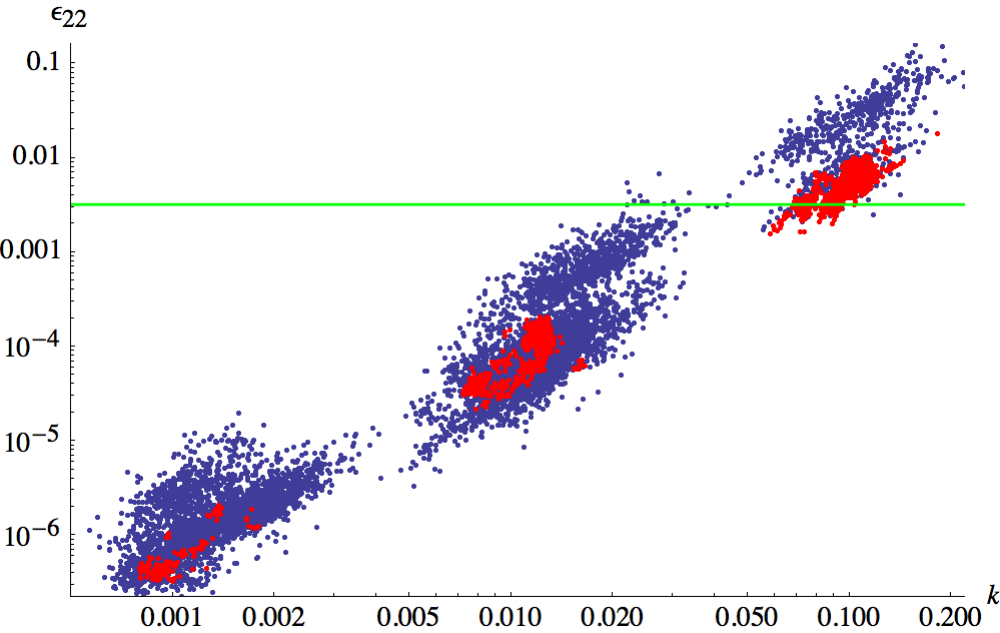}
\hspace*{2mm}&\hspace*{2mm}
\includegraphics[width=0.45\textwidth]{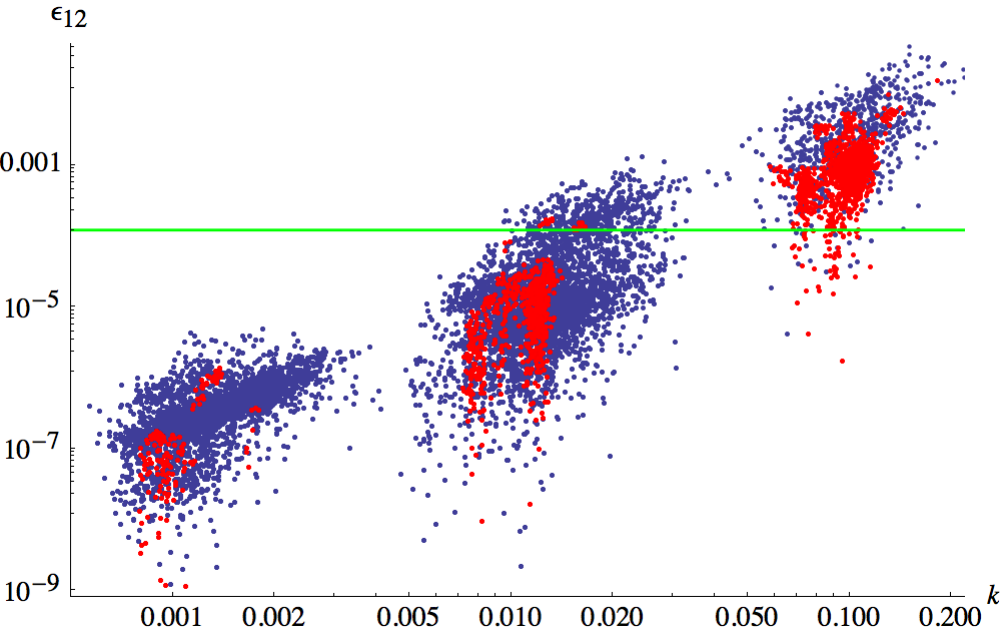}\end{tabular}
\caption{Examples of $\epsilon_{\alpha \beta}\equiv \left| \sum_{i=5}^8 U_{\alpha,i} \,U_{i,\beta}^\dagger \right|$
 entries, as  a function of an effective factor $k$ (generalisation of Eq.~(\ref{k}) for the ISS(2,3) model).
On the left, $ \epsilon_{22}$, for a  
mass regime in which the sterile neutrino masses are  between 1 GeV and $\Lambda_\text{EW}$; on the right, $ \epsilon_{12}$, in the regime where all sterile states are heavier than  $\Lambda_\text{EW}$.
The green lines indicate the corresponding upper bounds \cite{Antusch:2008tz}. Blue (red) points comply with oscillation data in the NH (IH) scheme. The scan details are summarised in the text.}
\label{nsi23gev}
\end{figure}

\subsubsection{LFV constraints: Br(${\mu \to e \gamma}$)}
For completeness, we illustrate the contributions of the new sterile states to rare LFV processes, in particular    considering Br($\mu \to e \gamma$), see Eq.~(\ref{Petcov:1976ff,Bilenky:1977du,Cheng:1980tp}). In Fig.~\ref{fig23Petcov:1976ff,Bilenky:1977du,Cheng:1980tp}, we display this observable as a function of the mass of the next-to-lightest sterile state, $m_5$. The investigated parameter space leads to contributions typically below the future experimental sensitivity. However, for $m_5$ in the range $[10^{2}, 10^{4}]$ GeV, one might observe a cLFV signal of the ISS(2,3) at MEG.

\begin{figure}[htb]
 \begin{center}
\includegraphics[width=0.6\textwidth]{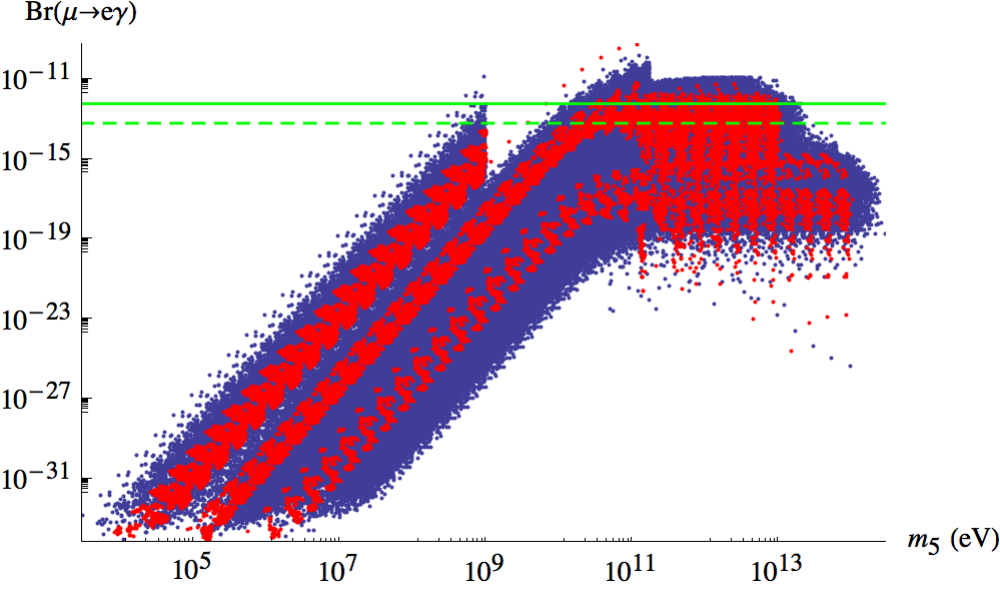}
\caption{Br(${\mu \to e \gamma}$) as a function of the mass of the next-to-lightest sterile state, $m_5$. 
The green full (dashed) horizontal lines denote MEG's current upper bound~\cite{Adam:2013mnn} (future sensitivity~\cite{Baldini:2013ke}); 
blue and red points  correspond to NH and IH solutions, respectively, and 
 pass all imposed constraints (oscillation data and NSI). Scan details as in Fig.~\ref{nsi23gev}.
}
\label{fig23Petcov:1976ff,Bilenky:1977du,Cheng:1980tp}
 \end{center}
\end{figure}

\subsubsection{An intermediate sterile scale}\label{lightsterile}
A fundamental difference between the "(2,2)" and the ISS(2,3) models is that, since in the latter case $\# s - \# \nu_R = 1$ (see  Section~\ref{generaliss}), the model has a third intermediate energy scale $\mathcal{O}(\mu)$, which corresponds to the mass of a sterile state.  It follows that if $\mu \approx $ eV this model can accommodate a $3\ +\ 1$-scheme that can potentially explain the (anti)-neutrino anomalies in  the short baseline, Gallium and reactor experiments. Should $\mu \approx $ keV, then the model can potentially provide a WDM candidate (see for example the analysis of~\cite{deVega:2013ysa}).

In Figure~\ref{nu_e-sterile_1_23}  we display the mixings of the light sterile state with $\nu_e$, as a function of  $m^2_{4}$. 
All points are in agreement with constraints from oscillation data,  
NSI, laboratory and LFV constraints.
As is clear from Fig.~\ref{nu_e-sterile_1_23}, the parameter space of the ISS(2,3) can provide solutions to either reactor anomaly. It can also provide a WDM candidate in the form of a sterile state of mass $\sim 1$ keV. 

\begin{figure}[htb]
 \begin{center}
\includegraphics[width=0.6\textwidth]{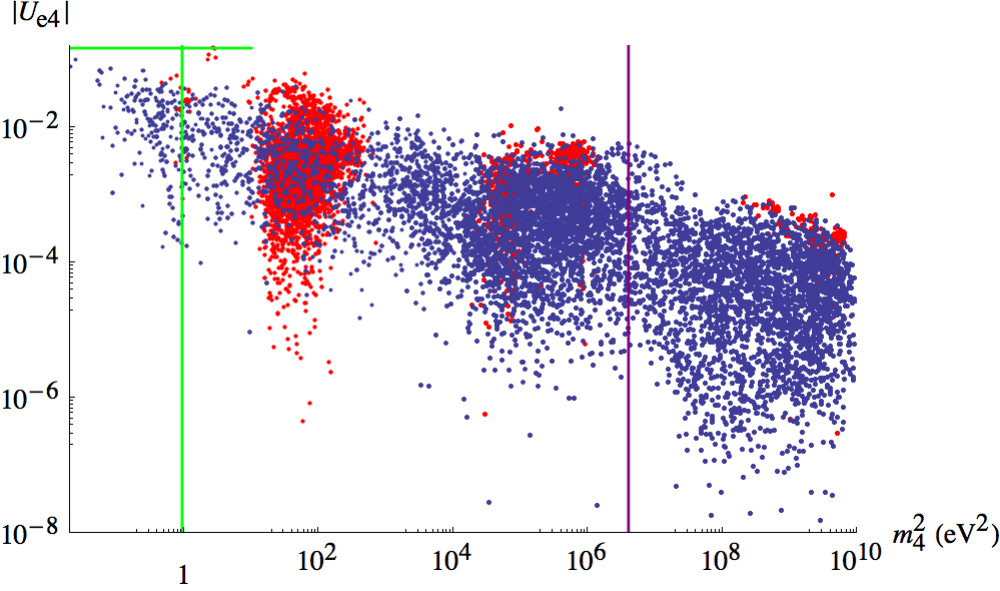}
\caption{Mixings between the electron neutrino and the lightest sterile state, as a function of the sterile squared mass $m_4^2$. The green lines indicate the best fit values of  $(\Delta m_{41}^2,|U_{e4}|)$ for the 3 + 1-scheme~\cite{Kopp:2013vaa}, while the purple vertical  line indicates the value $m_4^2=(2 \text{ keV})^2$, corresponding to the mass of the (warm) dark matter candidate suggested 
in~\cite{deVega:2013ysa}. 
Blue and red  points correspond to NH and IH solutions, respectively. The points displayed comply
with all imposed constraints (oscillation data, laboratory, NSI and Br(${\mu \to e \gamma}$)). Scan details as in Fig.~\ref{nsi23gev}.}
\label{nu_e-sterile_1_23}
 \end{center}
\end{figure}

\subsubsection{Neutrinoless double beta decay}\label{nu0bb-bis}
Due to the presence of the extra light sterile state, in the ISS(2,3) model there is an additional contribution to the effective mass derived in Eq.~(\ref{22bbdecay}). In our analysis we assumed the lightest sterile state to have a mass 
$m_4<100 \text{ keV} \ll |p| \approx 125 \text{ MeV}$, it contributes to the neutrinoless double beta  decay effective mass as 
\bee \label{23bbdecay}
m_\text{eff}^{\nu_e}&=&
\sum_{i=1}^8 U_{e,i}^2 \,p^2 \,\frac{m_i}{p^2-m_i^2} \,\simeq \left(\sum_{i=1}^4 U_{e,i}^2\, m_{\nu_i}\right) \non 
&&+ p^2\, \left(- U_{e,5}^2 \,\frac{|m_5|}{p^2-m_5^2}+U_{e,6}^2 \,\frac{|m_6|}{p^2-m_6^2}-U_{e,7}^2 \,\frac{|m_7|}{p^2-m_7^2}+U_{e,8}^2 \,
\frac{|m_8|}{p^2-m_8^2}\right)\,,
\eee 
trivially generalising Eq.~(\ref{22bbdecay}) and where above,  $p^2$ is again the virtual momentum of the propagating neutrino.

In Figure~\ref{bbdecay_23} we summarise our predictions for the effective electron neutrino mass as a function of $m_5$. Like in the previous case, by defining an "average" heavy sterile mass $m_s =\frac{m_5+m_6+m_7+m_8}{4}$,  one can  easily identify the three distinct regimes discussed in Section~\ref{nu0bb} for the ISS(2,2) scenario. Especially in regimes of heavier sterile masses (i.e., $m_5 \gtrsim 1 \ \text{GeV}$), the model 
is fairly predictive regarding the $0\nu2\beta$ decays: the value of the effective mass in ISS(2,3) scenario lies just below the current experimental bound and within the future sensitivity of ongoing experiments \cite{GomezCadenas:2011it}. 
Somewhat lighter sterile masses could also account for an effective mass within experimental reach, but these solutions are already excluded by the recent MEG bound and by laboratory constraints. 

\begin{figure}[htb]
 \begin{center}
\includegraphics[width=0.6\textwidth]{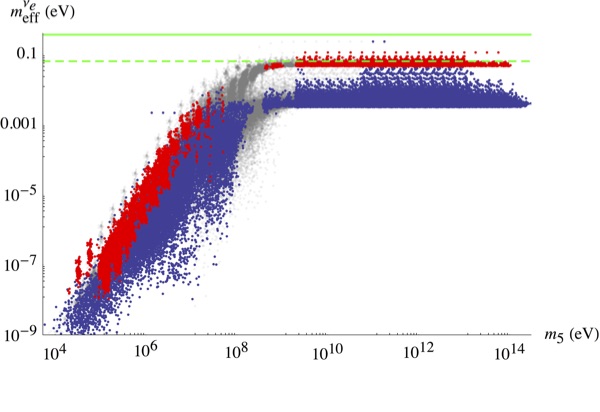}
\caption{Effective electron neutrino mass, $m_\text{eff}^{\nu_e}$, as a function of $m_5$. The green full and dashed horizontal lines denote the current upper bound and the expected future sensitivity \cite{GomezCadenas:2011it}; blue and red  points correspond to NH and IH solutions, respectively, and 
 pass all imposed constraints (oscillation data, NSI, Br($\mu \to e \gamma$) and laboratory direct searches), while grey points are excluded by laboratory bounds. Scan details as in Fig.~\ref{nsi23gev}.
}
\label{bbdecay_23}
 \end{center}
\end{figure}

\chapter{Dark Matter in the minimal Inverse Seesaw mechanism}\label{sec:DMMISS}

Sterile fermions are an intriguing and popular solution for the dark matter problem~\cite{Dodelson:1993je,Abazajian:2001nj,Dolgov:2000ew,Boyarsky:2009ix}. In particular, sterile neutrinos with masses around the keV can be viable Warm Dark Matter (WDM) candidates. They can potentially solve some tensions with structure formation observations, even if providing only a fraction of the total dark matter (DM) relic density~\cite{Klypin:1999uc,Moore:1999nt,Strigari:2010un,BoylanKolchin:2011de}. In addition, a sterile neutrino at this mass scale  could  in general decay into an ordinary neutrino and a photon which could be detected in cosmic rays. This last possibility has recently triggered a great interest in view of the indication, yet to be confirmed, of an unidentified photon line in galaxy cluster spectra at an energy $\sim 3.5$~keV~\cite{Bulbul:2014sua,Boyarsky:2014jta}.  

We have shown in Sections~(\ref{Sec:intro}-\ref{Sec:analysis}) that it is possible to construct several minimal distinct ISS scenarios that can 
reproduce the correct neutrino mass spectrum while fulfilling all phenomenological constraints. 
Based on a perturbative approach, we also shown that  
the mass spectrum of these minimal ISS realisations is characterised by either 2 or 3 different mass scales, 
corresponding to the one of the light active neutrinos $m_\nu$, that
corresponding to the heavy states $M_R$, and  an intermediate scale $\sim \mu$ 
only relevant 
when $\#s > \# \nu_R$. This allows to identify  two truly minimal ISS realisations (at tree level):
the first one, denoted ISS(2,2) model, corresponds to the SM extended by 
two RH neutrinos and two sterile states. It leads to a 3-flavour mixing scheme and prefers a normal hierarchy solution for the light neutrinos, while its full spectrum is characterised by only two mass scales (the light neutrino masses, $m_\nu$ and the RH neutrino masses, $M_R$).  
The second one, the ISS(2,3) realisation, corresponds to an extension of the SM by 
two RH neutrinos and three sterile states, and  allows to accommodate both hierarchies for the light neutrino spectrum (with the  inverse hierarchy  only marginally allowed), in a  3+1-mixing scheme. The mass of the lightest sterile neutrino can vary over a large interval: 
depending on its regime, the 
ISS(2,3) realisation can offer an explanation for the 
short baseline (reactor/accelerator) anomaly~\cite{Mueller:2011nm,Huber:2011wv,Mention:2011rk,Aguilar:2001ty,AguilarArevalo:2007it,AguilarArevalo:2010wv,Aguilar-Arevalo:2013pmq,Acero:2007su,Giunti:2010zu} (for a mass of the lightest sterile state around the eV), or provide a DM candidate (for a mass of the lightest sterile state in the keV range). 

In this chapter, we investigate in detail this last possibility, conducting a thorough analysis of the relic abundance of the dark matter candidate, taking into account all available phenomenological, astrophysical and cosmological constraints. The conventional DM production mechanism, the so called Dodelson-Widrow mechanism~\cite{Dodelson:1993je}, results in a tension with observational constraints from DM Indirect Detection (ID) and structure formation, since it can only  account for at most~$\sim 50\%$ of the total DM abundance. A sizeable DM density can  nonetheless be achieved when one considers the decay of the heavy pseudo-Dirac neutrinos. This possibility is realised in a  restricted region of the parameter space, $m_h < M_R \lesssim 1$ TeV, where $m_h$ is the Higgs boson mass. An extension of the model is thus needed in order to account for a viable DM in a broader portion of the parameter space.       

The rest of the chapter is organised as follows:  in Section~\ref{Sec:Model}, after  a recap of  the model - the  ISS(2,3) realisation -,  we address  the prospects of the lightest sterile state as a viable DM candidate, which are  stability,  indirect detection and the dark matter generation mechanism. In Section~\ref{Sec:Requirements}, we  consider all the relevant  different astrophysical and cosmological  constraints taking into account the effect of the heaviest sterile neutrinos (DM production from  decays of heavy sterile states or possible entropy injection effects from a scenario with lighter sterile neutrinos) accounting as well for the indication of the monochromatic 3.5 keV observed line.
Section~\ref{Sec:NISS} is devoted to an economical and motivated extension of the model which succeeds in providing   the observed  dark matter relic abundance in a larger region of the parameter space.  
The numerical details regarding the production and evolution of the sterile neutrinos can be found in 
the Appendix~\ref{sec:app_ISSDM}. 

\section{Description of the model}\label{Sec:Model}

\subsection{The ISS(2,3) framework}
The phenomenology of ISS mechanism has been discussed in Sections~(\ref{Sec:intro}-\ref{Sec:analysis}). We recall here that, depending on the number of fields, a generic ISS($\#\nu_R,\# s$) realisation is characterised by the following mass spectrum
\begin{itemize}
\item 3 light active states with masses of the form\begin{equation}\label{inv.ss}
m_\nu\approx \mathcal{O}(\mu) \frac{k^2}{1+k^2}\,, \,\,\,\,\,\,k\simeq\frac{\mathcal{O}(d)}{\mathcal{O}(n)}\ .
\end{equation}
This set must contain at least three different masses, in agreement with the two oscillation mass frequencies (the solar and the atmospheric ones). 
\item $ \# s- \# \nu_R$ light sterile states (present  only if $ \# s> \# \nu_R$) with masses $\mathcal{O}(\mu)$.
\item $\# \nu_R$ pairs of pseudo-Dirac heavy neutrinos with masses $\mathcal{O}\left(n\right)+\mathcal{O}\left(d\right)$.
\end{itemize}

The ISS(2,3) is the minimal viable realisation that accommodates a light sterile fermion.\footnote{It is worth mentioning that  a realisation of the ISS with 3 RH neutrinos and 4 sterile states fulfilling all possible constrains has been recently found in the context of conformal EW symmetry breaking~\cite{Lindner:2014oea}.}

Remarkably, and in order to comply with all constraints from neutrino oscillation and laboratory experiments,  
the coupling of this new state  to the active neutrinos must be highly suppressed, thus leading to a dominantly  sterile state, with a mass ranging from $\mathcal{O}(\text{eV})$ to several tens of keV.\footnote{Light sterile neutrinos, i.e. with masses ranging between the eV and keV scale also appear in the so called Minimal Radiative Inverse See-Saw~\cite{Dev:2012bd}.} As consequence of its very weak interactions, the lifetime of the lightest sterile neutrino largely exceeds the lifetime of the Universe and  it can  thus play a relevant r\^ole in cosmology. 

In this analysis we will focus on the possibility that this sterile neutrino accounts, at least partially, for the Dark Matter component of the Universe, identifying the viable regions of the parameter space with respect to DM phenomenology of the ISS(2,3) model. 

We point out that the heaviest sterile states might be involved in a  broad variety of particle physics processes and have then to comply with several laboratory bounds and electroweak precision tests (these bounds have been analysed in Sections~(\ref{22realisation},\ref{23realisation}) for the ISS(2,2) and ISS(2,3) realisations). On recent times the possibility of production of heavy neutrinos at collider has been as well considered. The most peculiar signatures of the ISS scenario are, as a consequence of the large Yukawa couplings of the right-handed neutrinos, additional decay channels of the Higgs boson into a heavy and an ordinary neutrino, if kinematically allowed, or into three SM fermions through an off-shell neutrino. These decay modes can be searched both directly, in particular the ones with leptonic final states~\cite{BhupalDev:2012zg,Das:2012ze,Cely:2012bz,Bandyopadhyay:2012px}, and indirectly, in global fits of the Higgs data, by looking at deviations from the SM prediction in the branching ratios of the observed channels~\cite{BhupalDev:2012zg}. 
Direct searches of decay channels of the Higgs provide bounds on the Yukawa couplings of the pseudo-Dirac neutrinos with masses ranging from approximately 60 GeV (at lower masses possible signals do not pass current analysis cuts employed by experimental collaborations) to 200 GeV which can be as strong as $\sim 10^{-2}$ while global analysis of Higgs data provide a limit, for the same mass range, as strong as $\sim 3 \times 10^{-3}$ but can be effective in a broader mass range.
Alternatively heavy sterile neutrinos can be looked in dilepton~\cite{Kersten:2007vk} or dilepton+dijet processes~\cite{Atre:2009rg}, which are sensitive to their coupling to the W boson, that is related to the mixing between the active and the sterile neutrinos and thus provide bounds on the the elements of the mixing matrix $U$.
In the low mass region, namely $\lesssim \mathcal{O}(\mbox{GeV})$, heavy neutrinos can be detected in decays of mesons~\cite{Atre:2009rg,Gorbunov:2007ak,Ruchayskiy:2011aa}. In this analysis we consider ISS(2,3) realisations satisfying the above experimental constraints. We remark that a sensitive improvement of these constraints in the low mass region is expected from the recently proposed SHiP~\cite{Bonivento:2013jag}. 

\subsection{Light sterile neutrino as Dark Matter}

Before the  analysis
we will briefly summarise the main issues that should be addressed in order for the lightest sterile neutrino to be a viable dark matter candidate.

\smallskip
\noindent {\bf Stability and Indirect Detection}:\\
The most basic requirement for a DM candidate is its stability (at least on cosmological scales).
All the extra neutrinos of the ISS model have a non zero mixing with ordinary matter. As a consequence, the lightest one is not totally stable  and can decay into an active neutrino and a photon $\gamma$. On the other hand, as already pointed out, its very small mixing makes the decay rate negligible with respect to cosmological scales. Nonetheless, a residual population of particles can decay at present times producing the characteristic signature of a monochromatic line in X-rays. This kind of signature is within reach of satellite detectors like CHANDRA and XMN which have put strong limits on the couplings between sterile and active neutrinos (due to  the lack of detection of this kind of signal). Recently,  the existence of an unidentified line in the combined spectrum of a  large set of X-ray galactic clusters has been reported~\cite{Bulbul:2014sua} and independently, in the combined observation of the Perseus Cluster and the M31 Galaxy~\cite{Boyarsky:2014jta}. These  observations can be compatible with the decay of a sterile neutrino with a mass of approximately 7 keV. 
 Confirmation of the latter result requires further observation, and most probably, higher resolution detectors like the forthcoming \emph{Astro-H}. 
 As we will show in the analysis, the ISS(2,3) model can account for this intriguing possibility; however, we will only impose that the sterile neutrino  lifetime does not exceed current observational limits. 
   
\smallskip
\noindent{\bf DM generation mechanism}:\\
The second issue to address is to provide a DM generation mechanism accounting for the experimental value of its abundance. In the pioneering work by Dodelson-Widrow (DW)~\cite{Dodelson:1993je}, it has been shown that the DM abundance can be achieved through active-sterile neutrino transitions.\footnote{The popular WIMP mechanism cannot be effective in our case  since sterile neutrinos could not exist in thermal equilibrium in the Early Universe due to  their suppressed interactions with ordinary matter.} This kind of production is always present provided that there is a non-vanishing  mixing between active and sterile neutrinos; as a consequence,  it is possible to constrain the latter as function of the neutrino mass by  imposing that the DM relic abundance does not exceed the observed value. The ISS(2,3) framework allows for an additional production mechanism, consisting in the decay of the heavy pseudo-Dirac states.
We will  discuss this point at a subsequent stage.    
 
\smallskip
\noindent{\bf Limits from structure formation}:\\
 Sterile neutrinos in the mass range relevant for the ISS(2,3) model are typically classified as warm dark matter. This class of candidates is subject to  strong constraints from structure formation, which typically translate into lower bounds on the DM mass.
We notice however, that the warm nature of the DM is actually related to the production mechanism determining the DM distribution function. Sterile neutrinos - with masses at the keV scale - produced by the DW mechanism can be considered as WDM; this may not be the case for other production mechanisms.

In the next section we will investigate whether the ISS(2,3) can provide a viable DM candidate.

\section{Dark matter production in the ISS(2,3)}\label{Sec:Requirements}
In this section we address the impact of the combination of three kinds of requirements on the DM properties on the ISS(2,3) parameter space.
The results presented below rely on the following hypothesis: a standard cosmological history is assumed with the exception of possible effects induced by the decays of heavy neutrinos; only the interactions and particle content of the ISS(2,3) extension of the SM are assumed.

Regarding DM production we will  not strictly impose that the relic abundance reproduces the observed  relic abundance, $\Omega_{\rm DM} h^2 \approx 0.12$~\cite{Ade:2013zuv},  but rather determine the maximal allowed DM fraction $f_{\rm WDM}$ within the framework of the ISS(2,3) parameter space.

The main production mechanism for DM is the DW, which is present as long as mixing with ordinary matter is switched on. In addition,  the DM could also be produced by the decays of the pseudo-Dirac neutrinos. However, a sizeable contribution can only be  obtained  if at least one of the pseudo-Dirac states lies in the mass range $130$ GeV - 1 TeV. Moreover, 
the pseudo-Dirac states can also have an indirect impact on the DM phenomenology since, under suitable conditions, they can release entropy at their decay, diluting the DM produced by active-sterile oscillations, as well as relaxing the bounds from structure formation. As will be shown below, this effect is also restricted to a  limited mass range for the pseudo-Dirac neutrinos.

For the  sake of simplicity and clarity, we  first discuss the case in which the heavy pseudo-Dirac neutrinos can be regarded as decoupled, and discuss at a second stage their impact on DM phenomenology. 

\subsection{Dark matter constraints without heavy neutrino decays}\label{sec:light_sterile}

We proceed to present the constraints from dark matter on the ISS(2,3) model, always  under the hypothesis that heavy neutrinos do not influence DM phenomenology. 

Regarding the relic density, for masses of the lightest-sterile neutrino with mass $m_s>0.1$~keV, we use the results\footnote{Notice that in~\cite{Asaka:2006nq},    the parametrisation $|M_D|_{\alpha 1} \equiv \theta_{\alpha s} m_s$ was used, while   in our work we use  $|U_{\alpha s}| \simeq \theta_{\alpha s}$ for  small mixing angles. } of~\cite{Asaka:2006nq}:
\begin{equation}\label{eq:Omega_s}
\Omega_{\rm DM} h^2 =1.1\times  10^7 \sum_\alpha  C_\alpha(m_s) \left|U_{\alpha s}\right|^2 \left(\frac{m_s}{\text{ keV}}\right)^2, \,\,\,\,\alpha=e,\mu,\tau\ .
\end{equation}
$C_\alpha$ are  active flavour-dependent coefficients\footnote{For DM masses of the order of 1 - 10~keV, the production peaks at temperatures of $\sim$150 MeV, 
corresponding  to the QCD phase transition in the primordial plasma. As a consequence,  
the numerical computation of the $C_\alpha$ coefficients is affected by uncertainties related to the determination of the rates of hadronic scatterings, and to the QCD equation of state.} which can be numerically computed by solving suitable Boltzmann equations. 
In the case of a sterile neutrino with mass $m_s<0.1$ keV, we have instead used the simpler expression~\cite{Abazajian:2001nj}:
\begin{equation}\label{eq:Omega_s_s}
\Omega_{\rm DM} h^2 = 0.3 \left(\frac{\sin^2 2 \theta}{10^{-10}}\right) \left(\frac{m_s}{100 \text{ keV}}\right)^2,
\end{equation}
where  $\sin^2 2 \theta = 4 \sum_{\alpha=e,\mu,\tau} |U_{\alpha s}|^2$, with  $|U_{\alpha s}|$  being the active-sterile leptonic mixing matrix element.
  We have then computed the DM relic density using Eqs.~(\ref{eq:Omega_s},\ref{eq:Omega_s_s})  for a set of ISS(2,3) configurations  satisfying data from neutrino oscillation experiment  and laboratory constraints. We have  imposed $f_{\rm WDM}=\Omega_{\rm DM}/\Omega_{\rm DM}^{\rm Planck} \leq 1$ thus obtaining constraints for  $m_s$ and $U_{\alpha s}$.  

The configurations with DM relic density not exceeding the experimental determination have been confronted with the limits coming from structure formation. 
There are several strategies to determine the impact of WDM on structure formation, leading to different constraints; in fact most of these constraints assume that the total DM component is accounted by WDM produced through the DW mechanism.  Notice that these constraints can  be relaxed when this hypothesis does not hold and we will address this point in a forthcoming section.

In the following, and when possible, we will thus reformulate the bounds from structure formation in terms of the quantity $f_{\rm WDM}$ which represents the amount of DM produced from active-sterile oscillation.\footnote{The results presented are in fact approximative estimates. A proper formulation would require detailed numerical studies, beyond the scope of this work.}   

The most solid bounds come from the analysis of the phase-space distribution of astrophysical objects. The WDM  free-streaming scale is of the order of the typical size of galaxies; as a consequence,  the formation of DM halos, as well as that of the associated galaxies is deeply influenced by the DM distribution function. According to this idea, it is possible to obtain  robust limits on the DM mass by requiring that the maximum of the dark matter distribution function inferred by  observation, the so called coarse grained phase space density, does not exceed the one of the fine-grained  density, which is theoretically determined and dependent on the specific DM candidate. 
Using this method, an absolute lower bound on the DM mass of around 0.3~keV, dubbed Tremaine-Gunn (TG) bound~\cite{Tremaine:1979we} was obtained by comparing the DM distribution from the observation of Dwarf Spheroidal Galaxies (Dphs) with the fine-grained distribution of a Fermi-gas. A devoted study of sterile neutrinos produced by DW mechanism has been presented in~\cite{Boyarsky:2008ju}, where a lower mass bound of the order of 2~keV was obtained. This limit can be evaded assuming that the WDM candidate is a subdominant component, while the DM halos are mostly determined by an unknown cold dark matter component. 
The reformulation of the limits in this kind of scenarios requires a dedicated study (an example can be found in~\cite{Anderhalden:2012qt}). In this work we conservatively rescale the results of~\cite{Boyarsky:2008ju} under the assumption that the observed phase-space density is simply multiplied by a factor $f_{\rm WDM}$. 
Moreover we have considered as viable the points of the ISS(2,3) model with $m_s < 2$~keV, featuring a value $f_{\rm WDM} \lesssim 1\%$, which corresponds approximatively to the current experimental uncertainty in the determination of the DM relic density.

For masses above 2~keV another severe bound is obtained from the analysis of the Lyman-$\alpha$ forest data. From these it is possible to indirectly infer the spectrum of matter density fluctuations, which are in turn determined by the DM properties. 
The Lyman-$\alpha$ constraint  is strongly model dependent and the bounds are related to the WDM production mechanism,  and to which extent this mechanism contributes to the total DM abundance. In order to properly take into account the possibility of only a partial contribution of the sterile neutrinos to the total DM abundance, we have adopted the results presented in~\cite{Boyarsky:2008xj} where the Lyman-$\alpha$ data have been considered in the case in which sterile neutrinos WDM account for the total DM abundance, as well as in the case in which they contribute only to a fraction (the remaining contribution being  originated by a cold DM component). More precisely, we have considered the most stringent 95\% exclusion limit\footnote{The limit considered actually relies on  data sets which are not  up-to-date. A more recent analysis~\cite{Viel:2013fqw} has put forward a stronger limit in the case of a pure WDM scenario, and thus the limits are  underestimated. As it will be clear in the following, the final picture is not  affected by this.}, expressed in terms of $(m_s, f_{\rm WDM})$, and translated it into an exclusion limit on the parameters of our model\footnote{Notice that the Lyman-$\alpha$ method is reliable for DM masses above 5~keV. For lower values there are very strong uncertainties and it is not possible to obtain solid bounds. In~\cite{Boyarsky:2008xj} it is argued  that the limit on $f_{\rm WDM}$ should not significantly change at lower masses with respect to the one obtained for neutrinos of 5~keV mass.}, namely the mass $m_s$ of the sterile neutrino and its effective mixing angle  with active neutrinos $\theta_s$.
We finally remark that WDM can be constrained also through other observations, as the number of observed satellites of the Milky way~\cite{Polisensky:2010rw,Horiuchi:2013noa,Boyarsky:2012rt}, giving a lower bound on the DM mass of approximately 8.8~keV. 
This last kind of limits however strictly relies on the assumption that the whole dark matter abundance is totally originated by a WDM candidate produced through the Dodelson-Widrow mechanism and cannot straightforwardly be reformulated in case of a  deviation from this hypothesis;  thus we have not been considered these limits in our study. 

The inverse Seesaw realisations passing the structure formation constraints have to be confronted to the limits from the X-ray searches, as reported in, for instance~\cite{Boyarsky:2012rt}. The corresponding constraints are  again given in the plane $(m_s,\theta_s)$, and can be schematically expressed by\footnote{Notice that the exclusion limit from X-rays is actually the combination of the outcome of different experiments and the dependence on the dark matter mass deviates in some regions from the one provided above. We have taken this effect into account in our analysis.}:
\begin{equation}
\label{eq:approx_X}
f_{\rm WDM} \sin^2 2\theta \lesssim  10^{-5} {\left(\frac{m_s}{1 \text{ keV}}\right)}^{\!-5}\ ,
\end{equation}    
where~\cite{Pal:1981rm,Boyarsky:2005us}:
\be 
\label{eq:pal}
\sin^2 2 \theta = \frac{16}{9} \sum_{i=1}^3 \left| \sum_{\alpha=e,\mu,\tau} U_{\alpha, s} \ U_{\alpha,i}^* F(r_\alpha)\right|^2, 
\,\,\,\,F(r_\alpha) = - \frac{3}{2} + \frac{3}{4} r_\alpha, \ r_\alpha=\left(\frac{m_\alpha}{M_W}\right)^2\ ,
\ee
with $i$ running over the different active neutrino mass eigenstates (3 different final states in the decay are possible). In the above expression we have again accounted for the possibility that the sterile neutrino contributes only partially to the DM component by rescaling the limit with a factor $f_{\rm WDM}$.\footnote{Notice that, contrary to the case of bounds from structure formation, this scaling is strictly valid only if the additional components does not decay into photons and thus it will not be applied in the next sections.} 

\begin{figure}[htb]
\begin{center}
\subfloat{\includegraphics[width=0.45\textwidth]{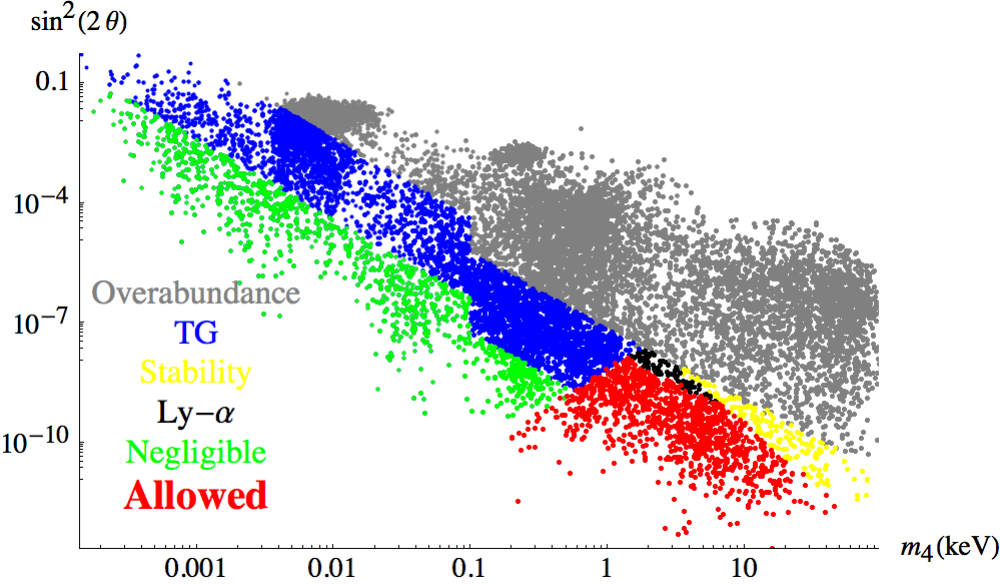}}\hspace*{0.5cm}
\subfloat{\includegraphics[width=0.45\textwidth]{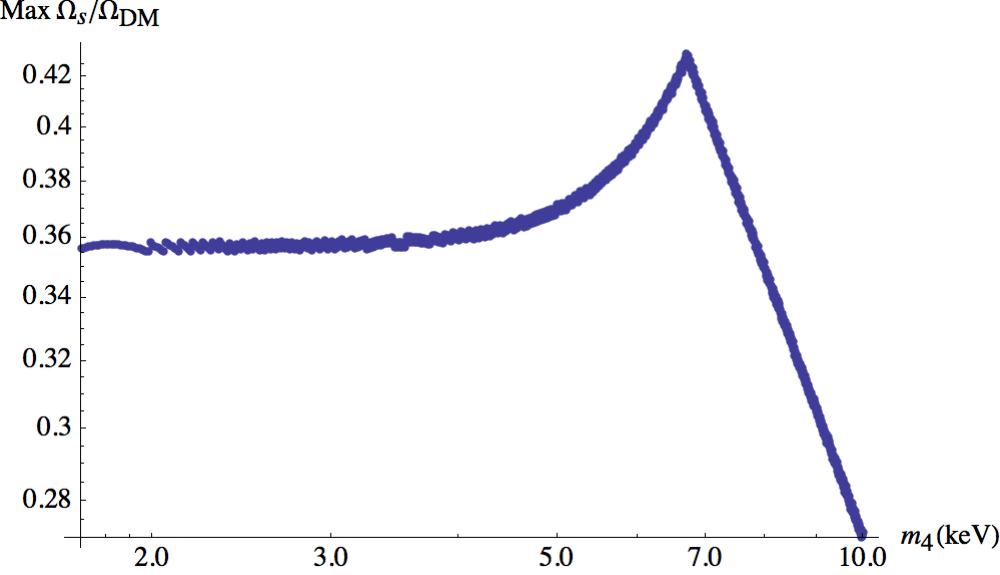}}
\end{center}
\caption{On the left panel, different regions of the lightest sterile neutrino parameter space $(m_4,\sin^2 2\theta$) identified by DM constraints. The grey region corresponds to a DM relic density exceeding the cosmological value. The blue, black and yellow regions are also excluded by phase space distribution, Lyman-$\alpha$ and X-ray searches constraints, respectively. The green region corresponds to configurations not excluded by cosmology but in which the lightest sterile neutrino contributes with a negligible amount to the DM relic density. Finally, the red region corresponds to the ISS(2,3)  configurations fulfilling  all the cosmological constraints, and for which the contribution to the dark matter relic density from the light sterile neutrino is sizeable. On the right panel, maximal value of $f_{\rm WDM}$ allowed by cosmological constraints as a function of the mass of the lightest sterile neutrino.}
\label{fig:summary_no_heavy}
\end{figure}

The result of the combination of the three kinds of constraints  applied in our analysis, namely dark matter relic density, structure formation and indirect detection, is reported in Figure~\ref{fig:summary_no_heavy}. As can be seen,  the requirement of a correct DM relic density has a very strong  impact, excluding a very large portion of the parameter space (grey region) at the highest values of the active-sterile mixing angles. Phase space density constraints rule-out most of   the configurations with mass of the lightest sterile neutrino below $\sim 2 \text{ keV}$ (blue region), a part a  narrow strip (green region) corresponding to $f_{\rm WDM} < 1 \%$. In this last region, and although not ruled out, the ISS(2,3) model cannot solve the Dark Matter puzzle, at least in its minimal realisation. In the large mass region, namely above 2~keV, a further exclusion comes form Lyman-$\alpha$ and indirect detection bounds (respectively black and yellow region) reducing the allowed active-sterile mixing. 
A sizeable contribution to the DM relic density can be thus achieved in a small localised region (in red)  of the parameter space, corresponding to masses of the lightest sterile neutrino in the range 2 - 50~keV and for active-sterile mixing angles $10^{-8} \lesssim \sin^2 2 \theta \lesssim 10^{-11}$.      
We show in the right panel of Figure~\ref{fig:summary_no_heavy} the maximal value of $f_{\rm WDM}$ allowed by the cosmological constraints as function of the DM mass. As  can be seen, the lightest sterile neutrino can only partially account for the DM component of the Universe with $f_{\rm WDM} \sim 0.43$ in the most favourable case. The maximal allowed DM fraction increases for the lowest values of the mass until a maximum at around 7~keV, after which it displays a  sharp decrease. This behaviour can be explained as follows: at lower masses, the Lyman-$\alpha$ bounds are the most effective and  become weaker as the mass of the sterile neutrino increases, thus allowing for larger $f_{\rm WDM}$. At the same time, the bounds from X-ray sources become stronger (since higher masses imply higher decay rates) thus reducing the allowed DM fraction as the mass increases.    

Notice that the above analysis is valid within the assumption that the production of the lightest sterile neutrino occurs in the  absence of a lepton asymmetry. Indeed, as firstly shown in~\cite{Shi:1998km}, the production of sterile neutrinos can be resonantly enhanced (as opposed to the conventional DW production usually called  non-resonant) in presence of a non-zero lepton asymmetry. In this case the correct dark matter abundance is achieved for much smaller active-sterile mixing angles, thus evading the limits from dark matter indirect detection; in addition, the resonant production alters the DM distribution function with respect to a non-resonant production, rendering  it  ``colder'' and thus compatible with Lyman-$\alpha$ constraints~\cite{Boyarsky:2008mt}. 

Interestingly a lepton asymmetry can be generated in frameworks featuring keV scale sterile neutrinos accompanied by heavier right-handed neutrinos. The entries of the active-sterile mixing matrix can in general be complex, and give rise to  CP-violating phases;  as a consequence, a lepton asymmetry can be generated by oscillation processes of the heavy neutrinos. In particular, it has been shown that a pair of quasi-degenerate right-handed neutrinos with masses of the order of a few GeV can generate a  lepton asymmetry before the EW phase transition (which is converted to the current baryon asymmetry of the Universe)  and then  at much later times, the lepton asymmetry needed to provide the correct relic density for a keV scale sterile neutrino~\cite{Shaposhnikov:2008pf,Laine:2008pg,Canetti:2012zc,Canetti:2012kh}. 
The ISS(2,3) model also features pairs of quasi-degenerate heavy neutrinos which can be of the correct order of mass. However, the lepton asymmetry needed to ensure the correct DM relic density, compatible with the bounds discussed, requires an extreme degeneracy in  the heavy neutrino spectrum, of the order of the atmospheric mass differences. Such an extreme degeneracy is not achievable for the ISS model since  the predicted degeneracy of the pair of heavy neutrinos is of ${\mathcal{O}}(\mu)$, corresponding to around 1~keV for the cases under consideration.\footnote{Notice that a mass degeneracy of ${\mathcal{O}}(\text{keV})$ is still feasible for baryogenesis through oscillations of the heavy right-handed neutrinos.} A sizeable lepton asymmetry can be, however, generated by oscillation of not-degenerate neutrinos in the so-called flavoured leptogenesis~\cite{Akhmedov:1998qx} where individual lepton asymmetries in the different flavours are generated due to oscillations but the total lepton number is not necessarily violated. This mechanism has been, indeed, proven to be successful in explaining baryogenesis via leptogenesis thanks to sphaleron interactions~\cite{Canetti:2012kh,Drewes:2012ma,Garbrecht:2014bfa,Canetti:2014dka}, provided that there are at least three neutrinos contributing to the generation of the lepton asymmetry, and might be also efficient in generating the correct lepton asymmetry in order to have a resonantly enhanced DM production. This scenario is particularly promising in the ISS(2,3) model since it features four pseudo-Dirac neutrinos, potentially contributing to the generation of a lepton asymmetry. A quantitative investigation is however beyond the scope of the present work and is left for a future study. 

\subsection{Impact of the heavy pseudo-Dirac states}

The picture presented above can be altered in some regions of the parameter space due to the presence of the heavy neutrinos. Indeed, contrary to the DM candidate, they can exist in sizeable abundances in the Early Universe owing to their efficient Yukawa interactions, and influence the DM phenomenology through their decays.
There are two possibilities. The first one is  direct DM production from decays mediated by Yukawa couplings.
The branching ratio of these processes is small when compared to that of other decay channels into SM states, since it is suppressed by the  small active-sterile mixing angle, an efficient DM production can nevertheless be achieved through the so called freeze-in mechanisms if the pseudo-Dirac neutrinos are heavier than the Higgs boson.  
Significantly lighter pseudo-Dirac neutrinos, namely with masses below~$\sim 20\,\mbox{ GeV}$, can instead indirectly  affect DM phenomenology. Indeed,  they can be  sufficiently long-lived such that they can dominate the energy density of the Universe,  injecting entropy at the moment of their decay. 
We will discuss separately these two possibilities in the next subsections.  

\subsubsection{Effects of entropy injection}   

The conventional limits on sterile neutrino DM can be in principle evaded in presence of an entropy production following the decay of massive states dominating the energy density of the Universe~\cite{Scherrer:1984fd}. A phase of entropy injection dilutes the abundance of the species already present in the thermal bath and, in particular, the one of DM if such an entropy injection occurs after its production. In addition, the DM momentum distribution gets redshifted - resembling a ``colder'' DM candidate - and   suffering weaker limits from Lyman-$\alpha$. This phase of entropy injection can be triggered in the ISS(2,3) model by the decay of the heavy pseudo-Dirac neutrinos if the following two conditions are realised:  
 Firstly, at least some of the heavy sterile neutrinos should be sufficiently abundant to dominate the energy budget of the Universe. Secondly they must decay after the peak of dark matter production, but before the onset of  Big Bang Nucleosynthesis (BBN). These two requirements will identify a  limited region of the parameter space outside which the results of the previous subsection strictly apply.  

All the massive eigenstates have Yukawa interactions with ordinary matter described by an effective coupling $Y_{\rm eff}$ which is defined by:
\begin{equation}
Y_{\alpha \beta} \,\overline{\ell_L^\alpha} \,\tilde{H }\, \nu_R^\beta = Y_{\alpha \beta} \, \overline{\ell_L}^\alpha \,\tilde{H }\, U_{\beta i} \nu_i
=Y_{\rm eff}^{\alpha i}\  \overline{\ell_L^\alpha} \,\tilde{H }\,  \nu_i\ .
\end{equation}
These interactions are mostly efficient at high temperature when scattering processes involving the Higgs boson and top quarks are energetically allowed;  in addition, they maintain the pseudo-Dirac neutrinos in thermal equilibrium until temperatures of the order of $\sim 100 \text{ GeV}$, provided that $Y_{\rm eff}^2 \gtrsim 10^{-14}$~\cite{Akhmedov:1998qx}. If this condition is satisfied, an equilibrium abundance of heavy pseudo-Dirac neutrinos existed at the early stages of the evolution of the Universe.   

The Yukawa interactions become less efficient as the temperature decreases. At low temperature  the transition processes from the light active neutrinos become important. For a given neutrino state, the rate of the transition processes reaches a maximum at around~\cite{Asaka:2006ek}:
\begin{equation}
\label{eq:T_N_D}
T_{\text{max},I} \simeq 130 {\left(\frac{m_I}{1\text{ keV}}\right)}^{\frac{1}{3}} \text{MeV}\ .
\end{equation}

The transition rate of each neutrino at the temperature $T_{\text{max},I}$ exceeds the Hubble expansion rate $H$ if~\cite{Asaka:2006ek}:
\begin{equation}
\label{eq:Heavy_equilibrium}
\theta > 5 \times 10^{-4} \left(\frac{1\text{ keV}}{m_I}\right)^{1/2}\ , 
\end{equation} 
and thus, if this condition is satisfied, the corresponding pseudo-Dirac neutrinos are in thermal equilibrium in an interval of temperatures around $T_{\text{max},I}$. 

Notice that the picture depicted above assumes that the production of sterile neutrinos from oscillations of the active ones is energetically allowed; as a consequence it is valid only for neutrino masses lower than $T_{\rm max}$:
\bee\label{eq:condition}
T_{\text{max},I} \simeq 130 {\left(\frac{m_I}{1\text{ keV}}\right)}^{\frac{1}{3}} \text{MeV} \geq m_{I} &\Rightarrow & m_I \leq m_{I,\text{max}} \approx  46.87 \text{ GeV}.
\eee
As will be made clear in the following, neutrinos heavier than $M_{I,\text{max}}$ have excessively large decay rates to affect DM production and  hence  will not be relevant in the subsequent analysis.

In Figure~\ref{heavy_unbroken}, we present the typical behaviour of the pseudo-Dirac neutrinos in the regimes of high and low temperatures, which are dominated, respectively, by Yukawa interactions and active-sterile transitions. In the left panel of Figure~\ref{heavy_unbroken}, we display  the values of the mass and effective Yukawa couplings ($Y_{\rm eff}$) of the lightest pseudo-Dirac state (the other heavy states exhibit an analogous behaviour), corresponding to a set of ISS(2,3) realisations compatible with laboratory tests of neutrino physics (red points). The green region translates the equilibrium condition for the Yukawa interactions. In the larger  mass region, i.e. for masses significantly larger than 10 GeV, the value of the effective Yukawa coupling $Y_{\rm eff}$ is always above the equilibrium limit and can even be of order one for  higher values of the mass. In this region,  the pseudo-Dirac neutrinos can have  a WIMP-like behaviour and can be in thermal equilibrium until low temperatures. As already mentioned neutrinos in this mass range have impact on Higgs phenomenology at the LHC; we have compared the configurations of Figure~\ref{heavy_unbroken} with the limits presented e.g. in~\cite{BhupalDev:2012zg} and found they all result viable. 
 In  the intermediate mass region, i.e. for masses  between 1 and  a few tens of GeV,  equilibrium configurations are still present. However the values of $Y_{\rm eff}$ are lower with respect to the previous case and the decoupling of the pseudo-Dirac neutrinos depends on the oscillation processes at low temperatures. Configurations for which $Y_{\rm eff}$ is too small to ensure the existence of a thermal population of pseudo-Dirac neutrinos in the early Universe (they can be nonetheless created by oscillations at lower temperatures) are also present. This last kind of configurations are the only ones corresponding to  masses below 0.1~GeV.
We emphasise that the outcome discussed here  is a direct consequence of the ISS(2,3) mechanism which allows to generate the viable  active neutrino mass spectrum for pseudo-Dirac neutrinos with masses of the order of the EW scale, and for large values of their Yukawa couplings.
For comparison, we display in the same plot the distribution of values of the effective Yukawa couplings of the WDM candidate as a function of the mass of next-to-lightest sterile state $m_5$  (blue points). As can be seen, the corresponding solutions are always far from thermal equilibrium due to the suppressed mixing $U_{\nu_R,4}$.  

\begin{figure}[htb]
 \begin{center}
\subfloat{\includegraphics[width=0.45\textwidth]{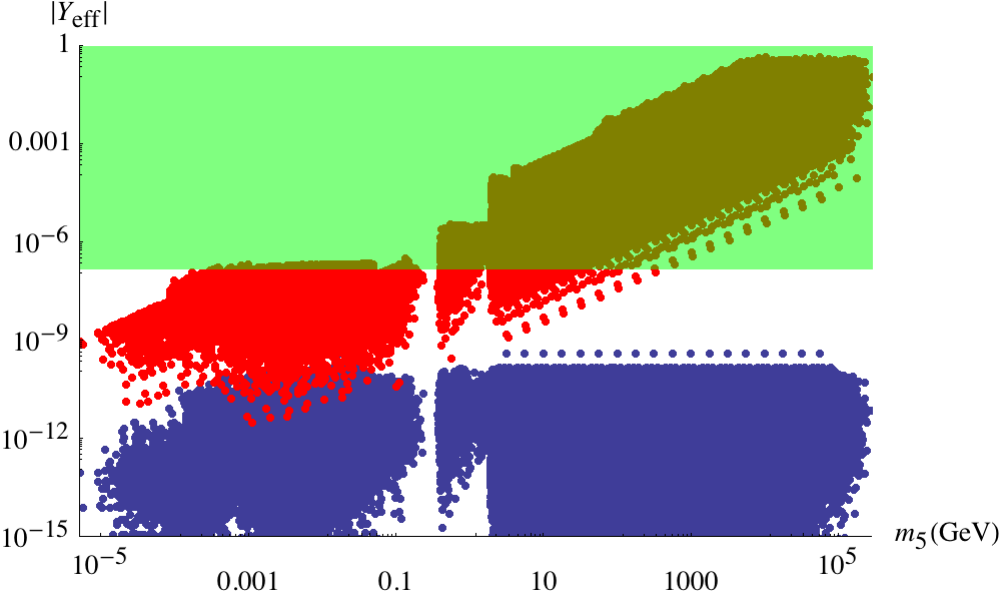}}\hspace*{0.6cm}
\subfloat{\includegraphics[width=0.45\textwidth]{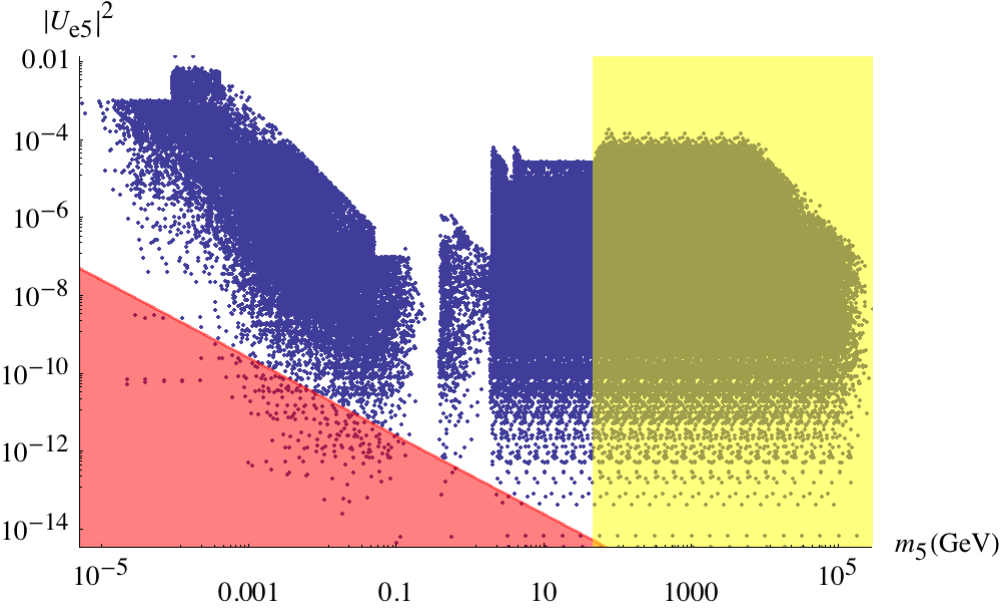}}
\end{center}
\caption{On the left panel: effective Yukawa couplings $Y_{\rm eff}$ for the neutrino DM candidate (blue~points) and of the lightest pseudo-Dirac particle (red points), as a function of the mass $m_5$.
The green region corresponds to   values $Y_{\rm eff}>\sqrt{2}\times 10^{-7}$, the limit above which the states are in thermal equilibrium. On the right panel: mixing of the electron neutrino with the lightest pseudo-Dirac state as a function of its mass. 
 The yellow region corresponds to the kinematically forbidden values of the sterile mass, see  Eq.~(\ref{eq:condition}).
 The red region denotes the solutions not in thermal equilibrium.}
\label{heavy_unbroken}
\end{figure}
In the right panel of Figure~\ref{heavy_unbroken}, we display  the mixing (for small angles it is possible to approximate $\theta_{e 5} \simeq U_{e 5}$) of the lightest pseudo-Dirac state with the electron neutrino as a function of  $m_5$  for the ISS realisations compatible with laboratory limits.  The yellow region corresponds to the values of the sterile mass for which the DW production mechanism is kinematically forbidden  (see Eq.~(\ref{eq:condition})).  The red region denotes the solutions which are not in thermal equilibrium.

Combining the results obtained from the  two panels of Figure~\ref{heavy_unbroken}, we can conclude that all the considered realisations in the relevant mass interval satisfy the equilibrium conditions.  Consequently we can always assume the presence of an equilibrium population of the pseudo-Dirac states up to temperatures of the order of $T_{\text{max},I}$. We stress that $T_{\text{max},I}$ is not the actual decoupling temperature that has been instead determined  in for instance~\cite{Shaposhnikov:2008pf}  and more recently in~\cite{Canetti:2012kh}, and which turns out to be lower than $T_{\text{max}}$; however this affects only marginally our discussion.

As already pointed out, we will be interested in masses of the pseudo-Dirac neutrinos not exceeding 10 - 20~GeV. For  such a mass range we can safely assume that the neutrinos decouple when they are relativistic (see Eq.~(\ref{eq:T_N_D})) and that their decay occurs at a much later stage, when they become non-relativistic, as described in~\cite{Asaka:2006ek}.

In this setup, the pseudo-Dirac states can dominate the energy budget of the Universe if their energy density, which is defined by
\begin{equation}
\rho_N(T) \equiv \sum_{I=5,8} m_I n_I(T),\,\,\,\,\,\,n_I(T)=\frac{g_{*}(T)}{g_{*}(T_D)}{\left(\frac{T}{T_D}\right)}^3 n_I^{\rm eq}(T_D)=\frac{g_{*}(T)}{g_{*}(T_D)} \frac{3 \zeta(3)}{2 \pi^2} T^3\ ,
\end{equation} 
exceeds the radiation energy density $\rho_r=\frac{\pi^2}{30}g_{*}(T)T^4$, where $g_{*}(T)$ represents the number of relativistic degrees of freedom at the temperature $T$. 
Provided that the pseudo-Dirac neutrinos are sufficiently long-lived, this occurs at a temperature $\overline{T}$ given by:
\begin{equation}
\label{eq:Tbar}
\overline{T} \approx 6.4 \, \mbox{MeV} \left(\frac{m_5}{1 \text{GeV}}\right)\left(\frac{\sum_{I} m_I Y_I}{m_5 Y_5}\right)\ ,
\end{equation}
where we have taken $g_{*}(T_D)=86.25$ and $m_5$ is the mass of the lightest pseudo-Dirac neutrino. In this scenario the decay of the pseudo-Dirac neutrinos is accompanied by a sizeable amount of entropy; the conventional radiation dominated era restarts at the reheating temperature $T_{r,I}$~\cite{Scherrer:1984fd},  
and the abundance of the species present in the primordial thermal bath is diluted by a factor $S$, which is defined as the ratio of the entropy densities of the primordial plasma at temperatures immediately below and above the reheating one.

Notice that the above discussion corresponds to a simplified limit: in general the four pseudo-Dirac neutrinos have different masses and different lifetimes. 
In the ISS(2,3) model the pseudo-Dirac states appear as pairs with the mass splitting in each pair much smaller than the masses of the corresponding states. Identifying the  mass scale of each pair as  $m$ and $M$, with $m < M$, we can write, to a  good approximation\footnote{The discussion of this section, as well as the expressions here presented, are valid in the so called "instantaneous reheating approximation" which assumes that the entropy injection occurs at the reheating temperature. In fact the entropy release is a continuous process and the quantities $T_{r,M/m}$ and $S_{M,m}$ are not determined analytically but  extrapolated from the numerical solution of suitable Boltzmann equations~\cite{Arcadi:2011ev}. Moreover at high temperatures, namely $T \gtrsim m_I$, the decay rate of massive states into radiation is altered by effects from, for example, thermal masses or quantum statistical effects~\cite{Kolb:2003ke,Yokoyama:2005dv,Yokoyama:2006wt,Bodeker:2006ij,Drewes:2013iaa} and the prediction for the reheating temperature might sensitively deviate from the prediction obtained in the instantaneous reheating approximation~\cite{Drewes:2014pfa}.      

In the setup under consideration we assume the pseudo-Dirac neutrinos decoupling at the temperature $T_{\rm max,I}$ defined in~(\ref{eq:T_N_D}). For the range of masses of pseudo-Dirac neutrinos for which the active-sterile transitions are effective $T_{\rm max,I} >m_I$ and increases while $m_I$ gets lower. In particular we have that $T_{\rm max}/m_I \sim 10$ for $m_I \sim 1\,\mbox{GeV}$. 

On the other hand, comparing the decay rates given in~(\ref{eq:Gammah}) and~(\ref{eq:Gammaz}), it results that the decay temperatures of the pseudo-Dirac neutrinos are lower than the masses of the neutrinos themselves at least for $m_I \lesssim 10\,\mbox{GeV}$ but they can be even lower by considering $Y_{\rm eff} \lesssim 10^{-3}$. The most relevant impact from the decays of the pseudo-Dirac neutrinos is obtained for very low decay temperatures, for which it is reasonable to neglect thermal corrections. Our main results can be thus described by the instantaneous reheating approximation.}:
\begin{equation}
 S=  S_{ m}\,  S_{M}\ ,
 \label{entropytot}
\end{equation}  
where $S_{ m}$ and $S_{M}$ are the dilution factors associated to the decays of the two pairs of pseudo-Dirac states, occurring at the two reheating temperatures $T_{r,M}$ and $T_{r,m}$, given by:
\begin{align}
\label{eq:entropy_release}
& S_M= {\left[1+2.95  {\left(\frac{2 \pi^2}{45}g_{*}(T_{r,M})\right)}^{1/3}{\left(\frac{\sum_{\alpha} m_\alpha Y_\alpha}{M\ Y_M}\right)}^{1/3}\frac{{(M\ Y_M)}^{4/3}}{(\Gamma_M M_{\rm PL})^{2/3}}\right]}^{3/4}\ ,  \nonumber\\
& S_{ m}= {\left[1+2.95  {\left(\frac{2 \pi^2}{45}g_{*}(T_{r,m})\right)}^{1/3}2^{1/3}\frac{{\left(\frac{m\ Y_m}{S_{ M}}\right)}^{4/3}}{(\Gamma_m M_{\rm PL})^{2/3}}\right]}^{3/4}\ ,
\end{align}
where $\Gamma_{m,M}$ is the decay rate of the heavy neutrinos. Notice that in the last term of each of the above equations, the effects of the first entropy dilution  have been included in the abundance of the lightest pair of heavy neutrinos.

The DM phenomenology is affected only when the pseudo-Dirac neutrinos dominate the Universe and decay after DM production. For keV scale DM,  this translates into the requirement $T_{r,m} \lesssim 150$ MeV. On the other hand,  a very late reheating phase would alter the population of thermal active neutrinos, leading to modifications of some quantities such as the primordial Helium abundance~\cite{Kawasaki:2000en} and the effective number of neutrinos $N_{\rm eff}$, and producing effects in structure formation  as well. By combining BBN and CMB data\footnote{There are also further cosmological constraints on heavy neutrinos derived using different approaches, see for instance~\cite{Dolgov:2000jw,Ruchayskiy:2012si,Hernandez:2014fha}.} it is possible to determine a solid bound $T_{r,m}>4$ MeV~\cite{Hannestad:2004px}. In addition, we have  considered  a (relaxed) limit of~$T_{r,m}>0.7$ MeV by  taking into account the possibility that this bound is evaded when the decaying state can produce ordinary neutrinos~\cite{Fuller:2011qy}. This choice is also motivated by the fact that after the decay of the neutrinos with mass $M$, the ratio $\rho_I/\rho_r$ between the energy densities of the remaining neutrinos and of the radiation is of order 3 - 5, implying that although subdominant, the radiation component is sizeable.  

The requirement $4 (0.7)\text{ MeV} \leq T_{r,m} \leq 150\text{ MeV}$ is satisfied only for a very restricted pseudo-Dirac neutrinos mass range. Indeed, sterile neutrinos can decay into SM particles through three-body processes mediated by the Higgs boson with a rate:
\begin{equation}
\label{eq:Gammah}
\Gamma_h= \frac{Y_{\rm eff}^2 \ m_I^5}{384 {\left(2 \pi\right)}^3 m_h^4} \sum_f y_f^2 \left(1-\frac{4 m_f^2}{m_I^2}\right)\ ,
\end{equation}
which implies an excessively  short lifetime for sterile neutrinos unless their masses are below (approximately) $\mathcal{O}(10\text{ GeV})$, in such a way that the decays into third generation quarks are kinematically forbidden and $Y_{\rm eff}$ can assume lower values. At these smaller masses, a sizeable contribution comes from $Z$ mediated processes with a rate:
\begin{equation}
\label{eq:Gammaz}
\Gamma_Z = \frac{G_F^2 m_I^5 \sin^2\theta_I}{192 \pi^3}\ , 
\end{equation}
where we have defined, for simplicity, effective mixing angles $\theta_I,I=m,M$ between the pseudo-Dirac neutrinos and the active ones. 

We have reported in Figure~\ref{fig:Tr_limits} the limit values of the lower reheating temperature $T_{\rm r, m}$ as a function of the mass scale $m$ and the effective mixing angle $\theta_m$, for three values of $Y_{\rm eff}$, namely 0.1, $10^{-3}$ and $10^{-6}$. The regions above the red curves correspond to an excessively large reheating temperature which does not affect DM production. 
The light-grey (dark-grey) regions below the blue curves represent values of the reheating temperature in conflict with the conservative (relaxed) cosmological limit of 4 (0.7) MeV. For ``natural'', i.e. $\mathcal{O}(1)$, values of the effective coupling, the decay rate of the heavy neutrinos is dominated by the Higgs channel and tends to be  too large except for  a narrow strip at masses of $1 - 2$ GeV. At lower values of $Y_{\rm eff}$ the size of the region corresponding to the interval 4 (0.7) - 150 MeV of reheating temperatures increases. The contours corresponding to the lower values of the reheating temperature are mildly affected by the values of $Y_{\rm eff}$ since,  at lower masses, the Higgs channel is suppressed by the Yukawa couplings of the first generation, in comparable amount with the $Z$ channel.
As can be seen from the left panel of  Figure~\ref{heavy_unbroken}, laboratory constraints favour values of $Y_{\rm eff} < 10^{-3}$ in the mass range $1 - 20$ GeV thus favouring  the possibility of an impact of the heavy neutrinos decays on the DM phenomenology,  in this region.

\begin{figure}[htb]
\begin{center}
\subfloat{\includegraphics[width=0.3\textwidth]{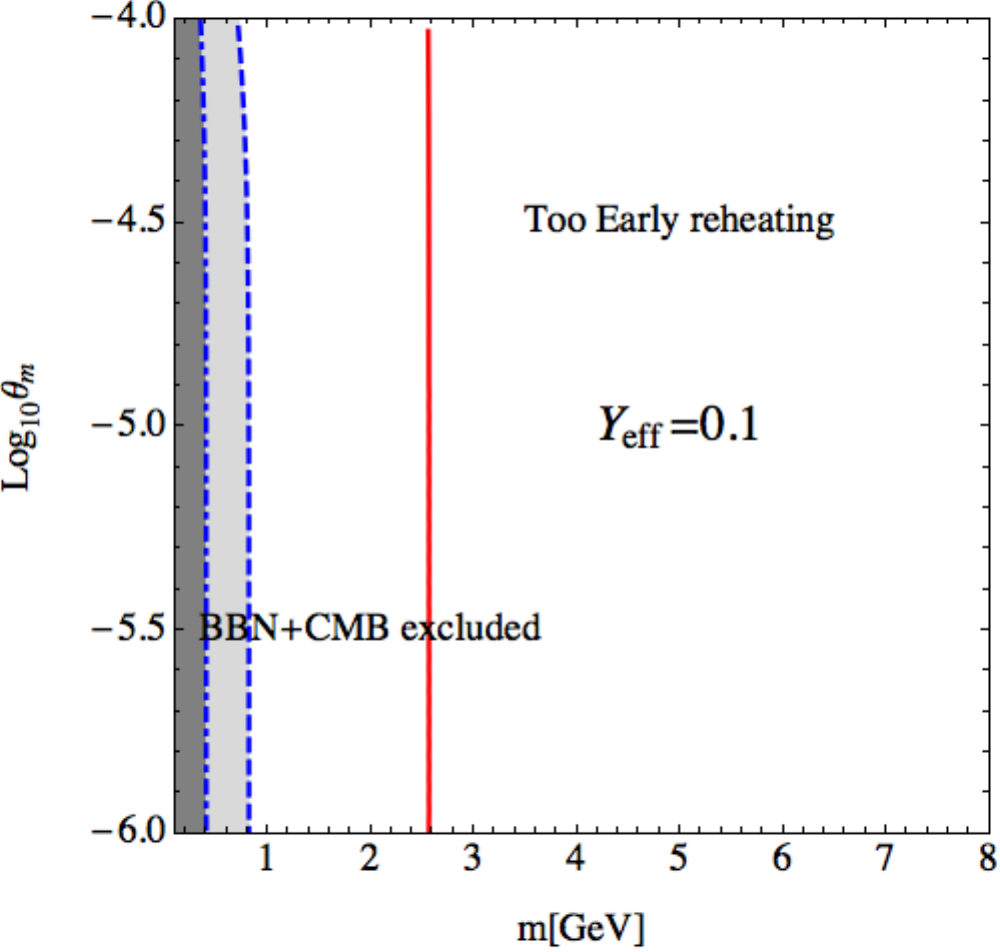}}\hspace*{0.4cm}
\subfloat{\includegraphics[width=0.3\textwidth]{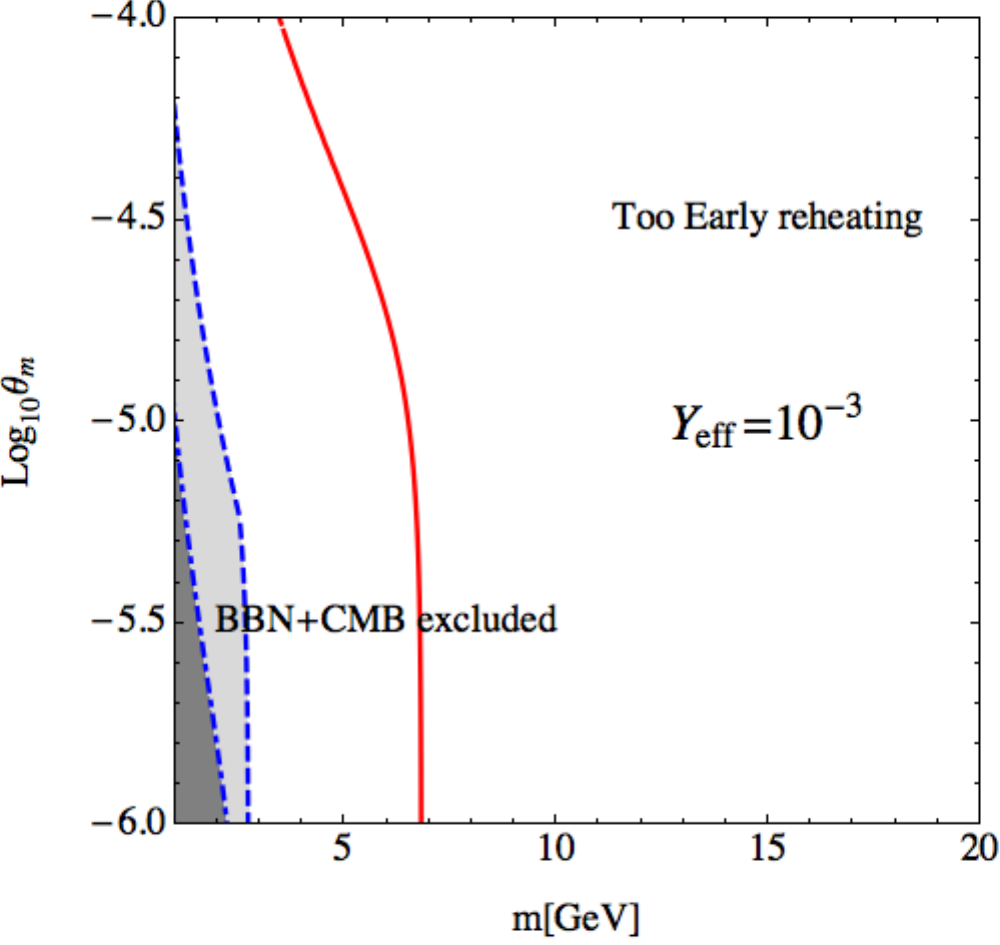}}\hspace*{0.4cm}
\subfloat{\includegraphics[width=0.3\textwidth]{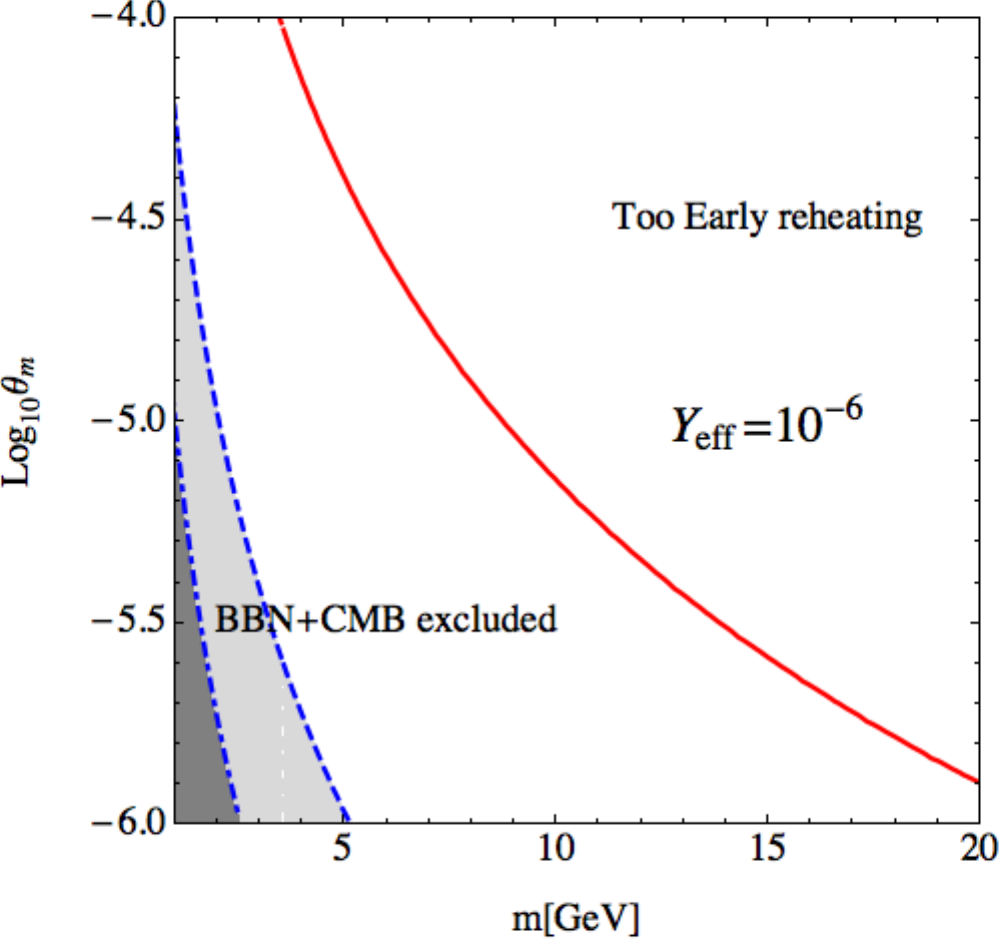}}
\end{center}
\caption{Region of the $(m,\theta_m)$ parameter space in which the decay of the pseudo-Dirac states can affect the DM phenomenology for three values of $Y_{\rm eff}$,  0.1, $10^{-3}$ and $10^{-6}$. The red lines in the panels represent $T_{r,m} = 150$ MeV. Above this line the reheating takes place before the DM production. The grey region below the dashed (dot-dashed) blue line is excluded by BBN/CMB combined constraints according to the limit $T_{r,m} > 4\ (0.7)$ MeV.}
\label{fig:Tr_limits}
\end{figure}

In Figure~\ref{fig:DeltaS} we have estimated the range of values of $S$  in the allowed parameter space, see Eq.~(\ref{entropytot}). In order to illustrate, we have chosen  to vary   $m$ and $ \theta_m$ fixing  $M$ to be $M=2\times m$ and $\theta_M=10^{-4}$, which corresponds to a  viable case for the ISS(2,3) mass spectrum; we have also  fixed the effective Yukawa coupling as $Y_\text{eff}=10^{-6}$ in order to maximise the phenomenologically relevant region of the parameter space.

\begin{figure}[htb]
\begin{center}
\includegraphics[width=0.55\textwidth]{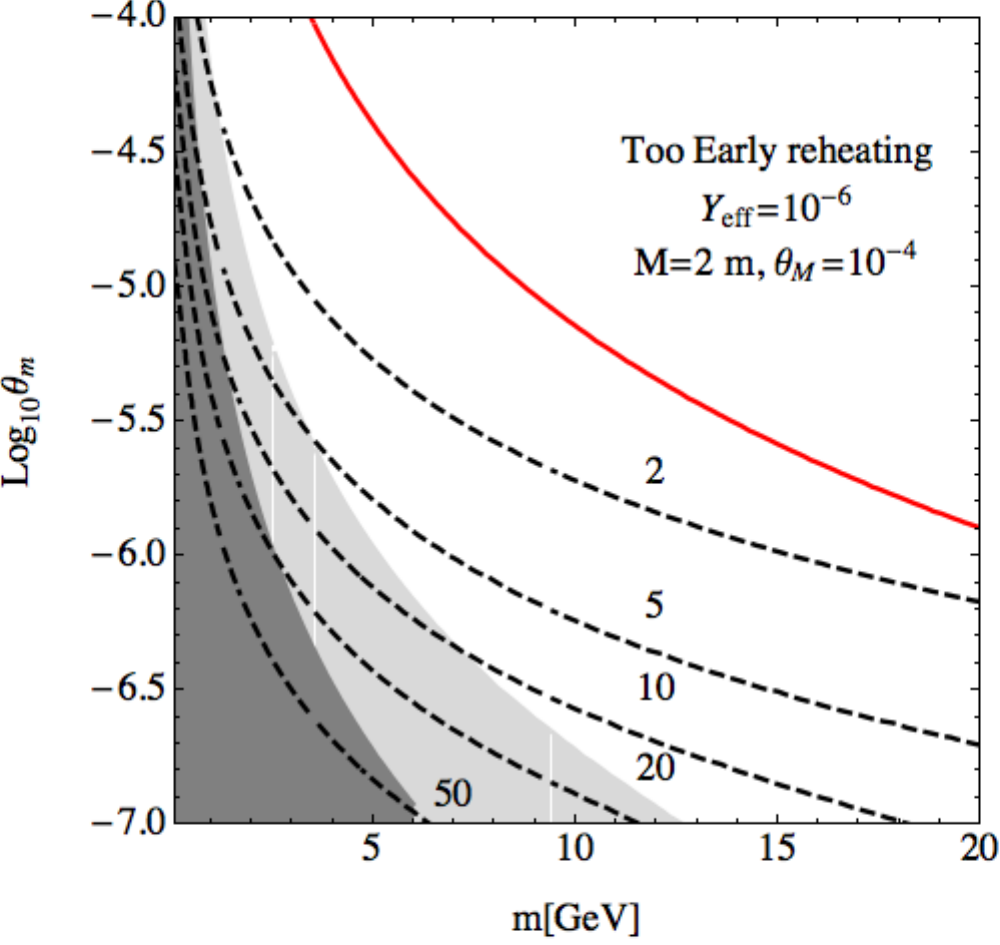}
\end{center}
\caption{Iso-curves of entropy injection in the plane $(m,\theta_m)$ with the remaining parameters $M$, $\theta_M$ and $Y_\text{eff}$ fixed according to the reported values. The grey (light grey) region corresponds to the lowest reheating temperature below 4 (0.7) MeV and is thus in tension with the cosmological bounds. In the region above the red curve the entropy injection takes place before the DM production.}
\label{fig:DeltaS}
\end{figure}

As can be seen from Figure~\ref{fig:DeltaS},  we have only very moderate values of the entropy injection when the conservative lower limit of 4 MeV  is imposed on the reheating temperature; values of $S$ up to around 20 can be achieved once a weaker bound is considered. 

Having determined the range of variation of the entropy dilution within the ISS(2,3) parameter space,  we have reformulated the limits on the DM mass and mixing angle, as presented in the previous section, for the case where $S>1$.  
The limits from DM relic density can be straightforwardly determined  by simply rescaling it by a factor $1/S$. The limits from Lyman-$\alpha$ are more difficult to  address since this would require a different analysis (as for example~\cite{Boyarsky:2008xj,Boyarsky:2008mt}), which is computationally demanding and lies beyond the scope of this work. 
To a good approximation, one can assume a redshift factor of $S^{1/3}$~\cite{Asaka:2006ek} for the DM momentum distribution and translate it into a modified limit for the DM mass given by $m_{s,Ly\alpha}^S \geq m_{s,Ly\alpha} /S^{1/3}$, where $m_{s,Ly\alpha}$ is the lower limit on the DM mass for a given value $f_{\rm WDM}$ of the DM fraction, in the case where  $S=1$.   

The X-ray limits remain  unchanged with respect to the previous section since these rely on the DM lifetime. Nevertheless, the entropy injection leads to an indirect effect since a given pair $(m_s,\theta)$ now corresponds in general to a lower relic density.   

\begin{figure}[htb]
\begin{center}
\subfloat{\includegraphics[width=0.45\textwidth]{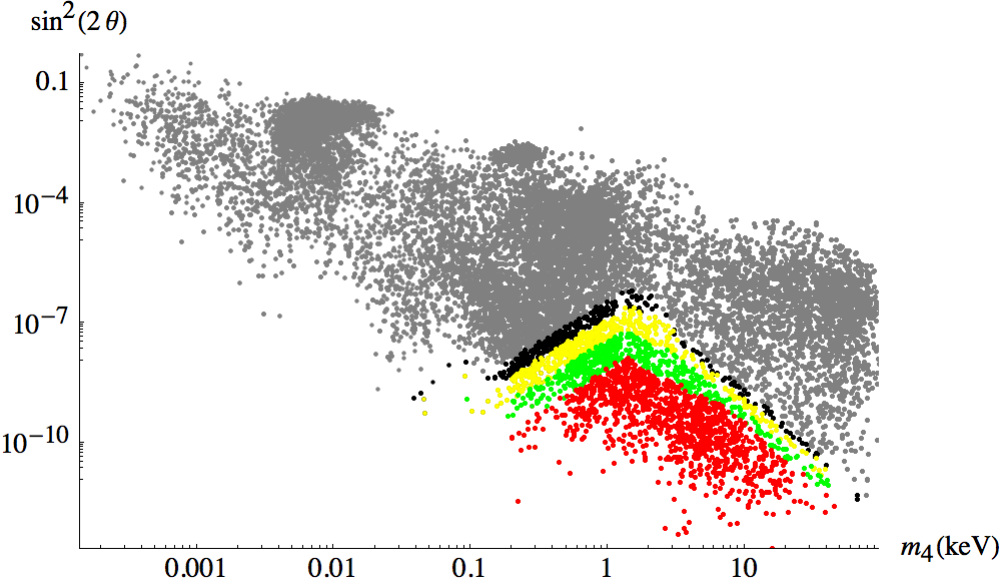}}\hspace*{0.4cm}
\subfloat{\includegraphics[width=0.45\textwidth]{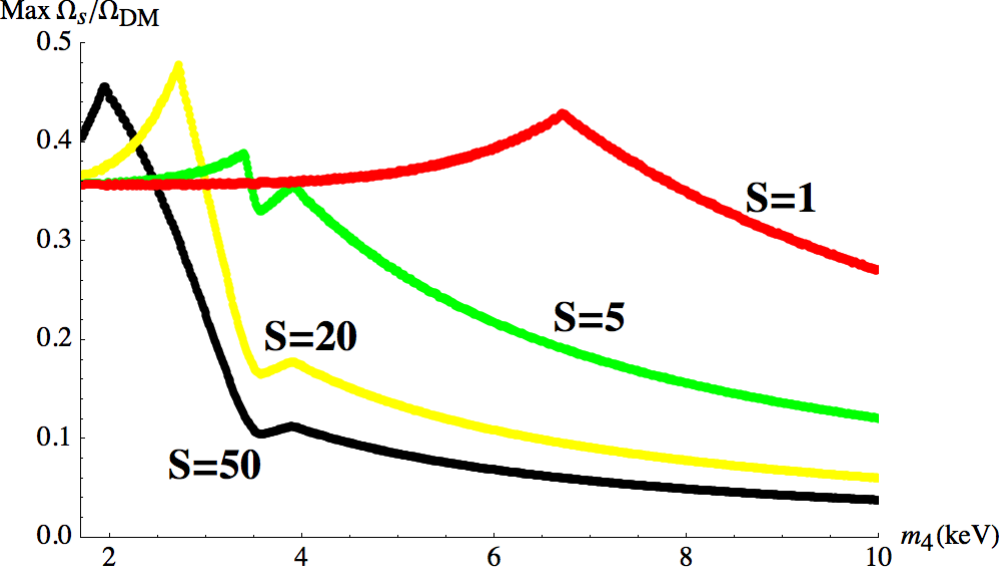}}
 \caption{\emph{Left panel}: Parameter space for the light sterile neutrino compatible with cosmological bounds in the hypothesis of an entropy injection, for values S=1 (red), S=5 (green), S=20 (yellow) and S=50 (black).  \emph{Right panel}: Maximal values of $f_{\rm WDM}$ allowed by the cosmological bounds as a function of the sterile neutrino mass in the hypothesis of an entropy injection $S$.}
\label{fig:entropy_abundance}
\end{center}
\end{figure}

The outcome of our analysis is summarised in Figure~\ref{fig:entropy_abundance}. 
In the left panel of Figure~\ref{fig:entropy_abundance},  we display the cosmologically favoured region for several values of  $S \leq 50$ as compared to the case $S=1$, represented by  red points. Values of $S$ larger than $\sim 20$ are not within reach in the framework of the present  model, but we have nonetheless extended our analysis up to these values in order to infer the maximal extension of the parameter space which could be achieved. The grey points are  excluded by DM constraints unless its relic density is negligible (see previous subsection). As one can see,  for $S>1$ we have a larger range of allowed values for the active-sterile mixing angle; interestingly, the augmentation of $S$ has a finite effect in enlarging the available parameter space.

On the right panel of Figure~\ref{fig:entropy_abundance}, we display  the maximal value of $f_{\rm WDM}$ for several values  of $S$ (we also display  for comparison the case corresponding to $S=1$). As one can see, there is only a marginal increase, namely from 0.43 to 0.48, of the maximal allowed value of $f_{\rm WDM}$. On the other hand, the maximal DM fraction is achieved for smaller  values of the  allowed DM mass, namely $\sim 2$~keV, as opposed to  values around $\sim7$~keV in the case where $S=1$. We finally notice that the maximal DM fraction is not an increasing function of $S$ but on the contrary, a maximum achieved at $S=20$ is followed by a sharp decrease. 
The reason of such  a behaviour is  mostly due to  the X-ray exclusion. Indeed, as already pointed out, any fixed value $f_{\rm WDM}$ imposes a condition on  $(m_s, \sin^2 2\theta)$ which is not sensitive to the mechanism accounting for the DM generation (more specifically, the value of $S$ in our case). 
Since the dark matter generation mechanism also depends on  $(m_s, \sin^2 2\theta)$, the interplay with the X-ray exclusion, as well as the effect of entropy injection,  favours larger mixing angles (thus maximising the  production of dark matter) and lower values of the mass (which in turn minimise the DM decay rate). 
Our analysis shows that  $f_{\rm WDM}=1$ could be achieved for  $S\lesssim 10$. This is not sufficient to relax the Lyman-$\alpha$ bound down to the value $m_s=2\text{ keV}$ because of the scaling of the latter limit as $S^{1/3}$. 

We finally point out that for very high values of $S$, sizeable values of $f_{\rm WDM}$ could be  achieved for very large mixing angles, already excluded by indirect dark matter  detection. This is at the origin of the saturation of the cosmologically favoured region observed in the left panel of 
Figure~\ref{fig:entropy_abundance}.

\subsubsection{Dark Matter Production from heavy neutrino decays and the 3.5 keV line}
As already mentioned, the pseudo-Dirac neutrinos can produce dark matter through their decays. These processes are mediated by  Yukawa interactions and the decay rate is proportional to $Y_{\rm eff} \sin\theta$, and thus suppressed with respect to the decay channels into only SM particles by the active-sterile mixing angle. A sizeable DM production can be nonetheless achieved through the so called freeze-in mechanism~\cite{Hall:2009bx,Chu:2011be,Chu:2013jja,Klasen:2013ypa,Blennow:2013jba}. It consists in the production of the DM while the heavy neutrinos are still in thermal equilibrium and, to be effective, requires that the rate of decay into DM is very suppressed, such that it results lower than the Hubble expansion rate. In our setup,  this condition can be expressed as: $Y_\text{eff} \sin\theta < 10^{-7}$. 

The dark matter relic density depends on  the decay rate of the pseudo-Dirac neutrinos into DM as follows:
\begin{equation}
\label{eq:general_fimp}
\Omega_{\rm DM}h^2 \simeq \frac{1.07 \times 10^{27}}{g_{*}^{3/2}}\sum_I g_I \frac{m_{\rm s} \Gamma\left(N_I \rightarrow \mbox{DM}+\mbox{ anything}\right)}{m_I^2}\ ,
\end{equation}
where the sum runs over the pseudo-Dirac states and $g_I$ represents the number of internal degrees of freedom of each state.
For pseudo-Dirac neutrinos lighter than the Higgs boson, DM production occurs through three-body processes whose rate is too suppressed to generate a sizeable amount of DM.  
On the other hand, the above analytical expression  is not strictly applicable for heavier pseudo-Dirac neutrinos since the mixing angle $\theta$ depends on the vacuum expectation value (VEV) of the Higgs boson and is thus zero above the EW phase transition temperature. To a good approximation, the correct DM relic density is determined by multiplying Eq.~(\ref{eq:general_fimp}) by  the function $\varepsilon(m_I)$ given by:
\begin{equation}
\varepsilon(m_I)=\frac{2}{3\pi}\int_{0}^{\infty} f(x_I) x_I^3 K_1(x_I) dx_I,\,\,\,\,\,x_I=\frac{m_I}{T}\, ,
\end{equation}  
with $f(x_I)$ describing the evolution of the Higgs VEV $v(T)$ with the temperature and which can be in turn approximated, according the results presented in~\cite{D'Onofrio:2014kta}, by:
\begin{equation}
\frac{v(T)}{v(T=0)}\ =\ \left \{
\begin{array}{cc}
1 & T < T_{\rm EW} \\
8 - \frac{m_I}{20 x_I} & T_{\rm EW} \leq T \leq 160\text{ GeV} \\
0 & T > 160 \text{ GeV}\ 
\end{array}
\right. \ ,
\end{equation} 
where $T_{\rm EW}\approx 140$ GeV is the temperature associated to the EW phase transition. 
As shown in Figure~\ref{fig:chi},  the function $\varepsilon(m_I)$ sharply decreases with the mass of the pseudo-Dirac neutrino since  most of the FIMP (Feebly Interacting Massive Particle) production occurs around the mass of the decaying particle. As a consequence, we can have sizeable production of DM only for masses of the decaying particles not too much above the scale of the electroweak  phase transition while DM production is  negligible for masses of the pseudo-Dirac neutrinos above the TeV scale.  
\begin{figure}[htb]
\begin{center}
\includegraphics[width=0.6\textwidth]{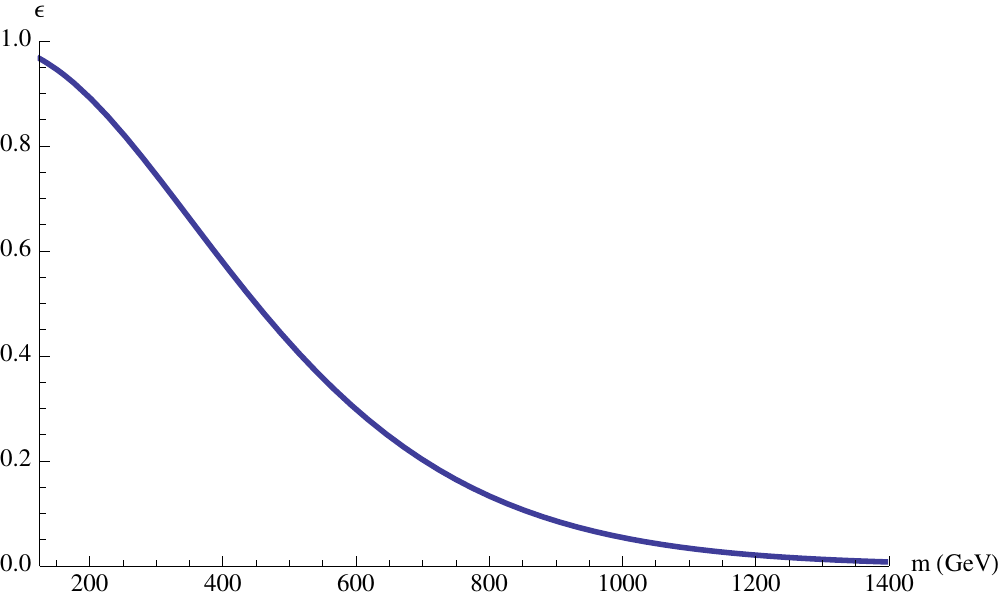}
\caption{Suppression factor in Eq.~(\ref{eq:heavy_fimp}) due to the electroweak symmetry restoration at high temperatures, as a function of the mass of the decaying particle.}
\label{fig:chi}
\end{center}
\end{figure} 
Using the expression of the rate associated to the process $N_I \rightarrow h \ +  \text{ DM}$:
\begin{equation}\label{eq:NhDM}
\Gamma\left(N_I \rightarrow h \ + \text{ DM}\right)=\frac{m_I}{16 \pi} Y^2_{\rm eff,I} \sin^2 \theta \left(1-\frac{m_h^2}{m_I^2}\right)\ , 
\end{equation}
the DM relic density is given by:
\begin{equation}
\label{eq:heavy_fimp}
\Omega_{\rm DM} h^2 \approx 2.16 \times 10^{-1} {\left(\frac{\sin\theta}{10^{-6}}\right)}^2 {\left(\frac{m_{\rm s}}{1 \text{ keV}}\right)} \sum_I g_I {\left(\frac{Y_{\rm eff,I}}{0.1}\right)}^2 {\left(\frac{m_I}{1 \mbox{TeV}}\right)}^{-1} \left(1-\frac{m_h^2}{m_I^2}\right)\varepsilon\left(m_I\right).
\end{equation}
It is then clear that the correct DM relic density can be achieved with a suitable choice of the parameters. It is worth noticing  that this production mechanism is  complementary to the DW one, which is always active provided that there is a nonzero active-sterile mixing. 

We have reported in Figure~\ref{fig:3.5kev} the (observed) value $\Omega_{\rm DM} h^2=0.12$ of the DM abundance, assuming for simplicity the same mass  $m_5$  and effective Yukawa couplings $Y_\text{eff}$ for the 4 heavy pseudo-Dirac states, 
for  different values of the DM mass and considering the maximal value of $\sin\theta$ allowed by cosmological constraints - including thus the corresponding contribution from DW production  mechanism-.
The displayed red points correspond to configurations of the ISS(2,3) model in agreement with all  laboratory constraints. Those configurations corresponding to pseudo-Dirac states far from thermal equilibrium, and  thus not accounting  for a freeze-in production mechanism, are delimited by a blue region. 
The shape of the lines can be understood as follows: for pseudo-Dirac masses comparable with the Higgs one,  the kinematical suppression in Eq.~(\ref{eq:heavy_fimp}) is significant, requiring sizeable Yukawas; for $m_I \gtrsim 200\mbox{ GeV}$ the dependence on $m_I$ is weaker, and the curve reaches a plateau, while for $m_I \gtrsim 500\mbox{ GeV}$ the suppression due to the function $\varepsilon\left(m_I\right)$ becomes significant requiring larger Yukawas, eventually violating the freeze-in condition $Y_\text{eff}\  \sin\theta<10^{-7}$ for $m_I \gtrsim 1.2\text{ TeV}$.

\begin{figure}[htb]
\begin{center}
\includegraphics[width=0.6\textwidth]{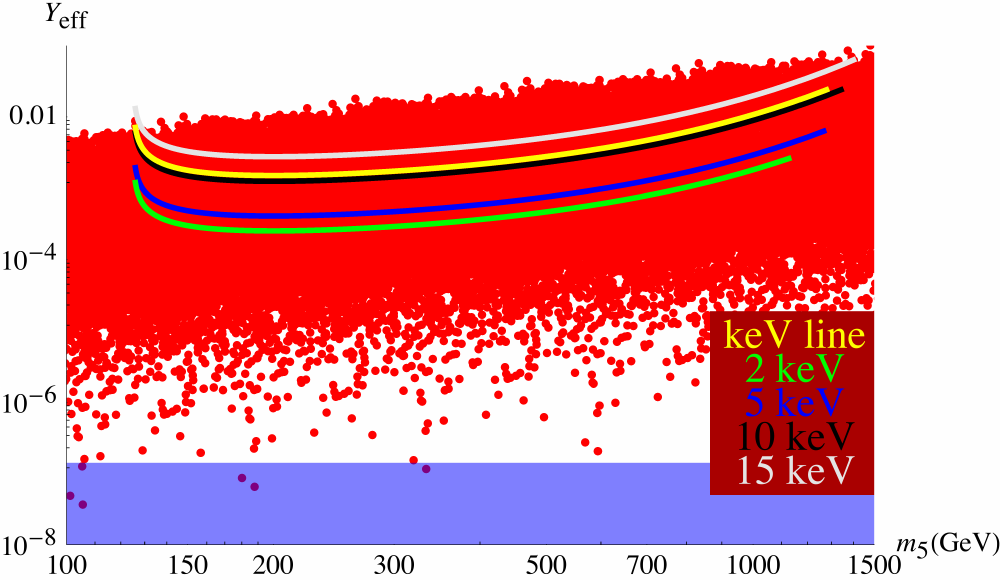}
\caption{
Viable configurations (continuous lines) for the heavy pseudo-Dirac masses $m_5$ and the corresponding effective Yukawa couplings $Y_\text{eff}$ accounting for the observed dark matter abundance of light sterile neutrinos via the freeze-in production mechanism, with masses and mixings for the sterile neutrinos compatible with cosmological bounds.
The red points denote different realisations of the ISS(2,3) model. In the blue region the production is not effective since the pseudo-Dirac states are out of thermal equilibrium. The lines end when the condition $Y_\text{eff}\  \sin\theta<10^{-7}$ is violated.
The yellow line accounts for the still unidentified monochromatic 3.5~keV line in galaxy cluster spectra~\cite{Bulbul:2014sua,Boyarsky:2014jta}.
}
\label{fig:3.5kev}
\end{center}
\end{figure}
The requirement of light  sub-eV  active neutrino masses together with $\mu \approx$ keV and $M_R \approx v$, implies values for the Yukawa couplings in the appropriate range to accounting for the observed DM abundance  ($m_\nu \approx \mu Y^2 v^2/M_R^2$, see Eq.~(\ref{inv.ss})). We emphasize here that this is not the case for a type-I Seesaw realisation  since in this case the relation $m_\nu\approx Y^2 v^2/M_R < 1$ eV implies $Y\lesssim10^{-6}$ if $M_R \approx v$, and thus the contribution from the freeze-in process is not sufficient to account for the total DM abundance.

Among the lines displayed in  Figure~\ref{fig:3.5kev}, we have highlighted in yellow  the one corresponding to the following  DM mass and mixing angle, 
\bee\label{eq:3.5kev}
m_s \simeq 7.1 \text{ keV},&\phantom{XXX}&
\sin^2 2 \theta \approx 7 \cdot 10^{-11},
\eee
which can account for the monochromatic 3.5~keV line  observed in the combined spectrum of several astrophysical objects~\cite{Bulbul:2014sua,Boyarsky:2014jta}. 

The results presented in Figure~\ref{fig:3.5kev} do not take into account the possible constraints  from structure formation.  As will be made clear in the next section, the  limits discussed  above should be sensitively relaxed since the DM produced through the freeze-in mechanism has a ``colder'' distribution with respect to the DW mechanism. A reformulation of the corresponding  limits  is beyond the scope of this work, especially in the case in which the DM production receives sizeable contributions both from DW and decay of the pseudo-Dirac neutrinos. 
We argue nonetheless  that the parameters accounting for the keV line can be compatible with bounds from structure formation since for this choice (of parameters),  the DM abundance is entirely determined by the decay of the heavy neutrinos (the DW contribution for that value of the mixing angle is less than 4\%) and the corresponding distribution function is  ``colder'' with respect to the one of a resonantly produced DM, which results compatible with the observational limits~\cite{Abazajian:2014gza}.

To summarise the results obtained and discussed in this section, one can  state that  in the  absence of effects from the heavy pseudo-Dirac neutrinos, the ISS(2,3) model can, in the most favourable case, account for  to approximatively the $\sim 43 \%$ of the total DM density for a mass of approximatively 7~keV. This percentage slightly increases up to $48\%$, for a DM mass of around 2~keV, once accounting for an entropy dilution factor of 5 - 20 which can be possible for masses of the pseudo-Dirac neutrinos of 3 - 10~GeV. The total DM component can be accounted for only in the region $m_h < m_I< 1.4\text{ TeV}$, when the DM can be produced through the freeze-in mechanism, although the compatibility with structure formation should be still  addressed. 
In order to also reproduce the correct relic density  for masses of the sterile neutrinos below  the Higgs boson  mass, it is necessary to extend the particle field content  of the model;  for this purpose, we will propose  in the following section a minimal  extension of the ISS(2,3) model.  

\section{Dark Matter Production in minimal extension of the ISS(2,3) model}
\label{Sec:NISS}
In order  to achieve the correct dark matter  relic density in  the pseudo-Dirac states low mass regime,  we consider a minimal extension of the ISS(2,3) model. This  consists in the introduction of a scalar field $\Sigma$, singlet under the  SM gauge group, interacting only with the sterile fermionic states and the Higgs boson. There are of course several other possibilities, see for instance~\cite{Bazzocchi:2012de,Fabbrichesi:2014qca,Molinaro:2014lfa}. In this minimal extension, the part of the  Lagrangian where the new singlet scalar field is involved reads:  
\begin{equation}
\label{eq:Sigma_lagrangian}
\mathcal{L}=\frac{1}{2}\partial_\mu \Sigma \partial^\mu \Sigma-\frac{h_{\alpha \alpha}}{2}\Sigma\overline{s^c}_{\alpha}s_{\alpha}+V(H,\Sigma).
\end{equation} 
We consider that the field $\Sigma$ has a non-vanishing VEV $\langle \Sigma \rangle$ that would be at the origin of  the Majorana mass coupling  $\mu$ which can  thus be expressed as:
\begin{equation}\label{mumu}
\mu \simeq 1 \text{ keV} \left(\frac{\langle \Sigma \rangle}{100 \text{ GeV}}\right) \left(\frac{h_{\alpha\alpha}}{10^{-8}}\right)\ .
\end{equation} 
For simplicity we will limit the scalar potential to the following terms (see e.g.~\cite{Petraki:2007gq} for a more general discussion):
\begin{equation}
V(H,\Sigma)=-\mu^2_{H} |H|^2-\frac{1}{2} \mu_{\rm \Sigma}^2 \Sigma^2 +2\lambda_{\rm H\Sigma}|H|^2 \Sigma^2\ .
\end{equation}
Following a pure phenomenological approach, we will consider values of the portal coupling $\lambda_{H\Sigma}$ from order of $10^{-2}$, corresponding to limits from effects on the Higgs width~\cite{Kamenik:2012hn}, down to very low values, i.e. ${\cal{O}}\left(10^{-8}\ \text{or}\  10^{-9}\right)$ (see for instance ~\cite{Merle:2013wta} and references therein for some examples of theoretically motivated models with extremely suppressed $\lambda_{H\Sigma}$).  

We will assume for simplicity that the scalar singlet field is  heavier that the Higgs boson, $m_\Sigma > 200\text{ GeV}$, and assume  $m_\Sigma \le \langle \Sigma \rangle$ in order to avoid non perturbative values of $\lambda_{\rm H\Sigma}$.

The DM density is generated by the decay of $\Sigma$  and is thus tied to the abundance of the latter, which in turn depends on the efficiency of the process $\Sigma\Sigma \leftrightarrow hh$ triggered by the portal like coupling $\lambda_{\rm H\Sigma}$ (by this we implicitly assume that, in case of very suppressed values of $\lambda_{\rm H\Sigma}$, the abundance of $\Sigma$ in the early stages of the evolution of the Universe is negligible). 
A proper description of the DM density requires the resolution of a system of coupled Boltzmann equations for the DM number density, as well as for the abundance of the $\Sigma$ field and possibly for the heavy pseudo-Dirac neutrinos, which also interact with $\Sigma$ -  also  including effects of entropy release. Further details of this computation can be found in the appendix. 

In the following we will present  analytical expressions which describe, to a good approximation, the  DM production mechanism. For simplicity we will assume that the pseudo-Dirac neutrinos are in thermal equilibrium (the case of non-equilibrium configurations substantially coincides with the studies already presented in~\cite{Petraki:2007gq,Merle:2013wta}) and with lifetimes such that the effects of entropy injection are not relevant.  As will be made clear, pseudo-Dirac neutrinos have a non trivial impact on DM production. We will thus for definiteness discuss two specific mass regimes, namely the case in which 
all the pseudo-Dirac neutrinos are lighter than $\Sigma$ and the case in which they  have instead similar or greater masses.

At high enough values of $\lambda_{H\Sigma}$,  the pair annihilation processes $\Sigma \Sigma \leftrightarrow hh$ maintain the field $\Sigma$ into thermal equilibrium.\footnote{Pair annihilation processes into fermion pairs are as well possible. For $m_\Sigma > m_h$, as assumed in this work, the relative rate is subdominant, being suppressed at least by a factor $v^2/m_\Sigma^2$.} Indeed,  
by comparing the $2 \rightarrow 2$ rate, associated to the thermally averaged cross-section $\langle \sigma v \rangle \sim 10^{-2}\times  \frac{\lambda_{\rm H\Sigma}^2}{m_\Sigma^2}$, with the Hubble expansion rate, the field $\Sigma$ can be considered to be  in thermal equilibrium in the Early Universe for $\lambda_{\rm H\Sigma} \ge \overline{\lambda}_{\rm H\Sigma}$ where:
\begin{equation}\label{lambdaFSIGMABAR}
 \overline{\lambda}_{\rm H\Sigma} \equiv 10^{-6} {\left(\frac{m_\Sigma}{100 \mbox{ GeV}}\right)}^{1/2}\ .
\end{equation}
On the contrary, its decay rate into DM is always suppressed compared to the Hubble rate due to the low value of the couplings $h_{\alpha\alpha}$ (see Eq.~(\ref{mumu})). The DM can thus be produced through the freeze-in mechanism from the decays of $\Sigma$ and its corresponding  abundance can be expressed as:
\begin{equation}
Y_{\rm DM}^{\rm FI}=\frac{135}{128 \pi^4} \sum_{I=5,8}\frac{|h_{\rm eff,I4}|^2}{g_{*}(T_{\rm prod})m_\Sigma}\left(1-\frac{m_I^2}{m_\Sigma^2}\right){\left(\frac{45 M_{\rm Pl}^2}{4 \pi^3 g_{*}}\right)}^{1/2}\ ,
\end{equation} 
where:
\begin{equation}
h_{\rm eff,I4}=\sum_{\alpha=1,3} U_{\rm I \alpha}^{T}\,h_{\alpha \alpha}\,U_{\alpha 4},
\end{equation}  
is an effective coupling taking into account all the decays $\Sigma \rightarrow N_I\, + \text{DM},\,\,\,I=4,8$ which are kinematically open.\footnote{Notice that since the scalar singlet field $\Sigma$ couples with all neutrinos, it can also decay into pseudo-Dirac states. However, the latter are in thermal equilibrium and thus no corresponding freeze-in production mechanism is possible.} The contribution to the relic DM density reads:
\begin{equation}
\Omega_{\rm DM}^{\rm FI} \approx 0.2 \sum_I {\left(\frac{|h_{\rm eff,I1}|}{10^{-8}}\right)}^2 \left(1-\frac{m_I^2}{m_\Sigma^2}\right) {\left(\frac{m_\Sigma}{200 \text{ GeV}}\right)}^{-1} \left(\frac{m_{s}}{1\text{ keV}}\right)\ . 
\end{equation}
On general grounds, out-of-equilibrium - i.e. after chemical decoupling - decays of $\Sigma$ may also contribute to DM production and 
the corresponding contribution 
to the DM density can be schematically expressed as:
\begin{equation}
\label{eq:ooe}
Y^{\rm SW}=B \ Y_\Sigma(T_{\rm f.o.})\,, \text{ where } B\equiv \sum_I b_I \text{Br}\left(\Sigma \rightarrow N_I N_1\right). 
\end{equation} 
In the above equation, $T_{\rm f.o.}$ is the standard freeze-out temperature of $\Sigma$ and  $b_I$ represents the number of DM particles produced per decay  for a given  decay channel. The branching ratio of the decay of $\Sigma$ into DM is given by:
\begin{equation}
\text{Br}\left(\Sigma \rightarrow N_I N_1\right)={{\sum_{I=1,5}^{}|h_{\rm eff,I1}|^2\left(1-\frac{m_I^2}{m_\Sigma^2}\right)}\over{\sum_{I,J=1,5}|h_{{\rm eff},IJ}|^2\left(1-\frac{{\left(m_I+m_J\right)}^2}{m_\Sigma^2}\right)+y_f^2 \sin^2 \alpha \left(1-\frac{4 m_f^2}{m_\Sigma^2}\right)}}\ ,
\end{equation}
where $\sin\alpha \propto \lambda_{H \Sigma}$ represents the mixing between $\Sigma$ and the Higgs boson.\footnote{This mixing exists only when the VEV of the SM Higgs doublet is different from zero. Analogously to what we did for  the active-sterile mixing angle,  we have adopted in our computation a temperature dependent scaling function.} Due to the very low couplings  $h_{\alpha \alpha}$, the branching ratio of the  decay of $ \Sigma$ into DM is very suppressed  with respect to the branching ratio of the decay  into two fermions induced by the mixing with the Higgs boson, even for low values of the mixing itself. Furthermore,  the total lifetime of the scalar field is comparable with the freeze-out  timescale.
Consequently,  the out-of-equilibrium production is sizeable for $\lambda_{H\Sigma} \sim \overline{\lambda}_{H\Sigma}$ when the scalar field features an early decoupling, i.e. $x_{\rm f.o.}=m_\Sigma/T_{\rm f.o.}=1\ -\ 3$~\cite{Petraki:2007gq}.

Finally, the DM relic density can be estimated as:
\begin{align}
& \Omega^{\rm SW}_\text{DM} \approx 0.11 \left(\frac{m_\text{s}}{2 \text{ keV}}\right) \left(\frac{m_\Sigma}{1000 \text{ GeV}}\right) \left(\frac{B}{0.01}\right) {\left(\frac{\lambda_{\rm H\Sigma}}{10^{-6}}\right)}^{-2} \ .
\end{align}

In the case in which $\lambda_{\rm H\Sigma} \ll \overline{\lambda}_{\rm H\Sigma}$ (see Eq.~(\ref{lambdaFSIGMABAR})), the field $\Sigma$ is too feebly interacting to be in thermal equilibrium in the Early Universe. Assuming, for simplicity, a negligible abundance at early times, it can be nonetheless produced in sizeable quantities by freeze-in and then decay through out-of-equilibrium processes~\cite{Merle:2013wta}. The field $\Sigma$ is produced by the $2 \rightarrow 2$ processes mediated by the portal interactions as well as by the $2 \rightarrow 1$ processes, $N_I N_I \rightarrow \Sigma,\,\,I=5,8$ if the pseudo-Dirac neutrinos are lighter than $\Sigma$. \\The abundance of $\Sigma$ can be expressed as:
\begin{align}
\label{eq:SFI}
& Y_\Sigma^{\rm SFI}\approx \frac{M_{\rm Pl}}{1.66 g_{*}(T_{\rm prod})}\frac{1}{m_\Sigma} \left(\sum_{I,J} \frac{135}{128\pi^4} |h_{{\rm eff},IJ}|^2+\frac{45}{1024 \pi^6} \lambda_{\rm H\Sigma}^2\right) \ ,\\ \nonumber \text{  with}\\ \nonumber
& h_{{\rm eff,} IJ}=U^{T}_{I\alpha}h_{\alpha \alpha}U_{\alpha J}\ . 
\end{align} 
The two terms inside the parenthesis refer to the contributions from the  $2 \rightarrow 1$ (where the sum over $I,J$ runs over the pseudo-Dirac neutrinos in thermal equilibrium) and the $2 \rightarrow 2$ processes, respectively. The DM abundance is given, analogously to Eq.~(\ref{eq:ooe}), by $B \ Y_{\Sigma}^\text{SFI}$.

 In the case where the pseudo-Dirac neutrinos are heavier than $\Sigma$,  the DM generation process in the regime $\lambda_{H\Sigma} \geq \overline{\lambda}_{H\Sigma}$ proceeds along the same lines as described before. There is however the additional contribution from the decays $N_I \rightarrow h\, DM$, given by Eq.~(\ref{eq:heavy_fimp}) as well as a further freeze-in contribution from the decays $N_I \rightarrow \Sigma DM$ given by:
\begin{equation}
\label{eq:Sigma_FIMP}
\Omega^{\rm FI}_{\Sigma} \approx 2.16 \times 10^{-3} \sum_{I=5,8} g_I {\left(\frac{m_{\rm s}}{1 \text{ keV}}\right)} {\left(\frac{|h_{\rm eff, I\,4}|}{10^{-8}}\right)}^{2} {\left(\frac{m_{\rm I}}{1 \mbox{TeV}}\right)}^{-1}     \ .
\end{equation}

In the regime where  $\lambda_{H\Sigma} \leq \overline{\lambda}_{H\Sigma}$,  the $2 \rightarrow 1$ production channel for the field $\Sigma$ is replaced by the production from the decays of the pseudo-Dirac neutrinos through the processes, if kinematically open, $N_I \rightarrow N_J \Sigma,\,\,\,I=5,8,\,J=4,I-1$. In this scenario the abundance of $\Sigma$ reads:
\begin{equation}
\label{eq:SFI_bis}
Y_\Sigma^{\rm SFI}\approx \frac{M_{\rm Pl}}{1.66 g_{*}(T_{\rm prod})}\left[\frac{1}{m_\Sigma}\frac{45}{1024 \pi^6} \lambda_{\rm H\Sigma}^2+\sum_{I,J} \frac{135}{64\pi^4} \frac{|h_{{\rm eff},IJ}|^2}{m_I}\left(1-\frac{{\left(m_\Sigma+m_J\right)}^2}{m_I^2}\right)\right]\ .
\end{equation}

\begin{figure}[htb]
\begin{center}
\subfloat{\includegraphics[width=0.45\textwidth]{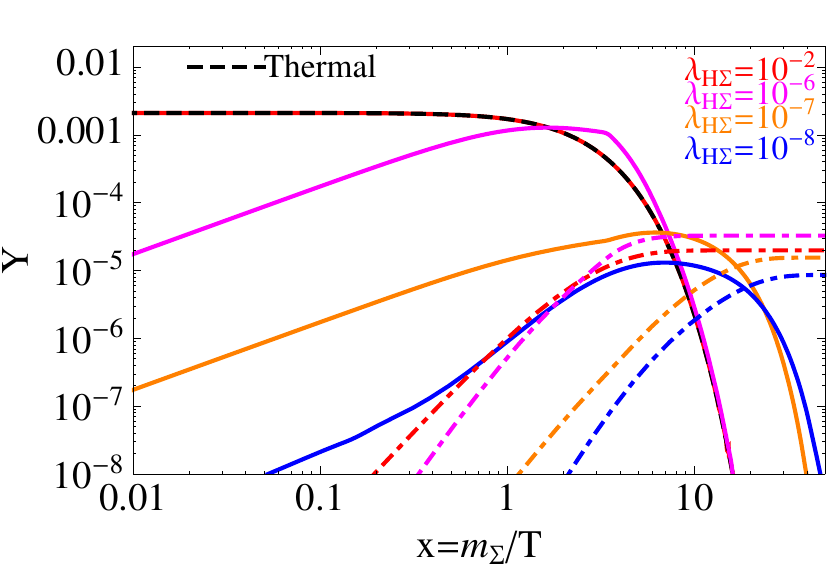}}\hspace*{0.5cm}
\subfloat{\includegraphics[width=0.45\textwidth]{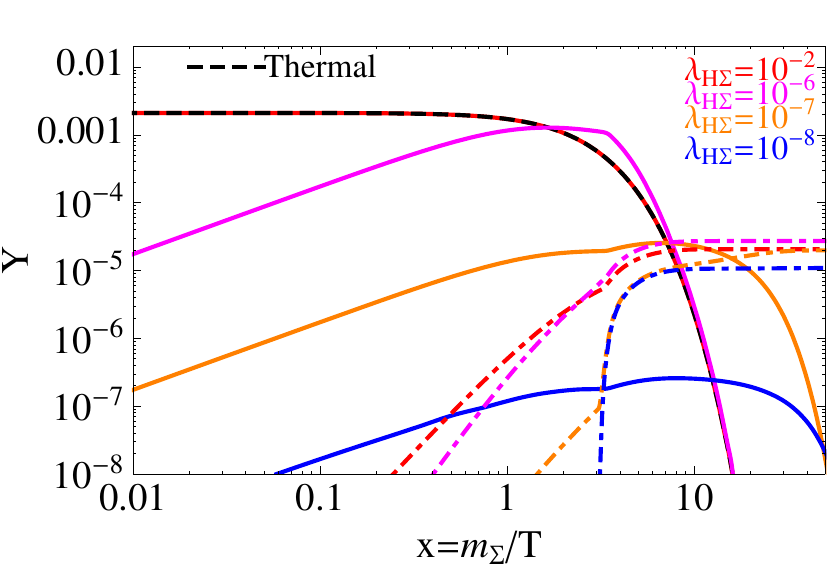}}
\caption{
Evolution of the abundance  of the $\Sigma$ field (solid lines) and of the DM (dot-dashed lines) for the four values of $\lambda_{H\Sigma}$ reported on the plot. The scalar field $\Sigma$ and the DM masses have been set to 500 GeV and 5 keV, respectively. The VEV $\langle \Sigma \rangle$ has been fixed to 1 TeV. The masses of two the pseudo-Dirac pairs are, respectively 10 and 20 GeV (left panel) and 500 and 1000 GeV (right panel). In both cases, DM production through $N_I \rightarrow h + \text{DM}$ decays and the effects of entropy production are negligible.
 }
\label{fig:sterile_production_boltzmann}
\end{center}
\end{figure} 
 
The validity of the assumptions leading to the analytical approximations above has been confirmed by (and complemented) by numerically solving the Boltzmann equations for the system $\Sigma-$DM abundances.
We display in Figure~\ref{fig:sterile_production_boltzmann} the evolution of the abundances  of $\Sigma$ and of the DM, for a set of values of the the coupling $\lambda_{\rm H\Sigma}$ and for fixed values of $m_{\rm s}$, $m_\Sigma$ and $\langle \Sigma \rangle$ to $5 \text{ keV}$, $500 \mbox{ GeV}$  and $1 \mbox{ TeV}$, respectively.
 On the left panel the masses of the pseudo-Dirac pairs have been fixed to, respectively, 10 and 20 GeV, while on the right panel the chosen values are 500~GeV and 1~TeV. For this last case we have fixed the coupling $Y_\text{eff}$ of the pseudo-Dirac neutrinos with the Higgs boson and the DM-active neutrino mixing angle $\theta$ to, respectively, 0.01 and $10^{-6}$ in such a way that  the freeze-in production from the decays $N_I \rightarrow h +{\text{ DM}}$ gives a subdominant contribution, not exceeding 30$\%$. 

At higher values of $\lambda_{H\Sigma}$,  the abundance of $\Sigma$ traces its equilibrium value and the DM production occurs prevalently through the freeze-in mechanism. At lower values of $\lambda_{H\Sigma}$ the out-of-equilibrium production becomes important thus increasing the total DM relic density. 

For $\lambda_{H\Sigma} < \overline{\lambda}_{H\Sigma}$, the abundance of $\Sigma$ does not follow the equilibrium value but is an increasing function of time (as consequence of the freeze-in production from the $2 \rightarrow 2$ scatterings as well as from the $2 \rightarrow 1$ processes, for pseudo-Dirac neutrinos lighter than $\Sigma$, or from decays of the pseudo-Dirac neutrinos themselves in the opposite case) until its decay, which occurs at  later timescales compared to the case of high values of $\lambda_{H\Sigma}$.

\begin{figure}[htb]
\begin{center}
\subfloat{\includegraphics[width=0.4\textwidth]{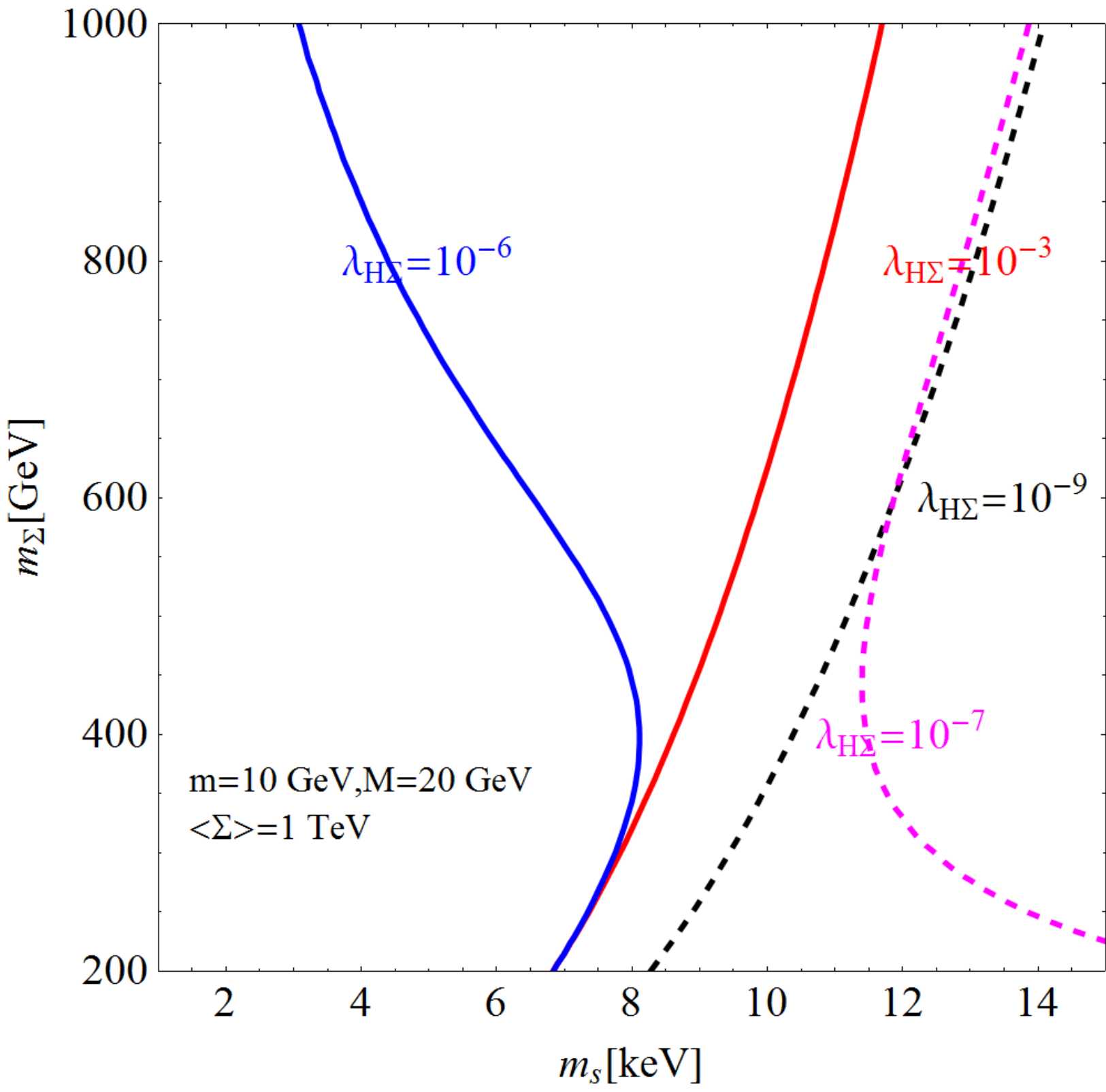}}
\hspace{5 mm}
\subfloat{\includegraphics[width=0.4\textwidth]{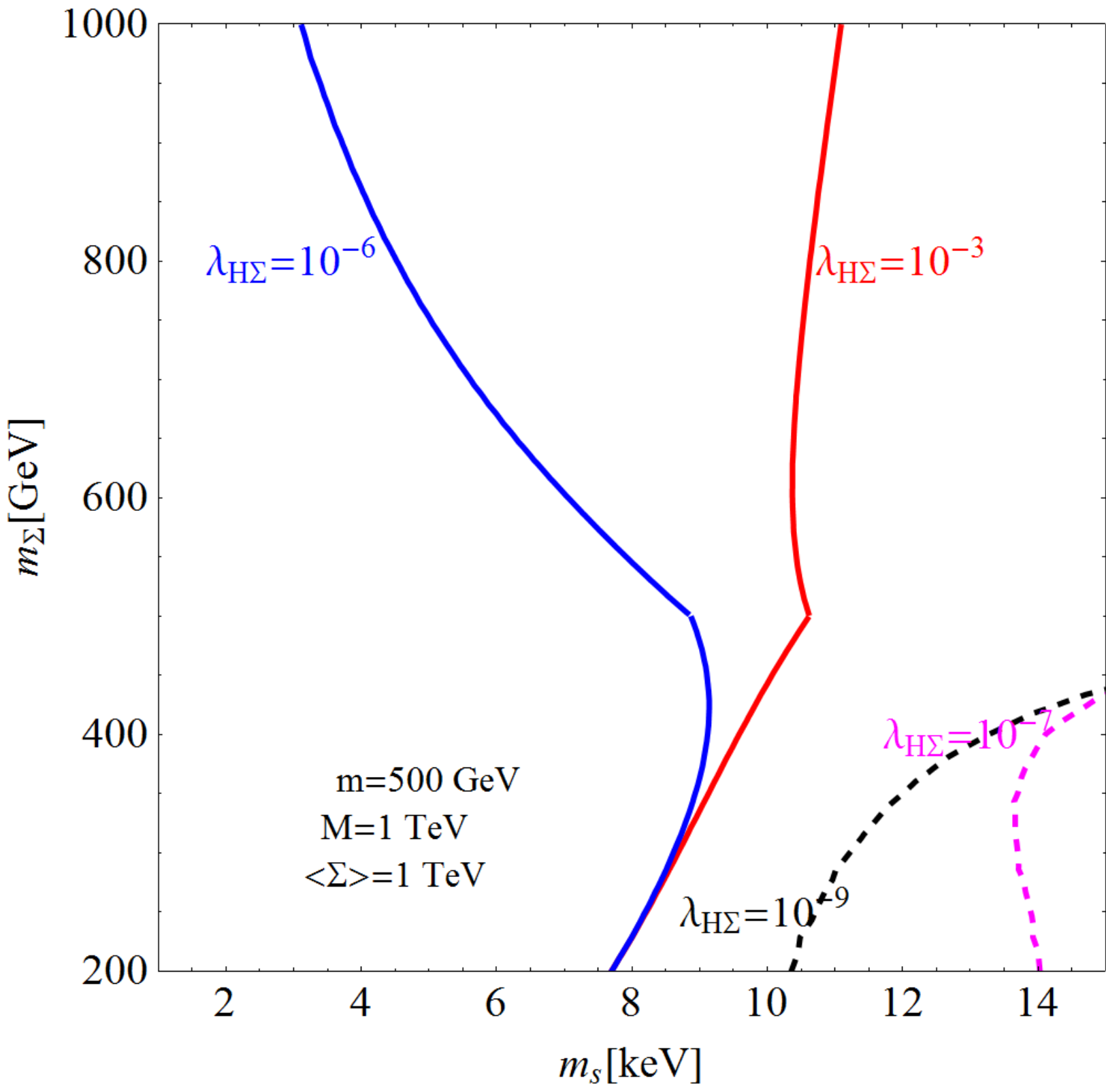}}
\caption{Contours of the cosmological value of the DM relic density in the plane $(m_s,m_\Sigma)$, for the values of $\lambda_{H\Sigma}$ reported in the plot. The other relevant parameters have been set as in Figure~\ref{fig:sterile_production_boltzmann}.
}
\label{fig:sterile_production_summary}
\end{center}
\end{figure} 

We report in Figure~\ref{fig:sterile_production_summary} the contours of the cosmological value of the DM relic density in the  $(m_s,m_{\rm \Sigma})$ plane for several  values of  the coupling $\lambda_{\rm H\Sigma}$ and for  the masses of the pseudo-Dirac neutrinos  considered in 
Figure~\ref{fig:sterile_production_boltzmann}. 
As already pointed out, for $\lambda_{H\Sigma}=10^{-3}$ the DM relic density is determined by the freeze-in mechanism and thus increases with the DM mass while decreasing with respect to $m_{\Sigma}$. For $\lambda_{H\Sigma}=10^{-6}$ the out-of-equilibrium production is instead the dominant contribution implying that $\Omega_s h^2 \propto m_s m_\Sigma$.
For $\lambda_{H\Sigma}=10^{-7}$ and $\lambda_{H\Sigma}=10^{-9}$, in the case of  light pseudo-Dirac neutrinos, the relic density is again proportional to the ratio $m_s/m_\Sigma$,  as expected from Eq.~(\ref{eq:SFI}).
 In the case of heavy pseudo-Dirac neutrinos, the regime $\lambda_{H\Sigma} \ll \overline{\lambda}_{H\Sigma}$ is substantially dominated by the freeze-in production of $\Sigma$ from the decays of   the pseudo-Dirac neutrinos and its subsequent out-of-equilibrium decays. The dependence on $m_\Sigma$ shown in Figure~\ref{fig:sterile_production_summary}  is due  to the kinematical factor in Eq.~(\ref{eq:SFI_bis}). 
We notice that the correct DM relic density, for the  chosen set of parameters, is achieved for DM masses between 1 - 15~keV. These results can be straightforwardly generalised in the case of entropy production from the pseudo-Dirac neutrinos. Indeed,  as can be seen from Figure~\ref{fig:sterile_production_boltzmann},  the DM production typically occurs at earlier stages compared to  the ones at which sizeable entropy production is expected (see previous section). As a consequence the instantaneous reheating approximation can be considered as valid and we can just rescale the DM relic density by a factor $S$. In this case,  the correct DM relic density is achieved for higher values of the DM masses. 

We emphasise, as already done in the previous section, that a complementary contribution to the DM relic density from DW production mechanism is in general present. The DM production related to the decays of $\Sigma$ allows to achieve the correct relic density without conflicting with the X-rays limits since it does not rely on the mixing with the active neutrinos;  this is not the case for the bounds from  structure formation.
However, applying  the limits on DM from structure formation is a very difficult task in our scenario since different DM production channels coexist, originating different dark matter distribution functions. A proper treatment would require to reformulate the bounds case by case by running suitable simulations, which lies beyond the scope of the present work. 
We will  nonetheless provide an approximate insight of how the latter bounds are altered, with respect to the conventional DW production mechanism, by taking  some representative examples.

In the following discussion, we consider  the case in which the DM is produced by the decays of the field $\Sigma$, either through freeze-in or through out-of-equilibrium decays.  An approximate reformulation of the limits from structure formation can be obtained by comparing  the average momentum of  DM at the keV scale with the one corresponding to DW production and by rescaling the lower limit on the DM mass with  the shift between these two quantities.  
The DM distribution function in the various cases of production from decay has been determined in e.g.~\cite{Petraki:2007gq} and~\cite{Boyanovsky:2008nc,Kamada:2013sh}). 
The dark matter produced through freeze-in is typically generated at temperatures of the order of the mass of $\Sigma$. Its average momentum depends only on the temperature and can be simply expressed, at temperatures of the order of the keV, as~\cite{Shaposhnikov:2006xi}:
\begin{equation}
\left.\left(\frac{\langle p \rangle}{T}\right)\right |_{T \sim \text{keV}}  \simeq 0.76  \ S^{-1/3}\ ,
\end{equation}
sensitively lower than the corresponding result  (of $\sim 2.83$) in the case of DW production. A similar result holds as well in the case of DM produced  through freeze-in from  the decays of the pseudo-Dirac neutrinos.\footnote{The DM distribution function can be obtained by solving the same Boltzmann equation as in~\cite{Petraki:2007gq} and  by replacing the Bose-Einstein distribution for the decaying state with a Fermi-Dirac function. 
The difference in the final result is of order one.} 
In the case in which the DM is prevalently produced out-of-equilibrium, the timescale of production varies with the lifetime of $\Sigma$ and the distribution function tends to be warmer as the latter increases.
\begin{figure}[htb]
\begin{center}
\includegraphics[width=0.6\textwidth]{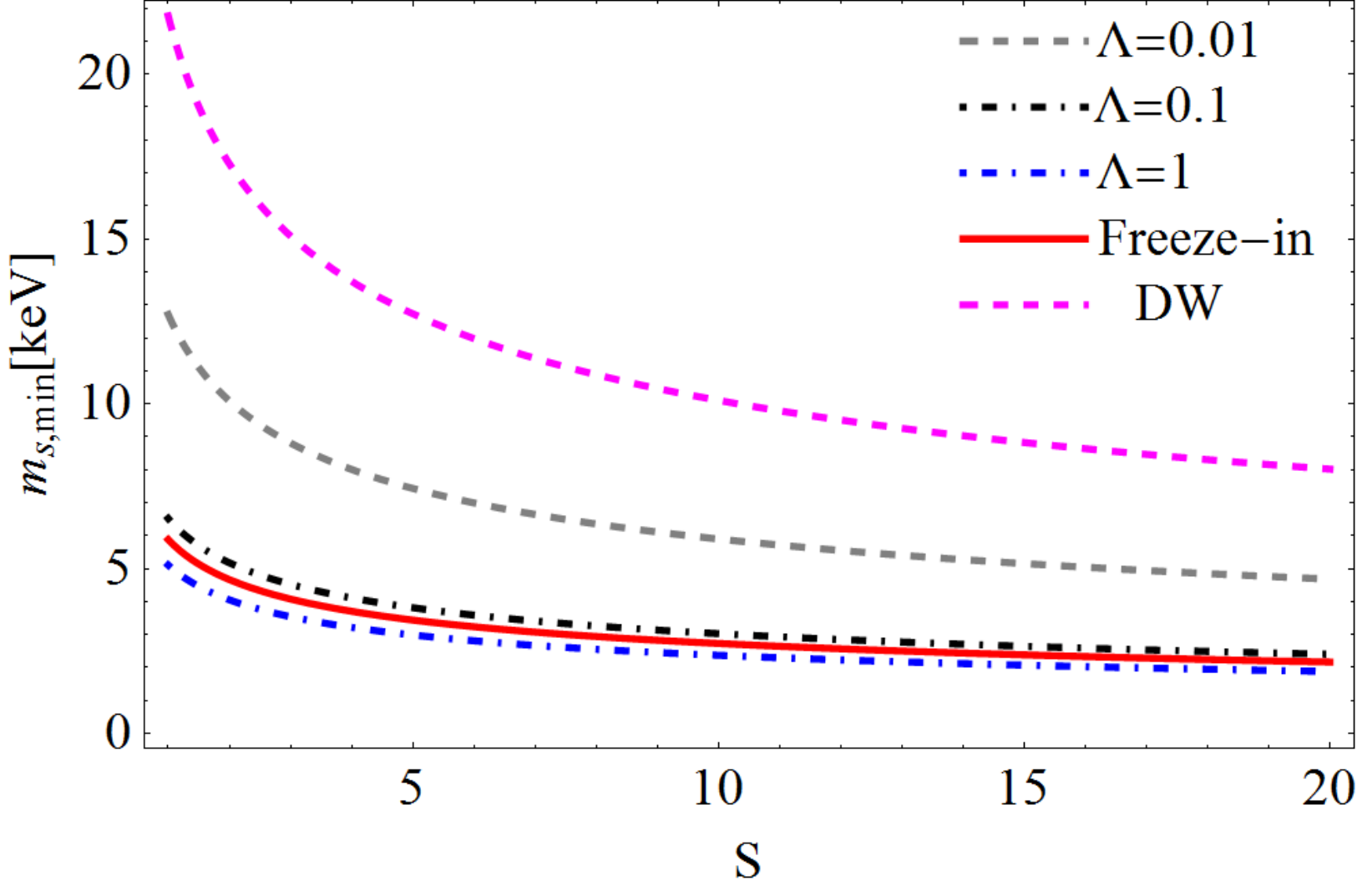}
\caption{Lower limit on the DM mass from Lyman-$\alpha$ 
 as a function of the entropy dilution factor $S$. The limit refers to the cases of dominant freeze-in production from the decay of the scalar field $\Sigma$ and dominant production from out-of-equilibrium decays  for the parameter $\Lambda= 0.01,\ 0.1,\   1$ (defined in Eq.~(\ref{def:Lambda})). We also report for comparison the   corresponding limit in the case of dominant DW production mechanism.
} 
\label{fig:S_Lyalpha}
\end{center}
\end{figure}

We report in Figure~\ref{fig:S_Lyalpha} the lower limit\footnote{Since we are here assuming that the lightest sterile neutrino is the only DM component, we adopt the most updated limit. This actually refers to a thermal  relic density. It can be reformulated in term of a limit on non-resonant DW production by using the formula given in~\cite{Viel:2005qj}.} from Lyman-$\alpha$ on the DM mass, obtained by applying our approximate rescaling to the limit presented in~\cite{Viel:2013fqw}, for some scenarios of DM production mechanism, namely freeze-in and out-of-equilibrium production for different decay rates of $\Sigma$ parametrised through the dimensionless quantity:
\begin{align}\label{def:Lambda}
& \Lambda=\frac{5 \overline{h}^2}{8 \pi m_\Sigma}{\left(\frac{45 M_{\rm Pl}^2}{4 \pi^3 g_{*}}\right)}^{1/2}\ , \nonumber\\
\text{where,}\nonumber\\
& \overline{h}^2=\sum_{I,J} |h_{{\rm eff},IJ}|^2 \left(1-\frac{{\left(m_I+m_J\right)}^2}{m_\Sigma^2}\right)+y_f^2 \sin\alpha^2 \left(1-\frac{4 m_f^2}{m_\Sigma^2}\right)\ . 
\end{align} 
As can be noticed,  in the most favourable cases, namely freeze-in or out-of-equilibrium production with $\Lambda \geq 1$ (corresponding to $m_\Sigma \lesssim 600\text{ GeV})$, the limit from Lyman-$\alpha$ is relaxed to approximately 5~keV, in absence of entropy injection, and to further low values for the case $S>1$. We remark again that these results assume that all the DM is produced by the decay of $\Sigma$ (into heavy neutrinos).

Interestingly, in the case in which  a sizeable contribution from DW production mechanism is also  allowed,  the DM distribution would feature a Warm and a ``colder'' component and thus the ISS(2,3) model  could potentially realise a mixed Cold + Warm DM scenario. This would constitute an  intriguing solution to some tensions with observation from structure formation (see e.g.~\cite{Weinberg:2013aya} and references therein). 
This possibility should be thoroughly investigated by means of   numerical simulations since the analytical estimates presented above are not valid for multi-component distributions. This will be thus left for a dedicated study.

\chapter{Lepton number violation as a key to low-scale leptogenesis}\label{introduction}

Within the Seesaw mechanism, it is in general possible to account for the BAU via thermal leptogenesis, see for instance~\cite{JosseMichaux:2008ix,Davidson:2008bu}. This kind of mechanism normally requires very high Seesaw  scales, above $10^8\,\mbox{GeV}$. An efficient thermal leptogenesis can be nonetheless achieved at a Seesaw  scale $\sim \,\mbox{TeV}$ in the presence of a resonant amplification~\cite{Pilaftsis:2003gt}. At even lower Seesaw  scales thermal leptogenesis is no longer at work and one must consider different mechanisms for the generation of any lepton asymmetry.
A viable possibility is provided by the mechanism first proposed in~\cite{Akhmedov:1998qx}, in which a lepton asymmetry is produced by the CP-violating oscillations of a pair of heavy neutrinos. This kind of mechanism has been successfully implemented in the so-called $\nu$MSM. Aiming at simultaneously
addressing the problems of neutrino mass generation, BAU and providing a viable DM
candidate, the $\nu$MSM is a truly minimal extension of the SM through the
inclusion of three RH neutrinos ($N_R^{1,2,3}$)~\cite{Asaka:2005an,Asaka:2005pn,Shaposhnikov:2008pf,Canetti:2012kh}. The lightest of these new states, with mass at the keV scale, is substantially sterile, i.e.\ with highly suppressed couplings both to active neutrinos and to the two other new states, and represents the Dark Matter candidate. The latter two heavy neutrinos are instead responsible for the light neutrino mass generation, as well as for lepton asymmetries both at early times, giving rise to the BAU, and at later times, accommodating the production of the correct amount of DM~\cite{Laine:2008pg}.  
For this to work, the spectrum in the additional sterile states requires a certain pattern, the two heaviest states $N_R^{2}$ and $N_R^{3}$ being almost degenerate. 
Notice however that this requirement can be relaxed by considering, relaxing the DM hypothesis, all the three right-handed neutrinos involved in the leptogenesis process, as it was shown in~\cite{Drewes:2012ma}.

The degeneracy between the heavy neutrinos, which is phenomenologically imposed in the $\nu$MSM, can be however naturally explained in frameworks in which the  smallness of (active) neutrino masses is directly linked to  a small violation of the total lepton number, cf. Section~\ref{sec:LNV_scale}. 
This can be achieved when, for instance, the Inverse Seesaw~\cite{Wyler:1982dd, Mohapatra:1986bd} mechanism  is embedded into the SM.
The mechanism consists in the addition of at least  two sets of additional sterile fermions 
with opposite lepton numbers, 
allowing for a small $\Delta L=2$ lepton number violating (LNV) mass parameter $\mu$ corresponding to a Majorana mass in the sterile sector. The masses of the mostly active neutrinos (light neutrinos) are proportional to $\mu$, while the remaining mostly sterile states are coupled into  pseudo-Dirac pairs with mass differences of the order of the LNV parameter $\mu$. 
In the limit where $\mu\rightarrow 0$,
lepton number conservation  is restored. However, in order to be phenomenologically viable, this mechanism requires at least four extra fermions (cf. Section~\ref{Sec:intro}). 
Another mechanism based on a small violation of the total lepton number is  the so called Linear Seesaw~\cite{Barr:2003nn,Malinsky:2005bi}.  It is similar to the ISS,  in the sense that it requires  the introduction of two types of fermion singlets (RH neutrinos and steriles) with opposite assignment for the total lepton number, and the smallness of the neutrino masses is linked to the small violation of the total lepton number conservation. The difference with respect to the ISS is that the lepton number violation (by two units) arises from the new LNV  Yukawa couplings of steriles with the left-handed neutrinos.

The present chapter focuses on the possibility of simultaneously having a very low  scale Seesaw mechanism - typically at $1 - 10$ GeV - at work for generating neutrino masses and mixings as well as an efficient leptogenesis at the electroweak scale, by considering a ``natural'' and minimal framework (with only two additional neutrinos) giving rise to the needed degeneracy in the spectrum of the sterile states.
For this, we consider the ISS and the LSS frameworks and revisit the mechanism of leptogenesis through oscillations~\cite{Akhmedov:1998qx,Asaka:2005an}.
More precisely, we have considered the minimal extension of the SM  by two sterile states with  couplings leading to an Inverse Seesaw  mass structure. Being insufficient to accommodate neutrino data, instead of adding further sterile states, we have completed the  scenario  with a Linear Seesaw mass term (see for instance ~\cite{Gavela:2009cd}), violating total lepton number also by two units. 
We have conducted a thorough analytical and numerical analysis investigating both neutrino mass hierarchies, normal (NH) and inverted (IH), for the neutrino mass spectrum.  To this end we have implemented and solved a system of Boltzmann equations 
 and have also derived an analytical expression for the baryon asymmetry in the weak washout regime, supporting our understanding of the behaviour of the numerical solutions. 
Our studies reveal that this scenario can incorporate a successful leptogenesis through oscillations between the two mostly sterile states while accommodating the observed neutrino data. 
We have also investigated if our scenario can be probed by SHiP~\cite{Alekhin:2015byh} or FCC-ee~\cite{Blondel:2014bra}.

In the second part of this chapter, we consider  the possibility of having the pure and minimal Inverse Seesaw mechanism with four or five extra neutrinos  - that is the ISS(2,2) (or ISS(2,3)) scenario with two RH neutrinos plus two (three) steriles states~(cf. Section~\ref{Sec:intro}) - at work for an efficient leptogenesis through  oscillations. Notice  that the ISS(2,3) scenario can provide in principle a DM candidate (cf. Chapter~\ref{sec:DMMISS}).
We find that the required mass splitting between the pseudo-Dirac pairs is too large to achieve a successful leptogenesis in the weak washout regime while accommodating neutrino data.

This chapter is organised as follows: 
in Section~\ref{sec_analytic} we first describe the idea by means of a toy model framework of one generation (one flavour) and two sterile states, with two sources of lepton number violation ($\Delta L=2$) corresponding  to a combined scenario of the ISS and the LSS. In the second part of this section, we  extend the toy model to the full three flavour case  and obtain estimates for the scales of our set of parameters.  Section~\ref{parametrisation} is devoted, in its first part, to an analytical derivation of the BAU in the weak washout regime, which has been confronted to a thorough numerical analysis we have conducted, taking into account the various constraints on the sterile states. 
The results derived in Section~\ref{parametrisation} are discussed in Section~\ref{Sec:Results.discussion}. Section~\ref{Sec:strong washout} is devoted to the scenario of strong washout regime for the Yukawa couplings.
We  dedicate Section \ref{ISS} to  the pure minimal ISS. 
The analytical determination of the BAU  and the analytical solution of the full system are presented in   Appendix~\ref{app_leptogenesis}. Finally, we provide  in   Appendix~\ref{app_benchmarks} the relevant numerical input parameters for all the solutions discussed in this analysis. 

\section{Leptogenesis and lepton number violation \label{sec_analytic}}

In Ref.~\cite{Akhmedov:1998qx}, followed up and refined e.g.\ in Refs.~\cite{Asaka:2005pn,Shaposhnikov:2008pf,Asaka:2010kk,Asaka:2011wq,Canetti:2012kh}, a compelling mechanism  accommodating neutrino data, the dark matter abundance and providing a successful mechanism for leptogenesis at the electroweak scale was proposed. In its simplest setup, this mechanism requires two additional heavy and nearly mass-degenerate neutrinos $N_R^{1,2}$ with sufficiently small Yukawa couplings to the SM, ensuring that these states have not yet reached thermal equilibrium at the electroweak phase transition. Starting from a zero initial abundance, these heavy states are produced thermally as the Universe approaches the electroweak phase transition. Oscillations between these two states produce a CP-asymmetry which induces particle-antiparticle asymmetries in the individual lepton flavours, which are produced in the decay of these states. These asymmetries in the active sector act as a background potential for the sterile flavours (similar to the MSW effect for neutrino oscillations in matter~\cite{Langacker:1986jv}), resulting in particle-antiparticle asymmetries for the sterile states. The two heavy states have opposite $CP$ and form a pseudo-Dirac pair, thus for negligible Majorana masses no total lepton asymmetry is induced,
 as the total lepton asymmetry in the active sector balances the one in the sterile sector. At the electroweak phase transition, $T \sim T_W$, the SM sphaleron processes freeze out, converting the asymmetry in the active sector (and only the active sector) to a net baryon asymmetry. To summarise, the Sakharov conditions~\cite{Sakharov:1967dj} necessary for a successful baryogenesis are fulfilled at $T \lesssim T_\text{W}$ because (i) the heavy states have not yet reached thermal equilibrium, (ii) the oscillations of the heavy states violate CP and (iii) sphaleron processes  violate baryon number. We review this mechanism in more detail in Appendix~\ref{app_leptogenesis}, deriving analytical expressions for the produced individual asymmetries.

A crucial ingredient for this mechanism is the small mass splitting between the two sterile states, with a relative degeneracy in the pair at the per mille level, $\Delta m/M \lesssim 10^{-3}$~\cite{Canetti:2012kh}. This small mass splitting is obtained naturally if there is a symmetry which imposes fully degenerate masses for the heavy states, the small mass splitting is then linked to the small breaking parameters of this symmetry. A simple choice is an additional global $U(1)$ symmetry with opposite charges assigned to the new fields. As a result, the two states $N_R^{1,2}$ form a Dirac spinor $\Psi= N_R^{1}+ (N_R^{2})^c$, and the $U(1)$ global symmetry mimics the lepton number one~\cite{Shaposhnikov:2006nn}. The small breaking of lepton number is moreover a promising source to explain the light neutrino masses, studied in detail in, for instance,  Refs.~\cite{Antusch:2015mia,Gavela:2009cd,Kersten:2007vk}.
In the following, we will thus focus on models with approximately conserved lepton number and investigate their viability for leptogenesis through neutrino oscillations. 

\subsection{An instructive toy model \label{sec_toymodel}}
Let us recapitulate the simplest implementation of this idea, by adding to the SM two sterile fermions with opposite lepton number, cf.\ also \cite{Gavela:2009cd}. In order to obtain clear analytical results we first consider a toy model with only one active neutrino. In this case the lepton conserving part of the mass matrix is, in the basis $(\nu_L, {N_R^{1}}^c, {N_R^{2}}^c)$,
\bee
M_0 &=& \left(\begin{array}{ccc} 0 & \frac{1}{\sqrt{2}}Y v & 0\\
\frac{1}{\sqrt{2}} Y v & 0 & \Lambda \\
0 & \Lambda & 0
\end{array}\right),
\eee
with $v$ denoting the vacuum expectation value of the Higgs boson, $v = 246$~GeV  after the EW phase transition, $Y$ denoting the Yukawa coupling of the sterile state with lepton number (+1) to the SM lepton and Higgs doublets and $\Lambda$ denoting a new mass parameter which will set the scale for the masses of the additional heavy states.
The mass spectrum resulting from this mass matrix is
\begin{equation}
m_\nu =0 \,,\quad
M_{1,2} = \sqrt{|\Lambda|^2 +\frac{1}{2}|Y v|^2}\ .
\label{eq_M}
\end{equation}

Let us consider now all possible patterns for breaking the global lepton number in $M_0$. A term in the $(1,1)$ entry breaks gauge invariance, and can only be generated in non minimal models, for example by adding an isospin triplet of Higgs fields. Since we are not interested in such a case, there are 3 possible patterns to perturb $M_0$
\begin{align}
\Delta M_{ISS}=\begin{pmatrix} 0 & 0 & 0\\
0 & 0 & 0\\
0 & 0 & \xi \, \Lambda
\end{pmatrix}, \quad 
\Delta M_{LSS}=\begin{pmatrix} 0 & 0 & \frac{1}{\sqrt{2}} \epsilon \, Y' v\\
0 & 0 & 0\\
\frac{1}{\sqrt{2}} \epsilon \, Y' v & 0 & 0 
\end{pmatrix}, \quad
\Delta M_{lp}=\begin{pmatrix} 0 & 0 & 0\\
0 & \xi' \, \Lambda & 0\\
0 & 0 & 0 
\end{pmatrix} \,.
\label{eq_DMs}
\end{align}
Here $\epsilon$, $\xi$ and $\xi'$ are small dimensionless parameters accounting for the breaking of lepton number and $Y' \sim Y$ is a new Yukawa coupling. Without loss of generality we can choose $|Y'| = |Y|$,  keeping $\epsilon$ as a free parameter\footnote{When considering three active flavours in the following, the six Yukawa couplings will be restricted to be of the same order of magnitude. \smallskip}.
The first possibility generates the usual Inverse Seesaw pattern~\cite{Wyler:1982dd, Mohapatra:1986bd}, the second one corresponds to the so called Linear Seesaw~\cite{Malinsky:2005bi}, while the third one does not generate neutrino masses at tree level but does it  at loop level\cite{Dev:2012sg,LopezPavon:2012zg}.
However loop corrections are only relevant in the regime of a large lepton number violation, $\xi' \gtrsim 1$, and since we focus on models with an approximate lepton number conservation we will concentrate in the first two possibilities, $\Delta M \equiv \Delta M_{ISS} + \Delta M_{LSS}$.\footnote{This structure can be obtained dynamically by extending the particle content of the SM, e.g.\ it is possible to generate a small $\Delta L=2$ mass $\sim  \xi \, \Lambda$ as in the general formulation of the Inverse Seesaw~\cite{Mohapatra:1986bd}, where the smallness of  $\xi$ was attributed to the supersymmetry breaking effects in a (superstring inspired) $E_6$ scenario. In the context of a non-supersymmetric $SO(10)$ model, which contains remnants of a larger $E_6$ symmetry, $ \xi \, \Lambda$ is generated at two-loop while $\xi' \, \Lambda$ is generated at higher loops, justifying its smallness compared to $\sim  \xi \, \Lambda$~\cite{Ma:2009gu}.} 
Here $M$  contains only a single physical complex phase (after absorbing  three complex phases by rotating the three fields of the toy model), which we will assign to $Y'$ in the following,  taking  $\Lambda $, $\xi$, $\epsilon$ and $Y$ to be real and positive.

Perturbatively diagonalising $M_0 + \Delta M$ yields expressions (at leading order in the small lepton number violating parameters $\epsilon,\xi \ll 1$) for the two quantities relevant for leptogenesis and neutrino mass generation: the mass-scale of the active neutrinos $m_\nu$ and the mass splitting between the two (heavy) states, $\Delta m^2$. For the Inverse Seesaw scenario, we find
\begin{align}
 m_\nu &= \xi \frac{(Y v)^2 \Lambda}{2 \Lambda^2 + (Y v)^2 } + {\cal O}({\xi^2}) \approx \xi \frac{(Y v)^2}{2 \Lambda } \,, \label{eq_mnuiss}\\
 \Delta m^2 &= 2 \xi \frac{\Lambda^3}{\sqrt{\Lambda^2 + \frac{1}{2}(Y v)^2 }} + {\cal O}({\xi^2})  \approx 2 \xi \Lambda^2 \,, \label{eq_M12iss}
\end{align}
whereas the Linear Seesaw yields
\begin{align}
  m_\nu &= 2 \epsilon \frac{Y^2 v^2 \Lambda}{2 \Lambda^2 + (Y v)^2 } + {\cal O}({\epsilon^2}) \approx \epsilon \frac{(Y v)^2}{\Lambda } \,,\label{eq_mnulss}\\
 \Delta m^2 &= 4 \epsilon \frac{(Y v)^2 \Lambda}{\sqrt{2 \Lambda^2 + (Y v)^2 }}  + {\cal O}({\epsilon^2})  \approx 2 \epsilon (Y v)^2 \,. \label{eq_M12lss}
\end{align}
Here after expanding in $\xi$ or in $\epsilon$, we that the heavy neutrinos cannot be fully thermalised for a viable leptogenesis scenario, implying an upper bound on their Yukawa couplings,
$ Y \, v \ll \Lambda$.

From these expressions we can draw several important conclusions. Firstly, comparing Eqs.~\eqref{eq_mnuiss} and \eqref{eq_mnulss} we note that both the Inverse and the Linear Seesaw realisations require the same degree of lepton number violation in order to reproduce the observed light neutrino masses,
\begin{equation}
\epsilon, \xi = (m_\nu M_{1,2})/m_D^2  \,,
\label{eq_epsmu}
\end{equation}
with $m_D = Y v/\sqrt{2} = |Y'| v/\sqrt{2}$. 

\noindent Secondly,  looking at Eqs.~\eqref{eq_M12iss} and \eqref{eq_M12lss}, we note that above the EW phase transition, where $\langle v \rangle = 0$, the mass splitting induced by the Linear Seesaw vanishes, contrary to the one induced by the Inverse Seesaw. This is a relevant detail since successful leptogenesis can occur only above the EW phase transition, where the sphalerons can efficiently convert a lepton asymmetry into a baryon asymmetry.  Thermal effects during the oscillation process might alleviate this difficulty~\cite{Canetti:2012kh}. For simplicity we will in the following however focus on the situation where the mass splitting is related to the Majorana mass term $\xi \Lambda$. In the pure Inverse Seesaw model this implies
\begin{equation}\label{eq:dmISS}
 (\Delta m^2)_{ISS}^{1/2} \simeq \left( \frac{2 \, m_\nu M_{1,2}}{m_D^2} \right)^{1/2} M_{1,2}\ .
\end{equation}
For example for $Y  < \sqrt{2} \times 10^{-7}$, $m_\nu = 0.05$~eV and $M_{1,2} = 1$~GeV, this yields ${\Delta m^2 \gtrsim (0.4~\text{GeV})^2}$ - a value far too large for a successful leptogenesis\footnote{This upper bound on $Y$ forces the heavy states to be out-of-equilibrium~\cite{Akhmedov:1998qx} and washout processes to be negligible. 
The numbers quoted here a priori only apply to the toy model discussed in this section, and not to realistic, more elaborate versions of the Inverse Seesaw mechanism. We will return to this point in Section~\ref{ISS}.}.

In conclusion, the minimal setup to accommodate acceptable light neutrino masses $m_\nu$ and a sufficiently small mass splitting $\Delta m^2$  is obtained  by simultaneously considering both  $\Delta M_{LSS}$ and  $\Delta M_{ISS}$, with $\epsilon > \xi$: the leading order contribution to the light neutrino masses stems from $\epsilon$, with the scale of $\epsilon$ determined by Eq.~\eqref{eq_epsmu}. Above the EW phase transition, the leading order contribution to the mass splitting is in turn set by Eq.~\eqref{eq_M12iss} and can be sufficiently small for $\xi \ll \epsilon$:
\begin{equation}
 m_\nu \simeq 2 \epsilon \frac{m_D^2}{M_{1,2}} \,, \quad \Delta m^2 \simeq 2 \xi M_{1,2}^2 \,.
 \label{eq_mueps}
\end{equation}

This analysis suggests that the minimal viable realisation of our ansatz is given by the mass matrix $M = M_0 +  \Delta M_{ISS} + \Delta M_{LSS}$.
Notice that the ordering of the second and third column/row  of Eqs.~\eqref{eq_M} and \eqref{eq_DMs} arises from the assignment of lepton number 1 and -1, respectively. Choosing $\epsilon > 1$ and $|Y| \simeq |Y'|$ correspondingly smaller, implies switching this assignment. Hence very large values of $\epsilon \gg 1$ also correspond to a small violation of lepton number, and there is an approximate symmetry under $\epsilon \rightarrow 1/\epsilon$ which becomes exact when $\xi, \xi' \rightarrow 0$. Accounting for solutions with $\epsilon \gg 1$ is equivalent to considering the mass matrix $M=M_0 +\Delta M_{LSS}+ \Delta M_{lp}$, which represents a minimal setup as well. The main difference between the two possibilities is that the Majorana mass term $\Delta M_{ISS}$ breaks the lepton number by $\Delta L=2$, i.e.\ by the same amount of the violation in the Yukawa sector given by $\Delta M_{LSS}$, while $\Delta M_{lp}$ carries $\Delta L=-2$. For simplicity we will focus on the case $\epsilon \ll 1$ in the remainder of this section, but our numerical study in Section~\ref{parameter_scan} will  cover the entire range for $\epsilon$. At leading order, the corresponding expressions for the perturbative expansion in $\epsilon' = 1/\epsilon \ll 1$ can be obtained by replacing $\epsilon \rightarrow 1/\epsilon$  and $Y \rightarrow \tilde{Y}\equiv \epsilon Y$ in the expressions below.

A further important parameter, which is  particularly relevant for leptogenesis, is the mixing between the two heavy neutrino mass eigenstates. To estimate this, we consider the effective potential for the heavy states arising from the interactions with the SM lepton and Higgs doublets in the surrounding hot thermal plasma~\cite{Asaka:2011wq},
\begin{equation}
 V_N = \frac{1}{8} T (Y^\text{eff})^\dagger Y^\text{eff}\,,
 \label{eq_VN}
\end{equation}
where $T$ denotes the temperature and $Y_\text{eff}$ are the Yukawa couplings of the mass eigenstates $m_j,\ j=1,2,3$,
\begin{equation}
Y^\text{eff}_{\alpha j} = Y_{\alpha I} U_{I j} \,.
\label{eq_Yeff}
\end{equation}
Here $U$ is the matrix which diagonalises $M = M_0 + \Delta M$, $U^T M U = \text{diag}(m_\nu, \, M_1, \, M_2)$, $\alpha$ runs over the active flavours (only one flavour in this toy model), $I$ denotes the sterile flavours and $Y_{\alpha 1} = Y_\alpha = Y$ and $Y_{\alpha 2} = \epsilon Y'_\alpha = \epsilon Y'$. Hence for the toy model of this section with $\beta \equiv \text{Arg}(\Lambda)$, $Y^\text{eff}_{\alpha j} = (0,  i\,  (e^{- i \beta} \,  Y - \epsilon Y'), e^{- i \beta} \, Y + \epsilon Y')/\sqrt{2} + {\cal O}(\xi/\Lambda)$, implying $|Y^\text{eff}| \simeq Y$. The eigenvectors of the potential $V_N$ corresponding to the two heavy states above the EW phase transition (i.e.\ for $v = 0$) are given by
\begin{equation}
 v_1 \simeq (0, 1 + e^{i \beta} \epsilon,\, 1 - e^{i \beta} \epsilon)\,, \quad v_2 \simeq (0, 1 + e^{i \beta} \epsilon,\, - 1 + e^{i \beta} \epsilon) \ ,
 \label{eq_sterile_mixing}
\end{equation}
up to corrections of order $\xi$, $\epsilon^2$. This indicates that in the parameter region of interest, which corresponds to $\xi \ll \epsilon$, maximal mixing between the heavy mass eigenstates with a mixing angle of $\theta_{PD} \simeq 45^\circ$ and hence particularly efficient oscillations are obtained for $\epsilon \rightarrow 0$.

In this regime, which we will refer to as ``perturbative'', since viable neutrino masses and a small enough splitting between the heavy states are obtained through a small violation of the lepton number, the condition for a successful leptogenesis can be casted as:
\begin{equation}
\label{eq:perturbative}
 |Y'| = Y = 10^{-7} \left(\frac{M_{1,2}}{1 \text{ Gev}}\right) \left( \frac{0.1}{\epsilon} \right)^{1/2} {\lesssim} \sqrt{2} \times 10^{-7},  \, \quad \xi < \frac{1}{2} \left( \frac{100 \text{ keV}}{M_{1,2}}\right)^2 \,,
\end{equation}
or, equivalently, for the flipped assignment of lepton charges, corresponding to $Y \rightarrow \tilde{Y} = \epsilon Y$:
\begin{equation}
 {|\tilde Y'| = \tilde Y = 10^{-7} \left(\frac{M_{1,2}}{1 \text{ Gev}}\right) \left( \frac{\epsilon}{10} \right)^{1/2}{\lesssim} \sqrt{2} \times 10^{-7}, \quad \xi < \frac{1}{2} \left( \frac{100 \text{ keV}}{M_{1,2}}\right)^2 \,.}
\end{equation}

In the more realistic model with three active flavours, the situation is more complicated, as cancellations in the matrix equations can arise. In particular, 
$\epsilon$ may be of order one and still yield viable solutions, though this in some sense goes against the spirit of our ansatz, linking the small mass splitting to an approximate symmetry. {In the following, we will refer to this latter type of viable parameter points, approximately identified by the condition $0.1 \lesssim \epsilon \lesssim 10 $, as ``generic'', as opposed to the ``{perturbative}'' solutions identified above.} This section served to clarify the  parameter region of interest which requires no matrix cancellations. We will proceed in the next section with a rigorous perturbative expansion of the full model in the perturbative region, before turning to a numerical study in Sections~\ref{parametrisation}, \ref{Sec:Results.discussion} and \ref{Sec:strong washout}. In Section~\ref{ISS} we will revisit the pure Inverse Seesaw scenario, and investigate if the conclusions above can be circumvented by considering the three active flavours.

\subsection{Perturbative expansion of the full model}

In the previous section, we illustrated a symmetry inspired ansatz for the neutrino mass matrix by means of a 3 $\times$ 3 toy model. In this section, we extend this analysis to a full model taking into account the three active flavours, confirming that the estimates for the scales which were obtained in the toy model (one flavour) remain  also valid in the full model.
Consider hence the following mass matrix,
\begin{equation}
 M =\ \Lambda \, \begin{pmatrix}
      0 & 0 & 0 & \frac{1}{\sqrt{2}}Y_1  v/ \Lambda & \frac{1}{\sqrt{2}} \epsilon Y'_1  v/\Lambda \\
      0 & 0 & 0 & \frac{1}{\sqrt{2}} Y_2  v/\Lambda & \frac{1}{\sqrt{2}} \epsilon Y'_2  v/\Lambda \\
      0 & 0 & 0 & \frac{1}{\sqrt{2}} Y_3  v/\Lambda& \frac{1}{\sqrt{2}} \epsilon Y'_3  v/\Lambda\\
      \frac{1}{\sqrt{2}} Y_1  v/\Lambda &  \frac{1}{\sqrt{2}} Y_2  v/\Lambda &  \frac{1}{\sqrt{2}} Y_3  v/\Lambda & 0 & 1 \\
    \frac{1}{\sqrt{2}}  \epsilon Y'_1  v/\Lambda &  \frac{1}{\sqrt{2}} \epsilon Y'_2  v/\Lambda  &   \frac{1}{\sqrt{2}} \epsilon Y'_3  v/\Lambda  & 1  & \xi
     \end{pmatrix}  \,. \label{eq_Mpertexp}
\end{equation}
In the parameter region of interest, as identified in Section~\ref{sec_toymodel}, all entries of this matrix except for the (4,5) and (5,4) entries are small,
\begin{equation}
 |\frac{1}{\sqrt{2}} Y_\alpha v/\Lambda|, \, |\frac{1}{\sqrt{2}}  \epsilon Y'_\alpha  v/\Lambda |, \, |\xi| \ll 1, \qquad \alpha=\{1,2,3\}\,,
\end{equation}
thus justifying a perturbative approach. In this setup we have two additional physical complex phases, whose assignment will be discussed in the next section. Expanding the eigenvalues of $M^\dagger M$ to fourth order in any combination of the perturbative parameters (including mixed terms), we can identify the leading order contributions to the decisive quantities,  $m_\nu$ and $\Delta m^2$. For the masses of the heavy states this yields
\begin{equation}
 M_{1,2}^2 \simeq |\Lambda|^2 \pm \frac{1}{2} |\xi| |\Lambda|^2 + \frac{1}{2}|\vec{Y}|^2 v^2 + \frac{1}{2} |\xi|^2 |\Lambda^2| + \frac{1}{2}|\epsilon|^2 |\vec{Y'}|^2 v^2 \ ,
\end{equation}
up to third order terms in $\{ \frac{1}{\sqrt{2}} |Y_\alpha v/\Lambda|, \,\frac{1}{\sqrt{2}} | \epsilon Y'_\alpha v/\Lambda |, \, |\xi| \}$. Here $|\vec{Y}|^2 \equiv \sum_{\alpha = 1}^3 |Y_\alpha|^2$. 
As in the toy model of Section~\ref{sec_toymodel}, the overall scale is hence determined by $|\Lambda|$ and the  leading order contribution to the mass splitting is $\Delta m^2 \simeq |\xi \Lambda^2|$.  

Proceeding to the light neutrino masses, we notice that one state remains exactly massless while the other two obtain small masses. This scale is given, up to fourth order in   $\{ \frac{1}{\sqrt{2}} |Y_\alpha v/\Lambda|, \, \frac{1}{\sqrt{2}}| \epsilon Y'_\alpha  v/\Lambda |, \, |\xi| \}$, by
\begin{equation}
 m_\text{tot}^2 \equiv \sum_{i = 1}^3 m_i^2  \simeq \frac{1}{2} |\epsilon|^2  \frac{v^4}{|\Lambda|^2} \left(\sum_{\alpha = 1}^3 |Y_\alpha|^2 |Y'_\alpha|^2  + \sum_{\alpha = 1}^3 \sum_{\beta = 1}^3 |Y_\alpha|^2 |Y'_\beta|^2     \right) \,,
\end{equation}
again in agreement with the expectation from the one flavour toy model studied in Section~\ref{sec_toymodel}.

\section{Computation of the baryon abundance \label{parametrisation}}
In this section we investigate the impact of  requiring  a successful leptogenesis on our scenario. To achieve this task we will compute the baryon abundance for a large set of model realisations (at low Seesaw scales), complying with the experimental constraints on the active neutrinos, as well as with limits from possible signatures of the extra sterile fermions in laboratory searches. In order to perform an efficient exploration of the parameter space, we adopt the parametrisation  of the neutrino mass matrix introduced in~\cite{Gavela:2009cd}, reviewed in detail in the next subsection.
An accurate determination of the baryon density would require the solution of a system of coupled Boltzmann equations, like the ones introduced in~\cite{Asaka:2005pn,Asaka:2011wq,Canetti:2012kh}, in the entire parameter space. Unfortunately this task is computationally demanding. For this reason we first focus our analysis on a subset of the parameter space, corresponding to very suppressed values of the Yukawa couplings of the new neutrinos (see below for details). We will refer to this scenario as ``weak washout'' regime. Here all the heavy neutrinos are far below thermal equilibrium during the entire leptogenesis process; as a consequence there is no depletion of the produced baryon asymmetry from washout processes. In this regime the system of Boltzmann equations can be perturbatively solved (see details on the derivation in Appendix~\ref{app_leptogenesis}), resulting in an analytical expression for the baryon abundance $Y_B$ well approximating the full numerical result.

We have further implemented the numerical solution of Boltzmann equations to validate and complement our analytical study,  extending our analysis beyond the reach of the analytical estimates, cf. Section~\ref{Sec:strong washout} for an analysis of the ``strong washout'' regime. This regime is characterised by higher values of the entries of the Yukawa matrix of the heavy neutrinos such that they reach thermal equilibrium at temperatures between the initial production of the lepton asymmetry (i.e. the temperature $T_L$ defined in Appendix~\ref{app_leptogenesis}) and $T_{\rm W}$. This entails a depletion of the lepton asymmetry.

For the sake of clearness we will present in the following the main results, while the details of both the analytical and numerical computations will be reviewed in the appendix.

\subsection{Parametrisation of the mass matrix}\label{sec:parametrisation:mass}

\noindent
We consider the neutrino mass matrix introduced in Eq.~\eqref{eq_Mpertexp}. As discussed in Section~\ref{sec_toymodel},
the lepton number violation is represented by the dimensionless parameters $\epsilon$ and $\xi$. The entries of the mass matrix associated to these parameters violate the lepton number by the same amount, namely $\Delta L=2$. Although in a ``natural'' scenario these two parameters would be expected to be of the same order of magnitude (and both small), we will stick to a more generic case taking them to be free and independent among each other. In particular we will here also allow for large lepton number violation $\epsilon, \xi \sim 1$, going beyond the perturbative expansion of Section~\ref{sec_toymodel}.

In order to identify a minimal set of parameters for a numerical scan, we have adopted 
the parametrisation introduced in~\cite{Gavela:2009cd}. The Yukawa matrices are expressed as a function of two free parameters $y$ and $y'$, of an additional parameter $\rho$ given by:
\begin{equation}
\rho=\frac{\sqrt{1+r}-\sqrt{r}}{\sqrt{1+r}+\sqrt{r}}\, , \quad r=\frac{|\Delta m_{\rm solar}^2|}{|\Delta m_{\rm atm}^2|}\ , 
\end{equation}
and of the elements of the PMNS matrix as:
\begin{align}
\label{eq:yukawa_parameters}
& Y_{\alpha}=\frac{y}{\sqrt{2}} \left[U^{*}_{\alpha 3} \sqrt{1+\rho} + U^{*}_{\alpha 2} \sqrt{1-\rho}\right] \ ,\nonumber\\
& Y_{\alpha }^{\prime}= \tilde{Y}_{\alpha }^{\prime}+\frac{k}{2}Y_{\alpha }\ ,\nonumber\\
& \tilde{Y}_{\alpha}^{\prime}=\frac{y^{\prime}}{\sqrt{2}}\left[U^{*}_{\alpha 3} \sqrt{1+\rho} - U^{*}_{\alpha 2} \sqrt{1-\rho}\right]\ , \nonumber\\
& k=\frac{\xi}{\epsilon}\ .
\end{align}
The three physical phases
in the mass matrix~\eqref{eq_Mpertexp} are conveniently assigned as follows: the Dirac phase $\delta_{CP}$ and the unique Majorana phase $\alpha$ of the PMNS matrix\footnote{The second Majorana phase in the PMNS matrix can be rotated away since in this case one neutrino is massless.} appear in $Y$ and $Y'$ through Eq.~\eqref{eq:yukawa_parameters}, the third `high-energy' phase is assigned to $\Lambda$. The parameters $\epsilon$, $\xi$, $v$, $y$ and $y'$ can then be taken to be real and positive and the $\Delta L=2$ Majorana mass term is taken equal to $\xi \left|\Lambda\right|$.
Using this parametrisation the mass eigenstates coincide with the expressions in the limit of pure Linear Seesaw (the Majorana mass parameter $\xi$ is encoded in the definition of the Yukawa matrices) and are thus given by~\cite{Gavela:2009cd}: 
\begin{align}
\label{eq:mass_eigen}
 |m_1|=0 \,, \qquad
 |m_2|= \frac{\epsilon y y^{'} \left(1-\rho\right) v^2}{2\left|\Lambda\right|}\,, \qquad
 |m_3|= \frac{\epsilon y y^{'} \left(1+\rho\right)v^2}{2\left|\Lambda\right|} \ ,
\end{align}
while in the limit $|\Lambda| \gg |Y| v, \epsilon |Y^{\prime}| v$, which is the one relevant for leptogenesis, the masses of the two mostly sterile states are given by:
\begin{equation}
M_{1,2}=|\Lambda| \left(1 \mp \xi \right)\ .
\end{equation}
Notice that this parametrisation generates neutrino masses only according to a normal hierarchy. An inverted hierarchical spectrum can be obtained by modifying the definition of $\rho$ as:
\begin{equation}
\rho=\frac{\sqrt{1+r}-1}{\sqrt{1+r}+1}\ ,
\end{equation}
and by replacing $U_{\alpha 3} \rightarrow U_{\alpha 2}$ and  $U_{\alpha 2} \rightarrow U_{\alpha 1}$ in Eq.~(\ref{eq:yukawa_parameters}).

\subsection{Parameter scan \label{parameter_scan}}

As seen in the previous section, the neutrino mass spectrum thus depends on 6 parameters, $y,y^{\prime},\Lambda, \epsilon, \xi, k$. These parameters are actually not independent among each other. We can reduce the number of free parameters by imposing the correct values for the neutrino masses (as can be inferred by the atmospheric and solar mass squared differences) through the expressions~(\ref{eq:mass_eigen}). We can, for example, determine $\epsilon$ by imposing a normal hierarchy for the light neutrino masses\footnote{For simplicity we are reporting just the case of a normal hierarchy regarding the light neutrino mass spectrum. An analogous procedure has been employed in the case of an  inverted hierarchy. \smallskip} leading to:
\begin{equation}
\label{eq:epsilon_neutrino}
\epsilon=\frac{2 \,  m_3 \,  \left|\Lambda\right|}{y y'\left(1+\rho\right) v^2}\ .
\end{equation}
The last line in Eq.~(\ref{eq:yukawa_parameters}) implies that only of the two parameters $\xi$ and $k$ is a free parameter, which we choose to be $k$.
We thus generate a set of models by scanning over $y,y',\Lambda,k$ within the following ranges:
\begin{align}
& 100\,\mbox{MeV} < {|}\Lambda{|} < 40\,\mbox{GeV}\ ,\nonumber\\
& 10^{-10} < y,y^{\prime} <1 \nonumber\\ 
& 10^{-10} < k <100 \ ,
\end{align}
where, in the spirit of the model, we chose $y$ and $y^{\prime}$ to be of the same order of magnitude for each generated realisation (see Section~\ref{sec_toymodel}).
All the generated points are required, besides complying with the correct neutrino mass and mixing pattern~\cite{Gonzalez-Garcia:2014bfa}, to satisfy bounds from direct laboratory searches of sterile fermions and BBN. The bounds used in our numerical study are based on Ref.~\cite{Alekhin:2015byh}.

Finally we calculate the baryon asymmetry generated in the oscillations of the sterile neutrinos (cf.\ Appendix~\ref{app_leptogenesis})\footnote{We follow the notation of Refs.~\cite{Asaka:2005pn,Asaka:2011wq,Canetti:2012kh} for our expression for the baryon asymmetry.\smallskip}:
\begin{equation}
\label{eq:baryo_analytical}
Y_{\Delta B}=\frac{n_{\Delta B}}{s}=\frac{945\, 2^{2/3}}{2528  \,\,  3^{1/3} \, \pi^{5/2}  \,  \Gamma(5/6)} \frac{1}{g_s\left(T_{\rm W}\right)}\sin^3 \phi \, \frac{M_0}{T_{\rm W}} \frac{M_0^{4/3}}{ \left(\Delta m^2\right)^{2/3}} \, Tr\left[ F^\dagger \delta F\right] \ ,
\end{equation}
where $F=Y^\text{eff} $ with $Y^\text{eff} $  defined in Eq.~(\ref{eq_Yeff}), 
 $\Delta m^2=M_2^2-M_1^2$ is the mass squared splitting of the heavy neutrinos, $T_{\rm W}$ is the temperature of the EW phase transition - set to 140 GeV, $M_0 \approx 7 \times 10^{17}\,\mbox{GeV}$, $\sin\phi \sim 0.012$ and $\delta = \text{diag}(\delta_\alpha)$ is the CP asymmetry in the oscillations defined as:
\begin{equation}\label{eq:deltaCP}
\delta_{\alpha}=\sum_{i >j} Im\left[F_{\alpha i} \left(F^{\dagger} F\right)_{ij} F^{\dagger}_{j\alpha}\right]\ .
\end{equation}
As before the index $\alpha$ corresponds to a flavour index, while the indices $i,j$ run over the sterile mass eigenstates. The derivation of this expression, firstly introduced in~\cite{Akhmedov:1998qx,Asaka:2005pn}\footnote{The expression~(\ref{eq:baryo_analytical}) differs by an $O(1)$ factor with respect to the one given in these references. The origin of these difference will be clarified in Appendix~\ref{app_leptogenesis}.}, is carefully revisited in the appendix. In the next section, this analytical determination will be confronted with the numerical solution of suitable Boltzmann equations, also detailed in the appendix.
This expression is valid under the assumption that the baryon asymmetry is produced with maximal efficiency, which is achieved if the heavy sterile neutrinos never reach thermal equilibrium during the generation process and, consequently, washout effects are always negligible. This requirement can be expressed, as rule of thumb, through the condition $\left|Y^\text{eff} _{\alpha i}\right| < \sqrt{2}\times 10^{-7}$~\cite{Akhmedov:1998qx} (the condition applies to all the elements of the matrix $Y^\text{eff} $).

We consider models as viable if Eq.~(\ref{eq:baryo_analytical}) yields a value for $Y_{\Delta B}$ such that $3 \times 10^{-11} \leq Y_{\Delta B} \leq 2.5 \times 10^{-10}$. The choice of this broad range, compared to the rather precise experimental determination~\cite{Ade:2015xua}, $Y_{\Delta B}= \left(8.6 \pm 0.01 \right) \times 10^{-11}$, is motivated by the need  to account for deviations with respect to the determination of $Y_{\Delta B}$ from the numerical solution of the Boltzmann equations. We expect, in particular, that the analytical expression~(\ref{eq:baryo_analytical}) overestimates the baryon asymmetry for values of $Y^\text{eff} $ close to the equilibrium value $\sqrt{2}\times 10^{-7}$, since in this case we have a late time equilibration of the heavy neutrinos with a reduction of the total baryon asymmetry.

\subsection{Comparison with numerical results}
\label{sec:numerical_ly}

In this section we compare the analytical expression~(\ref{eq:baryo_analytical}) for the baryon asymmetry in the weak washout regime with the numerical solution of the Boltzmann equations describing this process, cf.~Eq.~\eqref{eq:full_system} in Appendix~\ref{app_leptogenesis}, for a set of benchmark points.
In most of the cases we have found a good agreement, with deviations ranging between 5 and 15 $\%$. Larger deviations arise if the  entries of $Y^\text{eff} $ are very close to the out-of-equilibrium condition. 
An explicit comparison between the numerical and analytical determination of the baryon density is shown in Figs.~\ref{fig:bench_natural},~\ref{fig:bench_numsm} and~\ref{fig:bench_thermal}. The three corresponding benchmark points represent, respectively, a model satisfying the ``perturbative'' regime, featuring $\epsilon \sim 0.01$, a model in the ``generic'' regime, with $\epsilon \sim 1$, and, finally, a model with the entries of $Y^\text{eff} $ very close to the out-of-equilibrium condition. The relevant parameters, namely the mass scale $M=(M_1+M_2)/2$ of the heavy neutrinos and their mass splitting $\Delta m$ are reported in the fourth panel of each figure. The values of these parameters, together with the entries of the matrices $Y^{\rm eff}$, are also reported in Appendix~\ref{app_benchmarks}.

\begin{figure}[htb]
\begin{center}
\subfloat{\includegraphics[width=0.42\textwidth]{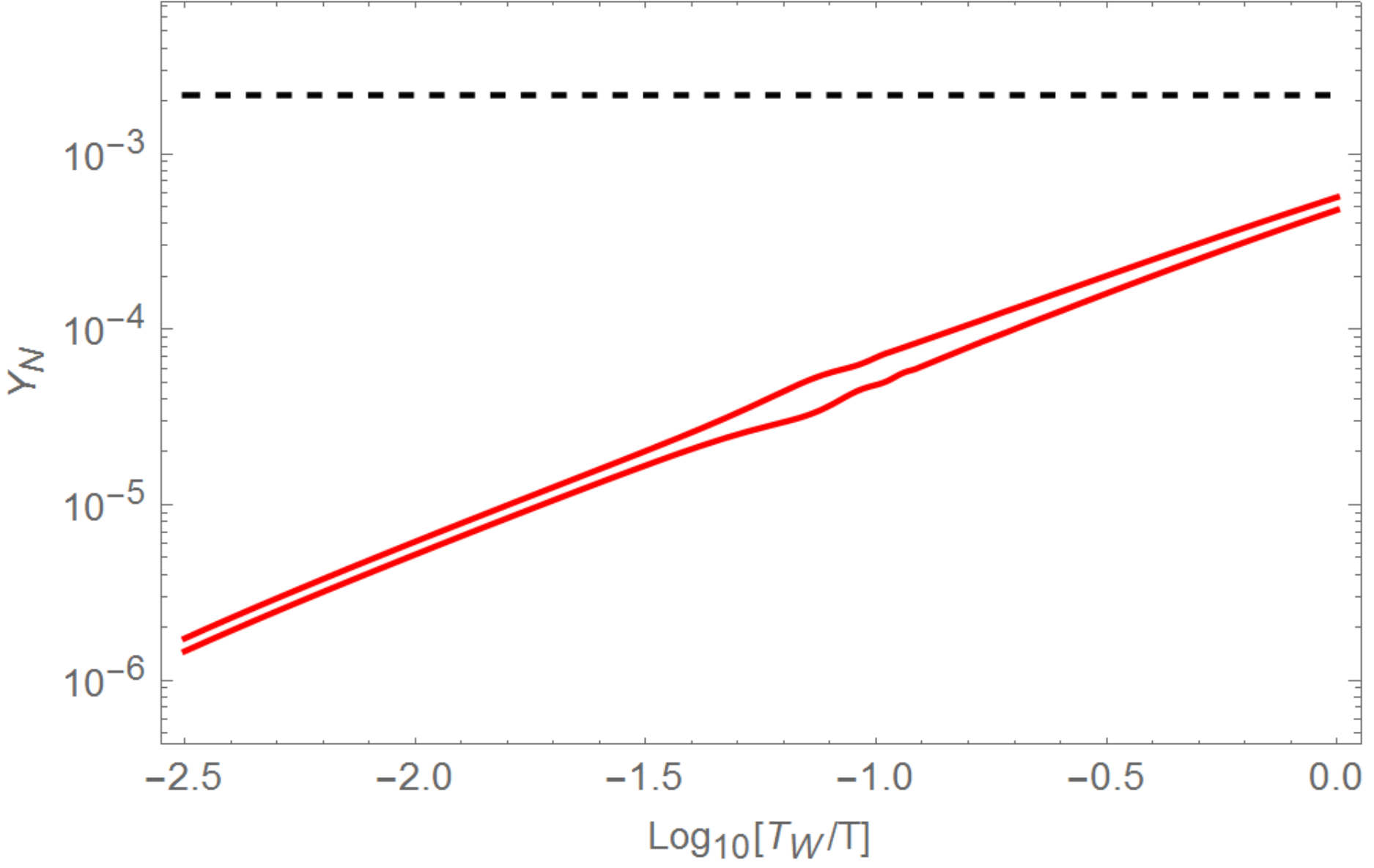}}\hspace{0.04\textwidth}
\subfloat{\includegraphics[width=0.42\textwidth]{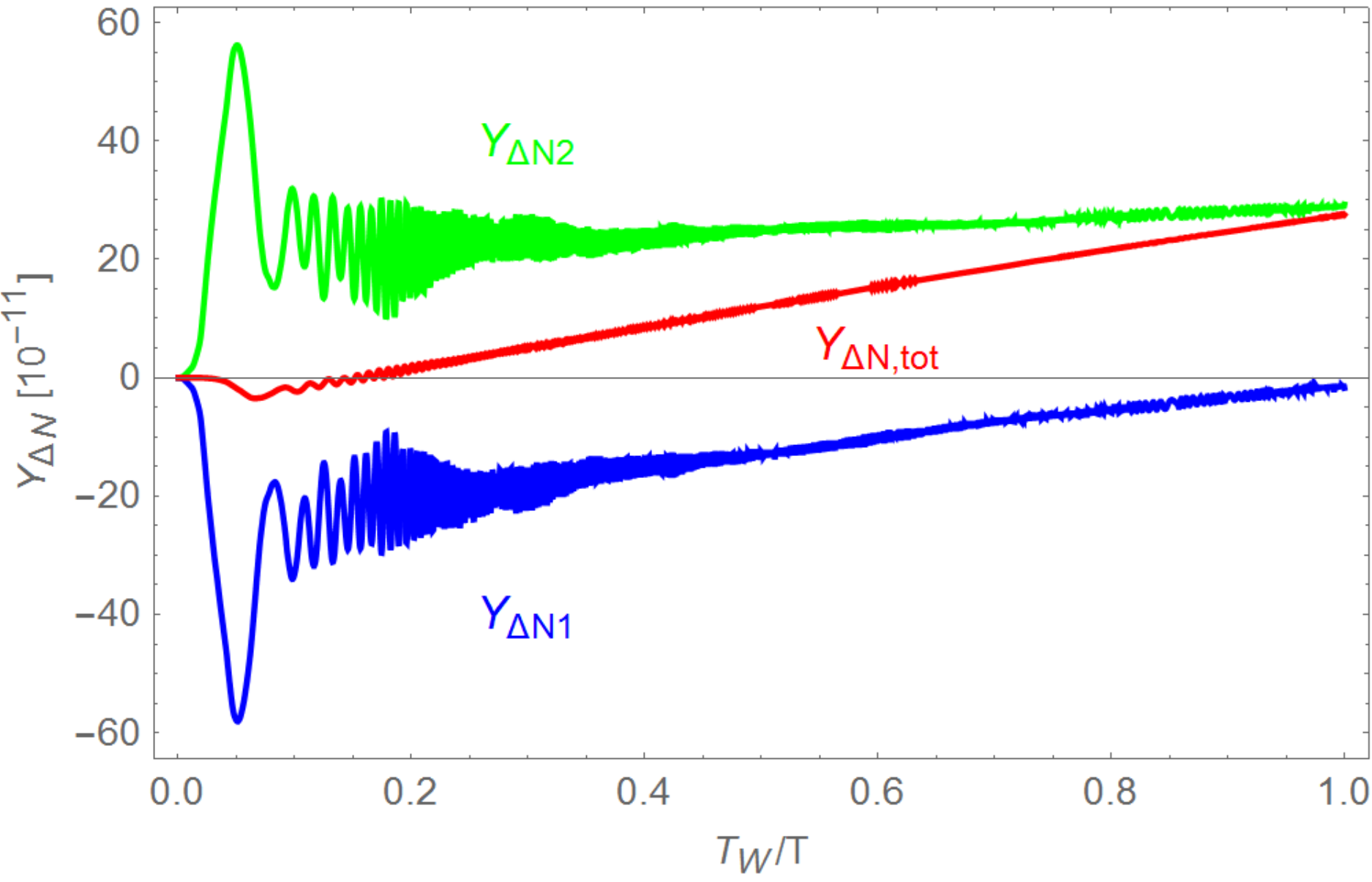}}\\
\subfloat{\includegraphics[width=0.45\textwidth]{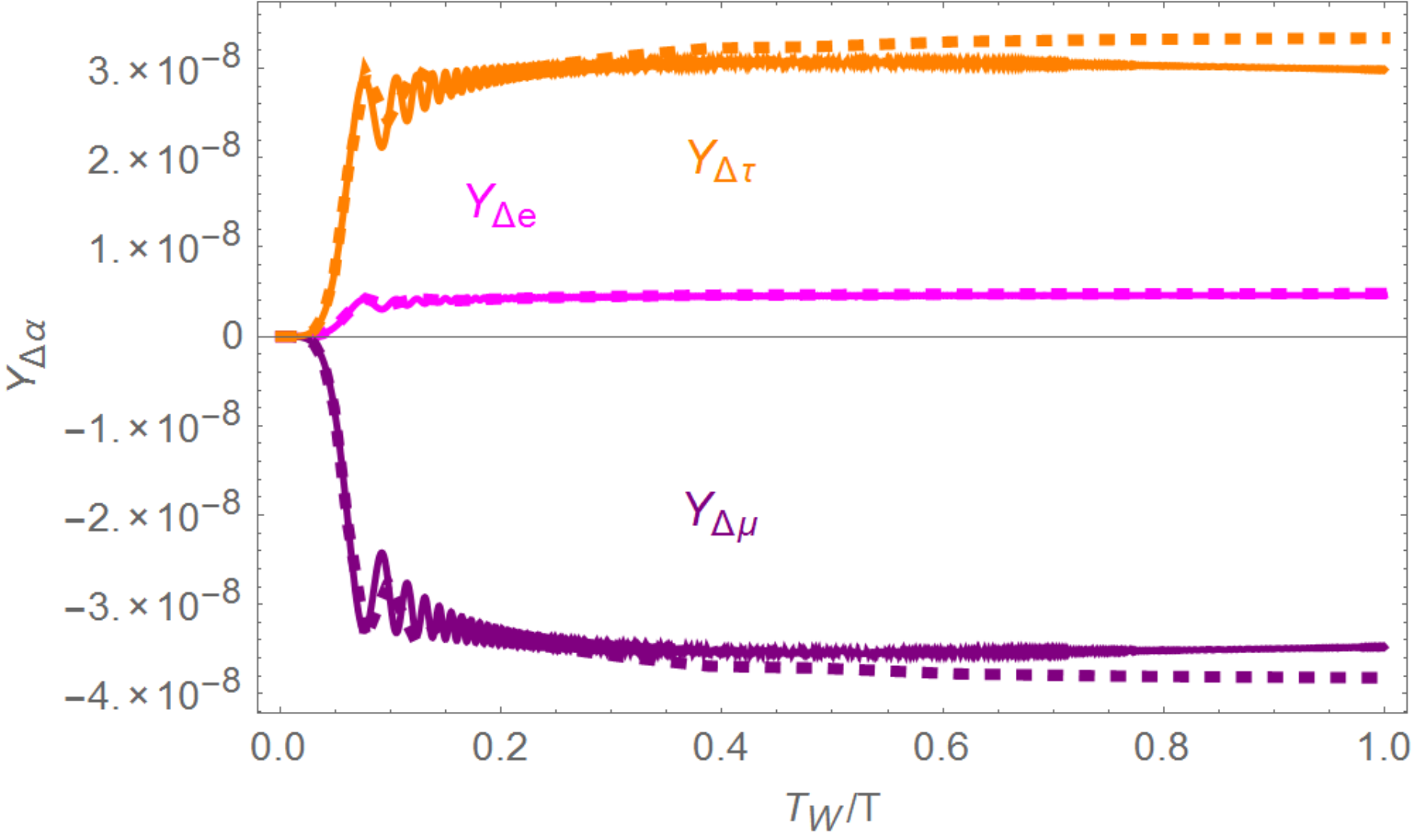}}\hspace{0.015\textwidth}
\subfloat{\includegraphics[width=0.45\textwidth]{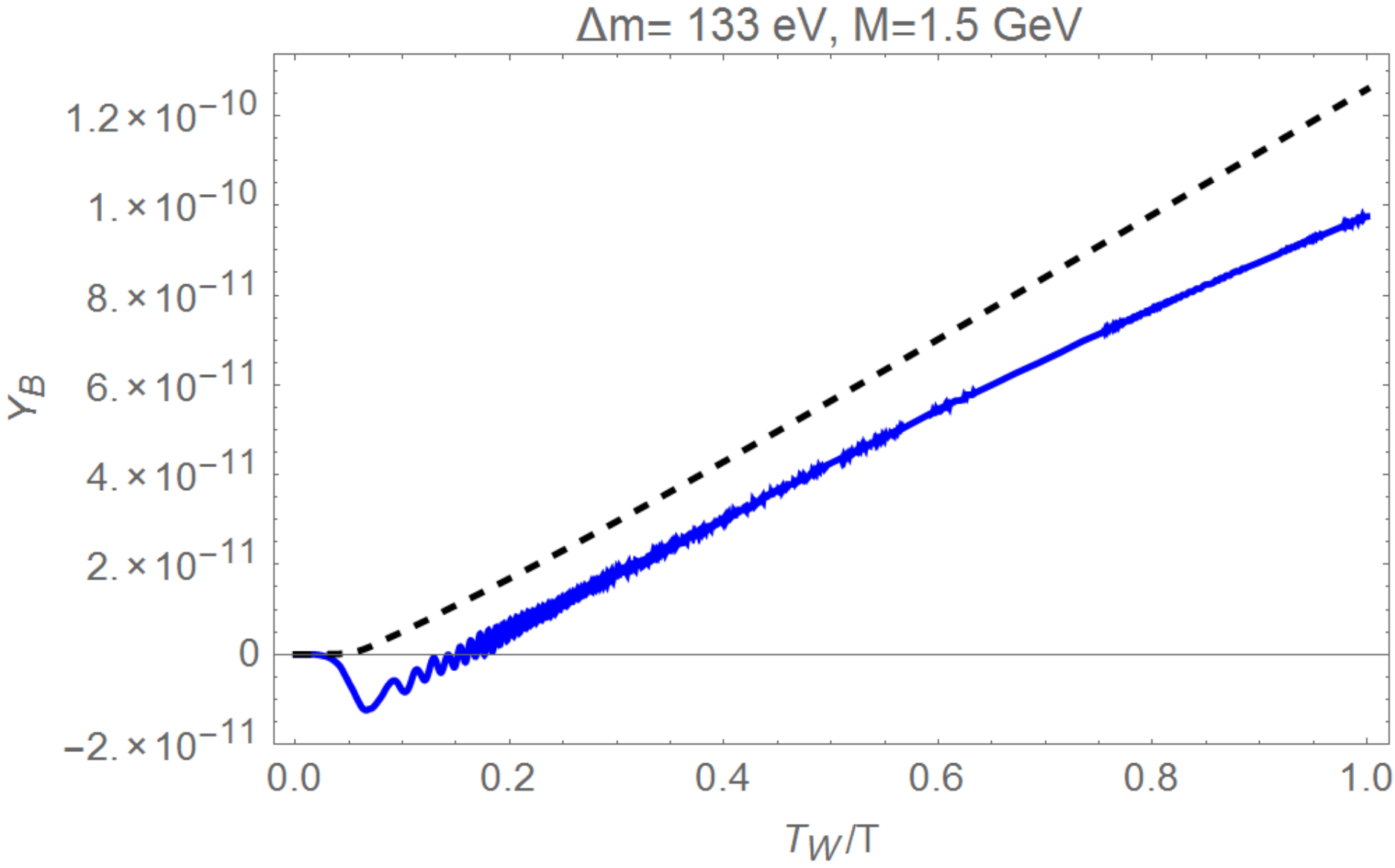}}
\caption{\footnotesize{Evolution of some relevant quantities as function of the temperature, obtained from our numerical treatment for a benchmark model with parameters $\epsilon$ and $\xi$ lying in the ``perturbative'' regime. Left top panel: Evolution of the abundance of the heavy neutrinos (red solid lines) compared with the equilibrium value (dashed black line) $Y_{{N0}}$, defined in the appendix. Right top panel: Evolution of the total (red line) and individual (blue and green lines) asymmetries in the two heavy neutrinos as function of the temperature. 
Left bottom panel: Evolution of the asymmetries in the active flavours according the numerical solution of the Boltzmann equations (solid lines) and the analytical estimate (dashed lines). 
Right bottom panel: Evolution of the baryon yield with temperature (blue line) compared with its analytical determination (dashed black line).}}
\label{fig:bench_natural}
\end{center}
\end{figure}

For each benchmark point we show a set of four plots describing the evolution of the abundance of the two heavy neutrinos, the individual and total asymmetries in the sterile sector, the individual asymmetries in the active sector and, finally, the baryon asymmetry $Y_B$. 
The baryon asymmetries are also compared with their analytical estimates, represented as dashed lines, whose derivation is described in detail in the appendix.
These plots illustrate the main features of the leptogenesis mechanism at work here: the abundance of sterile neutrinos (first panel) grows according to Eq.~\eqref{eq:first_step}, but does not reach the equilibrium value before sphaleron processes convert the lepton asymmetry into a baryon asymmetry at $T \sim T_\text{W}$, thus suppressing washout processes. The oscillations of these sterile states source an asymmetry in the individual active and sterile flavours (second and third panel), described by Eq.~\eqref{eq:secondstep} and \eqref{eq:third_step}, respectively. Using Eqs.~\eqref{Eqrho}, \eqref{eq:RN} and \eqref{eq:YN0} this translates into the abundances depicted in the second and third panel of Figs.~\ref{fig:bench_natural}, \ref{fig:bench_numsm} and~\ref{fig:bench_thermal}. We note that the asymmetries in the individual flavours, in particular in the active sector,  are typically much larger than the total asymmetry in that sector. We have confirmed that the analytical expressions presented in the appendix are well adapted to describe both the individual asymmetries, as well as the total asymmetries\footnote{Notice that the comparison between the numerical determination of the individual asymmetries in the active sector and the analytical determination~\eqref{eq:secondstep} should be regarded with care.  Besides deviations appearing at low temperatures in Figs.~\ref{fig:bench_numsm}-\ref{fig:bench_thermal} due to the fact that the heavy neutrinos get close to thermal equilibrium (see main text), the two determinations should not exactly coincide. Indeed in the analytical derivation a net baryon asymmetry appears, as a higher order effect, only in Eq.~\eqref{eq:third_step} (see also~\cite{Asaka:2005pn}). This effect, on the contrary, is already automatically encoded in the numerical determination of $Y_{\Delta L_{\alpha}}$.}. 
As expected from global lepton number conservation, the total asymmetries in the active and sterile sector are equal but with opposite sign, as demonstrated in Fig.~\ref{fig:numerical_cross_check} for the benchmark point of Fig.~\ref{fig:bench_natural}. The sphaleron processes however only act on the active flavours, yielding a total baryon asymmetry described by Eq.~\eqref{eq:baryo_analytical} and depicted in the fourth panel of Figs.~\ref{fig:bench_natural}, \ref{fig:bench_numsm} and~\ref{fig:bench_thermal}.

\begin{figure}[htb]
\begin{center}
\subfloat{\includegraphics[width=0.42\textwidth]{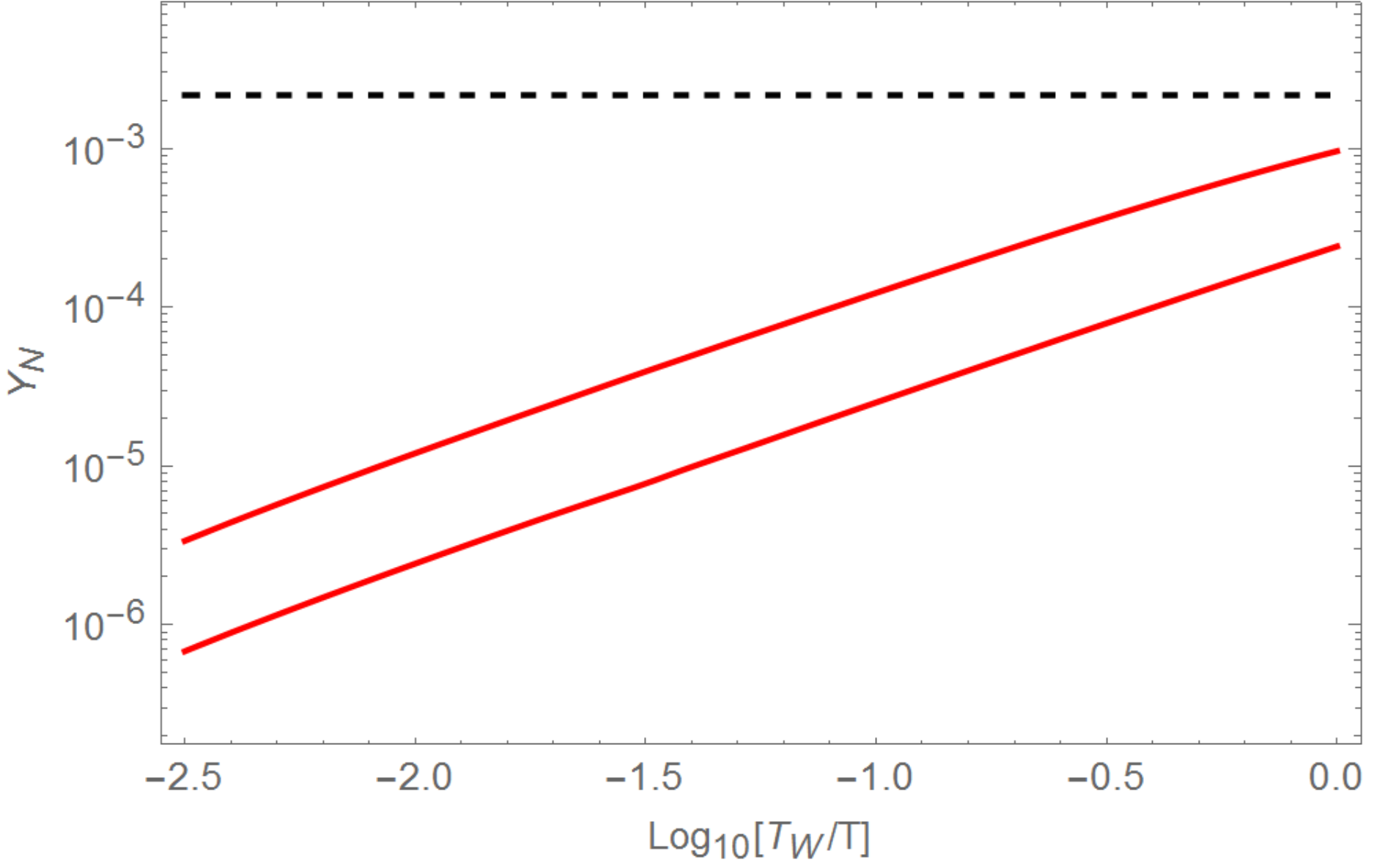}}\hspace{0.04\textwidth}
\subfloat{\includegraphics[width=0.42\textwidth]{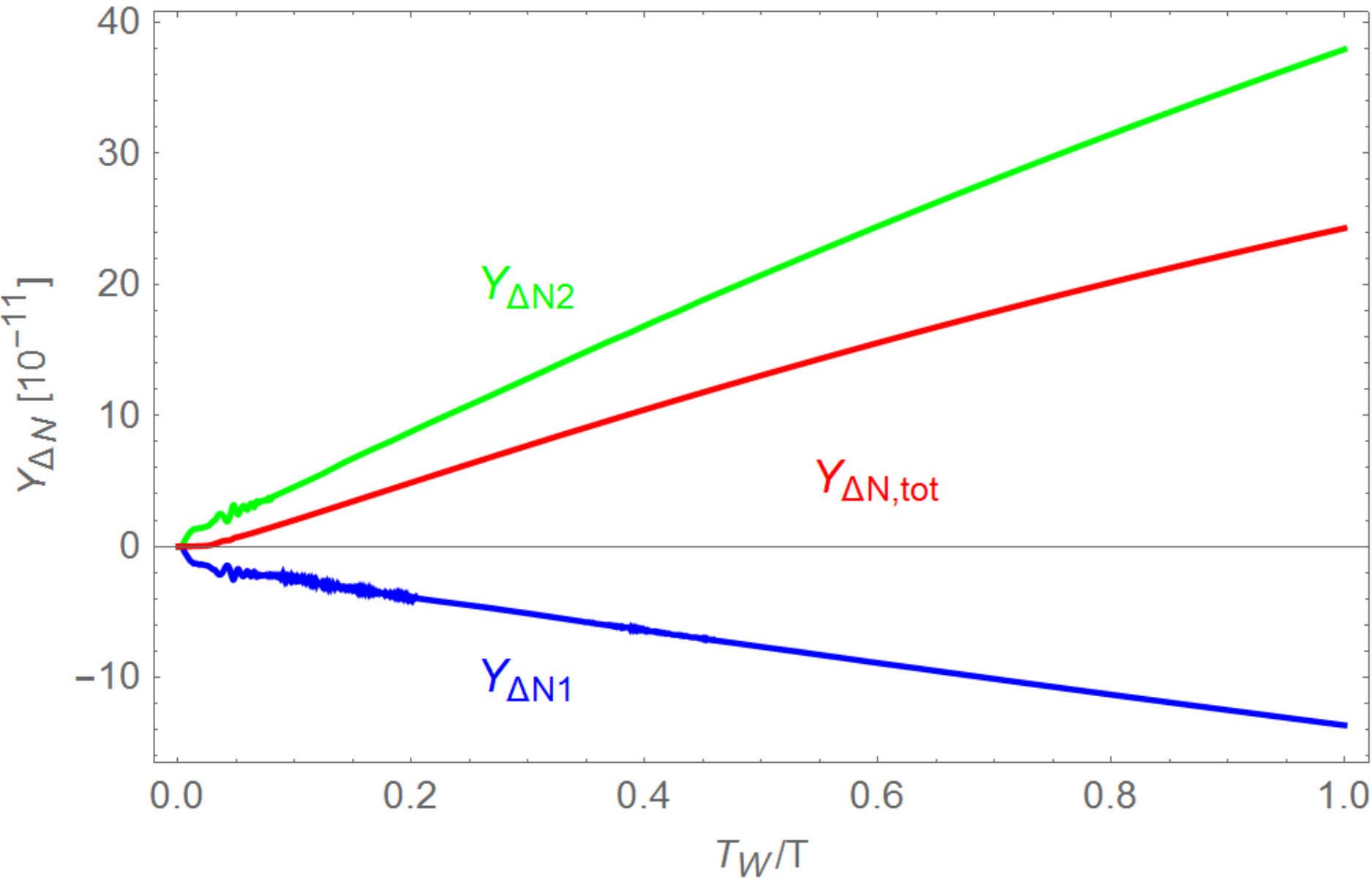}}\\
\subfloat{\includegraphics[width=0.45\textwidth]{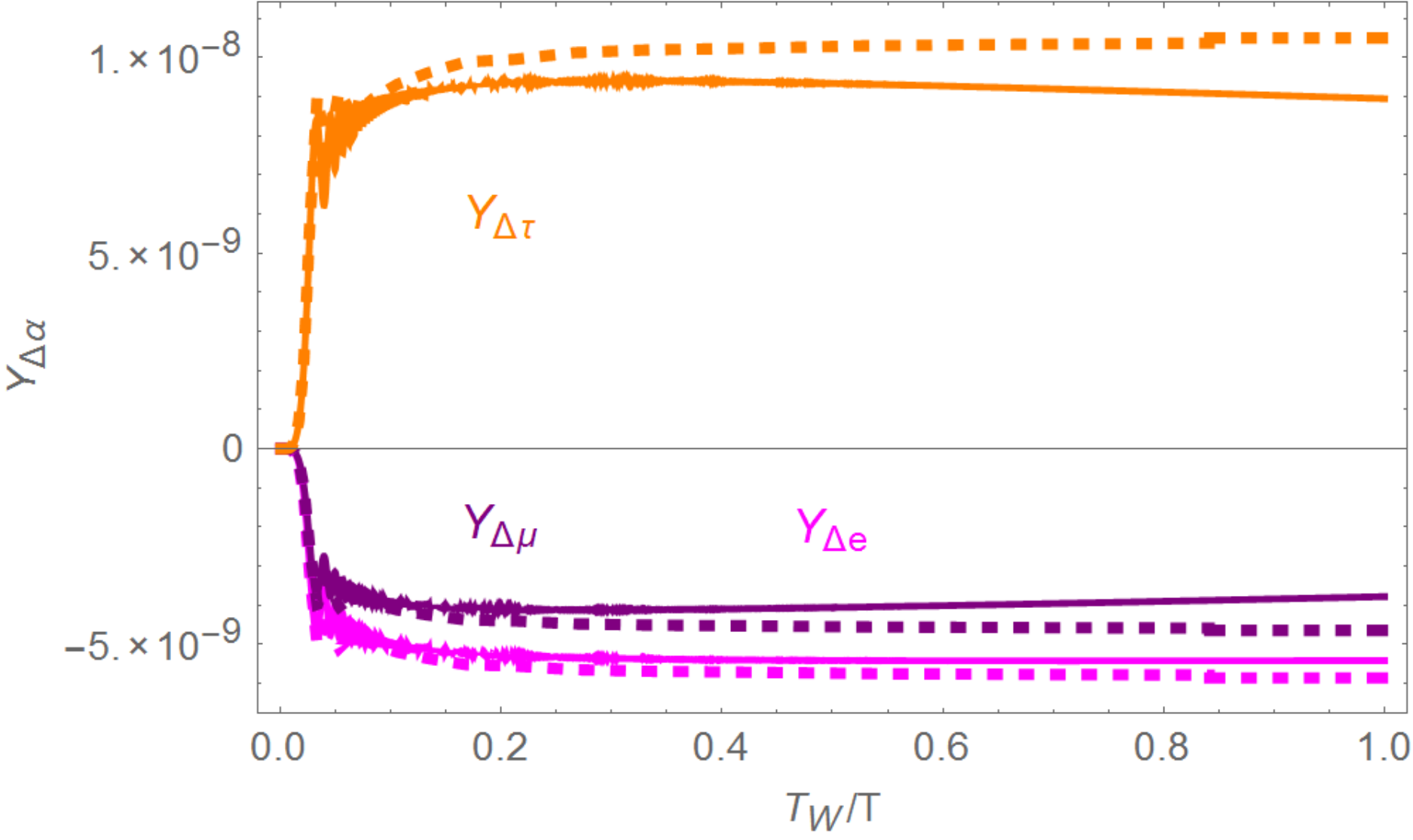}}\hspace{0.015\textwidth}
\subfloat{\includegraphics[width=0.45\textwidth]{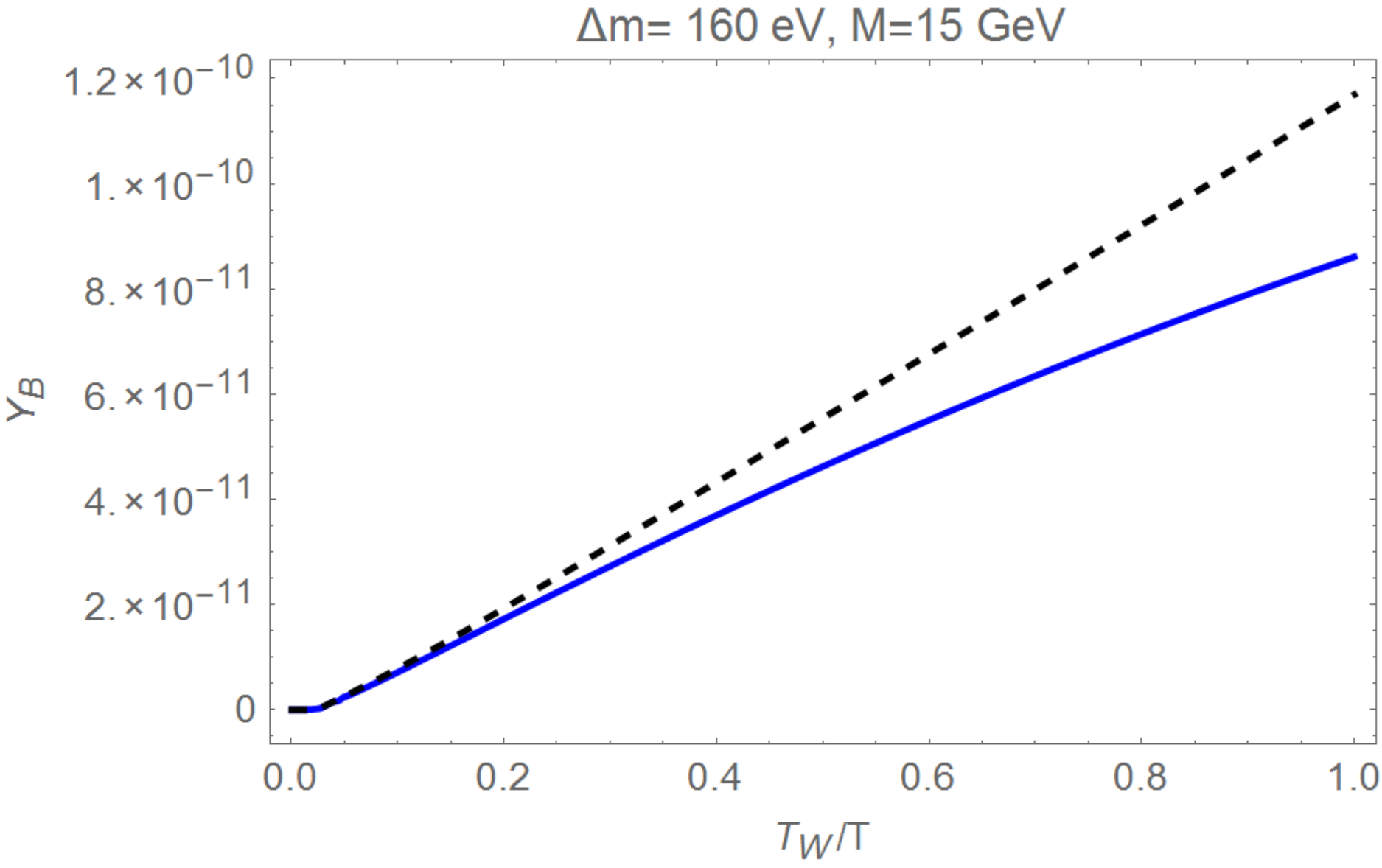}}
\caption{\footnotesize{As in Fig.~\ref{fig:bench_natural} but for a model with $\epsilon$ belonging to the ``generic'' regime.}}
\label{fig:bench_numsm}
\end{center}
\end{figure}

\begin{figure}[htb]
\begin{center}
\subfloat{\includegraphics[width=0.42\textwidth]{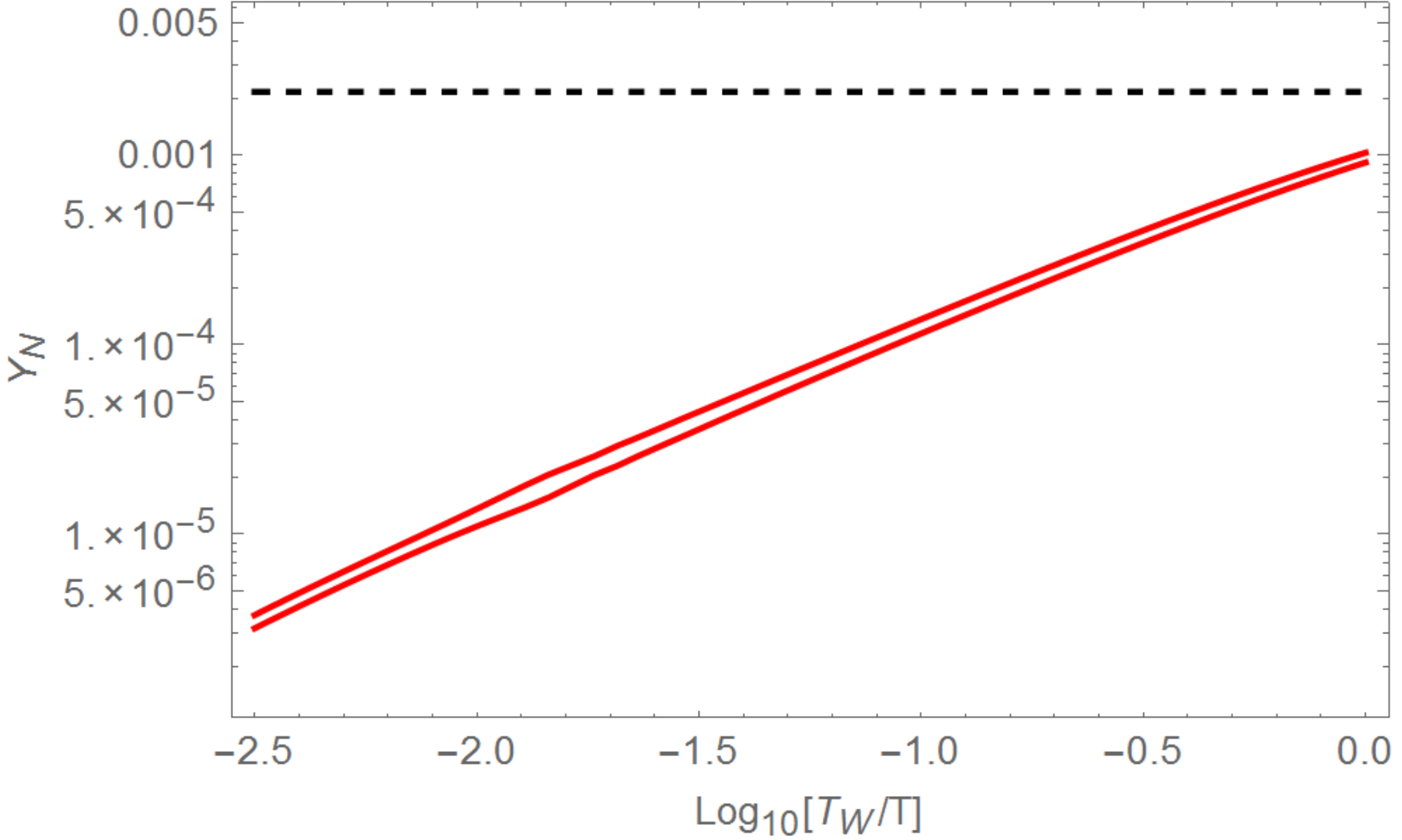}}\hspace{0.04\textwidth}
\subfloat{\includegraphics[width=0.42\textwidth]{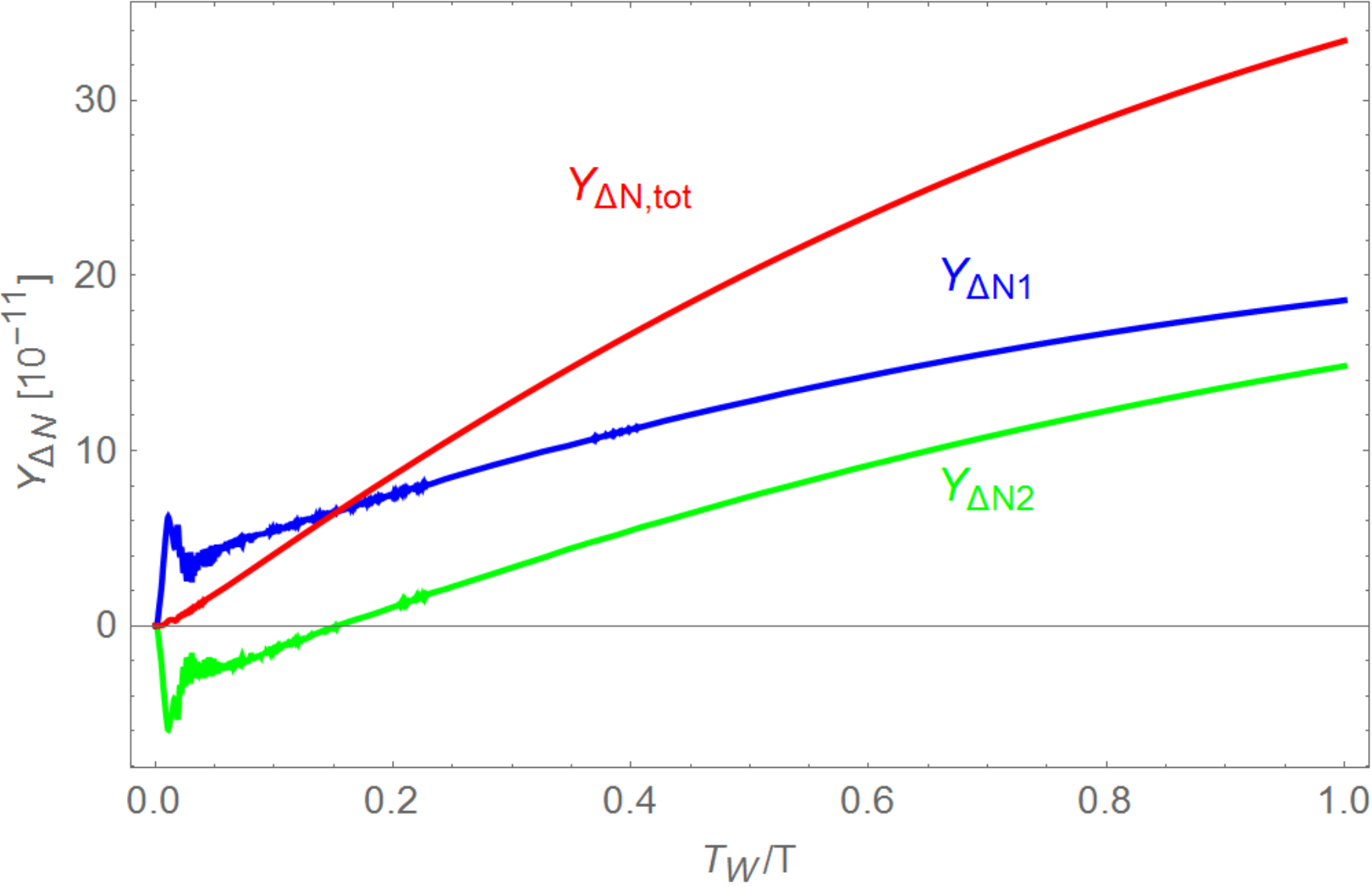}}\\
\subfloat{\includegraphics[width=0.45\textwidth]{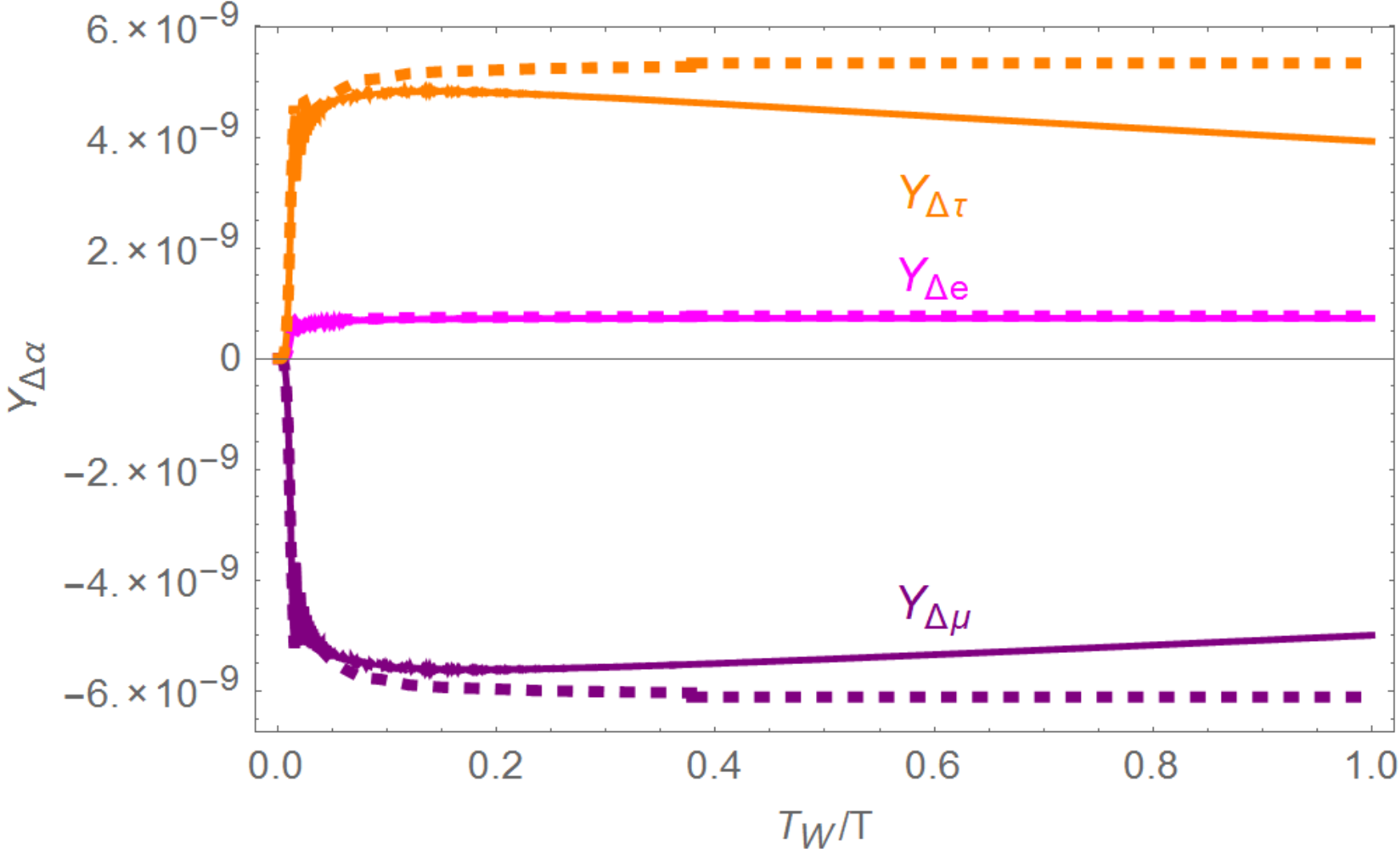}}\hspace{0.015\textwidth}
\subfloat{\includegraphics[width=0.45\textwidth]{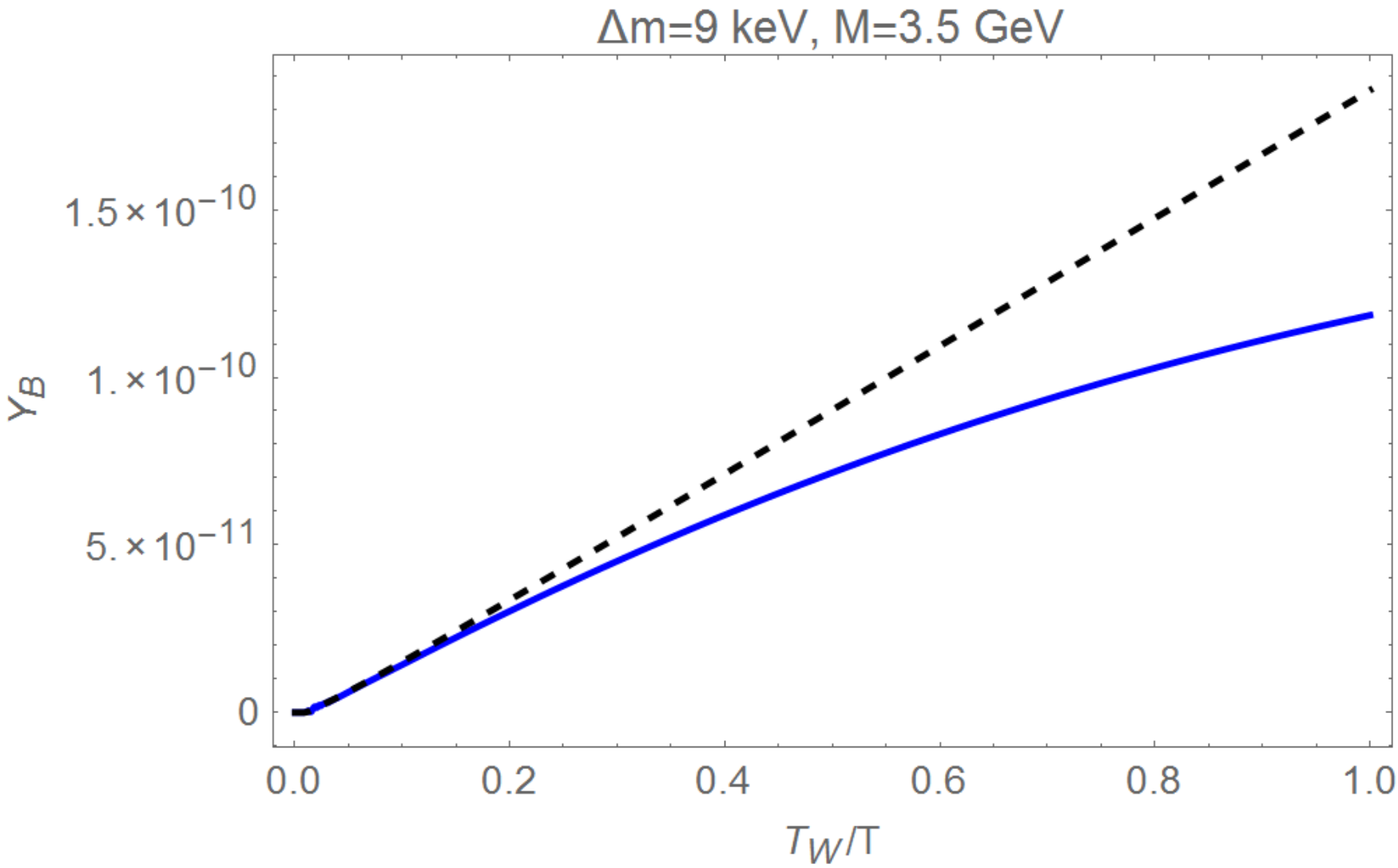}}
\caption{\footnotesize{As in Fig.~\ref{fig:bench_natural} but for a model in the ``perturbative'' regime with entries of $Y^\text{eff} $ close to the equilibration value. }}
\label{fig:bench_thermal}
\end{center}
\end{figure}

We will now discuss in more detail the main features of the each of the considered benchmarks. 
The model represented in Fig.~\ref{fig:bench_natural} is characterised by $\epsilon \sim 0.01$. As evident from the top left panel of the figure, and discussed in a more systematic way in the next subsection, this setup corresponds to a very strong, although not complete, superimposition of the two heavy neutrino states. This is the source of the nearly equal abundances of the two heavy states. In such a scenario essentially equal (large) and opposite in sign asymmetries are stored in the sterile states. A non-vanishing asymmetry arises at later times as a small difference between the individual asymmetries. This behaviour can be understood as follows. At very early times (corresponding to high temperatures) the heavy neutrino pair essentially behave as a single Dirac neutrino, thus carrying an approximately vanishing lepton asymmetry. A net asymmetry is created only after thermal/matter effects cause the oscillations to enter into the resonant regime. The net asymmetry increases at lower temperatures due to the not exact overlap between the neutrino states. As can be seen in  the right bottom panel, the analytical estimate does not provide a correct description of the early time behaviour of the numerical solution but provides nonetheless a good approximation of the total net asymmetry such that there is a relative difference of the order of 10 $\%$ between the numerical and analytical determination of $Y_B$.

Rather different is the case of the benchmark presented in Fig.~\ref{fig:bench_numsm}. This kind of benchmark has $\epsilon \sim 1$ and essentially resembles a $\nu$MSM realisation. As  can be seen on the top left plot of the figure, there is little overlap between the two heavy neutrinos. Contrary to the previous scenario, the two states acquire individual and uncorrelated net asymmetries which grow essentially monotonically in time. As evident from the plot there is a very good agreement with the analytical estimate at early times (high temperatures) while small deviations arise at later times since one of the two neutrinos gets very close to thermal equilibrium (see dashed black line in top left panel of Fig.~\ref{fig:bench_numsm}, causing a slight depletion of the asymmetries in the $\mu$ and $\tau$ flavours, which in this benchmark come with $Y^\text{eff}$ values close to the equilibrium value.

The late time depletion of the baryon asymmetry is more evident in Fig.~\ref{fig:bench_thermal}, where all the entries of the matrix $Y^{\rm eff}$ are close to the equilibrium value. This translates into a relative difference of approximately 40 $\%$ between the analytical and numerical solutions. With a value of $\epsilon \sim 0.01$ for this point, we notice  in the top left pane as in Fig.~\ref{fig:bench_natural} that the two heavy neutrinos appear strongly overlapped. Contrary to the first benchmark model, we observe here a good agreement between the analytical and numerical solutions at early times. This is due to the fact that the oscillations enter rather early in the resonant regime and thus only a small mismatch between the analytical and numerical solutions is originated. More generally we have found that in the regime of $Y^{\rm eff}$ close to the equilibrium value, the analytical determination of the baryon density can overestimate the correct value by up to a factor 3. This  motivates the choice of a broad range of allowed values for the baryon density in our scan of the parameter space.

Finally  we remark that, as already argued in~\cite{Drewes:2012ma}, the asymmetries in the individual active flavours are in general much higher in magnitude, with respect to the total net lepton asymmetry converted by the sphalerons. We notice in particular that larger (in magnitude) asymmetries are stored in the $\mu$ and $\tau$ flavours. This is because the considered benchmark model features a normal hierarchical neutrino spectrum. In such a case the matrix $Y^\text{eff} $ features as well a hierarchical structure with larger values of the entries associated to the $\mu$ and $\tau$ flavours, which thus achieve larger amounts of asymmetry.

\begin{figure}[htb]
\begin{center}
\includegraphics[width=0.55\textwidth]{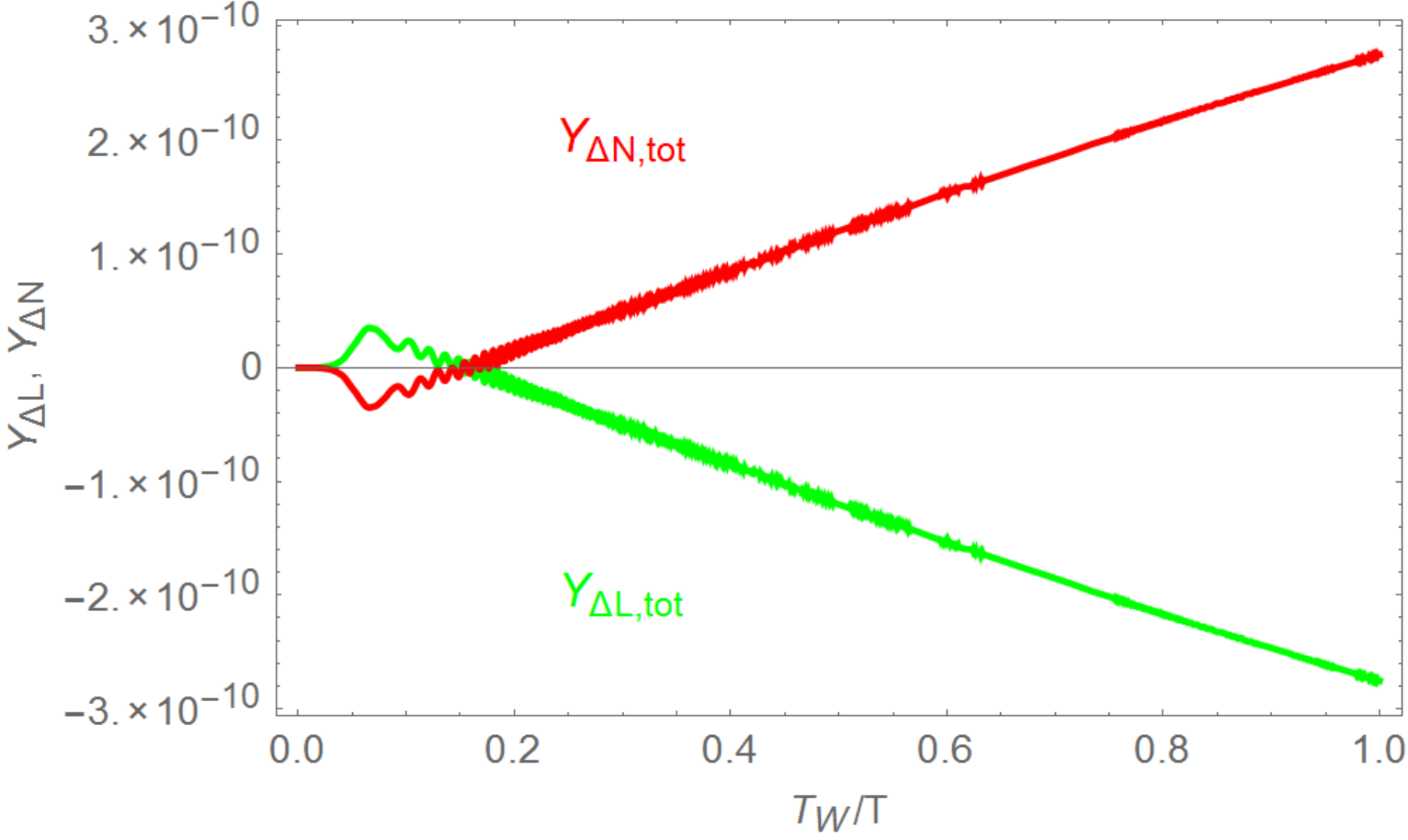}
\caption{\footnotesize{Total asymmetries in the sterile (red curve) and active (green curve) sectors as function of the temperature, for the first benchmark model (cf.\ Fig.~\ref{fig:bench_natural}). As expected from the conservation of the total lepton number, these are equal and opposite.}}
\label{fig:numerical_cross_check}
\end{center}
\end{figure}

\section{Discussion of the weak washout regime}\label{Sec:Results.discussion}

In this section we discuss the key results obtained from the dedicated parameter scan in the weak washout regime described in Section~\ref{parameter_scan}.

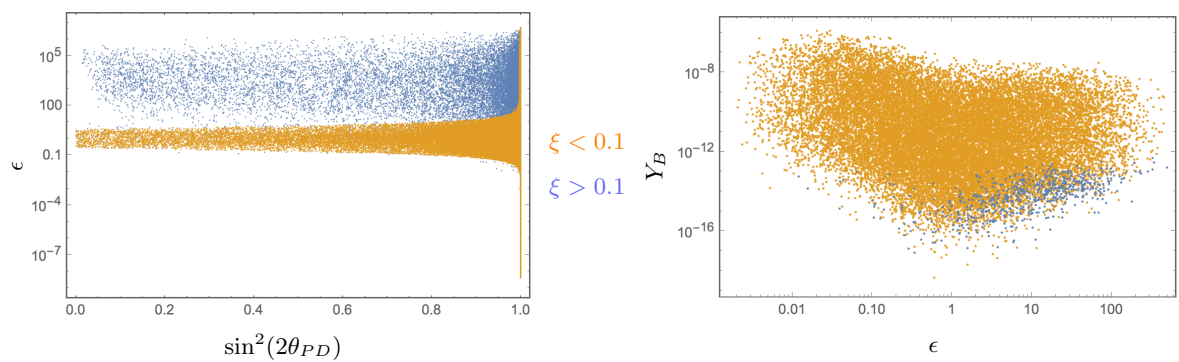
\begin{figure}[htb]
\begin{center}
\begin{tikzpicture}
 \node at (-4.5,0) {\includegraphics[width=0.45\textwidth]{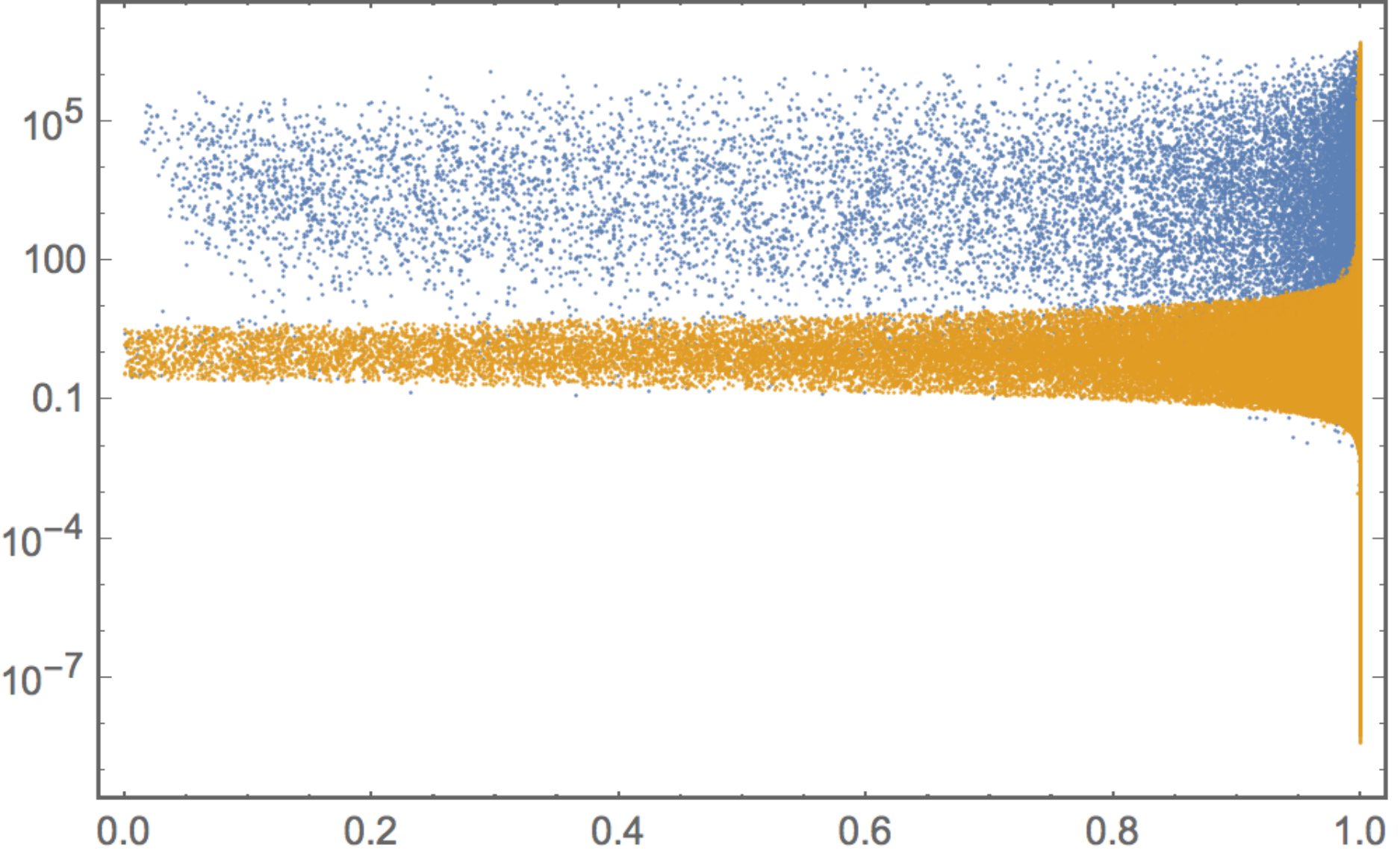}};
\node[rotate = 90] at (-8,0) {\footnotesize{$\epsilon$}};
\node at (-4.5, -2.4) {\footnotesize {$\sin^2(2 \theta_{PD}) $}};
\node at (-0.5, 0.3) {\textcolor{orange}{\footnotesize \bf $\xi < 0.1$}};
\node at (-0.5, -0.3) {\textcolor{mahtblue}{\footnotesize \bf $\xi > 0.1$}};
 \node at (4,0) {\includegraphics[width=0.45\textwidth]{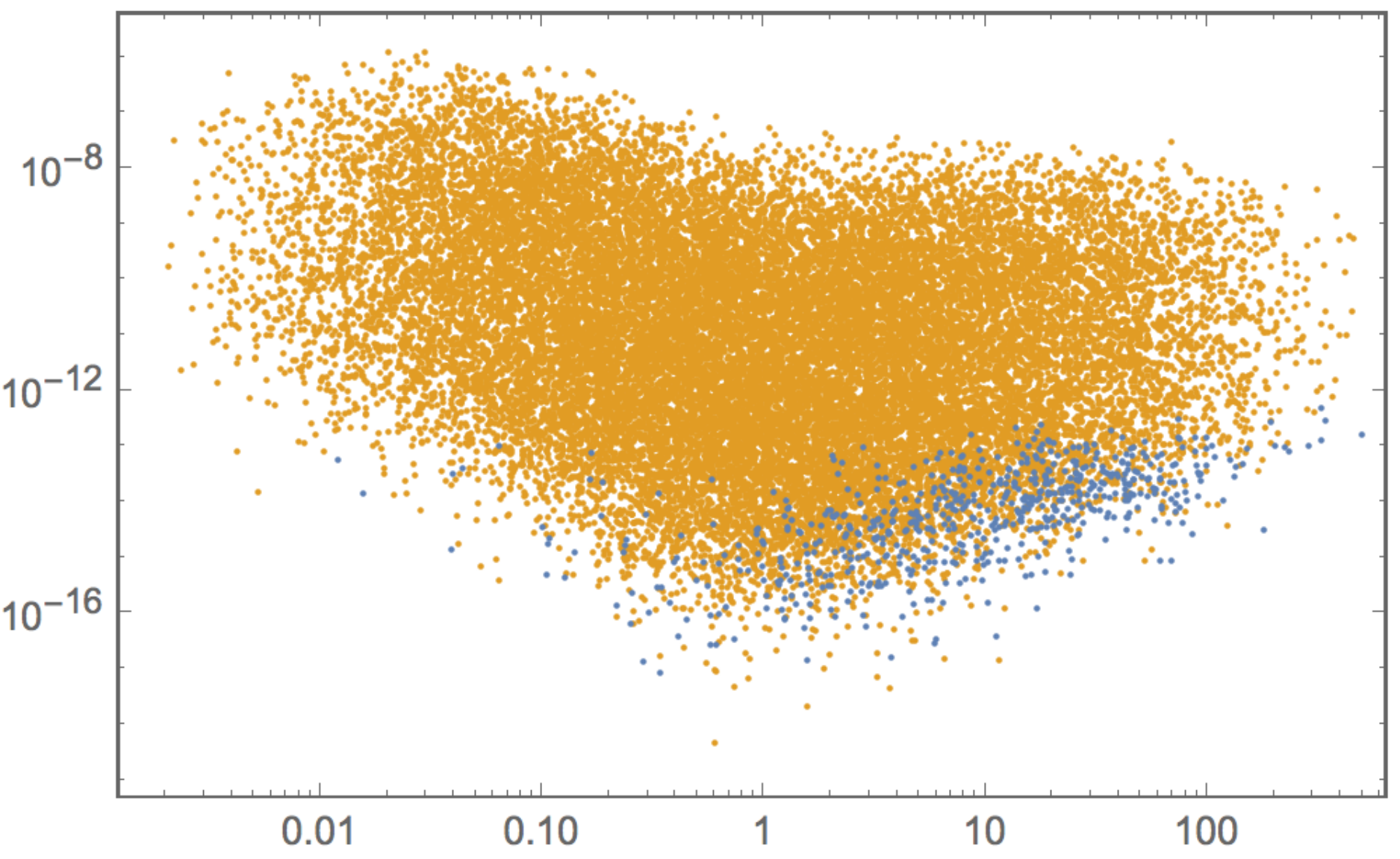}};
\node[rotate = 90] at (0.4,0) {\footnotesize{$Y_B$}};
\node at (4, -2.4) {\footnotesize{ $\epsilon$}};
\end{tikzpicture}
\caption{\footnotesize{Viable parameter points for $\xi > 0.1$ (in blue) and $\xi < 0.1$ (in orange).   Left Panel:  In the plane of the lepton number violating parameter $\epsilon$ and the mixing angle $\theta_{PD}$ between the two heavy mass eigenstates. Right panel:  In the plane of $\epsilon$ and the generated baryon asymmetry $Y_B$, after imposing the out-of-equilibrium condition $|Y^\text{eff}_{\alpha j}| < \sqrt{2} \times 10^{-7}$ but before imposing the constraint on the baryon asymmetry.}}
\label{fig:befYB}
\end{center}
\end{figure}

In Fig.~\ref{fig:befYB} we depict some instructive properties  of the solutions found in the scan over the parameter space before imposing the constraint on the baryon asymmetry. In the left panel, we show these solutions in the plane $(\epsilon, \sin^2(2 \theta_{PD}))$. Here $\theta_{PD}$ is the mixing angle between the two heavy mass eigenstates resulting from the potential in Eq.~\eqref{eq_VN}. Recall that a large mixing angle enhances the oscillations among the heavy states and hence the produced baryon asymmetry. As discussed in Section~\ref{sec_toymodel}, for small values of the lepton number violating parameter in the Majorana mass term, $\xi < 0.1$ (yellow points), the resulting distribution is approximately symmetric under the transformation $\epsilon \rightarrow 1/\epsilon$, corresponding to switching the lepton number assignments of the two additional states. In fact this will be the region we will focus on in the following, since large values of $\xi$ (blue points, $\xi > 0.1$) imply a mass splitting between the two heavy states too large to accomplish a successful leptogenesis, cf.\ Eq.~\eqref{eq_mueps}. Moreover, among the points with $\xi < 0.1$, we can distinguish two types of solutions. For $\epsilon < 0.1$ or $1/\epsilon < 0.1$ the mixing between the two heavy states is found to be close to maximal, $\sin^2(2 \theta_{PD}) \simeq 1$. This corresponds to the solutions found in the perturbative expansion of the toy model discussed in Section~\ref{sec_toymodel}, dubbed ``perturbative'' solutions. On the other hand, for $0.1 < \epsilon < 10$ any value of the mixing angle $\theta_{PD}$ can be obtained. This is what we referred to as ``generic'' solutions in Section~\ref{sec_toymodel}.

The right panel of Fig.~\ref{fig:befYB} shows the dependence of the resulting baryon asymmetry on the two lepton number violating parameters $\epsilon$ and $\xi$, after imposing the out-of-equilibrium condition $|Y^\text{eff}| < \sqrt{2} \times 10^{-7}$.  As anticipated from the previous figures, values of $\epsilon$ much larger or much smaller than one lead to a large mixing of the heavy states, rendering leptogenesis through the oscillations of these states very effective. Moreover, large values of $\xi$ imply a large mass splitting between the heavy states, rendering leptogenesis less effective (blue points versus yellow points). In fact above the EW phase transition, in the regime relevant for leptogenesis, the correlation between the mass splitting $\Delta m^2$ and $\xi$ is very well described by Eq.~\eqref{eq_mueps} in the regime $\xi \leq 1$.

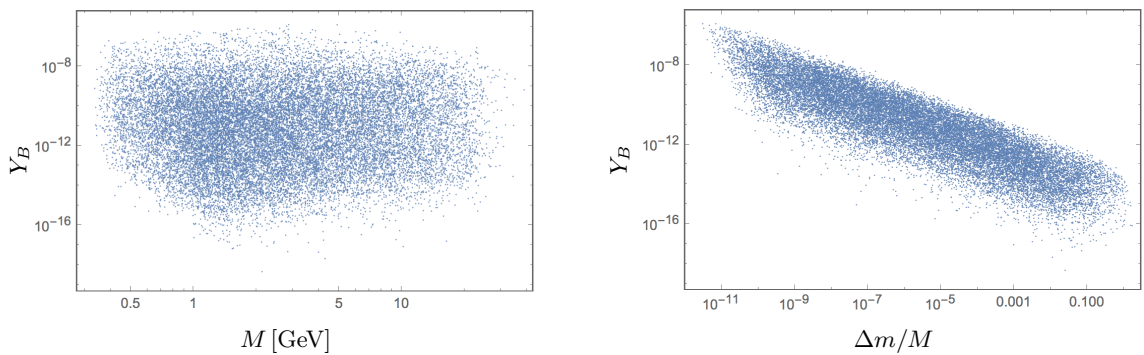
\begin{figure}[htb]
\begin{center}
\begin{tikzpicture}
 \node at (-4,0) {\includegraphics[width=0.45\textwidth]{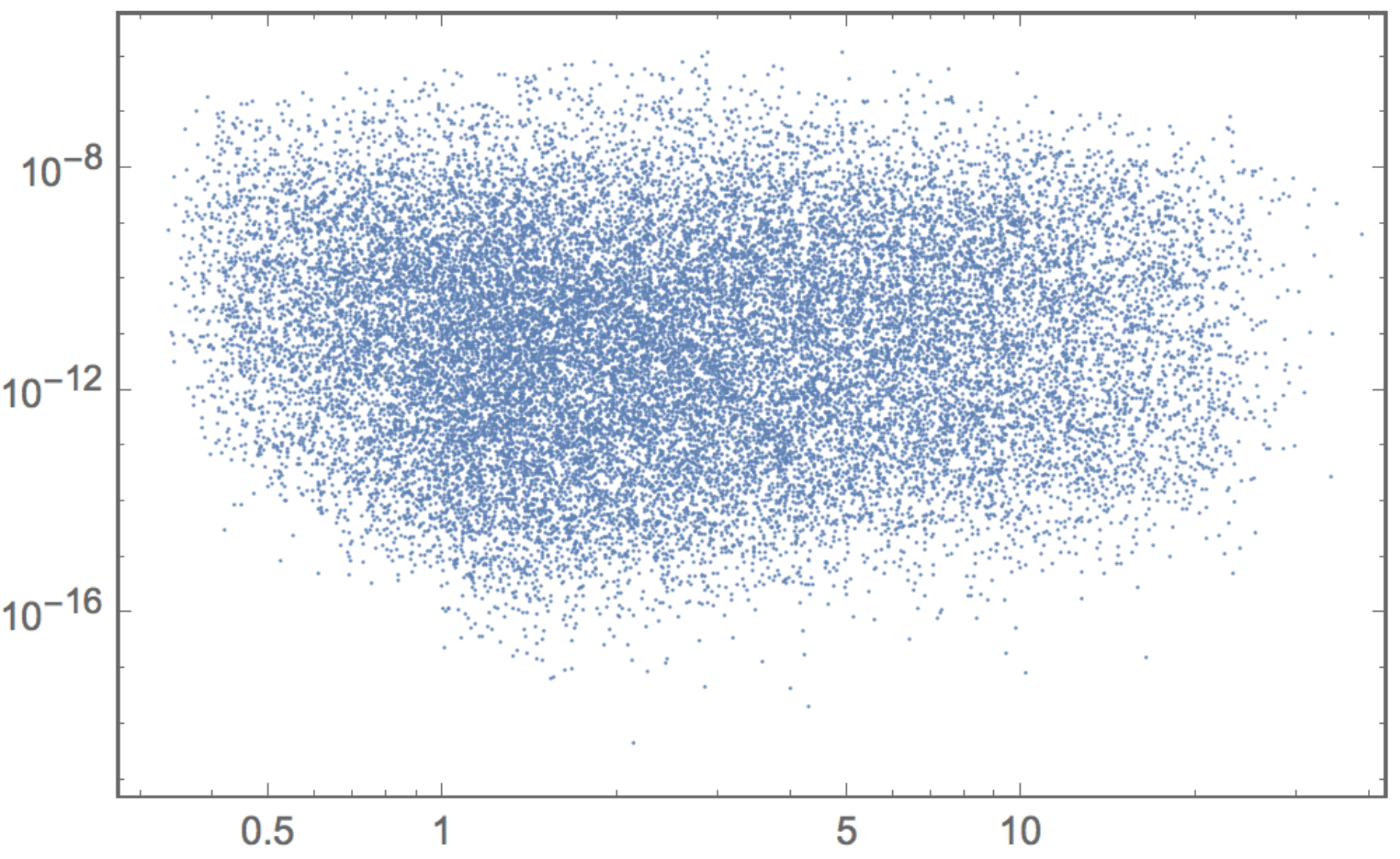}};
\node[rotate = 90] at (-7.5,0) {\footnotesize{$Y_B$}};
\node at (-4, -2.4) {\footnotesize{$M \, [ \text{GeV} ]$}};
 \node at (4,0) {\includegraphics[width=0.45\textwidth]{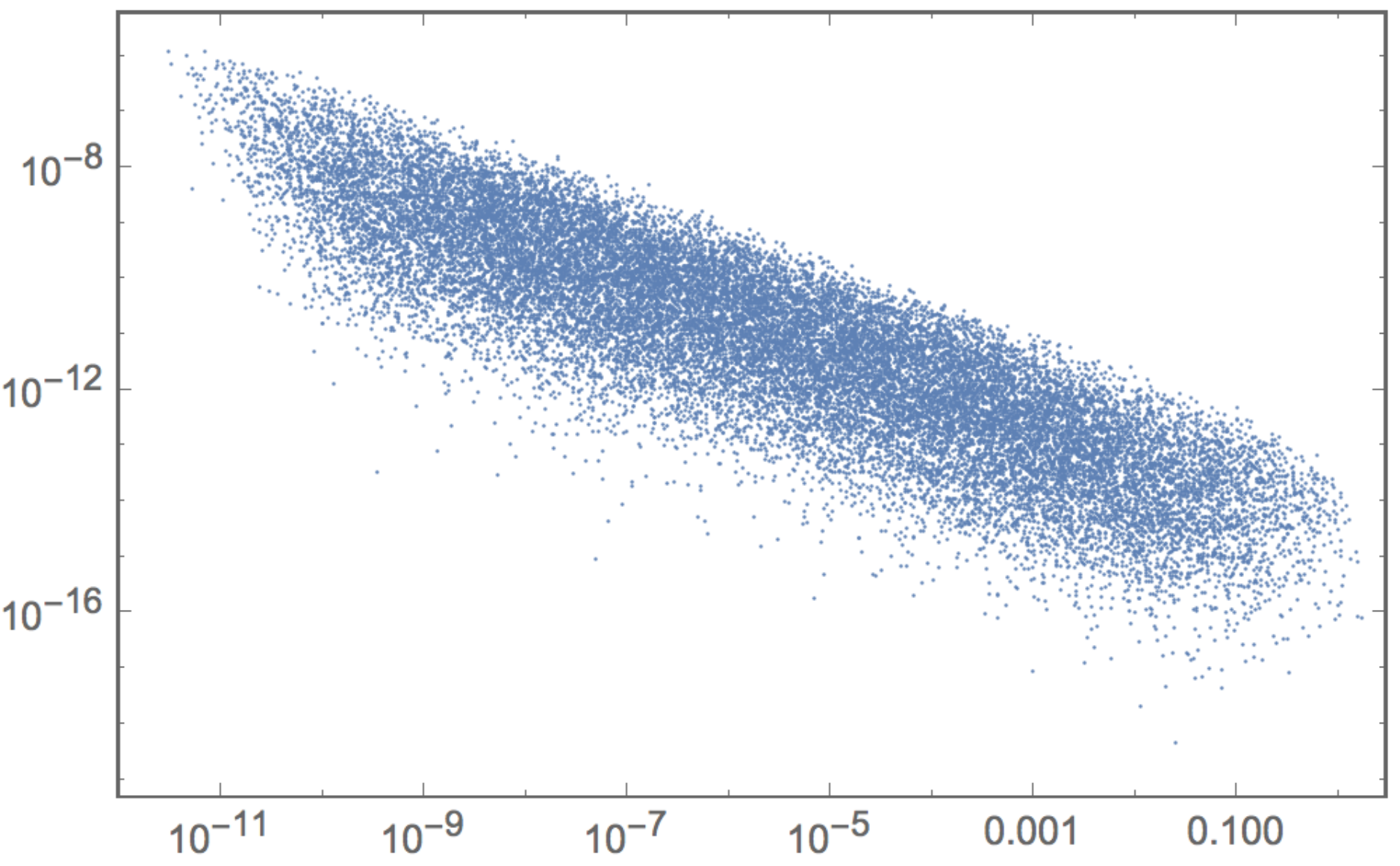}};
\node[rotate = 90] at (0.4,0) {\footnotesize{$Y_B$}};
\node at (4, -2.4) {\footnotesize{$\Delta m/M$}};
\end{tikzpicture}
\caption{\footnotesize{Viable parameter points  after imposing the out-of-equilibrium condition $|Y^\text{eff}_{\alpha j}| < \sqrt{2} \times 10^{-7}$ but before imposing the constraint on the baryon asymmetry. We show the baryon asymmetry $Y_B$ as function of the scale of the heavy neutrinos $M=(M_1+M_2)/2$ (left panel) and of the relative mass splitting $\Delta m/M$ between the two heavy states (right panel).}}
\label{fig:MdM}
\end{center}
\end{figure}

In Fig.~\ref{fig:MdM} our results have been re-expressed as function of the dimensionful parameter $M=(M_1+M_2)/2$, i.e. the mass scale of the heavy neutrinos, 
and of the relative mass splitting between the two heavy states, $\Delta m/M$, to give an impression of the viable parameter space. 
Here we show only solutions which obey the out-off equilibrium condition, $|Y^\text{eff}| < \sqrt{2} \times 10^{-7}$. We find solutions within the assumed viable range of values of the baryon asymmetry, i.e. $ 3 \times 10^{-11} < Y_B < 2.5 \times 10^{-10} $, for the heavy neutrino mass scale, $0.3~\text{GeV} \lesssim M \lesssim 35~\text{GeV}$ and relative mass splitting within  $10^{-11} \lesssim \Delta m/M \lesssim 10^{-3}$,  
with a lower bound on the mass splitting
  $\Delta m \gtrsim 10^{-2} \, \mbox{eV}$. \footnote{This lower limit on the mass splitting is not actually originated by the requirement of viable leptogenesis but has been imposed, as an additional constraint, in the parameter scan. It follows from the requirement $T_L > T_W$, where $T_L$ defined in Eq.~(\ref{eq:TL}) is the temperature at which the production of the lepton asymmetry is peaked. Although lower mass splittings are not excluded for a viable  leptogenesis~\cite{Canetti:2012kh}, since thermal effects can modify the value of the mass splitting inferred by the diagonalization of the mass matrix, we have imposed this lower bound in order to compare the numerical and the analytical determinations of the baryon abundance. As clarified in Appendix~\ref{app_leptogenesis}, the latter relies on the assumption that the temperature dependent mass splitting originated during oscillations is subdominant with respect to the one sourced by $\xi$.} 
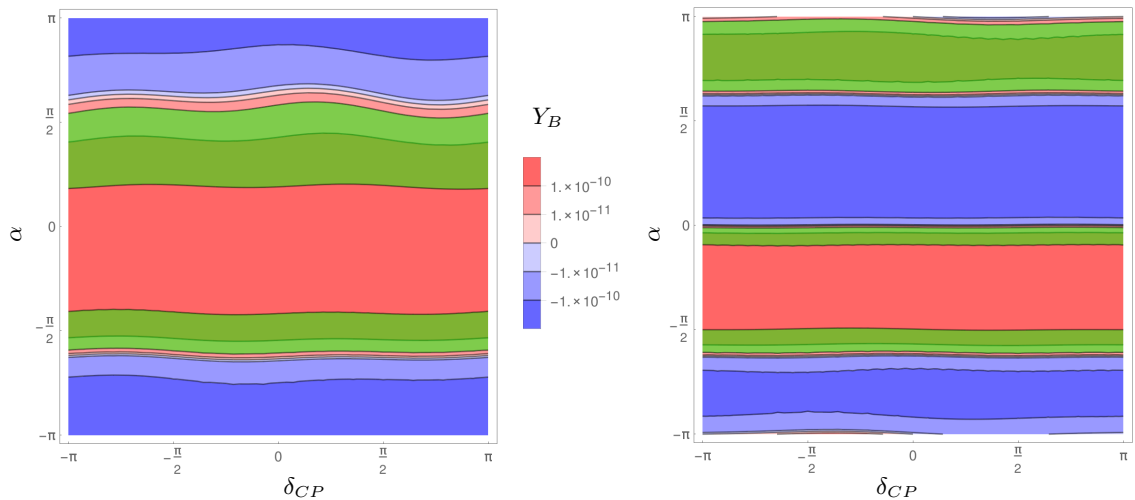
\begin{figure}[htb]
\begin{center}
\subfloat{
\begin{tikzpicture}
 \node at (0,0) {\includegraphics[width=0.55\textwidth]{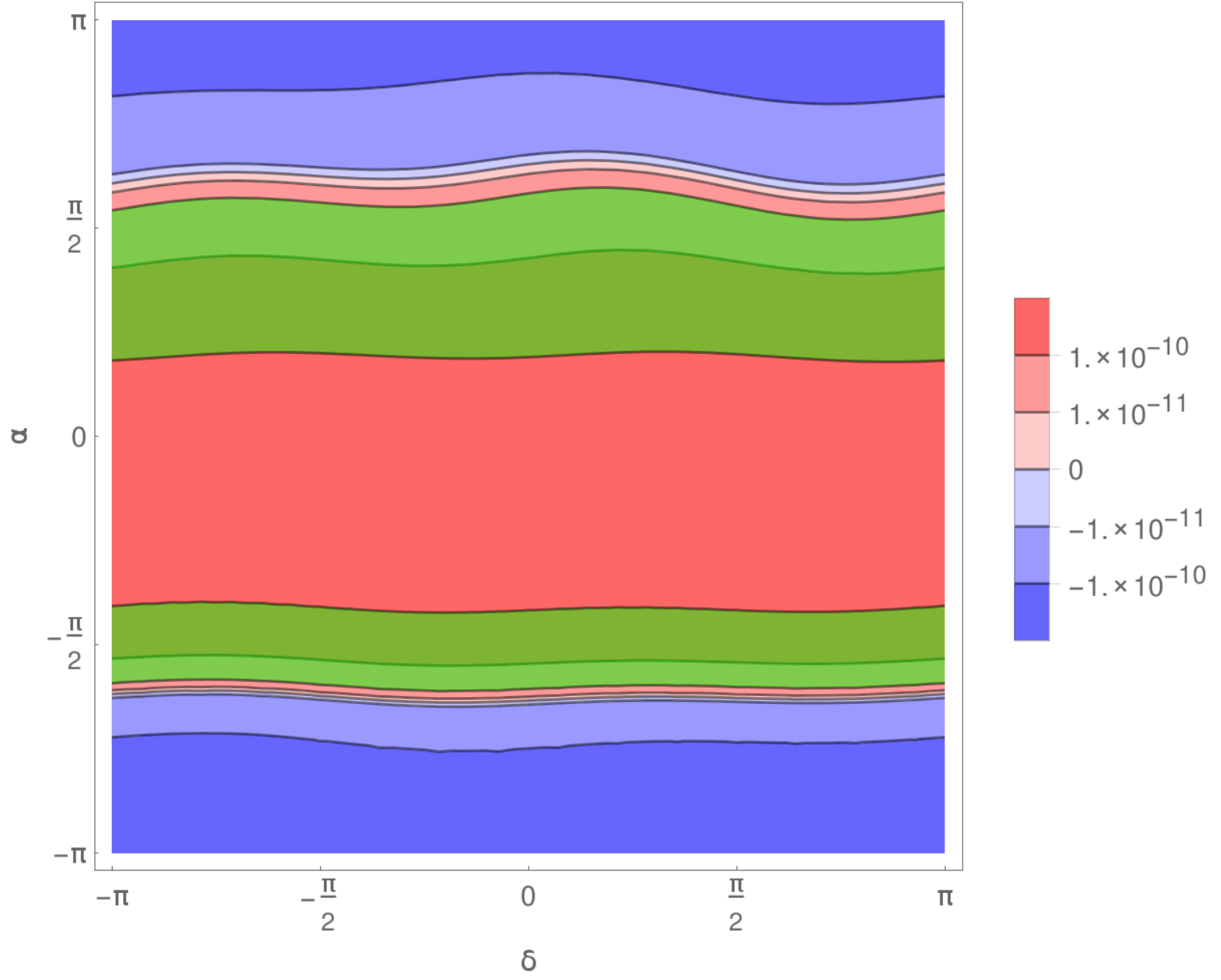}};
 \node[rotate = 90, fill = white] at (-4,0.25) {\footnotesize $\alpha$};
\node[fill = white] at (-0.2,-3.1) {\footnotesize $\delta_{CP}$};
\node at (3.0,1.8) {\footnotesize $Y_B$};
 \node at (7.5,0) {\includegraphics[width=0.435\textwidth]{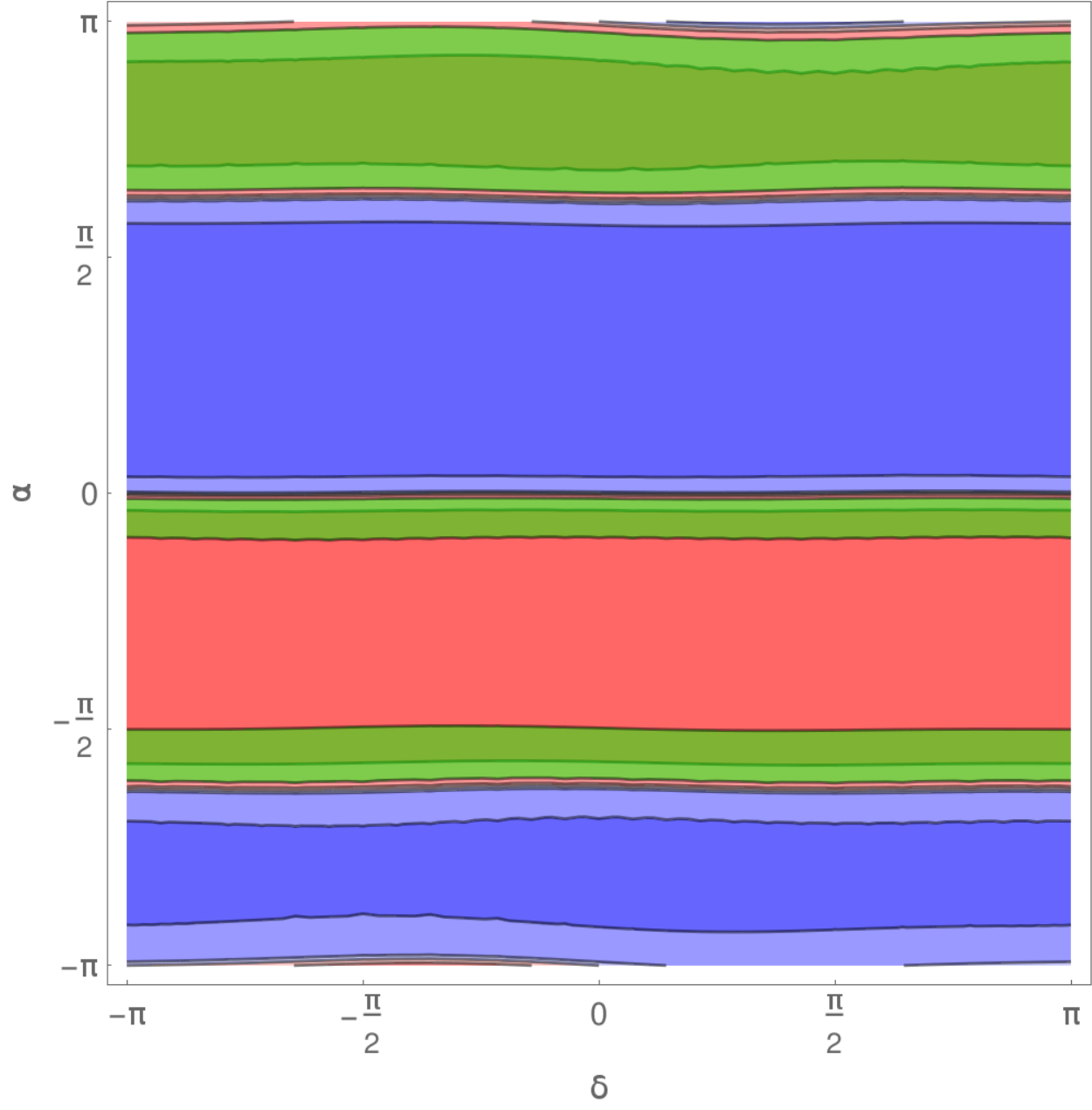}};
\node[rotate = 90, fill = white] at (4.4,0.25) {\footnotesize $\alpha$};
\node[fill = white] at (7.6,-3.1) {\footnotesize $\delta_{CP}$};
\end{tikzpicture}
} 
\caption{Contour plot of the baryon asymmetry $Y_B$ in terms of the Dirac phase $\delta_{CP}$ and the Majorana phase $\alpha$, for a fixed $(|Y_{\alpha i}|, \epsilon, \xi, |\Lambda|)$ parameter point, with a fixed non-zero phase assigned to $\Lambda$ (Arg($\Lambda) = 0.44 \,  \pi$, left panel) and a real value for $\Lambda$ (right panel). Negative (positive) asymmetry is marked in blue (red), overlayed by the green region marking a baryon asymmetry in agreement with observation, $ 3 \times 10^{-11} < Y_B < 2.5 \times 10^{-10} $.}
\label{fig:phases}
\end{center}
\end{figure}

In Fig.~\ref{fig:phases} we depict the impact of the Dirac phase $\delta_{CP}$ and the Majorana phase $\alpha$ (defined in Section~\ref{sec:parametrisation:mass}) on the determination of the baryon abundance $Y_B$.  To this purpose we have considered a fixed choice of the model parameters, namely  $\xi \simeq 6.7 \times 10^{-7}$, $\epsilon \simeq 0.075$, and $M_1 \simeq M_2 \simeq 4.4~\text{GeV}$, yielding  values for $Y_B$ in the allowed range. 
As can be seen, there is a very strong dependence on the Majorana phase while, on the contrary, the effect of the Dirac phase is negligible.
For values of $\epsilon$ close to one, the overall asymmetry is reduced (cf.\ Fig.~\ref{fig:befYB}), rendering the effect of the Dirac phase for determining the allowed regions more important. 
The different impact of the two phases $\delta_\text{CP}$ and $\alpha$ can be understood from the parametrisation of the PMNS matrix: the $\delta_{CP}$ phase multiplies the relatively small parameter $\sin \theta_{13} \simeq 0.15$ and a change of its value barely affects the amount of CP-violation encoded in the Yukawa couplings in Eq.~(\ref{eq:yukawa_parameters}); on the contrary the Majorana phase $\alpha$ multiplies an entire column of the PMNS matrix, thus a change in its value strongly affects the imaginary part of the Yukawa matrix. 
This qualitative picture  holds throughout the parameter space, the position of the allowed bands however varies significantly, since the third ``high-energy'' phase related to the parameter $\Lambda$ affects the value of $\delta_\alpha$ in Eq.~(\ref{eq:deltaCP}), rendering all values of the Majorana phase $\alpha$ possible when considering the entire parameter space. This third phase is also responsible for the non-zero value of the asymmetry even if $\delta_{CP}$ and $\alpha$ are zero, cf.\ left panel of Fig.~\ref{fig:phases}. { Vice versa, the correct value for the baryon asymmetry can also be obtained if this high-energy phase is zero, i.e.\ only through the phases of the PMNS matrix, as depicted in the right panel of Fig.~\ref{fig:phases}. }

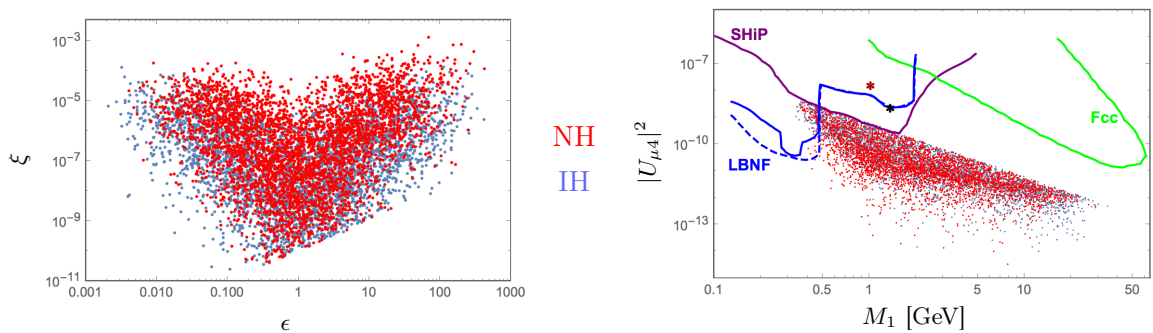
\begin{figure}[htb]
\begin{center}
\subfloat{
\begin{tikzpicture}
\node at (0,0) { \includegraphics[width=0.43\textwidth]{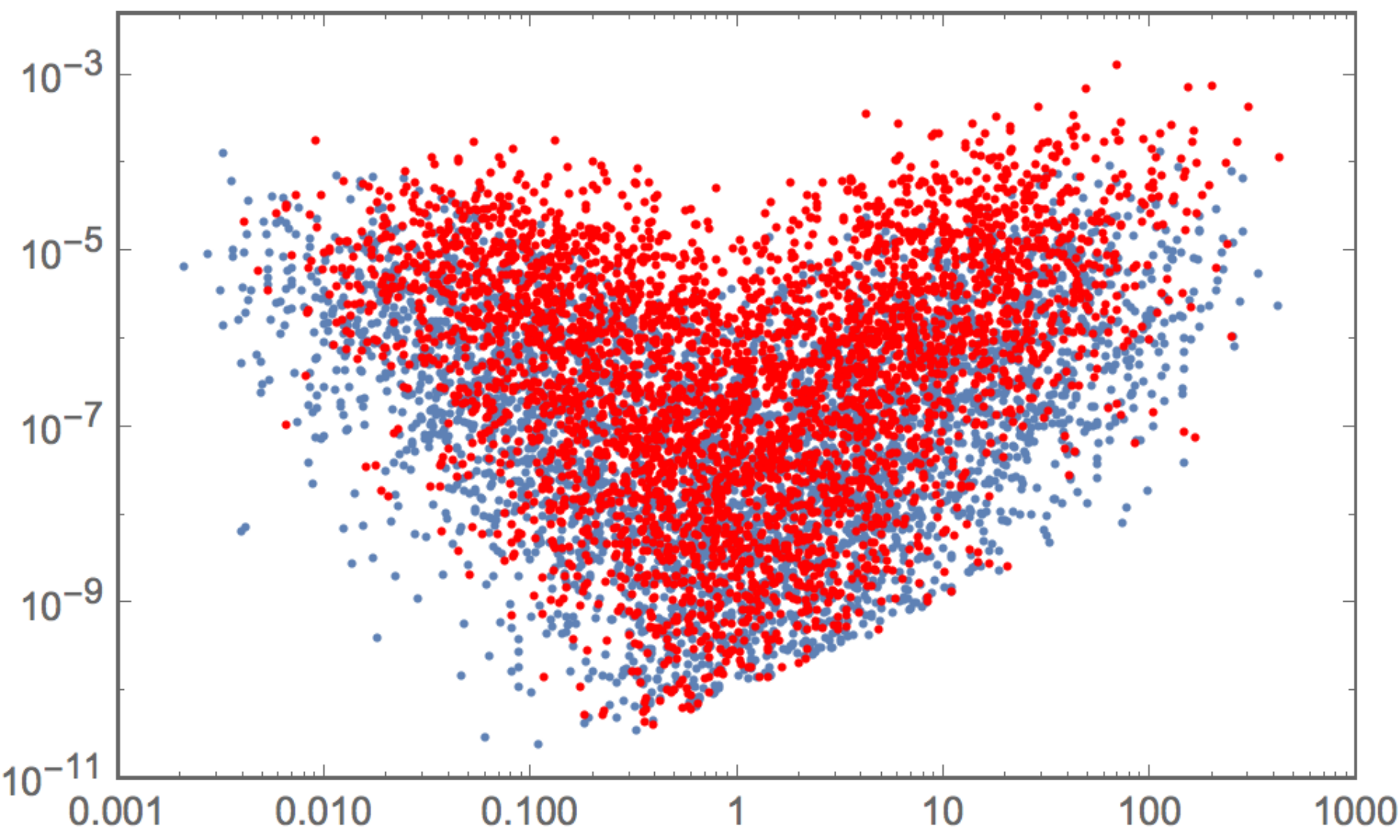}};
\node[rotate = 90] at (-3.5,0) {\footnotesize $\xi$};
\node at (0,-2.2) {\footnotesize $\epsilon$};
\node at (3.8, 0.3) {\textcolor{red}{NH}};
\node at (3.8, -0.3) {\textcolor{mahtblue}{IH}};
\end{tikzpicture}
} 
\subfloat{
\begin{tikzpicture}
\node at (0,0) { \includegraphics[width=0.43\textwidth]{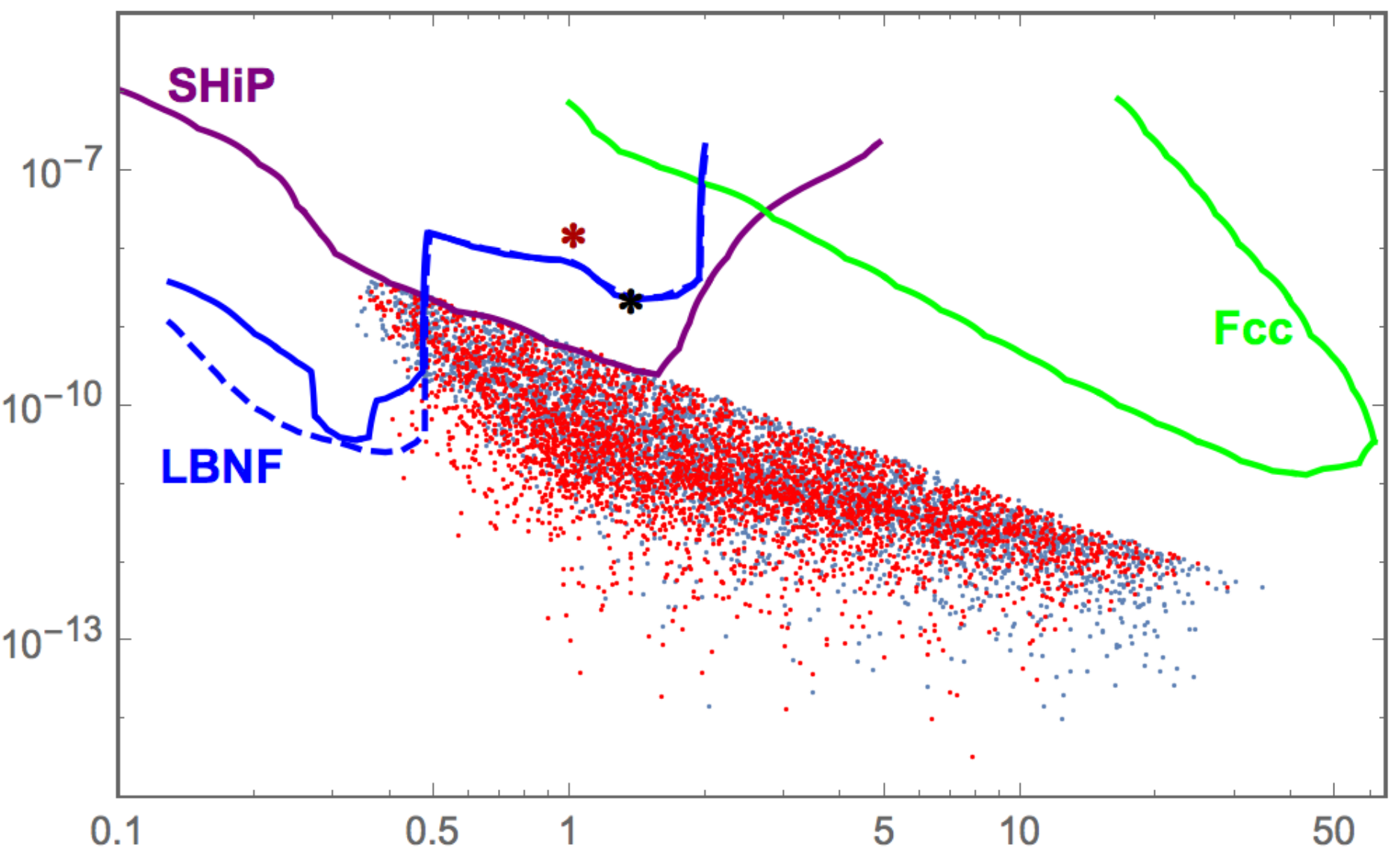}};
\node[rotate = 90] at (-3.5,0) {\footnotesize $|U_{\mu 4}|^2$};
\node at (0,-2.2) {\footnotesize $M_1$ [GeV]};
\end{tikzpicture}
}
\caption{\footnotesize{Left panel: Set of model points giving a viable baryon abundance in the weak washout regime, in the plane $(\epsilon, \xi)$. The red and blue points refer, respectively, to solutions with normal and inverted hierarchy for the active neutrino mass spectrum. Right panel: Models featuring a viable baryon abundance in the plane $\left(|U_{\rm \mu 4},M_1|\right)$ where $U_{\rm \mu 4}$ is the mixing between the lightest of the two exotic neutrinos with the $\mu$ flavour while $M_1$ is its mass. The color code is the same as in the left panel. The points are the result of a scan over the parameter space of weak washout regime. The asterisks refer to the benchmark solutions in the strong washout regime, characterised by the Yukawa couplings in Eqs.~(\ref{eq:benchSW1}) (red) and (\ref{eq:benchSW2}) (black).}}
\label{fig:aftYB}
\end{center}
\end{figure}

The results of our analysis are finally summarised in Fig.~\ref{fig:aftYB}, where the distribution of the parameter points featuring a viable baryon asymmetry is shown.
In the left panel of this figure we display the distribution of the viable parameter points in the plane of the lepton number violating parameters $\epsilon$ and $\xi$. The shape of this region can be well understood in terms of the toy model presented in Section~\ref{sec_toymodel}. Again we note the  approximate symmetry under $\epsilon \rightarrow 1/\epsilon$.
This figure demonstrates that  the parameter $\xi$ appearing in the Majorana mass term must be very small, in order to ensure a sufficiently small mass splitting between the two heavy neutrinos. Indeed all viable points are found to be within the range $\xi < 10^{-3}$. On the other hand, we find viable solutions for a large range of values of $\epsilon$, and moreover $\xi \lesssim 5\times  10^{-2}\ \epsilon$ in the entire parameter range.
This follows from the fact that the value of $\epsilon$ is inversely proportional to the size of the Yukawa couplings, cf. Eq~(\ref{eq:epsilon_neutrino}), and the requirement $\left| Y^\text{eff}\right| < 10^{-7}$ translates on the bound $10^{-3} \lesssim \epsilon \lesssim 10^3$. On the other hand the requirement of a sufficiently small mass splitting puts an upper bound on the possible values of $\xi$, resulting in the aforementioned bound on the ratio of the two parameters. Lower values of $\epsilon$, and consequently larger $\xi/\epsilon$ ratios, are nonetheless allowed for values of $Y^\text{eff} > 10^{-7}$, cf.\ Section~\ref{Sec:strong washout}. 

In the right panel of Fig.~\ref{fig:aftYB}, we show the mixing between the active and the sterile sector, parametrised by the mixing matrix element $|U_{\mu 4}|$, as a function  of $M_1$, the mass of the lighter of the two heavy states. The active-sterile mixing is a particularly interesting quantity, since it is in principle experimentally accessible through experiments such as SHiP~\cite{Anelli:2015pba,Alekhin:2015byh}, FCC-ee~\cite{Blondel:2014bra} and LBNF/DUNE~\cite{Adams:2013qkq}. Unfortunately, the viable parameter points for solutions in the weak washout regime are found to be below the expected sensitivity of these experiments, with the exception of a very small region of particularly light sterile states, $M_1 \lesssim 500$~MeV, which can be reached by LBNF/DUNE. We remark however that our study has been limited, up to now, to a subset of the parameter space, due to the limitation of the analytical expression of Eq.~(\ref{eq:baryo_analytical}). In the next subsection we will extend (at least partially) our analysis to regions characterised by higher values of $Y^\text{eff}$ and, consequently, higher values of the mixing between the heavy and the active neutrinos, which can be possibly probed in future facilities. We anticipate in Fig.~\ref{fig:aftYB} two solutions associated to  a viable leptogenesis in the strong washout scenario, whose active-sterile mixing is represented by the two asterisks. It is evident that these model realisations can be probed by both SHiP and LBNF/DUNE.

\section{Solutions in the strong washout regime}
\label{Sec:strong washout}

\begin{figure}[htb]
\begin{center}
\subfloat{\includegraphics[width=0.42\textwidth]{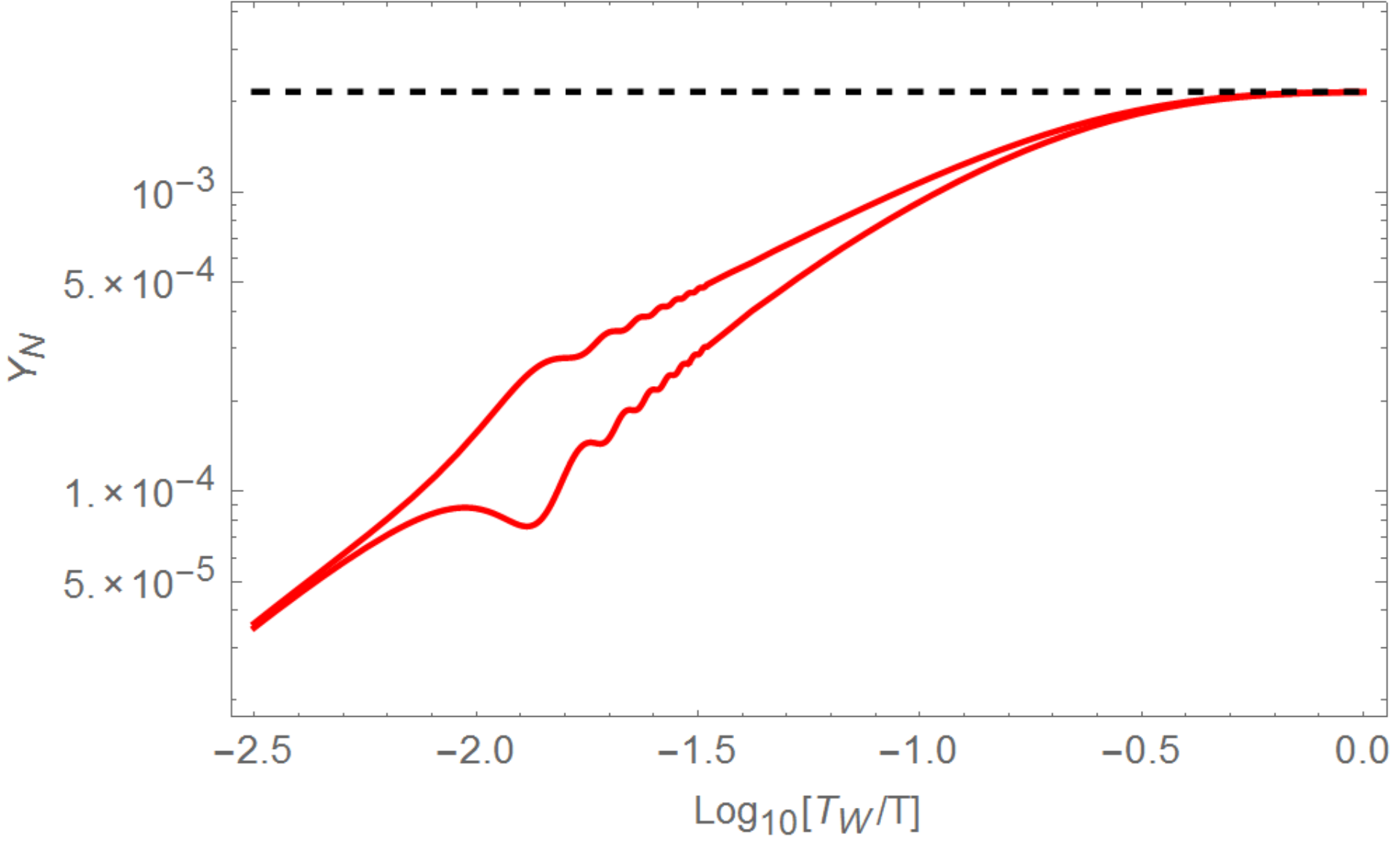}}\hspace{0.04\textwidth}
\subfloat{\includegraphics[width=0.42\textwidth]{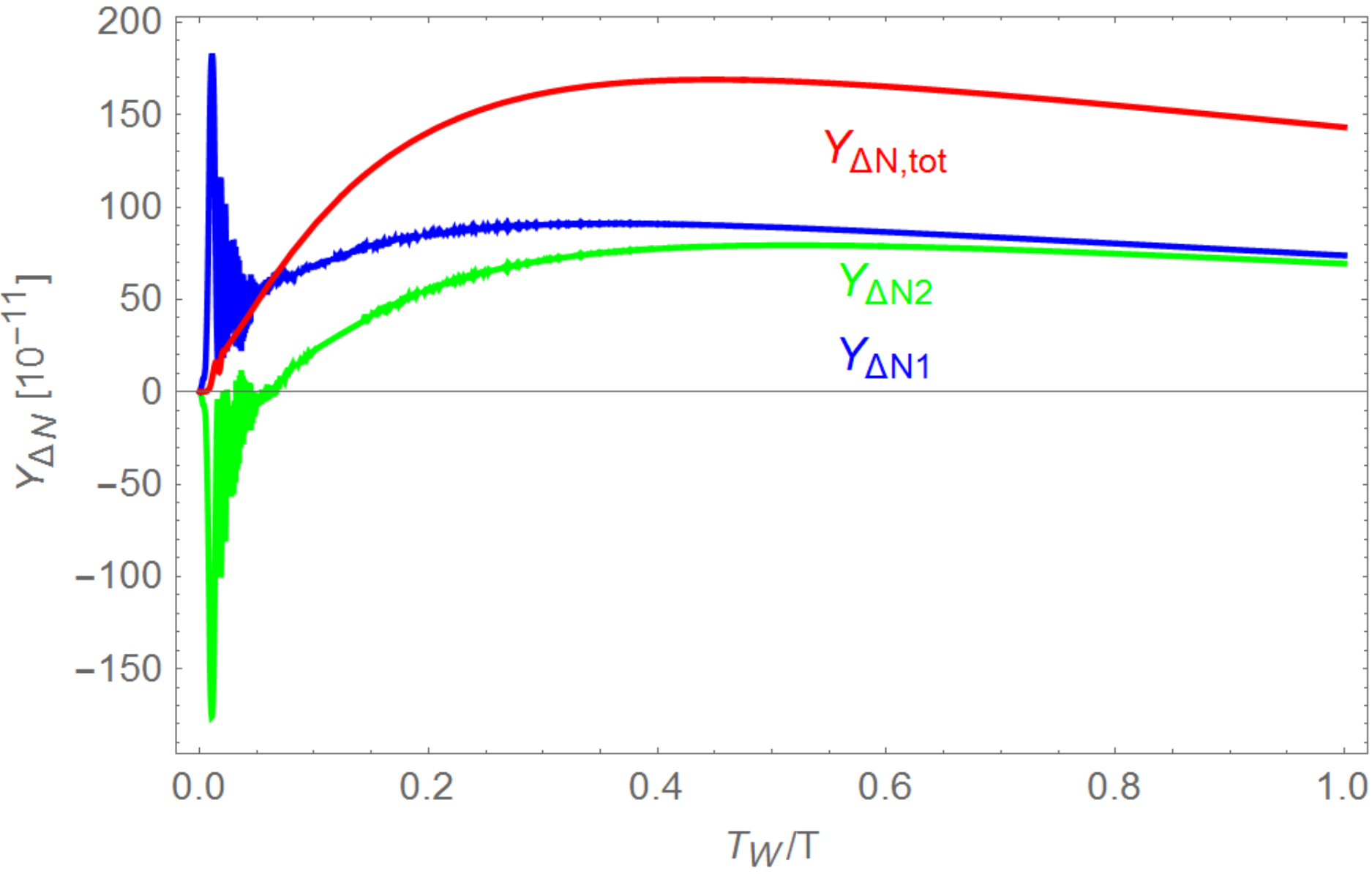}}\\
\subfloat{\includegraphics[width=0.45\textwidth]{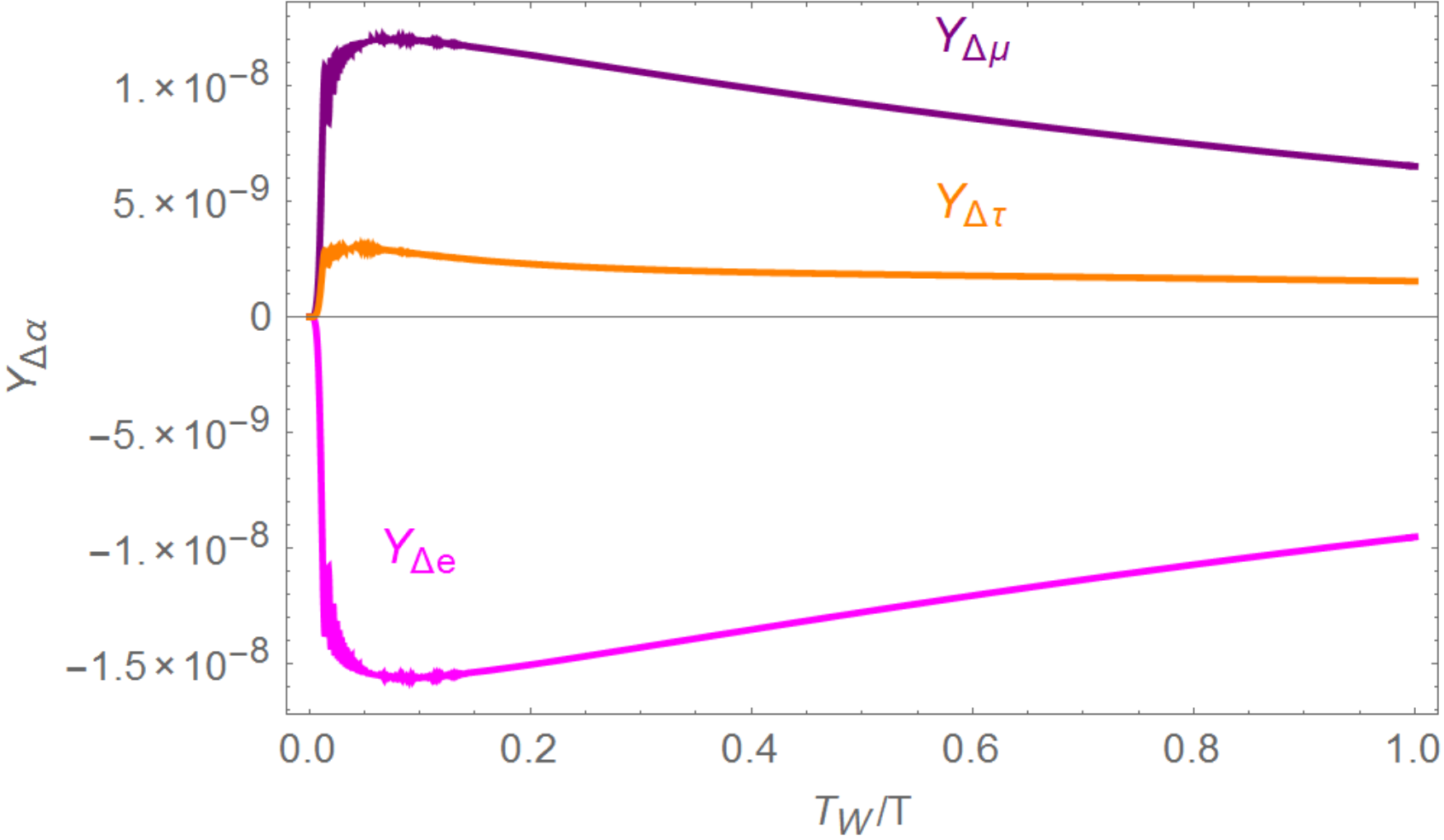}}\hspace{0.015\textwidth}
\subfloat{\includegraphics[width=0.45\textwidth]{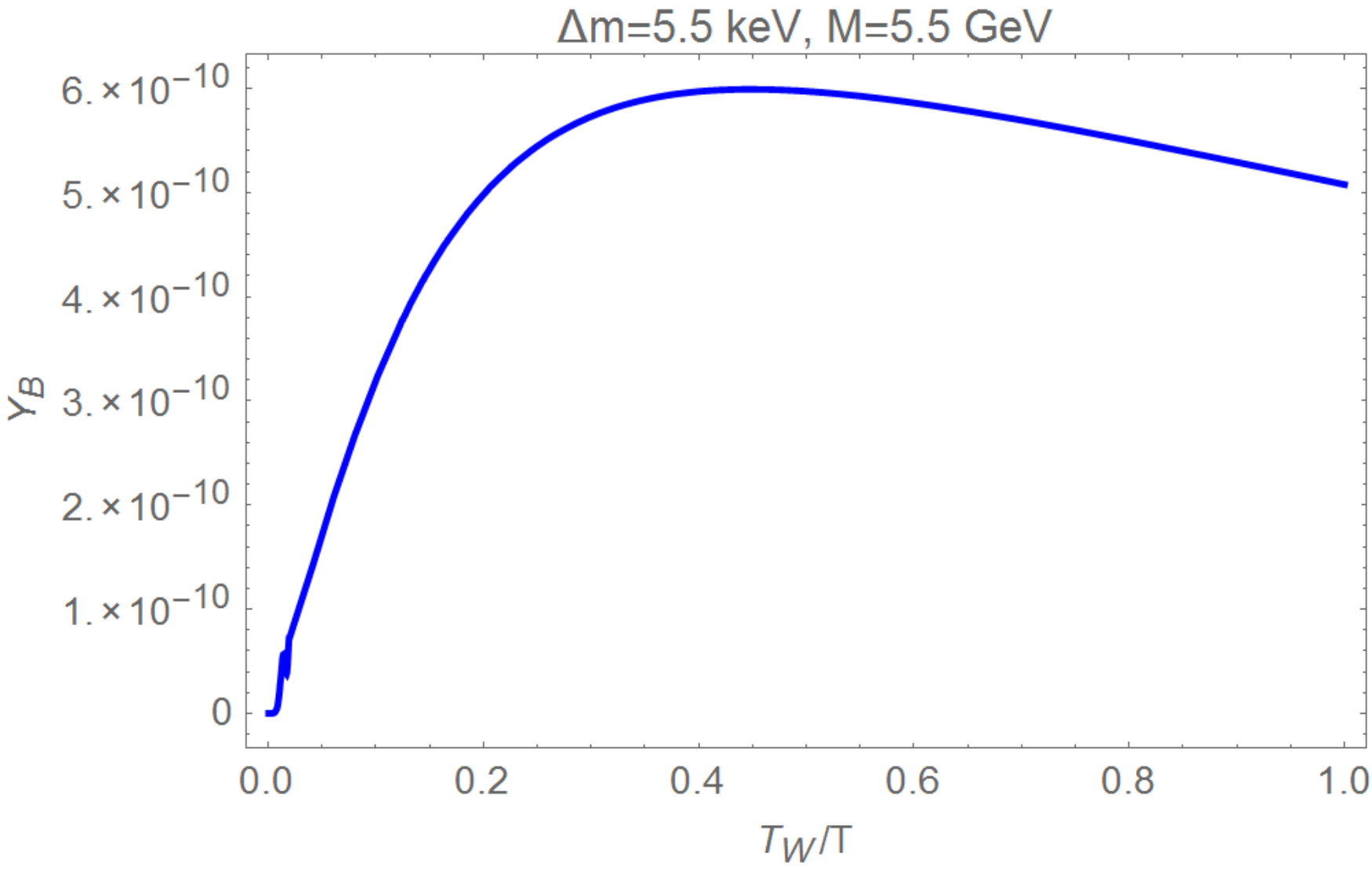}}
\caption{\footnotesize{Evolution of the baryon asymmetry (bottom right panel) as well as the individual lepton asymmetries in the active (bottom left panel) and sterile (upper right panel) sectors. Upper left panel: evolution of the abundances of the heavy neutrinos. In the model chosen the entries of the matrix $Y^\text{eff}$ exceed the equilibration value by ${\cal{O}}(1)$ amounts. The asymmetries are depleted at late times but a sizeable residual baryon asymmetry of the order of the observed value is nonetheless present. The values of the mass splitting of the heavy sterile neutrinos $\Delta m$ and of their mass scale $M=(M_1+M_2)/2$ are reported as well.}}
\label{fig:benchhy1}
\end{center}
\end{figure}

In this section we will investigate the possibility of achieving a successful leptogenesis in the case where the matrix $Y^\text{eff}$ has entries above the equilibration value $\sqrt{2} \times 10^{-7}$. In this situation the analytical solution of Eq.~(\ref{eq:baryo_analytical}) is not valid since we can no longer neglect the depletion of the baryon abundance when the heavy sterile neutrinos are in thermal equilibrium. At the same time higher entries of the $Y^\text{eff}$ correspond to a more efficient production of the sterile neutrinos which translates into an accordingly more efficient generation of a lepton asymmetry. The correct amount of the baryon abundance  might hence be in principle obtained even in a strong washout regime, provided that a sufficiently high initial lepton asymmetry is created. 

A full numerical exploration of the parameter space is computationally very demanding. We will thus limit our analysis to some relevant benchmark points which will be used to infer the main trends of the numerical solutions of the Boltzmann equation. We have, to this purpose, traced the evolution of the baryon abundance for three benchmarks characterised by increasing values of the entries of $Y^{\rm eff}$, ranging from $|Y^{\rm eff}_{\alpha i}| \sim \sqrt{2} \times 10^{-7}$ to $|Y^{\rm eff}_{\alpha i }| \approx 3 \times 10^{-6}$ (cf.\ Appendix~\ref{app_benchmarks}). Our results are reported in Figs.~\ref{fig:benchhy1}-\ref{fig:benchhy3}. Each figure reports the  same relevant quantities  as those chosen  in Section~\ref{sec:numerical_ly}. Before discussing the individual benchmarks we notice that all the plots show a very strong overlap between the neutrino states, except in very pronounced resonance regions. As can be inferred by Eq.~(\ref{eq:mass_eigen}), higher Yukawa couplings correspond to lower values of $\epsilon$ (higher values in the flipped regime $\epsilon > 1$). In the strong washout regime we thus expect the heavy neutrinos to typically form pseudo-Dirac pairs.

The first benchmark point, reported in Fig.~\ref{fig:benchhy1} has values of $|Y^\text{eff}_{\alpha i}|$ slightly above the equilibration condition. In this case the sterile neutrinos reach thermal equilibrium only at rather late time. The depletion of the lepton asymmetry is still limited and a value of $Y_B$ above the observed value is obtained, demonstrating the feasibility of leptogenesis in this regime.

\begin{figure}[htb]
\begin{center}
\subfloat{\includegraphics[width=0.42\textwidth]{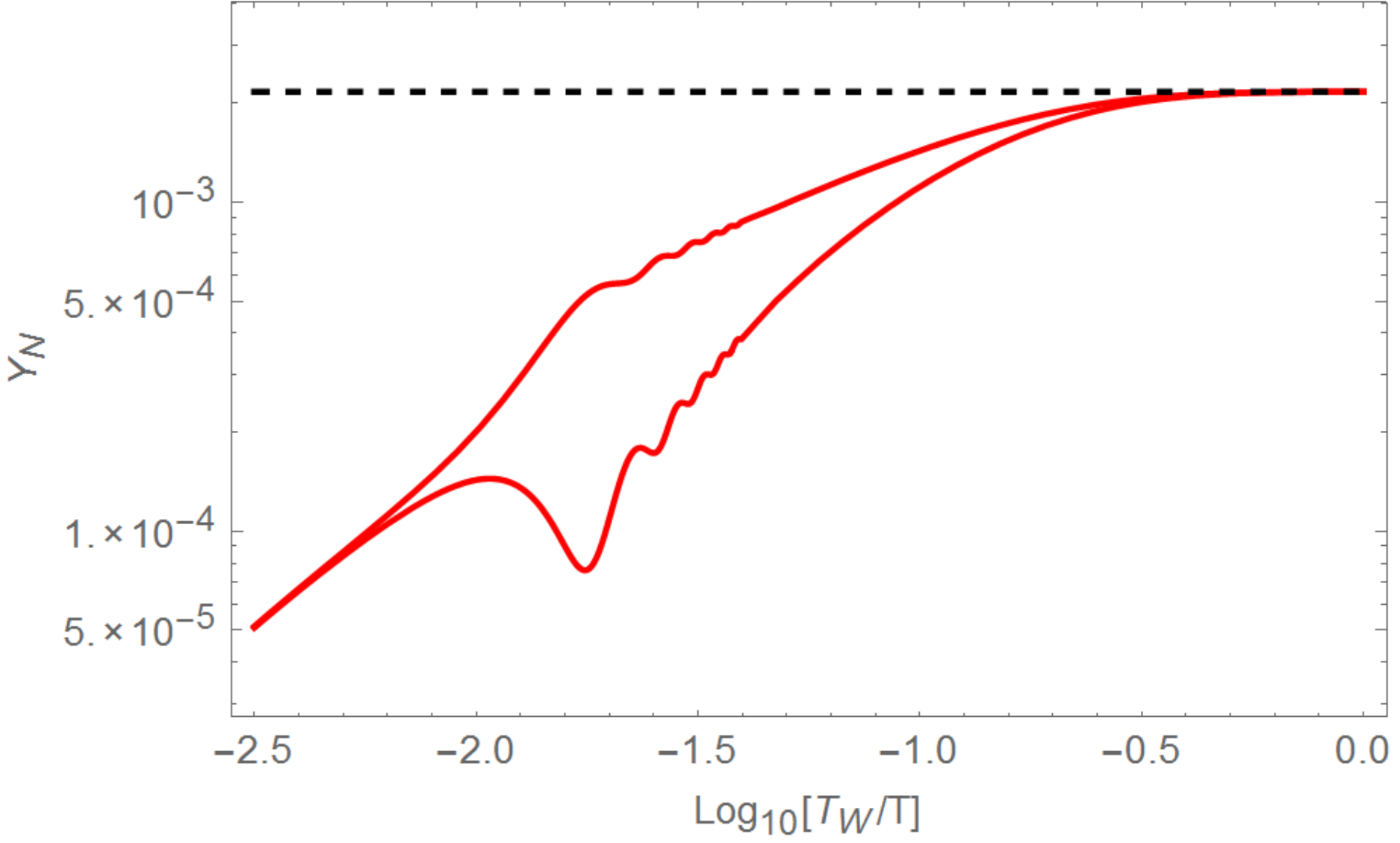}}\hspace{0.04\textwidth}
\subfloat{\includegraphics[width=0.42\textwidth]{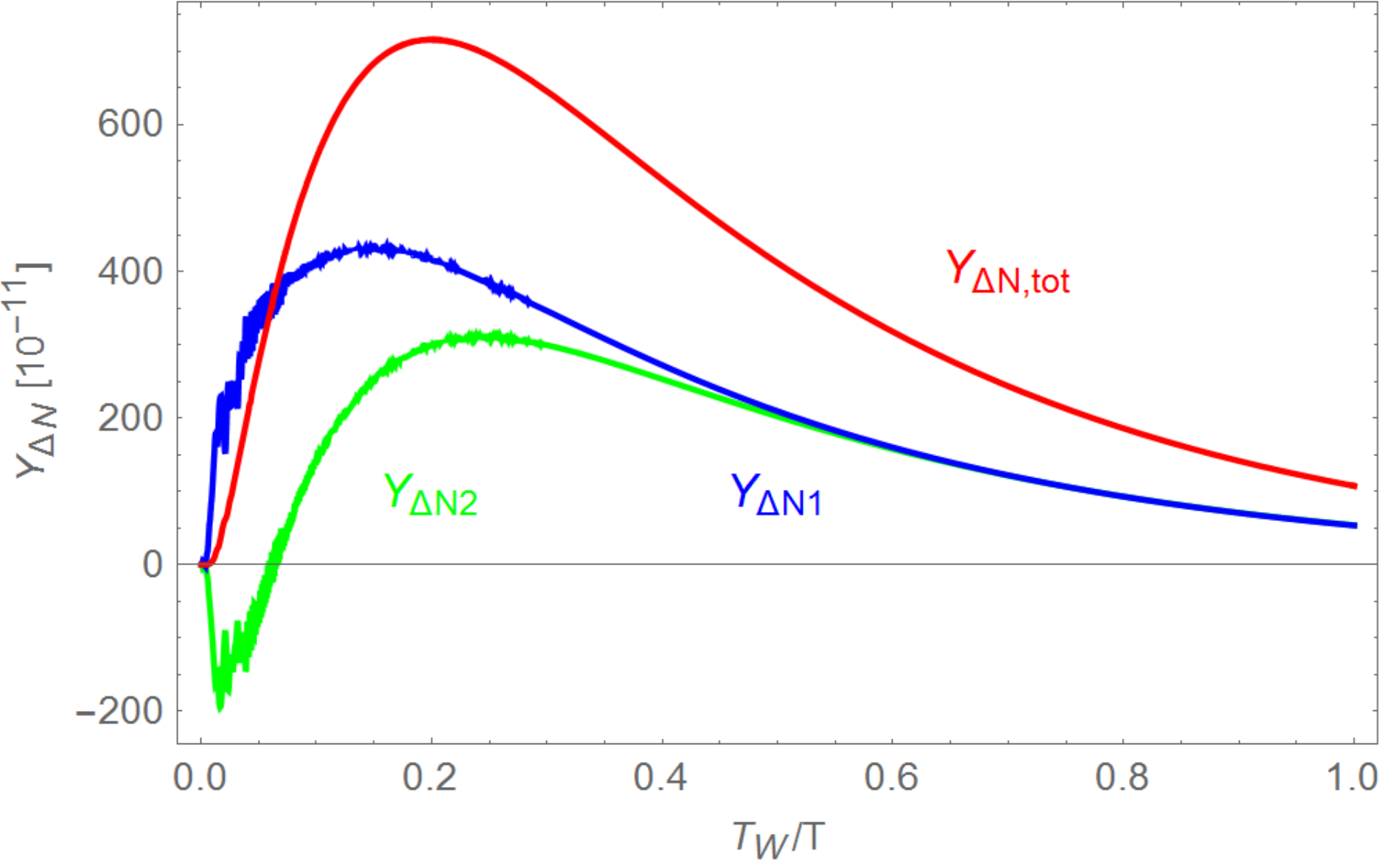}}\\
\subfloat{\includegraphics[width=0.45\textwidth]{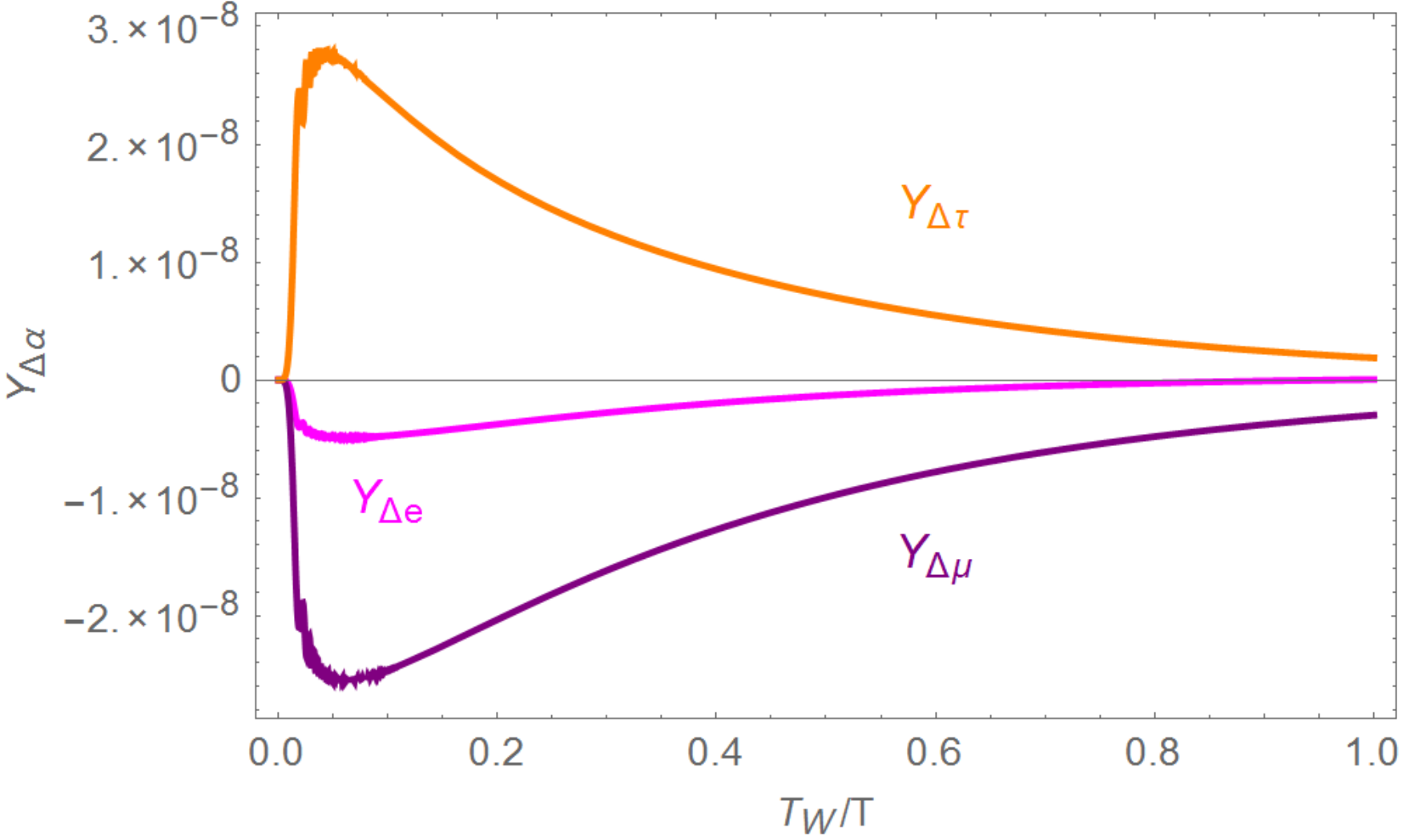}}\hspace{0.015\textwidth}
\subfloat{\includegraphics[width=0.45\textwidth]{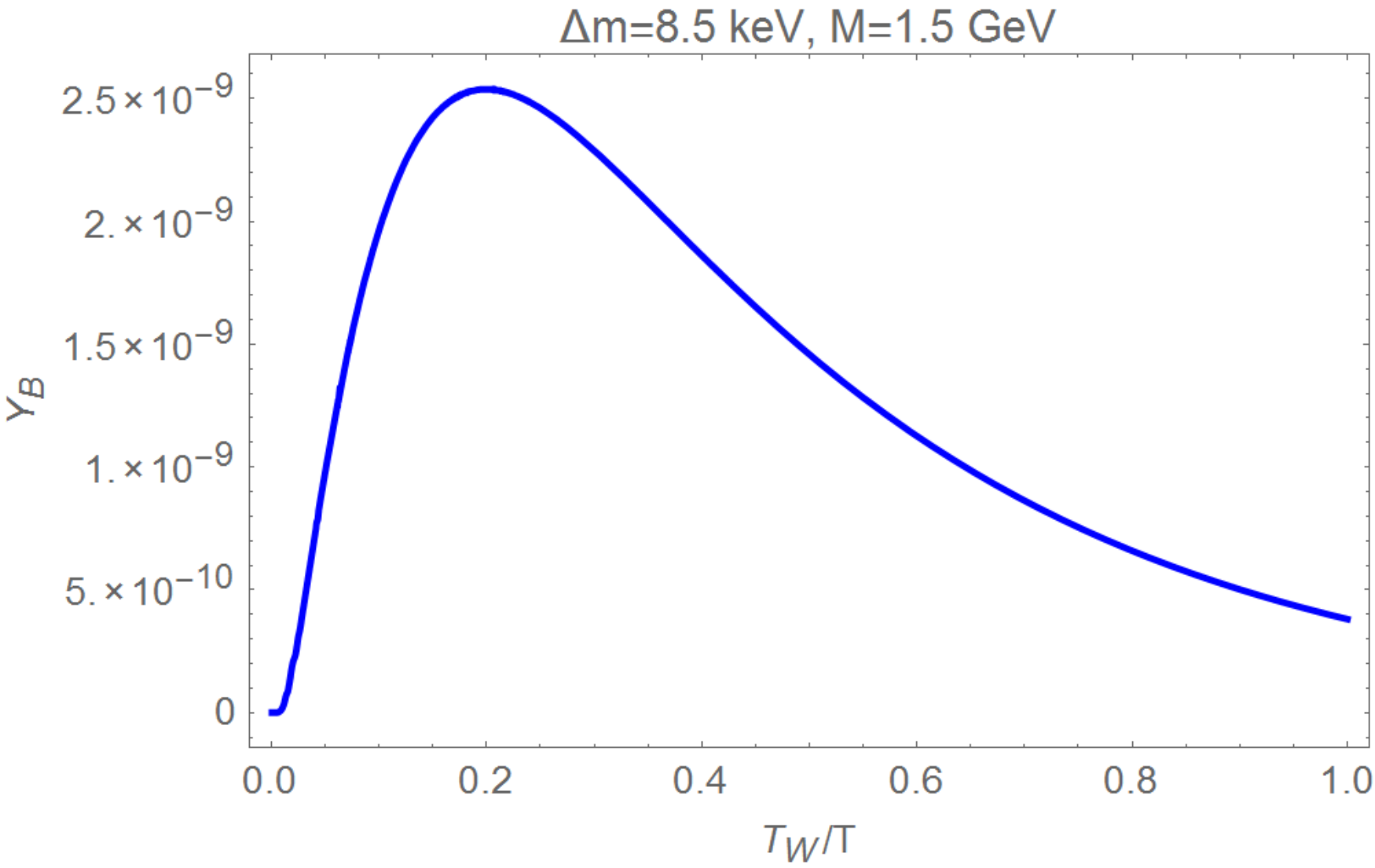}}
\caption{\footnotesize{As in Fig.~\ref{fig:benchhy1} but for a model with higher values of the entries of $Y_{\rm eff}$, but still below $10^{-6}$ (see Appendix~\ref{app_benchmarks} for details).}}
\label{fig:benchhy2}
\end{center}
\end{figure}

The second benchmark point, cf.\ Fig.~\ref{fig:benchhy2}, features higher entries of the Yukawa matrices but still exceeding  the equilibrium conditions by less than one order of magnitude, i.e. $\sqrt{2} \times 10^{-7} < |Y_{\alpha i}^\text{eff}| < 10^{-6}$. As already anticipated, the expected stronger washout effects are compensated by the higher initially produced lepton asymmetry leading to $Y_B(T_{\rm W}) \sim 4 \times 10^{-10}$, again sizeably above the observed value.

\begin{figure}[htb]
\begin{center}
\subfloat{\includegraphics[width=0.42\textwidth]{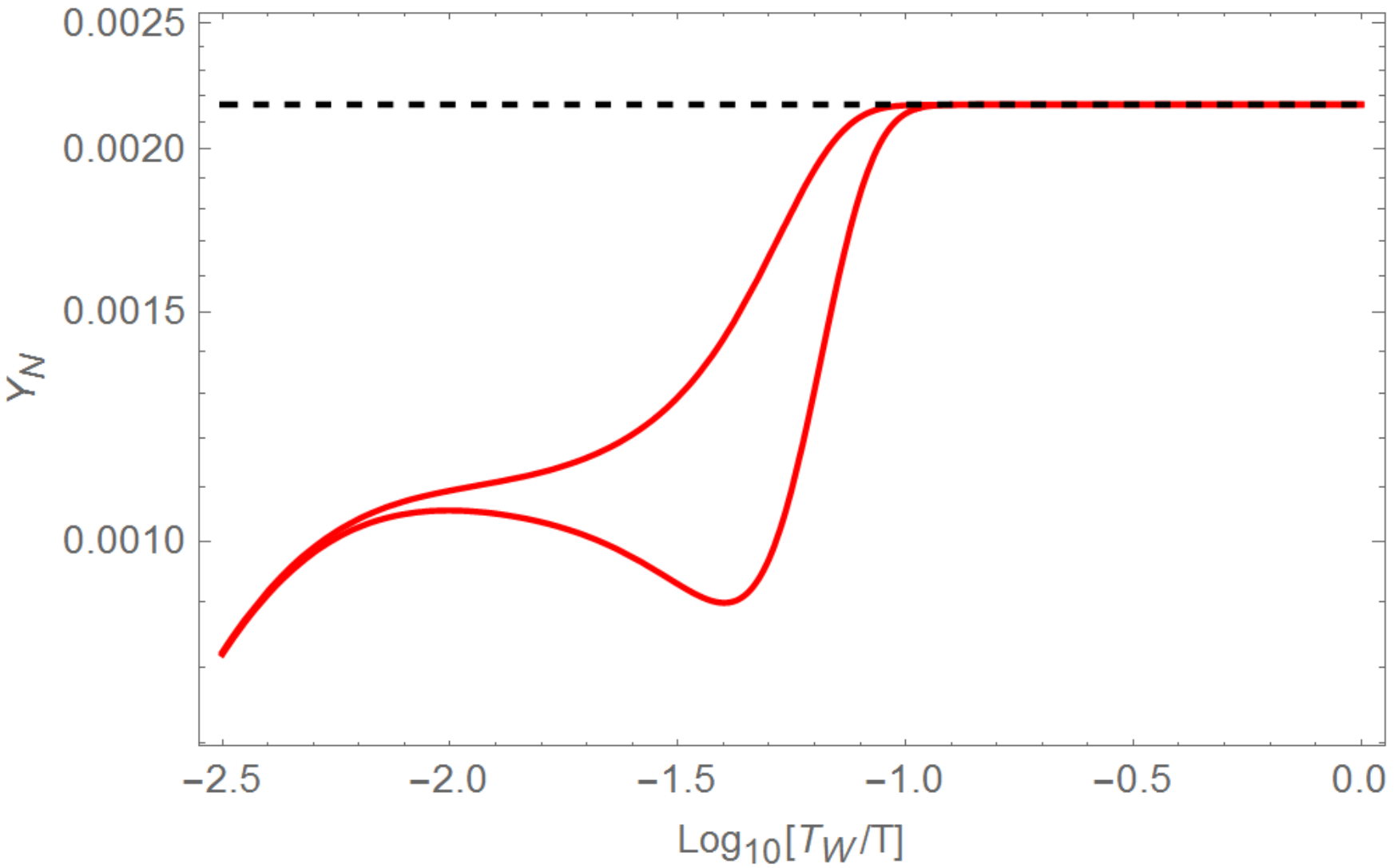}}\hspace{0.04\textwidth}
\subfloat{\includegraphics[width=0.42\textwidth]{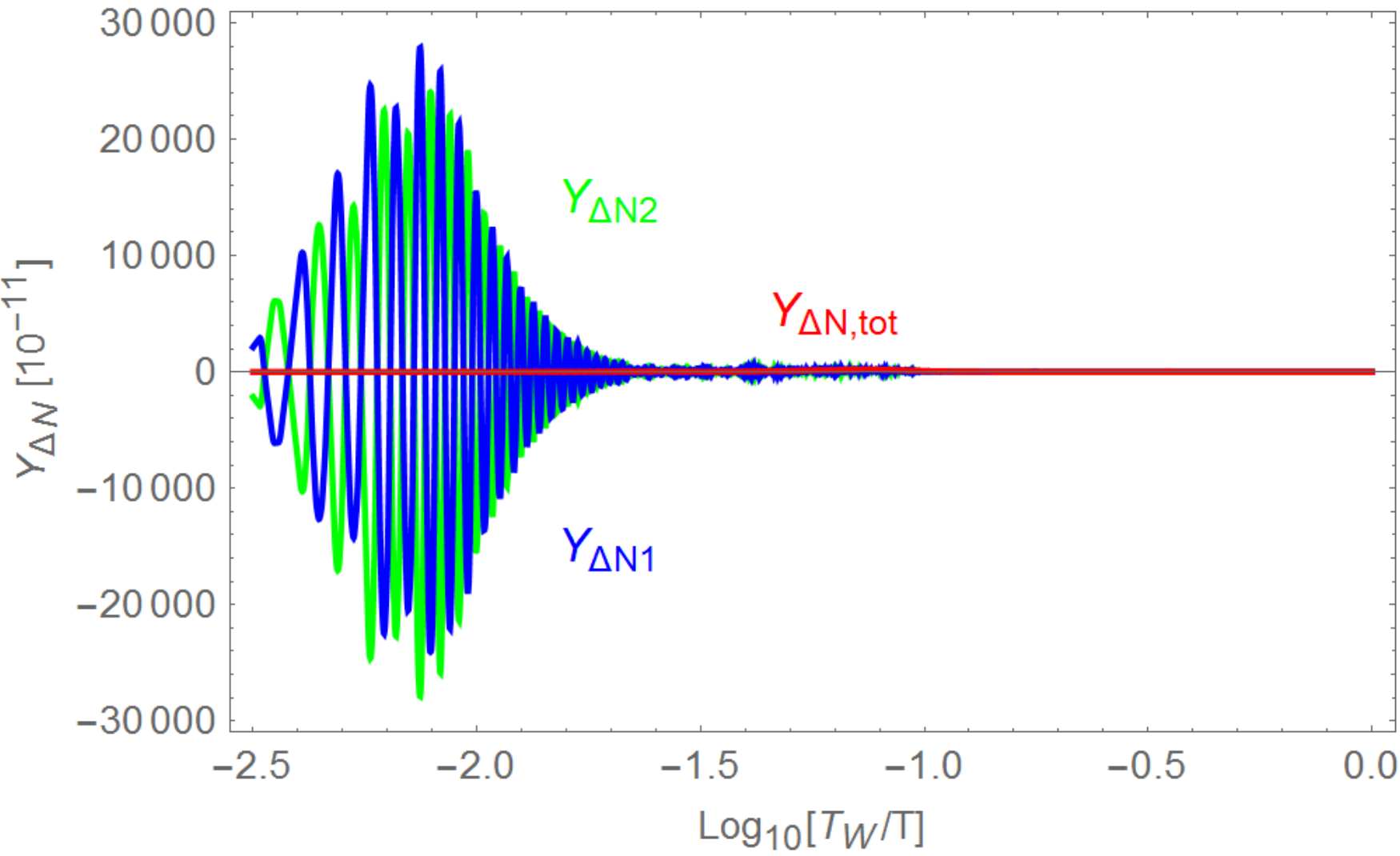}}\\
\subfloat{\includegraphics[width=0.45\textwidth]{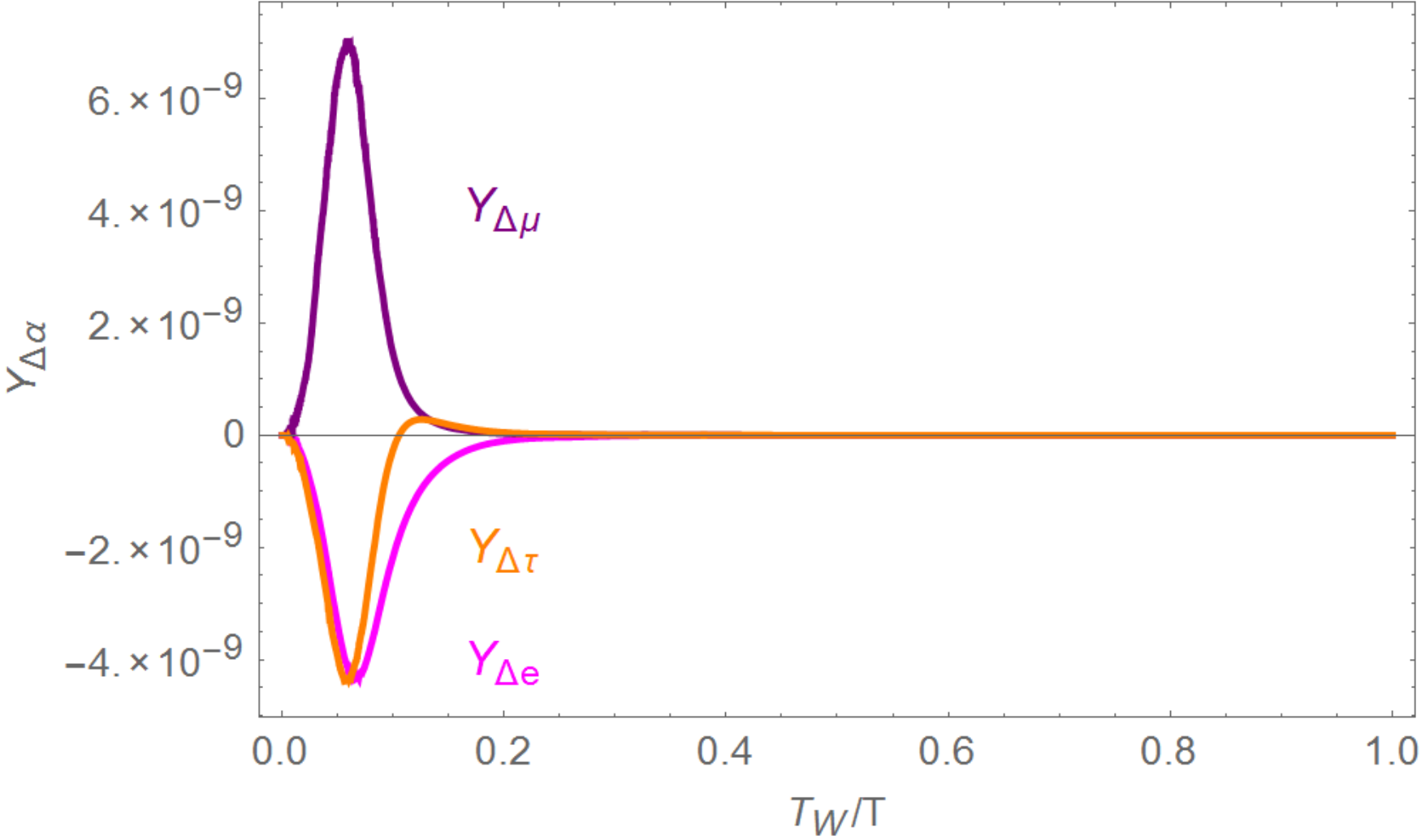}}\hspace{0.015\textwidth}
\subfloat{\includegraphics[width=0.45\textwidth]{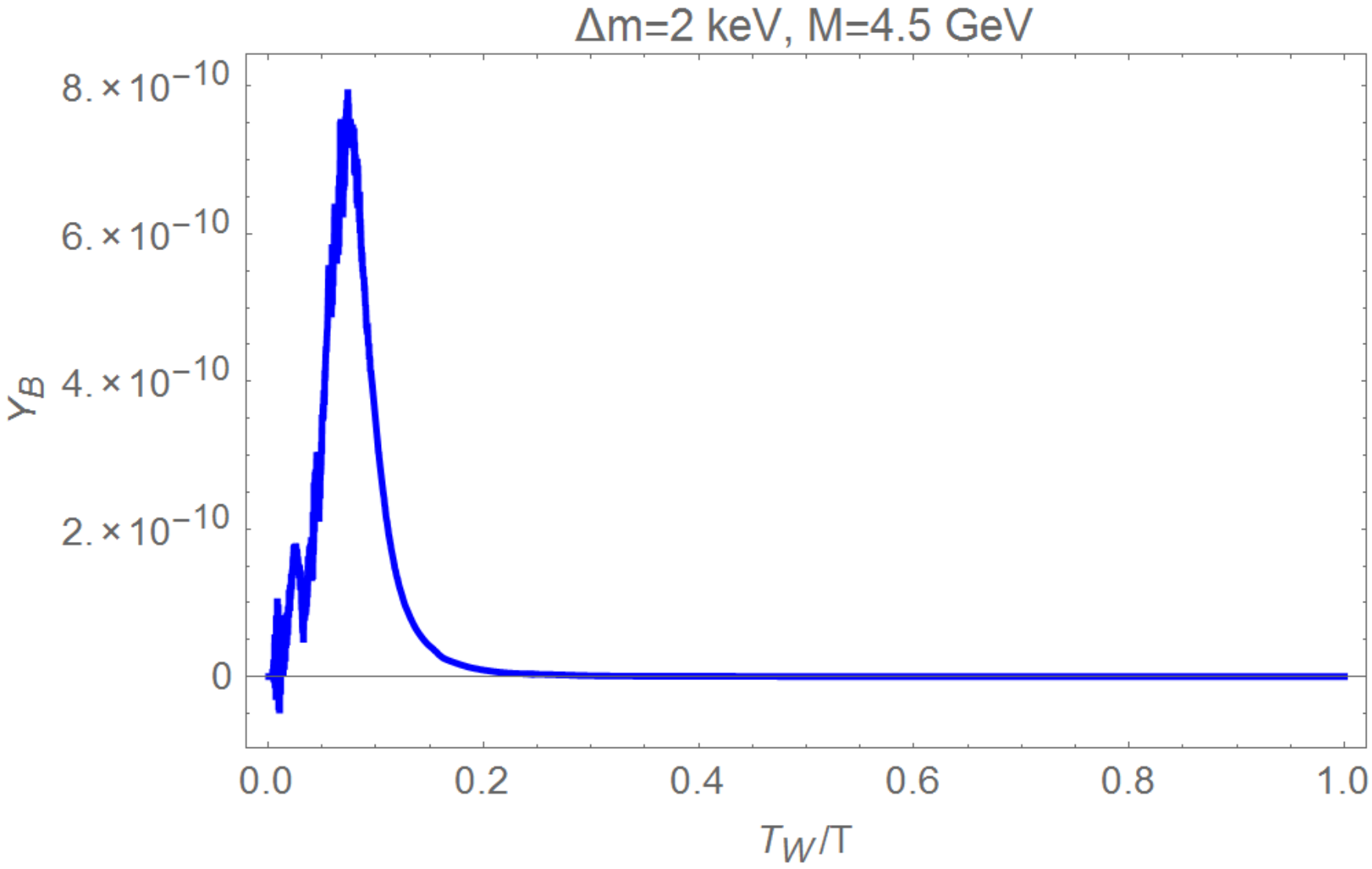}}
\caption{\footnotesize{As in Fig.~\ref{fig:benchhy1} but with entries of $Y^\text{eff}$ exceeding the equilibration value by more than one order of magnitude. In this case initially created asymmetries are completely depleted at later times and the final baryon abundance is negligible.}}
\label{fig:benchhy3}
\end{center}
\end{figure}

Finally we consider a benchmark point with $|Y^\text{eff}_{\alpha i}| \gtrsim 10^{-6}$, cf.\ Fig.~\ref{fig:benchhy3}. In this last case depletion effects are largely dominant and the final value of the baryon abundance is several orders of magnitude below the correct one, indicating that here the washout is too strong to allow for successful leptogenesis.

\begin{figure}[htb]
\begin{center}
\subfloat{\includegraphics[width=0.42\textwidth]{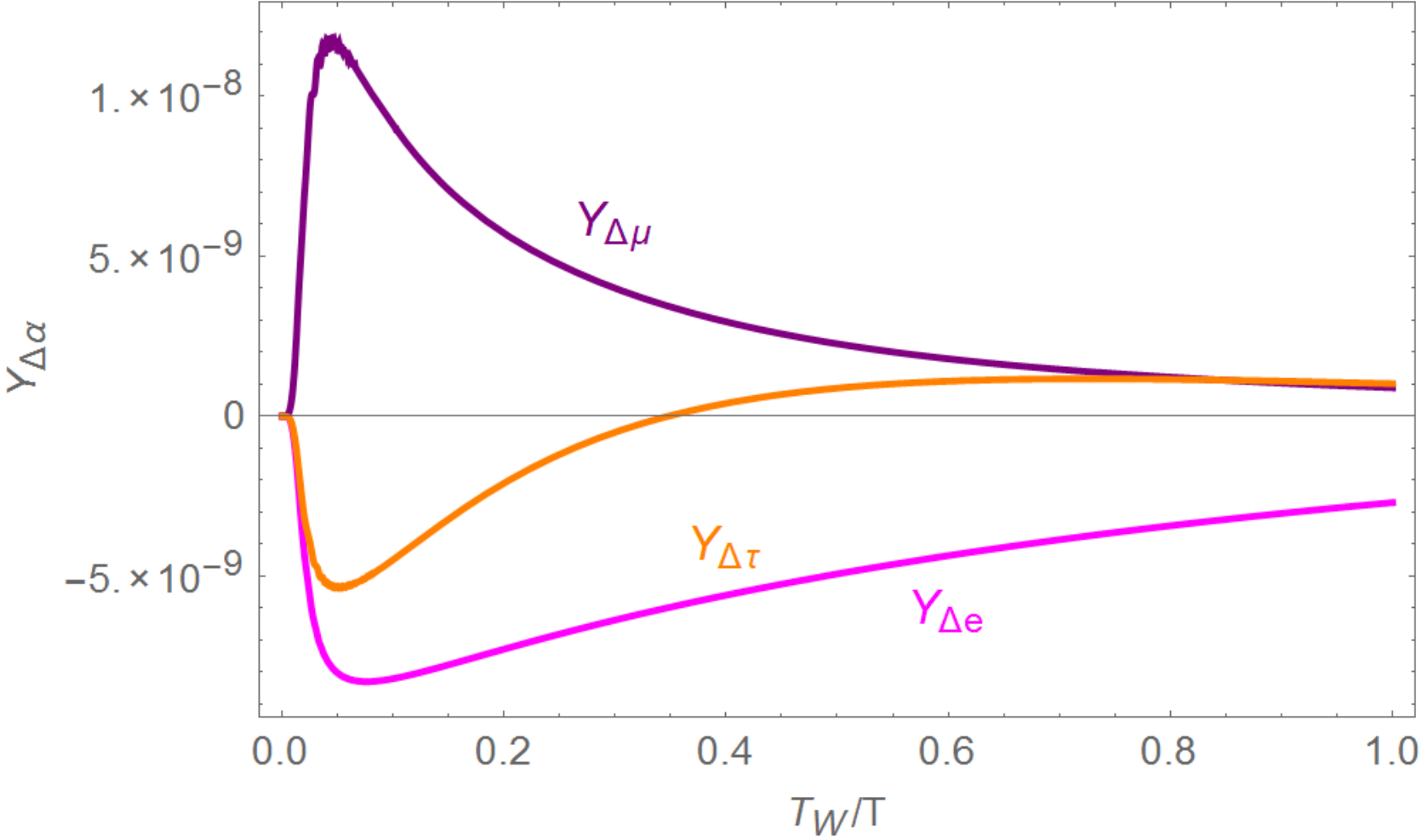}}\hspace{0.04\textwidth}
\subfloat{\includegraphics[width=0.42\textwidth]{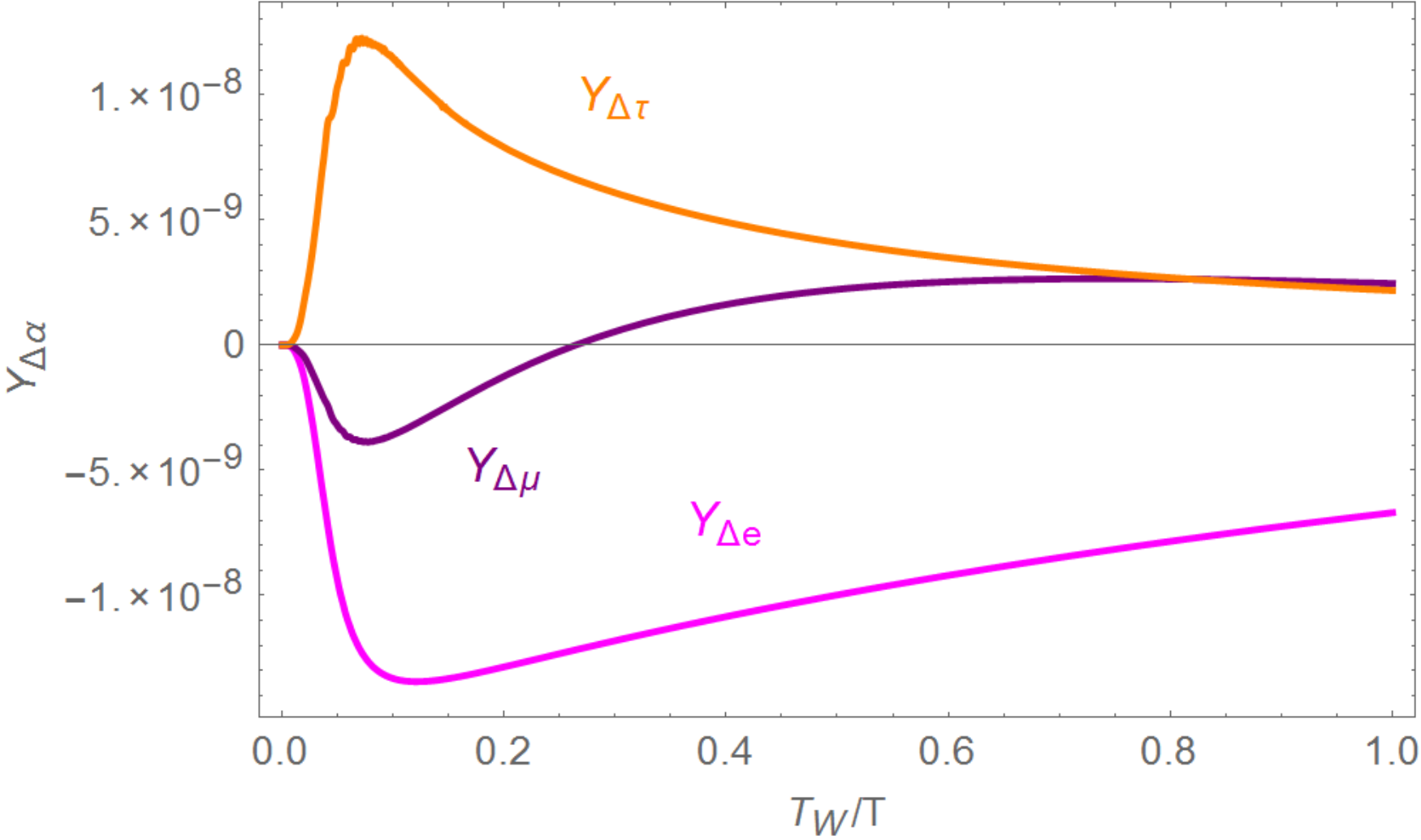}}\\
\subfloat{\includegraphics[width=0.45\textwidth]{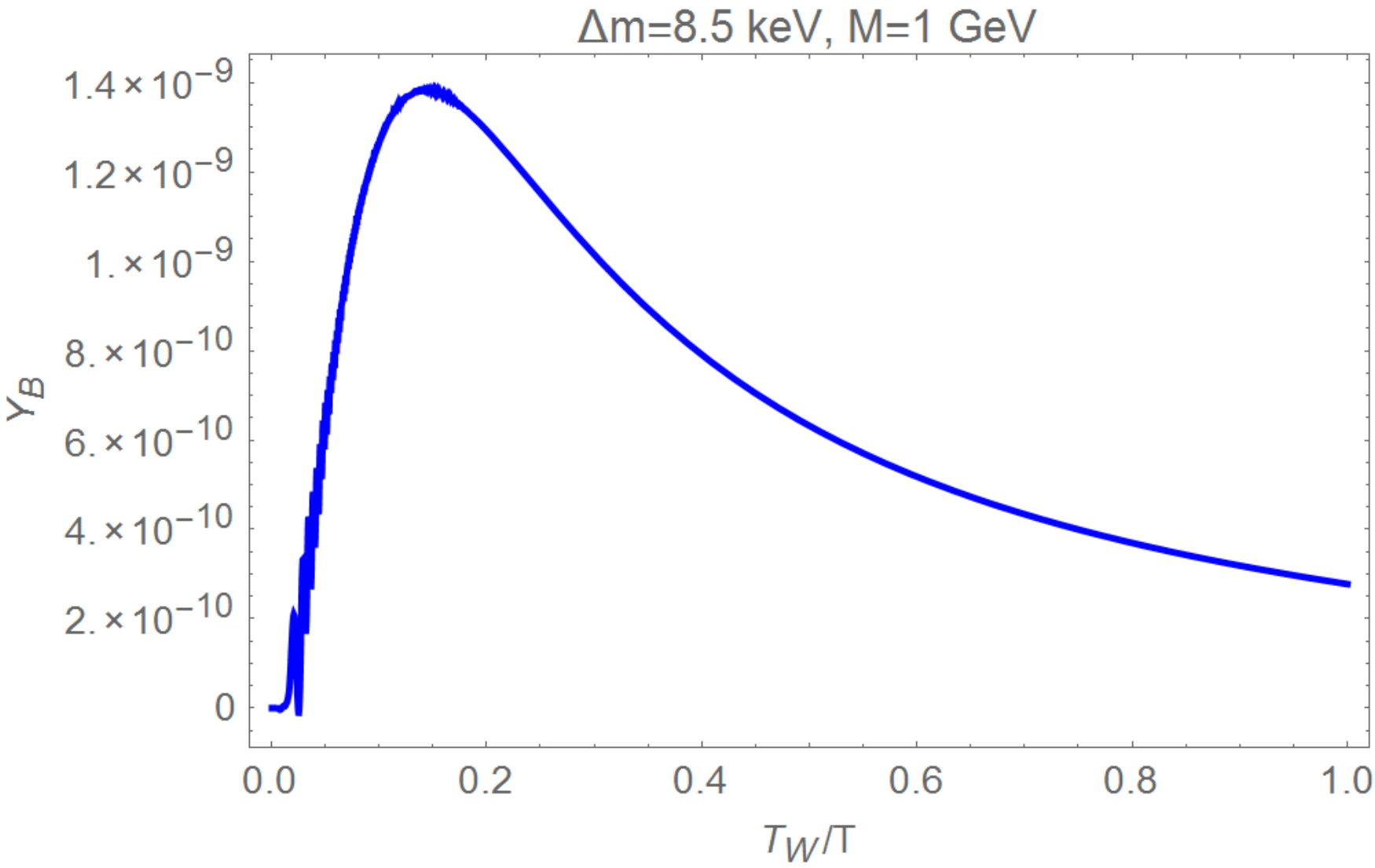}}\hspace{0.015\textwidth}
\subfloat{\includegraphics[width=0.45\textwidth]{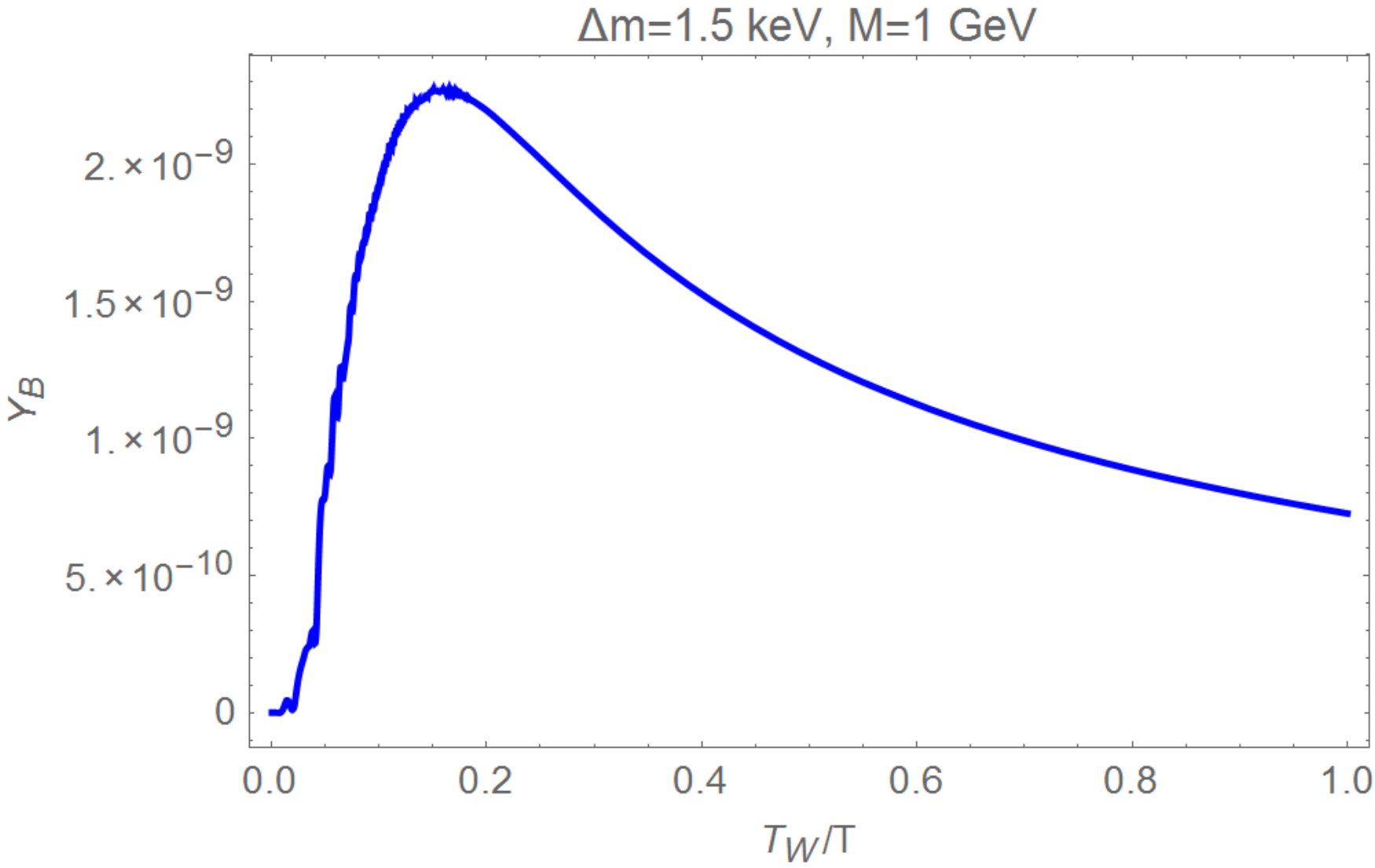}}
\caption{\footnotesize{Evolution of the asymmetries in the leptonic flavours (upper panels) and of the total baryon abundance (lower panels) for two benchmarks featuring a hierarchical structure in the matrix $Y^\text{eff}$: The entries $Y^{\rm eff}_{ei}$ are  below the equilibration value whereas the other entries are significantly above.
The value of $Y_B$ for the two benchmarks is $2.5 \times 10^{-10}$ and $7.5 \times 10^{-10}$, respectively.}}
\label{fig:bench_flavored}
\end{center}
\end{figure}

The benchmark points presented so far were characterised by matrices $Y^\text{eff}$ with entries of similar size (cf.\ Appendix~\ref{app_benchmarks}). This implies that the lepton asymmetry is generated with similar efficiency for all the three neutrino flavours. On the other hand, a viable neutrino spectrum can be obtained, in our scenario, also in the case of ``hierarchical'' structure, i.e.\ when there is a sizeable separation, possibly greater than one order of magnitude, between the entries of $Y^\text{eff}$ corresponding to different active flavours. In this case, it is possible to have realisations with $Y^\text{eff}$  entries below and above the equilibration condition. 

Two relevant examples are shown in Fig.~\ref{fig:bench_flavored}. They show two benchmark scenarios with  $Y^\text{eff}_{ei} \leq \sqrt{2} \times 10^{-7}$ and $Y^\text{eff}_{\mu (\tau) i } \geq \sqrt{2} \times 10^{-7}$. As can be seen in the upper panels of the figure, the asymmetry in the electronic flavour features a much weaker depletion than the other two flavours and tends, at late times, to become the dominant contribution  for the total baryon asymmetry. As evident from the bottom panels of the figure, in both cases the final value of $Y_B$ exceeds the experimental value, demonstrating the possibility of having an efficient baryon production in this kind of setup.  Interestingly the two benchmark points have values of $|U_{\mu i}|^2$ of, respectively, $10^{-8}$ and $2 \times 10^{-9}$, lying within the expected sensitivity of SHiP (cf. Fig.~\ref{fig:aftYB}). The plots shown refer to normally ordered spectra of active light neutrinos. Our result partially resembles the scenario of flavoured leptogenesis discussed in~\cite{Drewes:2012ma}. However contrary to the case discussed in this reference (where three right-handed neutrinos are involved in the generation of the BAU), in presence of only two heavy neutrinos, the hierarchy between the entries of $Y^{\rm eff}$ can hardly exceed one order of magnitude and the flavour effects are less efficient, still requiring approximate degeneracy between the heavy neutrinos. Although models with hierarchical structure for the Yukawa matrix $Y^{\rm eff}$ are present both for normal and  inverted hierarchy for the spectrum of active (light) neutrinos, we find that this kind of setup favours the normal hierarchy spectrum.

As evident from the analysis presented in this section, the viable parameter space is enlarged with respect to the one shown in Fig.~\ref{fig:aftYB}, towards larger values of the Yukawa couplings, at least $O\left(10^{-6}\right)$.
A conclusive statement on the extension of this parameter space requires a (computationally very demanding) numerical analysis. We have nevertheless shown that  our scenario can provide successful leptogenesis in the strong washout-out regime, with  values of the mixing between light active and heavy neutrinos within the sensitivity region of future facilities like SHiP, as has been found also in the three neutrino extension of the SM, cf.\ \cite{Canetti:2014dka}. Promising parameter points lie both in the region of hierarchical and non-hierarchical Yukawa couplings, e.g.\ $|U_{\mu i}|^2 \sim 10^{-8} - 10^{-9}$ for the points reported in Figs.~\ref{fig:benchhy2} and \ref{fig:bench_flavored}.

\section{A special case: the inverse Seesaw}\label{ISS}

A special case of the ansatz introduced in Section~\ref{sec_analytic} arises for $\epsilon \rightarrow 0$, referred to as the Inverse Seesaw.
As discussed in Section~\ref{Sec:intro}, minimal realisations of this scenario in agreement with neutrino oscillation data, laboratory and unitarity constraints, as well as  constraints from lepton flavour violating observables, require four or five additional heavy states (referred to as ISS(2,2) and ISS(2,3), respectively).  
 With respect to Eq.~\eqref{eq_Mpertexp}, the fourth row/column is extended to contain two ``right-handed neutrino'' fields and the fifth row/column is extended to two or three ``sterile'' fields, respectively. Schematically, the mass matrix can be written as
 \begin{equation}
  M = \begin{pmatrix}
   0  & \frac{1}{\sqrt{2}} Y v & 0 \\
  \frac{1}{\sqrt{2}} Y^T v & 0 & Z \Lambda \\
   0 & Z^T \Lambda & \xi \Lambda
  \end{pmatrix}.
 \end{equation}
Here in the ISS$(I,J)$ setup $Y$ and $\xi$ are understood as $3\times I$ and $J\times J$ matrices. $Z$ is a $I\times J$ matrix with entries of order unity. 
The entries of the $Z$ and $Y$ matrices are taken complex and the matrix $\xi$ can be taken real and diagonal, cf. Table~\ref{physpar}.
 
The new states form two heavy pseudo-Dirac pairs of mass ${\cal O}(\Lambda)$ with  squared mass-splittings of order $\mathcal{O}(\xi \Lambda^2)$,  the ISS(2,3) case features in addition a sterile state at the scale $\mathcal{O}(\xi \Lambda)$. In the mass ranges of eV or keV, the latter is an interesting candidate to address anomalies in the neutrino oscillation data or to explain dark matter (cf. Chapter~\ref{sec:DMMISS}), respectively. The two heavier pseudo-Dirac pairs are promising candidates for generating a baryon asymmetry, as discussed in Section~\ref{sec_toymodel}. Given that we are now dealing with a $7\times 7$ or $8\times 8$ mass matrix, there are clearly many possibilities for cancellations in the equations and we can no longer trust the simple estimates of Section~\ref{sec_toymodel}, which as we recall, lead us to disfavour the pure ISS due to a too large mass splitting of the heavy states. Indeed, a detailed scan of the ISS(2,3) parameter space reported in Chapter~\ref{sec:DMMISS} found solutions for the light sterile state in the sub-eV to 100 keV range, pointing to mass-splittings of order $\Delta m^2 \sim (10~\text{keV})^2 - (10~\text{MeV})^2$. In addition, suitable Yukawa couplings
 below the critical value of $\sqrt{2} \times 10^{-7}$ are indeed achievable for light neutrino masses and mixings in agreement with current observations. This renders this scenario very promising for a minimal low-energy setup to simultaneously explain neutrino masses, dark matter and leptogenesis. In the following we revisit this setup, clarifying that, despite the large number of parameters, a successful leptogenesis in the weak washout regime cannot be achieved.
 
We have shown in Section~\ref{sec_toymodel} that for the ISS toy model the requirements $Y<\sqrt{2}\times 10^{-7}$, $m_\nu = 0.05$~eV and $\Lambda = 1$~GeV, imply $\Delta m^2 \gtrsim (0.4~\text{GeV})^2$, a value too large for leptogenesis.
Let us now generalise this result using the full matrix equations and considering first an ISS(3,3) setup, for which all the relevant submatrices are invertible $3\times 3$ matrices.
 It is possible to define the $3\times3$ PMNS mixing matrix $N$ as\cite{Antusch:2009pm,Ibarra:2010xw}
\bee
N= (1+\eta)V,
\eee
where $V$ is a unitary matrix and $\eta$ parametrises the deviation from unitarity,
\bee
\eta\simeq -\frac{1}{4} \frac{v^2}{\Lambda^2} Y^* {Z^{-1}}^\dagger {Z^{-1}} Y^T,
\eee 
which is hermitian.
Retaining only the first order terms in the non-unitarity parameters, which are expected to be small, one has
\bee
NN^\dagger = (1+\eta)VV^\dagger (1+\eta)\simeq 1+2\eta.
\eee
In terms of the unitary $9\times 9$ leptonic mixing $U$ we have (no sum over $\alpha$) 
\bee
{\sum_{i = 4}^9 \,} |U_{\alpha i}|^2= 1-\sum_{k=1}^3 |U_{\alpha k}|^2= 1- \left(NN^\dagger\right)_{\alpha \alpha}\simeq -2\eta_{\alpha \alpha},
\eee
implying for the active-sterile mixing
\bee
{\sum_{i=4}^9} \sum_\alpha |U_{\alpha i}|^2\simeq  \frac{v^2}{2 \Lambda^2} \Tr{Y^*\ {Z^\dagger}^{-1}\ {Z^{-1}}\ Y^T}=\frac{v^2}{2 \Lambda^2} \Tr{\left|Y\ {Z^T}^{-1}\right| \left| {Z^{-1}}\ Y^T\right|},
\label{eq:Ubound}
\eee
since $\Tr{A^\dagger A} = \Tr{|A^T| |A|}$.
The  effective Yukawas are related to the mixing matrix and to the mass eigenstates by (no sum over i)
\bee
Y^\text{eff}_{\alpha i}= \sqrt{2}
U_{\alpha i}^* \frac{M_i}{v}\simeq \sqrt{2}
U_{\alpha i}^* \frac{\Lambda}{v}.
\eee
In the ISS scenario,  the neutrino mass matrix is given by
\bee\label{eq:mnuISS}
m_\nu \simeq - \frac{v^2}{2 \Lambda} Y\ {Z^{-1}}^T\ \xi \ {Z^{-1}}\ Y^T,
\eee
and in a basis in which $\xi$ is positive and diagonal
\bee
\max{\left[\xi_{jj}\right]} \frac{v^2}{2 \Lambda} \Tr{\left|Y\ {Z^{-1}}^T\right| \left|{Z^{-1}}\ Y^T\right|} &\geq&    \frac{v^2}{2 \Lambda}  \Tr{\left|Y\ {Z^{-1}}^T\ \xi\ {Z^{-1}}\ Y^T\right|}\non 
&=&\Tr{\left|m_\nu\right|}\ge 0.05 \text{ eV}\,.
\eee
With Eq.~\eqref{eq:Ubound}, the Yukawa couplings between the active flavours  and the heavy states are then bounded from below by
\begin{equation}
 {\sum_{i=4}^9}  \sum_{\alpha=1}^3 |Y^\text{eff}_{\alpha i}|^2 \geq 2 \,
 \frac{0.05~\text{eV}}{\text{max}|\xi_{jj}|} \frac{\Lambda}{v^2}\,.
 \label{eq_Ybound}
\end{equation}
Finally, imposing the lower bound in Eq.~(\ref{eq_Ybound}) to lie below the out-of equilibrium condition, $ \sum_{\alpha} |Y^\text{eff}_{\alpha i}|^2 < 2 \times 10^{-14}$ for all the heavy states $i$, implies $\max\left[ \xi_{jj}\right] \gtrsim 0.07 $ for $\Lambda = 1\text{ GeV}$, corresponding to a mass splitting $\mathcal{O}(\xi \Lambda^2) \gtrsim \left(0.25 \text{ GeV}\right)^2$, in good agreement with the estimation obtained in the toy model using Eq.~(\ref{eq:dmISS}). In conclusion, the ISS(3,3) model yields a mass splitting which is significantly too large for viable leptogenesis in the weak washout regime, which requires $\Delta m^2 \lesssim \text{MeV}^2$~\cite{Drewes:2012ma}. Moreover, the scale $\text{max}|\xi_{jj}| \Lambda$ which sets the scale of the  potential  DM candidate in the ISS(2,3) model  is found to be unpleasantly large: X-ray observations exclude sterile neutrinos heavier than about 100~keV contributing significantly to the DM abundance~\cite{Boyarsky:2005us}.

Since the lower bound in Eq.\ (\ref{eq_Ybound}) relies on the assumption $M_i \simeq \Lambda$ for every $i>3$, one may expect that the above conclusions are invalidated if a large mass difference among the different pseudo-Dirac pairs is present. In order to probe the feasibility of this configuration we performed a numerical scan of the the simpler phenomenologically viable realisation, the ISS(2,2). We generated the entries of the complex submatrices $Z\Lambda$ and $\xi \Lambda$ in the ranges $100 \text{ MeV} \le \left|Z_{ij}\right|\Lambda\le 40 \text{ GeV}$ and $1 \text{ eV} \le \left|\xi_{ij}\right|\Lambda\le 10 \text{ GeV}$, taking the different entries in each submatrix to be of the same order of magnitude.
The Dirac submatrix was generated using a modified version of the Casas-Ibarra parametrisation~\cite{Casas:2001sr} adapted for the ISS(2,2)
\bee
\frac{Y v}{\sqrt{2}} = N_\text{PMNS}^*\ \sqrt{\hat{m}_{\nu}}\ R\ \sqrt{\xi^{-1}}\ Z^T \Lambda,
\eee
where the ``orthogonal'' matrix $R$ is defined as
\bee
R\left(\theta\right) = \left( \begin{array}{cc} 1 & 0 \\ \cos \theta & \sin \theta \\ -\sin \theta & \cos \theta \end{array} \right),
\eee
and the complex angle $\theta$ is randomly varied in the range $0\le \left| \theta\right|\le 2\pi$. Each solution is required to accommodate neutrino oscillation data, laboratory bounds on direct searches of sterile fermions and BBN bounds.  The values of the effective Yukawas for the lightest sterile state, $Y_{4}^\text{eff}=\sum_{\alpha}\left|Y_{\alpha 4}^\text{eff}\right|$ as a function of the lightest sterile mass $M_1$ are reported in Fig~(\ref{Fig:ISS22}). The horizontal green line represents the out of equilibrium value $Y_{4}^\text{eff} = \sqrt{2}\times 10^{-7}$ while the colour code is related to the mass degeneracy in the lighter pseudo-Dirac pair
\bee
\delta_{45} \equiv 2\frac{M_2-M_1}{M_1+M_2} \,.
\label{eq:delta45}
\eee

\begin{figure}[htb]
\centering
\begin{tikzpicture}
 \node at (0,0) { \includegraphics[width=0.75\textwidth]{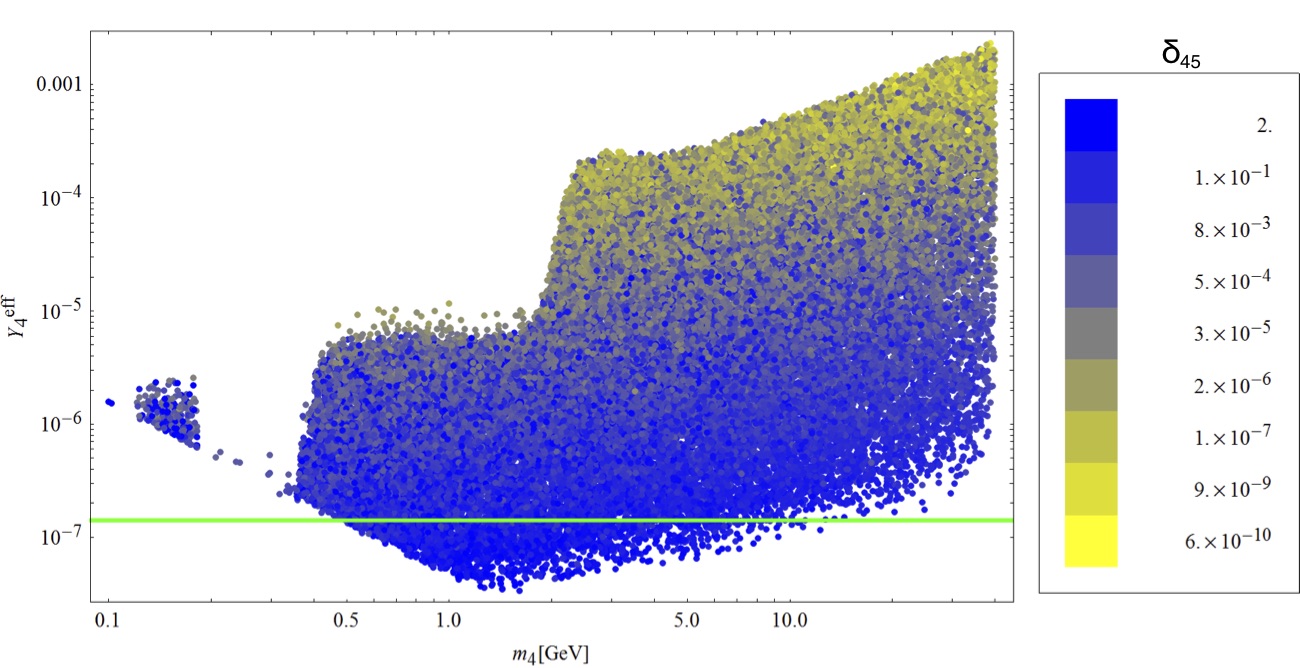}};
\node[fill = white] at (-0.5, -2.8) {\footnotesize $M_1  \, [\text{GeV}]$};
\node[fill = white, rotate = 90] at (-5.5, 0) {\footnotesize $Y_4^\text{eff}$};
\draw[fill = white, draw = white] (4.9, 1.5) rectangle (5.2, 2.1);
\node at (4.95,1.7) {\tiny $2 \times 10^0$};
\node[fill = white] at (4.5,2.5) {\footnotesize $\delta_{45}$};
\end{tikzpicture}
 \caption{Effective Yukawa coupling $Y^\text{eff}_4$ and mass  $M_1$ for the lightest sterile state in the ISS(2,2). The colour coding refers to the relative mass degeneracy $\delta_{45}$ with a high (low)  degeneracy marked in yellow (blue). }
 \label{Fig:ISS22}
\end{figure}
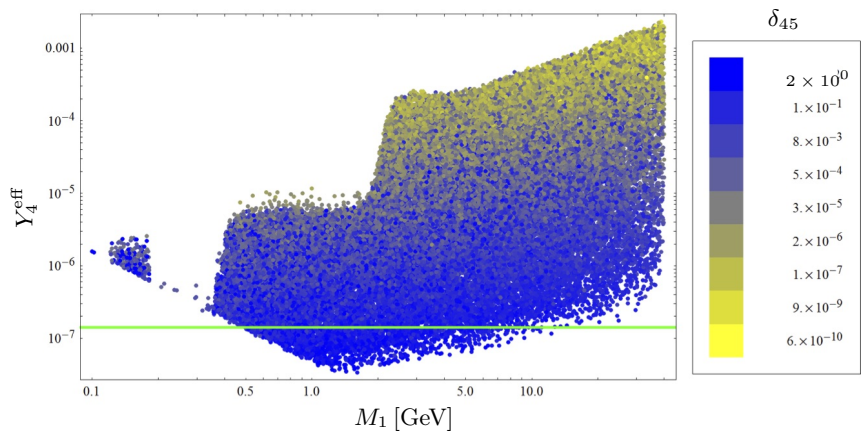

As is evident from this figure, smaller values of the Yukawa couplings are related to larger mass splittings in the pseudo-Dirac pair. Moreover the lower values for the Yukawa couplings are strongly limited by the BBN constraints in the region $M_1 \lesssim 1$ GeV, and by the Seesaw relation Eq.~(\ref{eq:mnuISS}) in the region $M_1 \gtrsim 1$ GeV, leaving only a small out-of-equilibrium region in the mass range $500 \text{ MeV} \lesssim M_1 \lesssim 10 \text{ GeV}$. Focusing now on the weak washout regime of Fig.~\ref{Fig:ISS22}, i.e.\  requiring $\left| Y^\text{eff}_{\alpha i}\right| < \sqrt{2}\times 10^{-7}$ for all $\alpha=e,\mu,\tau$, $i=4,\dots,7$, we depict in  Fig.~\ref{Fig:degISS22} the quantities governing the efficiency of leptogenesis:\footnote{Due to the larger Yukawa couplings, we generically expect the heavier pseudo-Dirac pair to yield the dominant contribution to leptogenesis in the weak washout regime.} the mass degeneracy $\delta_{67}$ and  effective Yukawa coupling $Y^\text{eff}_6$ of the heavier pseudo-Dirac pair (defined analogously to Eq.~\eqref{eq:delta45}).
As it is evident the condition $\delta_{67}<10^{-3}$ is not reached, and the parameter space of the model prefers the region $0.1 \lesssim \delta_{67} \lesssim 1$.
 Besides of being ineffective for leptogenesis, these realisations are clearly outside the ``natural'' region of the inverse Seesaw, in which $\xi \ll 1$. {We conclude that in the weak washout regime, and in particular in the regime of lepton number conservation, $\xi \ll 1$, solutions which are able to accommodate both neutrino oscillation data and a viable leptogenesis are hard, if not impossible, to find.}

\begin{figure}[htb]
\centering
 \includegraphics[width=0.55\textwidth]{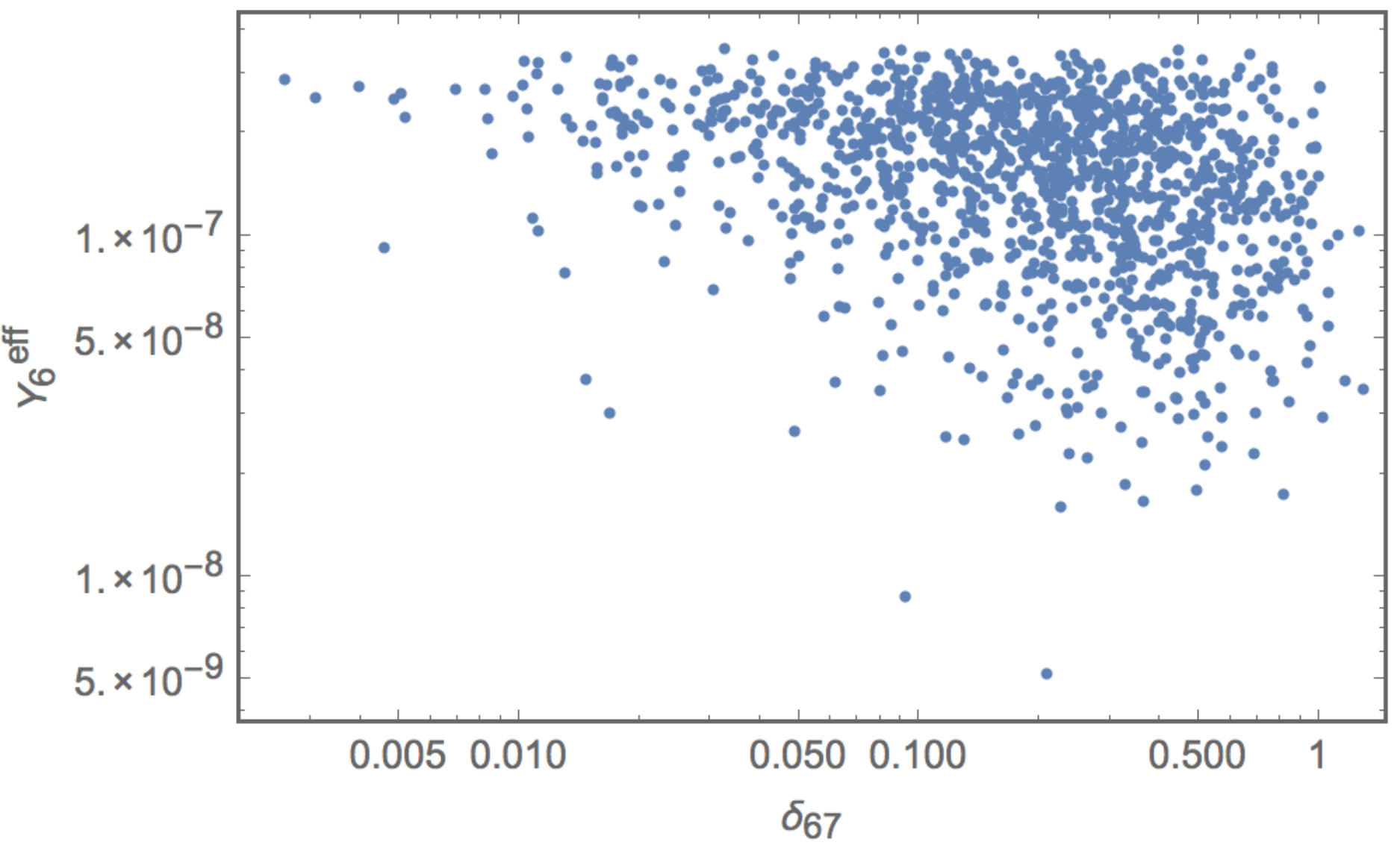}
 \caption{Effective Yukawa coupling $Y^\text{eff}_6$ and  relative mass degeneracy $\delta_{67}$ for the points in Fig.~\ref{Fig:ISS22} which lie below the equilibration value, $\left| Y^\text{eff}_{\alpha i}\right| < \sqrt{2}\times 10^{-7}$. In the weak washout regime a higher mass degeneracy cannot be obtained.}
 \label{Fig:degISS22}
\end{figure}

{Although we have conducted a detailed numerical study only for the ISS(2,2) scenario, we expect similar results to hold for the ISS(2,3) setup. The additional state in the ISS(2,3) with a mass of ${\cal O}(\xi \Lambda)$ comes with correspondingly suppressed Yukawa couplings, implying a negligible effect on leptogenesis, which relies only on the heavier states. Moreover, since the weak washout regime prefers large values of $\xi$, cf.\ Fig.~\ref{Fig:ISS22}, the particularly attractive parameter range of the ISS(2,3), which can simultaneously account for DM, is disfavoured since the potential DM candidate would be too heavy to comply with the aforementioned X-ray constraints.}

We remark that this however does not exclude the ISS as a viable setup for a low-scale leptogenesis, since solutions in the strong washout regime may be allowed. These region of the parameter space is however outside the range of validity of our analytical solution Eq.~(\ref{eq:baryo_analytical}). We leave the numerical exploration of the relevant parameter space of the model for a future study.

\chapter{Lepton flavour violating  decays of vector  quarkonia and of the $Z$ boson}\label{sec-0}

Neutrino oscillation experiments have firmly demonstrated that the individual lepton flavours are not preserved in the neutrino propagation, showing that the individual lepton numbers, eq.~(\ref{flavour_charges_sm}), are not conserved charges in Nature. Since this is a manifestation of BSM physics, it is natural to look for different lepton flavour violating (LFV) processes, whose experimental study can constrain possible neutrino mass generation models.

Currently, the search for manifestations of LFV constitutes a goal of 
several experimental facilities dedicated to rare lepton decays, such as $\ell_\alpha \to \ell_\beta \gamma$ and $\ell\to \ell_\alpha\ell_\beta \ell_\gamma$, and to the neutrinoless
$\mu-e$ conversion in muonic atoms. One of the most stringent bounds from these searches is the one derived by the MEG Collaboration, 
$\text{B}(\mu \to e \gamma) < 5.7 \times 10^{-13}$~\cite{Adam:2013mnn}, which is expected to be improved to a planned sensitivity of $6 \times 10^{-14}$~\cite{Baldini:2013ke}. 
Moreover, the bound  $\text{B}(\mu \to eee)
< 1.0 \times 10^{-12}$, set by the SINDRUM experiment~\cite{Bellgardt:1987du}, is expected to be improved by the Mu3e experiment where a sensitivity $\sim 10^{-16}$ is planned~\cite{Blondel:2013ia}. 
Limits on the $\tau$ radiative decays~\cite{Aubert:2009ag} and the three-body
decays of $\tau$~\cite{Hayasaka:2010np,Aushev:2010bq} appear to be less
stringent right now, but are likely to be improved at Belle II~\cite{Aushev:2010bq}, where the search for LFV decays of the $B$-meson will be made too~\cite{Bevan:2014iga}. 
The most promising  developments regarding LFV  are those related to the $\mu-e$ conversion in nuclei. The present bound for the  $\mu^- \mathrm{Ti} \rightarrow e^- \mathrm{Ti}$ conversion rate is  $4.3\times
10^{-12}$~\cite{Dohmen:1993mp}, and the planned sensitivity is $\sim10^{-18}$~\cite{Alekou:2013eta}. Similar is the case for gold and aluminium~\cite{Bertl:2006up,Kuno:2013mha}.

Searches for  LFV are also conducted in high-energy experiments and a first bound on the Higgs boson LFV decay $h \to \mu \tau$ has been reported by the CMS Collaboration
 \cite{CMS:2014hha}. The LHCb Collaboration, instead, reported the bound $\text{B}(\tau \to 3 \, \mu)< 8.0 \times 10^{-8}$~\cite{Aaij:2013fia}, which is likely to be improved in the near future~\cite{Agashe:2014kda}. 
 Notice also that they already improved the bounds on $\text{B}(B_{(s)} \to e \mu)$ by an order of magnitude~\cite{Aaij:2013cby}. \\

In this analysis we will focus on the indirect probes of new physics through the LFV processes of  neutral vector bosons, namely 
$V\to \ell_\alpha\ell_\beta$, with $\ell_{\alpha,\beta}\in \{e,\mu, \tau\}$, and $V\in \{ \phi, \psi^{(n)}, \Upsilon^{(n)}, Z\}$, where $\psi^{(n)}$ stands for $J/\psi$ and its radial excitations, and similarly for $\Upsilon^{(n)}$.  
Most of the research in this direction reported so far is related to the $Z\to \ell_\alpha\ell_\beta$ decay modes. 
More specifically, the experimental bounds,  obtained at LEP are found to be   $\mathrm{B}(Z\to e^\mp \mu^\pm) < 1.7 \times 10^{-6}$~\cite{Abreu:1996mj}, $\text{B}(Z\to \mu^{\mp}\tau^{\pm})< 1.2\times 10^{-5}$~\cite{Akers:1995gz,Abreu:1996mj}, 
and $\text{B}(Z\to e^{\mp}\tau^{\pm})< 9.8 \times 10^{-6}$~\cite{Akers:1995gz,Adriani:1993sy}. 
One of these bounds has been improved at LHC, namely $\text{B}(Z\to e^{\mp}\mu^{\pm})< 7.5\times 10^{-7}$~\cite{Aad:2014bca}. On the theory side, the $Z$ decays have been analysed in the extensions of the SM involving 
additional massive and sterile neutrinos that could mix with the standard (active) ones and thus give rise to the LFV decay rates~\cite{Mann:1983dv,Ilakovac:1994kj,Illana:1999ww}. A similar approach has been also adopted in Ref.~\cite{Abada:2014cca}, in the perspective of 
a Tera-$Z$ factory FCC-ee~\cite{Blondel:2013rn} for which a targeted sensitivity is expected to be $\text{B}(Z\to e^{\mp}\mu^{\pm}) \,\sim \, 10^{-13}$~\cite{Blondel:2014bra}.

Lepton flavour conserving decays of quarkonia have been measured to a high accuracy which can actually be used to fix the hadronic parameters (decay constants). Otherwise, one can use the results of 
numerical simulations of QCD on the lattice, which are nowadays accurate as well~\cite{Becirevic:2013bsa,Becirevic:2012dc,Donald:2012ga,Colquhoun:2014ica,Lewis:2012ir}. The experimentally established bounds for the simplest LFV decays of quarkonia are~\cite{Agashe:2014kda}:
\bea
&&\text{B}(\phi \to e \mu) < 2.0\times 10^{-6},   \qquad  \text{\cite{Achasov:2009en}}\nn \\
&&\hfill \nn\\
&&\text{B}(J/\psi \to e \mu) < 1.6\times 10^{-7}, \qquad \text{B}(J/\psi \to e \tau) < 8.3\times 10^{-6},  \nn \\
 &&\text{B}(J/\psi \to  \mu \tau) < 2.0\times 10^{-6} ,  \qquad  \text{\cite{Ablikim:2013qtm,Ablikim:2004nn}} \nn \\
&&\hfill \nn\\
&&\text{B}(\Upsilon \to  \mu \tau ) < 6.0\times 10^{-7} ,   \qquad \text{\cite{Love:2008ys}}   \nn \\
&&\hfill \nn\\
&& \text{B}(\Upsilon(2S) \to e \tau) < 8.3\times 10^{-6},\qquad  \text{B}(\Upsilon(2S) \to  \mu \tau) < 2.0\times 10^{-6}, \qquad \text{ \cite{Lees:2010jk}} \nn \\
&& \text{B}(\Upsilon(3S) \to e \tau) < 4.2\times 10^{-6},\qquad  \text{B}(\Upsilon(3S) \to  \mu \tau) < 3.1\times 10^{-6} , \qquad \text{ \cite{Lees:2010jk}}  \nn 
\eea
where each mode is to be understood as $\text{B}(V \to \ell_\alpha\ell_\beta) = \text{B}(V \to \ell_\alpha^+ \ell_\beta^-)+ \text{B}(V \to \ell_\alpha^- \ell_\beta^+)$.

Despite the appreciable experimental work on the latter observables, only a few theoretical studies have been carried out so far.  
The authors of Ref.~\cite{Nussinov:2000nm} applied a vector meson dominance approximation to $\mu \to 3 e$ and expressed the width of the latter process, 
$\Gamma(\mu \to ee e)=\Gamma(\mu\to V e )\Gamma(V\to ee)$. Since the values of $\Gamma(V\to ee)$ are very well known experimentally~\cite{Agashe:2014kda}, the experimental 
bound on $\Gamma(\mu \to 3 e)$ is then used to obtain an upper bound on the phenomenological coupling $g_{V\mu e}$, which is then converted to an upper bound on $\Gamma( V \to \mu e )$.  
A similar approach has been used in Ref.~\cite{Gutsche:2009vp} where instead of $\mu \to eee$, the authors considered the $\mu-e$ conversion in nuclei ($N$), which they described in terms 
of a product of couplings $g_{V\mu e}$ and $g_{V N N}$. 
The latter could be extracted from the experimentally measured $\Gamma(V \to p\bar p)$, and with that knowledge the experimental upper bound on  $\text{R}(\mu \text{Ti}\to e\text{Ti})$ results in an upper bound 
on $\Gamma( V \to \mu e )$. 
A more dynamical approach in modelling the $V \to \ell_\alpha\ell_\beta$ processes has been made in a supersymmetric extension of the SM with type I Seesaw~\cite{Sun:2012yq}. 

Sterile fermions were proposed in various neutrino mass generation mechanisms, but the interest in their properties was further motivated by the reactor/accelerator 
anomalies~\cite{Mueller:2011nm,Huber:2011wv,Mention:2011rk,Aguilar:2001ty,AguilarArevalo:2007it,AguilarArevalo:2010wv,Aguilar-Arevalo:2013pmq,Acero:2007su,Giunti:2010zu},  a possibility to offer a warm dark matter 
candidate~\cite{Dodelson:1993je,Abazajian:2001nj,Dolgov:2000ew,Boyarsky:2009ix}, and by indications from the large scale structure
formation~\cite{Klypin:1999uc,Moore:1999nt,Strigari:2010un,BoylanKolchin:2011de,Kusenko:2009up,Abazajian:2012ys}.

Incorporating neutrino oscillations (masses and mixing~\cite{Gonzalez-Garcia:2014bfa}) into the  SM implies that the  charged current 
is modified to 
\begin{equation}\label{eq:cc-lag1}
- \mathcal{L}_\text{cc} = \frac{g}{\sqrt{2}} U^{\alpha i} 
\bar{\ell}_\alpha \gamma^\mu P_L \nu_i  W_\mu^- + \, \text{c.c.}\,,
\end{equation} 
$U$ being the leptonic mixing matrix,  $\alpha$ the flavour of a charged lepton, and 
$i = 1, \dots, n_\nu$ denotes a physical neutrino state. If one assumes that  only three massive neutrinos are present,  
the matrix $U$ corresponds to the unitary Pontecorvo-Maki-Nakagawa-Sakata (PMNS) matrix. In that situation the GIM mechanism makes the decay 
rates B($V \to \ell_\alpha^\mp \ell_\beta^\pm$) completely negligible, $\lesssim10^{-50}$. That feature, however, can be drastically 
changed in the presence of a non-negligible mixing with heavy sterile fermions.
In what follows we will consider such situations, derive analytical expressions for B($V \to \ell_\alpha\ell_\beta$), and discuss the impact of the ISS(2,3) model (cf. Sections~\ref{Sec:intro}-\ref{Sec:analysis}).
We will also discuss a 
simplified model in which the effect of the heavy sterile neutrinos is described by one effective sterile neutrino state with non-negligible mixing with active neutrinos.\footnote{In this analysis, 
due to the tension between the most recent Planck results on extra light neutrinos (relics) and the reactor/accelerator  anomalies, we will consider the effect of (heavier) sterile neutrinos not 
contributing as light relativistic degrees of freedom~\cite{Ade:2013zuv}. We will require our models to be compatible with current experimental data and constraints and to fulfil the so-called perturbative 
unitary condition which puts a strong constraint on the models for the very heavy sterile fermion(s)~\cite{Chanowitz:1978mv}.} 
Despite several differences, our approach is similar to the one discussed in Ref.~\cite{Ilakovac:1999md}, where the SM has been extended by new, heavy, Dirac neutrinos, singlets under $SU(2)\times U(1)$, and 
applied to a number of low energy decay processes. Our sterile neutrinos are Majorana and we apply the approach to the leptonic decays of quarkonia for the first time. 

The remainder of this chapter is organised as follows: In Sec.~\ref{sec:obs} we formulate the problem in terms of a low energy effective theory of a larger theory which contains heavy sterile neutrinos, 
we derive expression for B($V \to \ell_\alpha  \ell_\beta $) and compute the Wilson coefficients.  In Sec.~\ref{sec:ext} we briefly describe the specific models with sterile neutrinos which are used in this paper to produce our results presented 
in Sec.~\ref{sec:results}.

\section{LFV decay of Quarkonia - Effective Theory\label{sec:obs}}

In this section we formulate a low energy effective theory of the LFV decays of quarkonia of type $V\to \ell_\alpha^\pm\ell_\beta^\mp$, and express the decay amplitude in terms of the quarkonium decay constants and the corresponding Wilson coefficients. 
The latter are then computed in the extensions of the SM which include the heavy sterile neutrinos.  We also derive the expression relevant to  $\Gamma(Z\to \ell_\alpha^\pm\ell_\beta^\mp)$.

\subsection{Effective Hamiltonian} 

Keeping in mind the fact that we are extending the SM by adding sterile fermions, without touching the gauge sector of the theory, the decays of vector quarkonia, 
$V (q) \to \ell_\alpha^\pm (p)\ \ell_\beta^\mp (q-p)$, can only occur through the photon and 
the $Z$-boson exchange at tree level. In the lepton flavour conserving processes the $Z$-exchange terms are very small with respect to those arising from the electromagnetic interaction and are usually neglected. The generic effective Hamiltonian can be written as
\begin{equation}\label{Heff:V}
	\mathcal{H}_{\text{eff}} = {\cal Q}_Q \frac{e^2 g^2}{2 m_V^2} \, \bar{Q} \gamma_\mu Q\, \cdot\, \bar{\ell}_\alpha \left[C_{VL} \gamma^\mu P_L +C_{VR} \gamma^\mu P_R +  \frac{p^\mu}{m_W} (C_R P_R + C_L P_L)\right] \ell_\beta,
\end{equation}
where $ {\cal Q}_Q$ is the electric charge of the quark $Q$, $m_V$ is the mass of quarkonium $V$ which is dominated by the valence quark configuration $\bar Q Q$,~\footnote{We remind the reader that the ground vector meson $\bar s s$, $\bar c c$, $\bar b b$ states are $\phi$, $J/\psi$, $\Upsilon$, respectively, and the corresponding charges are  $ {\cal Q}_{s,b}=-1/3$ and  $ {\cal Q}_c=2/3$.}  $C_{VL,VR,L,R}$ are the Wilson coefficients,  $p$ is the momentum of one of the outgoing leptons, and $P_{L/R}= \frac{1}{2}(1 \mp \gamma_5)$. Contributions to the scalar (left and right) terms are suppressed by $m_{\alpha,\beta}/m_W$, where $m_{\alpha,\beta}$ are the charged lepton masses. In this section we will keep such terms so that our expressions can be useful to approaches in which the scalar bosons are taken  in consideration. For our phenomenological discussion, however, it is worth emphasising that $C_{L,R,VR}\ll C_{VL}$. 
\begin{figure}[htb]
\begin{center}
\includegraphics[width=0.45\textwidth]{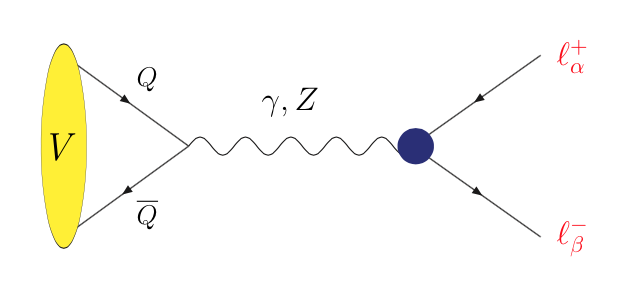}~\includegraphics[width=0.45\textwidth]{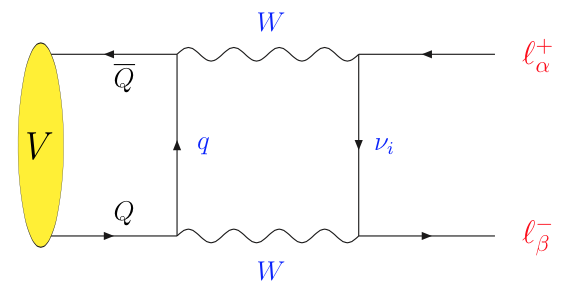}
\caption{{\footnotesize 
Diagrams contributing the LFV decay of quarkonia $V\to \ell_\alpha \ell_\beta$. The blob in the first diagram is related to the penguin loop that generates the LFV, and the box diagram is particularly important to be included in the case of 
$\Upsilon^{(n)}\to   \ell_\alpha \ell_\beta$ because of $V_{tb}\simeq 1$ and of the top quark mass, making the box diagram contribution to the Wilson coefficient significant.
 }}
\label{fig:1}
\end{center}
\end{figure}

Without entering the details of calculation it is easy to verify that the only relevant diagrams are those shown in Fig.~\ref{fig:1},
and therefore the structure of  the Wilson coefficients $C_i$ reads,
\begin{equation}
C_{i}=C^\gamma_{i} +  C^Z_{i} \frac{1}{\sin^2\theta_W\cos^2\theta_W}\frac{m_V^2}{m_V^2-m_Z^2} \frac{g_{V}^{\cal Q}}{{\cal Q}_Q}+C^{\mathrm{Box}}_{i} |V_{Qq}|^2\frac{1}{\sin^2\theta_W} \frac{m_V^2}{m_W^2} \frac{1}{{\cal Q}_Q},
\end{equation}
where $C^{\gamma,Z}_i$ are  the contributions arising from the photon and the $Z$-boson exchange, while $C_i^{\mathrm{Box}}$ comes from the box diagram that  involves the Cabibbo-Kobayashi-Maskawa 
coupling $V_{Qq}$.~\footnote{The box diagram contribution to $V \to \ell_\alpha\ell_\beta$ in the case of $V=\Upsilon$ is dominated by the top quark ($|V_{tb}|\simeq 1$); for $V= \psi$ it is negligible because 
the contribution of the $b$ quark is Cabibbo suppressed ($|V_{cb}|\simeq 0.004$) while the Cabibbo allowed one ($|V_{cs}|\simeq 0.99$) is suppressed by the strange quark mass; for $V=\phi$, the 
contributions of the charm and top quarks are comparable but overall smaller than in the $\Upsilon \to \ell_\alpha\ell_\beta$ case. 
} In the above expressions $g_V^Q=\frac{1}{2}I_3^Q-{\cal Q}_Q \sin^2\theta_W$.
The blob in the diagram shown in Fig.~\ref{fig:1} stands for the lepton loop diagrams that may contain one or two neutrino states and which, in the extensions of the SM involving a heavy neutrino sector, will give rise to the LFV decay  due to the effect of mixing which is parametrised by the matrix $U$ [see Eq.~(\ref{eq:cc-lag1})]. 
Separate contributions coming from different diagrams can be further reduced by factoring out the neutrino mixing matrix elements, namely
\begin{equation}
C_{i}^{\gamma,\mathrm{Box}}=\sum_{k=1}^{n_\nu} U_{\beta k} U_{\alpha k}^* C^{\gamma,\mathrm{Box}; k}_{i},  \quad \text{and} \quad C_{i}^Z=\sum_{k=1}^{n_\nu} U_{\beta k} U_{\alpha k}^* C^{Z, k}_{i}+\sum_{k=1}^{n_\nu}\sum_{j=1}^{n_\nu} U_{\beta k} U_{\alpha j}^* C^{Z, kj}_{i},
\end{equation}
where we see that the term involving two neutrino eigenstates appears only in the $Z$ coefficient because it is related to the vertex $Z \nu_k\nu_j$.
It is worth emphasising that the tensor structure in Eq.~(\ref{Heff:V}) 
can be easily obtained from the coefficients $C_{L,R}$  by applying the Gordon identity. 
Such contributions are $1/m_W$ suppressed, and thus completely negligible,  which is why we do not give explicit expressions for these coefficients.

Using the effective Hamiltonian (\ref{Heff:V}) and parameterising the hadronic matrix as 
\bea
\langle 0 \vert  \bar Q \gamma_\mu Q \vert V(q,\sigma) \rangle = f_V m_V \varepsilon_\mu^\sigma\,, 
\eea
where $f_V$ is the decay constant of a quarkonium $V$ with momentum $q$ and in a polarisation state $\sigma$, we can write the decay rate as,  
\begin{align}
\Gamma(V\to{\ell}_\alpha^- \ell_\beta^+ ) = \frac{8 \pi {\cal Q}_Q^2 \alpha^2}{3 m_V^3}  G_F^2 m_W^4 \left(\frac{f_V}{m_V}\right)^2  {\lambda^{1/2}(m_V^2,m_\alpha^2,m_\beta^2)} \phi_C,
\end{align}
with 
\begin{equation}
\lambda(a^2,b^2,c^2)=[a^2-(b-c)^2][a^2-(b+c)^2],\label{lambdadef}\end{equation} and 
\begin{align}
\phi_C=\left(-g^{\mu \nu}+\frac{q^\mu q^\nu}{m_V^2} \right) \mathrm{tr} \Big[(\slashed{q}-\slashed{p}+m_\beta) \cdot (C_{VL} \gamma^\mu P_L + C_{VR} \gamma^\mu P_R+C_L \frac{p^\mu}{m_W}P_L+C_R \frac{p^\mu}{m_W}P_R) \nonumber \\
\cdot (\slashed{p}-m_\alpha)\cdot (C_{VL}^* \gamma^\nu P_L + C_{VR}^* \gamma^\nu P_R+C_L^* \frac{p^\nu}{m_W}P_R+C_R^* \frac{p^\nu}{m_W}P_L)\Big],
\end{align}
which gives
\begin{align}
\phi_C=\frac{1}{4 m_V^2 m_W^2} &\Big{\lbrace}\lambda(m_V^2,m_\alpha^2,m_\beta^2)\Big[(m_V^2-m_\alpha^2-m_\beta^2)(|C_L|^2+|C_R|^2)-4\mathrm{Re} (C_L^* C_R) m_\alpha m_\beta \nn \\ 
&+4 m_W\mathrm{Re}(C_L^* (C_{VL} m_\beta+C_{VR} m_\alpha)+C_R^* (C_{VL} m_\alpha+ C_{VR} m_\beta))\Big] \nn\\
&+ 4 m_W^2 (|C_{VL}|^2+|C_{VR}|^2) \Big[2 m_V^4 - m_V^2(m_\alpha^2+m_\beta^2)- (m_\alpha^2-m_\beta^2)^2 \Big]\\
&+ 48 m_W^2 m_V^2 m_\alpha m_\beta   \mathrm{Re} (C_{VL}^* C_{VR}) \nn 
 \Big{\rbrace}.
\end{align}
As we mentioned above, we consider in our framework $C_{VL}\gg C_{VR,R,L}$, and therefore we can write 
\begin{align}
\Gamma (V\to \ell_\alpha^\pm\ell_\beta^\mp)=\frac{32\pi {\cal Q}_Q^2 \alpha^2}{3 m_V^3}  f_V^2 G_F^2 m_W^4   |C_{VL}|^2 
 \lambda^{1/2}(m_V^2,m_\alpha^2,m_\beta^2)\Big{[}1-  \frac{(m_\alpha^2+m_\beta^2)}{2 m_V^2}-\frac{(m_\alpha^2-m_\beta^2)^2}{2 m_V^4}\Big{]},
\end{align}
where $\lambda(a^2,b^2,c^2)$ is given in Eq.~(\ref{lambdadef}). In this last expression we also used $\Gamma (V\to \ell_\alpha^\pm\ell_\beta^\mp) = \Gamma (V\to \ell_\alpha^+\ell_\beta^-)+\Gamma (V\to \ell_\alpha^-\ell_\beta^+)$.

Besides quarkonia we will also revisit the issue of adding extra species of sterile neutrinos to the decay of $Z\to \ell_\alpha^\pm\ell_\beta^\mp$. In that case the effective Hamiltonian can be written as  
\begin{equation}
\mathcal{H}^Z_{\mathrm{eff}}= \frac{g^3}{2 \cos\theta_W} \bar{\ell}_\alpha \Big{[} D_{VL}\gamma^\mu P_L + D_{VR}\gamma^\mu P_R +D_{L}P_L+D_{R}P_R\Big{]}\ell_\beta Z^\mu, 
\end{equation}
where the Wilson coefficients are now denoted by $D_i$ and take the form  
\begin{equation}
	D_{i}=\sum_{k=1}^{n_\nu} U_{\beta k} U_{\alpha k}^* C^{Z, k}_{i}+\sum_{k =1}^{n_\nu}\sum_{j=1}^{n_\nu} U_{\beta k} U_{\alpha j}^* C^{Z, kj}_{i}\,.
\end{equation}
The decay rate in the similar limit, $D_{VL}\gg D_{VR,R,L}$, reads 
\begin{align}
\label{gammaz}
\Gamma(Z\to{\ell}_\alpha^- \ell_\beta^+ ) = \frac{8 \sqrt{2}}{3 \pi m_Z} \frac{G_F^3 m_W^6}{\cos^2\theta_W} |D_{VL}|^2 \lambda^{1/2}(m_Z^2,m_\alpha^2,m_\beta^2) \left[1-\frac{(m_\alpha^2+m_\beta^2)}{2 m_Z^2}-\frac{(m_\alpha^2-m_\beta^2)^2}{2 m_Z^4}\right].
\end{align}

\subsection{Wilson coefficients}
Concerning the computation of the Wilson coefficients we stress again that our results are obtained in a theory in which the Standard Model is extended to include extra species of sterile fermions, without changing the gauge sector. 
The origin  of the leptonic mixing matrix $U$ is model dependent and in order to be able to do a phenomenological analysis,  we will have to adopt a specific model which will be discussed in the next section. 

The blob in the diagram shown in Fig.~\ref{fig:1} stands for a series of diagrams such as those displayed in Fig.~\ref{fig:2}. All of them, including the box diagram in Fig.~\ref{fig:1}, have been computed in the Feynman gauge and the results are collected in Appendix~\ref{sec:app_quarkA}.
\begin{figure}[htb]
\hspace*{-11mm}\begin{tabular}{cccccc}
\includegraphics[width=0.3\textwidth]{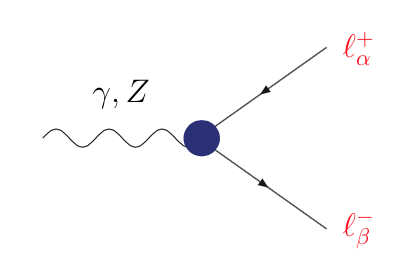}&$\raise 7.4ex \hbox{=}$&\includegraphics[width=0.3\textwidth]{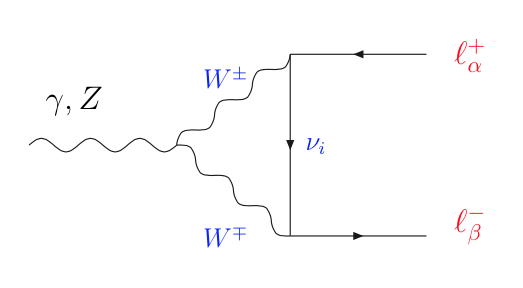}&$\raise 7.4ex \hbox{+}$&\includegraphics[width=0.3\textwidth]{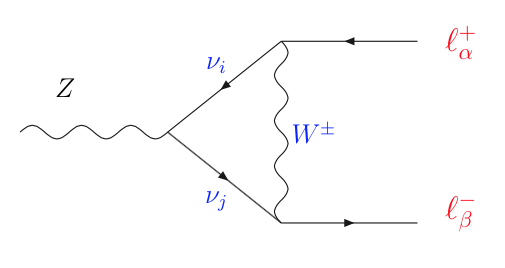}&$\raise 7.4ex \hbox{+~\dots}$
  \end{tabular}
  \caption{{\footnotesize 
Vertex diagrams contributing the LFV decays.
 }}
\label{fig:2}
\end{figure}
Here we focus on the most important contributions in the case of large masses of sterile (Majorana) neutrinos. Contributions to the Wilson coefficients coming from vertex diagrams can be divided into two pieces: those involving only one neutrino in the loop, $C_{VL}^{Z,\gamma}(x_i)$, where $x_i = m_i^2/m_W^2$, and those with two neutrinos in the loop, $C_{VL}^{Z }(x_i ,x_j)$. In the limit of large values of $x_{i,j} \gg 1$, we find the following behaviour 
\begin{align}\label{eq:Wils}
C_{VL}^{Z}(x_i)  \stackrel{x_{i}\gg 1}{\xrightarrow{\hspace*{9mm}}} &\,  \frac{5}{32\pi^2} \log x_i + \text{\small finite term} + {\cal O}(1/x_i) \sim \log x_i\, , \cr
 C_{VL}^{Z }(x_i ,x_i) \stackrel{x_{i}\gg 1}{\xrightarrow{\hspace*{9mm}}} &\,    \frac{C_{ii}}{64 \pi^2}\left\lbrace  \left( 2 x_i +3 -4 \log x_i\right)+x_i\left(\log x_i-\frac{7}{2}\right)\right\rbrace +\dots \cr  & \ \sim C_{ii}\ x_i \log x_i +  \dots 
\end{align}

To illustrate the relative contribution of the different diagrams we fix the values of the coefficients $C_{ij}\equiv \displaystyle{\sum_{\alpha=e,\mu,\tau}} U_{\alpha i}^\ast U_{\alpha j} = 10^{-5}$, and plot  $|C_{VL}(x_i)-C_{VL}(0)|$ and $|C_{VL}^{Z}(x_i,2\ x_i)-C_{VL}^{Z}(0,0)|$ for the case of $\Upsilon \to \mu\tau$, cf. Fig.~\ref{fig:3}.~\footnote{Due to the unitarity of the mixing matrix $U$, the terms in the Wilson coefficients that do not depend on neutrino masses give a vanishing contribution after summing over all neutrino states. We thus subtract the constant terms in the plots in order to better appreciate the dependence on the neutrino masses. Notice also that $C_{ij}= 10^{-5}$ is in agreement with all constraints discussed in the text when the neutrino masses are below $\mathcal{O}(100)$ TeV.}  We see that only for very large masses the diagrams with two neutrinos in the loop become more important than those with one neutrino state.  We should stress that each contribution to $C_{VL}(x_i)$, i.e.  $C_{VL}^{\rm Box}(x_i)$ and $C_{VL}^{Z}(x_i)$, scales as $\log x_i$  for large values of $x_i$, except for $C_{VL}^{\gamma}(x_i)$ which goes to a constant in the same limit. That can also be seen in Fig.~\ref{fig:3} where in the left panel we show the dependence of the total $C_{VL}(x_i)$ on $x_i$ and in the right panel we show  $C_{VL}^{\gamma}(x_i)$ and its dependence on the mass of the initial decaying meson, $\phi$, $J/\psi$, and $\Upsilon$.  The contribution of sterile neutrinos to the LFV decay of $\Upsilon$ is larger than the one to lighter mesons, since the Wilson coefficients are also proportional to the mass of the initial particle.

\begin{figure}[htb]
\begin{center}
\hspace*{-4mm}\includegraphics[width=0.45\textwidth]{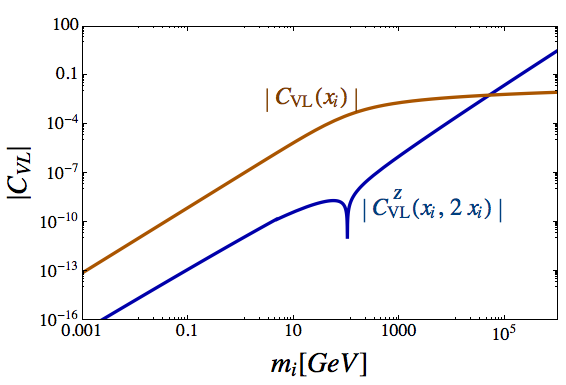}~\includegraphics[width=0.45\textwidth]{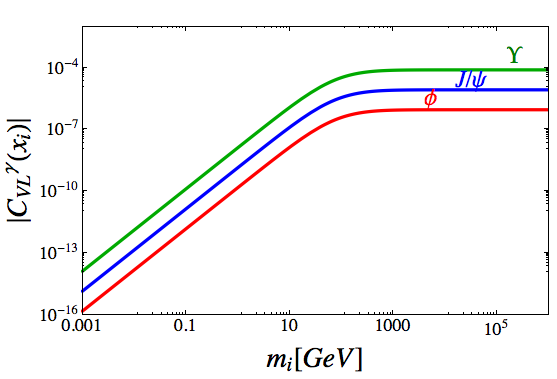}
\caption{{\footnotesize 
In the left panel are shown $C_{VL}(x_i)$ and $C_{VL}^Z(x_i ,x_j)$, for $x_j=2 x_i$,  as functions of $m_i =  m_W \sqrt{x_i}$,  the mass of the heavy sterile neutrino propagating in the loops. For illustration purpose, the couplings $C_{ij}$ were fixed to a common value, $10^{-5}$, and the example corresponds to the $\Upsilon \to \mu\tau$ decay. Right panel:  $C_{VL}^{\gamma}(x_i)$  is plotted as a function of $m_i$ for the case of $V\to e\mu$ in three specific cases $V\in \{\phi, J/\psi,\Upsilon\}$. In both cases the value of functions at $x_{i,j}=0$ have been subtracted away. }}
\label{fig:3}
\end{center}
\end{figure}
 Before closing this section we should reiterate that our  Wilson coefficients have  been computed in the Feynman gauge. Since  all divergencies cancel out,  our results are finite and gauge invariant, as was already  observed in Refs.~\cite{Mann:1983dv,Ilakovac:1994kj,Illana:1999ww,Abada:2014cca}.

\section{SM in the presence of sterile fermions \label{sec:ext} }
With the expressions derived above, we now have to specify a model for lepton mixing (couplings) $U_{\alpha i}$ in  the presence of heavy sterile neutrinos propagating in the loops. 
We opt for a minimal realization of the inverse Seesaw mechanism for the generation of neutrino masses, which is nowadays rather well constrained by the available experimental data. 
Furthermore, we will use a parametric model containing one effective sterile neutrino, which essentially mimics the behaviour at low energy scales of mechanisms involving heavy sterile fermions.

\subsection{The ISS(2,3) realization}
Among many possible realisations of accounting for massive neutrinos, we provide predictions for the LFV rates in the ISS(2,3) model, cf. Sections(\ref{Sec:intro}-\ref{Sec:analysis}). In fact the inverse Seesaw mechanism (ISS)~\cite{Mohapatra:1986bd,GonzalezGarcia:1988rw,Deppisch:2004fa} 
offers the possibility of accommodating
the smallness of the active neutrino masses for a comparatively  low Seesaw scale, but still with natural $\mathcal{O}(1)$
Yukawa couplings, rendering  this scenario phenomenologically appealing. Hence, depending on their masses and mixing with active neutrinos, the new  states can be
produced in collider and/or low energy experiments, and their
contribution to physical processes can be sizeable.

\subsection{A model with one effective sterile fermion}

Since the generic idea of obtaining a significant contribution to our observables applies to any model
in which the active neutrinos have sizeable mixing with some additional singlet states (sterile fermions),  
we can use an \textit{effective} model with three light active neutrinos plus one extra sterile neutrino.

The introduction of this extra state implies three new active-sterile mixing angles ($ \theta_{14}, \theta_{24}, \theta_{34}$), two extra Dirac $CP$ violating phases ($\delta_{14},\delta_{34}$) and one additional Majorana phase ($\phi_{41}$). 
The lepton mixing matrix is then a product of six rotations times the Majorana phases, namely
\begin{eqnarray} \label{eq:3+1rot}
U &=& R_{34}(\theta_{34},\delta_{34}) \cdot R_{24}(\theta_{24}) \cdot R_{14}(\theta_{14},\delta_{14}) 
\cdot R_{23} \cdot R_{13} \cdot R_{12}  \cdot \rm diag(\phi_{21},\phi_{31},\phi_{41}) \nonumber \\
&=& R_{34}(\theta_{34},\delta_{34}) \cdot R_{24}(\theta_{24}) \cdot R_{14}(\theta_{14},\delta_{14}) 
\cdot U_{\rm PMNS} \cdot \rm diag(\phi_{21},\phi_{31},\phi_{41})\,,
\end{eqnarray} 
where the rotation matrices $R_{34},R_{24},R_{14}$ can be defined as:
\begin{eqnarray} \label{eq:R}
R_{34}\ &=&\ \left( 
\begin{array}{cccc}
1 & 0 & 0 & 0 \\
0 & 1 & 0 & 0 \\
 0 & 0 & \rm cos \theta_{34} &\rm sin \theta_{34} \cdot e^{-i \delta_{34}}\\ 
 0 & 0 & -\rm sin \theta_{34} \cdot e^{i \delta_{34}}& \rm cos \theta_{34} 
\end{array}
\right)\,, \nonumber \\
R_{24}\ &=&\  \left( 
\begin{array}{cccc}
1 & 0 & 0 & 0 \\
0 & \rm cos \theta_{24}  & 0 & \rm sin \theta_{24}\\ 
0 & 0 & 1 & 0 \\
0 & - \rm sin \theta_{24}& 0 &  \rm cos \theta_{24} 
\end{array}
\right)\,, \nonumber \\
R_{14}\ &=& \left( 
\begin{array}{cccc}
\rm cos \theta_{14} & 0 & 0 & \rm sin \theta_{14} \cdot e^{-i \delta_{14}} \\
0 & 1 & 0 & 0 \\
0 & 0 & 1 & 0 \\
- \rm sin \theta_{14} \cdot e^{i \delta_{14}} & 0 & 0 &\rm cos \theta_{14} \\
\end{array}
\right) \,.
\end{eqnarray}

In the framework of the SM extended by sterile fermion states, which
have a nonvanishing mixing with active neutrinos, 
the Lagrangian describing the leptonic charged currents becomes
\begin{equation}\label{eq:cc-lag}
- \mathcal{L}_\text{cc} = \frac{g}{\sqrt{2}} U^{\alpha i} 
\bar{\ell}_\alpha \gamma^\mu P_L \nu_i  W_\mu^- + \, \text{c.c.}\,,
\end{equation}
where $i = 1, \dots, n_\nu$ denotes the physical neutrino states, 
and $\alpha = e, \mu, \tau$ are the flavours of the charged leptons. 
In the case of the SM with three neutrino generations,  $U$ is the  PMNS matrix, while in the case of $n_\nu \geq 4$, the 
$3\times 3$ submatrix ($\widetilde U_\text{PMNS}$) is not unitary anymore and one can parametrise it as
\begin{equation}\label{eq:U:eta:PMNS2}
U_\text{PMNS} \, \to \, \widetilde U_\text{PMNS} \, = \,(\mathbb{1} -\widetilde \eta)\, 
U_\text{PMNS}\,,
\end{equation}
where $\widetilde \eta$ is a matrix that accounts for the deviation of $\widetilde
U_\text{PMNS}$ from unitarity~\cite{Schechter:1980gr,Gronau:1984ct}, due to the presence of extra fermion states.
Many observables are sensitive to the 
active-sterile mixing and their current experimental values can be used to constrain the $\widetilde \eta$ matrix~\cite{Antusch:2014woa}. 

In order to express the deviation from unitarity in terms of a single parameter, we define
\bea
\eta = 1- |\det \widetilde U_\text{PMNS}|  \,,
\eea
which, in the case of the extension of the SM by only one sterile fermion and in terms of the mixing angles defined above, reads
\bea
\eta = 1 - \vert \cos\theta_{14} \cos\theta_{24}\cos\theta_{34}\vert \,.
\eea
\section{Results and discussion}\label{sec:results}

In this section we present and discuss our results.

Since the Wilson coefficients of the processes discussed here are proportional to the mass of the decaying particle, it is quite obvious that the most significant enhancement of B($V \to \ell_\alpha  \ell_\beta$) will occur for $V=\Upsilon$ and its radial excitations. 
For this reason we will present plots of our results for this decay channel. Plots for other channels are completely similar which is why we do not display them. Before we discuss the impact of the active-sterile neutrino mixing on the LFV decay rates further, we first specify the constraints on parameters of
both of our models.  
 
In Fig.~\ref{fig:4} (left panel), we plot the dependence of $\eta$ with respect to the  mass of the effective sterile neutrino $m_4$. 
\begin{figure}[htb] 
\begin{center}
\hspace*{-14mm}\begin{tabular}{cc}
\includegraphics[width=0.45\textwidth]{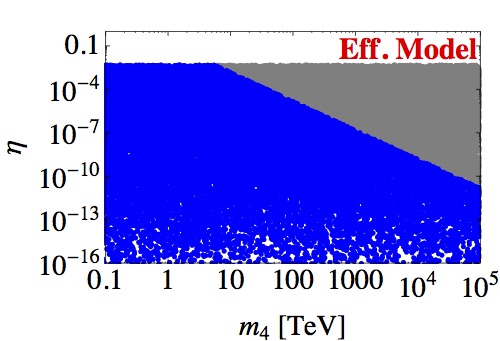}&\includegraphics[width=0.45\textwidth]{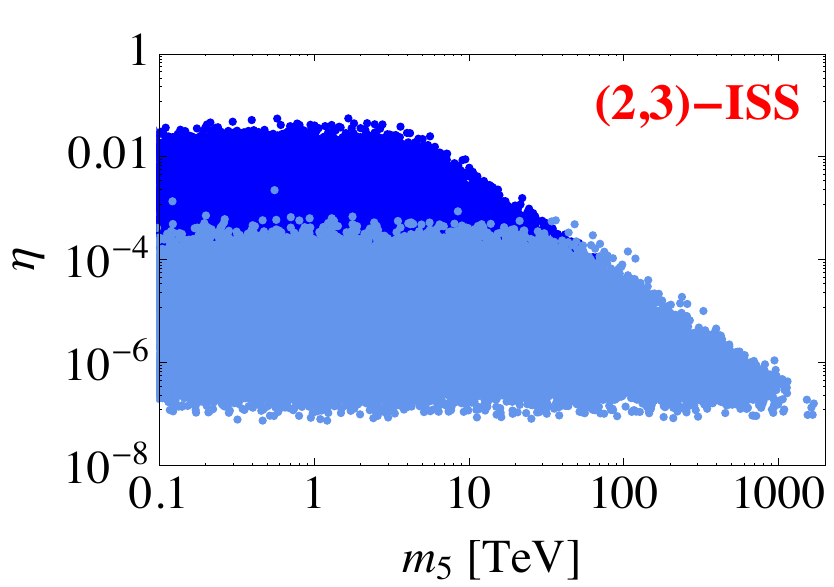} \cr
  \end{tabular}
  \caption{{\footnotesize 
$\eta$ parameter, which  parametrises the size of mixing between the active and heavy sterile states, is plotted vs the mass of the heavy sterile state. The grey points (left panel) correspond to solutions complying with all experimental data and constraints discussed in the text except for perturbative unitary condition~(\ref{pert-uni}), which we then applied to obtain  the region of dark-blue points. 
In the case of the ISS(2,3) model (right-panel), 
we  further imposed constraints of Ref.~\cite{Antusch:2014woa} on the matrix $\widetilde \eta$, as well as the bound $B(\mu\rightarrow eee)<10^{-12}$, resulting in the bright-blue region of points. 
 }}
\label{fig:4}
\end{center}
\end{figure}
Grey points in that plot are obtained by varying the mass of the lightest neutrino, $m_{\nu_e}\in (10^{-21}, 1)$~eV, and by 
imposing the following constraints: (i) Neutrino data (masses and mixing angles) respect the normal hierarchy, with $\Delta m_{21}^2 = 7.5(2) \times 10^{-5}$~eV, and $\Delta m_{31}^2 = 2.46(5) \times 10^{-3}$~eV~\cite{Gonzalez-Garcia:2014bfa}. We checked to see that our 
final results do not change in any significant manner if the inverse hierarchy is adopted. Furthermore, we vary the three mixing angles with the fourth neutrino by assuming $\theta_{i4}\in (0, 2\pi ]$, while keeping the other three mixing angles to 
their best-fit values, namely $\sin^2 \theta_{12} =0.30(1)$,  $\sin^2 \theta_{23} =0.47(4)$,  $\sin^2 \theta_{13} =0.022(1)$~\cite{Gonzalez-Garcia:2014bfa}. (ii) The selected points satisfy the upper bound ${\rm B}(\mu \to e\gamma)< 5.7 \times 10^{-13}$~\cite{Adam:2013mnn}. (iii) The results for 
$R_\pi = \Gamma(\pi \to e \bar \nu_e)/ \Gamma(\pi \to \mu \bar \nu_\mu )$, $R_K$, $\Gamma(W\to \ell\nu_\ell)$, and $\Gamma(Z\to \text{invisible})$, remain consistent with experimental findings. 
We see that for all (heavy) sterile neutrino masses the unitarity breaking parameter is $\eta \lesssim 0.005$. 
That parameter space is not compatible with the perturbative unitarity requirement, which for $m_{4}\gg m_W$ translates into~\cite{Ilakovac:1994kj},~\footnote{To write it in the form given in Eq.~(\ref{pert-uni}), we replaced $\alpha_W=g^2/(4\pi) = \sqrt{2} G_F m_W^2/\pi$. }
\bea\label{pert-uni}
{G_F m_4^2 \over \sqrt{2}\pi}\sum_\alpha \vert U_{\alpha 4}\vert^2  < 1\,.
\eea
The resulting region, i.e. the one that satisfies constraints (i), (ii), (iii) and Eq.~(\ref{pert-uni}), is depicted by blue points (the dark region) in Fig.~\ref{fig:4}, where we see that  the parameter $\eta$ is indeed diminishing with the increase of the heavy sterile mass $m_4$. In other words, the decoupling of a very heavy sterile neutrino entails the unitarity of the $3\times 3$ submatrix $\widetilde U_\text{PMNS}$. Decoupling from active neutrinos for very large masses was also explicitly emphasised in Ref.~\cite{Alonso:2012ji}.
We should mention that, besides the above constraints, we also implemented the constraint coming from ${\rm B}(\mu \to eee) < 10^{-12}$~\cite{Bellgardt:1987du}, but it turns out that the present experimental bound does not bring any additional improvement.  

By imposing the constraints (i) and Eq.~(\ref{pert-uni}) on the ISS(2,3) model, we get a similar region of allowed (blue) points in the right panel of Fig.~\ref{fig:4}. A notable difference with respect to the situation with one effective sterile neutrino is that the region of very small mixing angles is excluded due to relations between the active neutrino masses and the active-sterile neutrino mixing, cf. Section~\ref{ISS}. For very heavy $m_5$, on the other hand, the range of allowed $\eta$'s shrinks and eventually vanishes with $m_5\to \infty$.~\footnote{We recall that, in the ISS(2,3) model, $m_4$ stands for the mass of the light sterile state whose impact on the decays discussed here is negligible [as seen from Eq.~(\ref{eq:Wils})], while $m_5 > m_4$ can be large and is important for B($V\to \ell_\alpha\ell_\beta$).} 
Furthermore, we use the results of Ref.~\cite{Antusch:2014woa} which are derived in the minimal unitarity violation scheme in which the heavy sterile neutrino fields are integrated out, and therefore the observables computed in that scheme are functions of the deviation of PMNS matrix from unitarity only~\cite{Antusch:2006vwa,FernandezMartinez:2007ms,Antusch:2008tz}. We adapt and apply them to our ISS(2,3) model and get a region of the bright-blue points, as shown in the right panel of Fig.~\ref{fig:4}. 
To further constrain the parameter space we find it useful to account for the experimental bound on ${\rm B}(\mu \to eee)<1\times 10^{-12}$, as is discussed in Refs.~\cite{Ilakovac:1994kj,Deppisch:2005zm,Dinh:2012bp,Alonso:2012ji}.
This latter constraint appears to be superfluous in most of the parameter space, once the constraints of Eq.~(\ref{pert-uni}) and Ref.~\cite{Antusch:2014woa} are taken into account, except in the range $10\ \mathrm{TeV}\lesssim m_5 \lesssim 100\ \mathrm{TeV}$, where the bound ${\rm B}(\mu \to eee)<1\times 10^{-12}$ restricts the parameter space relevant to B($V \to e \mu$).

We also mention that we attempted implementing the constraints coming from various laboratory experiments, summarised in Ref.~\cite{Atre:2009rg}, but since those results only impact the region of relatively small sterile neutrino masses ($m_5 \lesssim 100$ GeV), they are of no relevance to the present study. 
\begin{figure}[htb]
\begin{center}
\hspace*{-11mm}\begin{tabular}{cc}
\includegraphics[width=0.45\textwidth]{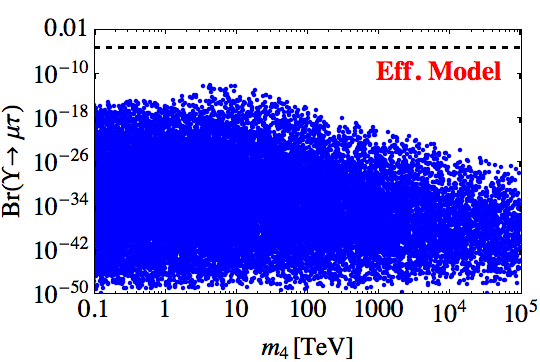}&\includegraphics[width=0.45\textwidth]{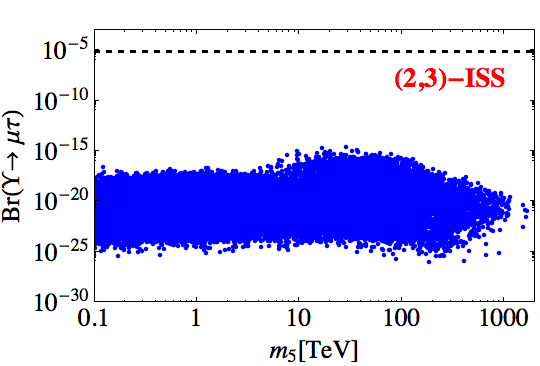} \cr
\includegraphics[width=0.45\textwidth]{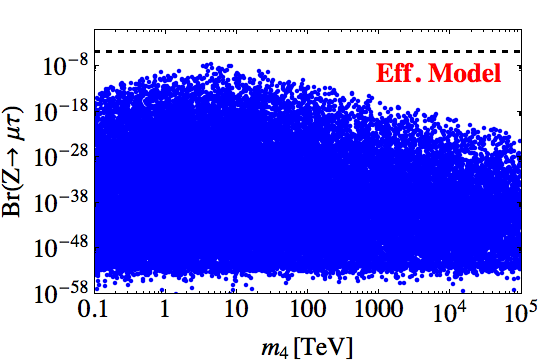}&\includegraphics[width=0.45\textwidth]{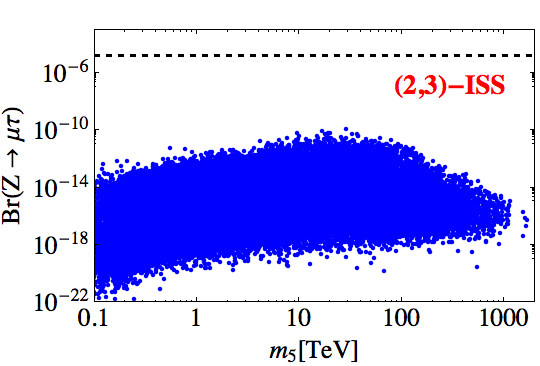} \cr
  \end{tabular}
  \caption{{\footnotesize 
${\rm B}(\Upsilon \to \mu \tau)$ and ${\rm B}(Z \to \mu \tau)$ are shown as functions of the heavy sterile neutrino(s) mass, and in both models considered in this paper. 
The points are selected in such a way that the models are consistent with the constraints discussed in the text and shown in Fig.~\ref{fig:4}.
Dashed horizontal lines correspond to the current experimental upper bounds for these decay rates. Notice again that the mass of the heavy sterile neutrino is denoted as $m_4$ in the effective model, and $m_5$ in the ISS(2,3) model because the latter contains a lighter sterile neutrino state, the impact of which  is negligible on the decay modes discussed here. 
 }}
\label{fig:5}
\end{center}
\end{figure}

\begin{table}[htb]
\renewcommand{\arraystretch}{1.5}
\centering{}
\resizebox{\textwidth}{!}{
\begin{tabular}{|cc|ccc|ccc|}
\hline 
$V$ & $\ell_\alpha \ell_\beta$ & $m_4 = 1$~TeV & $10$~TeV & $100$~TeV& $m_5 = 1$~TeV & $10$~TeV & $100$~TeV \\ \hline\hline
$\phi$ & $e\mu$  & $1 \times 10^{-24}$ & $5 \times 10^{-24}$ & $3 \times 10^{-24}$ & $1\times 10^{-23}$ & $6 \times 10^{-23}$ & $5 \times 10^{-23}$ \\ \hline
$J/\psi$ & $e\mu$  & $2 \times 10^{-21}$ & $3 \times 10^{-20}$ & $6 \times 10^{-21}$ & $2\times 10^{-20}$ & $9 \times 10^{-20}$ & $7 \times 10^{-20}$ \\  
             & $e\tau$  & $5 \times 10^{-18}$ & $8\times 10^{-17}$ & $2\times 10^{-19}$ & $1 \times 10^{-19}$ & $3\times 10^{-18}$ & $1 \times 10^{-19}$ \\  
             & $\mu\tau$  & $8 \times 10^{-18}$ & $6 \times 10^{-16}$ & $3 \times 10^{-20}$ & $4 \times 10^{-19}$ & $4 \times 10^{-18}$ & $8 \times 10^{-19}$ \\ \hline
$\psi (2S)$ & $e\mu$  & $9 \times 10^{-22}$ & $ 1.5 \times 10^{-20}$ & $3 \times 10^{-21}$ & $4 \times 10^{-21}$& $3\times 10^{-20}$ & $2 \times 10^{-20}$ \\  
             & $e\tau$  & $5 \times 10^{-18}$ & $2 \times 10^{-17}$ & $9 \times 10^{-21}$ & $4\times 10^{-20}$ & $1\times 10^{-18}$ & $4 \times 10^{-20}$ \\  
             & $\mu\tau$  & $8 \times 10^{-18}$ & $3 \times 10^{-17}$ & $1.2 \times 10^{-20}$ & $1 \times 10^{-19}$ & $1\times 10^{-18}$ & $2\times 10^{-19}$ \\ \hline
$\Upsilon$ & $e\mu$  & $7 \times 10^{-18}$ & $2 \times 10^{-17}$ & $6 \times 10^{-18}$ & $2 \times 10^{-19}$ & $2\times 10^{-17}$ & $2\times 10^{-17}$ \\  
             & $e\tau$  & $5 \times 10^{-14}$ & $2 \times 10^{-13}$ & $9 \times 10^{-17}$ & $6\times 10^{-18}$ & $4\times 10^{-16}$ & $5\times 10^{-17}$ \\  
             & $\mu\tau$  & $5\times 10^{-16}$  & $2.5\times 10^{-13}$ & $1.2\times 10^{-16}$ & $1\times 10^{-17}$& $8\times 10^{-16}$ &$3\times 10^{-16}$ \\ \hline
$\Upsilon(2S)$ & $e\mu$  & $5\times 10^{-18}$  & $5\times 10^{-18}$ & $1.5 \times 10^{-18}$ & $2 \times 10^{-19}$ & $2\times 10^{-17}$ & $2\times 10^{-17}$ \\  
             & $e\tau$  & $1.8 \times 10^{-14}$ & $3 \times 10^{-14}$ & $3\times 10^{-18}$ & $8\times 10^{-18}$ & $5 \times 10^{-16}$ & $5\times 10^{-17}$ \\  
             & $\mu\tau$  & $2 \times 10^{-16}$ & $2 \times 10^{-13}$ & $2 \times 10^{-17}$ & $2\times 10^{-17}$ & $8 \times 10^{-16}$ & $3 \times 10^{-16}$ \\ \hline
$\Upsilon(3S)$ & $e\mu$  & $1.5 \times 10^{-17}$ & $3 \times 10^{-17}$ & $1.5 \times 10^{-17}$ & $5 \times 10^{-19}$ & $5 \times 10^{-17}$ & $4\times 10^{-17}$ \\  
             & $e\tau$  & $5.5 \times 10^{-14}$ & $3 \times 10^{-14}$ & $4 \times 10^{-17}$ & $2\times 10^{-17}$ & $1\times 10^{-15}$ & $1 \times 10^{-16}$ \\  
             & $\mu\tau$  & $2 \times 10^{-15}$ & $2 \times 10^{-12}$ & $4 \times 10^{-17}$ &  $3 \times 10^{-17}$ & $2 \times 10^{-15}$ & $6\times 10^{-16}$ \\ \hline
$Z$ & $e\mu$  & $1.2\times 10^{-14}$ & $7\times 10^{-13}$ & $4 \times 10^{-13}$& $9\times 10^{-14}$& $8 \times 10^{-13}$ & $6 \times 10^{-13}$ \\  
             & $e\tau$  & $2\times 10^{-10}$ & $9 \times 10^{-9}$ & $4 \times 10^{-13}$ & $7\times 10^{-13}$ & $4\times 10^{-11}$ & $2 \times 10^{-12}$ \\  
             & $\mu\tau$  & $5.5\times 10^{-10}$  & $3.5\times 10^{-8}$ & $1.6\times 10^{-12}$ & $3\times 10^{-12}$& $6\times 10^{-11}$ &$1\times 10^{-11}$ \\ \hline
\end{tabular}
}
\caption{{\footnotesize{}\label{tab:results} Upper bound on ${\rm B}(V\to \ell_\alpha \ell_\beta)$ for three values of the mass $m_{4,5}$. The numbers in the three columns referring to $m_4$ are obtained by using the effective model discussed in the text, while the other three, referring to $m_5$, are results of the ISS(2,3) model (also discussed in the text). }}
\end{table}
After having completed the discussion on several constraints, we present our results for branching fractions B($V \to \mu  \tau$) depending on the mass of heavy sterile neutrino(s). In Fig.~\ref{fig:5} 
we plot our results for $V=\Upsilon$ and $V=Z$, for which the enhancement is more pronounced. Other cases of $V$ result in similar shapes but the upper bound becomes lower. 
In Table~\ref{tab:results} we collect our results for three values of the heavy sterile neutrino(s) mass. 

To better appreciate the enhancement of the LFV decay rates shown in Fig.~\ref{fig:5}, we emphasize that both of them are ${\rm B}(V \to \mu  \tau) < 10^{-50}$ in the absence of heavy sterile neutrinos. 
Current experimental bounds in both cases are shown by dashed lines. Since those bounds are expected to improve in the near future, a possibility of seeing the LFV modes discussed in this paper might become realistic. 
Conversely, an observation of the LFV modes $V \to \ell_\alpha  \ell_\beta$, with branching fractions significantly larger than the bounds presented in Table~\ref{tab:results} would be a way to disfavour many of
the models containing heavy sterile neutrinos as being the unique source of lepton flavour violation. In obtaining the bounds presented in Table~\ref{tab:results} we used masses and decay constants listed in Appendix~\ref{sec:app_quarkB}. 
In presenting our results (the upper bounds) for lepton flavour violating modes, we used the parameters from Ref.~\cite{Antusch:2014woa} which were determined at  90\% C.L. 
For that reason, we treated all  other input data to  2 $\sigma$ as well. Therefore, our final results  in Table~\ref{tab:results} are also obtained at 2 $\sigma$ level.

Finally, we compare in Table~\ref{tab:compare} our upper bounds for the modes for which we could find predictions in the literature.
\begin{table}[htb]
\renewcommand{\arraystretch}{1.5}
\centering{}
\hspace*{-4mm}\begin{tabular}{|cccccc|}
\hline 
Mode & Ref.~\cite{Nussinov:2000nm} & Ref.~\cite{Gutsche:2009vp} &  Ref.~\cite{Sun:2012yq}  & Eff. model & ISS(2,3) \\ \hline 
${\rm B}(\phi \to e\mu)$ & $<4\times 10^{-17}$  & $<1.3 \times 10^{-21}$ & $<5\times 10^{-20}$ & $< 5\times 10^{-24}$ & $<6\times 10^{-23}$ \\  
${\rm B}(J/\psi \to e\mu)$ & $<4\times 10^{-13}$  & $<3.5 \times 10^{-13}$ & $<1.9\times 10^{-18}$ & $< 3\times  10^{-20}$ & $< 9\times 10^{-20}$ \\
${\rm B}(J/\psi \to \mu \tau )$ & $-$ & $-$ & $< 1.6 \times 10^{-7}$ & $< 6\times  10^{-16}$  & $< 4\times 10^{-18}$ \\ 
${\rm B}(\Upsilon \to e\mu)$ & $< 2\times 10^{-9}$  & $<3.8 \times 10^{-6}$ & $< 3.6\times 10^{-18}$ & $< 2\times  10^{-17}$ & $< 2\times 10^{-17}$ \\
${\rm B}(\Upsilon \to \mu \tau )$ & $-$ & $-$ & $< 5.3 \times 10^{-7}$ & $< 2.5\times  10^{-13}$  & $< 8\times 10^{-16}$ \\ 
${\rm B}(Z \to e\mu)$ & $< 5\times 10^{-13}$  & $< 8 \times 10^{-15}$ & $-$ & $< 7\times  10^{-13}$ & $< 8\times 10^{-13}$  \\ \hline
\end{tabular}\caption{{\footnotesize{}\label{tab:compare} Upper bounds ${\rm B}(V\to \ell_\alpha \ell_\beta)$: Comparison of the results reported in the literature with the bounds 
obtained in this work by using two different models (the last two columns). The bounds for other similar decay modes that have not been discussed in the literature can be found in 
Table~\ref{tab:results}. }}
\end{table}

\chapter{Loop level constraints on Seesaw neutrino mixing}\label{sec:intro}

The extension of the SM field content by the addition of right-handed sterile fermions represents a minimal and viable solution to account for the origin of neutrino masses and mixing and, depending on the specific model, can also provide a simultaneous solution for the nature of DM and the origin of the BAU, cf. Chapters~\ref{sec:nu_mass_gen}, \ref{sec:DMMISS} and \ref{introduction}.

Given their singlet nature, a Majorana mass term for the right-handed neutrinos is directly allowed in the Lagrangian, thus inducing a new mass scale -the only one unrelated to electroweak (EW) symmetry breaking- to be determined by data. As discussed in Section~\ref{sec:LNV_scale}, depending on the size of this scale its phenomenological consequences are very different. 
One of the most appealing choices is that this new Majorana scale is high, leading to the well-known Seesaw mechanism~\cite{Minkowski:1977sc,Mohapatra:1979ia,Yanagida:1979as,GellMann:1980vs} and providing a rationale for the extreme smallness of neutrino masses when compared to the rest of the SM fermions and the EW scale. Values for the neutrino Yukawa couplings ranging between that of the electron and that of the top quark would lead to Majorana masses between the EW and the grand unification scale. Unfortunately, even for the lightest mass choice, any phenomenological consequence beyond neutrino masses tends to be hopelessly suppressed if the extra degrees of freedom only couple to the SM through their Yukawa interactions.
However, the smallness of neutrino masses could derive from symmetry arguments~\cite{Mohapatra:1986bd,Bernabeu:1987gr,Branco:1988ex,Buchmuller:1990du} rather than a hierarchy of scales. Indeed, the Weinberg operator~(\ref{weinbergmass}) leading to neutrino masses in Seesaw mechanisms is protected by the $B-L$ symmetry, conserved in the SM and violated in two units by the Weinberg operator. Thus, if the pattern of the Yukawa couplings and Majorana masses in a Seesaw realisation is such that it conserves $B-L$, the Weinberg operator will never be generated and the SM neutrinos will remain massless, even for $Y_\nu \sim 1$ and Majorana masses of the order of the EW scale. Small violations of $B-L$ in this pattern would thus induce the small neutrino masses observed in oscillation experiments. In this class of models fall the popular inverse~\cite{Mohapatra:1986bd,Bernabeu:1987gr} or linear~\cite{Malinsky:2005bi} Seesaw mechanisms which, contrary to the canonical type-I Seesaw, would lead to an extremely rich phenomenology through the large mixing allowed between the new extra sterile neutrinos and their SM siblings implying observable contributions in lepton flavour violating (LFV) processes, universality violation and signals in electroweak precision observables. It is then of interest to fit all these available data to determine the allowed values of the mixing of the heavy neutrinos with the SM charged leptons, examples of these constraints can be found in Refs.~\cite{Langacker:1988ur,Bilenky:1992wv,Nardi:1994iv,Tommasini:1995ii,Bergmann:1998rg,Loinaz:2002ep,Loinaz:2003gc,Loinaz:2004qc,Antusch:2006vwa,Antusch:2008tz,Alonso:2012ji,Abada:2012mc,Akhmedov:2013hec,Basso:2013jka,Abada:2013aba,Antusch:2014woa,Antusch:2015mia}.

When deriving such constraints on heavy-active neutrino mixing, it was recently pointed out in~\cite{Akhmedov:2013hec} that loop corrections involving the extra heavy neutrinos played an important role, obtaining qualitatively different results to those derived by staying at leading order. In particular, it was shown that corrections to the $T$ parameter~\cite{Peskin:1990zt,Peskin:1991sw} could be sizeable and that these, in turn, would affect the determination of $G_F$ through $\mu$ decay competing with the tree level effects. Since the value of $G_F$ from $\mu$ decay is generally in good agreement with the measured value of $M_W$ and other determinations of $\sin \theta_W$, in~\cite{Akhmedov:2013hec} it was found that the constraints stemming from these datasets could be weakened at loop level through partial cancellations between the tree level corrections and the $T$ parameter contribution. Furthermore, the invisible width of the $Z$, which is in slight tension with the SM prediction, is modified at tree level through the presence of extra heavy neutrinos, while the oblique corrections computed in~\cite{Akhmedov:2013hec} were found to be subleading. Thus, by accounting for these loop corrections, good fits with relatively large heavy-active mixing were found in~\cite{Akhmedov:2013hec}, since it is possible to alleviate the tension in the invisible width of the $Z$ without seriously affecting the determination of $G_F$ in $\mu$ decay through the partial cancellation of the tree and loop level contributions. 

However, when Ref.~\cite{Basso:2013jka} also investigated the relevance of the $T$ parameter the same cancellation was not reproduced and 
in~\cite{Antusch:2014woa} it was argued that loop contributions should always be negligible, since the heavy-active mixing that controls the strength 
of the couplings of the new degrees of freedom is bounded to be small ($\theta^2 \lesssim 10^{-2}$). Therefore, new tree-level bounds were derived instead 
through more updated fits to available data. While this argument is generally true, models based on an approximate $B-L$ symmetry are characterised by large 
Yukawas and EW-scale Majorana masses, thus, even if loop corrections through weak interactions further suppressed by $\theta^2$ are indeed negligible for all 
practical purposes, when the loop corrections are mediated by heavy neutrinos and/or the Higgs field or its Goldstones, the coupling involved in the vertex is 
no other than the large Yukawa coupling, so that loop corrections can indeed become relevant, as stated in~\cite{Akhmedov:2013hec}. However, not only the oblique 
corrections computed in~\cite{Akhmedov:2013hec} fall in this category, since the effect of the large Yukawa interactions does not vanish in the limit of massless 
neutrinos and charged leptons. Indeed, some vertex and box corrections involving large Yukawas are found not to vanish in the massless limit for light leptons 
(see e.g.~\cite{Kniehl:1996bd}).

In this analysis we clarify the importance of loop contributions to the determination of the heavy-active neutrino mixing including all loop corrections 
mediated by the potentially large Yukawa interactions. We find that, as discussed by~\cite{Akhmedov:2013hec}, the most relevant of these corrections are 
indeed the ones encoded through the oblique parameters but, in order to make them competitive with the tree-level contributions, EW scale Majorana masses and 
Yukawas on the very border of perturbativity are simultaneously required. Furthermore, we find that, as long as $B-L$ is conserved, the $T$ parameter is always
 positive, so that the partial cancellation discussed in~\cite{Akhmedov:2013hec} cannot take place in such a setup. Large violations of $B-L$ are thus required 
to achieve the negative and sizeable values of $T$ capable of reproducing the effect. But these large violations of $B-L$ would render the Weinberg operator 
unprotected and, in presence of the EW-scale Majorana masses and large Yukawas required for $T$, radiative corrections lead to unacceptably large contributions 
to the light neutrino masses, even if these are fixed to their correct value at tree level by means of the Casas-Ibarra parametrisation. Thus, when we impose an 
approximate $B-L$ symmetry with only 3 extra heavy right-handed neutrinos, we always find that loop corrections are irrelevant when deriving bounds on the 
heavy-active neutrino mixings. 

This chapter is organised as follows: In Section~\ref{sec:param} we detail the parametrisation employed for our study. In Section~\ref{sec:obs} we list the observables we analyse in our global fits. In Section~\ref{sec:res} we present our findings and discuss the importance of loop effects in the global fits as well as the necessity of large violations of $B-L$ in order to obtain partial cancellations between the tree and loop level contributions.
 
\section{Parametrization}
\label{sec:param}

In this work we explore the constraints that can be derived through various EW observables on the extra neutrino mass eigenstates mixing with charged leptons in a Seesaw scenario:
\begin{eqnarray}\label{eq:The3FormsOfNuMassOp}
\mathscr{L} &=& \mathscr{L}_\mathrm{SM} -\frac{1}{2} \overline{N_\mathrm{R}^i} (M_N)_{ij} N^{c j}_\mathrm{R} -(Y_{N})_{i\alpha}\overline{N_\mathrm{R}^i}  \phi^\dagger
\ell^\alpha_\mathrm{L} +\hc\; .
\end{eqnarray}
Here, $\phi$ denotes the SM Higgs field, which breaks the EW symmetry after acquiring its VEV $v_{\mathrm{EW}}$. We have also introduced 
the Majorana mass $M_N$ allowed for the right-handed neutrinos $N_\mathrm{R}^i$ as well as the Yukawa couplings between the neutrinos and the 
Higgs field. We will restrict our study to the extension of the SM by 3 right-handed neutrino fields. The VEV of the Higgs will induce Dirac masses 
$m_D = v_\text{EW} Y_N/\sqrt{2}$. Thus, the full $6\times 6$ mixing matrix $U$ is the unitary matrix that diagonalises the extended neutrino mass matrix: 
\begin{equation}
U^T \left(
\begin{array}{cc}
0 & m_D^T \\ m_D & M_N
\end{array}
\right) U = \left(
\begin{array}{cc}
m & 0 \\ 0 & M
\end{array}
\right), \label{eq:diag}
\end{equation}
where $m$ and $M$ are diagonal matrices containing respectively the masses of the 3 light $\nu_i$ and 3 heavy $N_i$ mass eigenstates. 
The diagonalising matrix $U$ can be written as~\cite{Blennow:2011vn}:
\begin{equation}
U = \left(
\begin{array}{cc}
\ c & s \\ -s^\dagger & \hat{c}
\end{array}
\right) \left(
\begin{array}{cc}
U_{\rm PMNS} & 0 \\ 0 & I 
\end{array}
\right), \label{eq:block}
\end{equation}
where
\begin{equation}
\left(
\begin{array}{cc}
\ c & s \\ -s^\dagger & \hat{c}
\end{array}
\right) \equiv \left(
\begin{array}{cc}
\displaystyle\sum\limits_{n=0}^\infty \frac{ \left(- \Theta \Theta^\dagger \right)^{n}}{(2n)!} & 
\displaystyle\sum\limits_{n=0}^\infty \frac{ \left(- \Theta \Theta^\dagger \right)^{n}}{\left(2n+1\right)!} \Theta  \\ 
-\displaystyle\sum\limits_{n=0}^\infty \frac{ \left(- \Theta^\dagger \Theta \right)^{n}}{\left(2n+1\right)!} \Theta^\dagger & 
\displaystyle\sum\limits_{n=0}^\infty \frac{ \left(- \Theta^\dagger \Theta \right)^{n}}{2n!}\end{array}
\right)
\label{eq:sincos}
\end{equation}
and $U_{\rm PMNS}$ is, \emph{approximately}, the PMNS matrix measured in neutrino oscillation experiments up to the non-Unitary (Hermitian) corrections from $c$. For alternative exact parametrisations of the full mixing matrix see Refs~\cite{Xing:2011ur,Donini:2012tt}. Indeed, due to this Hermitian correction, the actual PMNS matrix appearing in charge current interactions mixing the light neutrinos and charged leptons will, in general, not be Unitary and we will refer to it as $N$: 
\begin{equation}
N = c\,U_{\rm PMNS} 
\label{eq:N}
\end{equation}
The general matrix $\Theta$, representing the mixing between active ($\nu_e$, $\nu_\mu$ and $\nu_\tau$) and heavy ($N_1$, $N_2$ and $N_3$) neutrino states, and the mass eigenstates $m$ and $M$ are determined from Eq.~(\ref{eq:diag}) which leads to:
\begin{equation}
\label{eq:corr}
c^* U_{\rm PMNS} ^* m U_{\rm PMNS} ^\dagger c = - s^* M s^\dagger .
\end{equation}
In the Seesaw limit, that is $M_N \gg m_D$, these conditions reduce to the well-known results:
\begin{eqnarray}
\nonumber
\Theta &\simeq& m_D^\dagger M_N^{-1} \\
\nonumber
U_{\rm PMNS} ^* m U_{\rm PMNS} ^\dagger  &\simeq& - m_D^t M_N^{-1} m_D \equiv -\hat{m}\\
M &\simeq& M_N .
\label{eq:Seesaw}
\end{eqnarray}

Notice that, naively, the mixing between the active and heavy neutrinos $\Theta \Theta^\dagger \sim m/M$ and, given the smallness of neutrino masses $m$, the mixing effects we will study here would be unobservably small. However, in the context of Seesaw mechanisms with an approximate conservation of $B-L$ such as the inverse~\cite{Mohapatra:1986bd,Bernabeu:1987gr} or the linear~\cite{Malinsky:2005bi} Seesaws, this symmetry suppresses the neutrino mass $m$ while allowing a sizeable mixing. This approximate symmetry not only ensures an equally approximate cancellation in the combination $m_D^t M_N^{-1} m_D$ leading to the observed neutrino masses while allowing large -potentially observable- $\Theta \Theta^\dagger = m_D^\dagger M_N^{-2} m_D$, but also ensures the radiative stability and technical naturalness of the scheme~\cite{Kersten:2007vk}. 

When extending the SM Lagrangian by only 3 new singlet (right-handed neutrino) fields essentially the only neutrino mass matrix with an underlying $L$ symmetry that leads to 3 heavy massive neutrinos is~\cite{Abada:2007ux} (see also Ref.~\cite{Adhikari:2010yt}):
\begin{equation}
m_D = \frac{v_\text{EW}}{\sqrt{2}} \left(
\begin{array}{ccc}
Y_e & Y_\mu & Y_\tau \\ \epsilon_1 Y'_e & \epsilon_1 Y'_\mu & \epsilon_1 Y'_\tau \\ \epsilon_2 Y''_e & \epsilon_2 Y''_\mu & \epsilon_2 Y''_\tau
\end{array}
\right)
\qquad
\textrm{and}
\qquad
M_N = \left(
\begin{array}{ccc}
\mu_1 & \Lambda & \mu_3 \\ \Lambda & \mu_2 & \mu_4 \\ \mu_3 & \mu_4 & \Lambda'
\end{array}
\right), \label{eq:texture}
\end{equation}
with all $\epsilon_i$ and $\mu_j$ small lepton number violating parameters  (see also Ref.~\cite{Dev:2015vra} for a particular scenario where these small parameters arise naturally). Indeed, setting all $\epsilon_i=0$ and $\mu_j=0$, lepton number symmetry is recovered with the following $L$ assignments $L_e = L_\mu = L_\tau = L_1 = -L_2 = 1$ and $L_3 = 0$. In Eq.~(\ref{eq:Seesaw}) this leads to: $\hat{m}=0$ (3 massless neutrinos in the $L$-conserving limit), $M_1=M_2=\Lambda$ (a heavy Dirac pair) and $M_3=\Lambda'$ (a heavy decoupled Majorana singlet), but:
\begin{equation}
\Theta = \frac{v_\text{EW}}{2 \Lambda} \left(
\begin{array}{ccc}
-i Y_e^* & Y_e^* & 0 \\ -i Y_\mu^* & Y_\mu^* & 0 \\ -i Y_\tau^* & Y_\tau^* & 0
\end{array}
\right)
\equiv \frac{1}{\sqrt{2}} \left(
\begin{array}{ccc}
-i \theta_e & \theta_e & 0 \\ -i \theta_\mu & \theta_\mu & 0 \\ -i \theta_\tau & \theta_\tau & 0
\end{array}
\right)
\textrm{and}
\quad
\Theta \Theta^\dagger=  \left(
\begin{array}{ccc}
|\theta_e|^2 & \theta_e \theta_\mu^* & \theta_e \theta_\tau^* \\ \theta_\mu \theta_e^* & |\theta_\mu|^2 & \theta_\mu \theta_\tau^* \\ \theta_\tau \theta_e^* & \theta_\tau \theta_\mu^* & |\theta_\tau|^2
\end{array}
\right). \label{eq:theta}
\end{equation}

Thus, vanishing light neutrino masses can still be associated with arbitrarily large mixing between the heavy Dirac pair and active neutrinos and, for these kind of Seesaw scenarios, the bounds on the mixing we will explore are complementary and independent to the stringent constraints on the absolute light neutrino mass scale. 

The small $L$-violating parameters $\epsilon_i$ and $\mu_j$ will induce small non-zero neutrino masses and mixing among these light mass eigenstates but will only translate in negligible perturbations to the matrix $\Theta$. With the simple form in Eq.~(\ref{eq:theta}) for the heavy-active mixing, the series expansions in Eq.~(\ref{eq:sincos}) can be added exactly obtaining:
\begin{equation}
s= \frac{\sin \theta}{\theta} \Theta
\qquad
\textrm{and}
\qquad
c = I -\frac{1-\cos \theta}{\theta^2} \Theta \Theta^\dagger,
\label{eq:summed}
\end{equation}
with
\begin{equation}
\theta = \sqrt{|\theta_e|^2 + |\theta_\mu|^2 + |\theta_\tau|^2}.
\end{equation}

Regarding the role of the $\epsilon_i$ and $\mu_j$ parameters in the generation of the light neutrino masses and mixings observed in neutrino oscillations, 
all of them except $\mu_1$ and $\mu_3$ will lead to $\hat{m} \neq 0$ through Eq.~(\ref{eq:Seesaw}) when switched on:

\begin{eqnarray}
\hat{m}&=&\left( \mu_2 +\frac{\mu_4^2}{\Lambda'}\right)\mathbf{m^{\mathit{t}}_D}\Lambda^{-2}\mathbf{m_D}
-\epsilon_1 \mathbf{m'^{\mathit{t}}_D} \Lambda^{-1} \mathbf{m_D} - \epsilon_1 \mathbf{m^{\mathit{t}}_D} \Lambda^{-1} \mathbf{m'_D} 
+ \epsilon_2^2 \mathbf{m''^{\mathit{t}}_D} \Lambda'^{-1} \mathbf{m''_D}
\nonumber\\
&+&\epsilon_2 \frac{\mu_4}{\Lambda'} 
\left( \mathbf{m^{\mathit{t}}_D} \Lambda^{-1} \mathbf{m''_D}+ \mathbf{m''^{\mathit{t}}_D} \Lambda'^{-1} \mathbf{m_D} \right),
\label{eq:lightmass0}
\end{eqnarray}
with

\begin{equation}
\mathbf{m_D} \equiv \frac{v_\text{EW}}{\sqrt{2}} (Y_e, Y_\mu, Y_\tau) ,
\qquad
\mathbf{m'_D} \equiv \frac{v_\text{EW}}{\sqrt{2}} (Y'_e, Y'_\mu, Y'_\tau) 
\qquad
\textrm{and}
\qquad
\mathbf{m''_D} \equiv \frac{v_\text{EW}}{\sqrt{2}} (Y''_e, Y''_\mu, Y''_\tau). 
\end{equation}
Indeed, even though $\mu_1$ and $\mu_3$ do violate $L$, upon their inclusion the mass matrix in Eq.~(\ref{eq:diag}) does not increase its rank, 
which, in absence of the other $\epsilon_i$ and $\mu_j$, is only 3 and thus 3 massless eigenstates are still 
recovered\footnote{Notice that, even if $\mu_1$ and $\mu_3$ do not induce neutrino masses at tree level, the $L$ symmetry protecting them is 
now broken and loop contributions would appear instead~\cite{LopezPavon:2012zg}.}. The parameters $\mu_2$ and $\mu_4$ do contribute at tree level 
to generate light neutrino masses, however, their effect can be absorbed in a redefinition of the vectors $\mathbf{m'_D}$ and
 $\mathbf{m''_D}$ as follows: 
\begin{equation}
\epsilon_1 \mathbf{m'_D} \rightarrow \epsilon_1 \mathbf{m'_D} - \frac{\mu_2}{2 \Lambda} \mathbf{m_D}
\qquad
\textrm{and}
\qquad
\epsilon_2 \mathbf{m''_D} \rightarrow \epsilon_2 \mathbf{m''_D} - \frac{\mu_4}{\Lambda} \mathbf{m_D} 
\end{equation} 
up to contributions with two extra powers of the small $L$-violating parameters. Thus, in presence of non-zero $\epsilon_i$, 
it is enough to consider their contribution to the generation of neutrino masses which reads:
\begin{equation}
\hat{m} = \epsilon_1 \mathbf{m'^{\mathit{t}}_D} \Lambda^{-1} \mathbf{m_D} + \epsilon_1 \mathbf{m^{\mathit{t}}_D} \Lambda^{-1} \mathbf{m'_D} + \epsilon_2^2 \mathbf{m''^{\mathit{t}}_D} \Lambda'^{-1} \mathbf{m''_D} .
\label{eq:lightmass}
\end{equation}
Notice that the last term in Eq.~(\ref{eq:lightmass}) is suppressed by two powers of $\epsilon_2$ while the others only by one power of $\epsilon_1$. However, $\epsilon_2$ (and $\mu_3$ and $\mu_4$) violates $L$ by one unit while $\epsilon_1$ (and $\mu_1$ and $\mu_2$) by 2. Hence, if the source of $L$-violation is by one unit it is expected that $\epsilon_1 \sim \epsilon_2^2$. Thus, for full generality, we will keep the last term in Eq.~(\ref{eq:lightmass}). The six free parameters encoded in $\mathbf{m'_D}$ and $\mathbf{m''_D}$ allow to give mass to the three mass eigenstates observed in neutrino oscillations as well as the possibility of reproducing any mixing pattern including the, yet unknown, CP-violating phases of Dirac and Majorana types encoded in the PMNS matrix, while leaving $\mathbf{m_D}$, and hence $\Theta$, $s$ and $c$, mostly unconstrained~\footnote{In contrast, neglecting the last term in Eq.~(\ref{eq:lightmass}) would lead to the more constrained scenario explored in detail in Ref.~\cite{Gavela:2009cd}, with a massless neutrino and a mixing pattern in $\Theta$, $s$ and $c$ determined up to an overall factor from the observed neutrino oscillation parameters. This scenario has also been studied in Refs.~\cite{Zhang:2009ac,Malinsky:2009df,Ibarra:2010xw,Ibarra:2011xn,Cely:2012bz}}. One of the three elements of $\mathbf{m_D}$ is, however, fixed by the other two, the values of the light mass eigenstates and the elements of the PMNS matrix when solving for Eq.~(\ref{eq:lightmass}) obtaining the following relation:
\begin{equation}
\begin{split}
Y_\tau& \simeq \frac{1}{\hat{m}_{e \mu}^2 - \hat{m}_{ee} \hat{m}_{\mu \mu}}\left(Y_e\left(\hat{m}_{e \mu}\hat{m}_{\mu \tau}-\hat{m}_{e \tau}\hat{m}_{\mu \mu}\right)+\right.\\
&\left. Y_\mu\left(\hat{m}_{e \mu}\hat{m}_{e \tau}-\hat{m}_{ee}\hat{m}_{\mu \tau}\right)-\sqrt{Y_e^2\hat{m}_{\mu \mu}-2Y_eY_\mu \hat{m}_{e \mu}+Y_\mu^2\hat{m}_{ee}}\times \right.\\
&\left.\times\sqrt{\hat{m}_{e \tau}^2\hat{m}_{\mu \mu}-2\hat{m}_{e \mu}\hat{m}_{e \tau}\hat{m}_{\mu \tau}+\hat{m}_{ee}\hat{m}_{\mu \tau}^2+\hat{m}_{e \mu}^2\hat{m}_{\tau \tau}-\hat{m}_{ee}\hat{m}_{\mu \mu}\hat{m}_{\tau \tau}}\right) ,
\end{split}
\label{eq:Yt}
\end{equation}
where $\hat{m} = -U_{\rm PMNS} ^* m U_{\rm PMNS} ^\dagger$ is the mass matrix of the flavour eigenstates. Thus, in our numerical exploration of the parameter space in Section~\ref{sec:res} we will consider the 9 free parameters summarised in Table~\ref{tab:params}.  

An alternative parametrisation extensively used in the literature is the so-called Casas-Ibarra parametrisation~\cite{Casas:2001sr}. This parametrisation introduces the matrix $R = i M^{-1/2} m_D U_{\rm PMNS} m^{-1/2}$ exploiting the fact that, from \eq~(\ref{eq:Seesaw}), $R$ has to be (complex) orthogonal. The main advantage of this parametrisation is the ability to easily recover the Yukawa couplings through the heavy mass eigenvalues $M$ and the low energy observables $U_{\rm PMNS} $ and $m$ together with the elements of $R$ as $m_D = -i M^{1/2} R m^{1/2} U_{\rm PMNS} ^\dagger$. However, the physical range of the parameters contained in $R$ can be cumbersome and a physical interpretation of their values is not immediately transparent, see \Ref~\cite{Casas:2010wm} for a detailed discussion. Moreover, these relations only hold at tree level\footnote{See Ref.~\cite{Lopez-Pavon:2015cga} for a generalisation of the Casas-Ibarra approach to loop level.}. Thus, when values of $R$ are chosen so as to allow sizeable low energy phenomenology through large Yukawas and low $M$, it is important to check if the pattern displays an approximate $B-L$ symmetry. Otherwise, loop corrections to the unprotected Weinberg operator, that is to $U_{\rm PMNS} $ and $m$, will exceed present constraints even if their values were correct at tree level. For this reason we rather chose to perform the scan through the parameters summarised in Table~\ref{tab:params}.  

\begin{table}
\begin{center}
\begin{tabular}{|c||c|c|c|c|c|c|}
\hline
Parameter & $\left|Y_e\right| \times \left|Y_\mu\right| $ & $\left|Y_e\right| - \left|Y_\mu\right| $ & $m_1$ [eV] &  $\Lambda$ [GeV] & Phases
& Osc. data \rule{0pt}{2.6ex} \rule[-1.2ex]{0pt}{0pt}\\
\hline
Range &  $(0,10^{-4})$ & $(-0.1,0.1)$ & $(10^{-5},1)$ & $(10^{3},10^{4})$ & $(0,2\pi)$ & fixed \cite{Gonzalez-Garcia:2014bfa} \rule{0pt}{2.6ex} \rule[-1.2ex]{0pt}{0pt} \\
\hline
\end{tabular}
\caption{The 9 free parameters of our scan: the modulus and phase of the electron and muon Yukawas $|Y_e|$, $|Y_\mu|$, $\alpha_e$ and $\alpha_\mu$, the Majorana mass scale $\Lambda$, the absolute neutrino mass $m_1$ and the 3 yet unknown CP-violation phases (Dirac and Majorana) in the PMNS mixing matrix: $\delta$, $\alpha_1$ and $\alpha_2$. The PMNS mixing angles and mass splittings are fixed to their best fit from the global analysis in Ref.~\cite{Gonzalez-Garcia:2014bfa}. }
\label{tab:params}
\end{center}
\end{table}

At energies much below the masses of the heavy neutrinos $\Lambda$ and $\Lambda'$ the effects of their mixing $\Theta$ manifest dominantly through deviations from unitarity of the lepton mixing matrix $N$. Since any general matrix can be parametrised as the product of an Hermitian and a Unitary matrix, these deviations from unitarity have been often parametrised as~\cite{FernandezMartinez:2007ms}:
\begin{equation} 
N = (I - \eta) U_{\rm PMNS}
\end{equation} 
where the small Hermitian matrix $\eta$ (also called $\epsilon$ in other works) corresponds to the coefficient of the only dimension 6 operator obtained at tree level upon integrating out the heavy right-handed neutrinos in a Seesaw scenario~\cite{Broncano:2002rw} and, in our parametrisation it would be given from Eqs.~(\ref{eq:N}) and (\ref{eq:summed}) by: 
\begin{equation}
\eta = \frac{1-\cos \theta}{\theta^2} \Theta \Theta^\dagger .
\end{equation}

\section{Observables}
\label{sec:obs}

In this section we introduce the list of observables used for our analysis. While a more comprehensive set could be considered (see for example Ref.~\cite{Antusch:2014woa}), we have rather chosen the most representative of these observables since extending the analysis to the loop level for the whole set would be cumbersome and the dominant constraints as well as the main effects pointed out in~\cite{Akhmedov:2013hec} are contained in a smaller subset. We will thus present both the 1-loop contributions and the experimental constraints for a total of 13 observables. The loop amplitudes of the processes have been computed exploiting the Goldstone-boson equivalence theorem~\cite{Cornwall:1974km} under the assumption that the mass of the extra neutrinos $M_i$ is larger than the gauge boson masses; i.e. $M_i > M_{W,Z}$. Thus, we have made the simplifying assumption that the most relevant loop corrections are those were the loops are mediated by either the Higgs boson, $h$, the Goldstone bosons $\phi^\pm$ and $\phi^0$ or the heavy Majorana neutrinos. Indeed, this forces the vertexes to involve the potentially large Yukawa couplings (the only couplings that can be relevant at the loop level) and the corrections from including the transverse components are suppressed by $M_{W,Z}^2/M_N^2$. The set of 13 independent observables analysed in this study is composed of:

\begin{itemize}
\item 8 ratios constraining electroweak universality: $R^\pi_{\mu e}$, $R^\pi_{\tau \mu}$, $R^W_{\mu e}$, $R^W_{\tau \mu}$, $R^K_{\mu e}$, $R^K_{\tau \mu}$, $R^l_{\mu e}$, $R^l_{\tau \mu}$
\item The invisible $Z$ width
\item The $W$ mass $M_W$
\item 3 rare flavour-changing decays: $\mu\rightarrow e\gamma$, $\tau\rightarrow \mu\gamma$ and $\tau\rightarrow e\gamma$
\end{itemize}
All of them will be determined as a function of the three most precise electroweak measurements: $\alpha$, $M_Z$ and $G_\mu$ ($G_F$ as measured from $\mu$ decay)~\cite{Agashe:2014kda}:
\begin{eqnarray}
\alpha&=&\left(7.2973525698\pm0.0000000024\right)\times 10^{-3}, \nonumber\\
M_Z&=&\left(91.1876\pm0.0021\right) \text{ GeV}, \\
G_\mu&=&\left(1.1663787\pm0.0000006\right)\times 10^{-5} \text{ GeV}^{-2}. \nonumber
\end{eqnarray}

\begin{figure}[htb]
\begin{center}
\includegraphics[width=0.8\textwidth]{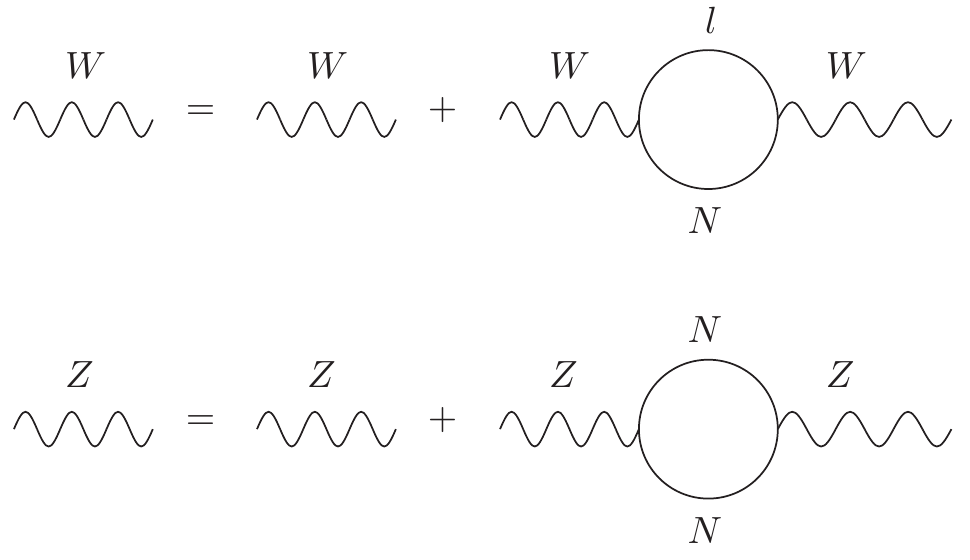}

\end{center}
\caption{1-loop correction of the new heavy neutrinos to $W$ and $Z$ propagators.}
\label{fig:propa}
\end{figure}

All observables will receive contributions from the loop corrections to the $W$ and $Z$ boson propagators through the diagrams in Fig.~\ref{fig:propa}. These contributions are encoded in the flavour-universal corrections $\delta_{W,Z}^\text{univ}$ that can be found in Eq.~(\ref{dW}) in the Appendix~\ref{sec:app_fits}. We now list the further corrections exclusive to each of the observables considered:

\subsection{$\mu$ decay, $G_F$ and $M_W$}

Our input value for $G_F$ is determined through $\mu$ decay, but this process will receive corrections both at the tree and the loop level (see Fig.~\ref{fig:GF}). Thus, the value determined from $\mu$ decay, $G_\mu$, is related to $G_F$ by:

\begin{figure}[htb]
\begin{center}
\includegraphics[width=0.8\textwidth]{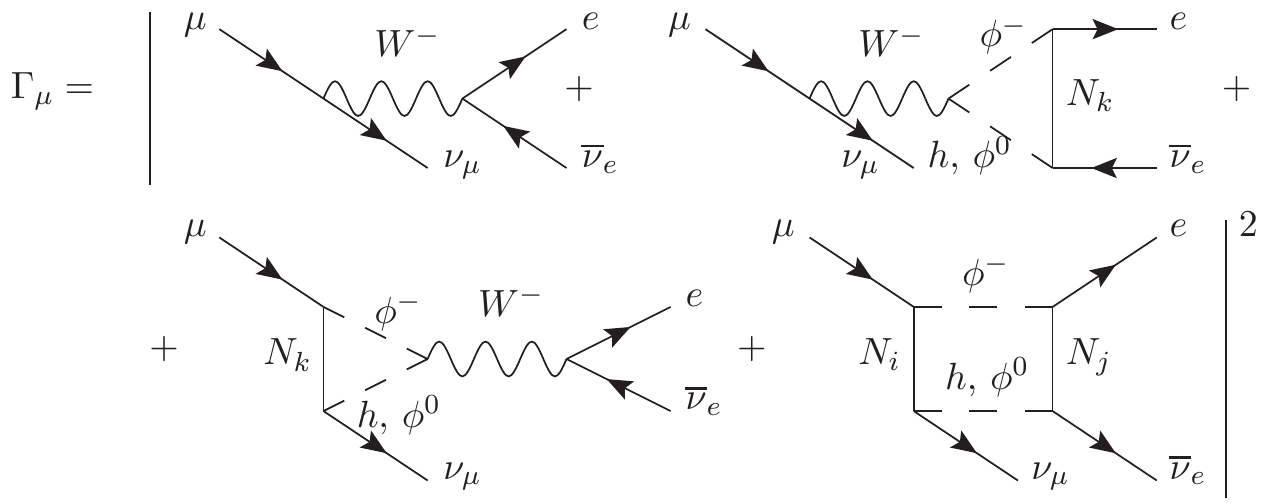}
\end{center}
\caption{1-loop corrections to $\mu$ decay.}
\label{fig:GF}
\end{figure}

\begin{equation}
\Gamma_\mu = \frac{m_\mu^5 G_F^2}{192 \pi^3}\left(1 - |\theta_e|^2 - |\theta_\mu|^2 + 2\delta^{\text{univ } N}_W+\delta G\right)
\equiv \frac{m_\mu^5 G_\mu^2}{192 \pi^3} ,
\label{mudecay}
\end{equation}
with

\begin{equation}
\delta G = 2Re[\mathcal{V}^W_e+\mathcal{V}^{W *}_\mu+\delta^{\text{CT } W}_{e}+\delta^{\text{CT } W *}_{\mu}+\mathcal{B}_{\mu e}]
\end{equation}
and where $\delta^{\text{univ } N}_W$ is the flavour-universal $W$ propagator correction, $\delta^{\text{CT } W}_{l}$ and $\mathcal{V}^W_{l}$ are the flavour-dependent lepton propagator and vertex contributions (see Eqs.~(\ref{eq:propscorr}) and~(\ref{eq:Wvertex}) in the Appendix~\ref{sec:app_fits}), and $\mathcal{B}_{\mu e}$ encodes the box diagram contribution computed in Eq.~(\ref{eq:box}) in the Appendix~\ref{sec:app_fits}.  

From Eq.~(\ref{mudecay}), we find:
\begin{equation}
G_\mu^2 = G_F^2 \left( 1 - |\theta_e|^2 - |\theta_\mu|^2 + 2\delta_W^\text{univ N}+\delta G\right)\,.
\label{Gmu}
\end{equation}

The second and third terms in Eq.~(\ref{Gmu}) correspond to the tree level correction, the fourth term is the universal 1-loop oblique correction 
which is given in Eq.~(\ref{dW}) of the Appendix~\ref{sec:app_fits}. This particular expression, when used in an observable mediated by the $Z$ and thus corrected 
through $2\delta_Z^\text{univ N}$, leads to a common correction to these observables given by $1 - |\theta_e|^2 - |\theta_\mu|^2 - 2 \alpha T$ 
(see Eqs.~(\ref{dW}) and (\ref{T})). This common dependence on the tree level and oblique corrections is the source of the cancellation analysed in Ref.~\cite{Akhmedov:2013hec}.

The the $W$ mass is also correlated to $G_F$ through
\begin{equation}
M_W^2 = \frac{\pi \alpha}{\sqrt{2} G_F s_\mathrm{W}^2 (1 - \Delta r)} , 
\end{equation}
with $\Delta r = 0.03639 \mp 0.00036 \pm 0.00011$~\cite{Agashe:2014kda}. Thus, the corrections induced at both the tree and loop levels by 
the heavy neutrinos from Eq.~(\ref{Gmu}) can be probed by the measurement of $M_W$ in LEP and Tevatron~\cite{Agashe:2014kda}:
\begin{equation}
M_W = 80.385 \pm 0.015 \quad \mathrm{GeV} .
\end{equation}

\subsection{Invisible $Z$ width}

The determination of the number of light active neutrinos by LEP through the invisible width of the $Z$ provides a constraint to heavy neutrino mixing already at the tree level. Additional loop corrections are induced through the diagrams in Fig.~\ref{fig:invZ} which lead to:
\begin{equation}
\Gamma_\text{inv}=\displaystyle\sum_{i,j=1}^3 \frac{G_F M_Z^3 \rho}{24 \sqrt{2} \pi}\left( \mathcal{Z}_{ij}+\mathcal{Z}_{ji}\right)\, ,
\label{eq:inv}
\end{equation}
where $\rho$ encodes the SM loop corrections to the process and 
\begin{equation}
\mathcal{Z}_{ij}=\vert C_{ij}\vert ^2\big(1+\delta_Z^\text{univ}\big)+2 Re\big[C_{ij}^*\left(\delta^{\text{CT } Z}_{ij}+\mathcal{V}^Z_{ij}\right)\big]\, ,
\end{equation}
with
\begin{equation}
C_{ij}=\sum_{\alpha=e,\mu,\tau} U_{\alpha i}^* U_{\alpha j}\,.
\end{equation}
and $\delta^{\text{CT } Z}_{ij}$ and $\mathcal{V}^Z_{ij}$ the lepton and vertex corrections shown in Eqs.~(\ref{eq:propscorrZ}) and~(\ref{eq:Zvertex}) 
in the Appendix~\ref{sec:app_fits}. 

Eq.~(\ref{eq:inv}) is often used to determine the number of active neutrinos $N_\nu$ lighter than $M_Z/2$ as:
\begin{equation}
\Gamma_\text{inv}=  \frac{G_F M_Z^3 \rho N_\nu}{12 \sqrt{2} \pi}\, ,
\label{eq:invi}
\end{equation}
The measurement by LEP of $\Gamma_\text{inv}=\left(0.4990\pm0.0015\right) \text{ GeV}$ combined with Eq.~(\ref{eq:invi}) leads to~\cite{Agashe:2014kda}:
\begin{equation}
N_\nu = 2.990 \pm 0.007 \, .
\end{equation}
We will exploit this result together with Eq.~(\ref{eq:inv}) to derive constraints on $C_{ij}$ and, hence, on the heavy neutrino mixings.

\begin{figure}[htb]
\begin{center}
\includegraphics[width=0.8\textwidth]{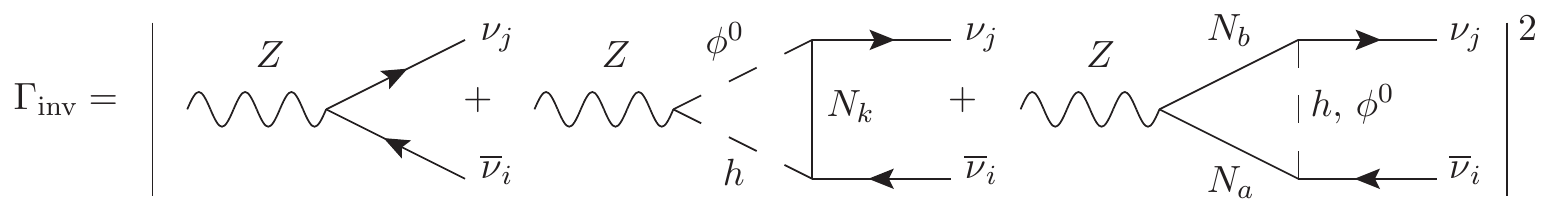}
\end{center}
\caption{1-loop corrections to the invisible decay of the $Z$.}
\label{fig:invZ}
\end{figure}

\subsection{Universality ratios}

Electroweak coupling universality is strongly constrained through ratios of leptonic decays of $K$, $\pi$, $W$ or charged leptons. In these ratios many uncertainties cancel and a clean constraint can be derived. These observables are corrected both at the tree and loop level, for instance, $R^\pi_{\mu e}=\Gamma\left(\pi^- \rightarrow \mu \overline{\nu}_\mu\right)/\Gamma\left(\pi^- \rightarrow e \overline{\nu}_e\right)$ is corrected by the diagrams in Fig.~\ref{fig:univ}.

\begin{figure}[htb]
\begin{center}
\includegraphics[width=0.8\textwidth]{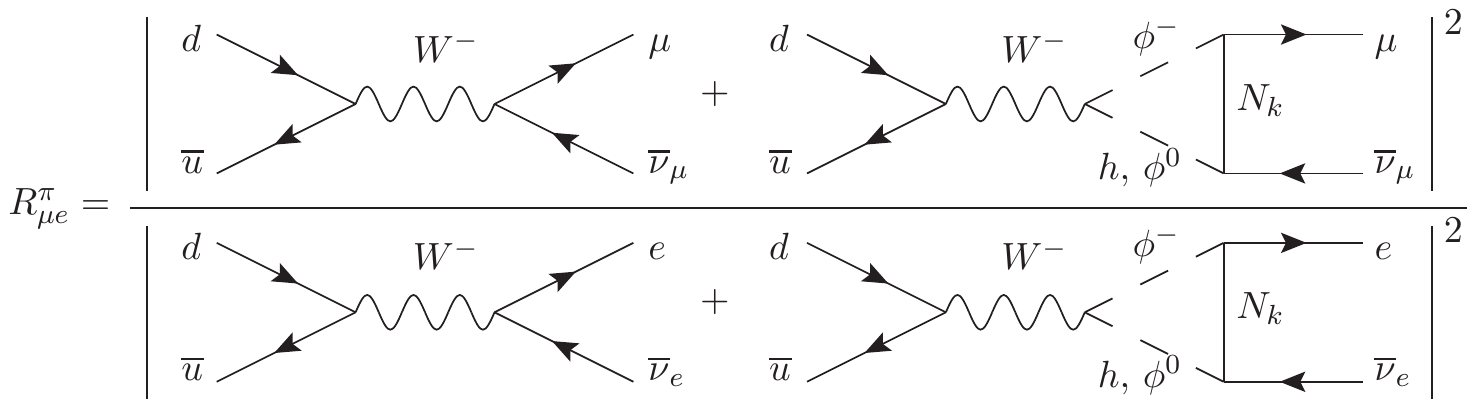}
\end{center}
\caption{1-loop corrections to weak universality ratios.}
\label{fig:univ}
\end{figure}

\begin{table}[htb]
\begin{center}
\begin{tabular}{c|c}
\hline
$BR\left(\pi^+\rightarrow e^+\nu_e\right)$& $\left(1.230\pm 0.004\right)\times 10^{-4}$\\
$BR\left(\pi^+\rightarrow \mu^+\nu_\mu\right)$& $\left(99.98770\pm 0.00004\right)\%$\\
$BR\left(\tau^-\rightarrow \pi^-\nu_\tau\right)$& $\left(10.83\pm 0.06\right)\%$\\
$BR\left(K^+\rightarrow e^+\nu_e\right)$& $\left(1.581\pm 0.008\right)\times 10^{-5}$\\
$BR\left(K^+\rightarrow \mu^+\nu_\mu\right)$& $\left(63.55\pm 0.11\right)\% 10^{-5}$\\
$BR\left(\tau^-\rightarrow K^-\nu_\tau\right)$& $\left(7.00\pm 0.10\right)\times 10^{-3}$\\
$BR\left(W^+\rightarrow e^+\nu_e\right)$& $\left(10.71\pm 0.16\right)\%$\\
$BR\left(W^+\rightarrow \mu^+\nu_\mu\right)$& $\left(10.63\pm 0.15\right)\%$\\
$BR\left(W^+\rightarrow \tau^+\nu_\tau\right)$& $\left(11.38\pm 0.21\right)\%$\\
$BR\left(\tau^-\rightarrow \mu^-\overline{\nu}_\mu \nu_\tau\right)$& $\left(17.41\pm 0.04\right)\%$\\
$BR\left(\tau^-\rightarrow e^-\overline{\nu}_e \nu_\tau\right)$& $\left(17.83\pm 0.04\right)\%$\\
\hline
$\tau_{\pi^\pm}$ & $\left(2.6033\pm 0.0005\right)\times 10^{-8}\text{ s}$\\
$\tau_{K^\pm}$ & $\left(1.2380\pm 0.0021\right)\times 10^{-8}\text{ s}$\\
$\tau_\tau$ & $\left(290.3\pm 5.0\right)\times 10^{-15}\text{ s}$\\
$\tau_\mu$ & $\left(2.1969811\pm 0.0000022\right)\times 10^{-6}\text{ s}$\\
\hline
$m_{\pi^\pm}$ & $139.57018\pm 0.00035 \text{ MeV}$ \\
$m_{K^\pm}$ & $493.677\pm 0.016 \text{ MeV}$ \\
$M_{W}$ & $80.385\pm 0.0015 \text{ MeV}$ \\
$m_{e}$ & $0.510998928\pm 0.000000011 \text{ MeV}$ \\
$m_{\mu}$ & $105.6583715\pm 0.0000035 \text{ MeV}$ \\
$m_{\tau}$ & $1776.82\pm 0.16 \text{ MeV}$ \\
\hline
$\delta R^\pi_{\mu e}$ & $\left(-0.374\pm 0.001\right)$ \\
$\delta R^\pi_{\mu \tau}$ & $\left(0.0016\pm 0.0014\right)$ \\
$\delta R^K_{\mu \tau}$ & $\left(0.0090\pm 0.0022\right)$ \\
\hline
\end{tabular}
\caption{Input values used for the constraints on weak universality from ratios of meson and charged lepton decays.}
\label{tab:exp-values}
\end{center}
\end{table}

Thus, the general expression for the ratio of lepton flavours $\alpha$ and $\beta$ is given by:

\begin{equation}
R_{\alpha \beta}=R^{SM}_{\alpha \beta}\frac{\displaystyle1 - |\theta_\alpha|^2 +2\text{Re}\left[\mathcal{V}^W_{\alpha}+\delta^{\text{CT }W}_{\alpha}\right]}{\displaystyle1 - |\theta_\beta|^2 +2\text{Re}\left[\mathcal{V}^W_{\beta}+\delta^{\text{CT }W}_{\beta}\right]} ,
\label{eq:univ1}
\end{equation}
where $R^{SM}_{\alpha \beta}$ is the SM value for this ratio, for example, for $\pi$ decay:
\begin{equation}
R^{\pi SM}_{\alpha \beta} = \left(\frac{ m_\alpha \left(m_\pi^2-m_\alpha^2\right)}{m_\beta\left(m_\pi^2-m_\beta^2\right)}\right)^2\frac{1}{1+\delta R_{\alpha \beta}^\pi}
\label{eq:univ2}
\end{equation}
and where $\delta R_{\alpha \beta}^\pi$ are the SM radiative corrections to this process~\cite{Decker:1994ea}. Notice that the flavour-universal contributions from the $W$ propagator cancel in the ratio. 

The predicted values of these ratios are computed through Eqs.~(\ref{eq:univ1}) and~(\ref{eq:univ2}) with data form \cite{Agashe:2014kda,Pich:2013lsa} and compared to the experimental measurements of the decay rates in our global fit. This data is summarised in Table \ref{tab:exp-values}.

\subsection{Rare decays}

The presence of extra heavy neutrinos beyond the three light ones participating in low energy weak processes induces deviations from unitarity in the PMNS matrix. Thus, the GIM cancellation~\cite{Glashow:1970gm} suppressing flavour-changing processes does not take place and strong constraints on the presence of these extra neutrinos can be derived. Moreover, the extra heavy neutrinos themselves also mediate the flavour-changing processes, such as radiative leptons decays $l_\alpha \to l_\beta \gamma$ in Fig.~\ref{fig:muegamma}. The contribution from both the heavy and light neutrinos is given by:
\begin{equation}
\frac{\Gamma\left(l_\alpha\rightarrow l_\beta \gamma \right)}{\Gamma\left(l_\alpha\rightarrow l_\beta \nu_\alpha \overline{\nu}_\beta \right)}=\frac{3 \alpha}{32 \pi}\frac{\Big| \displaystyle\sum_{k=1}^6 U_{\alpha k}U^\dagger_{k\beta}F(x_k)\Big|^2}{\left(U U^\dagger \right)_{\alpha\alpha}\left(U U^\dagger \right)_{\beta\beta}}
\label{eq:raredecaye}
\end{equation}
where $x_k\equiv \frac{M_{k}^2}{M_W^2}$, and $F(x_k)$ is given by:
\begin{equation}
F(x_k)\equiv\frac{10-43x_k+78x_k^2-49x_k^3+4x_k^4+18x_k^3\ln x_k}{3(x_k-1)^4}.
\end{equation}
Thus, for heavy neutrino masses much larger than $M_W$:
\begin{equation}
\frac{\Gamma\left(l_\alpha\rightarrow l_\beta \gamma \right)}{\Gamma\left(l_\alpha\rightarrow l_\beta \nu_\alpha \overline{\nu}_\beta \right)} \simeq \frac{3 \alpha}{32 \pi} |\theta_\alpha \theta^*_\beta|^2 (F(\infty) -F(0))^2 .
\label{eq:raredecay}
\end{equation}
The prediction from Eq.~(\ref{eq:raredecaye}) will be compared with the existing upper bounds from \cite{Agashe:2014kda}:
\begin{eqnarray}
BR_{\mu e}&<& 5.7\times 10^{-13}\, ,\\ 
BR_{\tau e}&< &3.3\times 10^{-8}\, ,\\
BR_{\tau \mu}&<&4.4\times 10^{-8} \,.
\end{eqnarray}
Notice that these bounds are quoted at the $90 \%$ CL so they will be rescaled to $1 \sigma$ to build the corresponding contribution to the $\chi^2$ function.

\begin{figure}[htb]
\begin{center}
\includegraphics[width=0.7\textwidth]{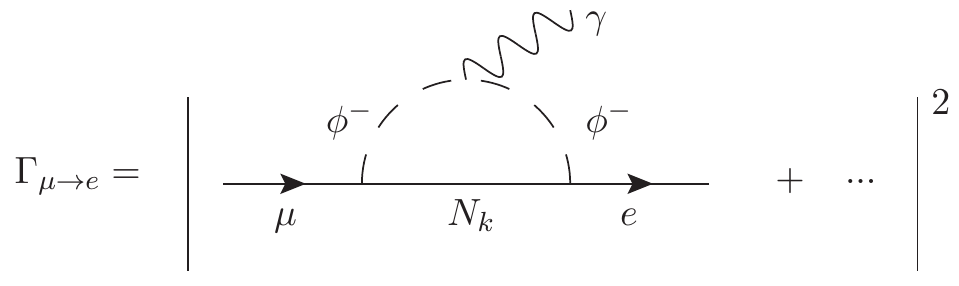}
\end{center}
\caption{Extra neutrino contributions to the $\mu \to e \gamma$ decay.}
\label{fig:muegamma}
\end{figure}

\section{Results}
\label{sec:res}

\subsection{Constraints from the global fit}

With the 13 observables discussed in Section~\ref{sec:obs} we build a $\chi^2$ function depending on the 9 parameters listed in Table~\ref{tab:params}. Given the large dimensionality of the parameter space, we make use of Markov chain Monte Carlo (MCMC) techniques for efficient parameter exploration. In particular, we implement importance sampling based on the Likelihood obtained from the observables through a Metropolis-Hastings algorithm. The range in which the 9 free parameters are varied is also summarised in Table~\ref{tab:params}. We have run simultaneously 5 different chains through the MCMC algorithm and have verified that good convergence (better than $R-1 < 0.035$~\cite{Gelman:1992zz}) for all parameters has been achieved. The results of the runs thus provide a good sample of the $\chi^2$ values in the preferred regions of the parameter space and have been used to marginalise over different subsets of the model parameters. In this way, we will present 2D and 1D frequentist contours on the more phenomenologically relevant parameters of the model. The post-processing of the chains to derive the allowed confidence regions has been performed with the MonteCUBES~\cite{Blennow:2009pk} user interface.

\begin{figure}[htb]
\centering
\includegraphics[width=0.46\textwidth]{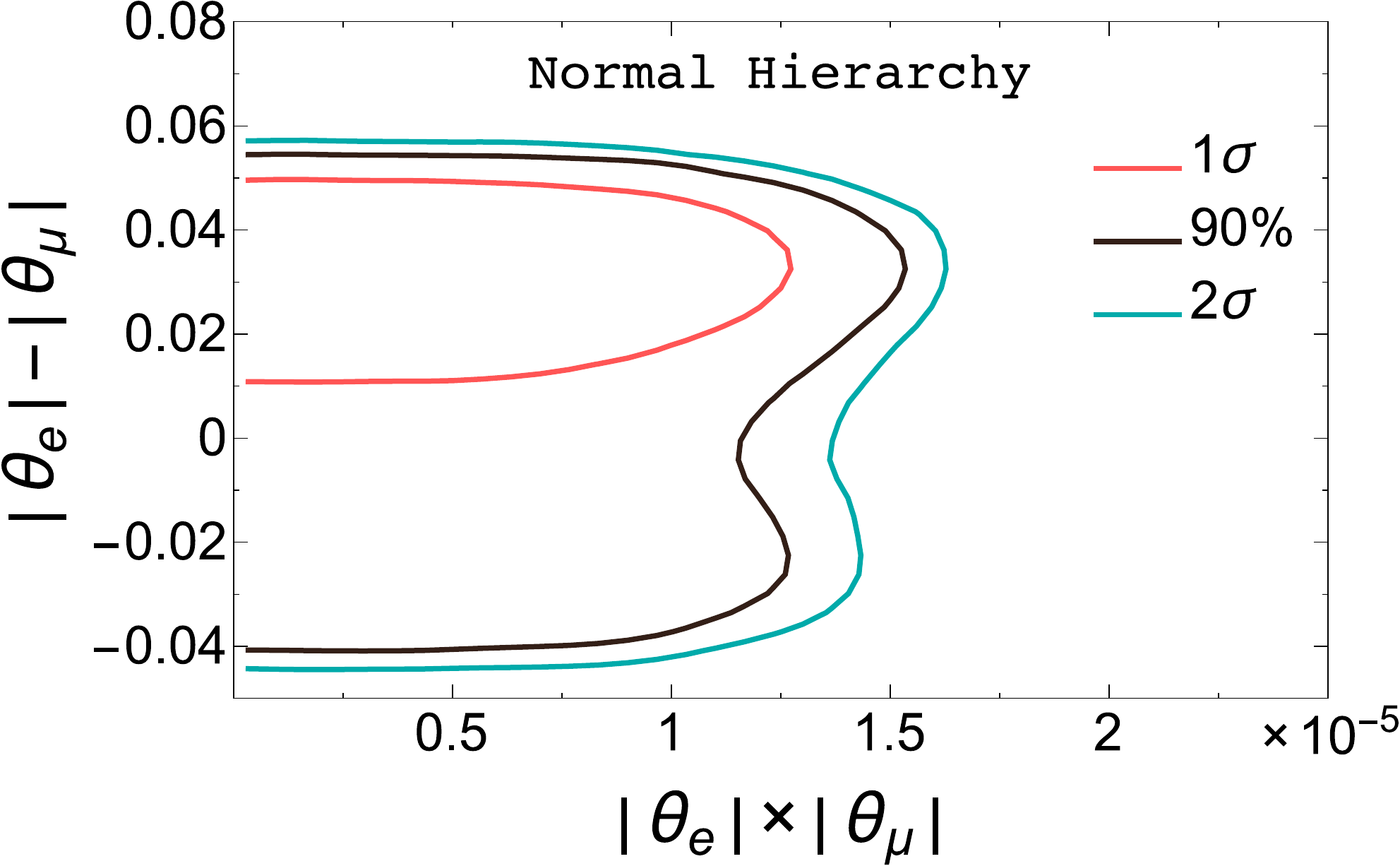}
\includegraphics[width=0.46\textwidth]{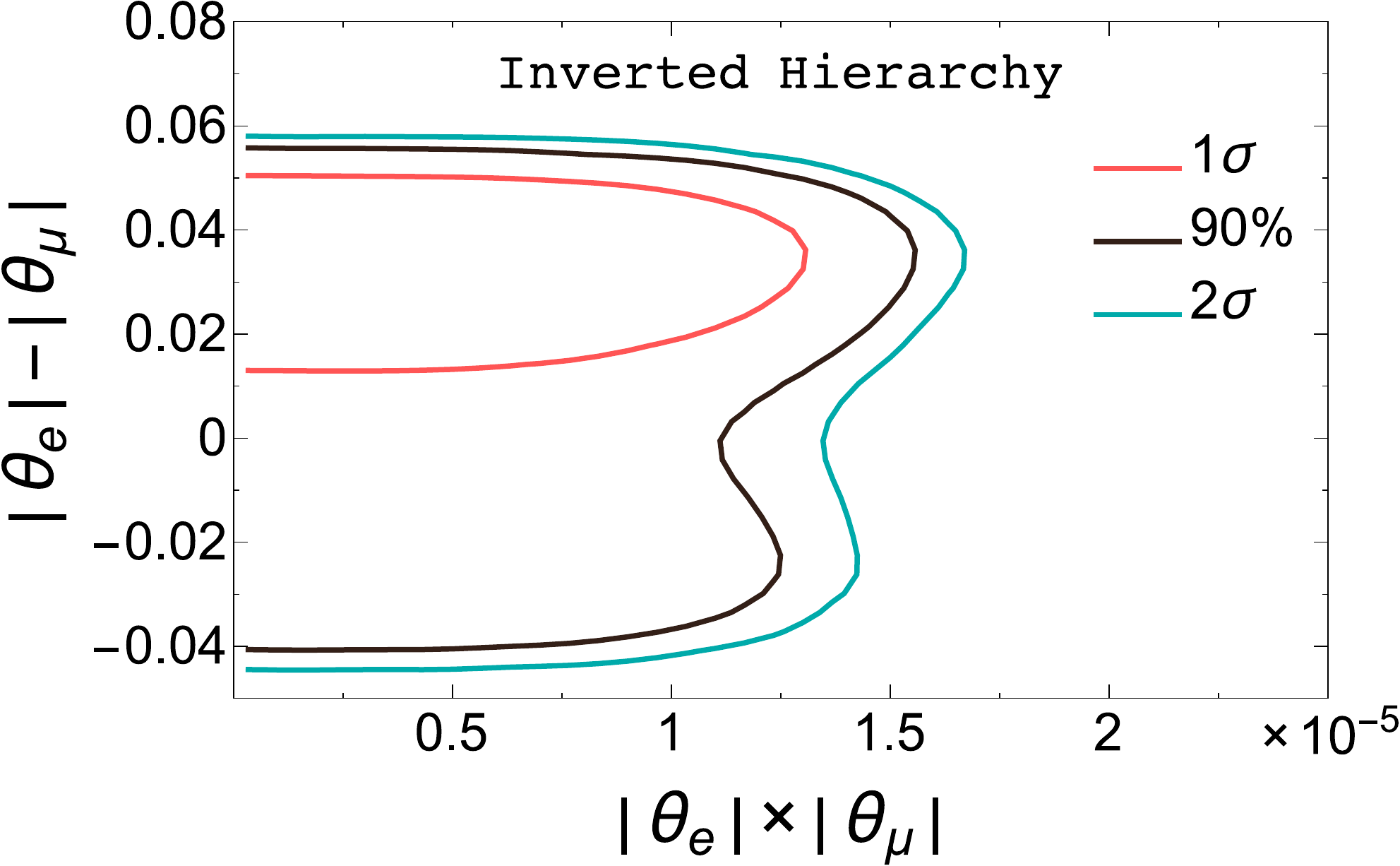}
\includegraphics[width=0.45\textwidth]{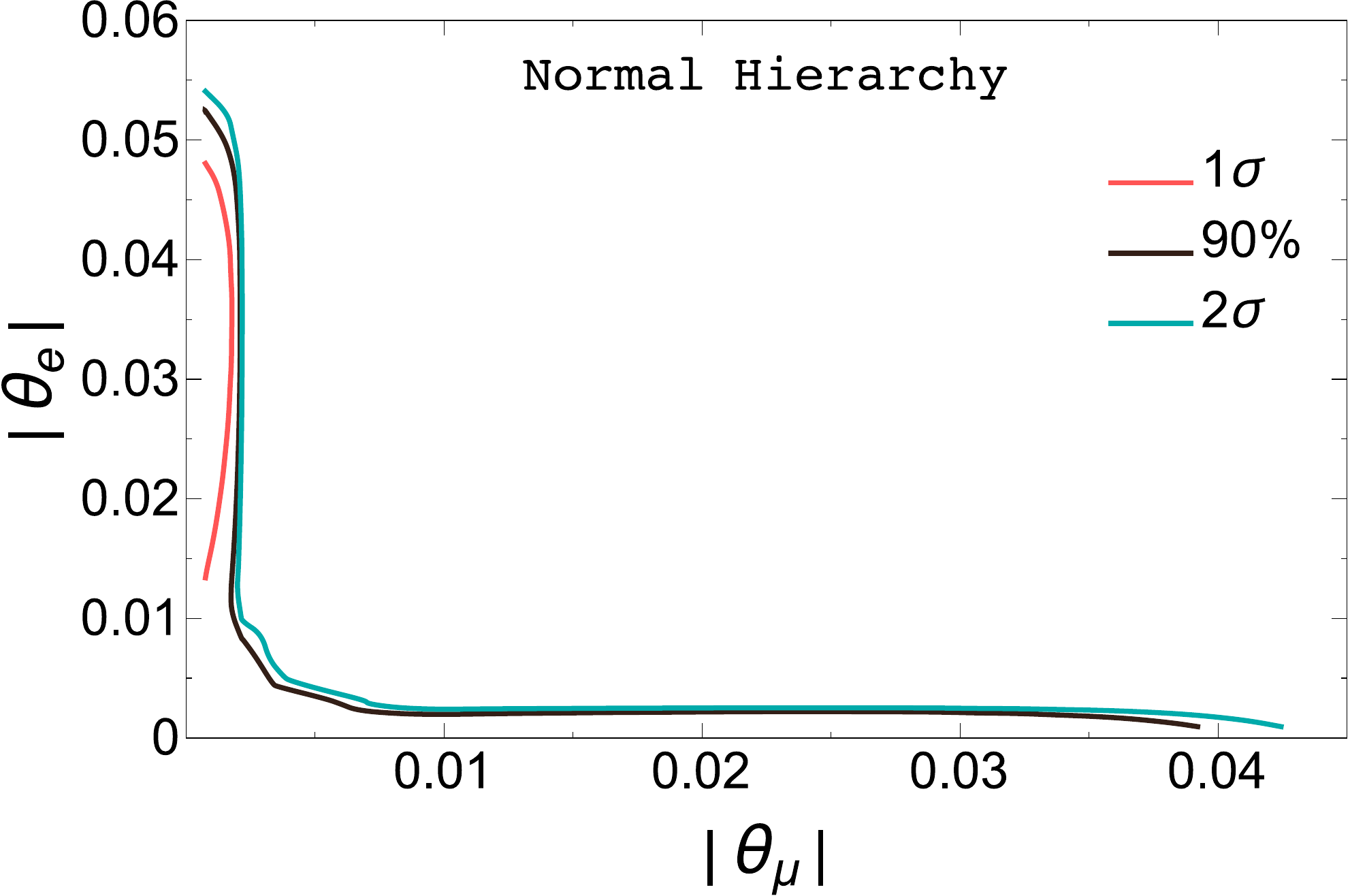}
\includegraphics[width=0.45\textwidth]{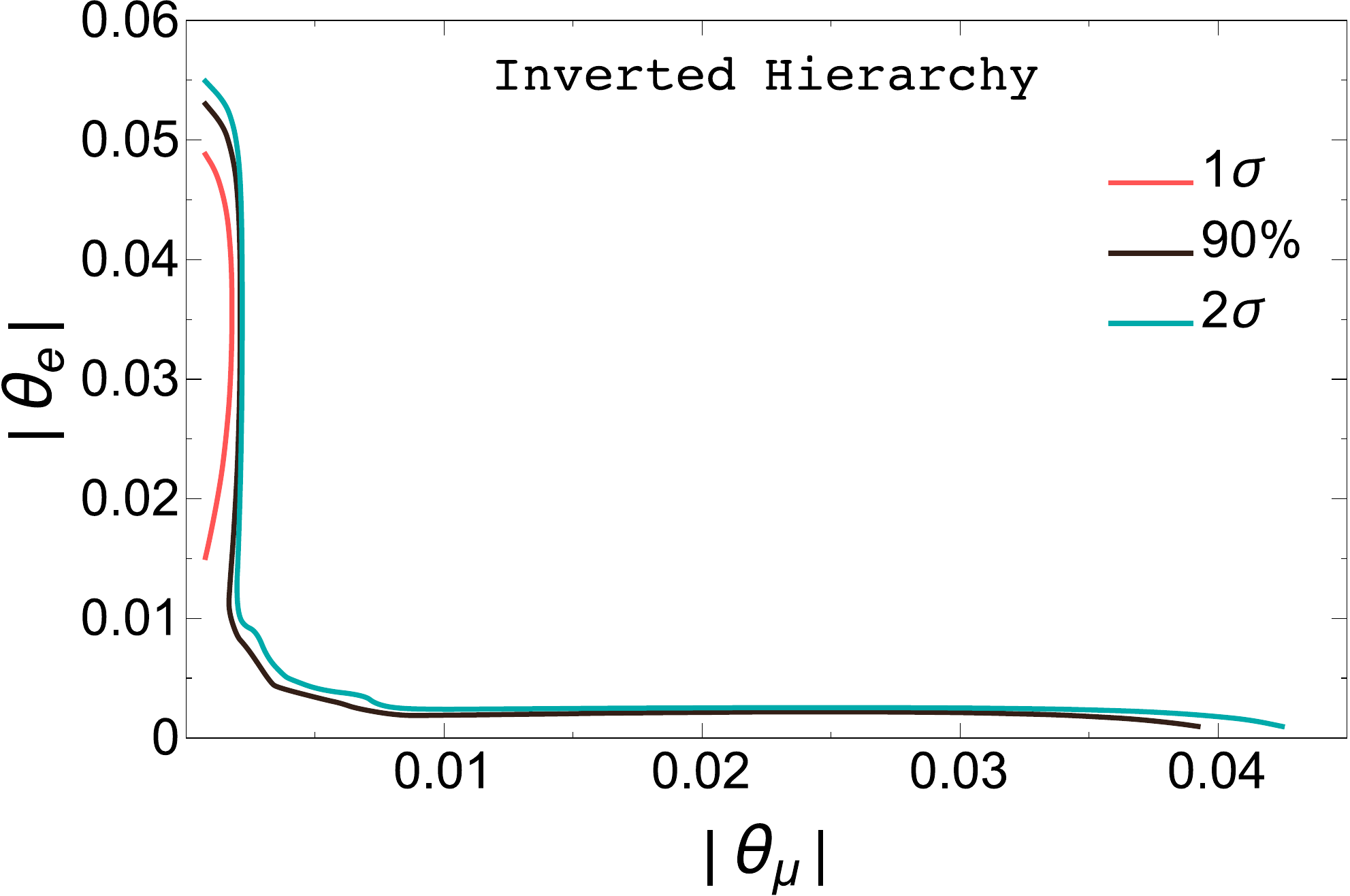}
\includegraphics[width=0.45\textwidth]{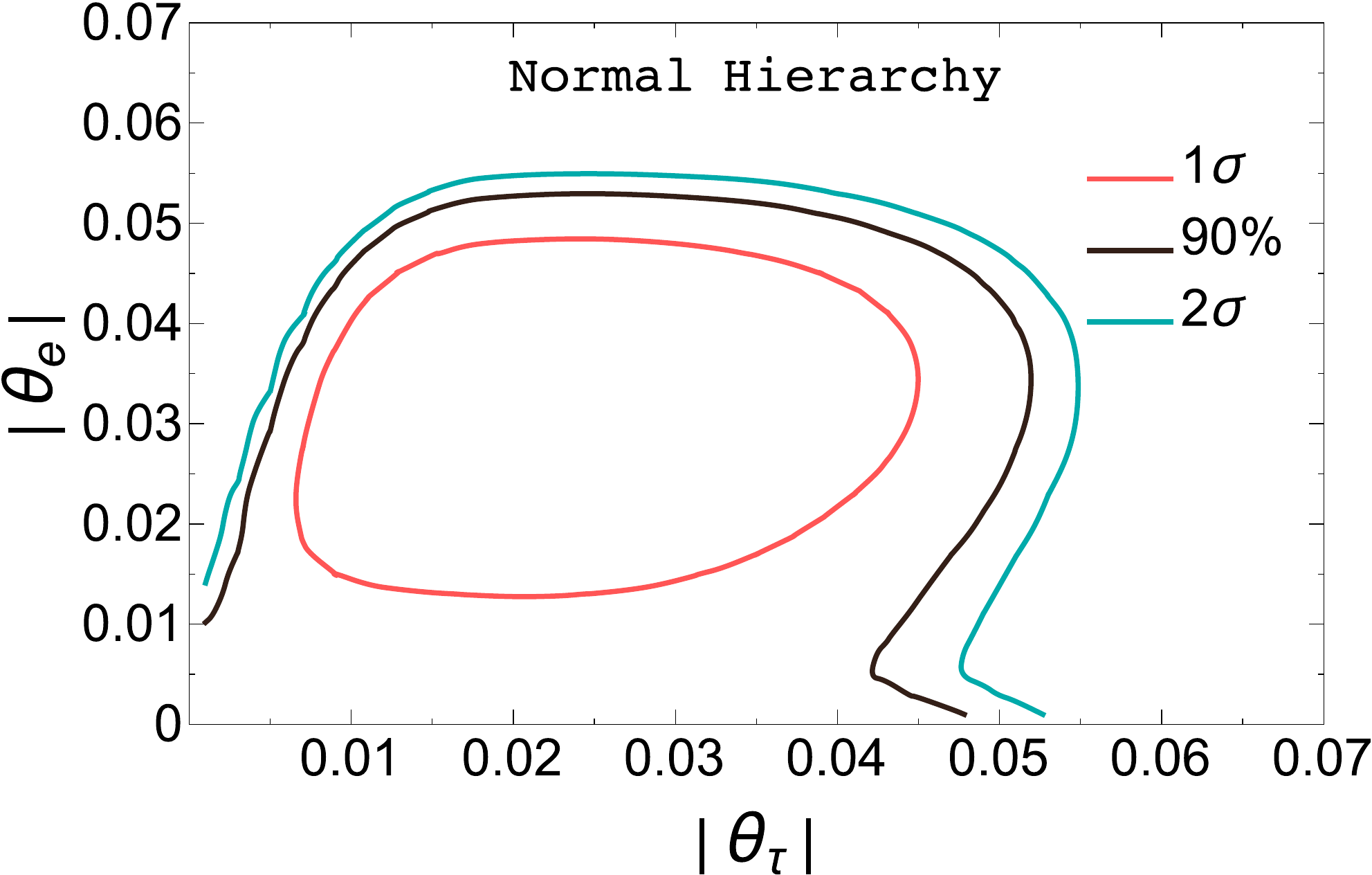}
\includegraphics[width=0.45\textwidth]{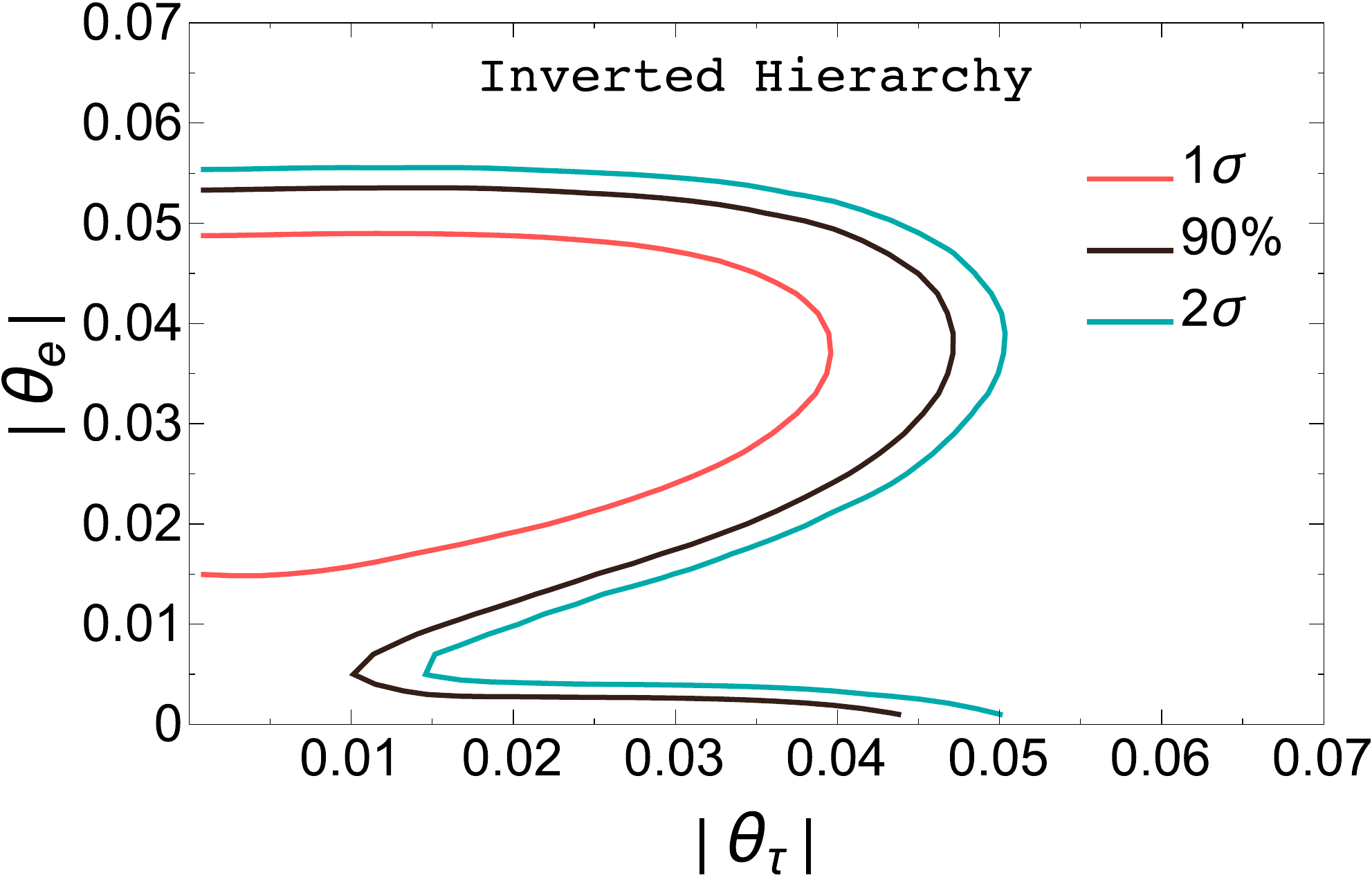}
\caption{Contours for $\theta_{e}$, $\theta_{\mu}$ and $\theta_{\tau}$ at $1\sigma$ (red), $90\%$ CL (black) and $2\sigma$ (blue). The left panels are obtained for normal hierarchy and the right for inverted.}
\label{fig:contours}
\end{figure}

In Fig.~\ref{fig:contours} we show the results of our MCMC scan for the 2 degrees of freedom constraints of different combinations of the heavy-active mixings $\theta_{\alpha}$ defined in Eq.~(\ref{eq:theta}). The contours correspond to the $1 \sigma$, $90 \%$ and $2 \sigma$ frequentist confidence regions. The upper panels show the bounds in the two combinations we choose to more directly sample (see Table~\ref{tab:params}): $|\theta_e| \times |\theta_\mu|$ and $|\theta_e| - |\theta_\mu|$. The rationale behind this is apparent upon inspection of Fig.~\ref{fig:contours}. Indeed, the constraints on the product are more than one order of magnitude smaller than those derived from the difference of the couplings $\sqrt{|\theta_e| \times |\theta_\mu|} \ll ||\theta_e| - |\theta_\mu||$, leading to a very pronounced hyperbolic degeneracy in the panels of the middle row, which contain the same information directly depicted as a function of $\theta_e$ and $\theta_\mu$. Thus, this particular choice of sampling parameters allowed to scan the hyperbolic degeneracy much more efficiently and speed the convergence of the MCMC. This very strong constraint in $|\theta_e| \times |\theta_\mu|$ stems from the strong bound on $\mu \to e \gamma$ from MEG that, from Eq.~(\ref{eq:raredecay}), sets a very stringent limit on $|\theta_\mu \theta_e^*|$. 

Finally, the lower panels of Fig.~\ref{fig:contours} contain the constraints derived for the mixing with the $\tau$ flavour $\theta_\tau$. Notice that $Y_\tau$, and hence $\theta_\tau$, was not a free parameter of the fit but was rather obtained from the other two Yukawas and the light neutrino masses and mixings from Eq.~(\ref{eq:Yt}). This is the source of the observed correlation between the values of $\theta_e$ and $\theta_\tau$. Notice also that, since the particular pattern of light neutrino masses plays an important role in Eq.~(\ref{eq:Yt}), the left (normal hierarchy) and right (inverted hierarchy) panels of Fig.~\ref{fig:contours} display different correlations. 

\begin{figure}[htb]
\centering
\includegraphics[width=0.45\textwidth]{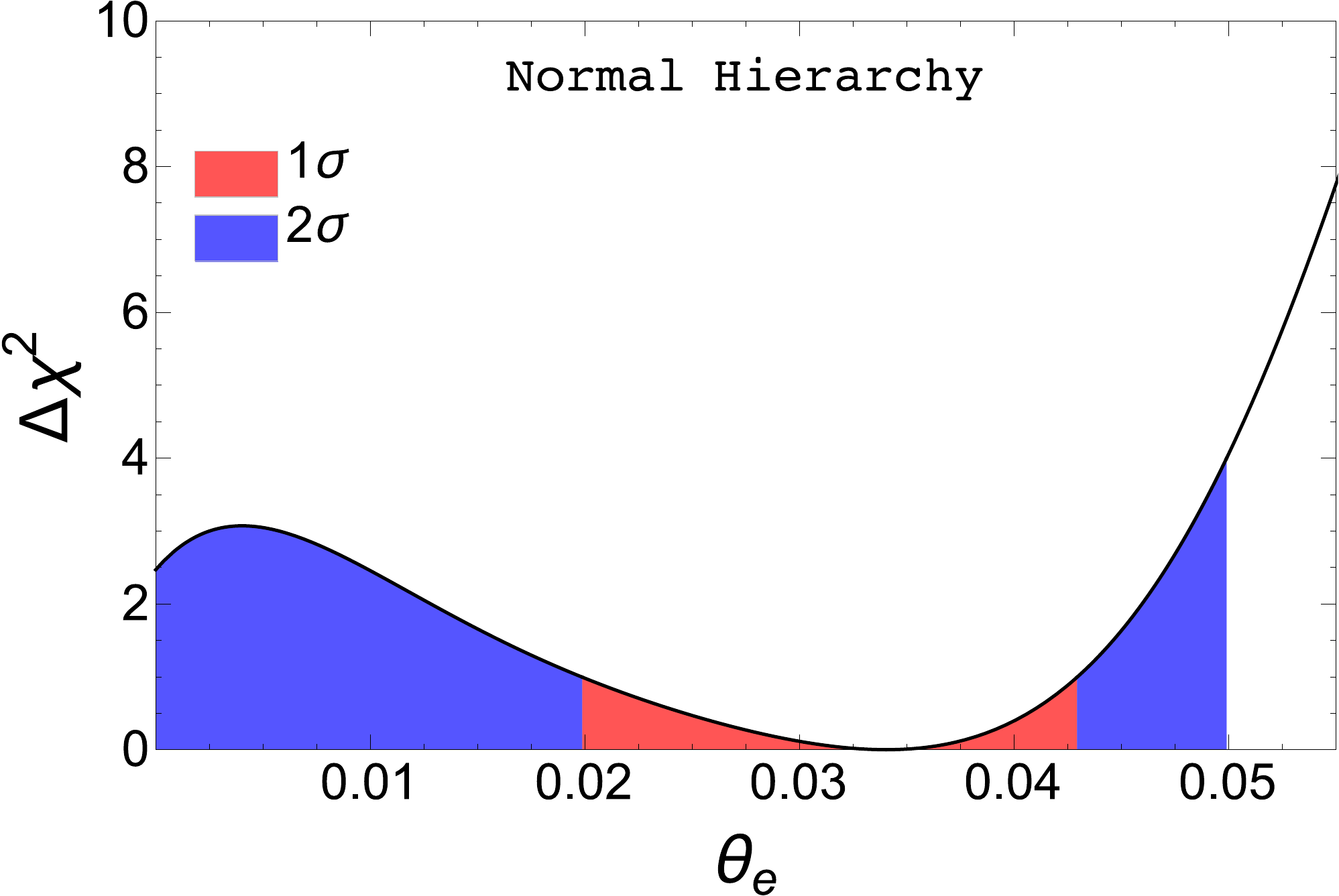}
\includegraphics[width=0.45\textwidth]{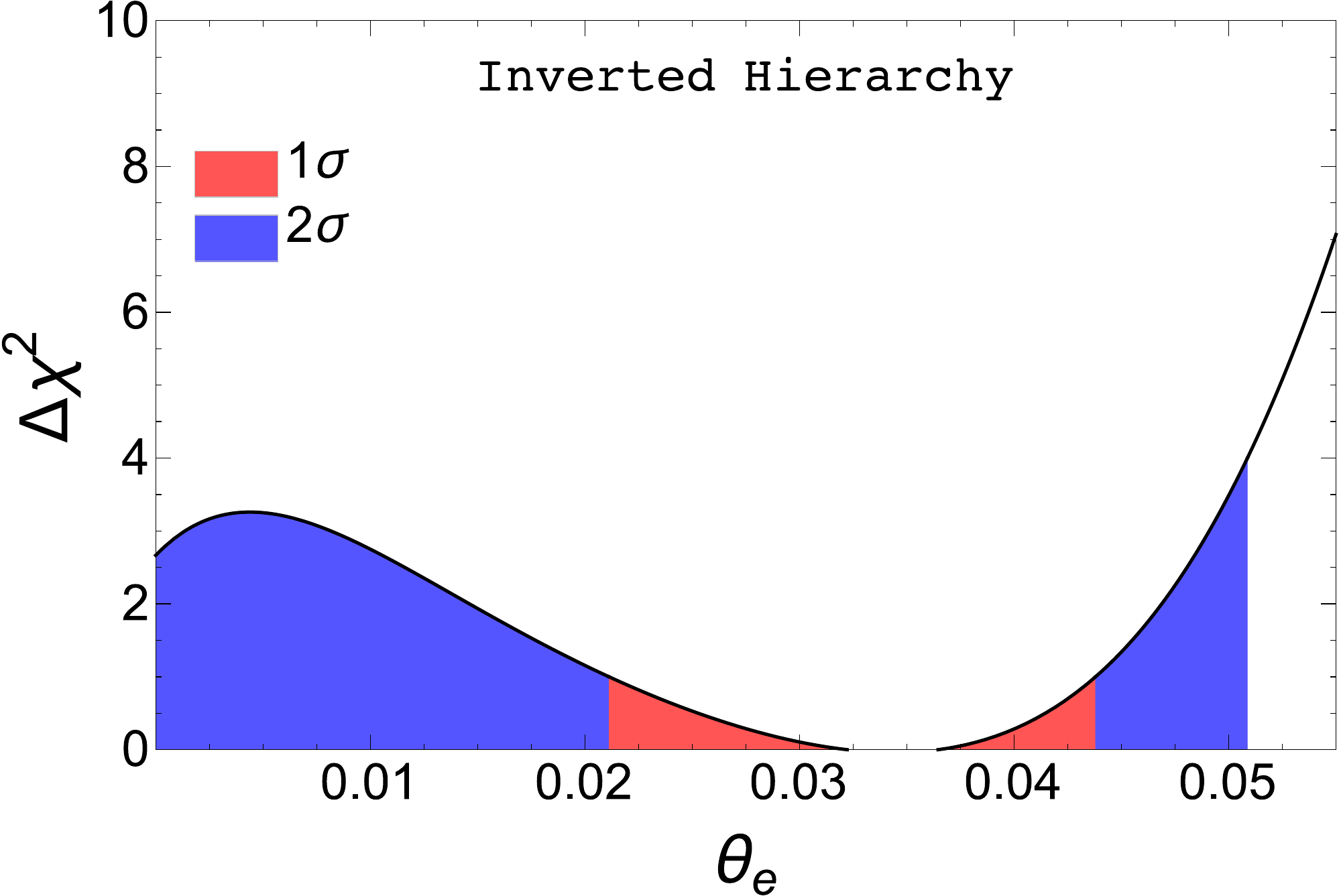}
\includegraphics[width=0.45\textwidth]{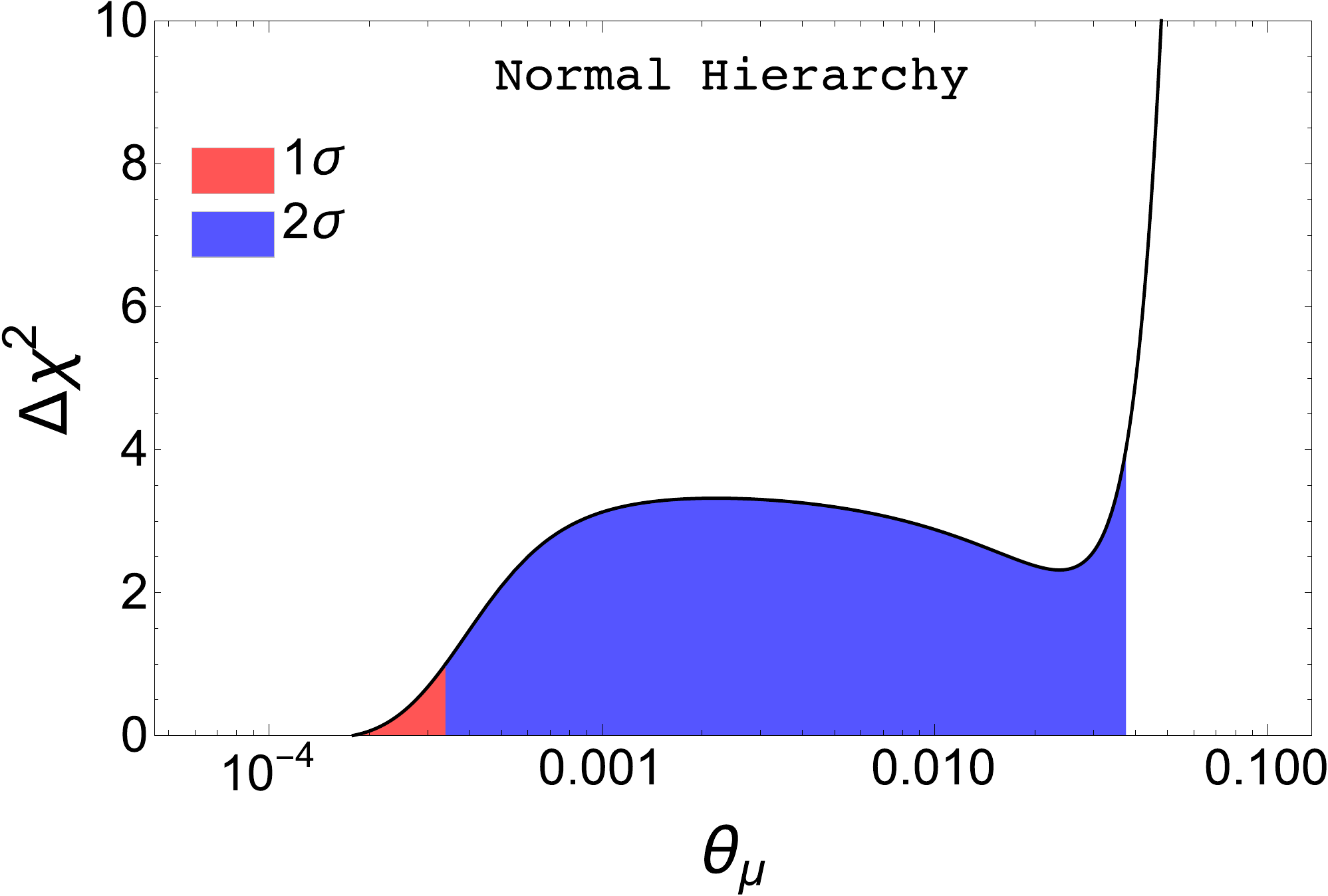}
\includegraphics[width=0.45\textwidth]{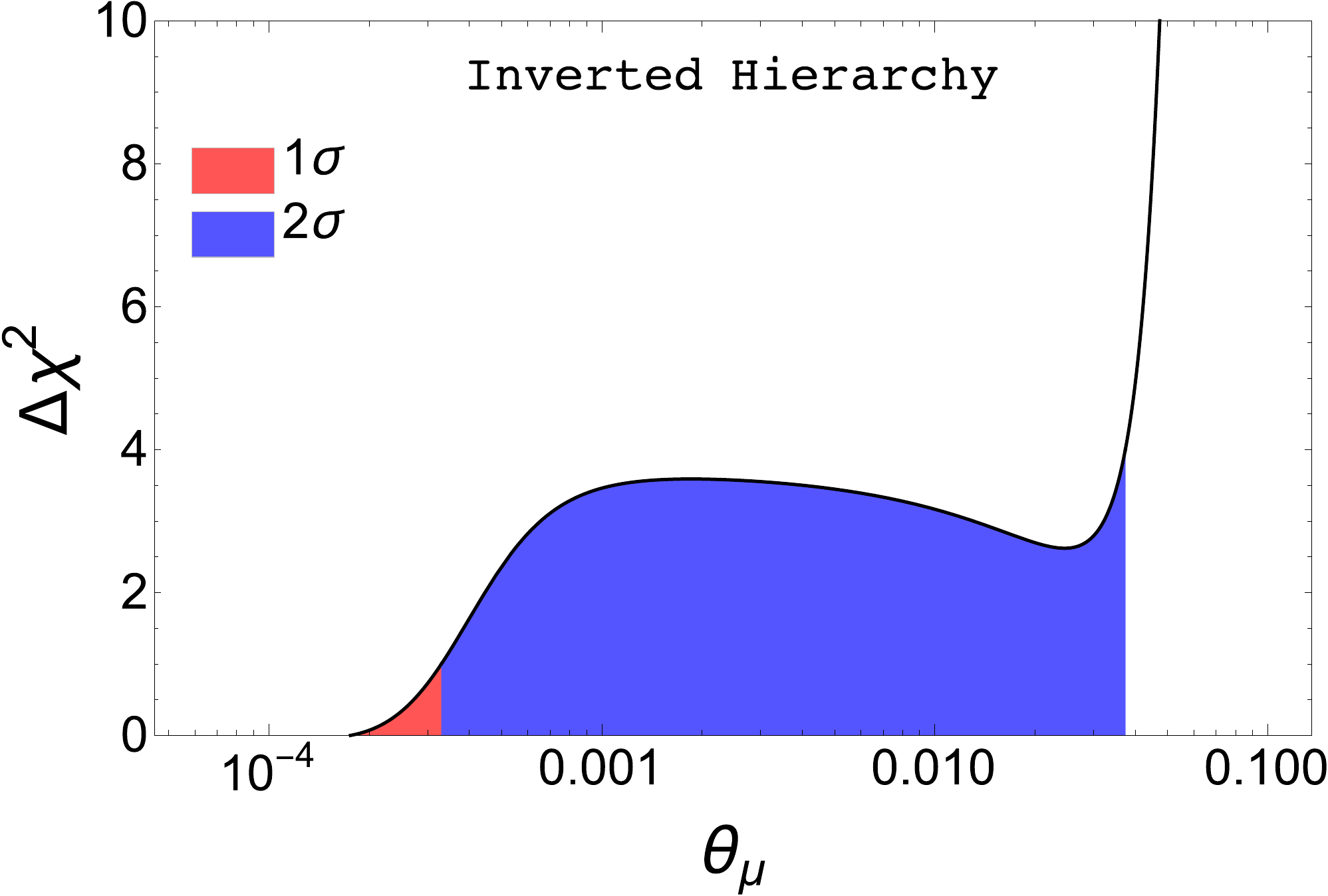}
\includegraphics[width=0.45\textwidth]{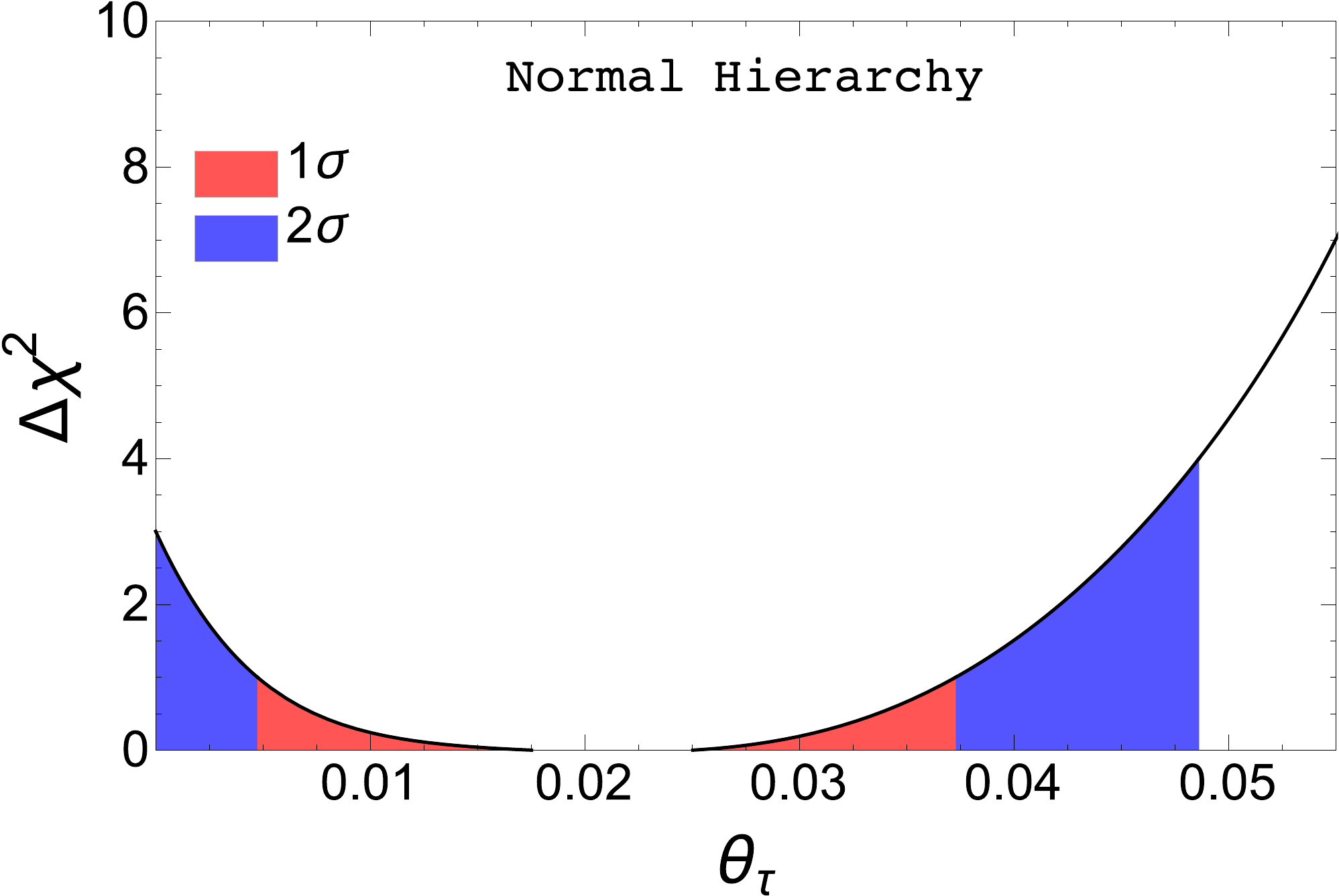}
\includegraphics[width=0.45\textwidth]{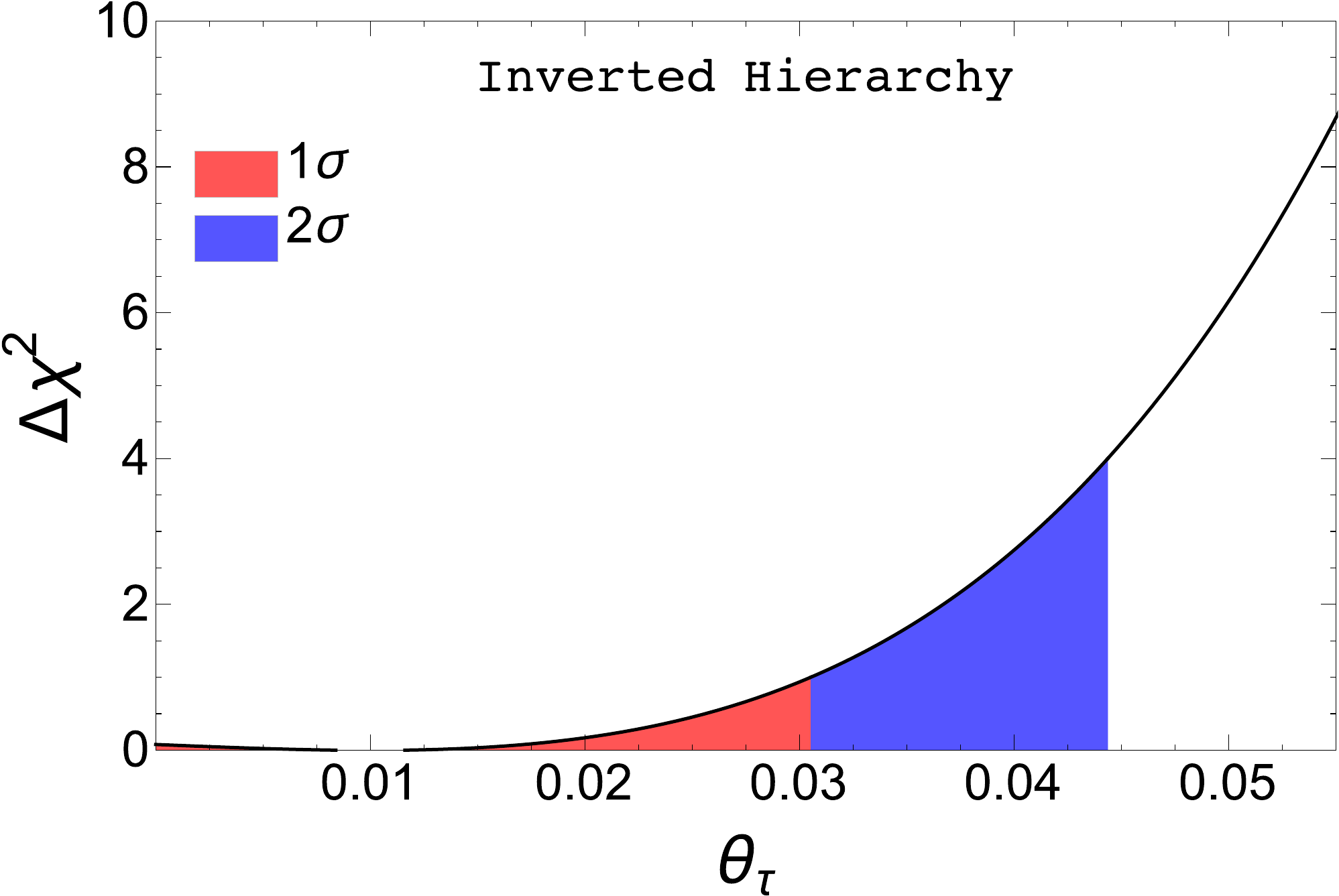}
\caption{$\Delta \chi^{2}$ (marginalised over all other parameters) for $\theta_{e}$, $\theta_{\mu}$ and $\theta_{\tau}$. Left panels show results for a normal hierarchy and right panels for inverted hierarchy.}
\label{fig:1D}
\end{figure}

In Fig.~\ref{fig:1D} we show the individual constraints that can be derived on $\theta_e$, $\theta_\mu$, and $\theta_\tau$ (from top to bottom) for a normal (left) and an inverted (right) hierarchy after marginalising over all other parameters. We generally find a slight improvement of the fit to the observables considered when some amount of mixing is present. In particular, we find that non-zero mixing with the electron is preferred at around the $90 \%$~CL by our dataset. Mixing with the tau flavour is also favoured for normal hierarchy due the correlations implied by Eq.~(\ref{eq:Yt}). At the $1 \sigma$ level, mixing with the $\mu$ flavour is significantly constrained due to the preference of some universality bounds (from $\pi$ and $\tau$ decays) for a slightly reduced coupling to the electron with respect to the muon. Thus, since universality constraints are corrected by $1-|\theta_\alpha|$ for each flavour, a non-zero $\theta_e$ is preferred in the fit while $\theta_\mu$ is kept at small values to satisfy the constraint from $\mu \to e \gamma$. Beyond the $1 \sigma$ level, the mixing with the electron is allowed to become small and thus the constraint on $\mu$ mixing at $2 \sigma$ is much weaker than naively expected from the $1 \sigma$ region. The limits of the 1 and $2 \sigma$ regions for the three mixing parameters are summarised in Table~\ref{tab:constraints}. 

\begin{table}[htb]
\centering
\begin{tabular}{|c|c|c|c|c|c|c|}
\hline
    & \multicolumn{2}{|c|}{$\theta_e$} & \multicolumn{2}{|c|}{$\theta_\mu$} & \multicolumn{2}{|c|}{$\theta_\tau$} \\
\hline
    & $1 \sigma$ & $2 \sigma$ & $1 \sigma$ & $2 \sigma$ & $1 \sigma$ & $2 \sigma$ 
 \\
\hline
NH & $0.034^{+0.009}_{-0.014}$ & $<0.050$ & $<3.2 \cdot 10^{-4}$ & $<0.037$ &$0.018^{+0.019}_{-0.013}$ & $<0.049$   \\
\hline
IH & $0.035^{+0.009}_{-0.014}$ & $<0.051$ & $<3.3 \cdot 10^{-4}$ & $<0.037$ & $<0.031$ & $<0.044$   \\
\hline
\end{tabular}
\caption{Constraints on $\theta_e$, $\theta_\mu$, and $\theta_\tau$ for normal and inverted hierarchy.}\label{tab:constraints}
\end{table}

In Fig.~\ref{fig:chi2} we show a comparison of the breakdown of the contributions of the different observables to the total $\chi^2$ for the SM (left panel) and 
our best fit (middle panel) as well as the difference of the two (right panel). It can be seen that some of the existing tension of the SM with the invisible width 
of the $Z$ can be alleviated by the presence of heavy neutrino mixings and also the agreement between the kinematic determination of $M_W$ and its SM value 
from $G_F$, $\alpha$ and $M_Z$ is improved. As already discussed, the universality constraints from $\pi$ and $\tau$ decays are also in better agreement when 
some mixing with the electron is present. On the other hand, universality tests from kaon decays rather point in the opposite direction. Thus, at the end, the 
preference for non-vanishing heavy-active mixing is mild and the final improvement of the $\chi^2$ with respect to the SM value is 3.7, not quite reaching the $2 \sigma$ level. Notice that, even if the number of free parameters in the fit is rather high, the observables actually depend 
on the combinations $|\theta_e|$, $|\theta_\mu|$ and $|\theta_\tau|$ only (and $\Lambda$ when loop corrections are relevant). Thus, the reduction by 3.7 of the 
$\chi^2$ should be attributed to the introduction of 3 (or 4) new parameters rather than 9.

\begin{figure}[htb]
\centering
\includegraphics[width=0.5\textwidth,angle=270]{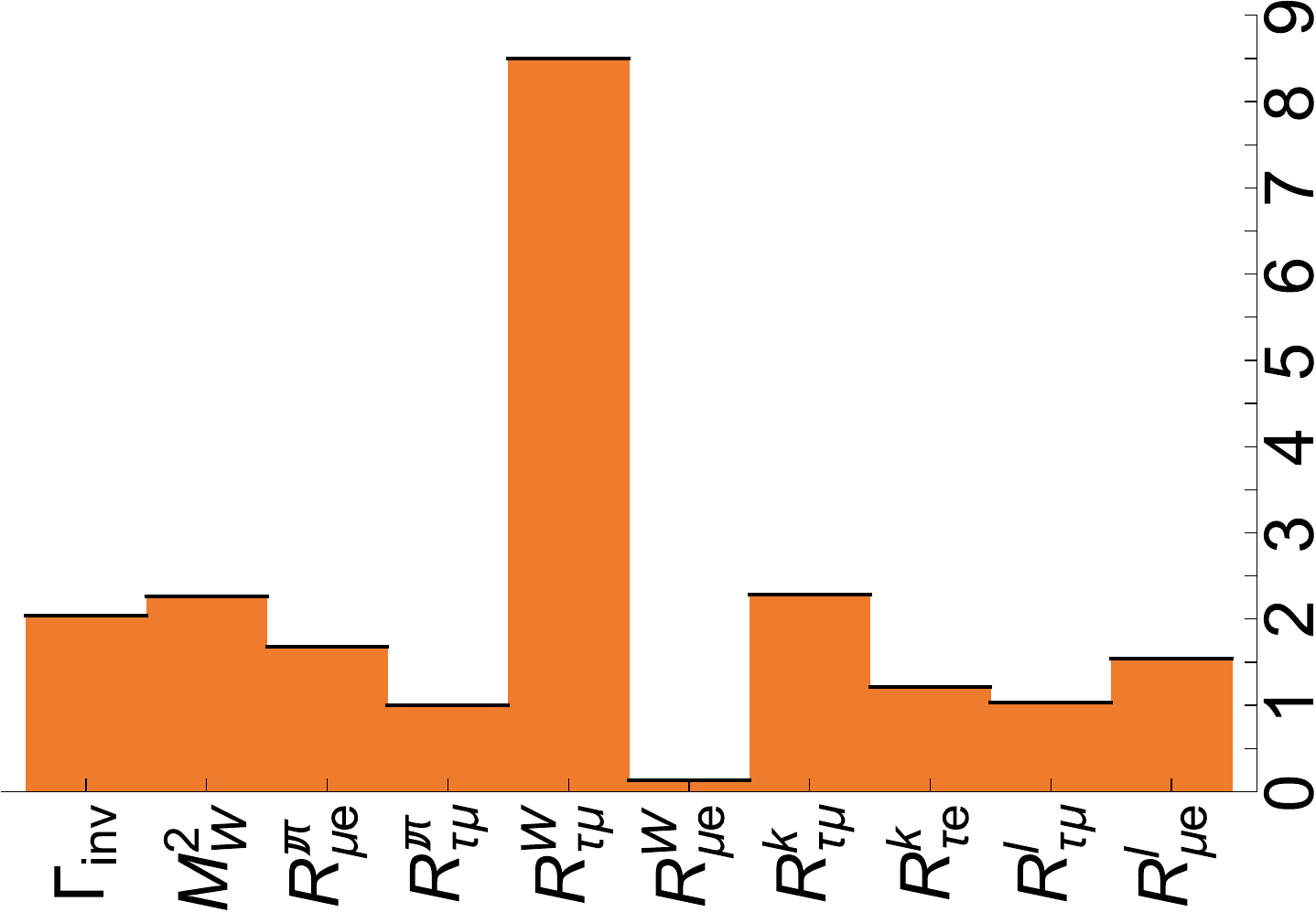}
\includegraphics[width=0.5\textwidth,angle=270]{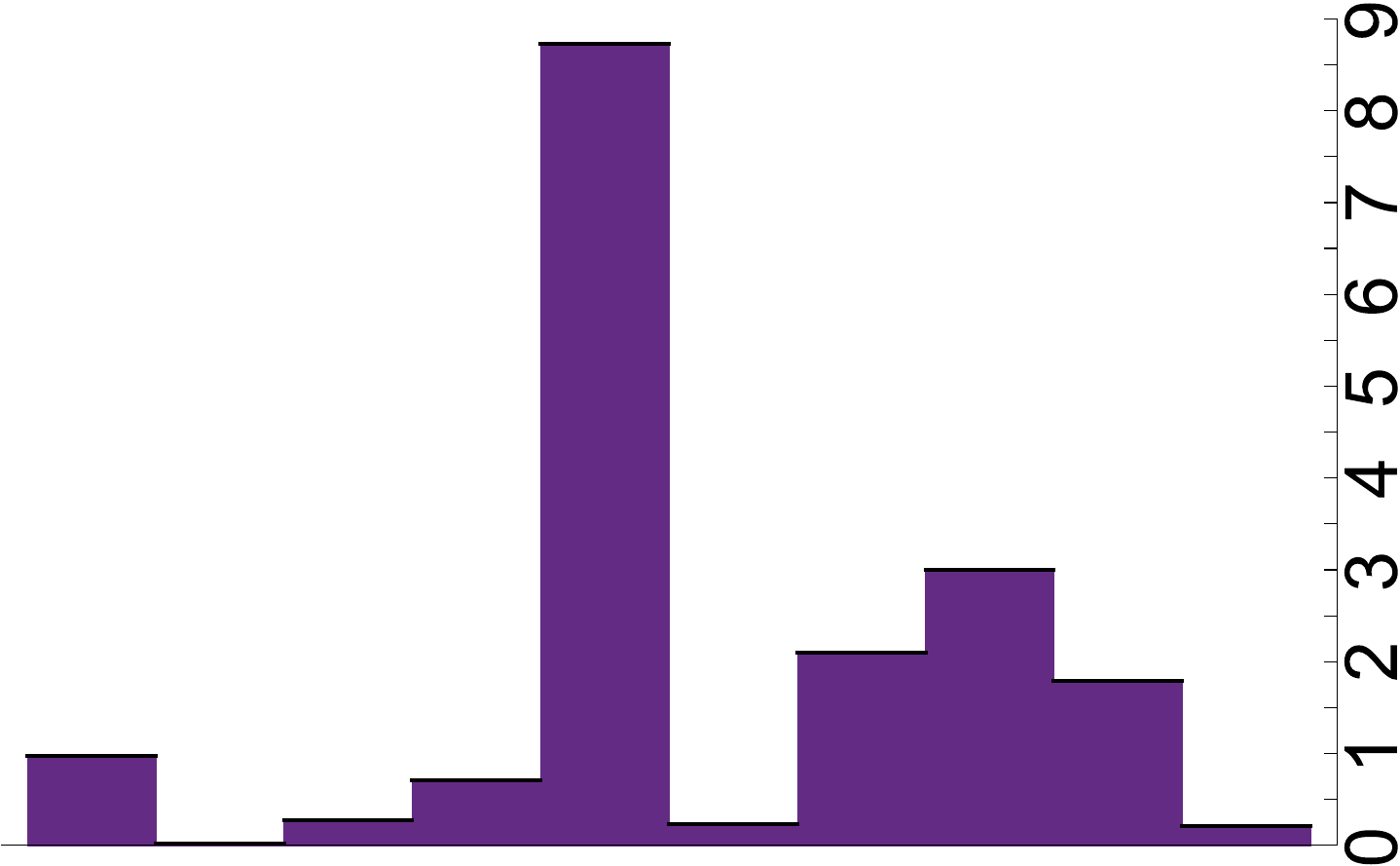}
\includegraphics[width=0.5\textwidth,angle=270]{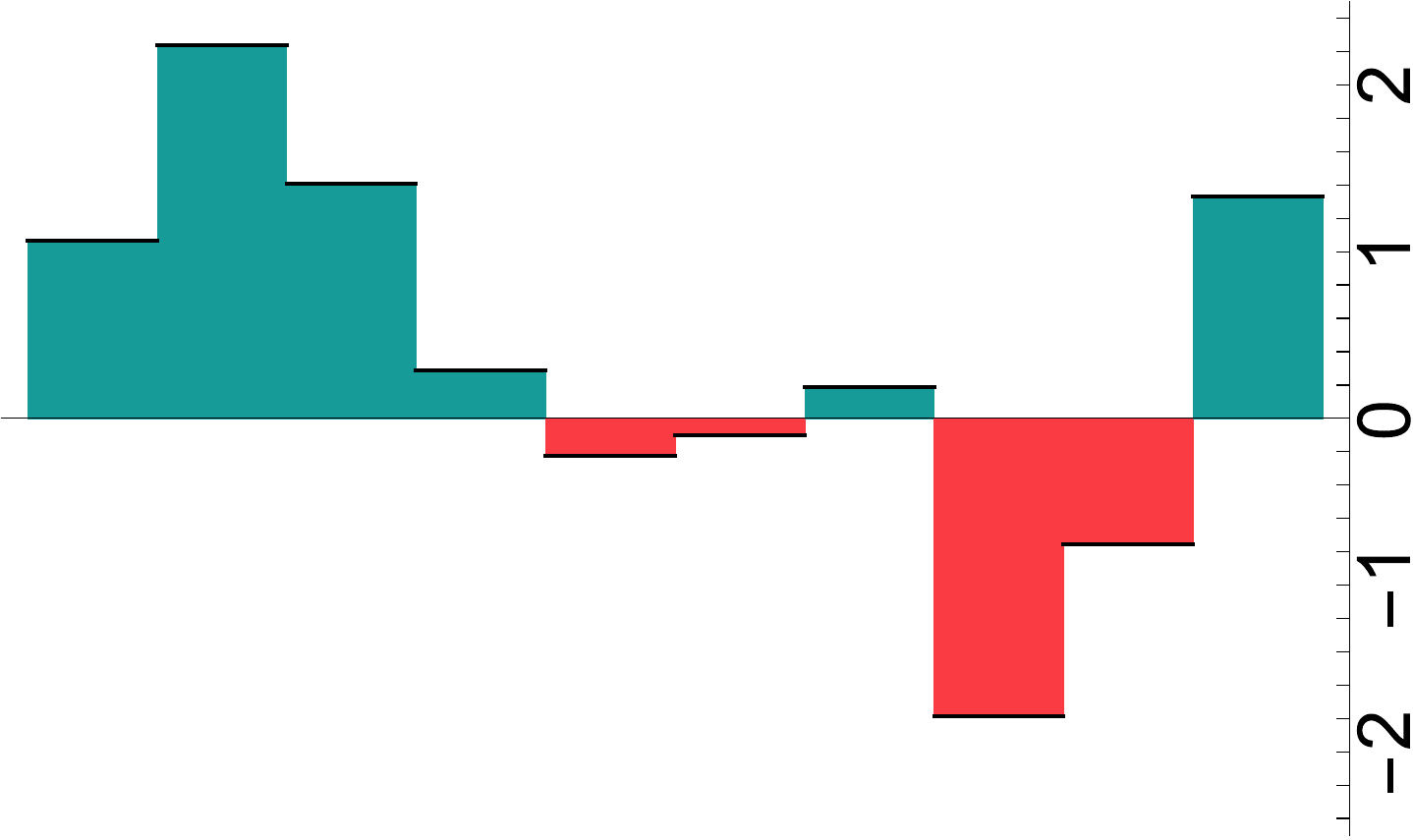}
${\chi^2(SM)}$ \hspace{3.5cm} $\chi^2(BF)$ \hspace{3.5cm} $\Delta\chi^{2}(SM)$
\caption{Contributions from the different observables to the $\chi^{2}$. Left plot shows the SM values. Middle plot shows the contributions from three right-handed neutrinos in the best-fit point. Right plot shows $\Delta\chi^2_i\equiv\chi^{2}_{i}(SM)-\chi^{2}_{i}(BF)$ for every observable $i$.}
\label{fig:chi2}
\end{figure}

Regarding the importance of the loop effects considered, we have performed a second set of MCMC runs where all loop corrections have been removed. The results of these simulations are essentially identical to the ones stemming from the full computation. By adding to the chain output also the value that the $T$ parameter took in the simulations, we find that its preferred values are $\sim 10^{-7}-10^{-6}$, negligible with respect to the best fit values of the tree level contributions. In order to understand this apparent lack of relevance of the loop corrections and the $T$ parameter in particular, in direct contrast to the results presented in~\cite{Akhmedov:2013hec}, we will now analyse in further detail the regions of the parameter space in which $T$ could be relevant and the necessary conditions for the cancellation with the tree level contributions to take place.

\subsection{The $T$ parameter}

The leading contributions (not suppressed by the light neutrino or charged lepton masses) to the $T$ parameter are given by~\cite{Akhmedov:2013hec}:
\begin{equation}
\alpha T = \frac{\alpha}{8 \pi s_\mathrm{W}^2 M^2_W}\left(  \sum_{\alpha,\beta,i,j}\left(U^*_{\alpha i} U_{\alpha j} U_{\beta i} U^*_{\beta j} f_{ij} +  U^*_{\alpha i} U_{\alpha j} U^*_{\beta i} U_{\beta j} g_{ij}\right) \right),
\label{eq:T}
\end{equation}
where
\begin{equation}
f_{ij} = \frac{M_i^2 M_j^2}{M_i^2-M_j^2}\ln{\frac{M_i}{M_j}}
\qquad
\mathrm{and}
\qquad
g_{ij} = \frac{2 M_i M_j^3}{M_i^2-M_j^2}\ln{\frac{M_i}{M_j}} ,
\end{equation}
and where $M_i$ are the neutrino mass eigenvalues. In~\cite{Loinaz:2002ep,Loinaz:2004qc} it was shown that several of the most constraining observables, notably the $Z$ decay to charged leptons and $\sin^2 \theta^{\rm eff}_w$~\cite{ALEPH:2002aa}, depended on the combination:
\begin{equation}
(NN^\dagger)_{ee} (NN^\dagger)_{\mu \mu} - 2 \alpha T \simeq 1 - |\theta_e|^2 - |\theta_\mu|^2 - 2 \alpha T .
\label{eq:cancellation}
\end{equation}
Since from Table~\ref{tab:constraints} $|\theta_e|^2 + |\theta_\mu|^2 \sim 10^{-3}$, $2 \alpha T$ must be of similar order so as to be competitive with the tree contribution. From Eq.~(\ref{eq:T})
\begin{equation}
2 \alpha T \simeq \frac{\alpha \Lambda^2 |\theta_\alpha|^4}{16 \pi s_\mathrm{W}^2 M^2_W},
\end{equation}
where $\Lambda$ is the mass scale of the heavy neutrinos and $\theta_\alpha/\sqrt{2}$ their mixing with the flavour states from Eq.~(\ref{eq:theta}). Thus, in order for $2 \alpha T \sim |\theta_\alpha|^2$ it is necessary that $\Lambda \sim 10-100$~TeV. And, since $|\theta_\alpha|^2 \sim |Y_\alpha|^2 v_\text{EW}^2/2 \Lambda^2 \sim 10^{-3}$, then $|Y_\alpha| \sim 1-10$, on the very limit of perturbativity but, a priori, an interesting possibility.

Furthermore, notice that the second term in Eq.~(\ref{eq:T}) has the typical structure in the elements of the mixing matrix $U$ of $L$-violating processes, such as, for example, neutrinoless double $\beta$ decay. Indeed, this term stems from the correction to the $Z$ propagator with two neutrinos running in the loop and a Majorana mass insertion and it is easy to see that it vanishes in the limit of exactly conserved Lepton number, taking all $\epsilon_i$ and $\mu_j$ to zero. Thus, if $B-L$ is approximately conserved, the first term in Eq.~(\ref{eq:T}) dominates the contribution to $T$. However, it can be shown that the matrix $f_{ij}$ is positive semi-definite for three extra heavy neutrinos or less\footnote{Preliminary explorations indicate that this argument can be generalised to more extra heavy neutrinos.} and can then be diagonalised as $f_{ij} = \sum_k V_{ik} \lambda_k V^*_{jk}$, where $V$ is a Unitary matrix and $\lambda_k \geq 0$. Thus, if $B-L$ is approximately conserved:
\begin{equation}
\alpha T  \sim  \frac{\alpha}{8 \pi s_\mathrm{W}^2 M^2_W}  \sum_{\alpha, \beta, i} \left| \sum_{k} U^*_{\alpha i} U_{\beta i}  V_{ik} \sqrt{\lambda_k} \right|^2 \geq 0 .
\end{equation}
But from Eq.~(\ref{eq:cancellation}) $T < 0$ is mandatory so as to have the cancellation between $T$ and $|\theta_\alpha|^2$ discussed in~\cite{Akhmedov:2013hec}. Thus, significant violations of $B-L$ are necessary so that the second term in Eq.~(\ref{eq:T}), which is allowed to be negative, can dominate over the first. 

\begin{figure}[htb]
\centering
\includegraphics[width=0.7\textwidth]{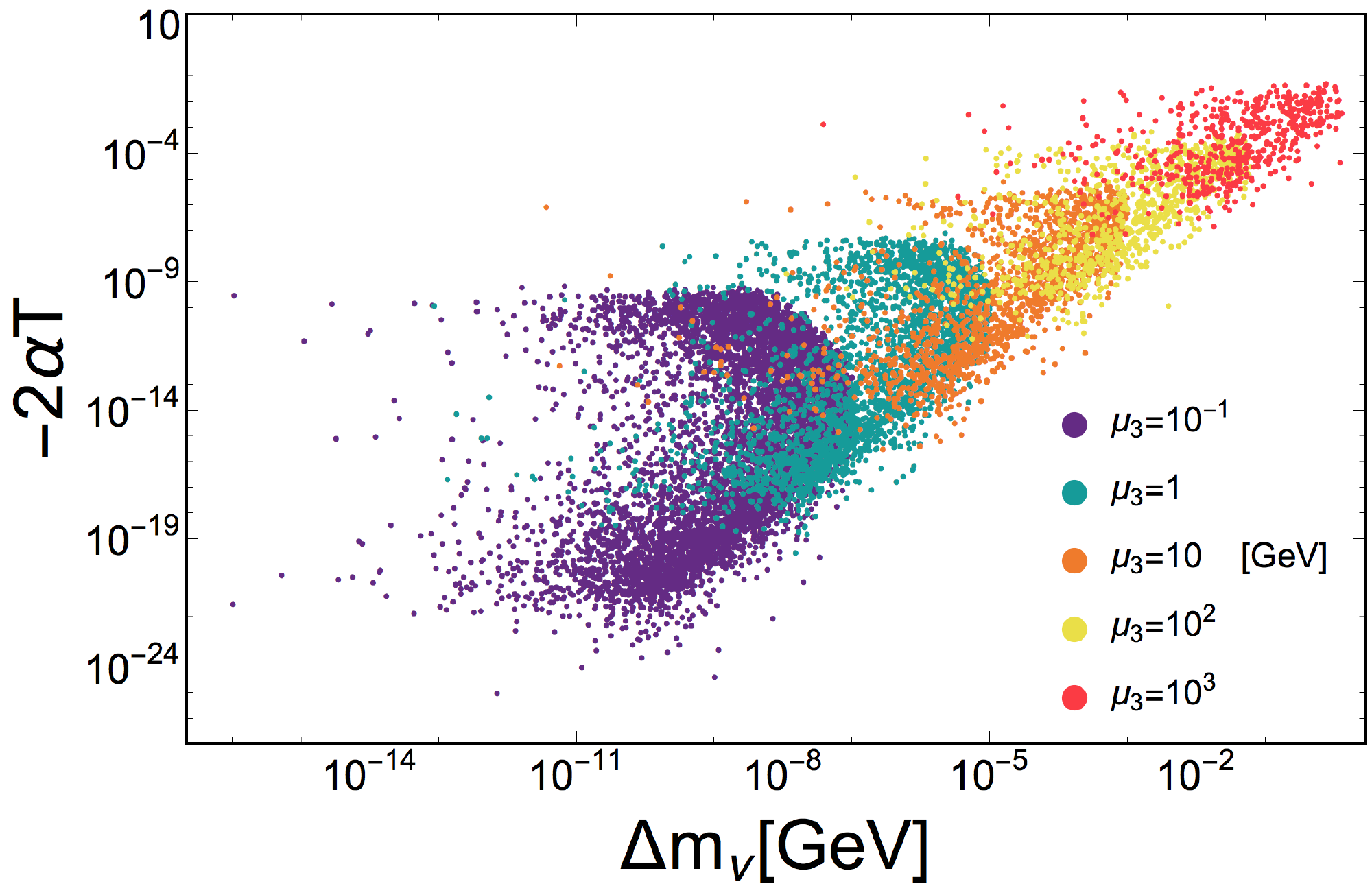}
\caption{$T$ parameter versus 1-loop correction to $m_\nu$ for different values of the $L$-violating parameters $\mu_1$ and $\mu_3$.}
\label{fig:T}
\end{figure}

Notice that, for arbitrary values of the $B-L$-violating parameters $\epsilon_i$ and $\mu_j$, Eq.~(\ref{eq:texture}) is a completely general parametrisation of 
a type-I Seesaw mechanism with three extra right-handed neutrinos. But, given Eq.~(\ref{eq:lightmass0}), only $\mu_1$ and $\mu_3$ are allowed to be sizeable 
given the present constraints on the light neutrino masses and mixings. If $|\mu_1| \gg \Lambda,\Lambda',\mu_3$ a negative $T$ can indeed be obtained:
\begin{equation}
T\simeq \frac{v_\text{EW}^4}{32\pi s_\mathrm{W}^2 M_W^2 \mu_1^2}\left(\displaystyle\sum_\alpha\left|Y_\alpha\right|^2\right)^2\left(3-4\log\left(\frac{\mu_1}{\Lambda}\right)\right) .
\end{equation}
If both $\mu_1$ and $\mu_3$ are simultaneously included and dominate over the $L$-conserving $\Lambda$ and $\Lambda'$ then $T$ is given by: 
\begin{equation}
T\simeq \frac{v_\text{EW}^4}{64\pi s_\mathrm{W}^2 M_W^2}\left(\displaystyle\sum_\alpha\left|Y_\alpha\right|^2\right)^2 \frac{6 \mu \mu_1 - \left( 3 \mu_1^2 + \mu^2 \right) \log\left(\frac{\mu + \mu_1}{\mu - \mu_1}\right)}{\mu^3 \mu_1}  ,
\label{eq:TLV}
\end{equation}
where $\mu = \sqrt{\mu_1^2 + 4 \mu_3^2}$. In this limit, negative values of $T$ are also easily accessible. However, the price to pay is high, the approximate $B-L$ symmetry protecting the Weinberg operator despite the Yukawas at the very border of perturbativity and the low Seesaw scale, has been strongly broken by $\mu_1$ and $\mu_3$. While this does not induce any dangerous corrections to neutrino masses at tree level, and hence when working with the Casas-Ibarra parametrisation as in Ref.~\cite{Akhmedov:2013hec} the correct masses and mixings seem to be recovered, loop corrections need to also be taken into account since no protecting symmetry can now suppress them. Indeed, the loop contributions mediated by $\mu_1$ and $\mu_3$ to the light neutrino masses are found to be~\cite{Pilaftsis:1991ug,Grimus:2002nk,AristizabalSierra:2011mn,LopezPavon:2012zg}:
\begin{equation}
\Delta m_{\nu_{\alpha\beta}}
= \frac{Y_\alpha Y_\beta}{32\pi^2\mu}\left(3M_Z^2f(M_Z)+M_h^2f(M_h)\right) \, ,
\end{equation}
with: 
\begin{equation}
f(M)=\frac{(\mu + \mu_1)^2\log{\left(\frac{\mu + \mu_1}{2M}\right)}}{\left( \mu + \mu_1 \right)^2-4M^2}-\frac{(\mu - \mu_1)^2\log{\left(\frac{\mu - \mu_1}{2M}\right)}}{\left(\mu - \mu_1 \right)^2-4M^2} \, .
\label{eq:f}
\end{equation}
These corrections can indeed be sizeable and in Fig.~\ref{fig:T} we show the values that the loop contribution to the light neutrino masses take in order to recover a given value for $-2 \alpha T$ for different values of $\mu_1$ and $\mu_3$. From inspection of Eq.~(\ref{eq:f}), the limit of vanishing $\mu_1$ would render $f(M) = 0$, keeping under control the loop corrections to neutrino masses\footnote{In this limit with $\mu_3 \gg \Lambda,\Lambda'$, $L$-symmetry is recovered with two degenerate neutrinos with mass $\mu_3$ that form a Dirac pair. Hence, the symmetry ensures the stability of $\nu$ masses at loop level but conversely drives $T$ to positive values.}. However, from Eq.~(\ref{eq:TLV}), $|\mu_1| > 1.9 |\mu_3|$ is necessary for $T < 0$. Indeed, as can be seen in Fig.~\ref{fig:T}, if $-2 \alpha T \sim 10^{-3}$ so as to implement the cancellation between tree and loop level contributions, corrections to the light neutrino masses ranging from $\sim 100$~keV to $\sim 100$~MeV, far exceeding present constraints, would be obtained. Thus, we conclude that, while the qualitatively important cancellations described in Ref.~\cite{Akhmedov:2013hec} can in principle take place and affect the constraints on the heavy-active neutrino mixing for $Y_\alpha \sim 1$ and $\Lambda \sim 10$~TeV, in practice large violations of the protecting $B-L$ symmetry would be required, leading to too large radiative corrections to light neutrino masses.

\chapter{Conclusion}

The neutrino mass generation mechanism, the nature of dark matter and the origin of the baryon asymmetry of the Universe are three  pressing questions in modern astroparticle physics that call for the introduction of physics beyond the Standard Model. Originally introduced to account for the observation of nonzero neutrino masses and mixings, sterile fermions have shown a strong potential in providing a simultaneous solution to the aforementioned observational problems. 

Any neutrino mass generation mechanism must account for the smallness of the neutrino mass scale when compared to the electroweak one. In this thesis we focused on the analysis of mechanisms characterised by an approximated $B-L$ symmetry, due to their potential to account for the observed neutrino masses with a relatively low new physics scale, $\mathcal{O}$(TeV) or lower, that makes them testable in present and future experimental facilities.

In~\cite{Abada:2014vea} we proposed a methodological approach to identify the most minimal Inverse Seesaw realisations
fulfilling all phenomenological requirements.
By adding extra sterile fermions to the SM (right-handed neutrinos, $\nu_R$, and sterile singlets, $s$)
whose number of generations were not fixed ($\# \nu_R$ not necessarily equal to $\# s$), 
we have shown that it is possible to construct several distinct ISS models that can 
reproduce the correct neutrino mass spectrum. 
Our general analysis has shown that the mass spectrum of an ISS realisation is characterised by either 2 or 3 different mass scales, 
corresponding to the one of the light active neutrinos, that
corresponding to the heavy states, and  an intermediate scale associated to $\#s - \# \nu_R$ sterile states (only relevant 
when $\#s > \# \nu_R$).
The approach we followed was based on  time-independent perturbation theory for linear operators, which allowed to analytically diagonalise the neutrino mass matrix. One can thus obtain analytic expressions for the neutrino eigenstates and their associated masses as a power series of the small parameters that violate the total lepton number. 
As a result, we were able to identify two classes of truly minimal ISS realisations that can successfully account for neutrino data.
The first, here denoted ISS(2,2) model, corresponds to the SM extended by 
two RH neutrinos and two sterile states. It leads to a 3-flavour mixing scheme, and requires only two scales (the one corresponding to the light neutrino masses, and the one corresponding to RH neutrino masses). Although considerably fine tuned, this ISS configuration still complies with all phenomenological constraints, and systematically leads to a Normal Hierarchy for the light neutrinos.  
The model could marginally give rise to an effective mass for $0\nu2\beta$ 
within experimental reach, but 
all these regions turn out to be excluded by current laboratory constraints and the MEG bound on $\mu \to e \gamma$ decays.
The second, named ISS(2,3) realisation, corresponds to an extension of the SM by 
two RH neutrinos and three sterile states. This class allows to accommodate both hierarchies for the light spectrum (although the IH  is only marginally allowed), in a  3 + 1-mixing scheme. Concerning $0\nu2\beta$ decays, the ISS(2,3) scenario leads to effective masses close to the current experimental bound and within future sensitivity of coming experiments. The mass of the lightest sterile neutrino can vary over a large interval: 
depending on its regime, the 
ISS(2,3) realisation can offer an explanation for the 
reactor anomaly (in this case, the lightest mostly sterile state has a mass $m_4 \sim$ eV), or provide a Warm Dark Matter candidate (for a mass of the lightest sterile state around the keV).
We have focused on the determination of the truly minimal inverse Seesaw  realisations. Our approach can be easily generalised  to probe the phenomenological viability and impact of any ISS extension of the SM (for an arbitrary number of RH states and sterile fermions).

The viability of the Dark Matter hypothesis in the ISS(2,3) was then analysed in~\cite{Abada:2014zra},  where we have considered the possibility of simultaneously addressing the dark matter problem and the neutrino mass generation mechanism.
We have conducted a comprehensive analysis, taking into account all the phenomenological and cosmological requirements and the several possibilities of neutrino mass spectra. We have found that in most of the parameter space the DM can be produced only through active-sterile transitions according to the DW production mechanism, accounting, in the most favourable case, for at most  $\sim43\%$ of the relic  DM abundance, without conflict with observational constraints.
 This situation can be improved  for two specific choices of the spectrum of the heavy pseudo-Dirac neutrinos. Firstly, one can consider the  case of moderately light, i.e. $\sim 1\ -\ 10$ GeV, pseudo-Dirac neutrinos. These states can dominate the energy density of the Universe and produce entropy at the moment of their decay, altering the impact of DM on structure formation. However the constraints from dark matter indirect detection  are still too severe and the allowed DM fraction is increased only up to $\sim 50\%$. The second possibility relies upon relatively heavy, $\sim 130\text{ GeV}-1\text{ TeV}$, pseudo-Dirac pairs, which can produce the correct amount of DM through their decays. In this kind of setup it is also possible for the ISS(2,3)  to account for the reported $3.5$ keV line in galaxy cluster spectra. 
In the final part of the work, we have proposed a minimal extension of the ISS(2,3) model with the addition of a scalar singlet (at the origin of the lepton number violating masses of the sterile fields) which allows to achieve the correct DM relic density for generic values of the masses of pseudo-Dirac neutrinos. The latter  can still participate, at various levels, to the production mechanism of DM.

In order to address the last issue, that is baryogenesis, we have proposed in~\cite{Abada:2015rta} a minimal extension of the Standard Model by the addition of two sterile fermions with opposite lepton number, forming a setup with an approximate lepton number conservation. The new fields form a pseudo-Dirac pair and are coupled to the active leptons via mixing terms.  The small mass splitting within this pair, as well as the smallness of the active neutrino masses, are due to two sources of lepton number violation with $\Delta L=2$,  corresponding to an Inverse Seesaw framework extended by a Linear Seesaw mass term. The main goal was to  study of the feasibility of simultaneously having a very low-scale Seesaw mechanism - typically at $1 - 10$ GeV - at work for generating neutrino masses and mixings as well as an efficient leptogenesis through oscillations at the electroweak scale within this ``natural'' and minimal framework. Here the naturally arising pseudo-Dirac state ensures a highly efficient leptogenesis due  to its small mass splitting. We have also considered the (pure) Inverse Seesaw mechanism in which several pseudo-Dirac states arise naturally.
We have conducted a comprehensive analytical and numerical analysis investigating both neutrino mass hierarchies, normal (NH) and inverted (IH), for the neutrino mass spectrum and exploring the different washout regimes for the baryon asymmetry of the Universe.
To this end we have implemented and solved a system of Boltzmann equations 
 and have additionally derived an analytical expression for the baryon asymmetry,  providing a better understanding of the behaviour of the solutions. 
Our studies reveal that our scenario (SM extended by two right-handed neutrinos
with two sources of lepton number violation 
by 2 units) 
is efficient to generate a successful leptogenesis through oscillations between the two mostly sterile states while complying with all available data. Our analytical expression is valid in the weak washout regime and agrees with the results obtained by numerically solving the system of Boltzmann equations. In the regime of strong washout, which is numerically very demanding, we have nevertheless proven that  our scenario can provide successful leptogenesis, with values of the active-sterile mixing that can be probed by future facilities such as SHiP.
We have conducted the same study for the pure Inverse Seesaw setup, in which case 
we find that the mass splitting between the states in the pseudo-Dirac pairs is too large to achieve a successful leptogenesis in the weak washout regime while accommodating the neutrino data. 
This analysis is however not conclusive to discard the ISS scenario since it relies on the (severe) condition that all the Yukawa couplings are below the equilibration value.
A complete analysis of the whole parameter space in this case is numerically very demanding, and will be the purpose of a future study.

In~\cite{Abada:2015zea}  we discussed the enhancement of the LFV decays of flavourless vector bosons,  $V \to \ell_\alpha  \ell_\beta$, with $V\in \left\{\phi, \psi^{(n)},\Upsilon^{(n)},Z\right\}$, induced by a mixing between the active and sterile neutrinos. 
The enhancement grows with the mass of the heavy sterile neutrino(s), as can be seen from the mass dependence of the Wilson coefficients that we explicitly calculated. We find that the most significant diagram 
that gives rise to the LFV decay amplitudes is the one coming from the $Z\nu \nu$ vertex, which suggests a steady growth of the decay rate with the mass of the sterile neutrino(s). In the physical amplitude, however, the region of very large 
mass of the sterile neutrino(s) is suppressed as the decoupling takes place, i.e. mixing between the active and sterile neutrinos rapidly falls. 
We illustrated the enhancement of ${\rm B}(V \to \ell_\alpha  \ell_\beta)$ in two scenarios: a model with one effective sterile neutrino that mimics the effect of a generic extensions of the SM including heavy sterile fermions, 
and in a minimal realisation of the inverse Seesaw scenario compatible with current observations (that is the ISS(2,3)). 
Our results for upper bounds on  ${\rm B}(V \to \ell_\alpha  \ell_\beta)$  [$V\in \left\{\phi,J/\psi, \psi(2S),\Upsilon(1S),\Upsilon(2S),\Upsilon(3S),Z\right\}$] are still considerably smaller than the current experimental bounds (when available), 
but that situation might change in the future as more experimental research will be conducted at Belle II, BESIII, LHC, and hopefully at FCC-ee (TLEP). 
If one of the decays studied is observed and turns out to have a branching fraction larger than the reported upper bounds, then sources of LFV other than those coming from mixing with heavy sterile neutrinos must be accounted for.

Finally in~\cite{Fernandez-Martinez:2015hxa} we have analysed in detail the importance of loop corrections when deriving constraints on the mixing between the SM flavour eigenstates and the new heavy neutrinos introduced in Seesaw mechanisms. Although naively the expectation is that radiative corrections involving these new states would be irrelevant given their weaker-than-weak interactions due to their singlet nature and, a priori, suppressed mixings with the SM neutrinos, Seesaw models may allow Yukawa couplings to be sizeable, even order one. Thus, loop corrections involving Yukawa vertices, when the loops involve the heavy neutrinos and the Higgs or the $W$ and $Z$ Goldstones, can indeed be sizeable as shown in Ref.~\cite{Akhmedov:2013hec}. In that work, it was shown that, for the low-scale Seesaw mechanisms characterised by large Yukawas and low (electroweak) Seesaw scale, the contribution of the new degrees of freedom to the oblique parameters could indeed become as important as the tree level effects in some regions of the parameter space. Moreover, it was observed that several observables shared a common dependence between the $T$ parameter and the tree level contribution, stemming from the modification by these effects of the muon decay through which $G_F$ is determined and subsequently used as input for other observables. Thus, a partial cancellation between these tree and loop level contributions can significantly relax the bounds derived from these observables. Indeed, in Ref.~\cite{Akhmedov:2013hec} a good fit with sizeable mixing was obtained in which the most stringent limits were avoided through this partial cancellation while standing tensions between the SM and some observables like the invisible width of the $Z$ were alleviated. 
We have extended the analysis performed in Ref.~\cite{Akhmedov:2013hec} to include also vertex corrections and not only oblique parameters, 
since the sizeable contributions from the heavy Yukawas do not vanish when taking the light neutrinos and charged lepton masses to zero. We have found that, 
all in all, the oblique parameters do tend to dominate over the other loop corrections and their contribution could be sizeable in some part of the parameter space. However, our MCMC scan shows no preference for any sizeable loop corrections and the partial cancellation found in~\cite{Akhmedov:2013hec} is not reproduced. We have then studied in detail the values of the $T$ parameter preferred by data through our MCMC scan and found that they were not only negligible, but always positive in our results, while, for the cancellation between tree level contributions and the $T$ parameter to take place, the latter must have negative values. We thus studied the necessary conditions for sizeable negative values of the $T$ parameter and realised that, not only sizeable Yukawas and relatively low Seesaw scales are required, but also large violations of $B-L$. We have then identified the only parameters in the mass matrix with three extra heavy neutrinos that could provide the necessary $B-L$ violation required for $T$ to be negative and competitive with tree level contributions, while keeping neutrino masses within their current bounds despite the large Yukawas, the low Seesaw scale and the loss of protecting $B-L$ symmetry. Finally, we have studied how these parameters would contribute to neutrino masses at loop level and found that, for the size of $T$ required for the cancellation to take place, light neutrino masses would range from 10~keV to 100~MeV, effectively ruling out this possibility.
We conclude that loop level corrections are only relevant in a small fraction of the Seesaw parameter space characterised by large Yukawa couplings and low Seesaw scale and that these corrections tend to strengthen the tree level contributions unless large deviations from $B-L$ are present. If $B-L$ is approximately conserved, data thus prefer regions of the parameter space where these loops are irrelevant. On the other hand, if $B-L$ is strongly violated, the cancellation discussed in Ref.~\cite{Akhmedov:2013hec} can indeed provide a good fit to data with a very relevant r\^ole of the loop contributions. However, these large violations of $B-L$ at loop level also lead to too large contributions to the light neutrino masses and hence this possibility is ruled out. We therefore conclude that loop corrections can safely be neglected in analyses of the heavy neutrino mixings in Seesaw mechanisms. 
Finally we have also obtained relevant constraints on these mixings when $B-L$ is an approximate symmetry, so as to recover the correct neutrino masses and mixings observed in neutrino oscillation searches. We find a mild ($\sim$ 90\% CL) preference for non-zero mixing with the $e$ flavour with a best fit at $\theta_e = 0.034^{+0.009}_{-0.014}$ or $\theta_e = 0.035^{+0.009}_{-0.014}$ for normal and inverted mass hierarchy respectively. In the case of normal hierarchy, this preference also induces non-zero mixing with the $\tau$ flavour $\theta_\tau = 0.018^{+0.019}_{-0.013}$ so as to recover the correct pattern of neutrino masses and mixings. On the other hand, small $\theta_\mu$ is preferred so as to keep $\mu \to e \gamma$ at acceptable levels in presence of non-zero $\theta_e$. At the $2 \sigma$ level the following upper bounds are found: $\theta_e < 0.051$, $\theta_\mu < 0.037$ and $\theta_\tau < 0.049$.

\newpage
\appendix

\chapter{Perturbative diagonalisation}\label{sec:min_ISS}

\section{Perturbative determination of the neutrino masses and of the leptonic mixing matrix}\label{AppendixA}
In the one generation ISS model, and in the basis defined by  $n_L \equiv \left( \nu_L,\nu_R^c,s \right)^T$, 
the neutrino mass matrix can be written as
\be\label{onegenmatrix}
M\,= \,\left( \begin{array}{ccc} 0 & d & 0 \\ d & m & n \\ 0 & n & \mu \end{array} \right),
\ee
where $d,m,n,\mu$ are complex numbers. 
This symmetric matrix can be diagonalised via~\cite{Schechter:1980gr}
\be\label{diagonalization-a}
U^T \,M \,U = \,\text{diag}(m_0,m_1,m_2)\,,
\ee
where $U$ is a unitary matrix and $m_{0,1,2}$ are the physical masses.
To obtain $U$, we use  the hermitian combination $M^\dagger M$ (or $M M^\dagger$),
\be\label{diagmsquare}
\text{diag}(m_0^2,m_1^2,m_2^2) \,= \,\left( U^T \,M \,U \right)^\dagger \,\left( U^T \,M \,U \right)\,  =\, 
U^\dagger \,M^\dagger \,M \,U\,,
\ee
so that the matrix $U$ diagonalising $M^\dagger \, M$ is the same as the one in Eq.~(\ref{diagonalization-a}).

In the following, we proceed to diagonalise the one-generation  squared mass matrix $M^\dagger\, M$ of Eq.~(\ref{onegenmatrix}), using 
perturbation theory for linear operators. We  also discuss the validity of the perturbative approach. 
The mass matrix $M$ can be decomposed as
\be
M= \underbrace{\left( \begin{array}{ccc} 0 & d & 0 \\ d & 0 & n \\ 0 & n & 0 \end{array} \right)}_{M_0} +\underbrace{ \left( \begin{array}{ccc} 0&0&0\\0&m&0\\0&0&\mu \end{array} \right)}_{\Delta M},
\ee
where $M_0$ is the zeroth order matrix and $\Delta M$ is the perturbation (which violates lepton number by two 
units). One can write $M^\dagger M$ as
\be
M^\dagger M \,=\,\underbrace{M_0^\dagger \,M_0}_{M^2_0} +\underbrace{ \Delta M^\dagger \,M_0 + M_0^\dagger \,\Delta M}_{M_I^2} + \underbrace{\Delta M^\dagger \,\Delta M}_{ M_{II}^2}\,,
\ee
where $M_I^2$ and $M_{II}^2$ are the components of the perturbation that are  homogenous functions of first and second order in the small parameters $ m$ and $\mu$ ($|m|,|\mu| \ll |d|,|n|$).

The perturbativity condition 
$||\Delta M|| \ll ||M_0||$  translates into conditions for the $M_0^2,M_I^2$ and $M_{II}^2$ matrices 
\begin{eqnarray}\label{condnormsI}
\frac{||M_I^2||}{||M_0^2||} &\leq& \frac{2 |m| |d| + 2|m| |n| + 2|\mu| |n|}{|d|^2 + |n|^2}\ll  1 \,,\nonumber \\
\frac{||M_{II}^2||}{||M_I^2||} &\le& \frac{|m|^2 + |\mu|^2}{|m| |n|} \ll 1\,.
\end{eqnarray} 
The perturbative determination of the mass eigenvalues is thus ensuring , 
provided that $|m|,|\mu| \ll |n|$.

For completeness, one must also determine perturbatively
 the matrix $U$ of  Eqs.~(\ref{diagonalization-a}, \ref{diagmsquare}), 
i.e. the leptonic mixing matrix (corresponding to the $U_\text{PMNS}$).
The eigenvalues of $M_0^2$ are given by 
\be
\label{onezeromasses}
\begin{array}{cc}
 {m_0^2}^{(0)} \,=\, 0\,, & {m_{1,2}^2}^{(0)} \,=\, |d|^2+|n|^2\,.
\end{array}
\ee
Denoting by $\mathbf{x}^{(0)}_{0}$ the normalised eigenvector associated to the null eigenvalue and by $\mathbf{x}^{(0)}_1$ and $\mathbf{x}^{(0)}_2$, an orthonormal combination of eigenvectors associated to the degenerate eigenvalue $|d|^2+|n|^2$,  the first order correction to $\mathbf{x}^{(0)}_0$ is given by 
\be\label{eigenvectorcorr}
\mathbf{x}^{(1)}_0 \,=\, \sum_{j=1,2} -\frac{{\mathbf{x}^{(0)}_j}^\dagger M_I^2 \  \mathbf{x}^{(0)}_0}{|d|^2 + |n|^2} \,\mathbf{x}^{(0)}_j\,.
\ee
Since $|\mu|,|m| \ll |n|$,   the coefficients in Eq.~(\ref{eigenvectorcorr}) verify
\begin{eqnarray}\label{condvecpert}
\left| \frac{{\mathbf{x}^{(0)}_j}^\dagger M_I^2\ \mathbf{x}^{(0)}_0}{|d|^2+|n|^2}\right| &\leq & \frac{||\mathbf{x}^{(0)}_j||\ ||M_I^2\ \mathbf{x}^{(0)}_0||}{|d|^2 +|n|^2} \ll 1\,.
\end{eqnarray}

Similar arguments apply to the first order corrections to $\mathbf{x}^{(0)}_{j=1,2}$; the second order 
eigenvector corrections are still subdominant, thus confirming the validity of the perturbative approach.

The lightest neutrino mass arises from perturbative corrections to the $m=0$ eigenvalue, while the two other states are massive and degenerate (pseudo-Dirac heavy neutrinos). The correction to  ${m_0^2}^{(0)}$  at second order is 
\begin{eqnarray}\label{1genneutrinomass}
{m_0^2}^{(2)} &=& \frac{|d|^4 |\mu |^2}{\left(|d|^2+|n|^2\right)^2},
\end{eqnarray}
which reduces to the usual inverse Seesaw result once the condition $|d| \ll |n|$ is assumed. 
As discussed in Section~\ref{Sec:towards}, in this approach the only assumption on the magnitude of the physical parameters is driven by the naturalness requirement, i.e.  $|m|,|\mu|\ll |d|,|n|$.

The eigenvector associated to ${m_0^2}^{(2)}$  is given  at zeroth order in the perturbative expansion by\footnote{The phases $\alpha_i$ cannot be fixed by diagonalising $M^\dagger M$ in (\ref{diagmsquare}). In fact, given an orthonormal basis of vectors, one  can freely change their phases and still have an orthonormal basis. They must be fixed using Eq.~(\ref{diagonalization-a}) and imposing that $m_i \geq 0$ for all $i$.}
\be\label{1genneutrinoeigenvec}
\mathbf{x}_0^{(0)} =e^{i \alpha_0} \left(
\begin{array}{c}
 -\frac{n d^*}{|d| \sqrt{|d|^2+|n|^2}} \\
 0 \\
 \frac{|d|}{\sqrt{|d|^2+|n|^2}}
\end{array}
\right),
\ee
and its  first order correction is
\be
\mathbf{x}_0^{(1)} =e^{i \alpha_0} \left(
\begin{array}{c}
 0 \\
 -\frac{\mu  |d| n^*}{\sqrt{\left(|d|^2+|n|^2\right)^3}} \\
 0
\end{array}
\right). 
\ee
The first order corrections to ${m_{1,2}^2}^{(0)}$ lift the degeneracy of the states and are given by 
\be
\begin{array}{cc}
 {m_1^2}^{(1)} = -\frac{\left|\mu ^* n^2+m |d|^2+m |n|^2\right|}{\sqrt{|d|^2+|n|^2}}, & {m_2^2}^{(1)}= \frac{\left|\mu ^* n^2+m |d|^2+m |n|^2\right|}{\sqrt{|d|^2+|n|^2}},
\end{array}
\ee
with zeroth order eigenstates
\be\label{toyend}
\mathbf{x}_1^{(0)}  =e^{i \alpha_1} \left(
\begin{array}{c}
 -\frac{d^* \left(m |d|^2+m |n|^2+n^2 \mu ^*\right)}{\sqrt{2} \sqrt{|d|^2+|n|^2} \left| {n^*}^2 \mu+m |d|^2+m
   |n|^2\right|} \\
 \frac{1}{\sqrt{2}} \\
 -\frac{n^* \left(m |d|^2+m |n|^2+n^2 \mu ^*\right)}{\sqrt{2} \sqrt{|d|^2+|n|^2} \left| {n^*}^2 \mu+m |d|^2+m
   |n|^2\right|}
\end{array}
\right),
\ee
\be
\mathbf{x}_2^{(0)} =e^{i \alpha_2} \left(
\begin{array}{c}
 \frac{d^* \left(m |d|^2+m |n|^2+n^2 \mu ^*\right)}{\sqrt{2} \sqrt{|d|^2+|n|^2} \left| {n^*}^2 \mu+m |d|^2+m
   |n|^2\right|} \\
 \frac{1}{\sqrt{2}} \\
 \frac{n^* \left(m |d|^2+m |n|^2+n^2 \mu ^*\right)}{\sqrt{2} \sqrt{|d|^2+|n|^2} \left| {n^*}^2 \mu+m |d|^2+m
   |n|^2\right|}
\end{array}
\right).
\ee

\section{Study of the ISS(2,2) realisation}\label{AppendixB}

Here, we use the perturbative approach described above to determine the neutrino spectrum and the leptonic mixing matrix. 
In this minimal model, the neutrino mass terms in the Lagrangian are 
\be
- \mathcal{L}_{m_\nu}\, =\, n_L^T\, C\, {M}\, n_L + \text{h.c.}\,,
\ee
where
\be
\begin{array}{cc}
n_L \,\equiv \, \left(  \nu_L^1,\, \nu_L^2,\,  \nu_L^3,\,  \nu_R^{c,1},\,  \nu_R^{c,2} ,\,  s^1 ,\, s^2 \right)^T,\hspace{1cm} & \text{and}\   C = i \gamma^2 \gamma^0. \end{array}
\ee
The  ISS(2,2) mass matrix ${M}$ is given by
\be\label{m22full}
{M} = \left( \begin{array}{ccccccc} 
0&0&0&d_{1,1}&d_{1,2}&0&0\\
0&0&0&d_{2,1}&d_{2,2}&0&0\\
0&0&0&d_{3,1}&d_{3,2}&0&0\\
d_{1,1}&d_{2,1}&d_{3,1}&m_{1,1}&m_{1,2}&n_{1,1}&n_{1,2}\\
d_{1,2}&d_{2,2}&d_{3,2}&m_{1,2}&m_{2,2}&n_{2,1}&n_{2,2}\\
0&0&0&n_{1,1}&n_{2,1}&\mu_{1,1}&\mu_{1,2}\\
0&0&0&n_{1,2}&n_{2,2}&\mu_{1,2}&\mu_{2,2}
 \end{array} \right)\, .
\ee
 Using Eq.~(\ref{nphys}), the number $n_p$ of physical parameters is $24$.  
 In  the following we choose\footnote{The mass matrix of Eq.~(\ref{m22full}) can be cast in such a form through the following  procedure: via a combination of  the transformations  in Eq.~(\ref{sym1.22}) and Eq.~(\ref{sym2.22}), one can always choose a basis in which the charged leptonic mass matrix ${\mathfrak{m}}$ is diagonal and real. With a combined transformation of Eq.~(\ref{sym3.22}) and Eq.~(\ref{sym4.22}) the matrix $n$ can be rendered real and diagonal; similar transformations allow to eliminate two phases form the matrix $\mu$ (for example those in the diagonal) while keeping $n$ real. Finally, another combined transformation of Eq.~(\ref{sym1.22}) and Eq.~(\ref{sym2.22}), allows to make 
one column  of the Dirac mass matrix, $d$, real (the first one, for example), while keeping ${\mathfrak{m}}$ real.} 
a  basis in which one has exactly 24 free parameters, as shown in Table~\ref{physpar22}. 
\begin{table}[htb]
\begin{center}
\begin{tabular}{|c|c|c|c|}
\hline
Matrix & \# of moduli & \# of phases &  Total \\
\hline
Diagonal and real ${\mathfrak{m}}$ & $3$ & $0$ & $3$ \\
$d$ with one real column & $6$ & $3$ & $9$ \\
 $m$ & $3$ & $3$ & 6 \\
Real and diagonal $n$ & $2$ & $0$ & 2 \\
$\mu$ with real diagonal & $3$ & $1$ & 4\\
\hline
Total & $17$ & $7$ & $24$\\
\hline 
\end{tabular}
\end{center}
\caption{Example of a basis in which the number of parameters matches the number of physical parameters.}
\label{physpar22}
\end{table}

\noindent 
In the chosen basis,  the mass matrices ${M}_0$ and  $ \Delta{M}$ (${M}={M}_0 + \Delta{M}$) are given by
\bea\label{22massmatrix}
\begin{tabular}{cc}
${M} _0= \!\!\left( \begin{array}{ccccccc} 
0&0&0&d_{1,1}&d_{1,2}&0&0\\
0&0&0&d_{2,1}&d_{2,2}&0&0\\
0&0&0&d_{3,1}&d_{3,2}&0&0\\
d_{1,1}&d_{2,1}&d_{3,1}&0&0&n_{1}&0\\
d_{1,2}&d_{2,2}&d_{3,2}&0&0&0&n_{2}\\
0&0&0&n_{1}&0&0&0\\
0&0&0&0&n_{2}&0&0
 \end{array} \right),$&$\!\!\!\!\Delta {M} = \!\!\left( \begin{array}{ccccccc} 
0&0&0&0&0&0&0\\
0&0&0&0&0&0&0\\
0&0&0&0&0&0&0\\
0&0&0&m_{1,1}&m_{1,2}&0&0\\
0&0&0&m_{1,2}&m_{2,2}&0&0\\
0&0&0&0&0&\mu_{1,1}&\mu_{1,2}\\
0&0&0&0&0&\mu_{1,2}&\mu_{2,2}
 \end{array} \right),$
\end{tabular}
\eea
where ($d_{i,1},n_i,\mu_{i,i}$) are real  and ($d_{i,2},\mu_{1,2},m_{i,j}$) are complex numbers.

\subsection{Massless eigenstate}
Having a massless eigenstate is an unavoidable feature of the minimal  ISS(2,2) and  ISS(2,3) realisations. 
In the minimal  ISS(2,2) realisation,  the massless eigenstate is given by 
\hskip -0.3cm\bee
\mathbf{v_1}&=&e^{i \left(\alpha_1-\phi_3\right)} \left( 
\tilde{\Delta_1},
-\tilde{\Delta_2},
\tilde{\Delta_3}, 0\,,0\,,0, 0
\right)^T,\  \tilde{\Delta_i}=\frac{\Delta_i}{\sqrt{|\Delta_1|^2+|\Delta_2|^2+|\Delta_3|^2}} =\left|\tilde{\Delta_i}\right|e^{i \phi_i},\eee
\bee
\text{with} \ \ \Delta_1=d_{2,1}d_{3,2}-d_{2,2}d_{3,1}, \quad
\Delta_2=d_{1,1}d_{3,2}-d_{1,2}d_{3,1}, \quad
\Delta_3=d_{1,1}d_{2,2}-d_{1,2}d_{2,1}, 
\eee
which is compatible with the constraints on the $U_\text{PMNS}$ matrix, in both cases of normal and inverted hierarchy.

\subsection{Perturbative diagonalization}\label{perturbative}
At  zeroth order, the (squared) masses of the system are given by the following set of eigenvalues of the matrix ${M}_0$ of Eq.~(\ref{22massmatrix})
\bee\label{22zeroth}
\lambda  =  \left\{ 0,0,0,\frac{f-\sqrt{f^2-4g}}{2},\frac{f-\sqrt{f^2-4g}}{2},\frac{f+\sqrt{f^2-4g}}{2},\frac{f+\sqrt{f^2-4g}}{2}\right\} ,
\eee

\bee\label{fg}
\text{where}\ \ f&\!\!=&\!\!|d_{1,2}|^2+|d_{2,2}|^2+|d_{3,2}|^2+d_{1,1}^2+d_{2,1}^2+d_{3,1}^2+n_{1,1}^2+n_{2,2}^2,\non
\text{and}\ \ g&\!\!=&\!\! |d_{1,2}|^2 \left(d_{2,1}^2+d_{3,1}^2+n_{1,1}^2\right)+|d_{3,2}|^2
   \left(d_{1,1}^2+d_{2,1}^2+n_{1,1}^2\right)+|d_{2,2}|^2 \left(d_{1,1}^2+d_{3,1}^2+n_{1,1}^2\right)\non 
   &&-d_{1,1} d_{2,1}
   d_{2,2} d_{1,2}^*-d_{1,1} d_{1,2} d_{3,1} d_{3,2}^*-d_{2,1} d_{2,2} d_{3,1} d_{3,2}^*-d_{1,1} d_{3,1} d_{3,2}
   d_{1,2}^*\non 
   && -d_{2,1} (d_{1,1} d_{1,2}+d_{3,1} d_{3,2}) d_{2,2}^*+d_{1,1}^2 n_{2,2}^2+d_{2,1}^2 n_{2,2}^2+d_{3,1}^2
   n_{2,2}^2+n_{1,1}^2 n_{2,2}^2\ .
\eee
Two of the three massless states receive perturbative contributions from 
$\Delta {M}$ of Eq.~(\ref{22massmatrix}) and, at second order in the perturbative expansion, the light neutrino  spectrum is given by
\bee\label{22issmasses}
{m_1^2}^{(2)}=0,\quad
{m_2^2}^{(2)}=\frac{b-\sqrt{b^2+4 c}}{2},\quad
{m_3^2}^{(2)}&=&\frac{b+\sqrt{b^2+4 c}}{2},
\eee
where the parameters $b$ and $c$ are expressed in terms of the entries of the (2,2) mass matrix given in Eq.~(\ref{m22full}) ($b$ and $c$ do not depend on the submatrix $m_{i,j}$). Due to the long and involved expressions for both parameters $b$ and $c$, we refrain from displaying the corresponding formulae here. 
Nevertheless, the compact expressions above allow to extract important information: the ISS(2,2) 
scenario strongly prefers the NH scheme. 

\chapter{Boltzmann equation for sterile neutrinos produced from decay}\label{sec:app_ISSDM}

In this appendix we briefly describe the numerical treatment used to validate and complement the results presented in Chapter~\ref{sec:DMMISS}. On general grounds one should solve a system of coupled Boltzmann equations for the abundance of the $\Sigma$ field as well as all the 5 extra neutrinos of the ISS scenario. As already
mentioned we will focus on the case in which the pseudo-Dirac neutrinos can be regarded in thermal equilibrium during the DM production phase. 
This allows to focus on a system of two coupled Boltzmann equations whose general form is:
\bee
\label{eq:num_sys}
 \frac{d n_{\Sigma}}{dt}+3 H n_{\Sigma}&=&-B \langle \Gamma \rangle n_{\Sigma} -(1-B) \langle \Gamma \rangle \left(n_{\Sigma}-n_{\Sigma, \rm eq}\right)\nonumber\\
&&+ \sum_I \tilde{B}_I \langle \Gamma_{N_I} \rangle n_{I}+\sum_I \left(1-\tilde{B}_I\right) \langle \Gamma_{N_I} \rangle \left(n_I-n_{I,\rm eq}\right) \nonumber\\ 
&&-\langle \sigma v \rangle \left(n_\Sigma^2-n_{\Sigma, \rm eq}^2\right), \nonumber\\
& \nonumber\\
 \frac{d n_{\rm DM}}{dt}+ 3 H n_{\rm DM}&=& B \langle \Gamma \rangle n_{\Sigma}\nonumber\\
&& +\sum_I \tilde{B}_I \langle \Gamma_{N_I} \rangle n_{I,\rm eq}+\sum_I \langle \Gamma\left(N_I \rightarrow h\, +\text{ DM}\right)\rangle n_{I,\rm eq} \nonumber\\
&& +DW.
\eee
The first equation traces the time evolution of the field $\Sigma$. The first row on the right-hand side represent the decay of $\Sigma$, respectively into at least one DM particle and only into pseudo-Dirac neutrinos, if kinematically allowed. Since the latter are assumed in thermal equilibrium this second term is balanced by a term accounting for inverse decays and thus vanishes if $\Sigma$ is in thermal equilibrium. On the contrary the first term is not balanced by an inverse decay term since the DM has too weak interactions to be in thermal equilibrium and then can be assumed to have a negligible abundance at early stages; this originates the freeze-in production channel. The second row represents the decays, if kinematical allowed, of the pseudo-Dirac neutrinos into $\Sigma$ and another neutrino. The factor $\left(n_I-n_{I,\rm eq}\right)$ assumes that $\Sigma$ is in thermal equilibrium and disappears if the pseudo-Dirac neutrinos are as well in thermal equilibrium. Here we have again distinguished the decay term into DM, which is non balanced by the inverse process, and the decay term into final thermal states (this distinction holds only if $\Sigma$ is in thermal equilibrium. In the regime $\lambda_{\rm H\Sigma} \ll \overline{\lambda}_{\rm H\Sigma}$ the second row should be replaced by the term $\sum_I \langle \Gamma_{N_I} \rangle n_{I}$). The last term finally represents the annihilation processes of $\Sigma$. $B$ and $\tilde{B}_I$ represent the effective branching fractions of decay of, respectively, $\Sigma$ and the pseudo-Dirac neutrinos. $\langle \Gamma \rangle$ and $\langle \sigma v \rangle$ represent the conventional definitions of the thermal averages~\cite{Gondolo:1990dk}:
\bee
 \langle \Gamma \rangle &=& \Gamma \frac{K_1(x)}{K_2(x)},\nonumber\\
 \langle \sigma v \rangle &=& \frac{1}{8 m_\Sigma^4 T K_2^2(m_\Sigma/T)} \int_{4 m_\Sigma^2}^{\infty} ds \sigma_{\rm ann} \left(s-4 m_\Sigma^2\right) \sqrt{s} K_1\left(\sqrt{s}/T\right),\,\,\,\,\,\,\sigma_{\rm ann} \propto \frac{\lambda_{H\Sigma}^2}{s} \nonumber \\
&=& \frac{\lambda_{H\Sigma}^2}{4 m_\Sigma^2 x^2}F(x),
\eee
where the function $F(x)$ is determined by numerically solving the integral above.

The second equation traces the DM number density. The first two rows represent the DM production from, respectively, $\Sigma$ and the pseudo-Dirac neutrinos. The term labelled $DW$ represents instead the contribution associated to production from oscillation processes. In the parameter space of interest the two production processes, decay and oscillations, occur at well separated time scales; as a consequence we can drop the DW term from the equations and possibly add its contribution to the final relic density.

In order to account possible effects of entropy injection from the decays of the pseudo-Dirac neutrinos the system above should be completed with a third equation accounting for the non conservation of the entropy (see e.g.~\cite{Arcadi:2011ev}). On the other hand it has been shown that the pseudo-Dirac neutrinos can dominate the energy budget of the Universe and inject sizeable amount of entropy only  at very late times, compared to the DM production from decay which occurs at temperature close to the mass scale of $\Sigma$ (a possible exception is the case $\lambda_{\rm H\Sigma} \ll \overline{\lambda}_{\rm H\Sigma}$). To a  good approximation we can thus stick on a system of the form~(\ref{eq:num_sys}) and apply a posteriori possible entropy effects.  

For simplicity we will describe two specific examples, namely all the pseudo-Dirac neutrinos lighter or heavier than $\Sigma$. In the first case all the source terms associated to the decays of the pseudo-Dirac neutrinos can be dropped. Moving to the quantities $Y_{\Sigma, \rm DM}=n_{\Sigma,\rm DM}/s$ and $x=m_\Sigma/T$ as, respectively, dependent and independent variables, the system reduces to:
\bee
\frac{dY_\Sigma}{dx} &=& -\frac{1}{16 \pi}\frac{\tilde{h}^2 m_\Sigma}{H x}\frac{K_1(x)}{K_2(x)} \left(Y_\Sigma -(1-B) Y_{\Sigma,eq}\right)\nonumber\\
&&-\frac{45 \lambda_{\rm H\Sigma}^2 m_\Sigma}{512 \pi^7 g_{*} H x^6}F(x)\left(Y^2_{\Sigma}-Y^2_{\Sigma,eq}\right),\nonumber\\
 \frac{dY_{\rm DM}}{dx} &=& \frac{1}{16 \pi}\frac{\tilde{h}^2 m_{\Sigma}}{H x}\frac{K_1(x)}{K_2(x)} B Y_\Sigma, \nonumber\\
\tilde{h}^2 &=& \sum_{IJ} |h_{{\rm eff},IJ}|^2 \left(1-\frac{{\left(m_I+m_J\right)}^2}{m_\Sigma^2}\right),\,\,\,\,\nonumber\\
 B &=& \frac{\sum_I |h_{\rm eff,I1}|^2\left(1-\frac{{\left(m_I\right)}^2}{m_{\Sigma}^2}\right)}{\sum_{I,J}|h_{\rm eff,I1}|^2\left(1-\frac{{\left(m_I+m_J\right)}^2}{m_{\Sigma}^2}\right)+y_f^2 \sin^2 \alpha \left(1-\frac{4 m_f^2}{m_\Sigma^2}\right)},
\eee
where $Y_{\Sigma,\rm eq}=\frac{45}{4 g_{*}\pi^4}x^2 K_2(x)$ and $H$ is the Hubble expansion rate $H=\sqrt{\frac{4 \pi^3 g_{*}}{45}}\frac{m_\Sigma^2}{x^2 M_{\rm Pl}}$. This last expression assumes that during the phase of DM generation the Universe is radiation dominated, this is reasonable since we have shown in the main text that the number density of the heavy neutrinos tends to dominate at  low temperatures.

The numerical solution of this system has been presented in the left panel of Figure~\ref{fig:sterile_production_boltzmann} for some sample values of the relevant parameters.

The analytic expressions provided in the text correspond instead to suitable limits in which this set of equations can be solved analytically.  In the regime $\lambda_{{\rm H}\Sigma} \geq \overline{\lambda}_{{\rm H}\Sigma}$ the right-hand side of the equation for the DM is dominated, at early times, by the annihilation term and we have simply $Y_{\Sigma}=Y_{\Sigma, \rm eq}$. In this regime we have only to solve the equation of the DM substituting $Y_{\Sigma, \rm eq}$ on the right-hand side. The equation can be straightforwardly integrated for:
\begin{equation}
Y_{\rm DM}= \frac{45}{1.66\,64 \pi^5 g_{*}^{3/2}}\frac{M_{\rm Pl}}{m_{\Sigma}}\sum_I |h_{\rm eff,I1}|^2\left(1-\frac{{\left(m_I\right)}^2}{m_{\Sigma}^2}\right) \int x^3 K_1(x) dx.
\end{equation}     
For late enough decays we can integrate the Bessel function from zero to infinity thus obtaining the freeze-in contribution to DM relic density:
\begin{equation}
Y^{\rm FI}_{\rm DM}= \frac{135}{1.66\,128 \pi^4 g_{*}^{3/2}}\frac{\sum_I |h_{\rm eff,I1}|^2 M_{\rm Pl}}{m_{\Sigma}},
\end{equation}
where we have neglected, for simplicity, the kinematical factors in this last expression.      
 At late times the only relevant terms in the equation are the decay terms, and the DM equation can be again integrated with initial condition $Y_{\Sigma}=Y_{\Sigma, \rm eq}(x_{f.o.})$, obtaining the SuperWimp contribution to the DM relic density.
 In the regime $\lambda_{\rm H\Sigma} < \overline{\lambda}_{\rm H\Sigma}$ instead the abundance of the $\Sigma$ field is always below the equilibrium value. We can thus drop the term proportional to $Y_{\Sigma}$ in the first Boltzmann equation which can be directly integrated over $x$. Assuming again enough late decays we can carry the integration until infinity obtaining Eq.~(\ref{eq:SFI}).   

In the case in which the pseudo-Dirac neutrinos are heavier than $\Sigma$ the system of Boltzmann equations is modified as:
\bee
\frac{dY_\Sigma}{dx} &=& -\frac{1}{16 \pi}\frac{\tilde{h}^2 m_\Sigma}{H x}\frac{K_1(x)}{K_2(x)} Y_{\Sigma}\nonumber\\
&& +\frac{1}{16 \pi}\sum_I \tilde{B}_I \frac{\overline{h}_I^2 m_I}{H x}\frac{K_1(x)}{K_2(x)}Y_{I} +\frac{1}{16 \pi}\sum_I \left(1-\tilde{B}_I\right) \frac{\overline{h}_I^2 m_I}{H x}\frac{K_1(x)}{K_2(x)}\left(Y_I-Y_{I,eq}\right)\nonumber\\
&&+\frac{45 \lambda_{\rm H\Sigma}^2 m_\Sigma}{512 \pi^7 g_{*} H x^6}F(x)\left(Y^2_{\Sigma}-Y^2_{\Sigma,eq}\right), \nonumber\\
 \frac{dY_{\rm DM}}{dx}&=&\frac{1}{16 \pi}\frac{\tilde{h}^2 m_{\Sigma}}{H x}\frac{K_1(x)}{K_2(x)} B Y_\Sigma \nonumber\\
&& +\frac{1}{16 \pi}\sum_I \tilde{B}_I \frac{\overline{h}_I^2 m_I}{H x}\frac{K_1(x)}{K_2(x)}Y_{I}+\frac{1}{16 \pi}\sum_I \tilde{B}_I \frac{Y^2_{\rm eff}\sin^2\theta m_I}{H x}\frac{K_1(x)}{K_2(x)}Y_{I},\nonumber\\
& \nonumber\\
\overline{h}_I^2 &=& \sum_{J} |h_{{\rm eff},IJ}|^2 \left(1-\frac{{\left(m_\Sigma+m_J\right)}^2}{m_I^2}\right),\non 
\tilde{B}_I &=& \frac{|h_{\rm eff,I4}|^2\left(1-\frac{{\left(m_\Sigma\right)}^2}{m_{I}^2}\right)}{\sum_{J}|h_{{\rm eff},IJ}|^2\left(1-\frac{{\left(m_\Sigma+m_J\right)}^2}{m_{I}^2}\right)}.
\eee
In the regime $\lambda_{\rm H\Sigma} > \overline{\lambda}_{\rm H\Sigma}$ the second row of the equation for $Y_\Sigma$ can be neglected and we can fix again $Y_{\Sigma}=Y_{\Sigma,\rm eq}$ and derive analytical solutions for the DM relic density through analogous steps as above. In the case $\lambda_{\rm H\Sigma} \ll \overline{\lambda}_{\rm H\Sigma}$ we have to replace the second row of the equation for $Y_\Sigma$ with $\frac{1}{16 \pi}\sum_I\frac{\overline{h}_I^2 m_I}{H x}\frac{K_1(x)}{K_2(x)}Y_{I},\,\,\,\,Y_I=Y_{I,\rm eq}$ and we can again fix $Y_\Sigma=0$ on the right-hand side. Analytical solutions are reliable if the timescales of production and decay of $\Sigma$ are well separated, otherwise one should refer to the numerical treatment.

\chapter{Leptogenesis equations and benchmark points}
\section{Analytical determination of the baryon asymmetry}\label{app_leptogenesis}

In this appendix we review the main details regarding the analytical and numerical determinations of the baryon asymmetry in Chapter~\ref{introduction}. The starting point is a system of coupled Boltzmann equations for the density matrices $\rho_{AB}$ with $A,B = \{N, \bar N, L , \bar L\}$ 
associated, respectively, to sterile neutrinos, active leptons and their anti-particles. This kind of system has been originally introduced in~\cite{Asaka:2005pn}, and has been reduced to a  system of ordinary differential equations.
A more refined version of this system, also adopted in this work, retaining the full dependence on the momentum $k$ of the density matrices, has been successively proposed in~\cite{Asaka:2011wq}. In this case one has to solve a system of integro-differential equations (cf. Eqs.~(12) and (14) of~\cite{Asaka:2011wq}) of the form:
\begin{align}
 \frac{d\rho_N}{dt} = & -i\left[H_N(k_N),\rho_N\right]
 -\frac{3}{2}\gamma_N^d(k_N) \left\{ F^\dagger\,F,\rho_N\right\}-\frac{1}{2}\gamma_N^d \left\{F^\dagger \left(A^{-1}-I\right)F,\rho_N\right\} \nonumber\\
& +3 \gamma_N^d(k_N) \rho_{\rm eq}(k_N) F^\dagger\,F+2 \gamma^d_N(k_N) \rho_{\rm eq}(k_N) \left[F^\dagger \left(A-I\right)F\right] \,,
\label{eq_N}
\end{align}
\begin{align}
 \frac{d\mu_{\alpha}}{dt}= & -\frac{3}{2}\gamma^d_\nu(T) FF^\dagger \tanh\mu_\alpha
 -\frac{\gamma_\nu^d (T)}{4}\left(1+\tanh\mu_\alpha\right) \int_0^\infty \frac{d k_N k_N}{T^2} F\left(\rho_{\bar N}^T -\rho_{\rm eq}\right)F^\dagger \nonumber\\
& +\frac{\gamma_\nu^d (T)}{4}\left(1-\tanh\mu_\alpha\right) \int_0^\infty \frac{d k_N k_N}{T^2} F^{*}\left(\rho_{N}^T -\rho_{\rm eq}\right)F^T \nonumber\\
& + \frac{\gamma_\nu^d (T)}{2 \cosh\mu_\alpha} \int_0^\infty \frac{dk_L}{T^2}\int_0^{k_L} dk_N \frac{\rho_{\rm eq}(k_L)}{\rho_{\rm eq}(k_N)}\left[F\rho_N F^\dagger-F^{*}\rho_{\bar N} F^T\right]\nonumber\\
&+\frac{\gamma_\nu^d (T)}{2 \cosh\mu_\alpha} \int_0^\infty \frac{dk_L}{T^2}\int_{k_L}^\infty dk_N \left[F\rho_N F^\dagger-F^{*}\rho_{\bar N} F^T\right]\nonumber\\
& -\frac{\gamma_\nu^d (T)}{2 \cosh\mu_\alpha} \int_0^\infty \frac{dk_L}{T^2}\rho_{\rm eq}(k_L)\int_0^\infty dk_N \left[F\rho_N F^\dagger-F^{*}\rho_{\bar N} F^T\right] \,.
\label{eq_mu}
\end{align}
Here we have taken the active leptons to be in thermal equilibrium, allowing us to trade the their equations for  equations of the chemical potentials $\mu_e,\mu_\mu,\mu_\tau$.
\begin{equation}\label{Eqrho}
\rho_{L}=N_D \, \rho_{\rm eq}(k) \, A,\,\,\,\,\,\,\rho_{\bar L}=N_D \, \rho_{\rm eq}(k) \, A^{-1},\,\,\,\,\,\rho_{\rm eq}=e^{-\frac{k}{T}} \,,
\end{equation}
with $A=\text{diag}(e^{\mu_e},e^{\mu_\mu},e^{\mu_\tau})$ representing a matrix of chemical potentials, $\rho_\text{eq}$  the equilibrium abundance of the mode with wavenumber $k$, determined by the temperature of the thermal bath $T$, and $N_D = 2$.

The first term on the rhs of Eq.~\eqref{eq_N} describes the oscillations of the heavy neutrinos in the presence of the effective Hamiltonian $H_N$, containing the free propagation and the effective potential induced by the medium effects. The following two terms describe the decay of the sterile states and the final two terms account for their production.  Both of these processes contain diagrams sensitive to the asymmetry in the active sector, leading to the terms proportional to $(A^{\pm 1} - I)$.  
The corresponding equation for the anti-particles $\rho_{\bar N}$  is straightforwardly obtained from Eq.~(\ref{eq_N}) by replacing $N \leftrightarrow \bar N$, $F \leftrightarrow F^{*}$ and $A \leftrightarrow A^{-1}$. For simplicity we will only show the equations for $\rho_N$ in the following.
Equation~\eqref{eq_mu} contains the decay and production of the active states, which in turn depend on the abundance and momentum of the sterile states. 
The functions $\gamma$ encoding the decay and production rates are defined as:
\begin{equation}
\gamma_N^d(k)=\frac{N_D N_C h_t^2}{64 \pi^3} \frac{T^2}{k},\,\,\,\,\,\,\gamma_\nu^d(k)=\frac{1}{N_D}\gamma_N^d(k) \,,
\end{equation}
with $N_C = 3$ and the top Yukawa coupling $h_t \simeq 1$.

 The abundances of the various species are given by
\begin{equation}
\label{eq:YN0}
Y_{N,L}=\frac{1}{s}\int \frac{d^3 k}{{\left(2 \pi\right)}^3} \, \rho_{N,L}(k) \,,
\end{equation}
where $s=\frac{2 \pi^2 g_{s}}{45}T^3$ denotes the entropy density of the thermal bath.
The system of integro-differential equations~(\ref{eq_N})-(\ref{eq_mu}) can be solved by specifying the masses of the heavy neutrinos $M_{1,2}$ and their Yukawa couplings $F_{\alpha i}$. The abundance of the heavy neutrino species and the asymmetry in the neutrino spectrum are given by:
\begin{align}
Y_{N,i}=\frac{1}{s}\int \frac{d^3 k}{{\left(2 \pi\right)}^3} \, {\left[\rho_{N}(k)\right]}_{ii} \,,\,\,\,\,i=1,2\nonumber\\
Y_{\Delta N,i}=\frac{1}{s}\int \frac{d^3 k}{{\left(2 \pi\right)}^3} \, {\left[\Delta\rho_{N}(k)\right]}_{ii}\,,\,\,\,\,\Delta \rho=\rho_N-\rho_{\bar N}
\end{align}
while the asymmetry in the leptonic flavour can be determined from the chemical potentials as:
\begin{equation}
Y_{\Delta L_\alpha}=\frac{45 N_D}{\pi^4 g_{s}} \sinh \mu_\alpha\,,\,\,\,\,\alpha=e,\mu,\tau
\end{equation}
According the conservation of the total (active plus sterile) lepton number, the baryon abundance $Y_B$ is given by:
\begin{equation} 
Y_B=-\frac{28}{79}\sum_{\alpha}Y_{\Delta L_\alpha}=\frac{28}{79} \sum_i Y_{\Delta N_i} \,.
\end{equation}
The properties of the system~(\ref{eq_N})-(\ref{eq_mu}) and of its solutions have been extensively studied in~\cite{Asaka:2011wq}. A useful simplification is to assume that the momentum distribution of the heavy neutrinos is proportional to the equilibrium one (this is equivalent to state the the heavy neutrinos are in kinetic equilibrium), i.e.:
\begin{equation}
\rho_{N,\bar N}=R_{N,\bar N}(t)  \, \rho_{\rm eq}(k)\ .
\label{eq:RN}
\end{equation}
With this substitution we can trace the evolution of the abundances of the heavy neutrinos through the only time dependent functions $R_{N,\bar N}$. The system of Boltzmann equations is then casted as:
\begin{align}
\label{eq:full_system}
 \frac{dR_N}{dt}= & -i\left[H(k_N),R_N\right] 
- \frac{3}{2}\gamma^d_N(k_N) \left\{F^\dagger F,R_N-I\right\}+2 \gamma_N^d(k_N) \left(F^\dagger\left(A-I\right)F\right) \nonumber\\
& -\frac{1}{2}\gamma^d_N(k_N)\left \{\left( F^\dagger\left(A^{-1}-I\right)F\right),R_N\right \}\nonumber\\
 \frac{d \mu_\alpha}{dt}= & -\frac{3}{2} \gamma_\nu^d(T){\left(FF^\dagger\right)}_{\alpha \alpha} \tanh\mu_\alpha 
 -\frac{\gamma_\nu^d(T)}{4}\left(1+\tanh\mu_\alpha\right)\left(F\left(R_{\bar N}^T-I\right)F^\dagger\right)_{\alpha \alpha} \\ \nonumber
&  +\frac{\gamma_\nu^d(T)}{4}\left(1-\tanh\mu_\alpha\right)(F^{*}\left(R_{N}^T-I\right)F^T)_{\alpha \alpha} 
+\frac{\gamma_\nu^d(T)}{2 \cosh\mu_\alpha}\left[F R_N F^\dagger-F^{*}R_{\bar N}F^T\right]_{\alpha \alpha}
\end{align}
As discussed in~\cite{Asaka:2011wq}, in very good approximation the system can be solved by reducing it to a system of ordinary equations for a single mode $k_*$, equivalent to the one presented~\cite{Asaka:2005pn}, by a suitable replacement of the type $k_* \sim T$. 
Notice that the choice of $k_{*}$ must maintain the system self-consistent, i.e.\ it should preserve lepton number. This condition can be stated as:
\begin{equation}
\Tr{\frac{dR_N}{dt}|_{k=k_{*}}-\frac{dR_{\bar N}}{dt}|_{k=k_{*}}+N_D \frac{d A}{dt}-N_D\frac{d A^{-1}}{dt}}=0
\end{equation}
and can be satisfied only for $k_{*}=2 \, T$, rather than for $k_{*}=3 \, T$, corresponding to the conventional thermal average.\footnote{In the notation used in this appendix this implies $\gamma(t) = \gamma(T) = \gamma(k_*/2) = 2 \, \gamma(k_*) $. }
The system~(\ref{eq:full_system}), with the substitution $k \rightarrow k_{*}=2 T$ is the one used in our study. Notice that, although very similar, the system~(\ref{eq:full_system}) does not exactly coincide with the one presented in~\cite{Asaka:2005pn}. In particular the coefficient of the third term of the right-hand side of the equation for $R_N$ differs by a factor 2/3. This is an important point since this term represent the connection term between the active and sterile sector which is mostly responsible of the generation of the lepton asymmetry.

As stated in the main text, despite the simplification discussed, an extensive numerical analysis is still very difficult. For this reason we have limited the numerical study to some relevant benchmarks, as reported e.g.\ in Figures~\ref{fig:bench_natural}, \ref{fig:bench_numsm} and \ref{fig:bench_thermal}  and have adopted, for the study of the parameter space, an analytical solution which is valid in the so-called weak wash-out regime. This analytical solution is derived following the procedure proposed in~\cite{Asaka:2005pn,Asaka:2010kk}. The final expression differs, however, by a $O(1)$ factor with respect to these references due to the different starting system, as mentioned above.

\subsection{Analytical solution in the weak washout regime}

An analytical expression of $Y_B$ can be obtained by solving Eq.~(\ref{eq:full_system}) perturbatively for small values of $\mu_\alpha$ and $F$.
Let us first consider the leading order in $\mu_\alpha$, i.e.\ we set tanh$\mu_\alpha \rightarrow 0$, cosh$\mu_\alpha \rightarrow 1$, $A - I \rightarrow 0$ and $A^{-1} - I \rightarrow 0$. The initial conditions are $R_{N,\bar N}(0)=0$, $\mu_{\alpha}=0$. 
The first step is to solve the equations for $R_{N,\bar N}$. First of all, one can perform the following transformation~\cite{Shaposhnikov:2008pf}:
\begin{equation}
\label{eq:transformation}
R_N=E(t) \tilde{R}_N E^{\dagger}(t) \,,
\end{equation}
with
\begin{equation}
E(t)=\exp\left[-i \int_0^t dt^{'} \Delta E \right],\quad \Delta E=\text{diag}(E_1,E_2)\ ,
\end{equation}
where $E_i$ denotes the energies of the two heavy neutrinos. This transformation encodes the oscillations processes in the sterile neutrino production term.  In this way we obtain
\begin{align}
\label{eq:diff2}
& \frac{d\tilde{R}_N}{dt}=-i\left[\tilde{H},\tilde{R}_N\right]-\frac{3}{2}\left\{\tilde{\Gamma}^d_N,\tilde{R}_N-I\right\} \,,
\end{align}
where we have dropped the terms proportional to $(A - I)$ and $(A^{-1} - I)$ and have defined
\begin{equation}
\tilde{\Gamma}{_N^d}(t)=E^\dagger(t)  \Gamma_N^d (t)  E(t)\,, \quad \Gamma_N^d=\frac{1}{2}\gamma_N^d(T) F^\dagger F \,.
\end{equation}
Physically, this corresponds to ignoring the back reaction of the asymmetry in the active sector on the production of the sterile neutrinos. For the size of asymmetries in the active sector which we are phenomenologically interested in, this is a very good approximation. The asymmetry in the active sector will however become important in the next order of our perturbative expansion, which we will need when determining the asymmetry in the sterile sector, as we will see below.

In the weak washout regime characterised by $R_N \ll 1$, we can solve Eq.~\eqref{eq:diff2} by dropping all the terms proportional to $R_N$,
\begin{equation}
\label{eq:first_step}
{\tilde{\textcolor{black}{R}}}_N =3\int_0^t dt_1 E^\dagger(t_1)  \Gamma_N^d (t_1)  E(t_1)\ .
\end{equation}
Let us now move to the equation for the chemical potential. At leading order in $\mu_\alpha$ and after inserting Eq.~\eqref{eq:first_step}, we find
\begin{align}
& \mu_\alpha= \frac{3}{4}\int_0^t dt_1 \gamma_\nu^d (t_1)  \int_0^{t_1} dt_2 \gamma_N^d(t_2) \left(\left[F E(t_1) E(t_2)^\dagger F^\dagger F  E(t_2)E(t_1)^\dagger  F^\dagger\right] \right.\nonumber\\
& \left. -  \left[F^{*}  E(t_1) E(t_2)^\dagger  F^T F^{*} E(t_2)E(t_1)^\dagger  F^T\right] \right)_{\alpha \alpha}\nonumber\\
& - \frac{3}{8}\int_0^t dt_1 \gamma_\nu^d (t_1)  \int_0^{t_1} dt_2 \gamma_N^d(t_2) \left(\left[F {\left[E(t_1) E(t_2)^\dagger  F^T F^{*} E(t_2)E(t_1)^\dagger\right]}^T  F^\dagger\right] \right.\nonumber\\
& \left. -  \left[F^{*}  {\left[E(t_1) E(t_2)^\dagger  F^\dagger F E(t_2)E(t_1)^\dagger\right]}^T  F^T\right] \right)_{\alpha \alpha} \, .
\end{align} 
After some manipulation, exploiting in particular
\begin{align}
&  E(t_1) E^\dagger (t_2)_{ij} = \text{diag}\left[\exp(i \int_{t_1}^{t_2} E_i)\right]  \,, \\
 & F_{\alpha i} (F^\dagger F)_{ij} (F^\dagger)_{j \alpha} - F^*_{\alpha i} (F^T F^*)_{ij} (F^T)_{j \alpha} = 2 \, \text{Im} ( F_{\alpha i} (F^\dagger F)_{ij} (F^\dagger)_{j \alpha})\,,
\end{align}
this expression can be simplified to
\begin{equation}
\label{eq:mualpha_asaka}
\mu_\alpha= {\frac{9}{2}} \, \delta_{\alpha}\int_0^t dt_1 \gamma_\nu^d (t_1) \int_0^{t_1} dt_2  \gamma_N^d (t_2) \sin\left(\int_{t_2}^{t_1} dt_3 E_2(t_3)-E_3(t_3)\right) \,,
\end{equation} 
with
\begin{equation}
\delta_{\alpha} \equiv \sum_{i >j} \text{Im}\left[F_{\alpha i} \left(F^{\dagger} F\right)_{ij} F^{\dagger}_{j\alpha}\right] \,.
\end{equation}
This result denotes the leading order asymmetry in the individual
flavours of the active sector induced by the sterile neutrino
oscillations. This asymmetry in turn generates an effective potential
for the sterile neutrino states, inducing an asymmetry in the sterile
sector, as we will discuss below. The backreaction of this asymmetry
in the sterile flavours will finally generate a net asymmetry (at next
order in the perturbative expansion) in the active sector.

Introducing
\begin{align}
& \int_{t_2}^{t_1} dt_3 (E_1(t_3)-E_2(t_3))=z(T_1)-z(T_2) \,, \nonumber\\
& z(T)=\int_0^t \frac{\Delta M^2_{12}}{2 T}=-\int_{T_0}^T \frac{M_0}{T^3} \frac{\Delta M^2_{12}}{4 T}=\frac{M_0 \Delta M^2_{12}}{12 T^3} \,,
\end{align}
the remaining integral can be computed by  changing the variables to $x_i \equiv \frac{T_L}{T_i}$ (with $dt_i=\frac{M_0}{T_L^2} {x_i} dx_i$), where
\begin{equation}\label{eq:TL}
T_L \equiv {\left(\frac{1}{12}M_0 \Delta M^2_{12}\right)}^{1/3}
\end{equation}
will turn out to be the characteristic temperature of the leptogenesis process.
From
\begin{align}
\gamma_N^d(t_i)=\frac{N_D N_C h_t^2}{64 \pi^3}T_i\,, \quad 
\gamma_\nu^d(t_i)=\frac{N_C h_t^2}{64 \pi^3}T_i \,, \quad \frac{N_C h_t^2}{64 \pi^3}=\frac{\sin\phi}{8} \,,
\end{align}
where $\sin\phi \simeq 0.012$ is defined in~\cite{Akhmedov:1998qx}\footnote{{To give a physical intuition, $\sin \phi$ roughly corresponds to the ratio of decay rate over effective potential for the sterile states, or correspondingly to the ratio of the imaginary over the real part of the one-loop diagram $N L \rightarrow N L$.}}, we find
\begin{align}
\label{eq:secondstep}
& \mu_\alpha={\frac{9}{64}} \sin^2\phi \frac{M_0^2}{T_L^2}\delta_\alpha J_{32}\left(\frac{T_L}{T}\right)\,, \\ 
& J_{32}(x)=\int_0^x dx_1 \int_0^{x_1}dx_2  \, \sin\left(x_1^3-x_2^3\right)\ .
\end{align}
The function $J_{32}$ has a very interesting behaviour. 
At early times, i.e. $x \ll 1$, $J_{32}(x)=\frac{3}{20}x^5$, while, after a sharp transition at $x \simeq 1$, it becomes constant. The asymptotic value for $x \gtrsim 1$ is given by:
\begin{equation}
J_{32}(x)=\frac{2^{1/3} \, \pi^{3/2}}{9 \, \Gamma(5/6)} \, .
\end{equation}
Given this behaviour, it is safe to assume that the lepton asymmetry, encoded in the chemical potential $\mu_{\alpha}$, is mostly generated at the temperature $T_L$.

The last step is to compute the asymmetry in the sterile sector. At leading order we have to compute:
\begin{align}
\frac{d \left(\Delta R\right)_{ii}}{dt}& = \gamma_N^d(t) \left[F^\dagger A F-F^T A^{-1} F^{*}\right]_{ii}
& =2 \gamma_N^d(t) \left[F^\dagger \sinh \mu_{\alpha} F\right]_{ii} \approx 2 \,\gamma_N^d(t) \left[F^\dagger \mu_\alpha F\right]_{ii}
\end{align}
Performing a direct integration this yields
\begin{equation}
\label{eq:third_step}
\left(\Delta R\right)_{ii} (T)=\frac{3\, 2^{2/3} \pi^{3/2}}{64 \,3^{1/3} \Gamma(5/6)}\sin^3 \phi\frac{M_0}{T} \frac{M_0^{4/3}}{\Delta M_{12}^{4/3}} \left(F^\dagger \delta_\alpha F\right)_{ii}
\end{equation} 
where we have profited from the asymptotic behaviour of the function $J_{32}$ to analytically solve the integral,  since the asymmetry in the sterile sector is generated mainly at $T < T_L$.
The asymmetry stored in the sterile sector is obtained as the trace of Eq.~(\ref{eq:third_step}). Since the total lepton number is conserved (recall that in the parameter space of interest the Majorana mass terms are much smaller than the temperature of the thermal bath), the same asymmetry but with an opposite sign is contained in the active flavours. SM sphaleron processes couple only to the active flavours, converting the asymmetry stored there into a baryon asymmetry,
\begin{equation}
 Y_{\Delta B} =  - \frac{28}{79} Y_{\Delta \alpha} = \frac{28}{79} Y_{\Delta N} = \frac{28}{79} Y_{N0} (\Delta R_{11}(T_{\rm W}) + \Delta R_{22}(T_{\rm W}))\,,
\end{equation}
with $Y_{N0} \simeq 0.0022$ denoting the equilibrium abundance, cf.\ Eq.~\eqref{eq:YN0}. Evaluating Eq.~\eqref{eq:third_step} at $T = T_{\rm W}$ demonstrates the strong enhancement $M_0/T_{\rm W}$ of the asymmetry, due to the separation of time-scales $ T_{\text W} < T_L \ll M_0$. 
We remark that at each step of the solution increasing powers of $\sin\phi$ and of the Yukawas are present, rendering the analytical procedure reliable.

\section{Numerical benchmark points}\label{app_benchmarks}

\subsection{Benchmarks in the weak wash-out regime}

\begin{itemize}
\item First benchmark (``perturbative'' model, Fig.~\ref{fig:bench_natural}):
\begin{align}
& M=1.5\, \mbox{GeV},\,\,\,\,\,\,\Delta m= 133\, \mbox{eV}\nonumber\\
& Y^{\rm eff}=\left(
\begin{array}{cc}
-3.35 \times 10^{-8} -i\, 1.27 \times 10^{-8} & -1.38 \times 10^{-8} + i\, 3.20 \times 10^{-8} \\
-2.89 \times 10^{-8} + i\, 5.89 \times 10^{-8} & 6.74 \times 10^{-8} + i\, 2.57 \times 10^{-8} \\
 2.30 \times 10^{-8} + i\, 6.99 \times 10^{-8} & 7.87 \times 10^{-8} - i\, 2.04 \times 10^{-8} 
\end{array}
\right)
\end{align}

\item Second benchmark (``generic'' model, Fig.~\ref{fig:bench_numsm}):
\begin{align}
& M=15\, \mbox{GeV},\,\,\,\,\,\,\Delta m= 163\, \mbox{eV}\nonumber\\
& Y^{\rm eff}=\left(
\begin{array}{cc}
4.91 \times 10^{-9} - i\, 3.67 \times 10^{-8} &  1.59 \times 10^{-8} - i\, 1.99 \times 10^-8 \\
6.23 \times 10^{-9} -  i\, 5.74 \times 10^{-8} & 1.13 \times 10^{-7} + i\, 1.09 \times 10^{-9} \\
-1.28 \times 10^{-8} + i\, 1.63 \times 10^{-8} & 1.10 \times 10^{-7} - i\, 2.24 \times 10^{-9}
\end{array}
\right)
\end{align}

\item Third benchmark (perturbative regime with large $Y^\text{eff}$, Fig.~\ref{fig:bench_thermal}):
\begin{align}
& M=3 \, \mbox{GeV},\,\,\,\,\,\,\Delta m= 9\, \mbox{keV}\nonumber\\
& Y^{\rm eff}=\left(
\begin{array}{cc}
-1.27 \times 10^{-8} -i\, 1.96 \times 10^{-8} & 1.87 \times 10^{-8} - i\, 6.78 \times 10^{-9}\\
-3.92 \times 10^{-8} + i\, 8.04 \times 10^{-8} & -9.30 \times 10^{-8} - i\, 3.50 \times 10^{-8}\\
 3.12 \times 10^{-8} + i\, 1.20 \times 10^{-7} & -1.31 \times 10^{-7} + i\, 2.78 \times 10^{-8}
\end{array}
\right)
\end{align}
\end{itemize}

\subsection{Benchmarks in the strong wash-out regime}

\begin{itemize}
\item First benchmark (Fig.~\ref{fig:benchhy1}):
\begin{align}
& M=5.5 \, \mbox{GeV},\,\,\,\,\,\,\Delta m= 5.5\, \mbox{keV}\nonumber\\
& Y^{\rm eff}=\left(
\begin{array}{cc}
 5.78 \times 10^{-8} + i\, 1.39 \times 10^{-7} & -1.37 \times 10^{-7} + i\, 5.92 \times 10^{-8}\\
 -1.79 \times 10^{-8} - i\, 1.90 \times 10^{-7} & 1.98 \times 10^{-7} - i\, 1.76 \times 10^{-8}\\
 -4.54 \times 10^{-9} - i\, 4.58 \times 10^{-7} & 4.63 \times 10^{-7} - i\, 4.45 \times 10^{-9}
\end{array}
\right)
\end{align}

\item Second benchmark (Fig.~\ref{fig:benchhy2}):
\begin{align}
& M=1.5 \, \mbox{GeV},\,\,\,\,\,\,\Delta m= 8.5\, \mbox{keV}\nonumber\\
& Y^{\rm eff}=\left(
\begin{array}{cc}
-1.65 \times 10^{-7} -i\, 1.26 \times 10^{-7} & 1.26 \times 10^{-7} - i\, 1.65 \times 10^{-7} \\
-8.65 \times 10^{-8} + i\, 2.59 \times 10^{-7} & -2.61 \times 10^{-7} -i\, 8.62 \times 10^{-8} \\
 9.37 \times 10^{-8} + i\, 5.16 \times 10^{-7} & -5.18 \times 10^{-7} + i\, 9.34 \times 10^{-8}
\end{array}
\right)
\end{align}

\item Third benchmark (Fig.~\ref{fig:benchhy3}):
\begin{align}
& M=4.5 \, \mbox{GeV},\,\,\,\,\,\,\Delta m= 2\, \mbox{keV}\nonumber\\
& Y^{\rm eff}=\left(
\begin{array}{cc}
 7.61\times 10^{-7} +i\, 6.89 \times 10^{-7} & -6.87 \times 10^{-7} + i\, 7.62 \times 10^{-7}\\
 4.00 \times 10^{-7} +i\, 2.79 \times 10^{-6} & -2.79 \times 10^{-6} +i\, 4.00 \times 10^{-7}\\
 -2.39 \times 10^{-7} +i\, 1.60 \times 10^{-6} & -1.60 \times 10^{-6} -i\, 2.39 \times 10^{-7}
\end{array}
\right)
\end{align}

\item First ``flavoured'' benchmark (left panels of Fig.~\ref{fig:bench_flavored}):
\begin{align}\label{eq:benchSW1}
& M=1 \, \mbox{GeV},\,\,\,\,\,\,\Delta m= 8.5\, \mbox{keV}\nonumber\\
& Y^{\rm eff}=\left(
\begin{array}{cc}
 -1.51 \times 10^{-7} - i\, 1.30 \times 10^{-7} & -1.30 \times 10^{-7} + i\, 1.51 \times 10^{-7}\\
 -6.69 \times 10^{-8} - 7.07 \times 10^{-7} & -7.07 \times 10^{-7} + i\, 6.68 \times 10^{-8}\\
 2.57 \times 10^{-8} - 3.92 \times 10^{-7} & -3.93 \times 10^{-7} - i\,2.57 \times 10^{-8}
\end{array}
\right)
\end{align}

\item Second ``flavoured'' benchmark (right panels of Fig.~\ref{fig:bench_flavored}):
\begin{align}\label{eq:benchSW2}
& M=1.4 \, \mbox{GeV},\,\,\,\,\,\,\Delta m= 1.6\, \mbox{keV}\nonumber\\
& Y^{\rm eff}=\left(
\begin{array}{cc}
1.20 \times 10^{-7} +i\, 1.08 \times 10^{-7} & 1.08 \times 10^{-8} - i\, 1.20 \times 10^{-8} \\
 1.28 \times 10^{-8} - i\, 3.60 \times 10^{-7} & -3.61 \times 10^{-7} - i\, 1.28 \times 10^{-8}\\
-4.41 \times 10^{-8} - i\, 8.29 \times 10^{-7} &  -8.30 \times 10^{-6} +i\, 4.40 \times 10^{-8}
\end{array}
\right)
\end{align}

\end{itemize}

\chapter{LFV operators and constraints}
\section{Wilson Coefficients}\label{sec:app_quarkA} 

In this Appendix we present detailed expressions for the Wilson coefficients discussed in Chapter~\ref{sec-0}.  
All computations have been made in the Feynman gauge. 
Contributions coming from the penguin and self-energy diagrams are shown in Fig.~\ref{fig:AA}, whereas the box diagrams are shown in Fig.~\ref{fig:B}.

\begin{figure}[htb]
\begin{center}
\begin{tabular}{cc}
\includegraphics[width=0.45\textwidth]{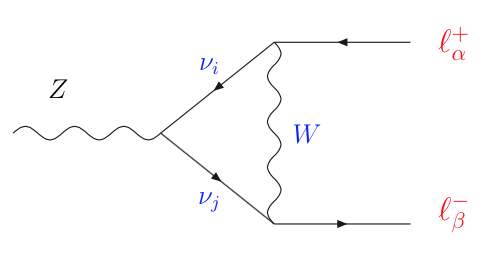}&\includegraphics[width=0.45\textwidth]{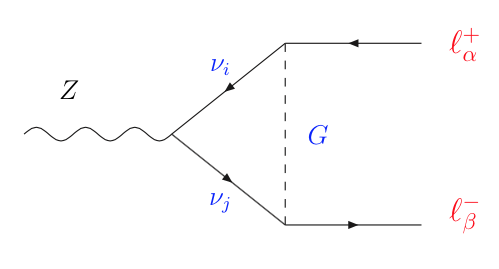} \cr
\includegraphics[width=0.45\textwidth]{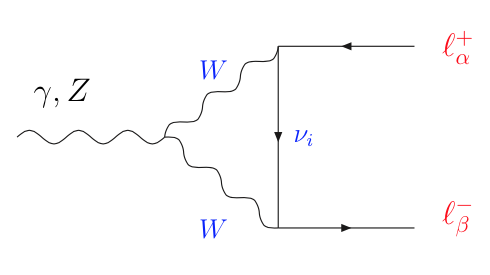}&\includegraphics[width=0.45\textwidth]{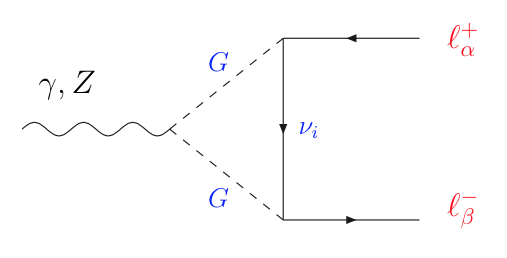} \cr
\includegraphics[width=0.45\textwidth]{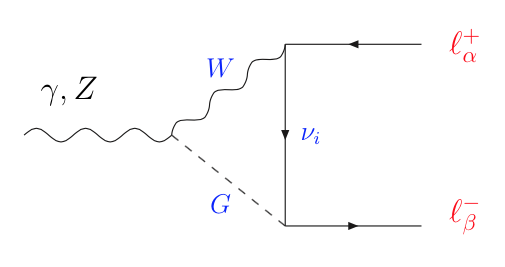}&\includegraphics[width=0.45\textwidth]{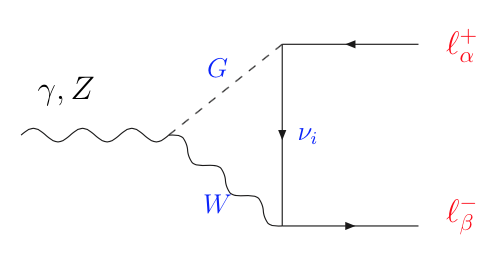} \cr
\includegraphics[width=0.45\textwidth]{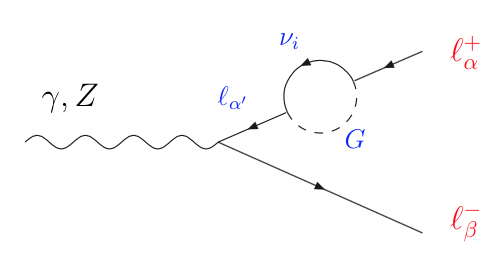}&\includegraphics[width=0.45\textwidth]{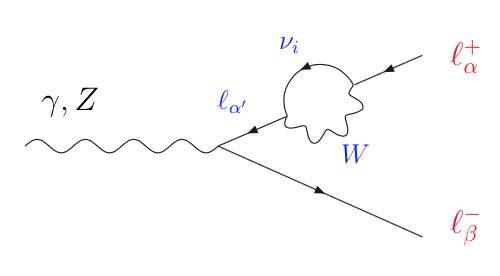} \cr
\includegraphics[width=0.45\textwidth]{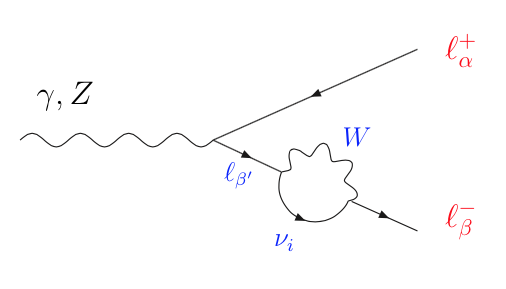}&\includegraphics[width=0.45\textwidth]{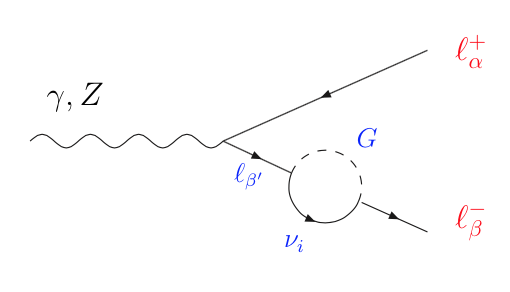} \cr
  \end{tabular}
  \caption{{\footnotesize 
Penguin and self-energy diagrams contributing the LFV decay in Feynman gauge. 
 }}
\label{fig:AA}
\end{center}
\end{figure}

\begin{figure}[htb]
\begin{center}
\begin{tabular}{cc}
\includegraphics[width=0.45\textwidth]{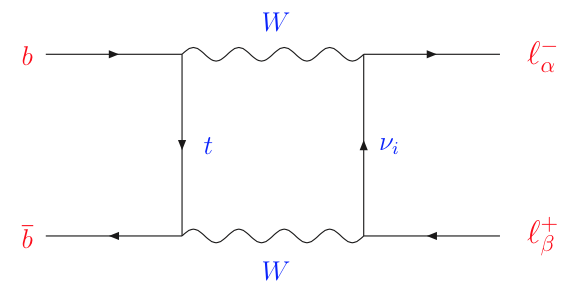}&\includegraphics[width=0.45\textwidth]{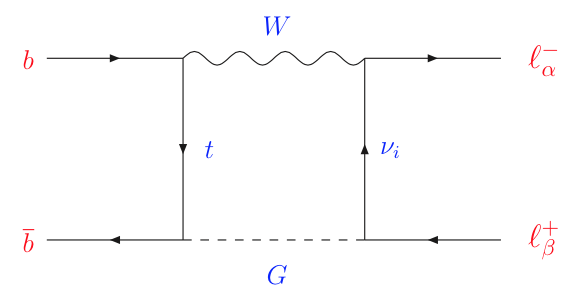} \cr
\includegraphics[width=0.45\textwidth]{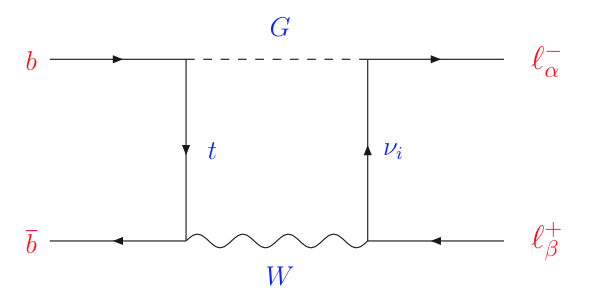}&\includegraphics[width=0.45\textwidth]{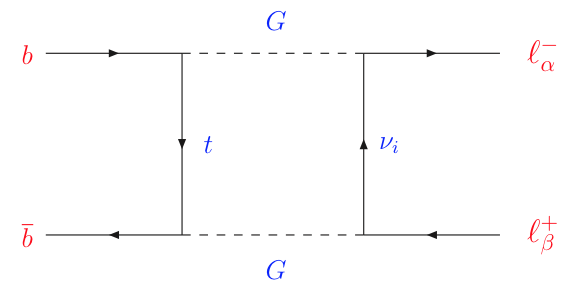} \cr
  \end{tabular}
  \caption{{\footnotesize 
Box diagrams contributing the LFV decay $\Upsilon^{(n)}\to \ell_\alpha\ell_\beta$ in Feynman gauge. 
 }}
\label{fig:B}
\end{center}
\end{figure}

We use the standard notation, $x_i=m_i^2/m_W^2$, $x_t=m_t^2/m_W^2$, $x_q=q^2/m_W^2= m_V^2/m_W^2$, and write 
\begin{equation}
C_{VL}^r=\displaystyle\sum_{i,j=1}^{n_\nu} U_{\beta i} U_{\alpha j}^* C^{r, ij}_{VL} (x_i,x_j),
\end{equation}
where $r\in\lbrace\gamma,Z,\text{box}\rbrace$. The coefficients $C^{r, ij}_{VL}$ related to $\gamma$ and the box contributions are diagonal, $C^{r, ij}_{VL}= \delta_{ij}C^{r, i}_{VL}$, while 
those related to the $Z$ penguins can also involve a coupling to two different neutrinos, since the $3\times 3$ mixing matrix is no longer unitary. 
We therefore separate the diagonal and nondiagonal parts of the corresponding coefficient $C^{Z, ij}_{VL}=\delta_{ij} C^{Z,i}+\widehat{C}^{Z,ij}$, where the second term depends on the parameter $C_{ij}$ defined by
\begin{align}
C_{ij} = \sum_{\alpha={e,\mu,\tau}} U_{\alpha i}^* U_{\alpha j},
\end{align}
which, in the presence of sterile neutrinos, is generally different from $\delta_{ij}$.
Furthermore, from the plots presented in the body of the present paper we see that the region of $m_{4,5}\gg m_V$ is particularly interesting because there occurs the enhancement 
of the LFV decay rate. For the sake of clarity we thus expand our expressions in $x_q$ and present here only the dominant terms. We also neglected, in the denominators of the loop integrals, the external momenta since they are 
negligible with respect to heavy neutrino masses.
Therefore, up to terms ${\cal O}(x_q^2)$, our results read:
\begin{align}
\label{cvgamma}
	C_{VL}^{\gamma,i} (x_i) = -\frac{1}{16\pi^2} + x_q\frac{-43 x_i^3 + 108 x_i^2 +6 (5 x_i-6)x_i^2 \log x_i-81 x_i+16}{288 \pi^2 (x_i-1)^4} ,
\end{align}

\begin{align}
	C_{VL}^{Z,i}(x_i) =  &  \frac{-1+12 x_i-11 x_i^2+10 x_i^2 \log x_i }{64 \pi^2 (x_i-1)^2} +   \cos^2 \theta_W   C_{VL}^{\gamma,i} (x_i)  ,
\end{align}

\begin{align}
	\widehat{C}_{VL}^{Z,ii}&(x_i,x_i) =C_{ii} \frac{(x_i-2) \left(3 \left(x_i^2-1\right)+2
   (x_i-4) x_i \log x_i\right)}{128 \pi^2 (x_i-1)^2}\\
	&-x_q C_{ii}\frac{(x_i-1) (x_i (x_i (2 x_i-47)+25)+14)+6 (x_i (12 x_i-13)+2) \log x_i}{1152 \pi^2 (x_i-1)^4},\nonumber
\end{align}

\begin{align}
	\widehat{C}_{VL}^{Z,ij}(x_i,x_j)&= \frac{\sqrt{x_i x_j} C_{ij}^*}{64 \pi^2} \Big{[}\frac{x_i (x_i-4)}{(x_i-1)(x_i-x_j)}\log x_i+\frac{x_j (x_j-4)}{(x_j-1)(x_j-x_i)}\log x_j-\frac{3}{2}\Big{]} \\
	&+\frac{C_{ij}}{64 \pi^2}\Big{[} \frac{2 x_i^2 (x_j-1)}{(x_i-1)(x_i-x_j)}\log x_i+\frac{2 x_j^2 (x_i-1)}{(x_j-1)(x_j-x_i)}\log x_j+3\Big{]}\nonumber \\
	&+\frac{x_q}{192 \pi^2} \Big{\lbrace} \sqrt{x_i x_j} C_{ij}^*\Big{[} \frac{x_i^2(x_i-3 x_j+2 x_i x_j)}{(x_i-1)^2(x_i-x_j)^3}\log x_i+\frac{x_j^2(x_j-3 x_i+2 x_i x_j)}{(x_j-1)^2(x_j-x_i)^3}\log x_j\nonumber\\
	&-\frac{x_i^3(x_j-1)-x_j^3(x_i-1)+x_i x_j (x_i-x_j)}{(x_i-1)(x_j-1)(x_i-x_j)^3}\Big{]}+ 2 C_{ij} \Big{[} \frac{x_i^2(3 x_i^2 +3x_j^2+3x_j-x_i-8x_i x_j)}{(x_i-1)^2(x_i-x_j)^3}\log x_i \nonumber \\
	&+\frac{x_j^2(3 x_j^2 +3x_i^2+3x_i-x_j-8x_i x_j)}{(x_j-1)^2(x_j-x_i)^3}\log x_j - \frac{8 x_i^2 x_j-8 x_i x_j^2-x_i^3 x_j+x_i x_j^3-2x_i^3+2x_j^3}{(x_i-1)(x_j-1)(x_i-x_j)^3}\Big{]}\Big{\rbrace},\nonumber
\end{align}

\begin{align}
\label{cvbox}
	C_{VL}^{\mathrm{Box},i} = \frac{1}{256 \pi^2} 
	 \left\lbrace \frac{[ x_i(x_t-8)+4]x_t^2 \log x_t}{(x_t-1)^2(x_i-x_t)}+\frac{[ x_t(x_i-8)+4]x_i^2 \log x_i}{(x_i-1)^2 (x_t-x_i)}+\frac{7 x_i x_t-4}{(x_i-1)(x_t-1)}\right\rbrace.
\end{align}

\section{Formulas and hadronic quantities}\label{sec:app_quarkB} 

In this Appendix we collect the expressions used to constrain the parameters of the models discussed in Chapter~\ref{sec-0}, as well as the values of the masses and decay constants used in our numerical analysis.
In the expressions below we used the value of $G_F=G_\mu = 1.166 \times 10^{-5} \ \text{GeV}^{-2}$, as extracted from $\mu \to e\nu_\mu\bar\nu_e$. 
In our scenarios, in which we extended the neutrino sector by adding heavy sterile neutrinos, the Fermi constant becomes $G_F=G_\mu/\sqrt{ \sum_{i,j} |U_{ei}|^2  |U_{\mu j}|^2}$. For the models used in this paper, we checked to see that 
$G_F = G_\mu$ remains an excellent approximation.  
 
\begin{itemize}
\item \underline{$\mu \to e\gamma$}: We use the experimentally established upper bound ${\rm B}(\mu\to e\gamma) < 5.7\times 10^{-13}$, and the expression~\cite{Ilakovac:1994kj}
\begin{align}
{\rm B}(\mu \to e\gamma) &= {\sqrt{2} G_F^3 s_W^2 m_W^2 \over 128 \pi^5 \Gamma_\mu} m_\mu^5 \vert U_{\mu 4}^\ast U_{e4} G_\gamma (x_4)\vert^2\,, \cr
G_\gamma (x) & = - { 2 x^3+5x^2 -x\over 4(1-x)^3 }
- {  3 x^3  \over 2(1-x)^4 } \log x\ ,
\end{align} 
to get one of the most significant constraints in this study. Notice that we use $s_W^2=1-m_W^2/m_Z^2$, and we kept the dominant contribution with $x_4$. 
\item \underline{$W \to \ell_\alpha \nu$}: Combining the measured ${\rm B}(W\to e\nu )=0.1071(16)$ and ${\rm B}(W\to \mu\nu)=0.1063(15)$, with the expression
\begin{align}
{\rm B}(W \to \ell_\alpha \nu) = {\sqrt{2} G_F m_W \over 24 \pi  \Gamma_W} \sum_{j=1}^4 \lambda(m_\alpha^2,m_j^2,m_W^2) \left( 
2 - { m_\alpha^2+m_j^2\over m_W^2}- { (m_\alpha^2-m_j^2)^2\over m_W^4} 
\right) \vert U_{\alpha j}^2\vert  ,
\end{align} 
we further restrain the possible values of $m_4$ while varying the mixing angles in the largest possible range.  
\item  \underline{$\Delta r_{K,\pi}= R_{K,\pi}^{\rm exp.}/R_{K,\pi}^{\rm SM} - 1$}: The ratio of the leptonic decay widths of a given meson $P$, $R_P=\Gamma(P\to e \nu_e )/\Gamma(P\to \mu \nu_\mu )$ was  recently shown to be quite restrictive on 
the possible values of $m_{4,5}$ and $\eta$~\cite{Abada:2013aba}.  The most significant constraints actually come from $\Delta r_{\pi}= 0.004(4)$ and  $\Delta r_{K}= -0.004(3)$, and the corresponding formula reads,
\begin{align}
\Delta r_P= - 1 + {m_\mu^2 (m_P^2-m_\mu^2)^2\over m_e^2 (m_P^2-m_e^2)^2} {\displaystyle{\sum_i} \vert U_{ei}\vert^2 \left[ m_P^2 (m_{\nu_i}^2 + m_e^2) -  (m_{\nu_i}^2 - m_e^2)^2 \right] \lambda^{1/2}(m_P^2,m_{\nu_i}^2 ,m_e^2) 
\over \displaystyle{\sum_i} \vert U_{\mu i}\vert^2 \left[ m_P^2 (m_{\nu_i}^2 + m_\mu^2) -  (m_{\nu_i}^2 - m_\mu^2)^2 \right] \lambda^{1/2}(m_P^2,m_{\nu_i}^2 ,m_\mu^2)
}.
\end{align} 
\item \underline{$Z \to \nu \nu$}: To saturate the experimental ${\Gamma}(Z\to \text{invisible} )=0.499(15)$~GeV, we sum over the kinematically available channels involving active and sterile neutrinos,
\begin{align}
\Gamma(Z \to \nu \nu) =& \sum_{i,j} \left(1-\frac{\delta_{ij}}{2}\right)  {G_F \over 12\sqrt{2} \pi m_Z}\lambda^{1/2}(m_Z^2,m_i^2,m_j^2) |C_{ij}|^2 \nn\\
                                              &\times  \left[  2 m_Z^2-m_i^2-m_j^2 - 6m_im_j - {(m_i^2-m_j^2)^2\over m_Z^2}\right]\,.
\end{align} 
\item \underline{$\mu \to e e e$}: We use the experimental upper bound ${\rm B}(\mu \to e e e) <1 \times 10^{-12}$~\cite{Bellgardt:1987du}, and the expression~\cite{Ilakovac:1994kj}
\bee
{\rm B}(\mu \to eee) &=& \frac{G_F^4 m_W^4 }{6144 \pi^7}\frac{m^5_\mu}{\Gamma_\mu}\non 
&&  \left\{ 2 \left|\frac{1}{2}F^{\mu eee}_{\rm Box}+F^{\mu e}_Z-2\sin^2\theta_W (F^{\mu e}_Z-F^{\mu e}_\gamma)\right|^2+4 \sin^4\theta_W \left|F^{\mu e}_Z-F^{\mu e}_\gamma\right|^2 \right. \non 
&& \left.		+ 16 \sin^2\theta_W \mathrm{Re} \left[	(F^{\mu e}_Z +\frac{1}{2}F^{\mu eee}_{\rm Box})	{G^{\mu e}_\gamma}^* 			\right]		- 48 \sin^4\theta_W \mathrm{Re}\left[	(F^{\mu e}_Z-F^{\mu e}_\gamma)	{G^{\mu e}_\gamma}^* 			\right] \right. \non 
&& \left.
		+32 \sin^4\theta_W |G^{\mu e}_\gamma|^2\left[		\ln \frac{m^2_\mu}{m^2_{e}} -\frac{11}{4}		\right]		\right\},
\eee
with the loop functions $F^{\mu eee}_{\rm Box},F^{\mu e}_Z,F^{\mu e}_\gamma,G^{\mu e}_\gamma$ defined in~\cite{Alonso:2012ji}.

\end{itemize}

Finally, the values of hadronic quantities not discussed in the body of the paper but used in our numerical analysis are listed in Table~\ref{tab:costs}.~\footnote{
Notice that the ratio of decay constants $f_{\psi(2S)}/f_{J/\psi}$ has been obtained from the corresponding (measured) electronic widths and the expression $\Gamma (\psi_n\to e^+e^-)= 16 \pi \alpha_{\rm em}^2 f_{\psi_n}^2/(27 m_{\psi_n}^2) $.
 }

\begin{table}[htb]
\renewcommand{\arraystretch}{1.5}
\centering{}
\hspace*{-4mm}\begin{tabular}{|ccc|ccc|}
\hline 
Quantity & Value & Ref. & Quantity & Value & Ref.  \\ \hline 
$m_\phi$ & $1.0195$~GeV  & \cite{Agashe:2014kda} & $f_\phi$ & $241(18)$~MeV &  \cite{Donald:2013pea} \\ 
$m_{J/\psi}$ & $3.0969$~GeV  & \cite{Agashe:2014kda} & $f_{J/\psi}$ & $418(9)$~MeV &  \cite{Becirevic:2013bsa,Becirevic:2012dc} \\ 
$m_{\psi(2S)}$ & $3.6861$~GeV  & \cite{Agashe:2014kda} & $f_{\psi(2S)}/f_{J/\psi}$ & $0.713(16)$ &  \cite{Agashe:2014kda} \\ \hline
$m_{\Upsilon}$ & $9.460$~GeV  & \cite{Agashe:2014kda} & $f_{\Upsilon}$ & $649(31)$~MeV &  \cite{Colquhoun:2014ica} \\ 
$m_{\Upsilon(2S)}$ & $10.023$~GeV  & \cite{Agashe:2014kda} & $f_{\Upsilon(2S)}$ & $481(39)$~MeV &  \cite{Colquhoun:2014ica} \\ 
$m_{\Upsilon(3S)}$ & $10.355$~GeV  & \cite{Agashe:2014kda} & $f_{\Upsilon(3S)}$ & $539(84)$~MeV &  \cite{Lewis:2012ir} \\ \hline 
\end{tabular}\caption{{\footnotesize{}\label{tab:costs} Masses and decay constants used in numerical analysis. }}
\end{table}

\chapter{Loop level corrections}\label{sec:app_fits}

In this Appendix we list the self-energies, counterterms and diagrams that enter in the renormalisation of the observables studied in Section \ref{sec:obs}. 

\subsection*{Lepton-flavour-dependent counterterms: $\delta^{\text{CT } W}_\alpha$ and $\delta^{\text{CT } Z}$ }

The unrenormalised charged lepton fields $l^0_{L\alpha}$ can be written in terms of the renormalised $\hat{l}_{L\alpha}$ ones as
\begin{equation}
l^0_{L\alpha}=\left(\delta_{\alpha\beta}+\frac{1}{2}\delta Z^{\text{l}}_{\alpha\beta}\right)\hat{l}_{L\beta}.
\end{equation}

The most general expression for the $l_{\beta}\rightarrow l_{\alpha}$ transition amplitude between fermionic Dirac states can 
be written as follows:

\begin{equation}
\Sigma_{\alpha \beta}^{\text{lep}}\left(\slashed{p}\right)=\slashed{p}P_{L}\Sigma_{\alpha \beta}^{L}\left(p^2\right)+
\slashed{p}P_{R}\Sigma_{\alpha \beta}^{R}\left(p^2\right)+P_{L}\Sigma_{\alpha \beta}^{D}\left(p^2\right)+P_{R}
\Sigma_{\alpha \beta}^{D*}\left(p^2\right)\,,
\end{equation}
where $\Sigma^{L}=\Sigma^{L\dagger}$ and $\Sigma^{R}=\Sigma^{R\dagger}$. In the physical observables 
only the Hermitian part of $\delta Z^{\text{l}}$ appears and it is given by
\begin{equation}
\begin{split}
\delta Z^{\text{lep}}_{\alpha \beta}\equiv &\frac{1}{2}\left(\delta Z^{\text{l}}_{\alpha \beta}+\delta Z_{\beta \alpha}^{\text{l}*}\right) \\
=& -\Sigma_{\alpha \beta}^{L}\left(m_{\beta}^{2}\right)-m_{\beta}\Big[ m_{\beta}\Big(\Sigma_{\alpha \beta}^{L\prime}
\left(m_{\beta}^{2}\right)+\Sigma_{\alpha \beta}^{R\prime}\left(m_{\beta}^{2}\right)\Big)+
\Big(\Sigma_{\alpha \beta}^{D\prime}\left(m_{\beta}^{2}\right)+
\Sigma_{\alpha \beta}^{D*\prime}\left(m_{\beta}^{2}\right)\Big) \Big] \, ,
\end{split}
\end{equation}
with $\Sigma^\prime\left(p^{2}\right)\equiv\text{d}\Sigma\left(p^{2}\right)/\text{d}p^{2}$. Therefore, the heavy neutrino  
contribution to $\delta Z^{\text{lep}}$ can be obtained simply computing

\begin{figure}[h]
\includegraphics[width=0.8\textwidth]{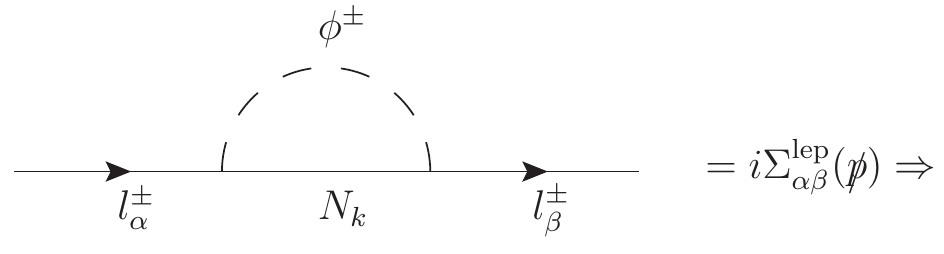}
\end{figure} 
\begin{equation}
\begin{split}
\Sigma_{\alpha \beta}^{\text{lep}}(\slashed{p})=&-\frac{\alpha}{8\pi s_\mathrm{W}^2 M_W^2}\sum_{k=4}^6 \bigg\lbrace M_{k}^2 U_{\beta k} U^*_{\alpha k} \Big[\left( P_L m_{\beta}+P_R m_{\alpha}\right) B_0(p^2,M_{k}^2,M_W^2) \\
&+\slashed{p} \left(P_R \frac{m_{\alpha} m_{\beta}}{M_{k}^2}+P_L\right) B_1(p^2,M_{k}^2,M_W^2) \Big]\bigg\rbrace \, ,
\end{split}
\end{equation}
where $B_i$ (and later $B_{ij}$, $C_{ij}$, $D_i$ and $D_{ij}$) are the Passarino-Veltman integrals~\cite{Passarino:1978jh} using the notation from Ref.~\cite{Ellis:2011cr}.

Similarly, the unrenormalised neutrino fields $\nu^0_{Lj}$ can also be written in terms of the renormalised ones 
$\hat{\nu}_{Lj}$ as
\begin{equation}
\nu^0_{Li}=\left(\delta_{ij}+\frac{1}{2}\delta Z^{\nu}_{ij}\right)\hat{\nu}_{Lj}.
\end{equation}

The transition amplitude between two Majorana states reads

\begin{equation}
\Sigma_{ij}^\text{neu}\left(\slashed{p}\right)=\slashed{p}P_{L}\Sigma_{ij}^{L}\left(p^2\right)+\slashed{p}P_{R}
\Sigma_{ij}^{L*}\left(p^2\right)+P_{L}\Sigma_{ij}^{M}\left(p^2\right)+P_{R}\Sigma_{ij}^{M*}\left(p^2\right) \, ,
\end{equation}
where $\Sigma^{L}=\Sigma^{R *}$ and $\Sigma^{M}=\Sigma^{M t}$. In the Majorana case, the Hermitian part of $\delta Z^{\nu}$ can be 
written as
\begin{equation}
\begin{split}
\delta Z^{\text{neu}}_{ij}\equiv&\frac{1}{2}\left(\delta Z^{\nu}_{ij}+\delta Z^{\nu *}_{ji}\right) \\
=&-\Sigma_{ij}^{L}\left(m_{j}^{2}\right)-m_{j}\Big[ m_{j}\Big(\Sigma_{ij}^{L\prime}\left(m_{j}^{2}\right)+\Sigma_{ij}^{L*\prime}\left(m_{j}^{2}\right)\Big)+\Big(\Sigma_{ij}^{M\prime}\left(m_{j}^{2}\right)+\Sigma_{ij}^{M*\prime}\left(m_{j}^{2}\right)\Big) \Big]\, .
\end{split}
\end{equation}
Analogously to the charged lepton case, $\delta Z^{\text{neu}}$ can thus be obtained from the heavy neutrino contribution 
to the neutrino self energy:

\begin{figure}[h]
\includegraphics[width=0.8\textwidth]{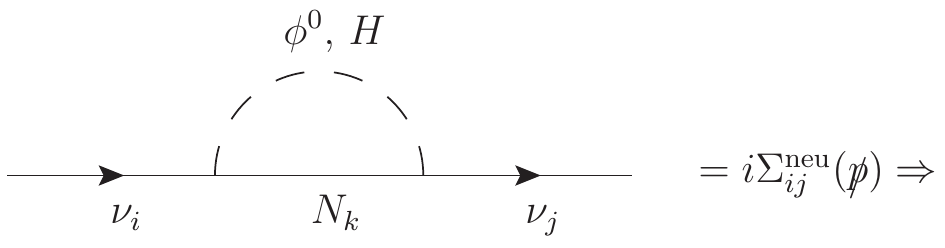}
\end{figure} 
\begin{equation}
\begin{split}
\Sigma_{ij}^{\text{neu}}(\slashed{p})=&-\frac{\alpha}{16\pi s_\mathrm{W}^2 M_W^2} \sum_{k=4}^6 \bigg\lbrace \slashed{p} P_L \left(M_j C_{jk}^*+M_k C_{jk}\right) \left(M_iC_{ki}^*+M_k C_{ki}\right)  \\
& \times \Big[B_1(p^2,M_k^2,M_Z^2)+B_1(p^2,M_k^2,M_h^2)\Big]  \\
& +\slashed{p} P_R  \left(M_j C_{jk}+M_k C^*_{jk} \right) \left(M_iC_{ki}+ M_kC^*_{ki}\right)  \\
&  \times \Big[B_1(p^2,M_k^2,M_Z^2)+B_1(p^2,M_k^2,M_h^2)\Big]\\
& +P_L M_k \left(M_jC_{jk}+M_kC^*_{jk}\right)\left(M_kC_{ki}+M_iC^*_{ki}\right)  \\
& \times \Big[B_0(p^2,M_k^2,M_Z^2)-B_0(p^2,M_k^2,M_h^2)\Big]\\
& +P_R M_k\left(M_j C^*_{jk}+M_kC_{jk}\right)\left(M_kC^*_{ki} +M_iC_{ki}\right)  \\
& \times \Big[B_0(p^2,M_k^2,M_Z^2)-B_0(p^2,M_k^2,M_h^2)\Big] \bigg\rbrace \, .
\end{split}
\end{equation}

Finally, the lepton-flavour-dependent combinations that will correct and cancel the divergences of 1-loop corrections to the vertex $W\nu l_\alpha$ and $Z\nu\nu$ are respectively:

\begin{eqnarray}
\label{eq:propscorr}
\delta^{\text{CT } W}_\alpha &=& \displaystyle\sum_{i=1}^3{\frac{U_{\alpha i}}{2}\left( \displaystyle\sum_{\beta=1}^3{\delta Z^{\text{lep}}_{\beta \alpha}U^*_{\beta i}}+\displaystyle\sum_{j=1}^6{U^*_{\alpha j}\delta Z^{\text{neu}}_{i j}}\right)} \\
\delta^{\text{CT } Z} &=& \displaystyle\sum_{k=1}^6 {\left( \delta Z^{\text{neu}}_{i k}C_{kj}+\delta Z^{\text{neu}}_{k j}C_{i k} \right)}
\label{eq:propscorrZ}
\end{eqnarray}

\subsection*{Vertex interferences: $\mathcal{V}_\alpha^W$ and $\mathcal{V}_{ij}^Z$}
\begin{figure}[h]
\includegraphics[width=0.8\textwidth]{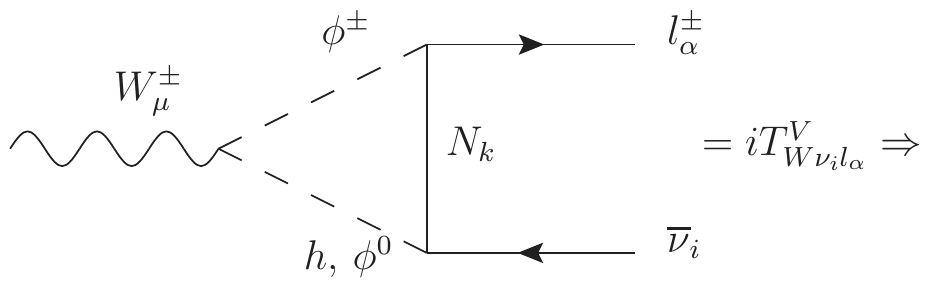}
\end{figure} 
\begin{equation}
\begin{split}
\mathcal{V}_\alpha^W\equiv&\displaystyle \frac{\sum_{i=1}^3 T_{0}^*T^{V}_{W\nu_il_\alpha}}{\sum_{i=1}^3 |T_0|^2} \\
=&\frac{\alpha}{8\pi s_\mathrm{W}^2 M_W^2}\displaystyle\sum_{i=1}^3\displaystyle\sum_{k=4}^6{M_{k}^2U_{\alpha i}U^*_{\alpha k}C^*_{ki}\left[ C_{00}(0,0,M_h^2,M_{k}^2,M_W^2)+C_{00}(0,0,M_Z^2,M_{k}^2,M_W^2)\right]} ,
\end{split}
\label{eq:Wvertex}
\end{equation}
up to higher order corrections and where $T_0$ is the corresponding tree level amplitude.

\begin{figure}[h]
\includegraphics[width=0.9\textwidth]{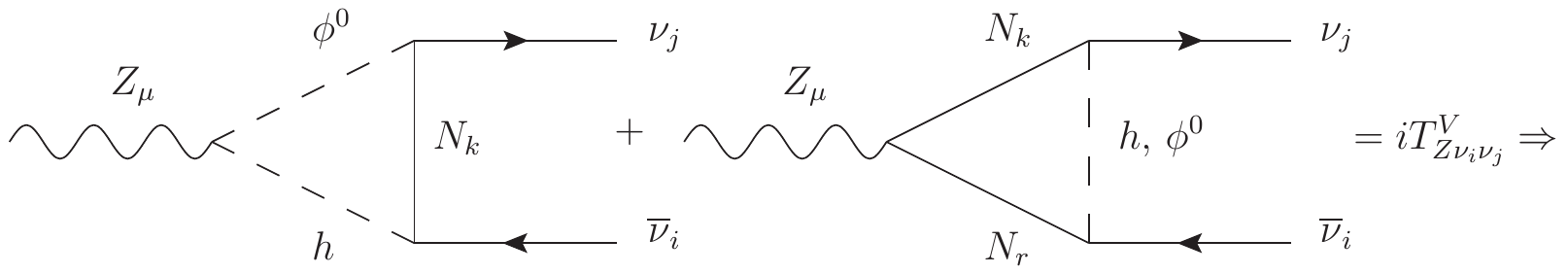}
\end{figure}
\begin{equation}
\begin{split}
\mathcal{V}_{ij}^Z \equiv & \frac{T_{0}^*T^{V}_{Z\nu_i\nu_j}}{|T_0|^2}\\
= & \frac{\alpha}{16\pi s_\mathrm{W}^2 M_W^2}\Bigg[\displaystyle\sum_{k,r=4}^6 \Bigg\lbrace -2 C_{kj}C_{ir}M_k M_r \bigg(C_{rk}M_kM_r\Big[C_0(0,M_Z^2,M_h^2,M_k^2,M_r^2)\\
&+C_0(0,M_Z^2,M_Z^2,M_k^2,M_r^2)\Big]+C_{kr}\Big[M_Z^2\big(C_{22}(0,M_Z^2,M_h^2,M_k^2,M_r^2)\\
&+C_{22}(0,M_Z^2,M_Z^2,M_k^2,M_r^2)-C_{21}(0,M_Z^2,M_h^2,M_k^2,M_r^2)-C_{21}(0,M_Z^2,M_Z^2,M_k^2,M_r^2)\big)\\ 
&+2\big(C_{00}(0,M_Z^2,M_Z^2,M_k^2,M_r^2)+C_{00}(0,M_Z^2,M_h^2,M_k^2,M_r^2)\big)\Big]\bigg)\Bigg\rbrace\\
&+\displaystyle\sum_{k=4}^6\Big[4 C_{kj}C_{ik}M_k^2\big(C_{00}(0,M_Z^2,M_k^2,M_h^2,M_Z^2)+C_{00}(0,M_Z^2,M_k^2,M_Z^2,M_h^2)\big)\Big]\Bigg]\, ,\end{split}
\label{eq:Zvertex}
\end{equation}
up to higher order corrections and where $T_0$ is the corresponding tree level amplitude.

\subsection*{Box contribution to $\mu$ decay: $\mathcal{B}_{\alpha \beta}$}

\begin{figure}[h]
\includegraphics[width=0.6\textwidth]{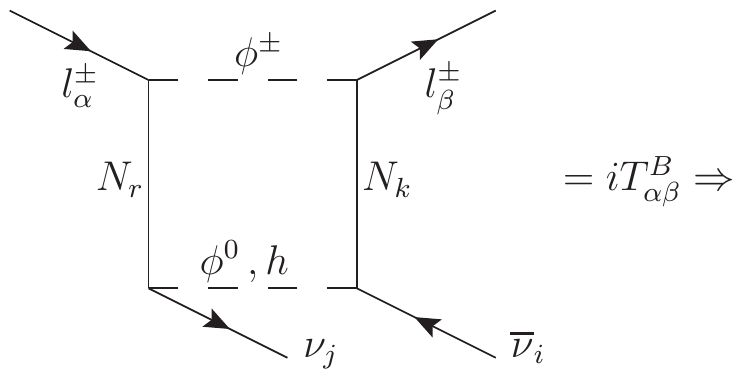}
\end{figure} 
\begin{equation}
\begin{split}
\mathcal{B}_{\alpha \beta} \equiv & \frac{\sum_{i,j=1}^3T_{0}^*T^{B}_{\alpha \beta}}{\sum_{i,j=1}^3|T_{0}|^2} \\
= & \frac{1}{5}\frac{g^2}{(16\pi)^2 M_W^2} \displaystyle\sum_{i,j=1}^3\displaystyle\sum_{k,r=4}^6{C_{ik}C_{jr}U_{\beta k}U^*_{\beta i}U^*_{\alpha r}U_{\alpha j}M^2_r M^2_k} \bigg\lbrace 20\Big[D_{00}(M_h^2)+D_{00}(M_Z^2)\Big] \\
&+m_\alpha^2\Big[3\big(D_{12}(M_h^2)+D_{12}(M_Z^2)\big)+2\big(D_{13}(M_h^2)+D_{13}(M_Z^2)\big)\\
&+3\big(D_{2}(M_h^2)+D_{2}(M_Z^2)\big) +2\big(D_{3}(M_h^2)+D_{3}(M_Z^2)\big)\Big] \bigg\rbrace\, ,
\end{split}
\label{eq:box}
\end{equation}
up to higher order corrections and where $T_0$ is the corresponding tree level amplitude and using the simplified notation $D_{ij}(M^2) \to D_{ij}(0,0,0,M_r^2,M^2,M_k^2,M_W^2)$. Apart from the explicit sum over final state neutrinos in Eq.~(\ref{eq:box}), the integral over the phase space is to be understood in both the numerator and denominator.

\subsection*{Flavour-universal corrections to the gauge boson propagators: $\delta_W^\text{univ N}$ and $\delta_Z^\text{univ N}$}

We label $\Sigma_{WW}$ and $\Sigma_{ZZ}$ the terms proportional to $g^{\mu\nu}$ in the $W$ and $Z$ self-energies respectively. Notice that the SM contribution has been subtracted from the total self-energy, as we are interested in the contribution stemming from the new extra neutrinos.

\begin{figure}[h]
\includegraphics[width=0.8\textwidth]{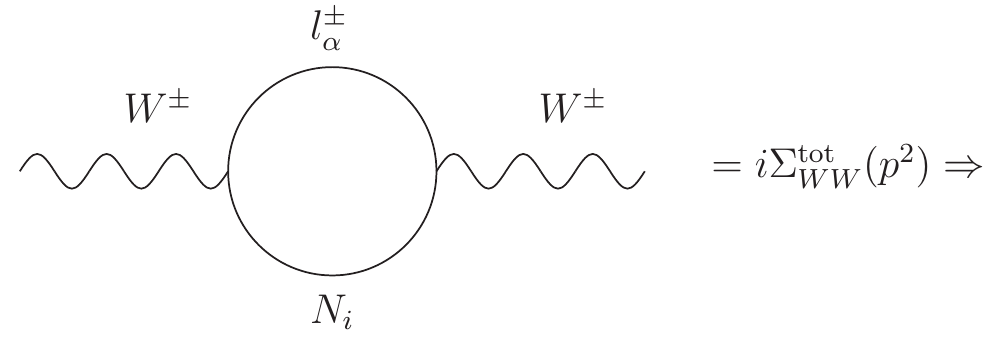}
\end{figure}
\begin{equation}
\begin{split} 
\Sigma_{WW}^N(p^2)\equiv&\Sigma_{WW}^\text{tot}(p^2)-\Sigma_{WW}^\text{SM}(p^2) \\
=&-\frac{\alpha}{4\pi s_\mathrm{W}^2}\displaystyle\sum_{\alpha=e,\mu,\tau}\Bigg\lbrace \displaystyle\sum_{i=1}^6 \vert U_{\alpha i}\vert^2 \bigg[ 2 B_{00}(p^2,M_i^2,m_\alpha^2) +p^2\Big( B_1(p^2,M_i^2,m_\alpha^2)  \\
 &+ B_{11}(p^2,M_i^2,m_\alpha^2)\Big) \bigg] - 2 B_{00}(p^2,0,m_\alpha^2)-p^2\Big(B_1(p^2,0,m_\alpha^2)+B_{11}(p^2,0,m_\alpha^2)\Big)\Bigg\rbrace 
\end{split}
\label{eq:sigmaW}
\end{equation}

\begin{figure}
\includegraphics[width=0.8\textwidth]{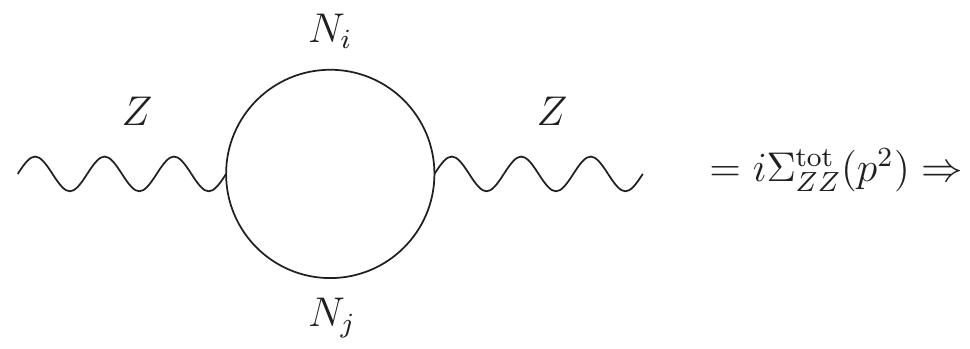}
\end{figure}

\begin{equation}
\begin{split}
\Sigma_{ZZ}^N(p^2)\equiv&\Sigma_{ZZ}^\text{tot}(p^2)-\Sigma_{ZZ}^\text{SM}(p^2)\\
=&-\frac{\alpha}{8\pi s_\mathrm{W}^2 c_\mathrm{W}^2}\Bigg\lbrace\displaystyle\sum_{\alpha, \beta}\displaystyle\sum_{i,j=1}^6 \bigg[ U_{\alpha i}U_{\alpha j}^*U_{	\beta j}U_{\beta i}^*M_iM_jB_0(p^2,M_i^2,M_j^2) + U_{\alpha j}U_{\alpha i}^*U_{\beta i}U_{\beta j}^*\\
&\times\Big( 2 B_{00}(p^2,M_i^2,M_j^2)+p^2\big(B_1(p^2,M_i^2,M_j^2)+B_{11}(p^2,M_i^2,M_j^2)\big)\Big) \bigg] \\
&- 3\Big[ 2 B_{00}(p^2,0,0)+p^2\big(B_1(p^2,0,0)+B_{11}(p^2,0,0)\big)\Big]\Bigg\rbrace
\end{split}
\label{eq:sigmaZ}
\end{equation}
Notice that both in Eq.~(\ref{eq:sigmaW}) and in Eq.~(\ref{eq:sigmaZ}) the sums run over all neutrino mass eigenstates (heavy and light) so here $M_i$ can represent both the heavy or the light neutrino masses.

The oblique universal corrections to the electroweak observables can be written as a combination of the three following 
independent parameters~\cite{Peskin:1990zt,Peskin:1991sw}:

\begin{eqnarray}
\label{S}
\alpha S&=&\frac{4s_\mathrm{W}^{2}c_\mathrm{W}^{2}}{M_{Z}^{2}}\Bigl[\hat{\Sigma}_{ZZ}^N(0)
+\hat{\Sigma}_{\gamma \gamma}^N(M_{Z}^{2})
-\frac{c_\mathrm{W}^{2}-s_\mathrm{W}^{2}}{c_\mathrm{W}s_\mathrm{W}}\hat{\Sigma}_{Z\gamma}^N(M_{Z}^{2})\Bigr]\,,\\
\label{T}
\alpha T&=&\frac{\hat{\Sigma}_{ZZ}^N(0)}{M_{Z}^{2}}-\frac{\hat{\Sigma}_{WW}^N(0)}{M_{W}^{2}}\,,\\
\label{U}
\alpha U&=&4s_\mathrm{W}^{2}c_\mathrm{W}^{2}\biggl[\frac{1}{c_\mathrm{W}^{2}}\frac{\hat{\Sigma}_{WW}^N(0)}{M_{W}^{2}}-
\frac{\hat{\Sigma}_{ZZ}^N(0)}{M_{Z}^{2}}
+\frac{s_\mathrm{W}^{2}}{c_\mathrm{W}^{2}}\frac{\hat{\Sigma}_{\gamma \gamma}^N(M_{Z}^{2})}{M_{Z}^{2}}-
\frac{2s_\mathrm{W}}{c_\mathrm{W}}\frac{\hat{\Sigma}_{Z\gamma}^N(M_{Z}^{2})}{M_{Z}^{2}}\biggr]\,.
\end{eqnarray}
and the renormalised self energies are given by:

\begin{eqnarray}
\hat{\Sigma}_{WW}^N\left(p^2\right)&=&\Sigma_{WW}^N\left(p^2\right)-\Sigma_{WW}^N\left(M_W^2\right)+(p^2-M_W^2)\left[
\frac{c_\mathrm{W}^2}{s_\mathrm{W}^2}\mathcal{R} -\Sigma_{\gamma\gamma}^{N\prime}(0)\right],\nonumber\\
\hat{\Sigma}_{ZZ}^N\left(p^2\right)&=&\Sigma_{ZZ}^N\left(p^2\right)-\Sigma_{ZZ}^N\left(M_Z^2\right)+(p^2-M_Z^2)\left[
\left(\frac{c_\mathrm{W}^2}{s_\mathrm{W}^2}-1\right)\mathcal{R}-\Sigma_{\gamma\gamma}^{N\prime}(0)\right],\nonumber\\
\hat{\Sigma}_{Z\gamma}^N\left(p^2\right)&=& \Sigma_{Z\gamma}^N\left(p^2\right)-\Sigma_{Z\gamma}^N\left(0\right)-p^2\frac{c_\mathrm{W}}{s_\mathrm{W}}\mathcal{R},\nonumber\\
\hat{\Sigma}_{\gamma\gamma}^N\left(p^2\right)&=& \Sigma_{\gamma\gamma}^N\left(p^2\right)-p^2\Sigma_{\gamma\gamma}^{N\prime}\left(0\right),
\end{eqnarray}
with 
\be
\mathcal{R}=\frac{\Sigma_{ZZ}^N\left(M_Z^2\right)}{M_Z^2}-\frac{\Sigma_{WW}^N\left(M_W^2\right)}{M_W^2}-\frac{2s_\mathrm{W}}{c_\mathrm{W}}
\frac{\Sigma_{Z\gamma}^N\left(0\right)}{M_Z^2}
\ee
Notice that, in the on-shell renormalisation scheme $\hat{\Sigma}_{WW}^N\left(M_W^2\right)=\hat{\Sigma}_{ZZ}^N\left(M_Z^2\right)=\hat{\Sigma}_{Z\gamma}^N\left(0\right)
=\hat{\Sigma}_{\gamma\gamma}^N\left(0\right)=0$. Moreover, there is no contribution to the propagator of the photon 
from the extra heavy neutrinos and therefore $\Sigma_{\gamma\gamma}^N$ and $\hat{\Sigma}_{\gamma\gamma}^N$ can be set to zero in the previous equations. In addition,
there is no correction to $\Sigma_{Z\gamma}$ either, so that $\Sigma_{Z\gamma}^N$ can be set to zero too. The universal oblique counterterms presented in Sec.~\ref{sec:obs} can thus be written as:

\begin{eqnarray}
\delta_W^\text{univ N}&=&\frac{\Sigma_{WW}^N\left(0 \right)-\Sigma_{WW}^N\left(M_W^2\right)}{M_W^2}-\frac{c_\mathrm{W}^2}{s_\mathrm{W}^2}\mathcal{R}=
\frac{\hat{\Sigma}_{WW}^N\left(0\right)}{M_W^2}\nonumber\\
\label{dW}
&=& \frac{1}{2s_\mathrm{W}^2}\alpha S-\frac{ c_\mathrm{W}^2}{s_\mathrm{W}^2}\alpha T-\frac{\cos 2\theta_W}{4s_\mathrm{W}^4}\alpha U\nonumber\\
\delta_Z^\text{univ N}&=&\frac{\Sigma_{ZZ}^N\left(0 \right)-\Sigma_{ZZ}^N\left(M_Z^2\right)}{M_Z^2} + 
\frac{1}{2}\left(1-\frac{c_\mathrm{W}^2}{s_\mathrm{W}^2}\right)\mathcal{R}=\frac{\hat{\Sigma}_{ZZ}^N\left(0\right)}{M_Z^2}\nonumber\\
\label{dZ}
&=&\frac{1}{2s_\mathrm{W}^2}\alpha S+\left(1-\frac{c_\mathrm{W}^2}{s_\mathrm{W}^2}\right)\alpha T-\frac{\cos 2\theta_W}{4s_\mathrm{W}^4}\alpha U .
\end{eqnarray}

\newpage
\addcontentsline{toc}{chapter}{References}

\bibliographystyle{JHEP}
\bibliography{thesis}

\end{document}